\RequirePackage[l2tabu]{nag}		
\documentclass[a4paper,11pt,reqno,openbib,oldfontcommands]{memoir} 
\usepackage{datetime}
\usepackage{ifpdf}
\ifpdf
	\fi
\ifdraftdoc 
	\usepackage{draftwatermark}				
	\SetWatermarkScale{0.3}
	\SetWatermarkText{\bf Draft: \today}
\fi
\newsubfloat{figure}
\newsubfloat{table}
\settrimmedsize{297mm}{210mm}{*}
\settypeblocksize{634pt}{448.13pt}{*} 
\setulmargins{4cm}{*}{*} 
\setlrmargins{*}{*}{1.5} 
\setmarginnotes{17pt}{51pt}{\onelineskip} 
\setheadfoot{\onelineskip}{2\onelineskip} 
\setheaderspaces{*}{2\onelineskip}{*} 
\checkandfixthelayout
\usepackage{fouriernc}
\usepackage[T1]{fontenc}
\usepackage{enumitem}
\OnehalfSpacing 
\setsecnumdepth{subsection} 
\maxsecnumdepth{subsubsection}
\usepackage{calc,soul,fourier}
\makeatletter 
\newlength\dlf@normtxtw 
\newsavebox{\feline@chapter} 
\newcommand\feline@chapter@marker[1][4cm]{%
	\sbox\feline@chapter{%
		\resizebox{!}{#1}{\fboxsep=1pt%
			\colorbox{gray}{\color{white}\thechapter}%
		}}%
		\rotatebox{90}{%
			\resizebox{%
				\heightof{\usebox{\feline@chapter}}+\depthof{\usebox{\feline@chapter}}}%
			{!}{\scshape\so\@chapapp}}\quad%
		\raisebox{\depthof{\usebox{\feline@chapter}}}{\usebox{\feline@chapter}}%
} 
\newcommand\feline@chm[1][4cm]{%
	\sbox\feline@chapter{\feline@chapter@marker[#1]}%
	\makebox[0pt][c]{
		\makebox[1cm][r]{\usebox\feline@chapter}%
	}}
\makechapterstyle{daleifmodif}{

	\renewcommand\printchapternum{\null\hfill\feline@chm[2.5cm]\par}

} 
\makeatother 
\chapterstyle{daleifmodif}
\makepagestyle{myvf} 
\makeoddfoot{myvf}{}{\thepage}{} 
\makeevenfoot{myvf}{}{\thepage}{} 
\makeheadrule{myvf}{\textwidth}{\normalrulethickness} 
\makeevenhead{myvf}{\small\textsc{\leftmark}}{}{} 
\makeoddhead{myvf}{}{}{\small\textsc{\rightmark}}
\makepagestyle{intro} 
\makeoddfoot{intro}{}{\thepage}{} 
\makeevenfoot{intro}{}{\thepage}{} 
\makeheadrule{intro}{\textwidth}{\normalrulethickness} 
\makeevenhead{intro}{\small\textsc{INTRODUCTION}}{}{} 
\makeoddhead{intro}{}{}{\small\textsc{INTRODUCTION}}
\makepagestyle{concl} 
\makeoddfoot{concl}{}{\thepage}{} 
\makeevenfoot{concl}{}{\thepage}{} 
\makeheadrule{concl}{\textwidth}{\normalrulethickness} 
\makeevenhead{concl}{\small\textsc{CONCLUSION AND OUTLOOK}}{}{} 
\makeoddhead{concl}{}{}{\small\textsc{CONCLUSION AND OUTLOOK}}
\newcommand{\clearemptydoublepage}{\newpage{\thispagestyle{empty}\cleardoublepage}}
\makeindex
\usepackage{import}

\usepackage{lipsum}					
\usepackage{amsfonts} 					
\usepackage[centertags]{amsmath}			
\usepackage{amssymb}					
\usepackage{amsthm, mathtools}	
\usepackage{braket}
\usepackage{framed}
\usepackage{mdframed}
\usepackage{cases}
\usepackage{newlfont}					
\usepackage{dsfont}
\usepackage{layouts}					
\usepackage{graphicx}					
\usepackage{longtable,rotating}			
\usepackage[utf8]{inputenc}			
\usepackage[table]{xcolor}
\usepackage{bm}
\usepackage{float}						
\usepackage{verbatim}					
\usepackage{upgreek }					
\usepackage{latexsym}					
\usepackage{url}						
\usepackage[spanish,english]{babel}		
\usepackage{color}                    				
\usepackage{memhfixc}					
\usepackage{enumerate}					
\usepackage{footnote}					
\usepackage{microtype}					
\usepackage{rotfloat}					
\usepackage{alltt}						
\usepackage{diagbox}
\usepackage{hhline}
\usepackage[version=0.96]{pgf}			
\usepackage{tikz}						
\usetikzlibrary{arrows,shapes,snakes,
		       automata,backgrounds,
		       petri,topaths}				
\usepackage[colorlinks=true,
		     allcolors=black]{hyperref}              
{%
    \MakeFramed{\hsize#1\advance\hsize-\width\FrameRestore}%
}
{\endMakeFramed}

\DeclareMathAlphabet{\mymathbb}{U}{BOONDOX-ds}{m}{n}
\DeclareMathAlphabet{\mathbbm}{U}{bbm}{m}{n}

\DeclareMathOperator{\poly}{poly\,}

\DeclareMathOperator{\Tr}{Tr}
\DeclareMathOperator{\Sym}{Sym}
\DeclareMathOperator{\Per}{Per}
\DeclareMathOperator{\Haf}{Haf\,}
\DeclareMathOperator{\lHaf}{lHaf\,}
\DeclareMathOperator{\Det}{Det}

\newcommand{\be}{\begin{equation}}
\newcommand{\ee}{\end{equation}}
\newcommand{\ba}{\begin{aligned}}
\newcommand{\ea}{\end{aligned}}

\newcommand{\R}{\mathbb{R}}
\newcommand{\bc}{\begin{center}}
\newcommand{\ec}{\end{center}}
\newcommand{\beq}{\begin{equation}}
\newcommand{\eeq}{\end{equation}}
\newcommand{\beqq}{\begin{equation*}}
\newcommand{\eeqq}{\end{equation*}}
\newcommand{\beqa}{\begin{align}}
\newcommand{\eeqa}{\end{align}}
\newcommand{\barr}{\begin{array}}
\newcommand{\earr}{\end{array}}
\newcommand{\bi}{\begin{itemize}}
\newcommand{\ei}{\end{itemize}}
\newcommand{\Hi}{\mathcal{H}}

\newcommand{\pgftextcircled}[1]{                                                                    
    \setbox0=\hbox{#1}%
    \dimen0\wd0%
    \divide\dimen0 by 2%
    \begin{tikzpicture}[baseline=(a.base)]%
        \useasboundingbox (-\the\dimen0,0pt) rectangle (\the\dimen0,1pt);
        \node[circle,draw,outer sep=0pt,inner sep=0.1ex] (a) {#1};
    \end{tikzpicture}
}
\newcommand{\blackged}{\hfill$\blacksquare$}
\newcommand{\whiteged}{\hfill$\square$}
\newcounter{proofcount}
\renewenvironment{proof}[1][\proofname.]{\par
 \ifnum \theproofcount>0 \pushQED{\whiteged} \else \pushQED{\blackged} \fi%
 \refstepcounter{proofcount}
 \normalfont 
 \trivlist
 \item[\hskip\labelsep
       \itshape
   {\bf\em #1}]\ignorespaces
}{%
 \addtocounter{proofcount}{-1}
 \popQED\endtrivlist
}
\let\oldsqrt\sqrt
\def\sqrt{\mathpalette\DHLhksqrt}
\def\DHLhksqrt#1#2{%
\setbox0=\hbox{$#1\oldsqrt{#2\,}$}\dimen0=\ht0
\advance\dimen0-0.2\ht0
\setbox2=\hbox{\vrule height\ht0 depth -\dimen0}%
{\box0\lower0.4pt\box2}}
\newcommand{\mycaption}[2][\@empty]{
	\captionnamefont{\scshape} 
	\changecaptionwidth
	\captionwidth{0.9\linewidth}
	\captiondelim{.\:} 
	\indentcaption{0.75cm}
	\captionstyle[\centering]{}
	\setlength{\belowcaptionskip}{10pt}
	\ifx \@empty#1 \caption{#2}\else \caption[#1]{#2}
}
\newcommand{\mysubcaption}[2][\@empty]{
	\subcaptionsize{\small}
	\hangsubcaption
	\subcaptionlabelfont{\rmfamily}
	\sidecapstyle{\raggedright}
	\setlength{\belowcaptionskip}{10pt}
	\ifx \@empty#1 \subcaption{#2}\else \subcaption[#1]{#2}
}
\usepackage{lettrine}
\newcommand{\initial}[1]{%
	\lettrine[lines=3,lhang=0.33,nindent=0em]{
		\color{gray}
     		{\textsc{#1}}}{}}
\theoremstyle{plain}
\newtheorem{theo}{Theorem}[chapter]
\theoremstyle{plain}

\theoremstyle{plain}

\theoremstyle{plain}
\theoremstyle{definition}
\newtheorem{defi}{Definition}[chapter]
\theoremstyle{plain}
\newtheorem{lem}{Lemma}[chapter]
\theoremstyle{plain}
\newtheorem{coro}{Corollary}[chapter]
\theoremstyle{plain}

\newtheorem{pb}{Problem}
\newtheorem{conj}{Conjecture}
\begin{document}
%
%
%
%
%
\frontmatter
\pagenumbering{roman}
%
%
%
%
%
\begin{titlingpage}
\begin{SingleSpace}
\calccentering{\unitlength} 
\begin{adjustwidth*}{\unitlength}{-\unitlength}
\vspace*{13mm}
\begin{center}
\rule[0.5ex]{\linewidth}{2pt}\vspace*{-\baselineskip}\vspace*{3.2pt}
\rule[0.5ex]{\linewidth}{1pt}\\[\baselineskip]
{\huge Continuous Variable Quantum Advantages}\\[4mm]
{\Large \textit{and Applications in Quantum Optics}}\\
\rule[0.5ex]{\linewidth}{1pt}\vspace*{-\baselineskip}\vspace{3.2pt}
\rule[0.5ex]{\linewidth}{2pt}\\
\vspace{6.5mm}
{\large By}\\
\vspace{6.5mm}
{\large\textsc{Ulysse Chabaud}}\\
\vspace{16mm}
\hspace{9mm}
\includegraphics[scale=0.3]{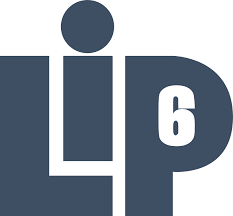}
\hspace{12mm}
\includegraphics[scale=0.12]{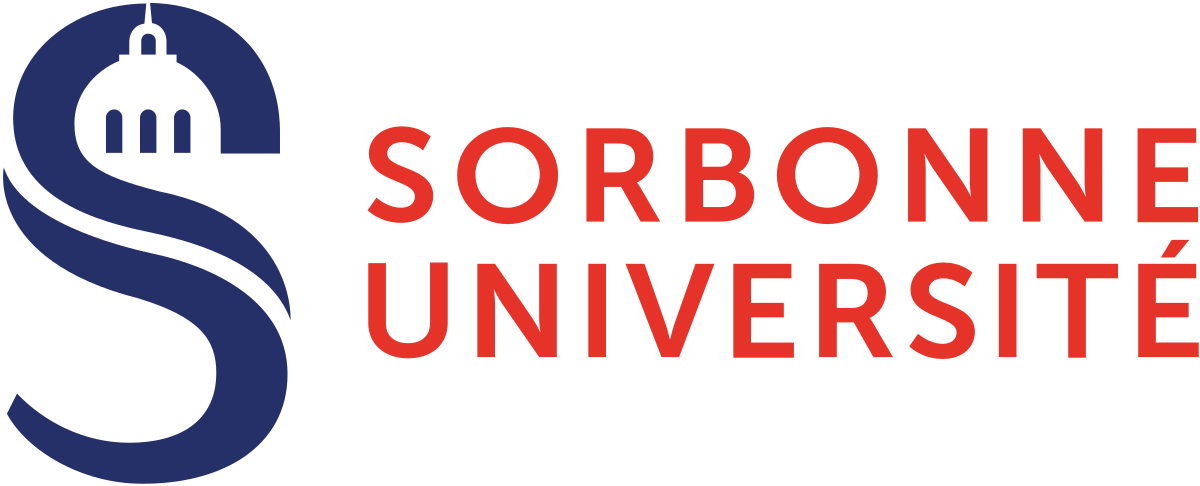}\\
\vspace{6mm}
{\large Laboratoire d'Informatique de Paris 6\\
\textsc{Sorbonne Universit\'e}}\\
\vspace{16mm}
\begin{minipage}{10cm}
A dissertation submitted to Sorbonne Universit\'e in accordance with the requirements of the degree of \textsc{Doctor of Philosophy}, under the supervision of {Damian Markham} and {Elham Kashefi}.\\

Members of the jury: Anthony Leverrier, Andreas Winter, S\'ebastien Tanzili, Perola Milman and Gerardo Adesso.
\end{minipage}\\
\vspace{12mm}
{\large\textsc{July 2020}}
\vspace{12mm}
\end{center}
\end{adjustwidth*}
\end{SingleSpace}
\end{titlingpage}
\clearemptydoublepage
\chapter*{Foreword}
\begin{SingleSpace}
\initial{T}he three years leading up to the writing of this dissertation have been incredibly rich. Intellectually, I found a stimulus that I had been missing for years. I also met and interacted with so many outstanding people! It has been truly an incredible time for which I feel lucky and grateful. 
My heartfelt thanks to my doctoral advisors Damian and Elham, who were incredibly supportive and distilled the perfect blend of guidance and freedom that allowed me to flourish as a researcher. I am also thankful to the members of my jury: Andreas Winter, Perola Milman, Sébastien Tanzilli, Gerardo Adesso and especially Anthony Leverrier for his continued support.

I share the love of continuous variable quantum information with Fr\'ed\'eric and we have had many exciting and inspiring discussions from his first ``coffee-break attack'', which have always been a pleasure. The friends and colleagues of the QI team have provided the best environment I could think of and I have to thank them all for their openness and passion, as I have been able to interact with them with pleasure, both in research and outside the lab: Eleni, Tom and Shane, Pierre-Emmanuel, Raja, Luka, Robert and Alisa---the members of the ``cool guys office''---Clément, Francesco, Nathan, Léo, Luis, Matthieu, Anu, Niraj, Andrea, Shraddha, Rhea, Victor, Federico, Adrien, Simon, Verena, Yao, Shouvik, Dominik, G\"ozde, Damien, Harold, Matteo and Cyril. My discussions with the LKB experimental team: Gana\"el, Mattia, Nicolas, Valentina, who agreed to listen to a mad theorist, were also great times and it was always a blast to join the Edinburgh team during our wild retreats : Alex, Atul, Ellen, Daniel, Brian, Rawad, Mina, Mahshid, Theodoros and Petros. I would like to thank especially Andru for his kindness and thoughtfulness. 

These years have been marked by exciting trips around the world and I am deeply grateful to Thomas Vidick, Scott Aaronson, Andrew Childs, Aram Harrow and Kae Nemoto for hosting my visits. I also had the opportunity and pleasure of interacting and working with Giulia Ferrini, Raul Garc\'ia-Patr\'on, Peter van Loock, Antoine Joux, Iordanis Kerenidis, Jens Eisert, Dominik Hangleiter, Nathan Walk, Ingo Roth and Adel Sohbi. I am also thankful to all the researchers I had the opportunity to meet at the lab, during workshops or at bigger conferences, for sharing their time and knowledge with me.

Thanks to my long-time friends Pierre, Safia, Oscar, Jim, Thomas, Aur\'elien, Alice, Vincent, Ziyad, Ga\"el, Balthazar, Maxence, Alexia, François, Nicolas, Alexandre, Arthur and Baptiste for making the moments outside of the quantum information universe extremely enjoyable, together with my brother, my parents and my grandfathers for their curiosity and interest, despite my sometimes foggy explanations.

My unlimited thanks to my wife Léonie for her love and support through all these years. Thank you for being in my life and helping me to become a better person. I can't wait to see what's next together with you!

\end{SingleSpace}
\clearpage
\clearemptydoublepage
%
%
%
%

\chapter*{Abstract}
\begin{SingleSpace}
\lettrine[lines=3,lhang=0.2,nindent=0em,lraise=0.1]{{\color{gray}\textsc{Q}}}{}uantum physics has led to a revolution in our conception of the nature of our world and is now bringing about a technological revolution. The use of quantum information promises indeed applications that outperform those of today's so-called classical devices. Continuous variable quantum information theory refers to the study of quantum information encoded in continuous degrees of freedom of quantum systems. This theory extends the mathematical study of quantum information to quantum states in Hilbert spaces of infinite dimension. It offers different perspectives compared to discrete variable quantum information theory and is particularly suitable for the description of quantum states of light. Quantum optics is thus a natural experimental platform for developing quantum applications in continuous variable.

This thesis focuses on three main questions: where does a \textit{quantum advantage}, that is, the ability of quantum machines to outperform classical machines, come from? How to ensure the proper functioning of a quantum machine? What advantages can be gained in practice from the use of quantum information? These three questions are at the heart of the development of future quantum technologies and we provide several answers within the frameworks of continuous variable quantum information and linear quantum optics.

Quantum advantage in continuous variable comes in particular from the use of so-called \textit{non-Gaussian} quantum states. We introduce the stellar formalism to characterize these states. We then study the transition from classically simulable models to models universal for quantum computing. We show that \textit{quantum computational supremacy}, the dramatic speedup of quantum computers over their classical counterparts, may be realised with non-Gaussian states and Gaussian measurements.

Quantum certification denotes the methods seeking to verify the correct functioning of a quantum machine. We consider certification of quantum states in continuous variable, introducing several protocols according to the assumptions made on the tested state. We develop efficient methods for the verification of a large class of multimode quantum states, including the output states of the \textit{Boson Sampling} model, enabling the experimental verification of quantum supremacy with photonic quantum computing.

We give several new examples of practical applications of quantum information in linear quantum optics. Generalising the \textit{swap test}, we highlight a connection between the ability to distinguish two quantum states and the ability to perform universal programmable quantum measurements, for which we give various implementations in linear optics, based on the use of single photons or coherent states. Finally, we obtain, thanks to linear optics, the first implementation of a quantum protocol for \textit{weak coin flipping}, a building block for many cryptographic applications.
\end{SingleSpace}
\clearpage
\clearemptydoublepage
%
%
\renewcommand{\contentsname}{Table of Contents}
\maxtocdepth{subsection}
\tableofcontents*
\addtocontents{toc}{\par\nobreak \mbox{}\hfill{\bf Page}\par\nobreak}
\clearemptydoublepage
%
%
%
%
\mainmatter
%
%
\let\textcircled=\pgftextcircled
\chapter*{Introduction}
\label{chap:intro}
\addcontentsline{toc}{chapter}{Introduction}
\chaptermark{Introduction}
\pagestyle{intro}

\lettrine[lines=3,lhang=0.2,nindent=0em,lraise=0.1]{{\color{gray}\textsc{Q}}}{}uantum mechanics has deepened our understanding of the world. It has led us to rethink the very notion of reality---how can a cat be neither dead nor alive?---by putting forth intriguing properties such as \textit{entanglement} and \textit{superposition}. 
Nowadays, new information processing devices using quantum properties are being developed, such as quantum computers, and it is fascinating and maybe incumbent to see whether and to what extent these quantum technologies may outperform conventional technologies.

\section*{Motivation and context}
\addcontentsline{toc}{section}{Motivation and context}  

While classical mechanics, as opposed to quantum, has been quite successful in describing the world at our scale, quantum mechanics has proven to be a very powerful tool for understanding the world at the particle scale. Interesting effects appear at this scale, and the challenge posed by the development of quantum technologies is not only to understand these effects but also to harness them. Quantum information---that is, information encoded in quantum degrees of freedom of physical systems---promises advantages over classical information notably for computing, communication, cryptography and sensing.
That the use of quantum mechanics may provide an advantage over classical mechanics for information processing is an exciting perspective, which raises the following question:

\begin{center}
\textit{What leads to a quantum advantage?}
\end{center}

\noindent This profound question has attracted enormous attention and so far has only partial answers. 
From a foundational point of view, this question asks what differentiates the quantum from the classical and what makes nature fundamentally nonclassical. While shedding light on the very nature of our world, answering this question also enables the development of new technologies exploiting quantum properties to gain an advantage over classical machines. 

\medskip

\noindent In order to understand the possible origins of a quantum advantage it is worthwhile to highlight some of the differences between quantum and classical information and in particular quantum features that are inherently nonclassical. 

Properties of quantum systems are intrinsically random prior to being measured and this randomness is lost whenever the quantum system is measured---hence the infamous Schr\"odinger's cat thought experiment,
in which a cat is locked in a box with a device that kills the animal with some probability: before opening the box, the cat is neither dead nor alive, but rather in a superposition of these two states, and opening the box collapses the state of the cat to either dead or alive. 
In a more general fashion, the state of a quantum system can be mathematically described by a wave function consisting of complex-valued probability amplitudes. The probabilities for the possible results of measurements made on the system can be derived from these amplitudes. As their name indicates, wave functions behave qualitatively like mechanical waves: they satisfy a linear wave equation and may interfere. This interference of probability amplitudes is a striking example of nonclassical phenomena. A quantum computer outperforming its classical counterpart would crucially interfere various branches of a computation. 

The linear evolution of probability amplitudes also has striking consequences: it implies that an arbitrary quantum state cannot be perfectly cloned \cite{wootters1982single}. This contrasts with the fact that classical information is trivial to copy. This quantum no-cloning property can also be derived from the uncertainty principle, which asserts that complementary quantum observables---such as position and momentum---cannot be simultaneously measured with arbitrary precision: measuring one of the two collapses the state of the measured quantum system such that the value of the other becomes uniformly random. If one was able to perfectly clone a quantum state, one could measure the position of the first copy and the momentum of the second and infer both quantities for the original state, thus contradicting the uncertainty principle. While uncertainty and no-cloning may be seen as limitations of quantum information, quantum advantage in cryptography notably comes from exploiting these properties to hide information from a possible eavesdropper \cite{bennett1984quantum}.

These quantum properties may be witnessed already for a single system. On the other hand, multiple systems may display correlations and it turns out that quantum systems may be correlated in a way classical systems cannot, as a consequence of entanglement. A quantum state over multiple subsystems is said to be entangled if it cannot be separated into the individual states of its subsystems. An important consequence of entanglement is the nonlocality of quantum theory, i.e., the fact that correlations displayed by spatially separated quantum systems cannot be reproduced locally by classical means \cite{bell1964einstein}---what Einstein famously described as ``spooky action at a distance''. While these nonclassical correlations may be exploited for the so-called quantum teleportation \cite{bennett1993teleporting}, they do not allow for superluminal communication, as a consequence of the no-signaling principle~\cite{peres2004quantum}.

\medskip

\noindent In theory, the nonclassical properties previously described may allow quantum devices to outperform their classical counterparts for a variety of information processing tasks, and in particular to demonstrate quantum computational supremacy \cite{harrow2017quantum}---a quantum computer performing efficiently a computational task which is provably intractable for classical computers---which marks a key milestone in the development of quantum technologies \cite{arute2019quantum}. 

However, a major obstacle to the use of the nonclassical properties of quantum information for technological applications is decoherence, i.e., the loss of coherence of the information encoded in a physical system, due to the interaction of that system with its environment. Quantum devices will inevitably interact with their environment and suffer the effect of noise. How to mitigate the consequences of decoherence is an active domain of research~\cite{preskill1998fault}. In theory, quantum computations may be performed fault-tolerantly, even though this results in a huge overhead in terms of physical systems needed for the computation. It is also not obvious how one can mitigate noise in other quantum information processing tasks, such as sensing or simulations, where the fault-tolerant quantum computing approach is not natural.
Hence, another question that arises when looking for an advantage using a quantum device is the following:

\begin{center}
\textit{How do we check the correct functioning of a quantum device?}
\end{center}

\noindent Answering this second question is a timely problem in the absence of fault-tolerant mechanisms, for benchmarking existing and upcoming quantum devices. It has also attracted a lot of attention \cite{eisert2020quantum}, under different names: validation, benchmarking, certification, verification. We shall use certification in the following when no context is precised.
The task of certification may indeed vary depending on the context: fundamental research, industrial quantum device, or even delegated quantum computing and quantum cryptography. In all these cases, what may vary is the level of trust one wants to guarantee, as well as the assumptions one is ready to make on the device being tested.

The challenge posed by the certification of quantum devices therefore depends on this context. What is more, the very properties of quantum information---entanglement, unclonability---add uniquely quantum challenges to the task of certification, and the way the information is encoded in physical systems also matters.

\medskip

\noindent Information, both classical and quantum, may be encoded using either discrete degrees of freedom of a physical system---such as the presence or absence of an electrical signal, or the spin of an electron---or continuous degrees of freedom---such as the position of a particle, or quadratures of the electromagnetic field. 

A great part of the theory already developed for discrete variable quantum information is still missing for continuous variable quantum information. The latter is based on the beautiful mathematics of quantum mechanics in infinite-dimensional Hilbert spaces and gives different perspectives on quantum information~\cite{Braunstein2005}. In addition, continuous variable quantum information has an exciting experimental status, thanks to quantum optics in particular, which enables the scalable generation of large entangled quantum states \cite{yokoyama2013ultra} and provides high efficiency measurements. Moreover, some continuous variable quantum technologies---such as continuous variable quantum key distribution~\cite{grosshans2002continuous}---compete with their discrete variable counterparts \cite{jouguet2013experimental}. This implies that the question of certification is of great importance for continuous variable quantum devices, which also allow for outperforming classical devices and demonstrating quantum computational supremacy \cite{Aaronson2013,Hamilton2016}.

\medskip

\noindent The demonstration of quantum supremacy, that is the convincing demonstration of a quantum computation beating what is possible classically, is however only a milestone, and what is at stake in the development of quantum technologies is to obtain advantages for real-world applications. 
It is thus natural to ask the following question:

\begin{center}
\textit{What useful advantages can we obtain from the use of quantum information?}
\end{center}

\noindent Depending on the application considered, a quantum advantage may take different forms: to obtain the result of a computation faster~\cite{shor1994algorithms,Grover98}, to communicate more messages within the same physical system~\cite{bennett1992communication} or in a more secured fashion~\cite{BB84}, or to perform a measurement with a better precision~\cite{giovannetti2011advances}, for example. Answering this third question amounts to developping new theoretical quantum algorithms as well as deriving realistic implementations for existing ones, for example with linear quantum optics and quantum states of light.

\medskip

\noindent This section has provided an overview of the different contexts on which the work of this thesis is based. Motivated by the three very general questions above---origin of quantum advantage, certification of quantum devices and useful quantum advantages---this dissertation explores various directions, with particular emphasis on continuous variable quantum information theory and optical quantum information processing. 
The next section presents a technical summary of the content of the thesis.

\section*{Summary of results}
\addcontentsline{toc}{section}{Summary of results}  
\label{sec:sec01}

\begin{figure}[h!]
	\centering
	\includegraphics[width=0.9\columnwidth]{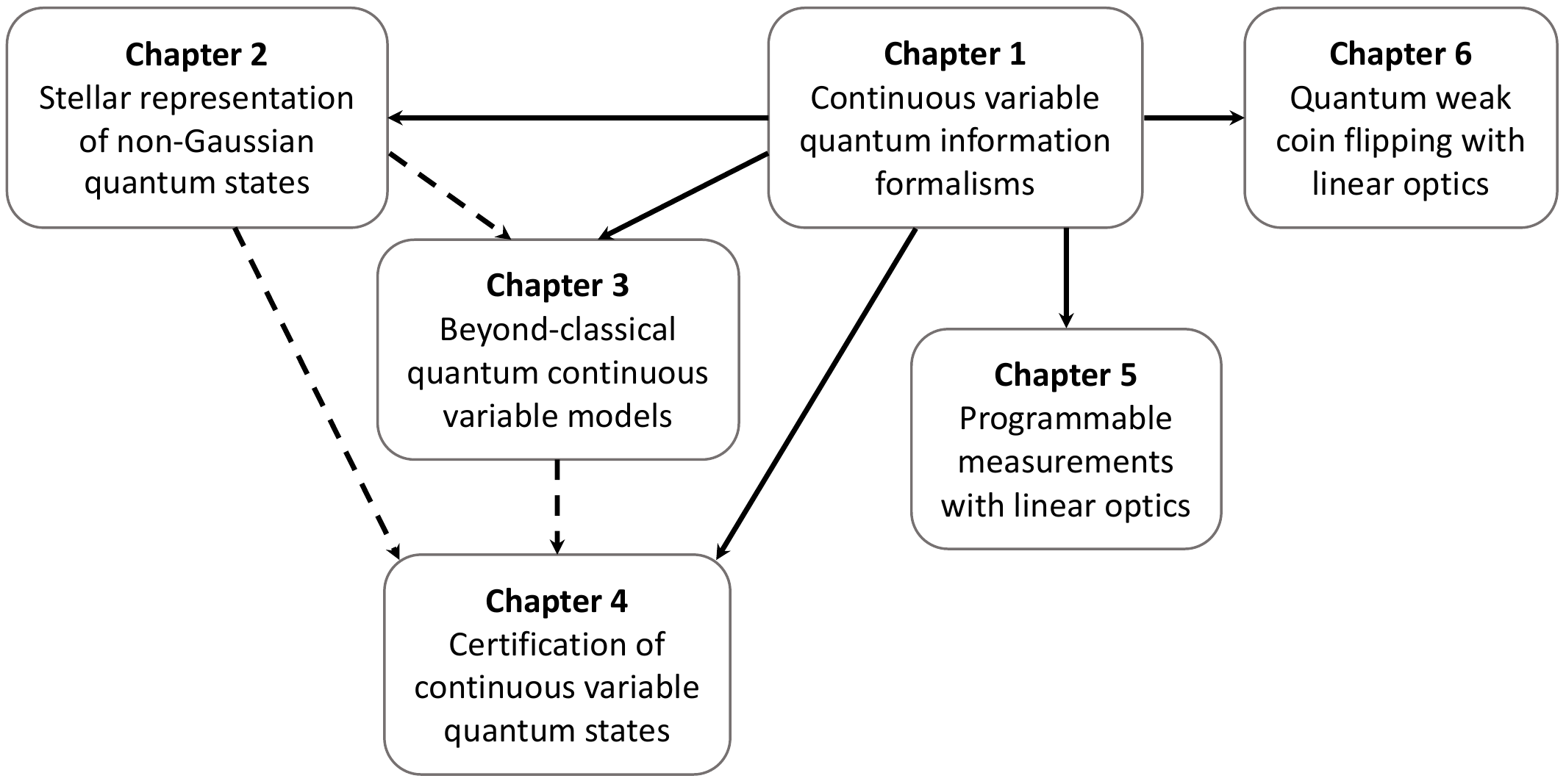}
	\mycaption{Dependencies between the chapters of this thesis. The solid arrows mean that one chapter depends on an other, while the dashed arrows indicate a partial dependence.}
	\label{fig:chaptersdep}
\end{figure}

After a preliminary chapter \ref{chap:CVQI}, chapters \ref{chap:stellar} and \ref{chap:CVS} deal with continuous variable quantum information theory and computing. Chapters \ref{chap:certif} and \ref{chap:prog} consider the probems of quantum state certification and testing, in the continuous variable regime and using quantum optics. Chapter \ref{chap:WCF} discusses the implementation of a quantum cryptography protocol with quantum optics. 
We detail the content of each chapter in what follows. The dependencies between the chapters are indicated in Fig.~\ref{fig:chaptersdep}.

\medskip

\noindent\textbf{Chapter \ref{chap:CVQI}.} After briefly introducing preliminary material on quantum information theory, this chapter presents the formalisms of continuous variable quantum information theory used in this thesis. Phase-space formalism is discussed. A description of Gaussian states and processes follows, together with the symplectic formalism. Then, quantum linear optics is presented with an exposition of Boson Sampling~\cite{Aaronson2013}. Finally, the Segal--Bargmann formalism is introduced.

\medskip

\noindent\textbf{Chapter \ref{chap:stellar}.} This chapter investigates the origin of quantum advantage for continuous variable quantum computing.
Continuous variable quantum states are separated into two broad families: Gaussian and non-Gaussian. While Gaussian states feature interesting properties such as entanglement, non-Gaussian states are crucial for a variety of quantum information tasks \cite{eisert2002distilling,fiuravsek2002gaussian,giedke2002characterization,wenger2003maximal,garcia2004proposal,ghose2007non,niset2009no,adesso2009optimal,barbosa2019continuous}. Characterizing and understanding the properties of these states is thus of major importance \cite{takagi2018convex,zhuang2018resource,albarelli2018resource,lami2018gaussian}. This chapter applies the Segal--Bargmann formalism to derive the stellar representation of single-mode non-Gaussian states. We define and study the stellar rank, using properties of holomorphic functions. This rank induces a hierarchy among continuous variable quantum states. The stellar representation is used to derive a criterion for Gaussian convertibility of states with finite stellar rank within this hierarchy. Its topology with respect to the trace norm is investigated and we show that the hierarchy is robust to small deviations and how to compute the robustness.
The main result of this chapter is a classification of single-mode continuous variable quantum states with respect to their non-Gaussian properties, which can be experimentally witnessed and has consequences for non-Gaussian quantum state engineering. 

\medskip

\noindent\textbf{Chapter \ref{chap:CVS}.} In this chapter, we explore the quantum advantage transition for continuous variables, i.e., the boundary between classically simulable quantum computational models and models capable of outperforming their classical counterparts, in terms of non-Gaussian resources. We give classical simulation algorithms for several quantum models and computational tasks, including linear optics with adaptive measurements and Gaussian circuits with non-Gaussian input states.
Then, we introduce a subuniversal family of continuous variable circuits related to Boson Sampling called Continuous Variable Sampling from photon-added or photon-subtracted squeezed states (CVS) circuits. We show that the continuous output probability densities of these circuits are on average hard to sample exactly classically, by relating their output probabilities to permanents of $(0,1)$-matrices. The main results of this chapter are classical simulation algorithms for Gaussian circuits with weakly non-Gaussian input states, as well as showing how quantum supremacy may be achieved with non-Gaussian states, together with Gaussian operations and measurements.

\medskip

\noindent\textbf{Chapter \ref{chap:certif}.} This chapter considers the certification of continuous variable quantum states. 
Determining an unknown quantum state is difficult especially in continuous variables, where it is described by possibly infinitely many complex parameters. Existing methods like homodyne quantum state tomography require many different measurement settings and heavy classical post-processing~\cite{lvovsky2009continuous}. This chapter shows how continuous variable quantum states can be efficiently verified: we introduce a reliable method for performing continuous variable quantum state state tomography using a single Gaussian measurement, namely heterodyne detection, which can be implemented with quantum optics ; then, we show how this tomography method may be promoted to a state certification protocol under the i.i.d.\@ assumption, by adding an energy test. We also derive a similar protocol for continuous variable quantum state verification, making no assumption whatsoever on the state preparation method, using a de Finetti reduction for infinite-dimensional systems~\cite{renner2009finetti}. We further show that this protocol extends to the multimode case and allows us to efficiently verify output states of Boson Sampling and CVS interferometers. The main result of this chapter is a flexible protocol for building trust for a large class of multimode mode continuous variable quantum states with Gaussian measurements, which provides analytical confidence intervals and allow for a reliable verification of quantum computational supremacy with photonic quantum computing.

\medskip

\noindent\textbf{Chapter \ref{chap:prog}.} On top of being a promising candidate for the demonstration of quantum supremacy with Boson Sampling, quantum optics provides an exciting experimental platform for near-term quantum applications, as well as for probing quantum behaviours. This chapter discusses the relations between quantum state discrimination, quantum state identity testing and universal programmable projective measurements and proposes implementations in linear optics. A generalisation of the swap test~\cite{buhrman2001quantum} is introduced, together with its implementation in linear optics using single-photon encoding. We show how this allows us to construct universal quantum-programmable projective measurements, based on a simple classical post-processing of samples from number-resolving or parity detectors. In order to simplify the experimental requirements, an alternative scheme is derived which uses a coherent state encoding, a simpler interferometer and single-photon threshold detectors, with applications to optical quantum communication protocols.

\medskip

\noindent\textbf{Chapter \ref{chap:WCF}.} Cryptographic protocols are built from a selection of simpler functionalities, called primitives. Remarkably, quantum mechanics allows for the implementation of some primitives with information-theoretic security which can only be achieved with conditional security classically, i.e., by relying on computational assumptions.
The so-called coin flipping by telephone~\cite{blum1983coin}, or weak coin flipping, is one of such cryptographic primitives. It refers to the cryptographic scenario in which two mistrustful and distant parties want to agree on a random bit, while they favor opposite outcomes. The use of quantum mechanics allows for achieving better security than classical mechanics. However, even though various quantum weak coin flipping protocols have been theorised~\cite{SR:PRL02,KN:IPL04,M:PRA05,M:arx07,ARV:arx19}, no practical implementation has been proposed so far.
This chapter introduces an implementation in linear optics of quantum weak coin flipping. 
The proposed implementation relies on adapting a theoretical protocol for quantum weak coin flipping~\cite{SR:PRL02} to linear optics, using the so-called dual-rail encoding, i.e., encoding a qubit with a photon in two spatial modes. The protocol can be implemented with current technology and may display quantum advantage over any classical protocol for the same task.

\section*{Additional remarks}
\addcontentsline{toc}{section}{Additional remarks}  
\label{sec:sec02}

This thesis is intended to be accessible to a reader familiar with the basics of quantum information and computing with discrete and continuous variables. Good introductions to the field of quantum information theory include~\cite{preskill1998lecture} and~\cite{nielsen2002quantum}, while~\cite{Braunstein2005} provides a comprehensive review of quantum information with continuous variables. Pointers to the relevant literature are also displayed throughout the thesis.

\medskip

\noindent This thesis is based on several previous works.
\begin{itemize}
\item
\textbf{Chapter~\ref{chap:stellar}.} This chapter is mainly based on a joint work with D.\@ Markham and F.\@ Grosshans \cite{chabaud2020stellar}, and section~\ref{sec:computeR} is based on a joint work with with G.\@ Roland, M.\@ Walschaers, F.\@ Grosshans, V.\@ Parigi, D.\@ Markham and N.\@ Treps \cite{inprepaLKB}. 
\item
\textbf{Chapter~\ref{chap:CVS}.} Section~\ref{sec:ALO} is based on a joint work with A.\@ Sohbi and D.\@ Markham~\cite{inprepaMLALO}, sections~\ref{sec:NGinput} and~\ref{sec:StrongsimuNG} on a joint work with G.\@ Ferrini, F.\@ Grosshans and D.\@ Markham~\cite{inprepaSimuNG}, and section~\ref{sec:CVS} on a joint work with T.\@ Douce, D.\@ Markham, P.\@ van Loock, E.\@ Kashefi and G.\@ Ferrini \cite{chabaud2017continuous}. 
\item
\textbf{Chapter~\ref{chap:certif}.} Section~\ref{sec:trust} is based on a joint work with J.\@ Eisert, D.\@ Hangleiter, N.\@ Walk, I.\@ Roth, D.\@ Markham, R.\@ Parekh, and E.\@ Kashefi \cite{eisert2020quantum}, sections~\ref{sec:certif1} to \ref{sec:certif3} on a joint work with T.\@ Douce, F.\@ Grosshans, D.\@ Markham and E.\@ Kashefi \cite{chabaud2019building}, section~\ref{sec:applicationscertif} on a joint work with with G.\@ Roland, M.\@ Walschaers, V.\@ Parigi, F.\@ Grosshans, D.\@ Markham and N.\@ Treps \cite{inprepaLKB}, and section~\ref{sec:applicationscertif3} on a joint work with F.\@ Grosshans, D.\@ Markham and E.\@ Kashefi \cite{chabaud2020efficient}. 
\item
\textbf{Chapter~\ref{chap:prog}.} Sections~\ref{sec:testing} and~\ref{sec:univ1} are based on a joint work with E.\@ Diamanti, D.\@ Markham, E.\@ Kashefi and A.\@ Joux \cite{chabaud2018optimal}, and section~\ref{sec:univ2} on a joint work with N.\@ Kumar, E.\@ Kashefi, D.\@ Markham, and E.\@ Diamanti \cite{inprepa1}. 
\item
\textbf{Chapter~\ref{chap:WCF}.} This chapter is based on a joint work with M.\@ Bozzio, E.\@ Diamanti and I.\@ Kerenidis \cite{bozzio2020quantum}. 
\end{itemize}

\clearemptydoublepage
\pagestyle{myvf}
%
%
\let\textcircled=\pgftextcircled
\chapter{Continuous variable quantum information formalisms}
\label{chap:CVQI}

\initial{C}ontinuous variable quantum information theory refers to the study of information encoded in quantum physical systems with continuous degrees of freedom. The approach of the work presented in this dissertation for studying continuous variable quantum information is to use different mathematical formalisms as different ways of gaining intuition. Juggling several representations of the same mathematical object is indeed an excellent way to get insights about this object. In this chapter, we briefly review the formalisms for continuous variable quantum information theory used throughout the rest of the thesis. These include phase space formalism for continuous variable quantum states and operators, symplectic formalism for Gaussian states, quantum optics and Boson Sampling, and Segal--Bargmann formalism for continuous variable quantum states.


\section{Preliminary material}
\label{sec:preli}

\subsection{Notations}
\label{sec:multiindex}

The sets $\mathbb N$, $\mathbb R$ and $\mathbb C$ are the usual sets of natural, real and complex numbers, with a $*$ exponent when $0$ is removed from the set. The size of a set $\mathcal X$ is denoted by $|\mathcal X|$. The natural logarithm is denoted $\log$.

We write complexity classes with sans serif font: \textsf{P}, \textsf{NP}...

The number of subsystems or modes will generally be denoted by $m\in\mathbb N^*$. Hilbert spaces are denoted by $\mathcal H$ or $\mathcal K$.  
The expressions $\ket\phi$, $\ket\psi$ denote pure states, and $\rho$ and $\sigma$ denote density operators of possibly mixed quantum states.

For vectors and operators, we denote by a $*$ exponent the complex conjugate, by a $T$ exponent the transpose and by a $\dag$ exponent the transpose complex conjugate (adjoint). Matrices are denoted by capital letters and covariance matrices will be denoted by $\bm V$. Operators are indicated by a hat, with the exception of density operators, positive-operator valued measure elements and identity operator $\mathbb1$. In particular, $\hat a$ and $\hat a^\dag$ denote the annihilation and creation operators and $\hat q$ and $\hat p$ denote the position-like and momentum-like quadrature operators. The identity matrix is also denoted $\mathbb1$, sometimes with an index indicating its size. The zero matrix is similarly denoted $\mymathbb0$. The trace is denoted by $\Tr$ and the determinant by $\Det$. 

$\Pr$ denotes a probability, while $\mathbb E$ denotes an expected value.
A function $\delta$ may stand for the Kronecker symbol or a Dirac delta, depending on the context. 
The letters $\alpha$, $\beta$ and $\gamma$ are used for coherent state amplitudes or complex amplitudes, while the letters $\xi$ and $\zeta$ are used for squeezing parameters. The letter $z$ denotes a complex variable.

We write $\otimes$ and $\oplus$ for the tensor product and the direct sum, respectively. We use bold math for multimode states, vectors and multi-index notations. Let $m,n\in\mathbb N^*$. We define $\bm0=(0,\dots,0)$ and $\bm1=(1,\dots,1)$, and we write $\bm0^n=(0,\dots,0)\in\mathbb N^n$ or $\bm1^n=(1,\dots,1)\in\mathbb N^n$ to avoid ambiguity. For all $k\in\{1,\dots,m\}$, we also define $\bm1_k=(0,\dots,0,1,0,\dots,0)$, where the $k^{th}$ entry is $1$ and all the other $m-1$ entries are $0$. For all $\bm z=(z_1,\dots,z_m)\in\mathbb C^m$, all $\bm z'=(z_1',\dots,z_m')\in\mathbb C^m$ and all $\bm p=(p_1,\dots,p_m)\in\mathbb N^m$ we write
\be
\ba
\,&\bm z^*=(z_1^*,\dots,z_m^*)\\
\,&-\bm z=(-z_1,\dots,-z_m)\\
\,&\bm{\tilde z}=\bm z\oplus\bm z^*=(z_1,\dots,z_m,z_1^*,\dots,z_m^*)\\
\,&\ket{\bm z}=\ket{z_1\dots z_m}\\
\,&\|\bm z\|^2=|z_1|^2+\cdots+|z_m|^2\\
\,&\bm z^{\bm p}=z_1^{p_1}\dots z_m^{p_m}\\
\,&\bm z+\bm z'=(z_1+z'_1,\dots,z_m+z'_m)\\
\,&\bm z\le\bm z'\Leftrightarrow z_k\le z'_k\quad\forall k\in\{1,\dots,m\}\\
\,&\bm p!=p_1!\dots p_m!\\
\,&|\bm p|=p_1+\cdots+p_m\\
\,&\partial^{\bm p}=\partial_1^{p_1}\dots\partial_m^{p_m}\\
\,&\left(\frac\partial{\partial\bm z}\right)^{\bm p}=\frac{\partial^{|\bm p|}}{\partial z_1^{p_1}\cdots\partial z_m^{p_m}}.
\ea
\label{multiindex}
\ee
We will use for brevity the notations $c_\chi=\cosh\chi$, $s_\chi=\sinh\chi$ and $t_\chi=\tanh\chi$, for all $\chi\in\mathbb C$. The commutator is denoted by $[\,,\,]$ and the anticommutator by $\{\,,\,\}$. Finally we adopt the convention $\hbar=1$ and use canonical conventions rather than optical ones.

\medskip

\noindent Note that proofs of intermediate technical results will be indicated by a vertical bar running along the side of the page, with a square symbol marking the end of the proof.

\subsection{Basics of quantum information theory}
\label{sec:DVQI}

The presentation given here is very succinct and good introductions to the field of quantum information theory include~\cite{preskill1998lecture} and~\cite{nielsen2002quantum}.

\medskip

\noindent In quantum information theory, we identify two notions of randomness. On the one hand, there is an inherent randomness in the formalism of quantum measurements, which we call quantum randomness. On the other hand, classical randomness corresponds to the usual notion of randomness to which we refer, for example, when we draw a card from a shuffled deck of cards or when we roll a die.
In practice, a quantum system can manifest both classical and quantum randomness.

\medskip

\noindent The properties of a quantum system are described by its quantum state. Quantum states with no classical randomness are called pure states. These pure quantum states are represented mathematically as normalised vectors in a separable Hilbert space $\mathcal H$. We adopt Dirac bra-ket notation~\cite{dirac1981principles} in what follows: a column vector $\psi$ is represented as the ket $\ket\psi$ and its adjoint (transpose complex conjugate) line vector is represented as the bra $\bra\psi$. In particular, the projector onto $\ket\psi$ is expressed as $\ket\psi\!\bra\psi$ and the inner product of two states $\ket{\phi}$ and $\ket\psi$ is denoted by $\braket{\phi|\psi}$, with $\braket{\phi|\psi}=\braket{\psi|\phi}^*$. The quantity $|\braket{\phi|\psi}|^2$ is referred to as the overlap of the states $\ket{\phi}$ and $\ket\psi$.

The simplest nontrivial example is a Hilbert space of dimension $2$. In that case, quantum states are referred to as qubit states, states in a Hilbert space of finite dimension $d>2$ being referred to as qudit states. Given an orthonormal basis $(\ket0,\ket1)$ of a Hilbert space of dimension $2$, a qubit state $\ket\psi$ is expressed as
\be
\ket\psi=a\ket0+b\ket1,\quad\quad|a|^2+|b|^2=1,
\label{basic1}
\ee
for $a,b\in\mathbb C$, with $\bra\psi=a^*\!\bra0+b^*\!\bra1$. The coefficients $a$ and $b$ are the complex amplitudes of the qubit state $\ket\psi$. If $a\neq0$ and $b\neq0$, the state $\ket\psi$ is said to be in a superposition of the states $\ket0$ and $\ket1$. 

The basis $(\ket0,\ket1)$ is referred to as the computational basis. On the other hand, setting $\ket\pm=\frac1{\sqrt2}(\ket0\pm\ket1)$, the states $(\ket+,\ket-)$ also form an orthonormal basis, referred to as the diagonal basis.

Observable physical quantities, or simply observables, are represented mathematically by self-adjoint (hermitian) operators $\hat O=\hat O^\dag$. Such operators have an orthonormal basis of eigenvectors and measuring the observable $\hat O$ gives an outcome sampled from the list of its eigenvalues. The probability of each outcome is determined by the Born rule:
\be
\Pr[\lambda]=\braket{\psi|\Pi_\lambda|\psi},
\label{basic2}
\ee
where $\lambda$ is the eigenvalue, $\ket\psi$ is the state of the measured quantum system and $\Pi_\lambda$ is a projector onto the eigenvector corresponding to the eigenvalue $\lambda$. Equivalently, we say that we measure in a specific orthonormal basis to say that we measure an observable which has this basis as an eigenbasis. In particular, Eq.~(\ref{basic1}) may be interpreted as follows: $\braket{\psi|\Pi_0|\psi}=\braket{\psi|0}\!\braket{0|\psi}=|\braket{0|\psi}|^2=|a|^2$ (resp.\@ $|b|^2$) is the probability that we obtain the outcome $0$ (resp.\@ $1$) when measuring the state $\ket\psi$ in the $(\ket0,\ket1)$ basis. The two probabilities sum to $1$, corresponding to the fact that the measurement will yield an outcome, either $0$ or $1$. The measurement outcome is random when $a\neq0$ and $b\neq0$, i.e., quantum randomness manifests when the measured state is in a superposition of eigenvectors of the observable. 
Measuring a quantum state collapses the state onto the eigenvector corresponding to the outcome obtained. In particular, any subsequent measurement of the same observable will yield the same result with probability $1$.

The most general notion of quantum measurement is captured by positive-operator valued measures (POVM). A POVM is a set of semidefinite operators $\{\Pi_i\}_{i\in\mathcal I}$ whose elements sum to the identity operator, indexed by a set of outcomes $\mathcal I$.
The operator $\Pi_i$ is associated to the measurement outcome $i\in\mathcal I$ and the probability for this outcome is given by Eq.~(\ref{basic2}), replacing $\lambda$ by $i$. The case where the operators $\Pi_i$ are projectors, as in Eq.~(\ref{basic2}), corresponds to projection-valued measures (PVM). 

\medskip

\noindent Quantum systems can also exhibit classical randomness. When that is the case, we refer to the quantum state as mixed. A mixed quantum state is represented mathematically by a so-called density operator, i.e., a hermitian operator with trace $1$ acting on a Hilbert space. The density operator for a pure state $\ket\psi$ is simply a projector $\ket\psi\!\bra\psi$. A mixed quantum state can be written as a convex combination, or mixture, of pure states. For example, the state obtained by flipping an unbiaised coin and choosing the state $\ket0$ for tails and $\ket1$ for heads is a mixed state expressed as $\frac12\ket0\!\bra0+\frac12\ket1\!\bra1$, which is different from the pure superposition $\frac1{\sqrt2}(\ket0+\ket1)$, whose density operator is given by $\frac12\ket0\!\bra0+\frac12\ket0\!\bra1+\frac12\ket1\!\bra0+\frac12\ket1\!\bra1$.
The Born rule for a mixed state $\rho$ reads
\be
\Pr[i]=\Tr\,(\Pi_i\rho),
\label{basic3}
\ee
where $\{\Pi_i\}_{i\in\mathcal I}$ is a POVM over a set of outcomes $\mathcal I$. Setting $\rho=\ket\psi\!\bra\psi$, we retrieve the Born rule for pure states in Eq.~(\ref{basic2}). Writing the semidefinite operator $\Pi_i=M_i^\dag M_i$, the state after a measurement with outcome $i\in\mathcal I$ is given by
\be
\rho^{(i)}=\frac{M_i\rho M_i^\dag}{\Tr\,(\Pi_i\rho)}.
\ee
Note that the choice of $M_i$ is not unique and this choice reflects different possible ways of physically implementing the same POVM. Given an observable $\hat O$, the quantity $\Tr\,(\hat O\rho)$ is the expectation of the operator $\hat O$ for the quantum state $\rho$ and is alternatively denoted $\left<\hat O\right>_\rho$.

\medskip

\noindent The global state of two independent quantum systems with states $\ket\phi$ and $\ket\psi$ in two Hilbert spaces $\mathcal H$ and $\mathcal H'$, respectively, lies in the tensor product $\mathcal H\otimes\mathcal H'$ and is obtained by taking the tensor product $\ket\phi\otimes\ket\psi$ of both states. We will usually write $\ket\phi\otimes\ket\psi=\ket{\phi\psi}$ when there is no ambiguity. The dimension of the Hilbert space $\mathcal H\otimes\mathcal H'$ is the product of the dimensions of the Hilbert spaces $\mathcal H$ and $\mathcal H'$, implying in particular that the computational basis of $n$-qubit states has size $2^n$.

Two quantum systems may not be independent and a pure quantum state which cannot be written as a tensor product of quantum states is called entangled. For example, the state $\frac1{\sqrt2}(\ket{00}+\ket{11})$
is entangled while the state $\frac12(\ket0+\ket1)\otimes(\ket0+\ket1)$ is separable.
A (mixed) quantum state is called separable if it can be written as a mixture of separable pure states, and entangled otherwise. 

Entanglement may be conceived as the quantum version of classical correlation~\cite{werner1989quantum}: the mixed quantum state $\frac12\ket{00}\!\bra{00}+\frac12\ket{11}\!\bra{11}$ is classically correlated---the measurements of each subsystem in the $(\ket0,\ket1)$ basis will always yield the same outcomes---but not entangled, since it is a mixture of product states. On the other hand, the pure state $\frac1{\sqrt2}(\ket{00}+\ket{11})=\frac1{\sqrt2}(\ket{++}+\ket{--})$ is entangled. This state is `more' correlated than the previous one in the following sense: not only the measurements of each subsystem in the $(\ket0,\ket1)$ basis will always yield the same outcomes but measuring each subsystem in the $(\ket+,\ket-)$ basis will also always yield the same outcomes.

Given a state $\rho$ over two subsystems in $\mathcal H$ and $\mathcal H'$, the reduced state of the first subsystem is obtained by tracing out, or taking the partial trace over, the second subsystem $\Tr_{\mathcal H'}(\rho)$. A separable state is fully described by the reduced states of its individual subsystems, while this is no longer the case for an entangled state.

\medskip

\noindent The simplest example of evolution of a quantum system is a unitary evolution over a time $t$, described by a unitary operator $\hat U$ with $\hat U^\dag\hat U=\mathbb1$, generated by a Hamiltonian $H$ with $H^\dag=H$, such that $\hat U=e^{-iHt}$. If the system is in a pure state $\ket\psi$, then the state after the evolution is a normalised pure state $\hat U\ket\psi$. If the system is in a mixed state $\rho$, then the state after the evolution is a mixed state with density operator $\hat U\rho\hat U^\dag$.

More general quantum evolutions are described by quantum channels, i.e., completely positive trace-preserving maps (CPTP). By Stinespring dilation theorem, CPTP maps can be expressed as unitaries acting on a larger space. Formally, if $\mathcal E$ is a CPTP map acting on a Hilbert space $\mathcal H$, then there exist a Hilbert space $\mathcal H'$ and a unitary operator $\hat U$ such that for all density operators $\rho$,
\be
\mathcal E(\rho)=\Tr_{\mathcal H'}[U(\rho\otimes\ket0\!\bra0)U^\dag].
\ee
In other words, any quantum channel can be obtained by tensoring with a second system in a fixed state, a unitary evolution and a reduction to a subsystem.
Naimark's theorem provides a similar result for decomposing a POVM as a unitary followed by a PVM on a larger space.

The most general physical evolutions are described by quantum operations, i.e., completely positive trace-decreasing maps (CPTD). These operations can be obtained as obtained by tensoring with a second system in a fixed state, a unitary evolution, a PVM and a reduction to a subsystem. 
Non-CPTD maps are referred to as unphysical operations. Such operations can be approximated by quantum operations, for example when they act as CPTD maps on a subset of the Hilbert space.

\medskip

\noindent A quantum computation is composed of the three following steps: input, evolution and measurement. With the above, one may conceive elaborate quantum computations as building a highly entangled state from a simple input product state via a unitary evolution and sampling from a probability distribution given by the Born rule and the choice of measurement.
Quantum computations can be looked at in the circuit picture, in which the unitary evolution is decomposed as a product of gates acting on at most two subsystems at a time.

\medskip

\noindent Discrimination of quantum states is a central element in many quantum information processing tasks~\cite{nielsen2002quantum} and various measures are available~\cite{fuchs1999cryptographic}. 
We review two measures used extensively in the thesis: the fidelity and the trace distance. The properties outlined are independent of the dimension of the Hilbert space.

\medskip

\noindent The fidelity between two states $\rho$ and $\sigma$ is defined as
\be
F(\rho,\sigma)=\Tr\left(\sqrt{\sqrt\sigma\rho\sqrt\sigma}\right)^2.
\ee
Note that the definition used here is the square of the definition in~\cite{fuchs1999cryptographic,nielsen2002quantum}. Even though it is not apparent with the above equation, the fidelity is symmetric in its arguments $\rho$ and $\sigma$. When at least one of the two states is a pure state, this expression reduces to
\be
F\left(\psi,\rho\right)=\Tr(\ket\psi\!\bra\psi\rho)=\braket{\psi|\rho|\psi}.
\label{fidepure}
\ee
In particular when both states are pure $F\left(\phi,\psi\right)=|\braket{\phi|\psi}|^2$.

\medskip

\noindent We write the Schatten $1$-norm of a bounded operator $T$ as
\be
\|T\|_1=\Tr\left(\sqrt{T^\dag T}\right)=\Tr(|T|).
\ee
The trace distance between two states $\rho,\sigma$ is defined as
\be
\ba
D(\rho,\sigma)&=\frac{1}{2}\|\rho-\sigma\|_1\\
&=\frac{1}{2}\Tr(|\rho-\sigma|).
\ea
\label{td0}
\ee
It is jointly convex in its two arguments. The fidelity is related to the trace distance by the Fuchs-van de Graaf inequalities~\cite{fuchs1999cryptographic}
\be
1-\sqrt{F(\rho,\sigma)}\le D(\rho,\sigma)\le\sqrt{1-F(\rho,\sigma)}.
\label{td1}
\ee
When one of the states is pure, the lower bound may be refined as
\be
1-F(\psi,\rho)\le D(\psi,\rho).
\label{td3}
\ee
When both states are pure, the upper bound in Eq.~(\ref{td1}) becomes an equality:
\be
\ba
D(\phi,\psi)&=\sqrt{1-F(\phi,\psi)}\\
&=\sqrt{1-|\braket{\phi|\psi}|^2}.
\ea
\label{td4}
\ee
The fidelity is nondecreasing under quantum operations and the trace distance is nonincreasing under quantum operations.
The total variation distance of two probability distributions $P$ and $Q$ over a sample space $\mathcal S$ is defined as
\be
\|P-Q\|_{tvd}=\frac12\sum_{s\in\mathcal S}{\left|P(s)-Q(s)\right|}.
\label{tvd}
\ee
A similar definition holds for probability densities over a continuous sample space, by replacing the discrete sum by a continuous sum. The trace distance verifies
\be
D(\rho,\sigma)=\max_{\hat O}{\|P_\rho^{\hat O}-P_\sigma^{\hat O}\|_{tvd}},
\label{td2}
\ee
where $P_\rho^{\hat O}$ (resp. $P_\sigma^{\hat O}$) is the probability distribution associated to measuring the observable $\hat O$ for the state $\rho$ (resp. $\sigma$) and where the maximum of the total variation distance is taken over all observables. The trace distance thus has an operational significance: if two states are close in trace distance, then any computation taking as input one of the two states is indistinguishable from the same computation taking as input the other state. Moreover, with Eq.~(\ref{td1}), lower bounds on the fidelity also give upper bounds on the total variation distance, which are tight when the states are pure, by Eq.~(\ref{td4}).

\medskip

In what follows, we consider the case of infinite-dimensional Hilbert spaces, allowing for the description of quantum systems with continuous degrees of freedom. Discrete variables can be encoded in continuous degrees of freedom and finite-dimensional Hilbert spaces may be embedded in infinite-dimensional ones. Despite its discrete character, we will also refer to the study of such embedded discrete variable quantum information in an infinite-dimensional Hilbert space as continuous variable quantum information theory, since the same mathematical formalisms are employed in both case.

\subsection{Continuous variable quantum information theory in a nutshell}
\label{sec:CVQI}

We refer the reader to the first chapters of~\cite{Braunstein2005,ferraro2005gaussian,adesso2014continuous} for a further introduction on the material presented in this section. While the presentation that follows is quite technical, it avoids many of the subtleties which appear when dealing with infinite-dimensional Hilbert spaces. The interested reader will find an example of a formal treatment in~\cite{de2005role}. 

\medskip

\noindent The continuous variable equivalent of a qubit or qudit is the qumode, or simply mode. Single-mode continuous variable quantum states are mathematically described as normalised complex vectors in an infinite-dimensional separable Hilbert space, with an infinite countable orthonormal basis $\{\ket n\}_{n\in\mathbb N}$ referred to as the Fock basis, or photon-number basis in the context of optical quantum information processing. In particular, $\ket0$ is referred to as the vacuum state and $\ket1$ as the single-photon state. A single-mode pure state $\ket\psi$ can be written in Fock basis as
\be
\ket\psi=\sum_{n\ge0}{\psi_n\ket n},
\ee
where $\psi_n\in\mathbb C$ for all $n\in\mathbb N$, with the normalisation condition $\sum_{n=0}^{+\infty}{|\psi_n|^2}=1$. The Fock basis comes with canonical adjoint operators $\hat a$ and $\hat a^\dag$ referred to as annihilation and creation operators, respectively, or photon subtraction and photon addition operators in the context of optical quantum information processing. These operators are defined by their action on the Fock basis as
\be
\ba
\,&\hat a\ket n=\sqrt n\ket{n-1},\quad\quad\quad\text{for }n\in\mathbb N^*,\\
\,&\hat a\ket 0=0,\\
\,&\hat a^\dag\ket n=\sqrt{n+1}\ket{n+1},\;\quad\text{for }n\in\mathbb N,
\ea
\ee
and follow the canonical commutation relation
\be
[\hat a,\hat a^\dag]=\mathbb 1,
\ee
where $\mathbb 1$ is the identity operator. The eigenstates of the annihilation operator are the coherent states $\{\ket\alpha\}_{\alpha\in\mathbb C}$, defined as
\be
\ket\alpha=e^{-\frac12|\alpha|^2}\sum_{n\ge0}{\frac{\alpha^n}{\sqrt{n!}}\ket n},
\ee
for all $\alpha\in\mathbb C$. 
Alternatively, defining the displacement operator as
\be
\hat D(\alpha)=e^{\alpha\hat a^\dag-\alpha^*\hat a},
\ee
for all $\alpha\in\mathbb C$, the coherent state of amplitude $\alpha\in\mathbb C$ is obtained from the vacuum state as
\be
\ket\alpha=\hat D(\alpha)\ket0.
\ee
The inner product of two coherent states $\ket\alpha$ and $\ket\beta$ is given by
\be
\braket{\alpha|\beta}=e^{\alpha^*\beta-\frac12(|\alpha|^2+|\beta|^2)},
\ee
for all $\alpha,\beta\in\mathbb C$. In particular, two coherent states have nonzero overlap. These states form an overcomplete family:
\be
\int_{\alpha\in\mathbb C}{\ket\alpha\!\bra\alpha\frac{d^2\alpha}\pi}=\mathbb 1,
\ee
where $d^2\alpha=d\Re(\alpha)d\Im(\alpha)$. The canonical position-like and momentum-like operators $\hat q$ and $\hat p$ are defined as
\be
\ba
\,&\hat q=\frac1{\sqrt2}(\hat a+\hat a^\dag),\\
\,&\hat p=\frac1{i\sqrt2}(\hat a-\hat a^\dag).
\ea
\ee
These hermitian operators, also referred to as quadrature operators in the context of optical quantum information processing, follow the canonical commutation relation
\be
[\hat q,\hat p]=i\mathbb 1.
\ee
They satisfy Heisenberg uncertainty principle~\cite{heisenberg1985anschaulichen}
\be
\sigma_{\hat q}\sigma_{\hat p}\ge\frac12,
\ee
where $\sigma_{\hat q}$ and $\sigma_{\hat p}$ denote the standard deviation of position and momentum, respectively, i.e., they cannot be measured both with arbitrary precision for the same quantum state: measuring one randomises the other. 

The eigenstates of $\hat q$ (resp.\@ $\hat p$) form a continuous family of unnormalisable states $\{\ket q\}_{q\in\mathbb R}$ (resp.\@ $\{\ket p\}_{p\in\mathbb R}$), thus technically lying outside of the Hilbert space. 
These states may be treated formally as an infinite uncountable basis of the Hilbert space, the so-called position basis (resp.\@ momentum basis). Expanding a single-mode pure state $\ket\psi$ in the position basis gives
\be
\ket\psi=\int_{q\in\mathbb R}{\psi(q)\ket qdq},
\ee
where $\psi(q)=\braket{q|\psi}$ is the position wave function of the state $\ket\psi$, with the normalisation condition for the position probability distribution $\int_{q\in\mathbb R}{|\psi(q)|^2dq}=1$. A similar expansion holds in the momentum basis with the momentum wave function. 
The position and momentum bases are related by a Fourier transform:
\be
\ket q=\frac1{\sqrt{2\pi}}\int_{p\in\mathbb R}{e^{-iqp}\ket pdp},
\ee
and
\be
\ket p=\frac1{\sqrt{2\pi}}\int_{q\in\mathbb R}{e^{iqp}\ket qdq}.
\ee
Note that the Fock state $\ket{n=0}$ and the coherent state $\ket{\alpha=0}$ are equal, but different from the position state $\ket{q=0}$ and the momentum state $\ket{p=0}$, themselves distinct.

\section{Phase space formalism}
\label{sec:phasespace}

\noindent We refer the reader to~\cite{cahill1969ordered,cahill1969density} for an introduction to the material presented in this section. In particular, we restrict to single-mode states and operators.

\medskip

\noindent The expectation values of the position and momentum operators lie in the so-called phase space, which is the quantum analogue of classical phase space. Continuous variable quantum states and operators can be alternatively described by a phase space representation. This formulation identifies a quantum state with a normalised distribution over phase space.

This allows for a simple and experimentally relevant classification of quantum states: those with a Gaussian phase space distribution are called Gaussian states and the others non-Gaussian states. By extension, operations mapping Gaussian states to Gaussian states are also called Gaussian. These Gaussian operations and states are the ones implementable with linear optics and quadratic non-linearities~\cite{Braunstein2005}, and are hence relatively easy to construct experimentally.

\medskip

\noindent Hereafter, we identify the single-mode phase space with $\mathbb C$, where the real part corresponds to expectation values of the position operator and the imaginary part to expectation values momentum operator. We adopt the convention $\alpha=\frac1{\sqrt2}(q+ip)\in\mathbb C$, with $\frac{d^2\alpha}\pi=\frac{d\Re(\alpha)d\Im(\alpha)}\pi=\frac{dqdp}{2\pi}$.

There exists a continuum of equivalent phase space distributions representing the same operator in phase space. This continuum of representations is parametrized by a real parameter $s\le1$.
For all $s\le1$, let us define the operator
\be
\hat T(\alpha,s):=\int_{\beta\in\mathbb C}{\hat D(\beta)\exp\left(\alpha\beta^*-\alpha^*\beta+\frac s2|\beta|^2\right)\frac{d^2\beta}\pi},
\ee
for all $\alpha\in\mathbb C$. The phase space representation with parameter $s$ of an operator $\hat O$ is defined as
\be
W_{\hat O}(\alpha,s)=\Tr\left[\hat T(\alpha,s)\,\hat O\right].
\label{Ws}
\ee
This expression should be treated formally for unbounded operators and the case $s=1$ should be understood as the limit $s\to1^-$. The same definition holds for density operators, in which case the representation is real-valued and corresponds to the expectation value of the operator $\hat T$. 
The phase space representations are normalised as
\be
\int_{\alpha\in\mathbb C}{W_\rho(\alpha,s)\frac{d^2\alpha}\pi}=\Tr\,(\rho),
\label{normWs}
\ee
for any density operator $\rho$ and any $s\le1$.
As the parameter $s$ decreases, the phase space representation smoothens. This is captured by the following relation:
\be
W(\alpha,s)=\frac2{t-s}\int_{\beta\in\mathbb C}{W(\beta,t)\exp\left(-\frac{2|\alpha-\beta|^2}{t-s}\right)\frac{d^2\beta}\pi},
\ee
for all $s<t\le1$, i.e., the representation with lower parameter is obtained from the representation with higher parameter by a Gaussian convolution. In particular, if one representation is a Gaussian function, then all representations are Gaussian. Moreover, for all operators $\hat O_1$ and $\hat O_2$,
\be
\Tr\left(\hat O_1\hat O_2\right)=\int_{\alpha\in\mathbb C}{W_{\hat O_1}(\alpha,-s)W_{\hat O_2}(\alpha,s)\frac{d^2\alpha}\pi},
\ee
for all $s\in[-1,1]$. This important property allows one to retrieve information about quantum systems by probing their phase space representation: if one of the two operators in the above equation is a density operator, the expectation value is obtained as
\be
\Tr\left(\hat O\rho\right)=\int_{\alpha\in\mathbb C}{W_{\hat O}(\alpha,-s)W_\rho(\alpha,s)\frac{d^2\alpha}\pi},
\label{OETgen}
\ee
for all $s\in[-1,1]$.

\medskip

\begin{figure}[h!]
	\begin{center}
		\includegraphics[width=0.5\columnwidth]{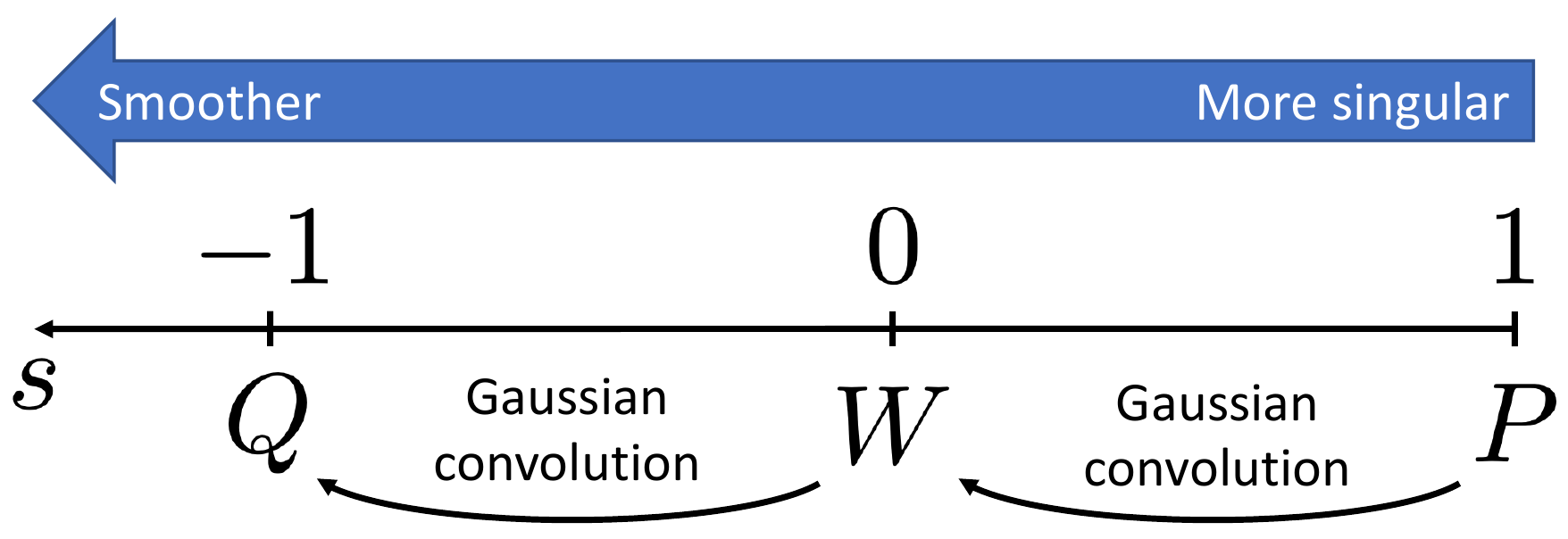}
		\caption{A pictorial representation of the continuum of phase space representations.}
		\label{fig:QPD}
	\end{center}
\end{figure}

\noindent In what follows, we detail some properties of the three most prominent representations in the literature: the Wigner $W$ function~\cite{wigner1997quantum}, the Glauber--Sudarshan $P$ function \cite{sudarshan1963equivalence,glauber1963coherent} and the Husimi $Q$ function~\cite{husimi1940some}, corresponding to the values $s=0$, $s=1$ and $s=-1$, respectively (Fig.~\ref{fig:QPD}). In particular, we will make extensive use of the Husimi representation throughout the first chapters of the thesis. We adopt the normalising conventions
\be
\ba
\,&W(\alpha)=\frac1\pi W(\alpha,0),\\
\,&P(\alpha)=\frac1\pi W(\alpha,1),\\
\,&Q(\alpha)=\frac1\pi W(\alpha,-1),
\ea
\ee
for all $\alpha\in\mathbb C$, so that the $W$, $P$ and $Q$ functions are normalised to $1$ for normalised states (note the difference of normalisation with~\cite{cahill1969density} for the Wigner and Husimi functions).\\

\subsection{Wigner $W$ function}

The Wigner function is a nonsingular distribution for all states and is referred to as a quasiprobability distribution, as it is a normalised distribution which can take negative values. This contrasts with classical phase space probability distributions. 

By virtue of Hudson's theorem~\cite{hudson1974wigner,soto1983wigner}, a pure quantum state is non-Gaussian if and only if its Wigner function has negative values. In other words, if a pure quantum state has a positive Wigner function, then it is a Gaussian state. Various notions relating to negativity of the Wigner function have been introduced for measuring how much non-Gaussian a quantum state is~\cite{kenfack2004negativity,albarelli2018resource}.

The Wigner function can be expressed as~\cite{royer1977wigner}
\be
W_{\hat O}(\alpha)=\frac2\pi\Tr\left[\hat D(\alpha)\hat\Pi\hat D^\dag(\alpha)\hat O\right],
\ee
for all $\alpha\in\mathbb C$ and for any operator $\hat O$, where $\hat\Pi=(-1)^{\hat a^\dag\hat a}=\sum_{n\ge0}{(-1)^n\ket n\!\bra n}$ is the parity operator and $\hat D(\alpha)=e^{\alpha\hat a^\dag-\alpha^*\hat a}$ is a displacement operator of amplitude $\alpha\in\mathbb C$. In particular, the Wigner function of a quantum state is related to the expectation value of displaced parity operators.

\subsection{Glauber--Sudarshan $P$ function}

The Glauber--Sudarshan $P$ function is the most singular phase space representation. For quantum states, it is actually always a singular distribution.

The $P$ function gives a convenient diagonal representation of a state in coherent state basis as
\be
\rho=\int_{\alpha\in\mathbb C}{P_\rho(\alpha)\ket\alpha\!\bra\alpha d^2\alpha},
\ee
and this representation is unique. The $P$ function can be expressed formally as~\cite{mehta1967diagonal}
\be
P_{\hat O}(\alpha)=\frac{e^{|\alpha|^2}}\pi\int_{\beta\in\mathbb C}{\braket{-\beta|\hat O|\beta}\exp\left(\alpha\beta^*-\alpha^*\beta+|\beta|^2\right)\frac{d^2\beta}\pi},
\ee
for all $\alpha\in\mathbb C$ and for any operator $\hat O$.

\subsection{Husimi $Q$ function}

The Husimi $Q$ function is a smoother version of the Wigner function and the Glauber--Sudarshan $P$ function. It is given by
\be
Q_{\hat O}(\alpha)=\frac1\pi\braket{\alpha|\hat O|\alpha},
\ee
for all $\alpha\in\mathbb C$ and for any operator $\hat O$, where $\ket\alpha$ is the coherent state of amplitude $\alpha\in\mathbb C$. 
The Husimi $Q$ function of a state thus is always nonnegative and normalised. However, it does not represent probabilities of mutually exclusive states since the overlap between two coherent states is always nonzero.

For any state $\rho$ and any operator $\hat O$ we have, with Eq.~(\ref{OETgen}), the so-called optical equivalence theorem for antinormal ordering:
\be
\Tr\left(\hat O\rho\right)=\pi\int_{\alpha\in\mathbb C}{Q_\rho(\alpha)P_{\hat O}(\alpha)\,d^2\alpha}.
\label{OETgen}
\ee
Hudson's theorem may be formulated as follows for the Husimi function~\cite{lutkenhaus1995nonclassical}: a pure quantum state is non-Gaussian if and only if its Husimi function has zeros. In other words, a pure quantum state is non-Gaussian if and only if it is orthogonal to at least one coherent state.

\section{Gaussian states and processes}
\label{sec:Gaussian}

Gaussian states and processes have been defined in the previous section, the former as the states having a Gaussian phase space representation and the latter as the processes mapping Gaussian states to Gaussian states. Ubiquitous in quantum physics, they are well understood theoretically~\cite{ferraro2005gaussian,weedbrook2012gaussian,adesso2014continuous} and routinely implemented experimentally~\cite{gerd2007quantum}.

\medskip

\noindent We review Gaussian processes and states in the following sections, restricting to pure states, unitary operations and projectors.

\subsection{Gaussian unitary operations}

The displacement operator of amplitude $\alpha\in\mathbb C$ has been introduced in the previous section and reads
\be
\hat D(\alpha)=e^{\alpha\hat a^\dag-\alpha^*\hat a}.
\ee
It satisfies the relations
\be
\ba
\,&\hat D^\dag(\alpha)=\hat D(-\alpha),\\
\,&\hat D(\alpha)\,\hat a\hat D^\dag(\alpha)=\hat a-\alpha\mathbb 1,\\
\,&\hat D(\alpha)\,\hat a^\dag\hat D^\dag(\alpha)=\hat a^\dag-\alpha^*\mathbb 1,\\
\,&\hat D(\alpha)\hat D(\beta)=e^{\frac12(\alpha\beta^*-\alpha^*\beta)}\hat D(\alpha+\beta),
\ea
\ee
for all $\alpha,\beta\in\mathbb C$. We denote a tensor product of $m$ single-mode displacements by $\hat D(\bm\alpha)=\bigotimes_{i=1}^m{\hat D(\alpha_i)}$ for all $\bm\alpha=(\alpha_1,\dots,\alpha_m)\in\mathbb C$.

The squeezing operator is defined as
\be
\hat S(\xi)=e^{\frac12(\xi\hat a^2-\xi^*\hat a^{\dag2})},
\ee
for all $\xi\in\mathbb C$. The parameter $\xi$ is called squeezing parameter. The squeezing operator satisfies the relations
\be
\ba
\,&\hat S^\dag(\xi)=\hat S(-\xi),\\
\,&\hat S(\xi)\,\hat a\hat S^\dag(\xi)=\cosh r\,\hat a+e^{-i\theta}\sinh r\,\hat a^\dag,\\
\,&\hat S(\xi)\,\hat a^\dag\hat S^\dag(\xi)=\cosh r\,\hat a^\dag+e^{i\theta}\sinh r\,\hat a,
\ea
\ee
for all $\xi=re^{i\theta}\in\mathbb C$. We denote a tensor product of $m$ single-mode squeezings by $\hat S(\bm\xi)=\bigotimes_{i=1}^m{\hat S(\xi_i)}$ for all $\bm\xi=(\xi_1,\dots,\xi_m)\in\mathbb C$.

The displacement and squeezing operators may be conceived as acting on a state by displacing and squeezing its phase space representation, respectively, as their name indicates. This geometrical intuition holds in particular for the Wigner quasiprobability distribution.

Any single-mode Gaussian unitary operation may be written as a squeezing and a displacement operator.
The ordering is only a convention, since the displacement and squeezing operators satisfy the braiding relation~\cite{nieto1997holstein}
\be
\hat D(\alpha)\hat S(\xi)=\hat S(\xi)\hat D(\gamma),\quad\quad\gamma=\alpha\cosh r+\alpha^*e^{-i\theta}\sinh r,
\ee
for all $\alpha\in\mathbb C$ and all $\xi=re^{i\theta}\in\mathbb C$. 

Passive linear transformation over $m$ modes are defined as the unitary transformations $\hat U$ which act unitarily on the creation operators of the modes $\hat a^\dag_1,\dots,\hat a^\dag_m$ as well as on the annihilation operators $\hat a,\dots,\hat a_m$. Any such transformation $\hat U$ is associated to an $m\times m$ unitary matrix $U$ which transforms the creation operators of the modes as
\be
\begin{pmatrix}\hat a^\dag_1\\ \vdots\\ \hat a^\dag_m\end{pmatrix}\rightarrow U\begin{pmatrix}\hat a^\dag_1\\ \vdots\\ \hat a^\dag_m\end{pmatrix},
\ee
and the annihilation operators of the modes as
\be
\begin{pmatrix}\hat a_1\\ \vdots\\ \hat a_m\end{pmatrix}\rightarrow U^*\begin{pmatrix}\hat a_1\\ \vdots\\ \hat a_m\end{pmatrix}.
\ee
These transformations map the multimode vacuum state onto itself.

Finally, Gaussian projectors are identified with projections onto Gaussian pure states, which we review in what follows.

\subsection{Single-mode Gaussian pure states}

General single-mode Gaussian pure states are obtained from the vacuum with a Gaussian unitary operation. They are the squeezed coherent states (or alternatively the displaced squeezed vacuum states):
\be
\hat S(\xi)\hat D(\alpha)\ket0,
\ee
for $\alpha,\xi\in\mathbb C$. Setting $\xi=0$ we obtain a coherent state of amplitude $\alpha\in\mathbb C$, while setting $\alpha=0$ we obtain a squeezed vacuum state with squeezing parameter $\xi\in\mathbb C$. 

\begin{figure}[h!]
	\begin{center}
		\includegraphics[width=0.4\columnwidth]{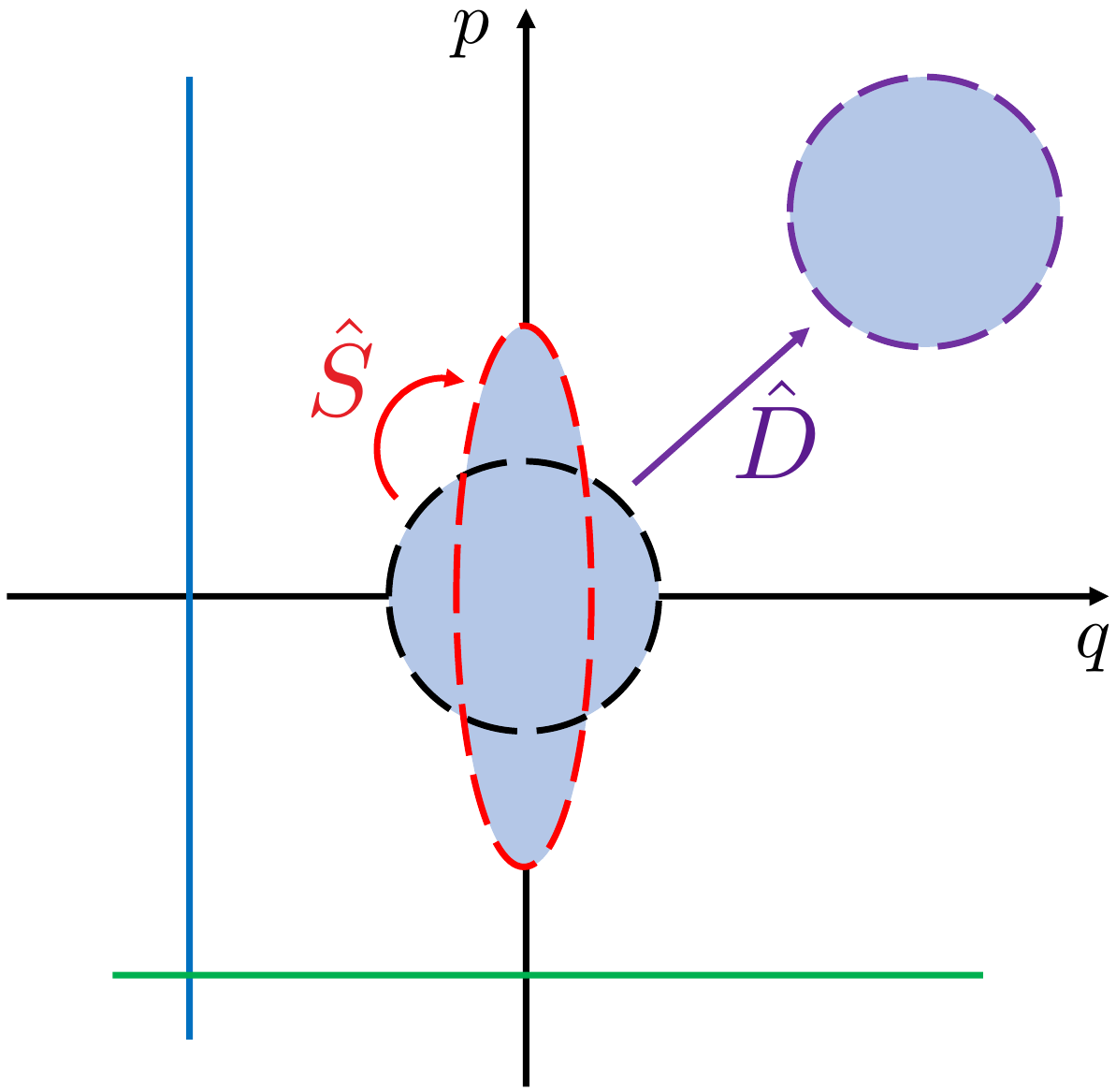}
		\caption{A pictorial representation of Gaussian states and processes in phase space. Circles are normalised Gaussian probability distributions---coherent states---viewed from the top and the ellipse represents a squeezed Gaussian probability distribution---a squeezed state. The vertical blue line and the horizontal green line correspond to position and momentum eigenstates, respectively.}
		\label{fig:phasespace}
	\end{center}
\end{figure}

The phase space representation of a coherent state is a Gaussian displaced in phase space, while the phase space representation of a squeezed vacuum state is a Gaussian centered at $0$, squeezed in a direction depending on the phase of the squeezing parameter. The strength of the squeezing depends on the modulus of the squeezing parameter (Fig.~\ref{fig:phasespace}). 
In particular, position and momentum eigenstates can be conceived formally as infinitely squeezed vacuum states, displaced by a finite amplitude~\cite{soto2013harmonic}.\\

\subsection{Multimode case: the symplectic formalism}
\label{sec:symplectic}

We present a short introduction to the symplectic formalism and refer to~\cite{adesso2014continuous} for a detailed exposition.

\medskip

\noindent Any $m$-mode Gaussian state $\bm\rho$ can be described by a $2m\times 2m$ covariance matrix $\bm V^{\mathbb R}$ containing its second canonical moments and a displacement vector $\bm d^{\mathbb R}$ of size $m$ containing its first canonical moments. The coefficients of the covariance matrix are defined, for $k,l\in\{1,\ldots,2m\}$, by 
$V_{kl}^{\mathbb R}=\frac{1}{2}\left<R_kR_l+R_lR_k\right>_\rho-\left<R_k\right>_\rho\left<R_l\right>_\rho$ where $\bm R=(\hat q _1,\dots,\hat q _m,\hat p _1,\dots,\hat p _m)$. The coefficients of the displacement vector are given by $d_j^{\mathbb R}=\left<R_j\right>_\rho$ for all $j\in\{1,\dots,2m\}$. Alternatively and more conveniently, one can describe covariance matrices and displacement vectors in the complex basis
$\bm\lambda=(\hat a_1,\dots,\hat a_m,\hat a_1^\dag,\dots,\hat a_m^\dag)$.
We write $\bm V$ and $\tilde{\bm d}$ the covariance matrix and displacement vector in that basis, with
\be
\bm V=\Omega\bm V^{\mathbb R}\Omega^\dag,\quad\quad\tilde{\bm d}=\Omega\,\bm d^{\mathbb R},
\ee
where
\be
\Omega=\frac1{\sqrt{ 2 }}\begin{pmatrix} \mathbb1_m & i\mathbb1_m \\ \mathbb1_m & -i\mathbb1_m \end{pmatrix}.
\label{omega}
\ee
The complex covariance matrix has the structure
\be
\bm V=\begin{pmatrix}
A&B\\B^*&A^*
\end{pmatrix},
\ee
with $A=A^\dag$ and $B=B^T$, so that $\bm V^\dag=\bm V$. The displacement vector has the structure
\be
\tilde{\bm d}=\begin{pmatrix}
\bm d\\ \bm d^*
\end{pmatrix}.
\ee
We will also refer to the above vector $\bm d$ as the displacement vector.

Gaussian multimode unitary operations are generated by Hamiltonians that are at most quadratic in the annihilation and creation operators of the modes. As a consequence, they induce affine transformations of the annihilation and creation operators which preserve their canonical commutation relations, i.e., symplectic linear transfomations, together with displacements. 
The evolution of a Gaussian state during a Gaussian evolution (excluding displacements) is described by a complex symplectic transformation of its complex covariance matrix and its complex displacement vector:
\begin{equation}
(\bm V,\tilde{\bm d})\rightarrow(S\bm VS^\dag,S\tilde{\bm d}),
\end{equation}
where a complex symplectic matrix $S$ satisfies 
\begin{equation}
S\Omega J\Omega^\dag S^{ \dag }=\Omega J\Omega^\dag,
\end{equation}
where $J=\begin{pmatrix} \mymathbb0_m & \mathbb1_m \\ -\mathbb1_m & \mymathbb0_m \end{pmatrix}$ and where the matrix $\Omega$ is defined in Eq.~(\ref{omega}). We will use the notations
\begin{equation}
S_{\bm\xi}\equiv\begin{pmatrix} D_c(\bm\xi) & D_s(\bm\xi) \\ D_s(\bm\xi) & D_c(\bm\xi) \end{pmatrix}  
\end{equation}
for all $\bm\xi=(\xi_1,\dots,\xi_m)\in\mathbb C^m$, with $D_c(\bm\xi)=\text{Diag}(c_{\xi_1},\ldots,c_{\xi_m})$ and $D_s(\bm\xi)=\text{Diag}(s_{\xi_1},\ldots,s_{\xi_m})$, where $c_\chi=\cosh\chi$ and $s_\chi=\sinh\chi$ for the symplectic matrices that implement squeezing and 
\begin{equation}
S_U\equiv\begin{pmatrix} U^* & \mymathbb0_m \\ \mymathbb0_m & U \end{pmatrix}
\end{equation}
for the symplectic matrix associated with a passive linear transformations with $m\times m$ unitary matrix $U$. A displacement does not affect the covariance matrix and only translates the displacement vector.

The so-called Bloch-Messiah or Euler decomposition implies that any $2m\times 2m$ complex symplectic matrix can be written as $S_US_{\bm\xi}S_V$ for some $m\times m$ unitary matrices $U$ and $V$ and some squeezing parameters $\bm\xi=(\xi_1,\dots,\xi_m)\in\mathbb C^m$.
In particular, any multimode Gaussian unitary operation can be decomposed as a passive linear transformation followed by a product of single-mode squeezings, followed by another passive linear transformation, together with single-mode displacements. 

Since any multimode Gaussian pure quantum state may be engineered from the vacuum with a Gaussian unitary operation, by virtue of Williamson decomposition, and since the vacuum is mapped onto itself by passive linear transformations, this means that any multimode Gaussian pure quantum state can be written as a tensor product of single-mode Gaussian states (displaced squeezed vacuum states) followed by a single passive linear transformation.

\section{Linear optics}
\label{sec:LO}

Linear optics covers the manipulation of light by unitary transformations whose exponent is at most quadratic in the field operator~\cite{walls2007quantum}, i.e., Gaussian unitaries. It induces transformations of quantum states of light which are divided in two categories, passive and active transformations, depending on whether these transformations change the total number of photons of the input state. In what follows, we review a few examples of quantum states of light and quantum optical measurements, and we detail passive linear optical transformations, implemented by unitary interferometers, with the examples of the Hong-Ou-Mandel effect~\cite{hong1987measurement} and its generalisation Boson Sampling~\cite{Aaronson2013}.

\subsection{Quantum states of light}

We briefly list single-mode quantum states that are common in the literature, some of which were already introduced in the previous sections, and which we will encounter in the following chapters.

\begin{itemize}
\item
Photon-number states: these states form the orthonormal Fock basis and are obtained from the vacuum as
\be
\ket n=\frac{(\hat a^\dag)^n}{\sqrt{n!}}\ket0,
\ee
for all $n\in\mathbb N$. Taking $n=0$ gives the vacuum state and photon-number states are non-Gaussian for $n>0$. They are the eigenstates of the photon-number operator $\hat n=\hat a^\dag\hat a$: for all $n\in\mathbb N$, $\hat n\ket n=n\ket n$.
\item
Coherent states: these Gaussian states are expressed as
\be
\ket\alpha=e^{-\frac12|\alpha|^2}\sum_{n\ge0}{\frac{\alpha^n}{\sqrt{n!}}\ket n},
\ee
for all $\alpha\in\mathbb C$. These states are a good approximation of the quantum state of a laser and are sometimes referred to as classical states, because their behaviour resembles that of a classical harmonic oscillator. They are the eigenstates of the annihilation operator $\hat a$: for all $\alpha\in\mathbb C$, $\hat a\ket\alpha=\alpha\ket\alpha$.
\item
Squeezed vacuum states: these Gaussian states are expressed as
\be
\ket\xi=\frac1{\sqrt{\cosh r}}\sum_{n\ge0}{(-e^{-i\theta}\tanh r)^n\frac{\sqrt{(2n)!}}{2^nn!}\ket{2n}},
\ee
for all $\xi=re^{i\theta}\in\mathbb C$. They display reduced variance for one quadrature, but increased variance for the conjugate quadrature, in accordance with the uncertainty principle.
\item
Photon-subtracted/added states: these states are obtained by applying the annihilation/creation operator to a state (and renormalising). These unphysical operations cannot be implemented determinisically and are implemented probabilistically in practice. For example, a photon subtraction may be implemented by mixing the input state with the vacuum on a beam splitter with near unity reflectance. Then, conditioned on a successful single-photon heralding of the transmitted light, the reflected state has been photon-subtracted.
\item
Cat states: named after Schrödinger's cat, these states are superpositions of two coherent states of equal amplitudes, $\ket\alpha$ and $\ket{-\alpha}$, like the cat in Schrödinger's thought experiment is in a superposition of two classical states, dead and alive. Varying the relative phase between the coherent states in the superposition gives different cat states. In particular, we introduce the cat$^+$ and cat$^-$ states:
\be
\ket{\text{cat}^\pm_\alpha}=\frac1{\sqrt{\mathcal N^\pm_\alpha}}(\ket\alpha\pm\ket{-\alpha}),
\ee
for all $\alpha\in\mathbb C$, where $\mathcal N^\pm_\alpha=2(1\pm e^{-2|\alpha|^2})$ is a normalisation factor.
\item
GKP states: finally, let us mention the Gottesman-Kitaev-Preskill (GKP) states which form a family of unphysical states with periodic wave functions~\cite{Gottesman2001}. These states are formal periodic superpositions of infinitely squeezed states and their physical approximations have applications for continuous variable quantum error correction~\cite{tzitrin2020progress}.
\end{itemize}

\subsection{Quantum optical measurements}
\label{sec:heterodynemeasurement}

We list various (idealised) single-mode measurements in what follows: homodyne detection, balanced heterodyne detection, unbalanced heterodyne detection, single-photon threshold detection, photon number parity detection and photon-number resolving detection. Detailed information on these detection methods can be found, e.g., in~\cite{ferraro2005gaussian}. We will only consider multimode detections that are tensor products of such single-mode detections.\\

\begin{figure}[h!]
	\begin{center}
		\includegraphics[width=2in]{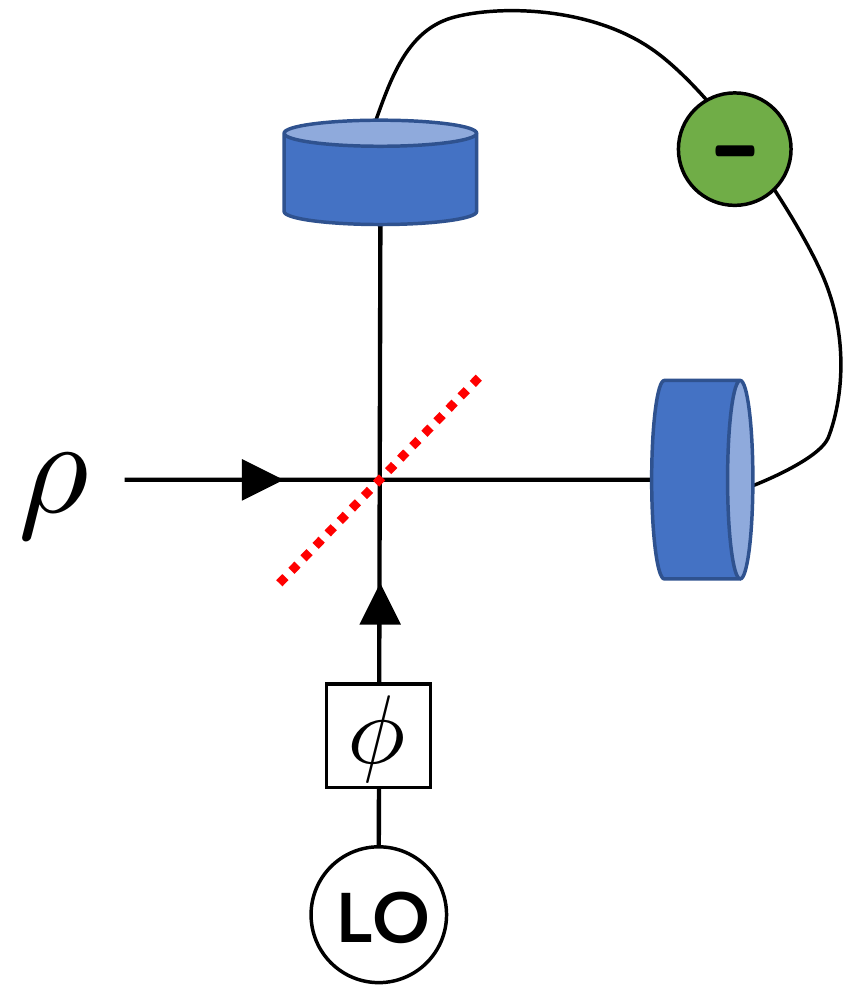}
		\caption{A schematic representation of homodyne detection of a state $\rho$. The dashed red line represents a balanced beamsplitter. LO stands for local oscillator, i.e., strong coherent state. The blue circles are photodiode detectors. Changing the phase $\phi$ of the local oscillator allows one to measure rotated quadratures.}
		\label{fig:homodyne}
	\end{center}
\end{figure}

\noindent Homodyne detection consists in a Gaussian measurement of a quadrature of the field, by mixing the state to be measured on a balanced beam splitter with a strongly excited coherent state, the local oscillator. Then, the intensities of both output arms are measured and their difference yields a value proportional to a quadrature of the input mode, rotated depending on the phase of the local oscillator (Fig.~\ref{fig:homodyne}). The POVM elements for homodyne detection with phase $\phi$ are given by
\be
\Pi_x^\phi=\ket x_\phi\!\bra x
\ee
for all $x\in\mathbb R$, where $\ket x_\phi$ is the eigenstate of the rotated quadrature operator $\hat x_\phi=\cos\phi\,\hat q+\sin\phi\,\hat p$ corresponding to the eigenvalue $x\in\mathbb R$.

\medskip

\noindent Balanced heterodyne detection, also called double homodyne or eight-port homodyne~\cite{ferraro2005gaussian}, consists in splitting the measured state with a balanced beam splitter and measuring both ends with homodyne detection. This corresponds to a joint noisy measurement of quadratures $\hat q$ and $\hat p$. 
This is a Gaussian measurement which yields two real outcomes, corresponding to the real and imaginary parts of $\alpha\in\mathbb C$. The POVM elements for balanced heterodyne detection are given by
\be
\Pi_\alpha=\frac1\pi\ket\alpha\!\bra\alpha,
\ee
for all $\alpha\in\mathbb C$, where $\ket\alpha$ is the coherent state of amplitude $\alpha\in\mathbb C$. Measuring a state with balanced heterodyne detection effectively amounts to sampling from its $Q$ function. \\
\begin{figure}[h!]
\begin{center}
\includegraphics[width=2.7in]{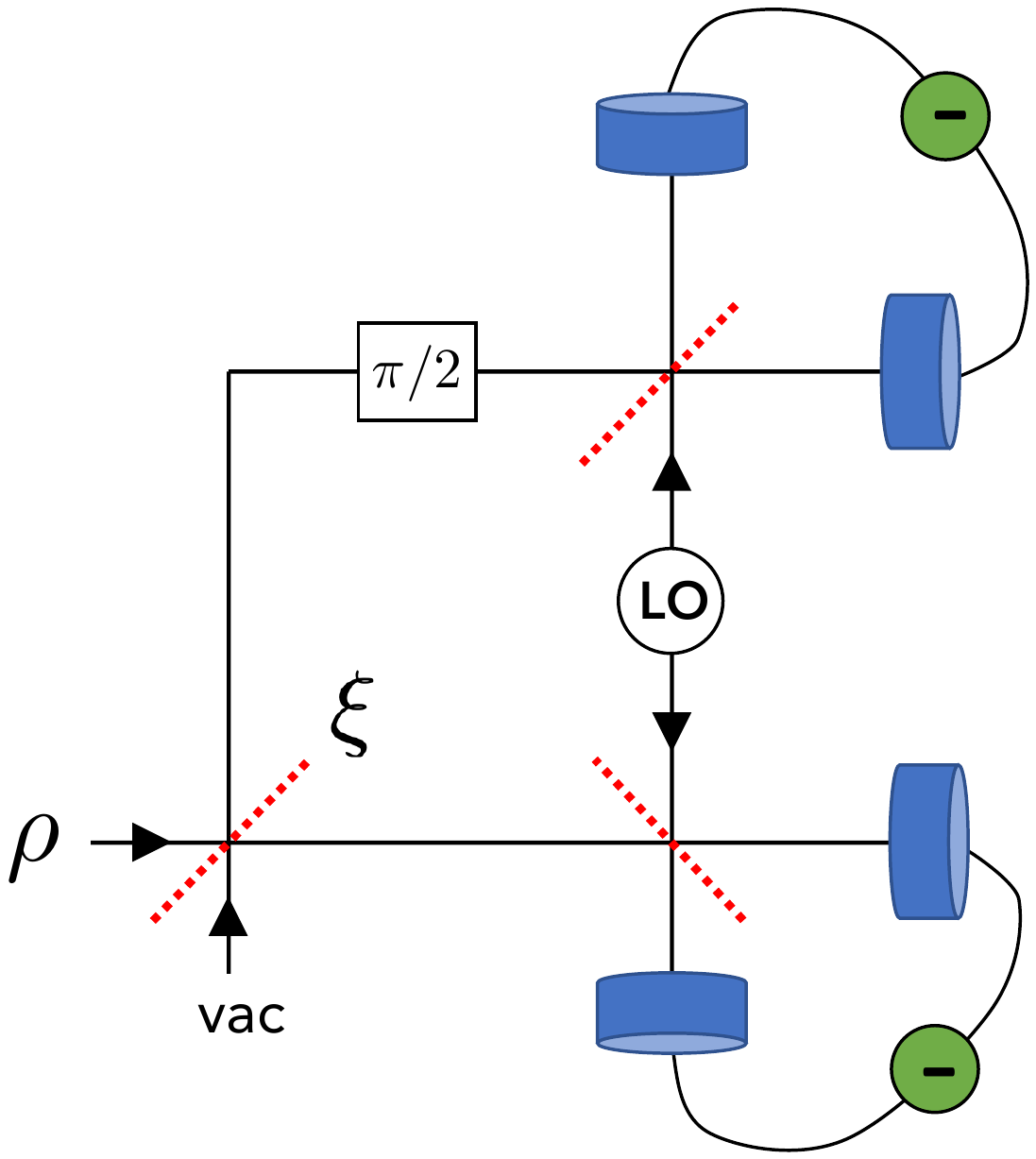}
\caption{Schematic representation of unbalanced heterodyne detection with unbalancing parameter $\xi=re^{i\theta}\in\mathbb C$. LO stands for local oscillator, i.e., strong coherent state. The blue circles are photodiode detectors. The $ \hat q$ and $\hat p$ measurements are each performed by balanced homodyne detection.}
\label{fig:2homodyne}
\end{center}
\end{figure}

\noindent A straightforward generalisation is unbalanced heterodyne detection (Fig.~\ref{fig:2homodyne}), where the input beam splitter is no longer balanced but characterized instead by a reflectance $R$ and a transmittance $T$, with $R^2+T^2=1$. The POVM elements for unbalanced heterodyne detection with unbalancing parameter $\xi\in\mathbb C$ are given by
\be
\Pi_\alpha^\xi=\frac1\pi\ket{\alpha,\xi}\!\bra{\alpha,\xi},
\ee
for all $\alpha\in\mathbb C$, where $\ket{\alpha,\xi}=\hat S(\xi)\hat D(\alpha)\ket0$ is a squeezed coherent state. Writing $\xi=re^{i\theta}$, the unbalalancing parameter is related to the optical setup by $r=\left|\log\left(\frac TR\right)\right|$, with $\theta$ being the phase of the local oscillator~\cite{chabaud2017continuous}. Measuring a state with unbalanced heterodyne detection effectively amounts to sampling from its squeezed $Q$ function. Setting $\xi=0$ gives balanced heterodyne detection, while sending $|\xi|=r$ to infinity gives homodyne detection. Any Gaussian measurement can thus be implemented by Gaussian unitary operations and heterodyne detection only, since it can be implemented by Gaussian unitary operations and homodyne detection only~\cite{giedke2002characterization,eisert2003introduction}.

\medskip

\noindent Additionnally, we introduce three non-Gaussian measurements, each giving more information about the photon number of the measured state. 
The first is single-photon threshold detection~\cite{H:NP09}, or simply threshold detection, whose POVM elements are given by
\be
\Pi_0=\ket0\!\bra0,\quad\Pi_1=\mathbb1-\ket0\!\bra0.
\ee
This binary measurement only distinguishes the vacuum state from other states. 
The second is photon number parity detection~\cite{haroche2007measuring}, or simply parity detection, whose POVM elements are given by
\be
\Pi_+=\sum_{n\ge0}{\ket{2n}\!\bra{2n}},\quad\Pi_-=\sum_{n\ge0}{\ket{2n+1}\!\bra{2n+1}}.
\ee
This is a binary measurement of the parity operator $\hat\Pi=(-1)^{\hat a^\dag\hat a}$ yielding, as its name indicates, the parity of the number of photons of the measured state.
The third is photon number-resolving detection~\cite{divochiy2008superconducting}, whose POVM elements are given by
\be
\Pi_n=\ket n\!\bra n,
\ee
for all $n\in\mathbb N$, i.e., projections onto Fock states.

\subsection{Linear interferometers}
\label{sec:linearinter}

Linear optical unitary interferometers are composed of beam splitters and phase shifters and implement passive linear transformations of the modes. In particular, any passive linear transformation $\hat U$ over $m$ modes with $m\times m$ unitary matrix $U$ can be implemented by a linear interferometer with at most $\frac{m(m-1)}2$ balanced beam splitters and $m$ phase shifters~\cite{reck1994experimental}.
The corresponding unitary interferometer is described by the same unitary matrix $U=(u_{ij})_{1\le i,j\le m}$. Unlike in the circuit picture, the matrix $U$ does not act on the computational basis, which in this case is the infinite multimode Fock basis, but rather describes the linear evolution of the creation operator of each mode. More precisely, 
\be
\begin{pmatrix} \hat a_1^\dag \\ \vdots\\ \hat a_m^\dag \end{pmatrix}\to U\begin{pmatrix} \hat a_1^\dag \\ \vdots\\ \hat a_m^\dag\end{pmatrix}=\begin{pmatrix} \sum_{k=1}^m{u_{1k}\hat a_k^\dag} \\ \vdots\\ \sum_{k=1}^m{u_{mk}\hat a_m^\dag}\end{pmatrix}.
\label{evocrea}
\ee
In that picture, the direct sum plays the role of the tensor product in the computational basis: taking the direct sum of two unitaries corresponds to putting linear optical elements in parallel, while multiplying unitaries corresponds to putting linear optical elements in sequence.

Multimode coherent states have a specific evolution through linear interferometers: they are mapped onto coherent states and do not become entangled, unlike other states. If $U$ is the unitary matrix describing an interferometer which implements a passive linear transformation $\hat U$, an input coherent state $\ket{\bm\alpha}$ is mapped to an output coherent state $\hat U\ket{\bm\alpha}=\ket{U\bm\alpha}$, where the vector of output amplitudes $U\bm\alpha$ is obtained by multiplying the vector of input amplitudes $\bm\alpha$ by the unitary matrix $U$.

Remarkable quantum effects may be witnessed when the input to linear optical unitary interferometers are single-photon Fock states instead of coherent states. The celebrated Knill--Laflamme--Milburn scheme~\cite{knill2001scheme} shows that single photons and linear optics are enough to achieve universal quantum computing together with adaptive measurements (making the rest of the computation depend on the result of intermediate measurements). Already without adaptive measurements, interesting effects can be observed. We give two notable examples in the following sections: the Hong--Ou--Mandel effect and Boson Sampling.

\subsection{Hong--Ou--Mandel effect}
\label{sec:HOM}

\begin{figure}[h!]
\begin{center}
\includegraphics[width=2.7in]{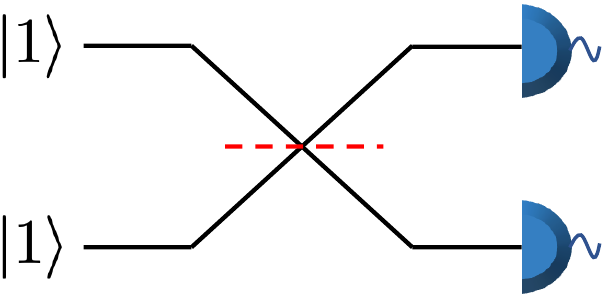}
\caption{Hong-Ou-Mandel effect. The dashed red line represents a balanced beam splitter. The number of photons is detected for both output arms. If the input single photons are indistinguishable, the outcomes $(20)$ and $(02)$ occur with the same probability $\frac12$ and the outcome $(11)$ never occurs.}
\label{fig:HOM}
\end{center}
\end{figure}

The Hong--Ou--Mandel effect, or photon bunching, refers to the bosonic behaviour of indistinguishable photons which bunch together when mixed on a balanced beamsplitter (Fig.~\ref{fig:HOM}). A balanced beam splitter is a unitary interferometer over two modes, with unitary matrix
\be
H=\frac1{\sqrt2}\begin{pmatrix}1&1\\1&-1\end{pmatrix}.
\ee

\noindent The input state is composed of two single photons, one in each mode. Labelling the modes $u$ and $d$, for `up' and `down', let $\hat a^\dag_u$, $\hat a^\dag_d$ and $\hat b^\dag_u$, $\hat b^\dag_d$ be the creation operators of the input and output modes, respectively. The balanced beam splitter acts on the input creation operators as
\be
\begin{pmatrix}\hat b^\dag_u\\ \hat b^\dag_d\end{pmatrix}=H\begin{pmatrix}\hat a^\dag_u\\ \hat a^\dag_d\end{pmatrix}.
\ee
The input state thus evolves as
\be
\ba
\ket{11}&=\hat a_u^\dag\hat a_d^\dag\ket{00}\\
&\overset H\rightarrow\frac12(\hat b_u^\dag+\hat b_d^\dag)(\hat b_u^\dag-\hat b_d^\dag)\ket{00}\\
&=\frac12(\hat b_u^{\dag2}-\hat b_d^{\dag2})\ket{00}\\
&=\frac1{\sqrt2}(\ket{20}-\ket{02}),
\ea
\ee
where we used $\hat b_u^\dag\hat b_d^\dag=\hat b_d^\dag\hat b_u^\dag$. In particular, measuring the photon number in both output modes will always yield $0$ for one of the modes: the outcome $(11)$  is never witnessed if the photons are indistinguishable, i.e., the photons have bunched together.\\

\subsection{Boson Sampling}
\label{sec:BosonSampling}

\begin{figure}[h!]
\begin{center}
\includegraphics[width=2.9in]{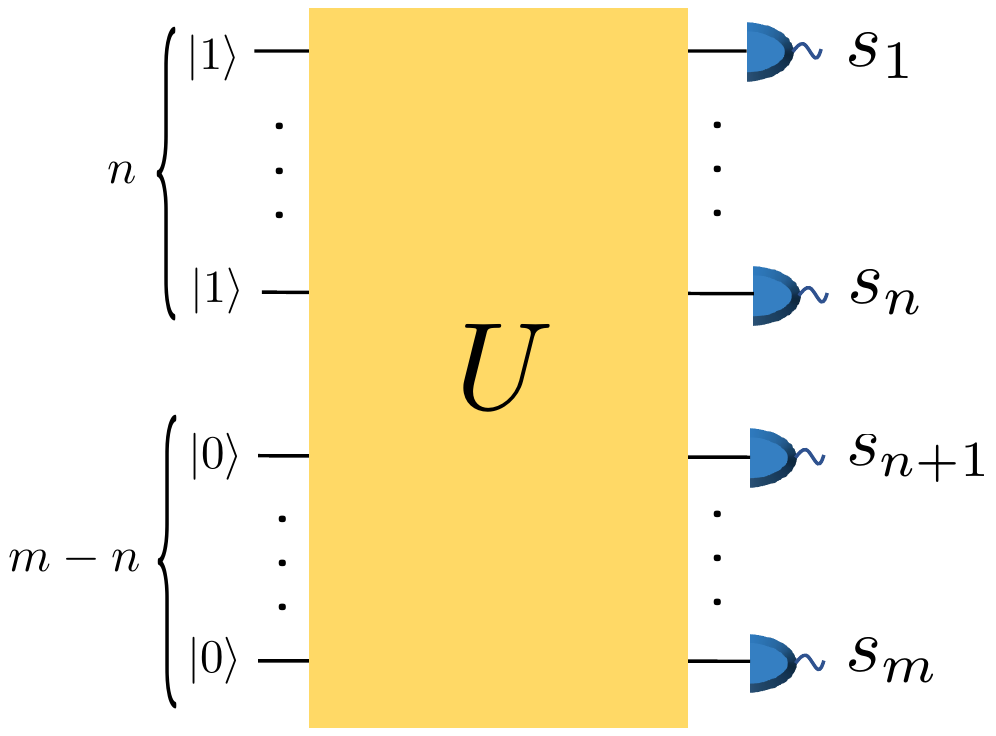}
\caption{BosonSampling with $n$ photons over $m$ modes. The outcomes $s_1,\dots,s_m$ denote the measured photon number for each mode.}
\label{fig:BosonSampling}
\end{center}
\end{figure}

Let $m\in\mathbb N^*$ and $n\in\mathbb N$, with $m\ge n$. Boson Sampling, introduced in~\cite{Aaronson2013}, is a generalisation of the Hong--Ou--Mandel setup, where the balanced beam splitter is replaced by a general unitary interferomer over $m$ modes with $m\times m$ unitary matrix $U$ and the input is composed of $n$ single photons in the first $n$ modes and vacuum in the remaining $m-n$ modes, the photon number of all output modes being measured (Fig.~\ref{fig:BosonSampling}). 

Even though Boson Sampling has been formulated for general bosonic particles, linear optics provides a convenient way of looking at it. Boson Sampling is a subuniversal model of quantum computation, believed to be hard to simulate by classical computers while not possessing the computational power of a universal quantum computer. We review this model in what follows and we refer to~\cite{Aaronson2013} for a detailed version of the material presented in this section. In particular, we do not discuss the theoretical use of postselection.

We denote photon number states over $m$ modes by
\be
\ket{\bm s}=\ket{s_1\dots s_m}=\frac{(\hat a_1^\dag)^{s_1}}{\sqrt{s_1!}}\cdots\frac{(\hat a_m^\dag)^{s_m}}{\sqrt{s_m!}}\ket0^{\otimes m},
\label{Fockm}
\ee
where $s_k$ and $\hat a_k^\dag$ are respectively the number of photons and the creation operator of the $k^{th}$ mode. We identify these states with $m$-tuples of integers $\bm s=(s_1,\dots,s_m)\in\mathbb N^m$ (see section~\ref{sec:multiindex} for multi-index notations). The input state with $n$ single photons in the first $n$ modes and vacuum in the other modes is denoted $\ket{\bm t}$, with $\bm t=(\bm1^n,\bm0^{m-n})$. We introduce, 
\be
\Phi_{m,n}:=\{\bm s\in\mathbb N^m,\text{ }|\bm s|=n\}.
\ee
This set corresponds to the $m$-mode Fock states with total number of photons equal to $n$. We have $|\Phi_{m,n}|=\binom{m+n-1}{n}$ and $\bm t\in\Phi_{m,n}$. \\

\noindent We consider a unitary interferometer of size $m$, described by an $m\times m$ unitary matrix $U=(u_{ij})_{1\le i,j\le m}$ acting on the creation and annihilation operators of the modes as in Eq.~(\ref{evocrea}). 
We write $\hat U$ the unitary action of the interferometer on the multimode Fock basis. Its entries are indexed by elements of $\Phi_{m,n}$, for all $n\in\mathbb N$.
Because the interferometer conserves the total number of photons, for all $p,q\in\mathbb N$, all $\bm s\in\Phi_{m,p}$ and all $\bm s'\in\Phi_{m,q}$,
\be
\braket{\bm s|\hat U|\bm s'}=0
\ee
whenever $p\neq q$. In particular, it may be written as the direct sum of its action on the various fixed energy subspaces. We write
\be
\hat U=\bigoplus_{n=1}^{+\infty}{\,\hat U_n},
\ee
where $\hat U_n$ is the $|\Phi_{m,n}|\times|\Phi_{m,n}|$ unitary submatrix of $\hat U$ obtained by only keeping the rows $\bm s$ and the columns $\bm s'$ for all $\bm s,\bm s'\in\Phi_{m,n}$. We have $\hat U_0=\begin{pmatrix}1\end{pmatrix}$, and $\hat U_1=U$ up to a reordering of the basis states.

Let $n\in\mathbb{N}$ and $\bm s,\bm t\in\Phi_{m,n}$. Combining Eq.~(\ref{evocrea}) and Eq.~(\ref{Fockm}) we obtain~\cite{Aaronson2013}
\be
\braket{\bm s|\hat U|\bm t}=\frac{\Per(U_{\bm s,\bm t})}{\sqrt{\bm s!\bm t!}},
\label{transf}
\ee
where $U_{\bm s,\bm t}$ is the $n\times n$ matrix obtained from $U$ by repeating $s_i$ times its $i^{th}$ row and $t_j$ times its $j^{th}$ column for $i,j=1,\dots,m$, and where the permanent of an $n\times n$ matrix $A=(a_{ij})_{1\le i,j\le n}$ is defined as
\be
\Per A=\sum_{\sigma\in S_n}{\prod_{i=1}^n{a_{i\sigma(i)}}},
\label{per}
\ee
where the sum is over the permutations of the set $\{1,\dots,n\}$.

We write $\Pr_{m,n}[.|\bm t]$ the probability distribution of the outputs over $\Phi_{m,n}$ of the unitary interferometer $U$ acting on the input $\ket{\bm t}$. With the previous notations we obtain, for all $\bm s,\bm t\in\Phi_{m,n}$,
\be
\text{Pr}_{m,n}[\bm s|\bm t]=\frac{\left|\Per\left(U_{\bm s,\bm t}\right)\right|^2}{\bm s!\bm t!}.
\label{proba}
\ee
With $\bm t=(\bm1^n,\bm0^{m-n})$ we have $\bm t!=1$ and thus
\be
\text{Pr}_{m,n}[\bm s|\bm t]=\frac1{\bm s!}\left|\Per\left(U_{\bm s,\bm t}\right)\right|^2.
\label{proba}
\ee
The output photon-number distribution of a Boson Sampling interferometer with $n$ photons over $m$ modes thus is related to the modulus squared of the permanent of an $n\times n$ matrix with complex entries. This matrix is obtained from the unitary matrix $U$ describing the interferometer by discarding its last $m-n$ columns and repeating its lines according to the detection pattern $\bm s$.

\medskip

\noindent The permanent defined in Eq.~(\ref{per}) is a `hard' quantity to compute. In order to appreciate this hardness, let us take a brief and informal detour through the realm of complexity theory~\cite{manson2001simplifying}. A formal introduction to the complexity classes presented here is given in~\cite{Aaronson2013}.

A complexity class is a set of computational problems. These problems may be of different types: in particular, a decision problem is a problem with yes or no answers, a function problem is a problem with more general answers (e.g., natural, real or complex numbers), and a sampling problem consists in outputting samples from a target probability distribution, either exactly or approximately.

In the language of complexity theory, an efficient algorithm is an algorithm which takes a number of steps which is polynomial in the size of its input (its number of bits), and the generic model for a classical computer is a deterministic Turing machine.

Given a complexity class \textsf{C}, a problem $p$ is said to be \textsf{C}-hard if any problem in \textsf{C} can be rephrased efficiently as an instance of the problem $p$. Roughly speaking, this means that the problem $p$ is harder than any of the problems in \textsf{C}. If the problem $p$ is also in \textsf{C}, it is referred to as \textsf{C}-complete.

The class of decision problems that can be solved efficiently by a classical computer is denoted \textsf{P}. 
The class of decision problems whose solution can be verified efficiently by a classical computer is denoted \textsf{NP}.
A great open problem in complexity theory is whether these two complexity classes are equal or if $\textsf{P}\neq\textsf{NP}$, the latter being widely believed. 

An oracle for a given computational problem is a black box which is able to produce a solution for any instance of this problem. An oracle for a complexity class is a black box which, given any problem in the complexity class, is able to produce a solution for any instance of this problem. The access to an oracle is denoted with an exponent. For example, a problem which can be solved efficiently when given access to an oracle for an \textsf{NP}-complete problem is in the class $\textsf P^{\textsf{NP}}$.

The polynomial hierarchy \textsf{PH} is a tower of complexity classes generalising \textsf{P} and \textsf{NP}. It can be defined inductively based on an oracle construction, where the level $0$ is \textsf P, the level $1$ contains \textsf{NP}, the level $2$ contains $\textsf{NP}^{\textsf{NP}}$, and so on. Each level is contained in the next one and if two consecutive levels are equal, then they are also equal to all of the above levels---we talk about a collapse of the polynomial hierarchy. The conjecture that the polynomial hierarchy does not collapse, i.e., that all levels within the hierarchy are distinct, is a stronger version of the $\textsf{P}\neq\textsf{NP}$ conjecture.

The class of decision problems that can be solved efficiently by a classical computer with access to a genuine random number source is denoted \textsf{BPP}. It lies at the second level of the polynomial hierarchy \textsf{PH}$_2$~\cite{lautemann1983bpp}.

The class of function problems which consist in counting the number of solutions of an \textsf{NP} problem is denoted \#\textsf P. Its equivalent complexity class of decision problems is denoted $\textsf P^{\#\textsf P}$ and by Toda's theorem~\cite{toda1991pp} we have $\textsf{PH}\subset\textsf P^{\#\textsf P}$, i.e., counting the solutions of \textsf{NP} problems is harder than any problem in the whole polynomial hierarchy of complexity classes.

\medskip

\noindent With these elements introduced, we are now in position to discuss the hardness of the permanent: computing exactly the permanent of matrices with $(0,1)$ entries is a \#\textsf P-complete problem~\cite{valiant1979complexity} and hence \textsf{PH}-hard. Moreover, approximating the permanent of real matrices up to multiplicative error, i.e., outputting an estimate $\tilde P$ such that $(1-1/\poly m)\,P\le\tilde P\le (1+/\poly m)\,P$ where $P$ is the permanent of a square matrix of size $m$ with real entries, is also \#\textsf P-hard~\cite{Aaronson2013}.

The computational problem `Boson Sampling' corresponds to the task of sampling from the output probability distribution in Eq.~(\ref{proba}), given the description $U$ of the Boson Sampling interferometer.

Making use of the hardness of the permanent and the connection between the output probabilities of a Boson Sampling interferometer and the permanent, two main results are derived in~\cite{Aaronson2013} about the hardness of classically solving two versions of the Boson Sampling problem, which we refer to as exact hardness and approximate hardness. 

Exact hardness corresponds to the following result: let $\mathcal O$ be an oracle which, given a unitary matrix $U$ and a random string as its unique source of randomness, samples exactly from the output probability distribution of the Boson Sampling interferometer $U$. Then $\textsf{PH}\subset\textsf{BPP}^{\textsf{NP}^{\mathcal O}}$. In particular, an efficient classical simulation of exact Boson Sampling collapses the polynomial hierarchy to its third level.

This result uses the fact that a single output probability of a Boson Sampling interferometer is hard to approximate up to multiplicative error and that being able to sample efficiently from a probability distribution allows one to obtain a multiplicative approximation of the probability of any outcome in $\textsf{FBPP}^{\textsf{NP}}$ (where \textsf{FBPP} is the class of function problems that can be solved efficiently using a \textsf{BPP} machine) thanks to Stockmeyer's approximate counting algorithm~\cite{stockmeyer1985approximation}. In that case, an oracle which samples from an exact Boson Sampling probability distribution is required.

On the other hand, approximate sampling refers to the task of sampling from a probability distribution which has a given constant total variation distance with a target distribution (see Eq.~(\ref{tvd})). Approximate hardness of Boson Sampling is more elaborate than exact hardness and relies on two plausible but unproven conjectures, even though the statement of the result is nearly identical: let $\mathcal O$ be an oracle which, given a unitary matrix $U$ and a random string as its unique source of randomness, samples approximately from the output probability distribution of the Boson Sampling interferometer $U$. Then $\textsf{PH}\subset\textsf{BPP}^{\textsf{NP}^{\mathcal O}}$. In particular, an efficient classical simulation of approximate Boson Sampling collapses the polynomial hierarchy to its third level.

Unlike for exact sampling, one cannot apply directly Stockmeyer's approximate counting algorithm in order to obtain multiplicative estimates of the probabilities of the target distribution. This is because the oracle now only outputs samples from an approximate probability distribution, i.e., a probability distribution which is very close to the correct one for most of the samples but not all samples. In the worst case, the probability that we are trying to estimate could be the probability of one of these `bad samples', and estimating this probability would merely give us a very bad estimate of the permanent, which is not hard to achieve. The trick to get around that problem is to hide the instance of the permanent that we wish to estimate into the probability of a random output of a Boson Sampling interferometer: given a classical machine which correctly performs the sampling for most of the samples, it would then correctly sample our instance with high probability. In the worst case, this effectively averages the constant total variation error over the sample space, allowing for a much more precise approximation of the permanent using Stockmeyer's algorithm.

This hiding procedure is based on the fact that small enough submatrices of random unitary matrices are very close to random complex Gaussian matrices. In order to restrict to matrices that do not have repeated lines, the so-called antibunching regime $n=O(\sqrt m)$ is chosen, which ensures a negligible probability of detecting more than one photon in the same output mode. The procedure outlined above then allows one to prove that the problem $|\text{GPE}|_\pm^2$ which consists in approximating up to additive error the square modulus of the permanent of random complex Gaussian matrices is in $\textsf{FBPP}^{\textsf{NP}^{\mathcal O}}$, where $\mathcal O$ is an oracle for approximate Boson Sampling.

The proof of approximate hardness then relies on two conjectures about the permanent of random complex Gaussian matrices in order to bridge the gap between additive approximations of the square modulus of the permanent of random complex Gaussian matrices and collapse of the polynomial hierarchy: the permanent of Gaussians conjecture and the permanent anti-concentration conjecture. The former conjecture states that the problem GPE$_\times$ which consists in approximating the permanent of random complex Gaussian matrices up to multiplicative error is \#\textsf P-hard. The latter conjecture states that with high probability the permanent of a randomly chosen complex Gaussian matrix is not too small. This implies in turn that the problem $|\text{GPE}|_\pm^2$ of additive approximation of the square modulus of the permanent of random complex Gaussian matrices is as hard as the problem GPE$_\times$ of multiplicative approximation of the permanent of random complex Gaussian matrices. With these two conjectures and the above argument, we obtain $\textsf{PH}\subset\textsf P^{\#\textsf P}\subset\text{GPE}_\times=|\text{GPE}|_\pm^2\subset\textsf{FBPP}^{\textsf{NP}^{\mathcal O}}$, which concludes the proof.

Assuming that the polynomial hierarchy does not collapse, Boson Sampling is hard to simulate exactly classically, and even approximately with additional mathematical conjectures. The approximate hardness of Boson Sampling is important since it opens the way for an experimental demonstration of quantum supremacy. Indeed, it is unrealistic to expect that an experimental Boson Sampling device would sample exactly from the ideal Boson Sampling distribution. Moreover, given the nature of the computational task at hand, i.e., outputting samples from a given probability distribution, there is no hope of being able to verify that an exact sampling has been performed. On the other hand, verifying that approximate Boson Sampling has been performed could be possible and indeed we derive such a verification protocol in chapter~\ref{chap:certif}.

\section{Segal--Bargmann formalism}
\label{sec:SegalBargmann}

The Segal--Bargmann formalism~\cite{bargmann1961hilbert,segal1963mathematical} associates to every quantum state an analytical function over the complex plane.
It has been used to study quantum chaos~\cite{leboeuf1990chaos,arranz1996distribution,korsch1997zeros,biswas1999distribution}, and the completeness of sequences of coherent states~\cite{perelomov1971completeness,bacry1975proof,boon1978discrete}. 
We give hereafter a quick introduction to this formalism. Further details may be found in chapter~\ref{chap:stellar} and in~\cite{vourdas2006analytic}.

\subsection{Definition}

We introduce below the analytical function, which we refer to as the stellar function. This function has been recently studied, in the context of non-Gaussian quantum state engineering~\cite{PhysRevA.99.053816}, in order to simplify calculations related to photon-subtracted Gaussian states.

\begin{defi}[Stellar function]
Let $\ket\psi=\sum_{n\ge0}{\psi_n\ket n}\in\mathcal H$ be a normalised state. The stellar function of the state $\ket\psi$ is defined as
\be
F_\psi^\star(z)=e^{\frac12|z|^2}\braket{z^*|\psi}=\sum_{n\ge0}{\psi_n\frac{z^n}{\sqrt{n!}}},
\ee
for all $z\in\mathbb C$, where $\ket z=e^{-\frac12|z|^2}\sum_{n\ge0}{\frac{z^n}{\sqrt{n!}}\ket n}$ is the coherent state of amplitude $z$.
\end{defi}

\noindent The stellar function is a holomorphic function over the complex plane, which provides an analytic representation of a quantum state.

\subsection{Properties of holomorphic functions}

A holomorphic function is a complex-valued function of one or more complex variables that is, at every point of its domain, complex differentiable in a neighbourhood of the point. As it turns out, the set of holomorphic functions is equal to the set of analytic functions, i.e., the functions that can be written as a convergent power series in a neighbourhood of each point of their domain. When their domain is the whole complex plane, they are called entire functions. In what follows we consider univariate entire functions.

These functions provide a natural extension of univariate complex polynomials and various properties of polynomials extend to entire functions. In particular, Liouville's theorem states that any bounded entire function is constant. The principle of permanence asserts that the zeros of an analytic function are isolated or this function is identically $0$. Furthermore, the number of zeros of an analytic function $f$ inside some contour is given by Cauchy's argument principle.

\begin{theo}[Cauchy's argument principle]\label{th:Cauchy}
Let $f$ be an analytic function and let $C$ be a contour in the domain of $f$. Then,
\be
Z_C(f)=\frac1{2i\pi}\oint_C{\frac{f'(z)}{f(z)}}dz,
\ee
where $Z_C(f)$ is the number of zeros of $f$ inside the contour $C$, counted with multiplicity.
\end{theo}

\noindent The growth of an analytic function is described by a pair of non-negative numbers $\rho,\sigma$ called the order and the type. They are defined as \cite{boas1954entire}
\be
\rho=\lim_{r\to+\infty}\sup\frac{\ln\ln M(r)}{\ln r},\quad\sigma=\lim_{r\to+\infty}\sup\frac{\ln M(r)}{r^\rho},
\ee
where $M(r)$ is the maximum value of the modulus of the funtion on the circle $|z|=r$. For polynomials, the growth is deeply related to the number of zeros---the degree. For entire functions, the growth is related to the density of zeros (see, e.g., \cite{stein2010complex} for more details). An entire function can also be factorized into a possibly infinite product involving its zeros, thanks to Weierstrass factorization theorem. For entire functions of finite order, this result is refined by Hadamard-Weierstrass factorization theorem.

\begin{theo}[Hadamard-Weierstrass factorization theorem]\label{th:Hadamard}
Let $f$ be an entire function of finite order $\rho$. Let $m\in\mathbb N$ be the multiplicity of $0$ as a root of $f$. Let $\{z_n\}n\in\mathbb N$ be the non-zero roots of $f$, counted with multiplicity. Then, there exist $p,q\in\mathbb N$, with $p,q\le\rho$, and a polynomial $P$ of degree $q$ such that, for all $z\in\mathbb C$,
\be
f(z)=z^me^{P(z)}\prod_{n=1}^{+\infty}{E_p\left(\frac z{z_n}\right)},
\ee
where
\be
E_p(z)=(1-z)\,e^{z+z^2/2+\dots+z^p/p}.
\ee
\end{theo}

\clearemptydoublepage
%
%

\addtocontents{toc}{\protect\pagebreak[4]}
	
\let\textcircled=\pgftextcircled
\chapter{Stellar representation of non-Gaussian quantum states}
\label{chap:stellar}

\initial{N}on-Gaussian states are crucial for a variety of quantum information tasks \cite{eisert2002distilling,fiuravsek2002gaussian,giedke2002characterization,wenger2003maximal,garcia2004proposal,ghose2007non,niset2009no,adesso2009optimal,barbosa2019continuous}. In particular, non-Gaussian states may be conceived as a resource for quantum computational advantage, Gaussian processes being classically simulable~\cite{Bartlett2002}. Hence, the characterisation of non-Gaussian states is of great importance and has attracted a lot of attention recently \cite{takagi2018convex,zhuang2018resource,albarelli2018resource,lami2018gaussian}. 

In this chapter, building on the Segal--Bargmann formalism, we introduce the stellar representation, which allows for the representation of the non-Gaussian properties of single-mode continuous variable quantum states by the distribution of the zeros of their Husimi $Q$ function in phase space. We use of this representation in order to derive an infinite hierarchy of single-mode states based on the number of zeros of the Husimi $Q$ function, the stellar hierarchy. We give an operational characterisation of the states in this hierarchy with the minimal number of single-photon additions needed to engineer them and derive equivalence classes under Gaussian unitary operations. We study in detail the topological properties of this hierarchy with respect to the trace norm, and discuss implications for the robustness of the states in the stellar hierarchy and for non-Gaussian state engineering.

This chapter is based on \cite{chabaud2020stellar,inprepaLKB}.


\section{The stellar function}
\label{sec:stellar1}

In continuous variable quantum information, quantum states are mathematically described by vectors in a separable Hilbert space of infinite dimension (see section \ref{sec:CVQI}). Alternatively, phase space formalism allows us to describe quantum states conveniently using generalised quasi-probability distributions~\cite{cahill1969density}, among which are the Husimi $Q$ function, the Wigner $W$ function, and the Glauber--Sudarshan $P$ function (see section \ref{sec:phasespace}). The states that have a Gaussian Wigner or Husimi function are called Gaussian states, while all the other states are called non-Gaussian. By extension, the operations mapping Gaussian states to Gaussian states are called Gaussian operations, and measurements projecting onto Gaussian states are called Gaussian measurements (see section \ref{sec:Gaussian}).

Hudson~\cite{hudson1974wigner} has notably shown that a single-mode pure quantum state is non-Gaussian if and only if its Wigner function has negative values and this result has been generalised to multimode states by Soto and Claverie~\cite{soto1983wigner}. 
This characterization is an interesting starting point for studying non-Gaussian states. From this result, one can introduce measures of a state being non-Gaussian using Wigner negativity, e.g., the negative volume~\cite{kenfack2004negativity}, that are invariant under Gaussian operations.
However, computing these quantities from experimental data is complicated in practice. 
Other measures and witnesses for non-Gaussian states have been derived~\cite{genoni2007measure,filip2011detecting,genoni2013detecting,hughes2014quantum}, which allow us to discriminate non-Gaussian states from mixtures of Gaussian states from experimental data, but they do not address the structure of non-Gaussian states and answer the question \textit{how much?} rather than \textit{how?}.

In order to adress the latter question, we will make use of another characterization of Gaussian states: the Wigner function having negative values is actually equivalent to the Husimi function having zeros, as shown by L\"utkenhaus and Barnett~\cite{lutkenhaus1995nonclassical}. Informally,

\begin{theo} \label{Husimi}
A pure quantum state is non-Gaussian if and only if its Husimi $Q$ function has zeros.
\end{theo}

\noindent Since the values of the $Q$ function are the overlaps with coherent states, this result may be understood as follows: a pure quantum state is non-Gaussian if and only if it is orthogonal to at least one coherent state. 

An interesting point is that for single-mode states, the zeros of the Husimi $Q$ function form a discrete set, as we will show in the next section. The non-Gaussian properties of single-mode states may thus be described by the distribution of these zeros in phase space. Based on this observation, we classify single-mode continuous variable quantum states with respect to their non-Gaussian properties in the following sections, using the so-called stellar representation, or Segal--Bargmann formalism (see section~\ref{sec:SegalBargmann}), its link with the Husimi $Q$ function and properties of holomorphic functions.

\subsection{Definition and uniqueness}

In what follows, $\mathcal H$ denotes a single-mode infinite-dimensional Hilbert space.
We recall the definition of the stellar function~\cite{bargmann1961hilbert,segal1963mathematical} and prove a few important properties.

\begin{defi}[Stellar function]
Let $\ket\psi=\sum_{n\ge0}{\psi_n\ket n}\in\mathcal H$ be a normalised state. The \textit{stellar function} of the state $\ket\psi$ is defined as
\be
F_\psi^\star(z)=e^{\frac12|z|^2}\braket{z^*|\psi}=\sum_{n\ge0}{\frac{\psi_n}{\sqrt{n!}}z^n},
\label{stellarfunc}
\ee
for all $z\in\mathbb C$, where $\ket z=e^{-\frac12|z|^2}\sum_{n\ge0}{\frac{z^n}{\sqrt{n!}}\ket n}\in\mathcal H$ is the coherent state of amplitude $z$.
\end{defi}

\noindent We now develop the formalism further, analysing the zeros of the stellar function to characterise states. The stellar function is a holomorphic function over the complex plane. For any normalised state $\ket\psi\in\mathcal H$ and all $z\in\mathbb C$,
\be
\ba
\left|F_\psi^\star(z)\right|^2&\le\left|\sum_{n\ge0}{\psi_n\frac{z^n}{\sqrt{n!}}}\right|^2\\
&\le\sum_{n\ge0}{|\psi_n|^2}\sum_{n\ge0}{\frac{|z|^{2n}}{n!}}\\
&=e^{|z|^2}
\ea
\ee
by Cauchy-Schwarz inequality. This implies that the stellar function of a normalised state is of finite order less or equal to $2$ and type less or equal to $\frac12$. 

\medskip

\noindent From the definition of the stellar function, for any state $\ket\psi\in\Hi$ we may write
\be
\ket\psi=\sum_{n\ge0}{\psi_n\ket n}=F_\psi^\star(\hat a^\dag)\ket0.
\label{Fadag}
\ee
From this equation one may understand the stellar function as an operational recipe for engineering a state from the vacuum, using the creation operator $\hat a^\dag$. This intuition will be made more precise in the following sections. An important property is that the stellar representation is unique:

\begin{lem} \label{lem:unique}
Let $\ket\phi$ and $\ket\psi$ be pure normalised single-mode states such that $F_\phi^\star=F_\psi^\star$. Then $\ket\phi=\ket\psi$.
Moreover, let $\ket\chi=f(\hat a^\dag)\ket0$ be a single-mode normalised pure state, where $f$ is analytic. Then $f=F_\chi^\star$.
\end{lem}

\begin{proof}
\begin{mdframed}[linewidth=1.5,topline=false,rightline=false,bottomline=false]

With the notations of the Lemma, $F_\phi^\star(z)=\sum_{n\ge0}{\phi_n\frac{z^n}{\sqrt{n!}}}$ and $F_\psi^\star(z)=\sum_{n\ge0}{\psi_n\frac{z^n}{\sqrt{n!}}}$. The functions $F_\phi^\star$ and $F_\psi^\star$ are analytic, so $F_\phi^\star(z)=F_\psi^\star(z)$ implies that $\phi_n=\psi_n$ for all $n\ge0$. Hence $\ket\phi=\ket\psi$.

\medskip

\noindent Now with $\ket\chi=\sum_{n\ge0}{\chi_n\ket n}=f(\hat a^\dag)\ket0$, let us write $f(z)=\sum_{n\ge0}{f_nz^n}$. We obtain
\be
\ba
\ket\chi&=\sum_{n\ge0}{f_n(\hat a^\dag)^n\ket0}\\
&=\sum_{n\ge0}{f_n\sqrt{n!}\ket n},
\ea
\ee
so $\chi_n=f_n\sqrt{n!}$ for all $n\ge0$. On the other hand, for all $z\in\mathbb C$,
\begin{align}
\nonumber F_\chi^\star(z)&=e^{\frac12|z|^2}\braket{z^*|\psi}\\ \displaybreak
&=\sum_{n\ge0}{\chi_n\frac{z^n}{\sqrt{n!}}}\\
\nonumber&=\sum_{n\ge0}{f_nz^n}\\
\nonumber&=f(z).
\end{align}
\end{mdframed}
\end{proof}

\noindent The stellar function of a state $\ket\psi\in\Hi$ is related to its Husimi $Q$ function, a smoothed version of the Wigner function~\cite{cahill1969density}, given by
\be
Q_\psi(z)=\frac1\pi|\braket{z|\psi}|^2=\frac{e^{-|z|^2}}\pi\left|F_\psi^\star(z^*)\right|^2,
\ee
for all $z\in\mathbb C$. The zeros of the Husimi $Q$ function are the complex conjugates of the zeros of $F_\psi^\star$. 
Hence, by Theorem~\ref{Husimi}, a single-mode pure quantum state is non-Gaussian if and only if its stellar function has zeros. These zeros form a discrete set, as the stellar function is a non-zero analytic function. 
The non-Gaussian properties of a single-mode pure state are then described by the distribution of the zeros over the complex plane. \\

\begin{figure}[h!]
	\begin{center}
		\includegraphics[width=\columnwidth]{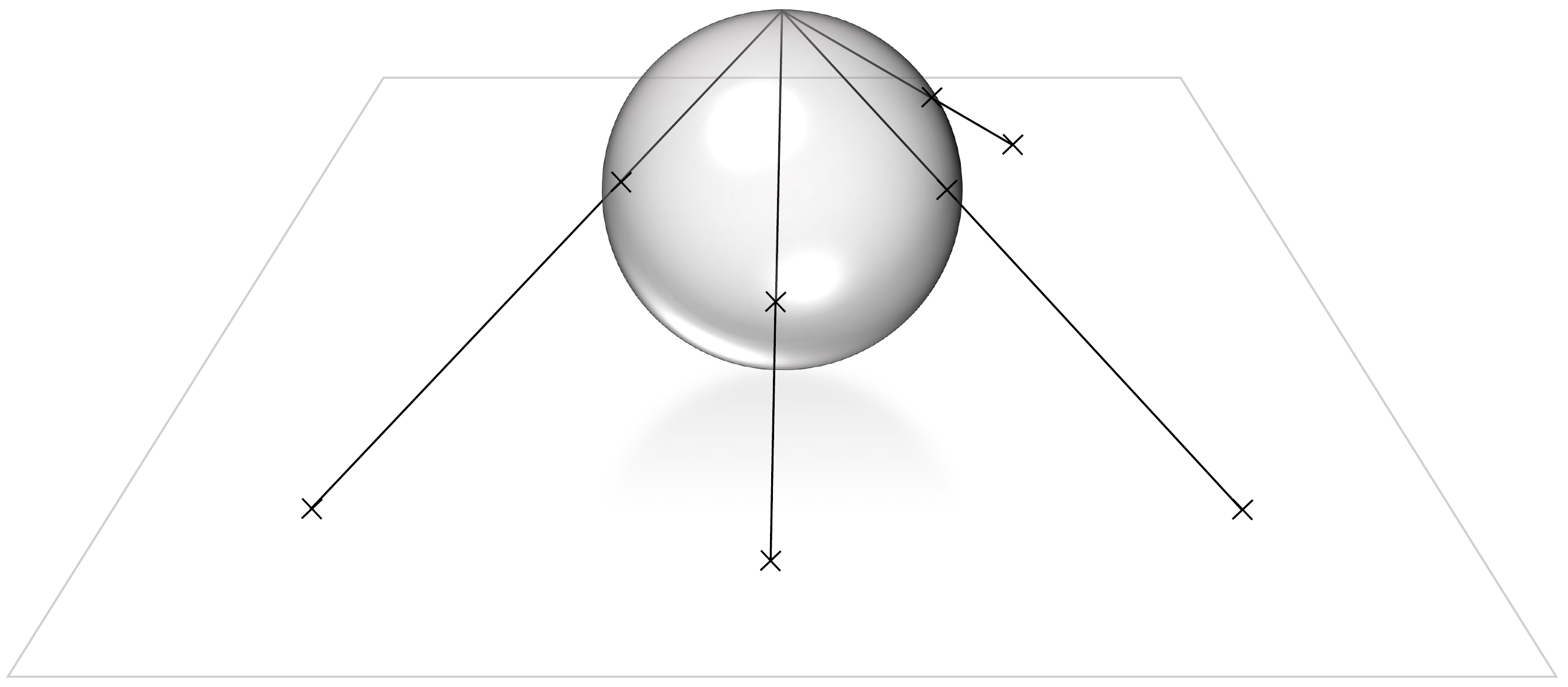}
		\caption{Antistereographic projection of four points onto the sphere}
		\label{fig:stellar}
	\end{center}
\end{figure}

\noindent Using anti-stereographic projection~\cite{sidoli2007arabic}, this amounts to describing the non-Gaussian properties of a pure state with a set of points on the sphere (Fig.~\ref{fig:stellar}), hence the name stellar representation, where the points on the sphere looked at from the center of the sphere are seen as stars on the celestial vault~\cite{tualle1995normal,korsch1997zeros}.

\medskip

\noindent In all the chapter we will use for brevity the notations $c_\chi=\cosh\chi$, $s_\chi=\sinh\chi$ and $t_\chi=\tanh\chi$, for all $\chi\in\mathbb C$.


\subsection{Examples}

In this section we give the stellar functions of various states and operators.

\subsubsection{Gaussian states and Fock states}

The displacement operator of amplitude $\alpha\in\mathbb C$ is given by $\hat D(\alpha)=e^{\alpha\hat a^\dag-\alpha^*\hat a}$. Its action on the vacuum state yields the coherent state $\ket\alpha$. The squeeze operator of parameter $\xi=re^{i\theta}\in\mathbb C$ is given by $\hat S(\xi)=e^{\frac12(\xi\hat a^2-\xi^*\hat a^{\dag2})}$. Its action on the vacuum state yields the squeezed state $\ket\xi$. 
All single-mode Gaussian operations may be decomposed as a squeezing operation and a displacement (see section~\ref{sec:Gaussian}).

For any single-mode Gaussian state $\hat S(\xi)\hat D(\alpha)\ket0$, where $\xi=re^{i\theta}$, the corresponding stellar function is~\cite{vourdas2006analytic}:
\be
G_{\xi,\alpha}^\star(z)=(1-|a|^2)^{1/4}e^{-\frac12az^2+bz+c},
\label{FSD}
\ee
where
\be
a:=e^{-i\theta}\tanh r,\text{ }b:=\alpha\sqrt{1-|a|^2}=\frac\alpha{\cosh r},\text{ }c:=\frac12a^*\alpha^2-\frac12|\alpha|^2.
\ee
In particular, we obtain
\be
G_{0,\alpha}^\star(z)=e^{\alpha z-\frac12|\alpha|^2},
\label{Fcoh}
\ee
for a coherent state of amplitude $\alpha\in\mathbb C$, and
\be
G_{\xi,0}^\star(z)=\frac1{\sqrt{\cosh r}}e^{-\frac12(e^{-i\theta}\tanh r)z^2},
\label{Fsqu}
\ee
for a squeezed vacuum state with squeezing parameter $\xi=re^{i\theta}\in\mathbb C$.

\medskip

\noindent For Fock states $\ket n$ with $n\in\mathbb N$, the stellar function is simply given by
\be
F_n^\star(z)=\frac{z^n}{\sqrt{n!}}.
\label{Fn}
\ee

\subsubsection{Cat states}

Let us define for $\alpha\in\mathbb C$ the cat$^+$ and cat$^-$ states:
\be
\ket{\text{cat}^\pm_\alpha}=\frac1{\sqrt{\mathcal N^\pm_\alpha}}(\ket\alpha\pm\ket{-\alpha}),
\ee
where $\ket\alpha$ is a coherent state, and $\mathcal N^\pm_\alpha$ is a normalisation factor. 

\begin{lem}\label{lem:Fcat}
The stellar functions of cat states are given by
\be
F^\star_{\text{cat}_\alpha^+}(z)=\frac{\cosh(\alpha z)}{\sqrt{\cosh(|\alpha|^2)}},
\label{Fcat+}
\ee
and
\be
F^\star_{\text{cat}_\alpha^-}(z)=\frac{\sinh(\alpha z)}{\sqrt{\sinh(|\alpha|^2)}},
\label{Fcat-}
\ee
for all $z,\alpha\in\mathbb C$.
\end{lem}

\begin{proof}
\begin{mdframed}[linewidth=1.5,topline=false,rightline=false,bottomline=false]

The overlap between two coherent states is given by
\be
\braket{z|\alpha}=e^{-\frac12(|z|^2+|\alpha|^2-2z^*\alpha)},
\ee
for $z,\alpha\in\mathbb C$. Hence, with $\braket{\text{cat}^\pm_\alpha|\text{cat}^\pm_\alpha}=1$ we have 
\be
\mathcal N^\pm_\alpha=2(1\pm e^{-2|\alpha|^2}).
\ee
We then obtain for $z,\alpha\in\mathbb C$,
\begin{equation}
\ba
F^\star_{\text{cat}_\alpha^\pm}(z)&=e^{-\frac12|z|^2}\braket{z^*|\text{cat}_\alpha^\pm}\\
&=\frac1{\sqrt{\mathcal N^\pm_\alpha}}\left(e^{-\frac12|\alpha|^2+\alpha z}\pm e^{-\frac12|\alpha|^2-\alpha z}\right)\\
&=\frac1{\sqrt{2\left(e^{|\alpha|^2}\pm e^{-|\alpha|^2}\right)}}\left(e^{\alpha z}\pm e^{-\alpha z}\right).
\ea
\end{equation}
We finally obtain
\be
F^\star_{\text{cat}_\alpha^+}(z)=\frac{\cosh(\alpha z)}{\sqrt{\cosh(|\alpha|^2)}},
\ee
and
\be
F^\star_{\text{cat}_\alpha^-}(z)=\frac{\sinh(\alpha z)}{\sqrt{\sinh(|\alpha|^2)}}.
\ee
\end{mdframed}
\end{proof}

\subsubsection{GKP states}

The set of Gottesman-Kitaev-Preskill (GKP) states have been proposed as a means for encoding a qubit in an oscillator, in a way which is fault-tolerant to small shifts in position and momentum \cite{Gottesman2001}. An example of such states is the simultaneous $+1$ eigenstate of the two commuting displacements operators $e^{-i\sqrt{2\pi}\hat p}$ and $e^{i\sqrt{2\pi}\hat q}$. The corresponding encoding may correct for comparable shifts in $\hat q$ and $\hat p$. An expression for this unphysical state (it has infinite norm) is given by
\be
\ba
\ket{\text{GKP}}&=\sum_{s\in\mathbb Z}{e^{-i\sqrt{2\pi}s\hat p}}\sum_{t\in\mathbb Z}{e^{i\sqrt{2\pi}t\hat q}}\ket0\\
&=\sum_{s,t\in\mathbb Z^2}{(-1)^{st}\hat D\left(2\sqrt\pi(s+it)\right)\ket0}\\
&=\sum_{s,t\in\mathbb Z^2}{(-1)^{st}\ket{2\sqrt\pi(s+it)}},
\ea
\ee
as an infinite superposition of coherent states. The stellar function of this state is then given by
\be
\ba
F^\star_{\text{GKP}}(z)&=\sum_{s,t\in\mathbb Z^2}{(-1)^{st}F^\star_{2\sqrt\pi(s+it)}(z)}\\
&=\sum_{s,t\in\mathbb Z^2}{(-1)^{st}e^{-2\pi(s^2+t^2)}e^{2\sqrt\pi(s+it)z}},
\ea
\ee
where we used Eq.~(\ref{Fcoh}) in the second line. This stellar function may be expressed as a Riemann theta function~\cite{riemann1857theorie}. Using properties of these functions, we obtain the following result:

\begin{lem}\label{lem:GKP}
$F^\star_{\text{GKP}}$ has an infinite number of zeros and has exactly $16$ zeros counted with multiplicity in each square region of the complex plane of size $4\sqrt\pi\times4\sqrt\pi$.
\end{lem}

\begin{figure}[h!]
	\begin{center}
		\includegraphics[width=0.5\columnwidth]{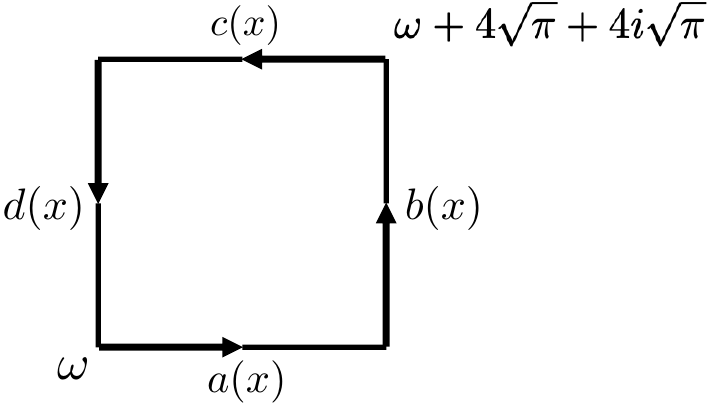}
		\caption{The contour used for the argument principle}
		\label{fig:contour}
	\end{center}
\end{figure}

\begin{proof}
\begin{mdframed}[linewidth=1.5,topline=false,rightline=false,bottomline=false]

We first derive a few invariance properties of $F^\star_{\text{GKP}}$ and conclude with the argument principle (Theorem~\ref{th:Cauchy}).
For all $z\in\mathbb C$, we have
\begin{eqnarray}
\label{GKProtation}
F^\star_{\text{GKP}}(iz)&=&\sum_{s,t\in\mathbb Z^2}{(-1)^{st}e^{-2\pi(s^2+t^2)}e^{2\sqrt\pi(-t+is)z}}\nonumber \\
&\underset{t\rightarrow -t}=&\sum_{s,t\in\mathbb Z^2}{(-1)^{st}e^{-2\pi(s^2+t^2)}e^{2\sqrt\pi(t+is)z}}\\
&\underset{s\leftrightarrow t}=&F^\star_{\text{GKP}}(z)\nonumber.
\end{eqnarray}
Now for all $z\in\mathbb C$,
\begin{eqnarray}
F^\star_{\text{GKP}}(z)&=&\sum_{s,t\in\mathbb Z^2}{(-1)^{st}e^{-2\pi(s^2+t^2)}e^{2\sqrt\pi(s+it)z}} \nonumber\\
&\underset{s\rightarrow s-2}=&\sum_{s,t\in\mathbb Z^2}{(-1)^{(s+2)t}e^{-2\pi((s+2)^2+t^2)}e^{2\sqrt\pi((s+2)+it)z}}\nonumber\\
&=&\sum_{s,t\in\mathbb Z^2}{(-1)^{st}e^{-2\pi(s^2+t^2)}e^{-8\pi s-8\pi}e^{2\sqrt\pi(s+it)z}e^{4\sqrt\pi z}}\\
&=&e^{4\sqrt\pi z-8\pi}\sum_{s,t\in\mathbb Z^2}{(-1)^{st}e^{-2\pi(s^2+t^2)}e^{2\sqrt\pi(s+it)(z-4\sqrt\pi)}}\nonumber\\
&=&e^{4\sqrt\pi z-8\pi}F^\star_{\text{GKP}}(z-4\sqrt\pi)\nonumber
\label{GKPRperiod}
\end{eqnarray}
Combining this with Eq.~(\ref{GKProtation}) we also obtain for all $z\in\mathbb C$,
\begin{align}
\nonumber F^\star_{\text{GKP}}(z)&=F^\star_{\text{GKP}}(iz)\\
\label{GKPIperiod}&=e^{4\sqrt\pi iz-8\pi}F^\star_{\text{GKP}}(iz-4\sqrt\pi)\\
\nonumber&=e^{4\sqrt\pi iz-8\pi}F^\star_{\text{GKP}}(z+4i\sqrt\pi).
\end{align}
This means that $F^\star_{\text{GKP}}$ is quasiperiodic along the horizontal and vertical directions in the complex plane, with period $4\sqrt\pi$. The functions $z\mapsto e^{4\sqrt\pi z-8\pi}$ and $z\mapsto e^{4\sqrt\pi iz-8\pi}$ do not vanish and it is thus sufficient for our purpose to prove that $F^\star_{\text{GKP}}$ has at least one zero: the quasiperiodicity implies that $F^\star_{\text{GKP}}$ would also vanish on the lattice with square cells of size $4\sqrt\pi$ containing this zero. 

\medskip

\noindent By Theorem~\ref{th:Cauchy}, the number of zeros of $F^\star_{\text{GKP}}$ inside a closed contour $C$ counted with multiplicity is given by
\be
Z_C\left(F^\star_{\text{GKP}}\right)=\frac1{2i\pi}\oint_C{\frac{\partial_zF^\star_{\text{GKP}}(z)}{F^\star_{\text{GKP}}(z)}}dz.
\label{argprinciple}
\ee
For all $\omega\in\mathbb C$, we consider the square contour $C_\omega$ with corners $\omega$, $\omega+4\sqrt\pi$, $\omega+4\sqrt\pi+4i\sqrt\pi$ and $\omega+4i\sqrt\pi$, parametrised by
\be
\ba
a(x)&=\omega+4\sqrt\pi x\\
b(x)&=\omega+4\sqrt\pi+4i\sqrt\pi x\\
c(x)&=\omega+4\sqrt\pi(1-x)+4i\sqrt\pi\\
d(x)&=\omega+4i\sqrt\pi(1-x),
\ea
\ee
for $x\in[0,1]$ (Fig.~\ref{fig:contour}). We have $a'(x)=4\sqrt{\pi}$, $b'(x)=4i\sqrt{\pi}$, $c'(x)=-4\sqrt{\pi}$ and $d'(x)=-4i\sqrt{\pi}$ for all $x\in[0,1]$. 

\noindent The quasiperiodicity of $F^\star_{\text{GKP}}$ may be rewritten as
\be
\begin{cases}
F^\star_{\text{GKP}}(z+4\sqrt{\pi})=e^{8\pi+4\sqrt{\pi}z}F^\star_{\text{GKP}}(z),\\
F^\star_{\text{GKP}}(z+4i\sqrt{\pi})=e^{8\pi-4i\sqrt{\pi}z}F^\star_{\text{GKP}}(z),
\end{cases}
\ee
for all $z\in\mathbb C$. Taking the derivative with respect to $z$ we obtain
\be
\begin{cases}
\partial_zF^\star_{\text{GKP}}(z+4\sqrt{\pi})=e^{8\pi+4\sqrt{\pi}z}\left[\partial_zF^\star_{\text{GKP}}(z)+4\sqrt{\pi}F^\star_{\text{GKP}}(z)\right],\\ 
\partial_zF^\star_{\text{GKP}}(z+4i\sqrt{\pi})=e^{8\pi-4i\sqrt{\pi}z}\left[\partial_zF^\star_{\text{GKP}}(z)-4i\sqrt{\pi}F^\star_{\text{GKP}}(z)\right],
\end{cases}
\ee
and thus
\be
\begin{cases}
\frac{\partial_zF^\star_{\text{GKP}}(z+4\sqrt{\pi})}{F^\star_{\text{GKP}}(z+4\sqrt{\pi})}=4\sqrt{\pi}+\frac{\partial_zF^\star_{\text{GKP}}(z)}{F^\star_{\text{GKP}}(z)},\\ 
\frac{\partial_zF^\star_{\text{GKP}}(z+4i\sqrt{\pi})}{F^\star_{\text{GKP}}(z+4i\sqrt{\pi})}=-4i\sqrt{\pi}+\frac{\partial_zF^\star_{\text{GKP}}(z)}{F^\star_{\text{GKP}}(z)}.
\end{cases}
\label{logderivGKP}
\ee
With Eq.~(\ref{argprinciple}), for all $\omega\in\mathbb C$,
\begin{align}
\nonumber Z_{C_\omega}\left(F^\star_{\text{GKP}}\right)&=\frac1{2i\pi}\oint_{C_\omega}{\frac{\partial_zF^\star_{\text{GKP}}(z)}{F^\star_{\text{GKP}}(z)}}dz\\
\label{zerosGKPinter}&=\frac1{2i\pi}\int_0^1\left[\frac{\partial_zF^\star_{\text{GKP}}(a(x))}{F^\star_{\text{GKP}}(a(x))}a'(x)+\frac{\partial_zF^\star_{\text{GKP}}(b(x))}{F^\star_{\text{GKP}}(b(x))}b'(x)\right.\\
\nonumber&\quad\quad\quad\quad\quad\quad\left.+\frac{\partial_zF^\star_{\text{GKP}}(c(x))}{F^\star_{\text{GKP}}(c(x))}c'(x)+\frac{\partial_zF^\star_{\text{GKP}}(d(x))}{F^\star_{\text{GKP}}(d(x))}d'(x)\right]dx.
\end{align}
Given that $c(x)=a(1-x)+4i\sqrt{\pi}$, we have
\be
\ba
\int_0^1{\frac{\partial_zF^\star_{\text{GKP}}(c(x))}{F^\star_{\text{GKP}}(c(x))}c'(x)dx}&=\int_0^1{-4\sqrt{\pi}\frac{\partial_zF^\star_{\text{GKP}}(a(1-x)+4i\sqrt{\pi})}{F^\star_{\text{GKP}}(a(1-x)+4i\sqrt{\pi})}}dx\\
&=\int_0^1{-4\sqrt{\pi}\left(-4i\sqrt{\pi}+\frac{\partial_zF^\star_{\text{GKP}}(a(1-x))}{F^\star_{\text{GKP}}(a(1-x))}\right)}\\
&=16i\pi-\int_0^1{\frac{\partial_zF^\star_{\text{GKP}}(a(1-x))}{F^\star_{\text{GKP}}(a(1-x))}a'(1-x)dx}\\
&=16i\pi-\int_0^1{\frac{\partial_zF^\star_{\text{GKP}}(a(x))}{F^\star_{\text{GKP}}(a(x))}a'(x)dx},
\ea
\ee
where we used Eq.~(\ref{logderivGKP}) in the second line with $z=a(1-x)$. Hence,
\be
\frac1{2i\pi}\int_0^1{\left[\frac{\partial_zF^\star_{\text{GKP}}(a(x))}{F^\star_{\text{GKP}}(a(x))}a'(x)+\frac{\partial_zF^\star_{\text{GKP}}(c(x))}{F^\star_{\text{GKP}}(c(x))}c'(x)\right]dx}=8.
\ee
Similarly $b(x)=d(1-x)+2\sqrt{2\pi}$ gives
\be
\frac1{2i\pi}\int_0^1{\left[\frac{\partial_zF^\star_{\text{GKP}}(b(x))}{F^\star_{\text{GKP}}(b(x))}b'(x)+\frac{\partial_zF^\star_{\text{GKP}}(d(x))}{F^\star_{\text{GKP}}(d(x))}d'(x)\right]dx}=8.
\ee
With Eq.~(\ref{zerosGKPinter}) we finally obtain
\be
Z_{C_\omega}\left(F^\star_{\text{GKP}}\right)=16.
\label{ZComega}
\ee
The function $F^\star_{\text{GKP}}$ thus has an infinite number of zeros. Since Eq.~(\ref{ZComega}) is independent of the choice of $\omega\in\mathbb C$, $F^\star_{\text{GKP}}$ has exactly $16$ zeros (counted with multiplicity) in each square region of the complex plane of size $4\sqrt\pi\times4\sqrt\pi$. Moreover, with the property of invariance under rotation in Eq.~(\ref{GKProtation}), it has exactly $4$ zeros in each square region of size $2\sqrt\pi\times2\sqrt\pi$ whose corners have coordinates in $2\sqrt\pi\mathbb Z$ (by considering the square region of size $4\sqrt\pi\times4\sqrt\pi$ centered on the origin).

\end{mdframed}
\end{proof}

\subsubsection{Operators}

While operators have their own treatment in the Segal--Bargmann formalism~\cite{vourdas2006analytic}, it is sufficient for our purpose to consider the following correspondences: the creation and annihilation operators have the stellar representations
\be
\hat a^\dag\rightarrow z,\quad \hat a\rightarrow\partial_z,
\label{castellar}
\ee
i.e., the operator corresponding to $\hat a^\dag$ in the stellar representation is the multiplication by $z$ and the operator in the stellar representation corresponding to $\hat a$ is the derivative with respect to $z$. This implies that the stellar function of a photon-added state $\hat a^\dag\ket\psi$ is given by $z\mapsto zF_\psi^\star(z)$, while the stellar function of a photon-subtracted state $\hat a\ket\psi$ is given by $z\mapsto\partial_zF_\psi^\star(z)$. In particular, photon-added states are always non-Gaussian, since $0$ is a root of their stellar function, while photon-subtracted states can be Gaussian (e.g., the Fock state $\ket1$, for which $\hat a\ket1=\ket0$, or the coherent states $\ket\alpha$, for $\alpha\in\mathbb C$, for which $\hat a\ket\alpha=\alpha\ket\alpha$). 

Any operator written as a power series in $\hat a^\dag$ and $\hat a$ thus has a stellar representation obtained by taking the same power series in the operator multiplication by $z$ and the operator derivative with respect to $z$, which corresponds to its effect on the stellar function of a state it is acting on. For example, the photon number operator $\hat n=\hat a^\dag\hat a$ acts on the stellar function as
\be
F_\psi^\star(z)\mapsto z\partial_zF_\psi^\star(z).
\ee
For various operators however, the corresponding stellar representation may be expressed more concisely than with a power series. We give a few examples in what follows.

\medskip

\noindent The displacement and squeeze operators satisfy the following commutation rules (see section~\ref{sec:Gaussian})
\be
\ba
\hat D(\alpha)\,\hat a^\dag\hat D^\dag(\alpha)&=\hat a^\dag-\alpha^*\\
\hat S(\xi)\,\hat a^\dag\hat S^\dag(\xi)&=c_r\hat a^\dag+s_re^{i\theta}\hat a,
\ea
\label{commutDS}
\ee
where $\alpha,\xi=re^{i\theta}\in\mathbb C$, with $c_r=\cosh r$ and $s_r=\sinh r$. For all $\ket\psi=\sum_{n\ge0}{\psi_n\ket n}$ we thus have
\be
\ba
\hat D(\alpha)\ket\psi&=\hat D(\alpha)F_\psi^\star(\hat a^\dag)\ket0\\
&=\sum_{n\ge0}{\frac{\psi_n}{\sqrt{n!}}\hat D(\alpha)\,(\hat a^\dag)^n\ket0}\\
&=\sum_{n\ge0}{\frac{\psi_n}{\sqrt{n!}}(\hat a^\dag-\alpha^*)^n\hat D(\alpha)\ket0}\\
&=F_\psi^\star(\hat a^\dag-\alpha^*)\ket\alpha\\
&=F_\psi^\star(\hat a^\dag-\alpha^*)e^{\alpha\hat a^\dag-\frac12|\alpha|^2}\ket0,
\ea
\ee
where we used Eq.~(\ref{Fadag}) in the first line, Eq.~(\ref{stellarfunc}) in the second line, Eq.~(\ref{commutDS}) in the third line and Eq.~(\ref{Fcoh}) in the last line. Hence, with Lemma~\ref{lem:unique}, the displacement operator $\hat D(\alpha)$ acts on the stellar function as
\be
F_\psi^\star(z)\mapsto e^{\alpha z-\frac12|\alpha|^2}F_\psi^\star(z-\alpha^*),
\label{dispstellar}
\ee
for all $\alpha\in\mathbb C$. Similarly, for all $\ket\psi$ we have
\be
\ba
\hat S(\xi)\ket\psi&=\hat S(\xi)F_\psi^\star(\hat a^\dag)\ket0\\
&=F_\psi^\star(c_r\hat a^\dag+s_re^{i\theta}\hat a)\hat S(\xi)\ket0\\
&=F_\psi^\star(c_r\hat a^\dag+s_re^{i\theta}\hat a)\ket\xi\\
&=\frac1{\sqrt{c_r}}F_\psi^\star(c_r\hat a^\dag+s_re^{i\theta}\hat a)e^{-\frac12e^{-i\theta}t_r(\hat a^\dag)^2}\ket0,
\ea
\ee
where $\xi=re^{i\theta}$ with $c_r=\cosh r$, $s_r=\sinh r$ and $t_r=\tanh r$, and where we used Eq.~(\ref{Fadag}) in the first line, Eq.~(\ref{commutDS}) in the second line and Eq.~(\ref{Fsqu}) in the last line. Hence, with Lemma~\ref{lem:unique}, the squeezing operator $\hat S(\xi)$ acts on the stellar function as
\be
F_\psi^\star(z)\mapsto\frac1{\sqrt{c_r}}F_\psi^\star(c_rz+s_re^{i\theta}\partial_z)e^{-\frac12e^{-i\theta}t_rz^2},
\label{squestellar}
\ee
for all $\xi=re^{i\theta}\in\mathbb C$.

\medskip

\noindent The POVM corresponding to a threshold detection is $\{\ket0\!\bra0,\mathbb1-\ket0\!\bra0\}$. The projector onto the vacuum acts as
\be
\ket\psi\mapsto\braket{0|\psi}\ket0,
\ee
so it maps the stellar function of a state $\ket\psi$ as
\be
F_\psi^\star(z)\mapsto\braket{0|\psi}F_{\ket0}^\star(z).
\ee
We have $\braket{0|\psi}=F_\psi^\star(0)$ and $F_{\ket0}^\star(z)=1$, so the projector $\ket0\!\bra0$ acts on the stellar function as
\be
F_\psi^\star(z)\mapsto F_\psi^\star(0),
\ee
while the projector $\mathbb1-\ket0\bra0$ acts as
\be
F_\psi^\star(z)\mapsto F_\psi^\star(z)-F_\psi^\star(0).
\ee
In particular, the click of a threshold detector projects the measured state onto a non-Gaussian state for which $0$ is a root of the stellar function. This is consistent with the fact that the measured state is orthogonal to the vacuum state---a coherent state of amplitude $0$---after being projected onto the support of $\mathbb1-\ket0\bra0$.

\medskip

\noindent Finally, the parity operator $\hat\Pi=(-1)^{\hat a^\dag\hat a}=e^{i\pi\hat n}$ maps the Fock state $\ket n$ to $(-1)^n\ket n$, for all $n\in\mathbb N$. Hence, it acts on the stellar function as
\be
F_\psi^\star(z)\mapsto F_\psi^\star(-z).
\ee
by Eq.~(\ref{stellarfunc}).

\section{The stellar hierarchy}

\subsection{The stellar rank}

The Hilbert space $\mathcal H$ is naturally partitioned into sets of states whose stellar functions---or equivalently Husimi $Q$ function---have the same number of zeros counted with multiplicity. We introduce the following related definition:

\begin{defi}[Stellar rank]
The \textit{stellar rank} $r^\star(\psi)$ of a pure single-mode normalised quantum state $\ket\psi\in\mathcal H$ is defined as the number of zeros of its stellar function $F_\psi^\star$, counted with multiplicity.
\end{defi}

\noindent By analogy with the Schmidt rank in entanglement theory~\cite{terhal2000schmidt}, we define the stellar rank of a mixed state $\rho$ as
\be
r^\star(\rho):=\inf_{p_i,\psi_i}\sup\,r^\star(\psi_i),
\label{rankmixed}
\ee
where the infimum is over the statistical ensembles $\{p_i,\psi_i\}$ such that $\rho=\sum_i{p_i\ket{\psi_i}\!\bra{\psi_i}}$. In particular, a mixed quantum state has nonzero rank if and only if it cannot be written as a mixture of Gaussian states.

\medskip

\begin{figure}[h!]
	\begin{center}
		\includegraphics[width=\columnwidth]{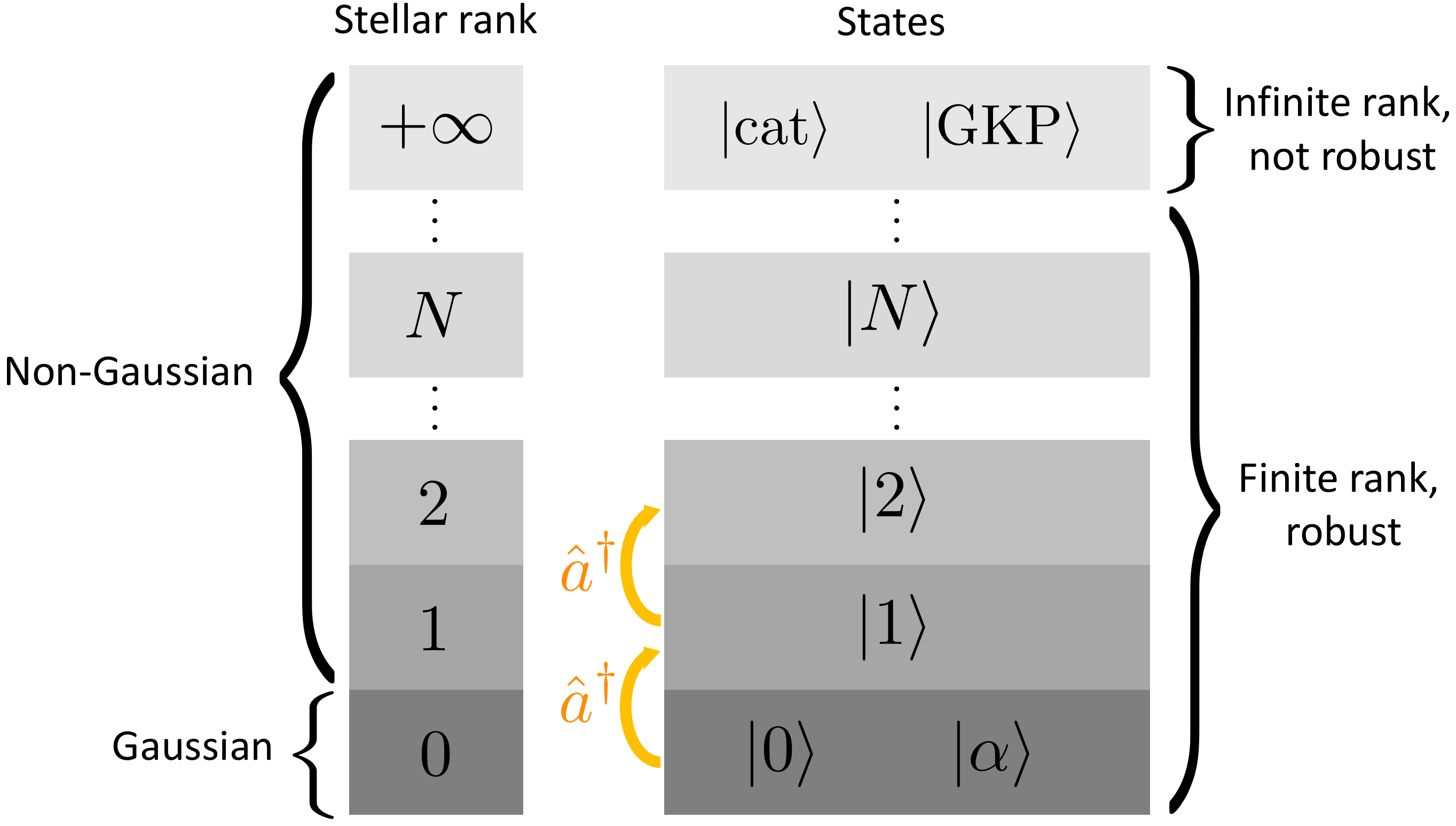}
		\caption{The stellar hierarchy of single-mode normalised quantum states. Each rank $N$ contains states obtained from the vacuum with $N$ single photon additions and Gaussian unitary operations (Theorem~\ref{finitez}). The states of finite rank are robust, while the states of infinite rank are not (section~\ref{sec:robust}).}
		\label{fig:stellarH}
	\end{center}
\end{figure}

\noindent We introduce hereafter the notation $\overline{\mathbb N}=\mathbb N\cup\{+\infty\}$, so that $r^\star(\psi)\in\overline{\mathbb N}$, and extend naturally the ordering from $\mathbb N$ to $\overline{\mathbb N}$, with the convention $N<+\infty\Leftrightarrow N\in\mathbb N$. For $N\in\overline{\mathbb N}$, we define
\be
R_N:=\{\ket\psi\in\Hi,\text{ }r^\star(\psi)=N\}
\label{finiterng}
\ee
the set of states with stellar rank equal to $N$. The \textit{stellar hierarchy} is the hierarchy of states induced by the stellar rank (Fig~\ref{fig:stellarH}).
The following properties are easily obtained:

\begin{itemize}
\item
By Lemma~\ref{lem:unique}, if $M\neq N$ then $R_M\cap R_N=\varnothing$, for all $M,N\in\overline{\mathbb N}$, so all the ranks in the stellar hierarchy are disjoint. 
\item
We have $\Hi=\bigcup_{N\in\overline{\mathbb N}}{R_N}$, i.e., the stellar hierarchy covers the whole space of normalised states, and the set of states of finite stellar rank is given by $\bigcup_{N\in\mathbb N}{R_N}$. 
\item
By Theorem~\ref{Husimi}, the rank zero of the stellar hierarchy $R_0$ is the set of single-mode normalised pure Gaussian states, and non-Gaussian states populate all higher ranks.
\item
For all $N\in\mathbb N$ ,the Fock state $\ket N$ is of stellar rank $N$, by Eq.~(\ref{Fn}), while cat states are of infinite stellar rank, by Lemma~\ref{lem:Fcat}, so all ranks are non empty.
\end{itemize}

\noindent In the following, we investigate further properties of the stellar hierarchy. We prove a first general decomposition result for pure states of finite stellar rank:

\begin{theo} \label{finitez}
Let $\ket\psi\in\bigcup_{N\in\mathbb N}{R_N}$ be a pure state of finite stellar rank. Let $\{\alpha_1,\dots,\alpha_{r^\star(\psi)}\}$ be the roots of the Husimi $Q$ function of $\ket\psi$, counted with multiplicity. Then,
\be
\ket\psi=\frac1{\mathcal N}\left[\prod_{n=1}^{r^\star(\psi)}{\hat D(\alpha_n)\,\hat a^\dag\hat D^\dag(\alpha_n)}\right]\ket{G_\psi},
\ee
where $\hat D(\alpha)$ is a displacement operator, $\ket{G_\psi}$ is a Gaussian state, and $\mathcal N$ is a normalisation constant. Moreover, this decomposition is unique up to reordering of the roots.
\end{theo}

\begin{proof}
\begin{mdframed}[linewidth=1.5,topline=false,rightline=false,bottomline=false]

We consider a state $\ket\psi$ of finite stellar rank $r^\star(\psi)\in\mathbb N$. 
Its stellar function is an analytic function over the complex plane of order less or equal to $2$, so by Hadamard-Weierstrass factorization theorem (Theorem~\ref{th:Hadamard}),
\be
F_\psi^\star(z)=z^k\left[\prod_{n=1}^{r^\star(\psi)-k}{\left(1-\frac z{z_n^*}\right)e^{\frac z{z_n^*}+\frac12\left(\frac z{z_n^*}\right)^2}}\right]e^{g_0+g_1z+g_2z^2},
\label{HW}
\ee
for all $z\in\mathbb C$, where $k\in\mathbb N$ is the multiplicity of $0$ as a root of $F_\psi^\star$, where the $\{z_n\}$ are the non-zero roots of $Q_\psi$ counted with multiplicity (i.e., the $\{z_n^*\}$ are the non-zero roots of $F^\star_\psi$ counted with multiplicity), and where $g_0,g_1,g_2\in\mathbb C$. Let us introduce for brevity $m=r^\star(\psi)-k\in\mathbb N$. Because the product in the above equation is finite, we need not worry about convergence of individual factors, and we may reorder the expression at will. We obtain
\be
\ba
F_\psi^\star(z)&=z^k\prod_{n=1}^m{\left(1-\frac z{z_n^*}\right)}\cdot\prod_{n=1}^m{e^{\frac z{z_n^*}+\frac12\left(\frac z{z_n^*}\right)^2}}\cdot e^{g_0+g_1z+g_2z^2}\\
&=z^k\prod_{n=1}^m{\left(1-\frac z{z_n^*}\right)}\cdot e^{g_0+\left(g_1+\sum_{n=1}^m{\frac1{z_n^*}}\right)z+\left(g_2+\frac12\sum_{n=1}^m{\frac1{(z_n^*)^2}}\right)z^2}\\
&=\frac{(-1)^m}{\prod_{n=1}^m{z_n^*}}\left[z^k\prod_{n=1}^m{\left(z-z_n^*\right)}\right]\cdot e^{g_0+\left(g_1+\sum_{n=1}^m{\frac1{z_n^*}}\right)z+\left(g_2+\frac12\sum_{n=1}^m{\frac1{(z_n^*)^2}}\right)z^2}.
\ea
\ee
With Eqs.~(\ref{Fadag}) and (\ref{commutDS}), we obtain, for all $\alpha\in\mathbb C$,
\begin{align}
\nonumber\ket\psi&=F_\psi^\star(\hat a^\dag)\ket0\\
&=\frac{(-1)^m}{\prod_{n=1}^m{z_n^*}}\left[(\hat a^\dag)^k\prod_{n=1}^m{\left(\hat a^\dag-z_n^*\right)}\right]\cdot e^{g_0+\left(g_1+\sum_{n=1}^m{\frac1{z_n^*}}\right)\hat a^\dag+\left(g_2+\frac12\sum_{n=1}^M{\frac1{(z_n^*)^2}}\right)(\hat a^\dag)^2}\ket0\\
\nonumber&=\frac{(-1)^m}{\prod_{n=1}^m{z_n^*}}\left[(\hat a^\dag)^k\prod_{n=1}^m{\hat D(z_n)\,\hat a^\dag\hat D^\dag(z_n)}\right]\cdot e^{g_0+\left(g_1+\sum_{n=1}^m{\frac1{z_n^*}}\right)\hat a^\dag+\left(g_2+\frac12\sum_{n=1}^m{\frac1{(z_n^*)^2}}\right)(\hat a^\dag)^2}\ket0.
\end{align}
Gouping the non-zero roots $\{z_n\}$ and the $k$ zero roots into the set of zeros counted with multiplicity $\{\alpha_n\}$, we obtain
\be
\ket\psi=\frac{(-1)^m}{\prod_{n=1}^m{z_n^*}}\left[\prod_{n=1}^{r^\star(\psi)}{\hat D(\alpha_n)\,\hat a^\dag\hat D^\dag(\alpha_n)}\right]\cdot e^{g_0+\left(g_1+\sum_{n=1}^m{\frac1{z_n^*}}\right)\hat a^\dag+\left(g_2+\frac12\sum_{n=1}^m{\frac1{(z_n^*)^2}}\right)(\hat a^\dag)^2}\ket0.
\ee
The state
\be
e^{g_0+\left(g_1+\sum_{n=1}^m{\frac1{z_n^*}}\right)\hat a^\dag+\left(g_2+\frac12\sum_{n=1}^m{\frac1{(z_n^*)^2}}\right)(\hat a^\dag)^2}\ket0
\ee
is a (non normalised) Gaussian state, by Eq.~(\ref{FSD}) and Lemma~\ref{lem:unique}. We finally obtain
\be
\ket\psi=\frac1{\mathcal N}\left[\prod_{n=1}^{r^\star(\psi)}{\hat D(\alpha_n)\,\hat a^\dag\hat D^\dag(\alpha_n)}\right]\ket{G_\psi},
\ee
where $\mathcal N$ is a normalisation constant, and $\ket{G_\psi}$ is a Gaussian state. The decomposition is unique by Lemma~\ref{lem:unique} (up to a reordering of the roots).

\end{mdframed}
\end{proof}

\noindent This decomposition implies that any state of finite stellar rank may be obtained from a Gaussian state by successive applications of the creation operator at different locations in phase space, given by the zeros of the Husimi $Q$ function. Experimentally, this corresponds to the probabilistic non-Gaussian operation of single-photon addition~\cite{zavatta2004quantum,marco2010manipulating,walschaers2018tailoring}. 
Using this decomposition, we obtain the following property for the stellar rank:

\begin{theo} \label{Ginv}
A unitary operation is Gaussian if and only if it leaves the stellar rank invariant.
\end{theo}

\begin{proof}
\begin{mdframed}[linewidth=1.5,topline=false,rightline=false,bottomline=false]

If a unitary operation leaves the stellar rank invariant, it maps in particular all pure states of stellar rank zero to pure states of stellar rank zero, i.e., all Gaussian states to Gaussian states, so it is a Gaussian operation.

\medskip

\noindent Reciprocally, let us show that Gaussian unitary operations leave the stellar rank invariant.
We first consider finite stellar rank pure states. Let $\ket \psi$ be such a state. By Theorem~2,
\be
\ket\psi=P_\psi(\hat a^\dag)\ket{G_\psi},
\ee
where $P_\psi$ is a polynomial of degree $r^\star(\psi)$ and $\ket{G_\psi}$ is a Gaussian state. By Eq.~(\ref{commutDS}) and by linearity we have
\be
\ket{\psi_\alpha}:=\hat D(\alpha)\ket\psi=\hat P_\psi(\hat a^\dag-\alpha^*)\hat D(\alpha)\ket{G_\psi},
\ee
and
\be
\ket{\psi_\xi}:=\hat S(\xi)\ket\psi=P_\psi(c_r\hat a^\dag+s_re^{i\theta}\hat a)\hat S(\xi)\ket{G_\psi},
\ee
where $\xi=re^{i\theta}$.
By Eq.~(\ref{dispstellar}), during a displacement of $\alpha$, the stellar function of $\ket\psi$ is modified as
\be
F_\psi^\star(z)\rightarrow F_{\psi,\alpha}^\star(z)=e^{z\alpha-\frac12|\alpha|^2}F_\psi^\star(z-\alpha^*)=P_\psi(z-\alpha^*)\,G_\alpha^\star(z),
\label{Fdisplaced}
\ee
where $G_\alpha^\star(z)$ is the Gaussian stellar function corresponding to the Gaussian state $\hat D(\alpha)\ket{G_\psi}$. Moreover, by Eq.~(\ref{squestellar}), during a squeezing of $\xi$, the stellar function of $\ket\psi$ is modified as
\be
\ba
F_\psi^\star(z)\rightarrow F_{\psi,\xi}^\star(z)&=P_\psi\left(c_rz+s_re^{i\theta}\partial_z\right)G_\xi^\star(z)\\
&=Q_{\psi,r}(z)\,G_\xi^\star(z),
\ea
\ee
where $G_\xi^\star(z)$ is the Gaussian stellar function corresponding to the Gaussian state $\hat S(\xi)\ket{G_\psi}$, and where $Q_{\psi,\xi}(z)=G_\xi^{\star-1}(z)\left[P_\psi\left(c_rz+s_re^{i\theta}\partial_z\right)G_\xi^\star(z)\right]$ is a polynomial. Let us compute the leading coefficient of $Q_{\psi,\xi}$. Writing $p$ the leading coefficient of $P_\psi$, and $N=r^\star(\psi)$ its degree for brevity, the leading coefficient of $Q_{\psi,\xi}$ is given by the leading coefficient of
\be
G_\xi^{\star-1}(z)\left[p\left(c_rz+s_re^{i\theta}\partial_z\right)^NG_\xi^\star(z)\right].
\ee
Let us write $G_\xi^\star(z)=e^{-\frac12az^2+bz+c}$, as in Eq.~(\ref{FSD}). The leading coefficient of $Q_{\psi,\xi}$ may then be obtained as the leading coefficient of
\be
e^{\frac12az^2}\left[p\left(c_rz+s_re^{i\theta}\partial_z\right)^Ne^{-\frac12az^2}\right].
\label{lead1}
\ee
For all $x,\lambda$, we have~\cite{wyss2017two}
\be
\left(x+\lambda\partial_x\right)^N=\sum_{n=0}^{\left\lfloor{\frac N2}\right\rfloor}{\frac{N!\lambda^n}{(N-2n)!n!2^n}\sum_{k=0}^{N-2n}{\binom{N-2n}{k}x^k\partial_x^{N-2n-k}}},
\ee
so the leading coefficient of $Q_{\psi,\xi}$ is equal to the leading coefficient of:
\be
pc_r^N\sum_{n=0}^{\left\lfloor{\frac N2}\right\rfloor}{z^{N-2n}(1-a)^{N-2n}\frac{N!t_r^ne^{in\theta}}{(N-2n)!n!2^n}},
\ee
where $t_r=\tanh r$. Finally, taking the leading coefficient in $z$ of this expression, corresponding to $n=0$, gives
\be
pc_r^N(1-a)^N.
\ee
It is non-zero unless $a=1$, which corresponds to an infinite value for the modulus $r$ of the squeezing parameter $\xi=re^{i\theta}$ by Eq.~(\ref{FSD}).
Hence the polynomials $P_\psi$ and $Q_{\psi,\xi}$ have the same degree.
This shows that a finite number of zeros is not modified by Gaussian operations.

Gaussian operations also map states with infinite number of zeros to states with infinite number of zeros. Indeed, assuming there exist a state $\ket\phi$ with an infinite number of zeros which is mapped by a Gaussian operation $\hat G$ to a state $\ket\psi$ with a finite number of zeros, then $\hat G^\dag$ would map $\ket\psi$ to $\ket\phi$, thus changing the (finite) number of zeros of $F_\psi^\star$, which would be in contradiction with the previous proof.
Hence Gaussian unitary operations leave the stellar rank of pure states invariant.

Now by Eq.~(\ref{rankmixed}), the stellar rank of a mixed state $\rho$ is given by
\be
r^\star(\rho)=\inf_{p_i,\psi_i}\sup\,r^\star(\psi_i),
\ee
where the infimum is over the statistical ensembles such that $\rho=\sum_i{p_i\ket{\psi_i}\!\bra{\psi_i}}$. 
For $\hat G$ a unitary Gaussian operation,
\begin{align}
\nonumber r^\star(\hat G\rho\hat G^\dag)&=\inf_{p_i,\psi_i}\sup\,r^\star(\hat G\psi_i)\\ \displaybreak
&=\inf_{p_i,\psi_i}\sup\,r^\star(\psi_i)\\
\nonumber&=r^\star(\rho),
\end{align}
where we used in the second line the fact that Gaussian unitary operations leave the stellar rank of pure states invariant. Hence, Gaussian unitary operations leave the stellar rank invariant.

\end{mdframed}
\end{proof}

\noindent An interesting consequence is that the number of single-photon additions in the decomposition of Theorem~\ref{finitez} is minimal. Indeed, if a quantum state is obtained from the vacuum by successive applications of Gaussian operations and single-photon additions, then its stellar rank is exactly the number of photon additions, because each single-photon addition increases by one its stellar rank---it adds a zero to the stellar function at zero---while each Gaussian operation leaves the stellar rank invariant by Theorem~\ref{Ginv}. Hence, the stellar rank is a measure of the non-Gaussian properties of a quantum state which may be interpreted as a minimal non-Gaussian operational cost, in terms of single-photon additions, for engineering the state from the vacuum.

\subsection{Gaussian convertibility}

Now that the first properties of the stellar hierarchy are laid out, we consider the convertibility of quantum states using Gaussian unitary operations:

\begin{defi}[Gaussian convertibility] \label{Gconvert}
Two states $\ket\phi$ and $\ket\psi$ are \textit{Gaussian-convertible} if there exists a Gaussian unitary operation $\hat G$ such that $\ket\psi=\hat G\ket\phi$.
\end{defi}

\noindent Note that this notion is different from the notion of Gaussian conversion introduced in~\cite{yadin2018operational}, which denotes the conversion of Gaussian states with passive linear optics, and a subclass of Gaussian measurements and feed-forward.

Gaussian convertibility defines an equivalence relation in $\mathcal H$. By Theorem~\ref{Ginv}, having the same stellar rank is a necessary condition for Gaussian convertibility. However, this condition is not sufficient. In order to derive the equivalence classes for Gaussian convertibility, we introduce the following definition:

\begin{defi}[Core state]
\textit{Core states} are defined as the single-mode normalised pure quantum states which have a polynomial stellar function.
\end{defi}

\noindent By Eq.~(\ref{Fadag}) and Lemma~\ref{lem:unique}, core states are the states with a bounded support over the Fock basis, i.e., finite superpositions of Fock states. These correspond to the minimal non-Gaussian core states introduced in~\cite{menzies2009gaussian}, in the context of non-Gaussian state engineering. 

\medskip

\noindent With this definition, we obtain our following result.

\begin{theo} \label{Gconv}
Let $\ket\psi\in\bigcup_{N\in\mathbb N}{R_N}$ be a state of finite stellar rank. Then, there exists a unique core state $\ket{C_\psi}$ such that $\ket\psi$ and $\ket{C_\psi}$ are Gaussian-convertible.

By Theorem~\ref{finitez}, $\ket\psi=P_\psi(\hat a^\dag)\ket{G_\psi}$, where $P_\psi$ is a polynomial of degree $r^\star(\psi)$ and $\ket{G_\psi}=\hat S(\xi)\hat D(\alpha)\ket0$ is a Gaussian state, where $\hat D(\alpha)=e^{\alpha\hat a^\dag-\alpha^*\hat a}$ is a displacement operator, and $\hat S(\xi)=e^{\frac12(\xi\hat a^2-\xi^*\hat a^{\dag2})}$ is a squeezing operator, with $\xi=re^{i\theta}$. Then,
\be
\ket\psi=\hat S(\xi)\hat D(\alpha)\ket{C_\psi}=\hat S(\xi)\hat D(\alpha)F_{C_\psi}^\star(\hat a^\dag)\ket0,
\label{decomp2}
\ee
where the (polynomial) stellar function of $\ket{C_\psi}$ is given by
\be
F_{C_\psi}^\star(z)=P_\psi\left(c_rz-s_re^{i\theta}\partial_z+c_r\alpha^*-s_re^{i\theta}\alpha\right)\cdot1,
\label{polycore}
\ee
for all $z\in\mathbb C$.
\end{theo}

\begin{proof}
\begin{mdframed}[linewidth=1.5,topline=false,rightline=false,bottomline=false]

Let $\ket\psi\in\bigcup_{N\in\mathbb N}{R_N}$ be a state of finite stellar rank. By Theorem~\ref{finitez},
\be
\ket\psi=P_\psi(\hat a^\dag)\ket{G_\psi},
\label{decomppsi}
\ee
where $P_\psi$ is a polynomial of degree $r^\star(\psi)$ and $\ket{G_\psi}=\hat S(\xi)\hat D(\alpha)\ket0$ is a Gaussian state, with $\xi=re^{i\theta}$. Let us define $\ket{C_\psi}=\hat D^\dag(\alpha)\hat S^\dag(\xi)\ket\psi$. The states $\ket\psi$ and $\ket{C_\psi}$ are Gaussian-convertible. Moreover, from the commutation relations in Eq.~(\ref{commutDS}) and by linearity we obtain
\be
\ba
\ket{C_\psi}&=\hat D^\dag(\alpha)\hat S^\dag(\xi)P_\psi(\hat a^\dag)\ket{G_\psi}\\
&=\hat D^\dag(\alpha)P_\psi\left(c_r\hat a^\dag-s_re^{i\theta}\hat a\right)\hat S^\dag(\xi)\ket{G_\psi}\\
&=P_\psi\left[c_r(\hat a^\dag+\alpha^*)-s_re^{i\theta}(\hat a+\alpha)\right]\ket0\\
&=P_\psi\left(c_r\hat a^\dag-s_re^{i\theta}\hat a+c_r\alpha^*-s_re^{i\theta}\alpha\right)\ket0,
\ea
\ee
where we used Eq.~(\ref{decomppsi}) in the first line. By Eq.~(\ref{castellar}), the stellar operator corresponding to $\hat a^\dag$ is the multiplication by $z$ and the stellar operator corresponding to $\hat a$ is the derivative with respect to $z$. Hence,
\be
F_{C_\psi}^\star(z)=P_\psi\left(c_rz-s_re^{i\theta}\partial_z+c_r\alpha^*-s_re^{i\theta}\alpha\right)\cdot1,
\ee
for all $z\in\mathbb C$, which is a polynomial function, so the state $\ket{C_\psi}$ is a core state.

\medskip

\noindent In order to conclude the proof, we need to show that $\ket{C_\psi}$ is the unique core state Gaussian-convertible to $\ket\psi$. Let $\ket C=P_C(\hat a^\dag)\ket0$ be a core state Gaussian-convertible to $\ket\psi$. The states $\ket{C_\psi}$ and $\ket C$ are Gaussian-convertible so there exist $\xi,\alpha\in\mathbb C$ such that
\be
\ba
\ket{C_\psi}&=\hat S(\xi)\hat D(\alpha)\ket C\\
&=\hat S(\xi)\hat D(\alpha)P_C(\hat a^\dag)\ket0\\
&=P_C(c_r\hat a^\dag+s_re^{i\theta}\hat a-\alpha^*)\hat S(\xi)\hat D(\alpha)\ket0,
\ea
\ee
where we used Eq.~(\ref{commutDS}). Hence,
\be
F_{C_\psi}^\star(z)=P_C(c_rz+s_re^{i\theta}\partial_z-\alpha^*)\,G_{\xi,\alpha}^\star(z).
\ee
With Eq.~(\ref{FSD}), this function may be expressed as a polynomial multiplied by a Gaussian function $G_{\xi,\alpha}^\star$. On the other hand $F_{C_\psi}^\star$ is a polynomial, since $\ket{C_\psi}$ is a core state. By comparison of the speed of convergence, this implies that the Gaussian function $G_{\xi,\alpha}^\star$ is constant, i.e., that
\be
e^{-i\theta}\tanh r=0\quad\text{and}\quad\alpha\sqrt{1-\tanh^2r}=0,
\ee
by Eq.~(\ref{FSD}). This in turn implies $\xi=\alpha=0$, and $\ket C=\hat S(\xi)\hat D(\alpha)\ket C=\ket{C_\psi}$.

\end{mdframed}
\end{proof}

\noindent This result has several consequences:

\begin{itemize}
\item
It implies a second general decomposition result, in addition to Theorem~\ref{finitez}: by Eq.~(\ref{decomp2}), any state of finite stellar rank can be uniquely decomposed as a finite superposition of equally displaced and equally squeezed number states. This shows that the stellar hierarchy matches the genuine $n$-photon hierarchy introduced in~\cite{lachman2018faithful}: a pure state exhibits \textit{genuine} $n$\textit{-photon quantum non-Gaussianity} if and only if it has a stellar rank greater or equal \mbox{to $n$}.
Formally, for all $N\in\mathbb N$, the set $R_N$ of states of stellar rank equal to $N$ is obtained by the free action of the group of single-mode Gaussian unitary operations on the set of core states of stellar rank $N$, which is isomorphic to the set of normalised complex polynomials of degree $N$.
\item
It shows that two different core states are never Gaussian-convertible, while any state of finite stellar rank is always Gaussian-convertible to a unique core state. This implies that equivalence classes for Gaussian convertibility for states of finite stellar rank correspond to the orbits of core states under Gaussian operations.
\item
It gives an analytic way to check if two states of finite stellar rank are Gaussian-convertible, given their stellar functions, by checking with Eq.~(\ref{polycore}) if they share the same core state.
\item
It shows that photon-subtracting a state of finite stellar rank, which amounts to derivating its stellar function, can either decrease its stellar rank by $1$, leave it invariant, or increase it by $1$, depending on whether the Gaussian operation which converts the state to its core state is either the identity, a displacement, or a Gaussian operation with nonzero squeezing parameter. In particular, this implies that the stellar rank is a lower bound on the number of photon subtractions necessary to enginer a state from the vacuum, together with Gaussian unitary operations.
\end{itemize}

\noindent We consider the following simple example to illustrate the use of Theorem~\ref{Gconv} for determining Gaussian convertibility: a photon-subtracted squeezed state, a photon-added squeezed state and a single-photon Fock state.
We write $\ket\phi=-\frac1{s_\xi}\hat a\ket{\xi}$ a normalised photon-subtracted squeezed vacuum state and $\ket\psi=\frac1{c_\xi}\hat a^\dag\ket\xi$ a normalised photon-added squeezed vacuum state, with $\xi\in\mathbb R^*$. We write also $\ket\chi=\ket1$ a single-photon Fock state. Using Eq.~(\ref{FSD}) and Eq.~(\ref{castellar}), we obtain for all $z\in\mathbb C$
\be
\ba
F_\phi^\star(z)&=-\frac1{s_\xi}\partial_z\left[e^{-\frac12t_\xi z^2}\right]\\
&=\frac z{c_\xi}e^{-\frac12t_\xi z^2},
\ea
\ee
and
\be
F_\psi^\star(z)=\frac z{c_\xi}e^{-\frac12t_\xi z^2},
\ee
where $c_\xi=\cosh\xi$ and $t_\xi=\tanh\xi$. Hence $F_\phi^\star=F_\psi^\star$, so the states $\ket\phi$ and $\ket\psi$ are actually equal. 

We also have $F_\chi^\star(z)=z$. With the notations of Theorem~\ref{Gconv}, we have $r_\phi=\xi$, $r_\chi=0$, $\theta_\phi=\theta_\chi=\alpha_\phi=\alpha_\chi=0$, $\hat G_\phi=\hat S(\xi)$, $\hat G_\chi=\mathbb1$, $P_\phi(z)=\frac z{c_\xi}$, and $P_\chi(z)=z$, so for all $z\in\mathbb C$,
\be
\ba
P_\phi\left(c_{r_\phi}z-s_{r_\phi}e^{i\theta_\phi}\partial_z+c_{r_\phi}\alpha_\phi^*-s_{r_\phi}e^{i\theta_\phi}\alpha_\phi\right)\cdot1&=\frac1{c_\xi}\left(c_\xi z-s_\xi\partial_z\right)\cdot1\\
&=z,
\ea
\ee
and
\be
\ba
P_\chi\left(c_{r_\chi}z-s_{r_\chi}e^{i\theta_\chi}\partial_z+c_{r_\chi}\alpha_\chi^*-s_{r_\chi}e^{i\theta_\chi}\alpha_\chi\right)\cdot1&=z\cdot1\\
&=z,
\ea
\ee
thus $\ket\phi$ and $\ket\chi$ share the same core state. By Theorem~\ref{Gconv}, this means that $\ket\phi$ and $\ket\chi$ are Gaussian-convertible, and we have $\ket\phi=\hat G_\phi\hat G_\chi^\dag\ket\chi$, where
\be
\hat G_\phi\hat G_\chi^\dag=\hat S(\xi).
\ee
Using Eq.~(\ref{commutDS}) confirms indeed that
\be
-\frac1{s_\xi}\hat a\ket{\xi}=\frac1{c_\xi}\hat a^\dag\ket\xi=\hat S(\xi)\ket1.
\label{Gconvexample}
\ee

\section{Robustness of non-Gaussian states}
\label{sec:robust}

The stellar hierarchy provides a ranking of non-Gaussian states, in terms of the minimal number of photons additions necessary to engineer them. However, for this hierarchy to be relevant in realistic experimental scenarios, it has to be robust to small deviations. We consider this formally in what follows and analyse the robustness properties of the stellar hierarchy.

\subsection{Definitions}

We introduce the following definition:

\begin{defi}[Stellar robustness]
Let $\ket\psi\in\Hi$. The \textit{stellar robustness} of the state $\ket\psi$ is defined as
\be
R^\star(\psi):=\inf_{r^\star(\phi)<r^\star(\psi)}{D(\phi,\psi)},
\ee
where $D$ denotes the trace distance and where the infimum is over all states $\ket\phi\in\Hi$ such that $r^\star(\phi)<r^\star(\psi)$.
\end{defi}

\noindent The stellar robustness quantifies how much one has to deviate from a quantum state in trace distance to find another quantum state of lower stellar rank: states with a positive stellar robustness will be referred to as \textit{robust}. The stellar robustness inherits the property of invariance under Gaussian operations of the stellar rank, because the trace distance between two states is invariant under unitary operations. Because of its operational properties, the choice of the trace distance is especially relevant in the context of non-Gaussian state engineering and quantum computing with non-Gaussian states.

A similar notion, though more restricted, is the quantum non-Gaussian depth~\cite{straka2018quantum} which quantifies the maximum attenuation applicable on a quantum state, after which quantum non-Gaussianity can still be witnessed.
A natural generalisation of the notion of stellar robustness is the following:

\begin{defi}[$k$-robustness]
Let $\ket\psi\in\Hi$. For all $k\in\overline{\mathbb N}^*$, the \textit{$k$-robustness} of the state $\ket\psi$ is defined as
\be
R^\star_k(\psi):=\inf_{r^\star(\phi)<k}{D(\phi,\psi)},
\ee
where $D$ denotes the trace distance and where the infimum is over all states $\ket\phi\in\Hi$ such that $r^\star(\phi)<k$. 
\end{defi}

\noindent For all $k\in\mathbb N^*$, the $k$-robustness quantifies how much one has to deviate from a quantum state in trace distance to find another quantum state which as a stellar rank between $0$ and $k-1$. States with a positive $R^\star_k$ will be referred to as \textit{robust with respect to states of stellar rank lower than $k$}. When $k=+\infty$, the $\infty$-robustness quantifies how much one has to deviate from a quantum state in trace distance to find another quantum state of finite stellar rank. Note that the stellar robustness satisfies $R^\star(\psi)=R^\star_{r^\star(\psi)}(\psi)$. We introduce the related definition:

\begin{defi}[Robustness profile]
Let $\ket\psi\in\Hi$. The \textit{robustness profile} of the state $\ket\psi$ is defined as
\be
\mathcal R(\psi):=\left(R^\star_k(\psi)\right)_{k\in\mathbb N^*}.
\ee
\end{defi}

\noindent The robustness profile is the sequence of $k$-robustnesses for all $k\in\mathbb N^*$. This profile describes how hard a non-Gaussian state is to produce experimentally, using photon additions. 

\medskip

\noindent A dual notion to the robustness is the following:

\begin{defi}[Smoothed non-Gaussianity of formation]
Let $\rho$ be a single-mode normalised state, and let $\epsilon>0$. The \textit{$\epsilon$-smoothed non-Gaussianity of formation} $\mathcal N\mathcal G\mathcal F_\epsilon(\rho)$ is defined as the minimal stellar rank of the states $\sigma$ that are $\epsilon$-close to $\rho$ in trace distance. Formally,
\be
\mathcal N\mathcal G\mathcal F_\epsilon(\rho):=\inf_\sigma{\left\{r^\star(\sigma),\text{ s.t. }D(\rho,\sigma)\le\epsilon\right\}},
\ee
where $D$ denotes the trace distance.
\label{NGformation}
\end{defi}

\noindent The infimum in the definition is also a minimum, since the set considered only contains integer values and is lower bounded by zero.
That minimum is not necessarily attained for the energy cut-off state (consider, e.g., a Gaussian pure state).

The smoothed non-Gaussianity of formation can be obtained directly from the robustness profile and gives a smoothed version of the stellar rank, dual to the robustness.
By Theorem~\ref{finitez}, it quantifies the minimal number of single-photon additions that need to be applied to a Gaussian state in order to obtain a state $\epsilon$-close to a target state, and provides an operational cost measure for non-Gaussian resource states, which is also invariant under Gaussian operations. 

\medskip

\noindent The robustness is related to the fidelity by the following result:

\begin{lem}\label{lem:Rfide}
Let $\ket\psi\in\mathcal H$. For all $k\in\overline{\mathbb N}^*$,
\begin{equation}
\sup_{r^\star(\rho)<k}F(\rho,\psi)=1-[R_k^\star(\psi)]^2,
\end{equation}
where $F$ is the fidelity.
\end{lem}

\begin{proof}
\begin{mdframed}[linewidth=1.5,topline=false,rightline=false,bottomline=false]
For any pure state $\ket\psi\in\Hi$, and any set of pure states $\mathcal X$, we have
\be
\ba
\sup_{\substack{\rho=\sum{p_i{\ket\phi_i}\bra{\phi_i}}\\ \sum{p_i}=1,\phi_i\in\mathcal X}}{F(\rho,\psi)}&=\sup_{\substack{\rho=\sum{p_i{\ket\phi_i}\bra{\phi_i}}\\ \sum{p_i}=1,\phi_i\in\mathcal X}}{\braket{\psi|\rho|\psi}}\\
&=\sup_{\sum{p_i}=1}\sup_{\phi_i\in\mathcal X}{\sum{p_i|\braket{\phi_i|\psi}|^2}}\\
&=\sup_{\phi\in\mathcal X}{|\braket{\phi|\psi}|^2}\\
&=\sup_{\phi\in\mathcal X}{F(\phi,\psi)}.
\ea
\ee
Hence, for $\mathcal X$ the set of pure states of stellar rank less than $k$,
\begin{align}
\nonumber R_k^\star(\psi)&=\inf_{r^\star(\phi)<k}{D(\phi,\psi)}\\
\label{robustnessinter}&=\inf_{r^\star(\phi)<k}{\sqrt{1-|\braket{\phi|\psi}|^2}}\\ \displaybreak
\nonumber&=\sqrt{1-\sup_{r^\star(\phi)<k}{F(\phi,\psi)}}\\
\nonumber&=\sqrt{1-\sup_{r^\star(\rho)<k}{F(\rho,\psi)}},
\end{align}
where $D$ denotes the trace distance, where we used the definition of the stellar rank for mixed states~(\ref{rankmixed}). We finally obtain
\be
\sup_{r^\star(\rho)<k}{F(\rho,\psi)}=1-[R_k^\star(\psi)]^2.
\ee
\end{mdframed}
\end{proof}

\noindent Certifying that a (mixed) state $\rho$ has a fidelity greater than $1-[R^\star_k(\psi)]^2$ with a given target pure state $\ket\psi$ thus ensures that the state $\rho$ has stellar rank greater or equal to $k$. However, this is only possible if the two following conditions are met:

\begin{itemize}
\item
The target state $\ket\psi$ is robust with respect to states of stellar rank less than $k$, i.e., $R^\star_k(\psi)>0$.
\item
The value of the $k$-robustness $R^\star_k(\psi)$ is known.
\end{itemize}

\noindent We consider these two problems in what follows. First, we determine for all $k\in\overline{\mathbb N}^*$ which states are robust with respect to states of stellar rank less than $k$. Then, we show how to compute their $k$-robustness.

\subsection{Topology of the stellar hierarchy}

Determining which states are robust amounts to characterizing the topology of the stellar hierarchy, with respect to the trace norm. Formally, this topology is summarised by the following result for states of finite stellar rank:

\begin{theo}\label{topology}
For all $N\in\mathbb N$,
\be
\overline{R_N}=\underset{0\le K\le N}{\bigcup}{R_K},
\ee
where $\overline X$ denotes the closure of $X$ for the trace norm in the set of normalised states of $\mathcal H$.
\end{theo}

\begin{proof}
\begin{mdframed}[linewidth=1.5,topline=false,rightline=false,bottomline=false]

Recall that the set of normalised pure single-mode states is closed for the trace norm in the whole Hilbert space, since it is the reciprocal image of $\{1\}$ by the trace norm, which is Lipschitz continuous---with Lipschitz constant $1$---hence continuous.

\medskip

\noindent For the proof, we fix $N\in\mathbb N$. We prove the theorem by showing a double inclusion. We first show that $\bigcup_{K=0}^N{R_K}\subset\overline{R_N}$,
and then that the set $\bigcup_{K=0}^N{R_K}$ is closed in $\mathcal H$ for the trace norm. Since the closure of a set $X$ is the smallest closed set containing $X$, and given that $R_N\subset\bigcup_{K=0}^N{R_K}$, this will prove the other inclusion and hence the result.

\medskip

\noindent We have $R_N\subset\overline{R_N}$. Let $\ket\psi\in\bigcup_{K=0}^{N-1}{R_K}$. There exists $K\in\{0,\dots,N-1\}$ such that $r^\star(\psi)=K$. By Theorem~4, there exists a core state $\ket{C_\psi}$, with a polynomial stellar function of degree $K$, and a Gaussian operation $\hat G_\psi$ such that $\ket\psi=\hat G_\psi\ket{C_\psi}$. We define the sequence of normalised states
\be
\ket{\psi_m}=\sqrt{1-\frac1m}\ket\psi+\frac1{\sqrt m}\hat G_\psi\ket N,
\ee
for $m\ge1$. We have
\be
\ket{\psi_m}=\hat G_\psi\left(\sqrt{1-\frac1m}\ket{C_\psi}+\frac1{\sqrt m}\ket N\right),
\ee
and the state $\sqrt{1-\frac1m}\ket{C_\psi}+\frac1{\sqrt m}\ket N$ is a normalised core state whose stellar function is a polynomial of degree $N$, hence $\ket{\psi_m}\in R_N$. Moreover, $\{\ket{\psi_m}\}_{m\ge1}$ converges to $\ket\psi$ in trace norm. This shows that $\bigcup_{K=0}^N{R_K}\subset\overline{R_N}$.

\medskip

\noindent We now prove that the set $\bigcup_{K=0}^N{R_K}$ is closed in $\mathcal H$ for the trace norm. For $N=0$ (i.e., showing that the set of Gaussian states is a closed set), this is already a nontrivial result, and a proof may be found, e.g., in~\cite{lami2018gaussian}. 

For all $N\ge0$, the sketch of the proof is the following: given a converging sequence in $\bigcup_{K=0}^N{R_K}$, we want to show that its limit has a stellar rank less or equal to $N$. We first use the decomposition result of Theorem~\ref{Gconv}, in order to obtain a sequence of Gaussian operations acting on a sequence of core states of rank less or equal to $N$. We make use of the compactness of this set of core states to restrict to a unique core state. Then, we show that the squeezing and the displacement parameters of the sequence of Gaussian operations cannot be unbounded. This allows us to conclude by extracting converging subsequences from these parameters.

The trace distance $D$ is induced by the trace norm. Let $\{\ket{\psi_m}\}_{m\in\mathbb N}\in\bigcup_{K=0}^N{R_K}$ be a converging sequence for the trace norm, and let $\ket\psi\in\mathcal H$ be its limit. By Theorem~\ref{Gconv}, there exist a sequence of core states $\{\ket{C_m}\}_{m\in\mathbb N}$, with polynomial stellar functions of degrees less or equal to $N$, and a sequence of Gaussian operations $\{\hat G_m\}_{m\in\mathbb N}$ such that for all $m\in\mathbb N$, $\ket{\psi_m}=\hat G_m\ket{C_m}$.

The set of normalised core states with a polynomial stellar function of degree less or equal to $N$ corresponds to the set of normalised states with a support over the Fock basis truncated at $N$, and is compact for the trace norm in $\mathcal H$ (isomorphic to the set of norm $1$ vectors in $\mathbb C^{N+1}$). Hence, the sequence $\{\ket{C_m}\}_{m\in\mathbb N}$ admits a converging subsequence $\{\ket{C_{m_k}}\}_{k\in\mathbb N}$. Let the core state $\ket C$, with a polynomial stellar function of degree less or equal to $N$, be its limit. Along this subsequence,
\be
\ket{\psi_{m_k}}=\hat G_{m_k}\ket{C_{m_k}},
\label{psimk}
\ee
and we have $\text{lim}_{k\rightarrow+\infty}D(\ket{\psi_{m_k}},\ket\psi)=0$ and $\text{lim}_{k\rightarrow+\infty}D(\ket{C_{m_k}},\ket C)=0$. Moreover, for all $k\in\mathbb N$,
\be
\ba
D(\hat G_{m_k}\ket C,\ket\psi)&\le D(\hat G_{m_k}\ket C,\ket{\psi_{m_k}})+D(\ket{\psi_{m_k}},\ket\psi)\\
&=D(\hat G_{m_k}\ket C,\hat G_{m_k}\ket{C_{m_k}})+D(\ket{\psi_{m_k}},\ket\psi)\\
&=D(\ket C,\ket{C_{m_k}})+D(\ket{\psi_{m_k}},\ket\psi),
\ea
\ee
where we used the triangular inequality in the first line, Eq.~(\ref{psimk}) in the second line, and the invariance of the trace distance under unitary transformations in the third line. Hence, the sequence $\{\hat G_{m_k}\ket C\}_{k\in\mathbb N}$ converges in trace norm to $\ket\psi$. This shows that we can restrict without loss of generality to a unique core state, with a polynomial stellar function of degree less or equal to $N$, instead of a sequence of such core states.

Let $\ket C$ thus be a core state, with a polynomial stellar function of degree $K$ less or equal to $N$. We write
\be
\ket C=P_C(\hat a^\dag)\ket0=\sum_{n=0}^K{\frac{p_n}{\sqrt{n!}}\ket n},
\label{coreC}
\ee
with $\sum_{n=0}^K{\frac{|p_n|^2}{n!}}=1$. Let us consider a converging sequence $\{\hat G_m\ket C\}_{m\in\mathbb N}$, where $\hat G_m$ are Gaussian operations, and denote $\ket\psi$ its limit. There exists two sequences $\{\xi_m\}_{m\in\mathbb N}$ and $\{\alpha_m\}_{m\in\mathbb N}$, such that for all $m\in\mathbb N$,
\be
\hat G_m=\hat S(\xi_m)\hat D(\alpha_m).
\ee
We write $\xi_m=r_me^{i\theta_m}$, with $r_m\ge0$, for all $m\in\mathbb C$. We may rewrite $\hat G_m=\hat D(\gamma_m)\hat S(\xi_m)$, where for all $m\in\mathbb N$,
\be
\gamma_m=c_{r_m}\alpha_m+s_{r_m}e^{i\theta_m}\alpha_m^*,
\ee
where $c_{r_m}=\cosh(r_m)$ and $s_{r_m}=\sinh(r_m)$. With these notations, we prove the following result:

\begin{lem}
The sequences $\{\xi_m\}_{m\in\mathbb N}$ and $\{\gamma_m\}_{m\in\mathbb N}$ are bounded.
\end{lem}

\begin{proof} 

We first compute an upper bound for the $Q$ function of the state $\hat G_m\ket C$, which we obtain in Eq.~(\ref{boundQ3}). For $m\in\mathbb N$, we have:
\be
\ba
Q_{\hat G_m\ket C}(z)&=Q_{\hat D(\gamma_m)\hat S(\xi_m)\ket C}(z)\\
&=Q_{\hat S(\xi_m)\ket C}(z-\gamma_m)\\
&=\frac{e^{-|z-\gamma_m|^2}}\pi\left|F_{\hat S(\xi_m)\ket C}^\star(z^*-\gamma_m^*)\right|^2,
\ea
\label{QGmC1}
\ee
for all $z\in\mathbb C$. 

\noindent We have
\be
\ba
\hat S(\xi_m)\ket C&=\hat S(\xi_m)P_C(\hat a^\dag)\ket0\\
&=P_C(c_{r_m}\hat a^\dag+s_{r_m}e^{i\theta_m}\hat a)\hat S(\xi_m)\ket0.
\ea
\ee
Hence, with Eq.~(\ref{FSD}) and (\ref{castellar}),
\be
\ba
F_{\hat S(\xi_m)\ket C}^\star(z)&=(1-|t_{r_m}|^2)^{1/4}P_C(c_{r_m}z+s_{r_m}e^{i\theta_m}\partial_z)\cdot e^{-\frac12t_{r_m}e^{-i\theta_m}z^2}\\
&=\frac1{\sqrt{c_{r_m}}}\sum_{n=0}^K{\frac{p_n}{\sqrt{n!}}(c_{r_m}z+s_{r_m}e^{i\theta_m}\partial_z)^n}\cdot e^{-\frac12t_{r_m}e^{-i\theta_m}z^2}.
\ea
\label{FSC0}
\ee
where $t_{r_m}=\tanh(r_m)$.

The Hermite polynomials~\cite{abramowitz1965handbook} satisfy the following recurrence relation
\be
\mathit{He}_{n+1}(z)=z\mathit{He}_n(z)-\partial_z\mathit{He}_n(z),
\label{recH}
\ee
for all $n\ge0$ and all $z\in\mathbb C$, and $\mathit{He}_0=1$. Setting
\be
f_n(z):=e^{\frac12t_{r_m}e^{-i\theta_m}z^2}(c_{r_m}z+s_{r_m}e^{i\theta_m}\partial_z)^n\cdot e^{-\frac12t_{r_m}e^{-i\theta_m}z^2},
\ee
we obtain $f_0(z)=1$, and 
\be
\ba
f_{n+1}(z)&=e^{\frac12t_{r_m}e^{-i\theta_m}z^2}(c_{r_m}z+s_{r_m}e^{i\theta_m}\partial_z)\left[e^{-\frac12t_{r_m}e^{-i\theta_m}z^2}f_n(z)\right]\\
&=\frac z{c_{r_m}}f_n(z)+s_{r_m}e^{i\theta_m}\partial_z f_n(z).
\ea
\ee
Hence, with Eq.~(\ref{recH}), for all $n\ge0$ and all $z\in\mathbb C$,
\be
f_n(z)=\lambda_m^{n/2}\mathit{He}_n\left(\frac z{c_{r_m}\sqrt{\lambda_m}}\right),
\ee
where we have set $\lambda_m=-e^{i\theta_m}t_{r_m}$. With Eq.~(\ref{FSC0}) we thus obtain
\be
\ba
F_{\hat S(\xi_m)\ket C}^\star(z)&=\frac1{\sqrt{c_{r_m}}}\sum_{n=0}^K{\frac{p_n}{\sqrt{n!}}f_n(z)}\cdot e^{-\frac12t_{r_m}e^{-i\theta_m}z^2}\\
&=\frac1{\sqrt{c_{r_m}}}\sum_{n=0}^K{\frac{p_n\lambda_m^{n/2}}{\sqrt{n!}}\mathit{He}_n\left(\frac z{c_{r_m}\sqrt{\lambda_m}}\right)}e^{-\frac12t_{r_m}e^{-i\theta_m}z^2}.
\label{FSC}
\ea
\ee
From this and Eq.~(\ref{QGmC1}) we deduce
\begin{align}
\nonumber Q_{\hat G_m\ket C}(z)&=\frac{e^{-|z-\gamma_m|^2}}{\pi c_{r_m}}\left|\sum_{n=0}^K{\frac{p_n\lambda_m^{n/2}}{\sqrt{n!}}\mathit{He}_n\left(\frac{z^*-\gamma_m^*}{c_{r_m}\sqrt{\lambda_m}}\right)}e^{-\frac12t_{r_m}e^{-i\theta_m}(z^*-\gamma_m^*)^2}\right|^2\\
&\le\frac{e^{-|z-\gamma_m|^2}}{\pi c_{r_m}}\left|e^{-\frac12t_{r_m}e^{-i\theta_m}(z^*-\gamma_m^*)^2}\right|^2\sum_{n=0}^K{\frac{|p_n|^2}{n!}}\cdot\sum_{n=0}^K{\left|\lambda_m^{n/2}\mathit{He}_n\left(\frac{z^*-\gamma_m^*}{c_{r_m}\sqrt{\lambda_m}}\right)\right|^2}\\
\nonumber&=\frac1{\pi c_{r_m}}e^{-|z-\gamma_m|^2-\frac12t_{r_m}[e^{i\theta_m}(z-\gamma_m)^2+e^{-i\theta_m}(z^*-\gamma_m^*)^2]}\sum_{n=0}^K{\left|t_{r_m}^{n/2}\mathit{He}_n\left(\frac{z-\gamma_m}{c_{r_m}\sqrt{\lambda_m^*}}\right)\right|^2},
\end{align}
where we used Cauchy-Schwarz inequality in the second line, $|\lambda_m|=t_{r_m}$ and the fact that the coefficients of $\mathit{He}_n$ are real in the third line. 

\noindent Setting
\be
\alpha_m(z):=-\frac{ie^{\frac12i\theta_m}}{c_{r_m}}(z-\gamma_m),
\label{zm}
\ee
for all $m\in\mathbb N$ and for all $z\in\mathbb C$, we obtain
\be
\ba
Q_{\hat G_m\ket C}(z)&\le\frac1{\pi c_{r_m}}e^{-|z-\gamma_m|^2-\frac12t_{r_m}[e^{i\theta_m}(z-\gamma_m)^2+e^{-i\theta_m}(z^*-\gamma_m^*)^2]}\sum_{n=0}^K{\left|t_{r_m}^{n/2}\mathit{He}_n\left(\frac{e^{-i\theta_m}\alpha_m(z)}{\sqrt{t_{r_m}}}\right)\right|^2}\\
&=\frac1{\pi c_{r_m}}e^{-c_{r_m}^2|\alpha_m(z)|^2+\frac12c_{r_m}s_{r_m}[\alpha_m^2(z)+\alpha_m^{*2}(z)]}\sum_{n=0}^K{\left|t_{r_m}^{n/2}\mathit{He}_n\left(\frac{e^{-i\theta_m}\alpha_m(z)}{\sqrt{t_{r_m}}}\right)\right|^2}\\
&=\frac1{\pi c_{r_m}}e^{-c_{r_m}(c_{r_m}-s_{r_m})x_m^2(z)}e^{-c_{r_m}(c_{r_m}+s_{r_m})y_m^2(z)}\sum_{n=0}^K{\left|t_{r_m}^{n/2}\mathit{He}_n\left(\frac{e^{-i\theta_m}\alpha_m(z)}{\sqrt{t_{r_m}}}\right)\right|^2},
\ea
\label{boundQ1}
\ee
where $\alpha_m(z)=x_m(z)+iy_m(z)$. For all $r\in\mathbb R$,
\be
c_r(c_r-s_r)=\frac12(1+e^{-2r})>\frac12,
\ee
and
\be
c_r(c_r+s_r)=\frac12(1+e^{2r})>\frac12,
\ee
so with Eq.~(\ref{boundQ1}) we obtain
\be
Q_{\hat G_m\ket C}(z)\le\frac1{\pi c_{r_m}}e^{-\frac12|\alpha_m(z)|^2}\sum_{n=0}^K{\left|t_{r_m}^{n/2}\mathit{He}_n\left(\frac{e^{-i\theta_m}\alpha_m(z)}{\sqrt{t_{r_m}}}\right)\right|^2}.
\label{boundQ2}
\ee
Finally, we obtain the following bound for all $n\in\{0,\dots,K\}$:
\be
\ba
\left|t_{r_m}^{n/2}\mathit{He}_n\left(\frac{e^{-i\theta_m}\alpha_m(z)}{\sqrt{t_{r_m}}}\right)\right|&=\left|t_{r_m}^{n/2}\sum_{k=0}^{\lfloor\frac n2\rfloor}{\frac{(-1)^kn!}{2^kk!(n-2k)!}\left(\frac{e^{-i\theta_m}\alpha_m(z)}{\sqrt{t_{r_m}}}\right)^{n-2k}}\right|\\
&\le\sum_{k=0}^{\lfloor\frac n2\rfloor}{\frac{n!}{2^kk!(n-2k)!}t_{r_m}^k|\alpha_m(z)|^{n-2k}}\\
&\le\sum_{k=0}^{\lfloor\frac n2\rfloor}{\frac{n!}{2^kk!(n-2k)!}|\alpha_m(z)|^{n-2k}},
\ea
\label{boundHe}
\ee
for all $m\in\mathbb N$ and all $z\in\mathbb C$. Let us define for brevity the polynomial
\be
T(X):=\sum_{n=0}^K{\left(\sum_{k=0}^{\lfloor\frac n2\rfloor}{\frac{n!}{2^kk!(n-2k)!}X^{n-2k}}\right)^2}.
\ee
Plugging Eq.~(\ref{boundHe}) in Eq.~(\ref{boundQ2}) yields
\be
Q_{\hat G_m\ket C}(z)\le\frac1{\pi c_{r_m}}e^{-\frac12|\alpha_m(z)|^2}T(|\alpha_m(z)|),
\label{boundQ3}
\ee
for all $m\in\mathbb N$ and all $z\in\mathbb C$.

\medskip

With this bound on the $Q$ function obtained, we may now prove that the sequences $\{\xi_m\}_{m\in\mathbb N}=\{r_me^{i\theta_m}\}_{m\in\mathbb N}$ and $\{\gamma_m\}_{m\in\mathbb N}$ are bounded.

Assuming that $\{r_m\}_{m\in\mathbb N}$ is unbounded implies that it has a subsequence $\{r_{m_k}\}_{k\in\mathbb N}$ going to infinity. Since the function $x\mapsto e^{-\frac12x^2}T(x)$ is bounded, $Q_{\hat G_{m_k}\ket C}(z)\rightarrow0$ for all $z\in\mathbb C$ when $k\rightarrow+\infty$ by Eq.~(\ref{boundQ3}). But $Q_{\hat G_{m_k}\ket C}(z)\rightarrow Q_\psi(z)$ for all $z\in\mathbb C$ when $k\rightarrow+\infty$, by property of the convergence in trace norm. This would imply $Q_\psi(z)=0$ for all $z\in\mathbb C$, which is impossible since $\ket\psi$ is normalised. Hence $\{r_m\}_{m\in\mathbb N}$ is a bounded sequence, and so is $\{\xi_m\}_{m\in\mathbb N}$.

With the same reasoning, if $\{|\alpha_m(z)|\}_{m\in\mathbb N}$ was unbounded for all $z\in\mathbb C$, this would imply by Eq.~(\ref{boundQ3}) that $Q_\psi(z)=0$ for all $z\in\mathbb C$, giving the same contradiction. Hence, there exists $z_0\in\mathbb C$ such that the sequence $\{|\alpha_m(z_0)|\}_{m\in\mathbb N}$ is bounded. By Eq.~(\ref{zm}), this implies that the sequence $\{\gamma_m\}_{m\in\mathbb N}$ is also bounded, since the sequence $\{r_m\}_{m\in\mathbb N}$ is bounded.

\end{proof}

\noindent The sequences $\{\xi_m\}_{m\in\mathbb N}$ and $\{\gamma_m\}_{m\in\mathbb N}$ being bounded, one can consider simultaneously converging subsequences $\{\xi_{m_k}\}_{k\in\mathbb N}$ and $\{\gamma_{m_k}\}_{k\in\mathbb N}$. We write $\xi=re^{i\theta}=\lim_{k\to\infty}{\xi_{m_k}}$ and $\gamma=\lim_{k\to\infty}{\gamma_{m_k}}$. On one hand, we have
\be
\ba
F_{\hat G_{m_k}\ket C}^\star(z)&=F_{\hat D(\gamma_{m_k})\hat S(\xi_{m_k})\ket C}^\star(z)\\
&=e^{\gamma_{m_k}z-\frac12|\gamma_{m_k}|^2}F_{\hat S(\xi_{m_k})\ket C}^\star(z-\gamma_{m_k}^*)\\
&=\frac1{\sqrt{c_{r_{m_k}}}}\sum_{n=0}^K{\frac{p_n\lambda_{m_k}^{n/2}}{\sqrt{n!}}\mathit{He}_n\left(\frac{z-\gamma_{m_k}^*}{c_{r_{m_k}}\sqrt{\lambda_{m_k}}}\right)}e^{-\frac12t_{r_{m_k}}e^{-i\theta_{m_k}}(z-\gamma_{m_k}^*)^2+\gamma_{m_k}z-\frac12|\gamma_{m_k}|^2},
\ea
\label{compF}
\ee
for all $k\in\mathbb N$ and all $z\in\mathbb C$, where we have used Eq.~(\ref{Fdisplaced}) in the second line, where $\lambda_{m_k}=-e^{i\theta_{m_k}}t_{r_{m_k}}$, and where we have used Eq.~(\ref{FSC}) in the last line. Setting $\lambda=-e^{i\theta}t_r$, we obtain
\be
\ba
\lim_{k\to\infty}{F_{\hat G_{m_k}\ket C}^\star(z)}&=\frac1{\sqrt{c_r}}\sum_{n=0}^K{\frac{p_n\lambda^{n/2}}{\sqrt{n!}}\mathit{He}_n\left(\frac{z-\gamma^*}{c_r\sqrt{\lambda}}\right)}e^{-\frac12t_re^{-i\theta}(z-\gamma^*)^2+\gamma z-\frac12|\gamma|^2}\\
&=F^\star_{\hat G\ket C}(z),
\ea
\label{limF1}
\ee
for all $z\in\mathbb C$, where $\hat G=\hat D(\gamma)\hat S(\xi)$, and where the second line comes from reversing the calculations of Eq.~(\ref{compF}). On the other hand, for all $z\in\mathbb C$,
\begin{align}
\nonumber\lim_{k\to\infty}{F_{\hat G_{m_k}\ket C}^\star(z)}&=e^{\frac12|z|^2}\lim_{k\to\infty}{\braket{z^*|\hat G_{m_k}|C}}\\
\label{limF2}&=e^{\frac12|z|^2}\braket{z^*|\psi}\\
\nonumber&=F_\psi^\star(z),
\end{align}
by property of the convergence in trace norm. 
Combining Eq.~(\ref{limF1}) and Eq.~(\ref{limF2}) yields
\be
F_\psi^\star(z)=F_{\hat G\ket C}^\star(z),
\ee
for all $z\in\mathbb C$. By Lemma~1, this implies that $\ket\psi=\hat G\ket C\in R_K$. This shows that $\overline{\bigcup_{K=0}^N{R_K}}=\bigcup_{K=0}^N{R_K}$, so $\overline{R_N}\subset\bigcup_{K=0}^N{R_K}$, which concludes the proof.

\end{mdframed}
\end{proof}

\noindent This result implies that the set $\bigcup_{0\le K\le N}{R_K}$, containing the states of stellar rank smaller or equal to $N$, is a closed set in $\Hi$ for the trace norm, for all $N\in\mathbb N$. In particular, since all ranks of the stellar hierarchy are disjoint, for any state of finite rank $N$, there is no sequence of states of strictly lower stellar rank converging to it. Each state of a given finite stellar rank is thus isolated from all the lower stellar ranks, i.e., there is a ball around it in trace norm which only contains states of equal or higher stellar rank. 

Moreover, each state of infinite stellar rank is isolated from states of finite stellar rank lower than $N$, for all $N\in\mathbb N^*$, i.e., there is a ball around it in trace norm which only contains states of stellar rank higher than $N$.

On the other hand, with the other inclusion, no state of a given finite stellar rank is isolated from any equal or higher stellar rank, i.e., one can always find a sequence of states of any higher rank converging to this state in trace norm. 

\medskip

\noindent We also prove the following density result: 

\begin{lem} \label{lem:dense}
The set of states of finite stellar rank is dense for the trace norm in the set of normalised pure single-mode states:
\be
\overline{\underset{N\in\mathbb N}{\bigcup}{R_N}}=\mathcal H,
\ee
where $\overline X$ denotes the closure of $X$ for the trace norm in the set of normalised states in $\mathcal H$.
\end{lem}

\begin{proof}
\begin{mdframed}[linewidth=1.5,topline=false,rightline=false,bottomline=false]

Recall that the set of normalised pure single-mode states is closed for the trace norm in the whole Hilbert space, since it is the reciprocal image of $\{1\}$ by the trace norm, which is Lipschitz continuous---with Lipschitz constant $1$---hence continuous.

Let $\ket\psi\in\mathcal H$ be a normalised state. We consider the sequence of normalised cut-off states
\be
\ket{\psi_m}=\frac1{\sqrt{\mathcal{N}_m}}\sum_{n=0}^m{\psi_n\ket n},
\ee
where $\mathcal{N}_m=\sum_{n=0}^m{|\psi_n|^2}$ is a normalising factor (non-zero for $m$ large enough). All the states $\ket{\psi_m}$ have a finite support over the Fock basis, so their stellar function is a polynomial. Hence $\{\ket{\psi_m}\}_{m\in\mathbb N}\in\bigcup_{N\in\mathbb N}{R_N}$.

Moreover, for all $m\in\mathbb N$,
\begin{align}
\nonumber D(\psi_m,\psi)&=\sqrt{1-|\braket{\psi_m|\psi}|^2}\\ \displaybreak
&=\sqrt{1-\sum_{n=0}^m{|\psi_n|^2}}\\
\nonumber&=\sqrt{\sum_{n\ge m+1}{|\psi_n|^2}},
\end{align}
where we used that $\ket\psi$ and $\ket{\psi_m}$ are pure states in the first line, and the fact that $\ket\psi$ is normalised in the third line. Furthermore, $\sum_{n\ge m+1}{|\psi_n|^2}\rightarrow0$ when $m\rightarrow+\infty$, because $\ket\psi$ is normalised. Hence, $\{\ket{\psi_m}\}_{m\in\mathbb N}$ converges in trace norm to $\ket\psi$, which concludes the proof.

\end{mdframed}
\end{proof}

\noindent This result implies that states of infinite stellar rank are not isolated from lower stellar ranks, unlike states of finite stellar rank. Given a state of infinite stellar rank, there always exists a sequence of states of finite stellar ranks converging to it. However, the ranks of the states in this sequence have to go to infinity, since by Theorem~\ref{topology} states of infinite stellar rank are isolated from states of finite stellar rank lower than $N$, for all $N\in\mathbb N^*$.

\medskip

\noindent The consequences of Theorem~\ref{topology} and Lemma~\ref{lem:dense} for the robustness are summarised with the following result:

\begin{coro}\label{coro:robust}
For all $\ket\psi\in\bigcup_{N\in\mathbb N}{R_N}$ and for all $k\in\overline{\mathbb N}^*$,
\begin{subnumcases}{}
   R_k^\star(\psi)>0 & for $k\le r^\star(\psi)$, \label{systemrobust1}
   \\
   R_k^\star(\psi)=0 & for $k>r^\star(\psi)$. \label{systemrobust2}
\end{subnumcases}
In particular, states of finite stellar rank are robust: for all states $\ket\psi\in\bigcup_{N\in\mathbb N}{R_N}$, we have $R^\star(\psi)=R^\star_{r^\star(\psi)}(\psi)>0$. 

\medskip

\noindent For all $\ket\psi\in R_\infty$ and for all $k\in\overline{\mathbb N}^*$,
\begin{subnumcases}{}
   R_k^\star(\psi)>0 & for $k\in\mathbb N$, \label{systemnotrobust1}
   \\
   R_k^\star(\psi)=0 & for $k=\infty$. \label{systemnotrobust2}
\end{subnumcases}
In particular, states of infinite stellar rank are not robust: for all states $\ket\psi\in R_\infty$, we have $R^\star(\psi)=R^\star_\infty(\psi)=0$.
\end{coro}

\noindent Eqs.~(\ref{systemrobust1}), (\ref{systemrobust2}) and (\ref{systemnotrobust1}) are deduced from Theorem~\ref{topology}, and Eq.~(\ref{systemnotrobust2}) is deduced from Lemma~\ref{lem:dense}. 

This result implies that the robust states (i.e., $R^\star>0$) are exactly the non-Gaussian states of finite stellar rank. When considering imperfect single-mode non-Gaussian state engineering, one may thus restrict to states of finite stellar rank, which by Theorem~\ref{finitez} are obtained uniquely by a finite number of single-photon additions to a Gaussian state. Alternatively, one may also describe such states using Theorem~\ref{Gconv} as finite superpositions of equally displaced and squeezed number states, or equivalently as Gaussian-convertible to core states. Engineering of such states has recently been considered in~\cite{su2019conversion}, by photon detection of Gaussian states.

Moreover, for $k\in\mathbb N^*$, the states that are robust with respect to states of stellar rank lower than $k$ (i.e., $R^\star_k>0$) thus are the states $\ket\psi$ such that $r^\star(\psi)\ge k$.

\subsection{Computing the robustness}
\label{sec:computeR}

Importantly, the $k$-robustness is state-dependent and the following result gives a simple expression. Let us define, for all $n\in\mathbb N$,
\be
\Pi_n=\sum_{m=0}^n{\ket m\!\bra m}
\ee
the projector onto the subspace spanned by the Fock states $\ket0,\dots,\ket n$.

\begin{theo} \label{th:robust}
Let $k\in\mathbb N^*$ and let $\ket\psi\in\Hi$. Then,
\be
R^\star_k(\psi)=\sqrt{1-\sup_{\hat G\in\mathcal G}{\Tr\left[\Pi_{k-1}\hat G\ket\psi\!\bra\psi\hat G^\dag\right]}},
\ee
where the supremum is over Gaussian unitary operations. Moreover, assuming the optimisation yields a Gaussian operation $\hat G_0$, an optimal approximating state is
\be
\hat G_0^\dag\left(\frac{\Pi_{k-1}\hat G_0\ket\psi}{\left\|\Pi_{k-1}\hat G_0\ket\psi\right\|}\right).
\ee
\end{theo}

\begin{proof}
\begin{mdframed}[linewidth=1.5,topline=false,rightline=false,bottomline=false]

From Lemma~\ref{lem:Rfide} and in particular Eq.~(\ref{robustnessinter}) we have
\be
R^\star_k(\psi)=\sqrt{1-\sup_{r^\star(\phi)<k}{|\braket{\phi|\psi}|^2}}.
\label{Rstarkinter}
\ee
By Theorem~\ref{Gconv}, for any pure state $\ket\phi$ such that $r^\star(\phi)<k$, there exist a normalised core state $\ket{C_\phi}$ of stellar rank lower than $k$ and a Gaussian operation $\hat G_\phi$ such that
\be
\ket\phi=\hat G_\phi\ket{C_\phi}.
\label{phiGC}
\ee
We obtain
\be
\ba
|\braket{\phi|\psi}|^2&=|\braket{C_\phi|\hat G_\phi^\dag|\psi}|^2\\
&=|\braket{C_\phi|\Pi_{k-1}\hat G_\phi^\dag|\psi}|^2\\
&\le|\braket{C_\phi|C_\phi}|^2|\braket{\psi|\hat G_\phi\Pi_{k-1}\hat G_\phi^\dag|\psi}|^2\\
&=\Tr\left[\Pi_{k-1}\hat G_\phi^\dag\ket\psi\!\bra\psi\hat G_\phi\right],
\ea
\label{boundoverlapPik}
\ee
where we used $\ket{C_\phi}=\Pi_{k-1}\ket{C_\phi}$ in the second line, since $\ket{C_\phi}$ is a core state of stellar rank lower than $k$ (hence its support is contained in the support of $\Pi_{k-1}$), Cauchy-Schwarz inequality in the third line and $|\braket{C_\phi|C_\phi}|^2=1$ in the last line. This upperbound is attained if
\be
\ket{C_\phi}=\frac{\Pi_{k-1}\hat G_\phi^\dag\ket\psi}{\sqrt{\Tr\left[\Pi_{k-1}\hat G_\phi^\dag\ket\psi\!\bra\psi\hat G_\phi\right]}},
\label{coreapprox}
\ee
which is indeed a normalised core state of stellar rank lower than $k$.

With Eqs.~(\ref{Rstarkinter}) and~(\ref{boundoverlapPik}), the robustness of the state $\ket\psi$ is then given by
\be
\ba
R^\star_k(\psi)&=\sqrt{1-\sup_{\hat G_\phi\in\mathcal G}{\Tr\left[\Pi_{k-1}\hat G_\phi^\dag\ket\psi\!\bra\psi\hat G_\phi\right]}}\\
&=\sqrt{1-\sup_{\hat G\in\mathcal G}{\Tr\left[\Pi_{k-1}\hat G\ket\psi\!\bra\psi\hat G^\dag\right]}},
\ea
\ee
where the supremum is over Gaussian unitary operations and where we used the fact that the set of Gaussian unitary operations is invariant under adjoint in the second line. With Eq.~(\ref{phiGC}), assuming the optimisation yields an optimal Gaussian unitary $\hat G_0$, an optimal approximating state is $\ket\phi=\hat G_\phi\ket{C_\phi}$, where $\hat G_\phi=\hat G_0^\dag$ and where $\ket{C_\phi}$ is given by Eq.~(\ref{coreapprox}). Namely,
\be
\ket\phi=\hat G_0^\dag\left(\frac{\Pi_{k-1}\hat G_0\ket\psi}{\left\|\Pi_{k-1}\hat G_0\ket\psi\right\|}\right),
\ee
i.e., $\hat G_0\ket\phi$ is the renormalised truncation of $\hat G_0\ket\psi$ at photon number $k-1$.

\end{mdframed}
\end{proof}

\noindent From Theorem~\ref{th:robust}, the robustness profile $\mathcal R(\psi)=\left(R^\star_k\right)_{k\in\mathbb N^*}$ is a non-increasing sequence for any state $\ket\psi$, and each term in the sequence may be obtained with an optimisation over two complex parameters. 

In particular, with the Hermite polynomials
\be
\ba
\mathit{He}_m(z)&=(-1)^me^{\frac12z^2}\partial_z^m e^{-\frac12z^2}\\
&=\sum_{p=0}^{\left\lfloor\frac m2\right\rfloor}{\frac{m!(-1)^p}{2^pp!(m-2p)!}z^{m-2p}},
\ea
\label{Hermite}
\ee
for all $m\in\mathbb N$ and all $z\in\mathbb C$, the robustness of cat states has the following expression:

\begin{coro}\label{coro:Rcat}
Let $k\in\mathbb N^*$ and let $\alpha\in\mathbb C$. Then, writing $c_x=\cosh x$, $s_x=\sinh x$, and $t_x=\tanh x$ for brevity,
\be
R^\star_k(\text{cat}_\alpha^+)=\sqrt{1-\sup_{\xi=re^{i\theta},\beta\in\mathbb C}{\frac{e^{-|\beta|^2}}{4c_rc_{|\alpha|^2}}\sum_{m=0}^{k-1}{\frac{t_r^m}{m!}\left|u_m^+(|\alpha|,\xi,\beta)\right|^2}}},
\ee
and
\be
R^\star_k(\text{cat}_\alpha^-)=\sqrt{1-\sup_{\xi=re^{i\theta},\beta\in\mathbb C}{\frac{e^{-|\beta|^2}}{4c_rs_{|\alpha|^2}}\sum_{m=0}^{k-1}{\frac{t_r^m}{m!}\left|u_m^-(|\alpha|,\xi,\beta)\right|^2}}},
\ee
where
\be
u_m^\pm(|\alpha|,\xi,\beta):=e^{-|\alpha|\beta^*+\frac12t_re^{i\theta}(|\alpha|+\beta)^2}\mathit{He}_m\left(\frac{|\alpha|+\beta}{\sqrt{c_rs_r}}e^{i\theta/2}\right)\pm e^{|\alpha|\beta^*+\frac12t_re^{i\theta}(\beta-|\alpha|)^2}\mathit{He}_m\left(\frac{\beta-|\alpha|}{\sqrt{c_rs_r}}e^{i\theta/2}\right).
\ee
\end{coro}

\begin{proof}
\begin{mdframed}[linewidth=1.5,topline=false,rightline=false,bottomline=false]

Let $\alpha\in\mathbb C$. We have
\be
\ket{\text{cat}^\pm_\alpha}=\frac1{\sqrt{\mathcal N^\pm_\alpha}}(\ket\alpha\pm\ket{-\alpha}),
\ee
where $\mathcal N^\pm_\alpha=2(1\pm e^{-2|\alpha|^2})$. By Theorem~\ref{th:robust},
\be
R^\star_k(\text{cat}_\alpha^\pm)=\sqrt{1-\sup_{\hat G\in\mathcal G}{\Tr\left[\Pi_{k-1}\hat G\ket{\text{cat}_\alpha^\pm}\!\bra{\text{cat}_\alpha^\pm}\hat G^\dag\right]}}.
\label{Rcatgen}
\ee
for all $k\in\mathbb N^*$, where $\Pi_{k-1}$ in the projector onto the Fock basis with less the $k-1$ photons. We have
\be
\Tr\left[\Pi_{k-1}\hat G\ket{\text{cat}_\alpha^\pm}\!\bra{\text{cat}_\alpha^\pm}\hat G^\dag\right]=\sum_{m=0}^{k-1}{\left|\braket{m|\hat G|\text{cat}_\alpha^\pm}\right|^2}.
\label{interRcat0}
\ee
Setting $\hat G=\hat S(\xi)\hat D(\beta)$, for $\xi=re^{i\theta},\beta\in\mathbb C$, we obtain
\be
\ba
\braket{m|\hat G|\text{cat}_\alpha^\pm}&=\frac1{\sqrt{\mathcal N^\pm_\alpha}}\left(\braket{m|\hat S(\xi)\hat D(\beta)|\alpha}\pm\braket{m|\hat S(\xi)\hat D(\beta)|-\alpha}\right)\\
&=\frac1{\sqrt{\mathcal N^\pm_\alpha m!}}\left(e^{\frac12(\alpha^*\beta-\alpha\beta^*)}\braket{0|\hat a^m\hat S(\xi)\hat D(\alpha+\beta)|0}\pm e^{\frac12(\alpha\beta^*-\alpha^*\beta)}\braket{0|\hat a^m\hat S(\xi)\hat D(\beta-\alpha)|0}\right).
\ea
\ee
Switching to the stellar representation we obtain
\be
\ba
\braket{m|\hat G|\text{cat}_\alpha^\pm}&=\frac1{\sqrt{\mathcal N^\pm_\alpha m!}}\left[\partial_z^m e^{\frac12(\alpha^*\beta-\alpha\beta^*)}G^\star_{\xi,\alpha+\beta}(z)\pm\partial_z^m e^{\frac12(\alpha\beta^*-\alpha^*\beta)}G^\star_{\xi,-\alpha+\beta}(z)\right]_{z=0}\\
&=\frac{e^{-\frac12(|\alpha|^2+|\beta|^2)}}{\sqrt{c_r\mathcal N^\pm_\alpha m!}}\Bigg[e^{-\alpha\beta^*+\frac12t_re^{i\theta}(\alpha+\beta)^2}\partial_z^m e^{-\frac12e^{-i\theta}t_rz^2+\frac{\alpha+\beta}{c_r}z}\\
&\quad\quad\quad\quad\quad\quad\quad\pm e^{\alpha\beta^*+\frac12t_re^{i\theta}(\beta-\alpha)^2}\partial_z^m e^{-\frac12e^{-i\theta}t_rz^2+\frac{\beta-\alpha}{c_r}z}\Bigg]_{z=0},
\ea
\label{interRcat}
\ee
where we used $\braket{0|\psi}=F^{\star}_\psi(0)$ and Eq.~(\ref{FSD}). With Eq.~(\ref{Hermite}) we have
\be
\left[\partial_z^m e^{-\frac12az^2+bz}\right]_{z=0}=a^{m/2}\mathit{He}_m\left(\frac b{\sqrt a}\right),
\ee
where $\mathit{He}_m$ is the $m^{th}$ Hermite polynomial.

\noindent With Eq.~(\ref{interRcat}) we obtain
\be
\ba
\left|\braket{m|\hat G|\text{cat}_\alpha^\pm}\right|^2&=\frac{e^{-(|\alpha|^2+|\beta|^2)}t_r^m}{c_r\mathcal N^\pm_\alpha m!}\Bigg|e^{-\alpha\beta^*+\frac12t_re^{i\theta}(\alpha+\beta)^2}\mathit{He}_m\left(\frac{\alpha+\beta}{\sqrt{c_rs_r}}e^{i\theta/2}\right)\\
&\quad\quad\quad\quad\quad\quad\quad\quad\pm e^{\alpha\beta^*+\frac12t_re^{i\theta}(\beta-\alpha)^2}\mathit{He}_m\left(\frac{\beta-\alpha}{\sqrt{c_rs_r}}e^{i\theta/2}\right)\Bigg|^2.
\ea
\label{interRcat2}
\ee
Combining Eqs.~(\ref{Rcatgen}), (\ref{interRcat0}) and (\ref{interRcat2}) yields
\be
R^\star_k(\text{cat}_\alpha^\pm)=\sqrt{1-\sup_{\xi=re^{i\theta},\beta\in\mathbb C}{\frac{e^{-(|\alpha|^2+|\beta|^2)}}{c_r\mathcal N^\pm_\alpha}\sum_{m=0}^{k-1}{\frac{t_r^m}{m!}\left|u_m^\pm(\alpha,\xi,\beta)\right|^2}}},
\ee
where we have set
\be
u_m^\pm(\alpha,\xi,\beta):=e^{-\alpha\beta^*+\frac12t_re^{i\theta}(\alpha+\beta)^2}\mathit{He}_m\left(\frac{\alpha+\beta}{\sqrt{c_rs_r}}e^{i\theta/2}\right)\pm e^{\alpha\beta^*+\frac12t_re^{i\theta}(\beta-\alpha)^2}\mathit{He}_m\left(\frac{\beta-\alpha}{\sqrt{c_rs_r}}e^{i\theta/2}\right).
\ee
Since the robustness is invariant under Gaussian operations, $R^\star_k(\text{cat}^\pm_\alpha)$ does not depend on the phase of $\alpha$, since one can map a cat state of amplitude $\alpha$ to a cat state of amplitude $e^{i\phi}\alpha$ through a Gaussian rotation (this corresponds to mapping $\beta$ to $e^{i\phi}\beta$ and $\theta$ to $\theta-2\phi$ in the previous expressions). Hence, we may assume without loss of generality that $\alpha\in\mathbb R$ and replace $\alpha$ by $|\alpha|$.
With
\be
c_{|\alpha|^2}=\frac{\mathcal N^+_\alpha}{4e^{-|\alpha|^2}}\quad\text{and}\quad s_{|\alpha|^2}=\frac{\mathcal N^-_\alpha}{4e^{-|\alpha|^2}},
\ee
this concludes the proof.

\end{mdframed}
\end{proof}

\begin{figure}[h!]
	\begin{center}
		\includegraphics[width=0.8\columnwidth]{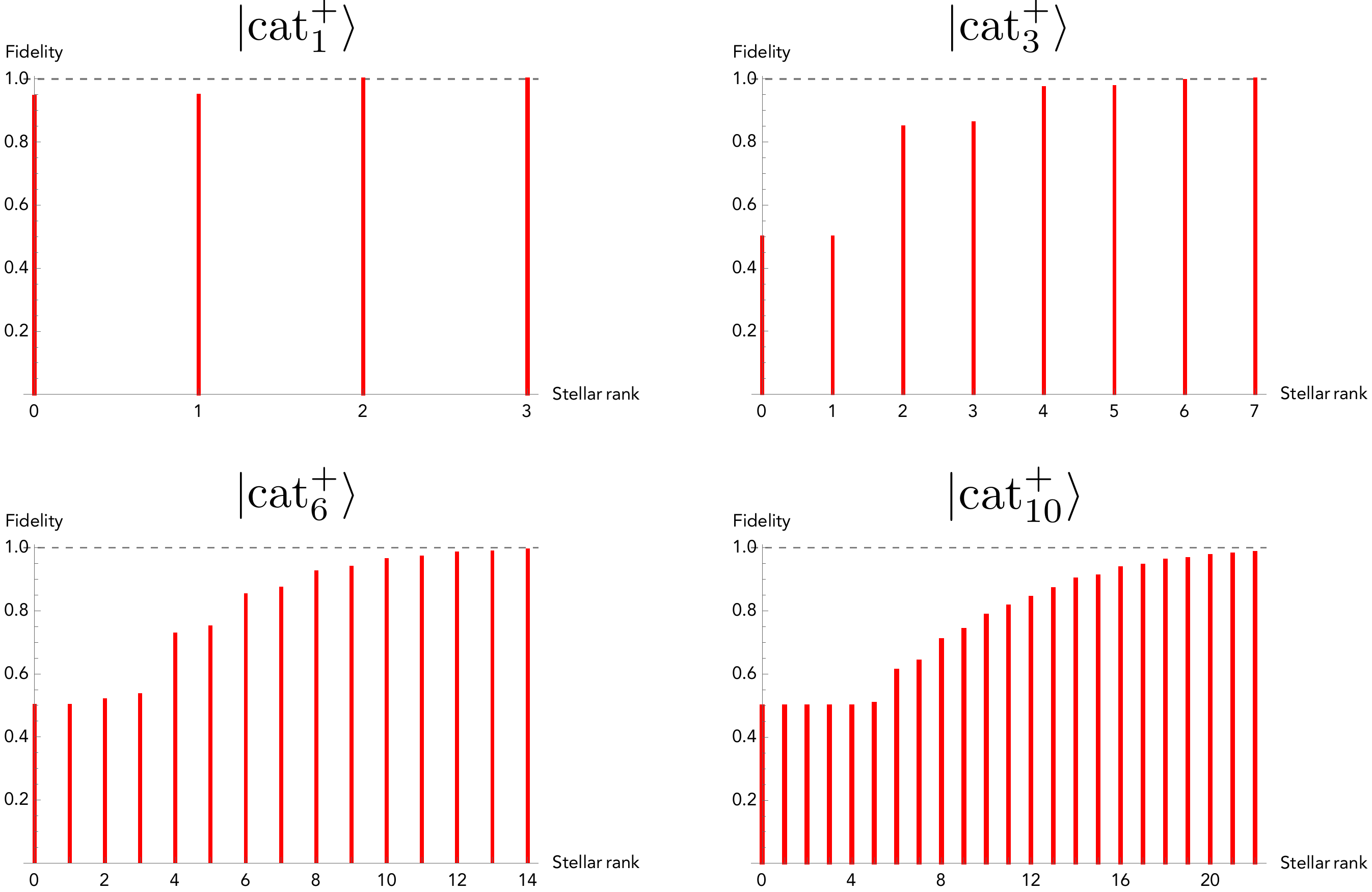}
		\caption{Achievable fidelities for target cat$^+$ states of amplitudes $1$, $3$, $6$ and $10$. For each rank $k\in\mathbb N^*$, the vertical line depicts the achievable fidelities between the target state and states of rank $k$.}
		\label{fig:Rcatp}
	\end{center}
\end{figure}

\begin{figure}[h!]
	\begin{center}
		\includegraphics[width=0.8\columnwidth]{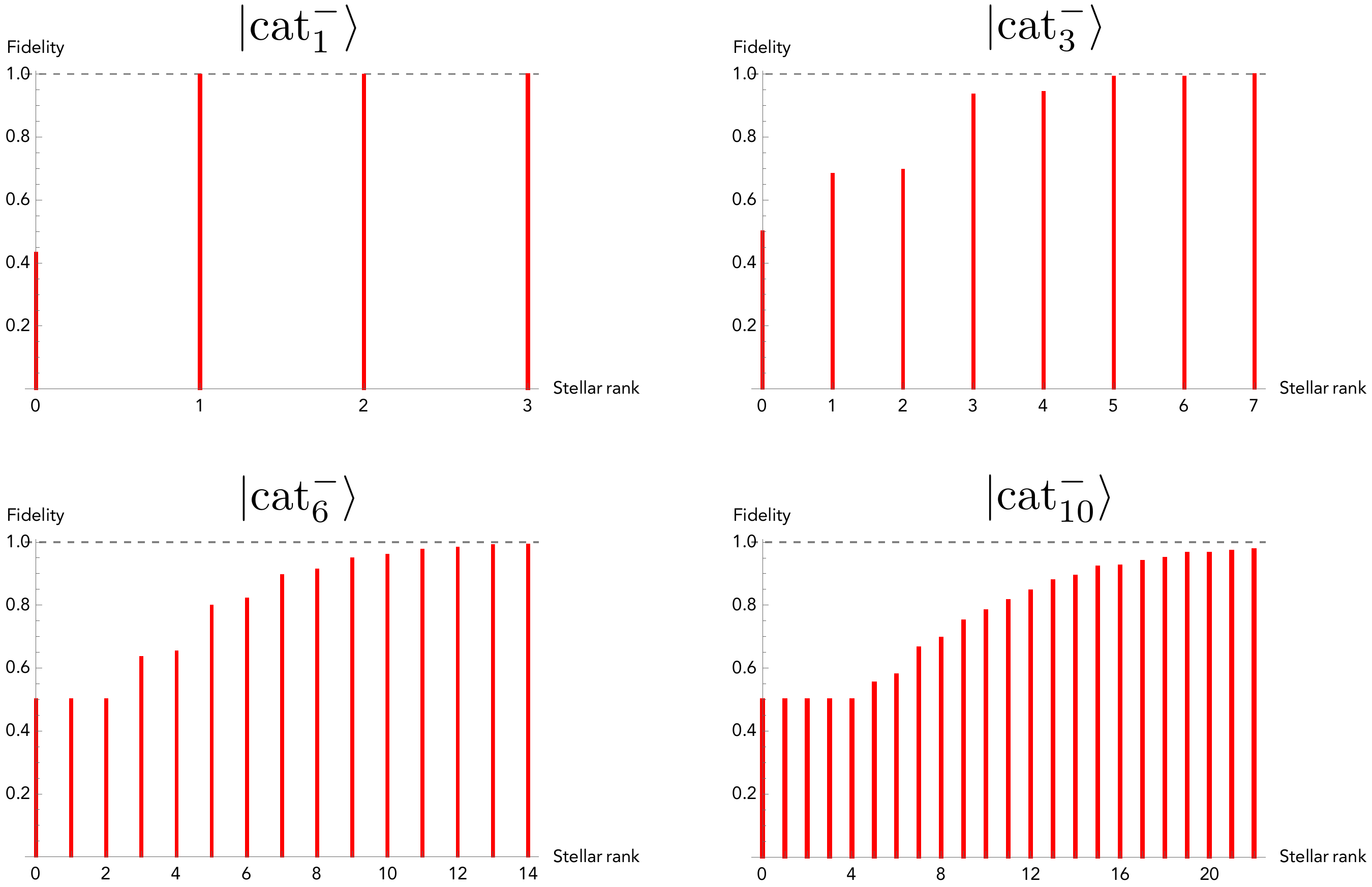}
		\caption{Achievable fidelities for target cat$^-$ states of amplitudes $1$, $3$, $6$ and $10$. For each rank $k\in\mathbb N^*$, the vertical line depicts the achievable fidelities between the target state and states of rank $k$.}
		\label{fig:Rcatm}
	\end{center}
\end{figure}

\noindent By Lemma~\ref{lem:Rfide}, the sequence of maximum achievable fidelities for each rank $k-1$ with a given target state is obtained from the robustness profile with $F=1-R^{\star2}_k$. We have computed numerically the values of $R^\star_k(\text{cat}_\alpha^+)$ and $R^\star_k(\text{cat}_\alpha^-)$ for different values of $k$ and $\alpha$ and the corresponding achievable fidelities are depicted in Fig.~\ref{fig:Rcatp} and Fig.~\ref{fig:Rcatm}. For each rank, if $\rho$ denotes a state for which the maximum fidelity is achieved, then any lower fidelity may be obtain by considering the states $\rho_p=p\ket{\perp}\bra{\perp}+(1-p)\rho$, for $0\le p\le1$, where $\ket{\perp}$ is a coherent state orthogonal to the target state (which exists by Theorem~\ref{Husimi}, since the target state is non-Gaussian).
From the numerics and the obtained profiles of the cat states, we make various observations:
\begin{itemize}
\item
The main difference between low amplitude cat$^+$ and cat$^-$ states is that the former are easier to approximate by Gaussian states than the latter: at low amplitude, cat$^+$ states are closer to the vacuum while cat$^-$ states are closer to the single photon Fock state.
\item
High amplitude cat states are `more non-Gaussian' than low amplitude cat states, in the sense that one needs more photon additions to approximate them to the same precision.
\item 
The maximum achievable fidelity increases more from odd to even ranks (resp.\@ even to odd ranks) than from even to odd ranks (resp.\@ odd to even ranks) for cat$^+$ states (resp.\@ cat$^-$ states). This is due to cat$^+$ states (resp.\@ cat$^-$ states) having support only on even (resp.\@ odd) Fock states.
\item
For each given amplitude, there is a critical stellar rank after which good approximation of the cat state becomes possible. Before that stellar rank, the best Gaussian operation in the optimisation of Corollary~\ref{coro:Rcat} is roughly a displacement of the amplitude of the cat. Past that rank, it is a smaller displacement combined with a squeezing.
\end{itemize}

\noindent If the state $\ket\psi$ is of finite rank, the $k^{th}$ term of the sequence $\mathcal R(\psi)$ is zero for all $k>r^\star(\psi)$ by Corollary~\ref{coro:robust}. For $k\le r^\star(\psi)$, an expression depending on the core state of $\ket\psi$ may be obtained for $R^\star_k(\psi)$:

\begin{coro} \label{coro:robustcomp}
Let $k\in\mathbb N^*$ and let $\ket\psi\in\Hi$ be a non-Gaussian pure state of finite stellar rank $r^\star(\psi)\ge k$, with core state $\ket{C_\psi}=\sum_{n=0}^{r^\star_\psi}{C_n\ket n}$. Then,
\be
R^\star_k(\psi)=\sqrt{1-\sup_{\xi,\alpha\in\mathbb C}{\sum_{m=0}^{k-1}{|u_m(\xi,\alpha)|^2}}},
\ee
where for all $m\in\{0,\dots,k-1\}$ and all $\xi=re^{i\theta},\alpha\in\mathbb C$, 
\be
u_m(\xi,\alpha)=\frac1{\sqrt{m!c_r}}\sum_{n=0}^{r^\star_\psi}{\frac{C_n^*}{\sqrt{n!}}\left[\partial_z^n(c_rz+s_re^{i\theta}\partial_z-\alpha^*)^me^{-\frac12e^{-i\theta}t_rz^2+\frac\alpha{c_r}z+\frac12e^{i\theta}t_r\alpha^2-\frac12|\alpha|^2}\right]_{z=0}},
\ee
with $c_r=\cosh r$, $s_r=\sinh r$ and $t_r=\tanh r$. Moreover, assuming the optimisation yields values $\xi_0,\alpha_0\in\mathbb C$, an optimal approximating state is
\be
\hat D^\dag(\alpha_0)\hat S^\dag(\xi_0)\left(\frac{\Pi_{k-1}\hat S(\xi_0)\hat D(\alpha_0)\ket{C_\psi}}{\left\|\Pi_{k-1}\hat S(\xi_0)\hat D(\alpha_0)\ket{C_\psi}\right\|}\right).
\ee
\end{coro}

\begin{proof}
\begin{mdframed}[linewidth=1.5,topline=false,rightline=false,bottomline=false]

Let $\ket\psi\in\Hi$ be a pure state of finite stellar rank $r^\star(\psi)\in\mathbb N^*$ with core state $\ket{C_\psi}=\sum_{n=0}^{r^\star_\psi}{C_n\ket n}$. By Theorem~\ref{Gconv}, there exist a Gaussian operation $\hat G_\psi$ such that $\ket\psi=\hat G_\psi\ket{C_\psi}$.
From Theorem~\ref{th:robust} we have
\be
\ba
R^\star_k(\psi)&=\sqrt{1-\sup_{\hat G\in\mathcal G}{\Tr\left[\Pi_{k-1}\hat G\ket\psi\bra\psi\hat G^\dag\right]}}\\
&=\sqrt{1-\sup_{\hat G'\in\mathcal G}{\Tr\left[\Pi_{k-1}\hat G'\ket{C_\psi}\bra{C_\psi}\hat G'^\dag\right]}}\\
&=\sqrt{1-\sup_{\xi,\alpha\in\mathbb C}{\Tr\left[\Pi_{k-1}\hat D^\dag(\alpha)\hat S^\dag(\xi)\ket{C_\psi}\bra{C_\psi}\hat S(\xi)\hat D(\alpha)\right]}}\\
&=\sqrt{1-\sup_{\xi,\alpha\in\mathbb C}{\sum_{m=0}^{k-1}\left|\braket{C_\psi|\hat S(\xi)\hat D(\alpha)|m}\right|^2}},
\ea
\label{robusku}
\ee
where we used the group structure of the Gaussian unitary operations in the second line and the fact that any single-mode Gaussian unitary operation may be decomposed as a squeezing and a displacement in the third line. Assuming the optimisation yields values $\xi_0,\alpha_0\in\mathbb C$, the optimal core state used in the approximation is
\be
\ket C=\frac{\Pi_{k-1}\hat S(\xi_0)\hat D(\alpha_0)\ket{C_\psi}}{\left\|\Pi_{k-1}\hat S(\xi_0)\hat D(\alpha_0)\ket{C_\psi}\right\|}.
\label{coreopti}
\ee
Now for all $m\in\{0,\dots,k-1\}$,
\be
\braket{C_\psi|\hat S(\xi)\hat D(\alpha)|m}=\sum_{n=0}^{r^\star_\psi}{C_n^*\braket{n|\hat S(\xi)\hat D(\alpha)|m}},
\ee
and for all $n\in\{0,\dots,r^\star_\psi\}$,
\be
\ba
\braket{n|\hat S(\xi)\hat D(\alpha)|m}&=\frac1{\sqrt{m!n!}}\braket{0|\hat a^n\hat S(\xi)\hat D(\alpha)(\hat a^\dag)^m|0}\\
&=\frac1{\sqrt{m!n!}}\braket{0|\hat a^n(c_r\hat a^\dag+s_re^{i\theta}\hat a-\alpha^*)^m\hat S(\xi)\hat D(\alpha)|0},
\ea
\label{FockSD}
\ee
where we used Eq.~(\ref{commutDS}) in the second line, with $c_r=\cosh r$, $s_r=\sinh r$. Hereafter we also set $t_r=\tanh r$. We have $\braket{0|\chi}=F^\star_\chi(0)$ for all states $\chi$, hence switching to the stellar representation Eq.~(\ref{FockSD}) rewrites
\be
\braket{n|\hat S(\xi)\hat D(\alpha)|m}=\frac1{\sqrt{m!n!c_r}}\left[\partial_z^n\left(c_rz+s_re^{i\theta}\partial_z-\alpha^*\right)^me^{-\frac12e^{-i\theta}t_rz^2+\frac\alpha{c_r}z+\frac12e^{i\theta}t_r\alpha^2-\frac12|\alpha|^2}\right]_{z=0},
\ee
where we used Eq.~(\ref{FSD}). Hence,
\be
\braket{C_\psi|\hat S(\xi)\hat D(\alpha)|m}=\sum_{n=0}^{r^\star_\psi}{\frac{C_n^*}{\sqrt{m!n!c_r}}\left[\partial_z^n\left(c_rz+s_re^{i\theta}\partial_z-\alpha^*\right)^me^{-\frac12e^{-i\theta}t_rz^2+\frac\alpha{c_r}z+\frac12e^{i\theta}t_r\alpha^2-\frac12|\alpha|^2}\right]_{z=0}}.
\ee
Setting, for all $m\in\{0,\dots,k-1\}$, $u_m(\xi,\alpha)=\braket{C_\psi|\hat S(\xi)\hat D(\alpha)|m}$, thus omitting the dependency in $\psi$, we finally obtain with Eq.~(\ref{robusku}),
\be
R^\star_k(\psi)=\sqrt{1-\sup_{\xi,\alpha\in\mathbb C}{\sum_{m=0}^{k-1}{|u_m(\xi,\alpha)|^2}}},
\ee
where for all $m\in\{0,\dots,k-1\}$ and all $\xi=re^{i\theta},\alpha\in\mathbb C$, 
\be
u_m(\xi,\alpha)=\frac1{\sqrt{m!c_r}}\sum_{n=0}^{r^\star_\psi}{\frac{C_n^*}{\sqrt{n!}}\left[\partial_z^n(c_rz+s_re^{i\theta}\partial_z-\alpha^*)^me^{-\frac12e^{-i\theta}t_rz^2+\frac\alpha{c_r}z+\frac12e^{i\theta}t_r\alpha^2-\frac12|\alpha|^2}\right]_{z=0}},
\ee
with $c_r=\cosh r$, $s_r=\sinh r$ and $t_r=\tanh r$. Moreover, assuming the optimisation yields values $\xi_0,\alpha_0\in\mathbb C$, an optimal approximating state is $\ket\phi=\hat D^\dag(\alpha_0)\hat S^\dag(\xi_0)\ket C$, where $\ket C$ is defined in Eq.~(\ref{coreopti}). Namely,
\be
\ket\phi=\hat D^\dag(\alpha_0)\hat S^\dag(\xi_0)\left(\frac{\Pi_{k-1}\hat S(\xi_0)\hat D(\alpha_0)\ket{C_\psi}}{\left\|\Pi_{k-1}\hat S(\xi_0)\hat D(\alpha_0)\ket{C_\psi}\right\|}\right),
\ee
i.e., $\hat S(\xi_0)\hat D(\alpha_0)\ket\phi$ is the renormalised truncation of $\hat S(\xi_0)\hat D(\alpha_0)\ket{C_\psi}$ at photon number $k-1$.

\end{mdframed}
\end{proof}

\noindent From this result, the value of the stellar robustness may be obtained analytically for low stellar rank states and numerically for all finite stellar rank states. In particular, we obtain:

\begin{lem}\label{lem:RFock} For the single photon Fock state $\ket1$ of stellar rank $1$ we have
\be
R^\star(1)=\sqrt{1-\frac{3\sqrt3}{4e}}.
\ee
The corresponding maximum achievable fidelity is given by
\be
\frac{3\sqrt3}{4e}\approx0.478.
\ee
\end{lem}

\begin{proof}
\begin{mdframed}[linewidth=1.5,topline=false,rightline=false,bottomline=false]

The single-photon Fock state $\ket1$ is a core state of stellar rank $1$ (its stellar function is given by $F^\star_1(z)=z$ for all $z\in\mathbb C$). Hence, by Corollary~\ref{coro:robustcomp},
\be
\ba
R_1^\star(1)&=R^\star(1)\\
&=\sqrt{1-\sup_{\xi=re^{i\theta},\alpha\in\mathbb C}{\frac1{c_r}\left|\left[\partial_z e^{-\frac{t_r}2e^{-i\theta}z^2+\frac\alpha{c_r}z+\frac{t_r}2e^{i\theta}\alpha^2-\frac12|\alpha|^2}\right]_{z=0}\right|^2}}\\
&=\sqrt{1-\sup_{\xi=re^{i\theta},\alpha\in\mathbb C}{\frac{|\alpha|^2}{c_r^3}\left|e^{\frac{t_r}2e^{i\theta}\alpha^2-\frac12|\alpha|^2}\right|^2}}\\
&=\sqrt{1-\sup_{\xi=re^{i\theta},\alpha\in\mathbb C}{\frac{|\alpha|^2}{c_r^3}e^{\frac{t_r}2(\alpha^2e^{i\theta}+\alpha^{*2}e^{-i\theta})-|\alpha|^2}}}.
\ea
\ee
Setting $\gamma=x+iy=i\alpha e^{i\theta/2}$ we obtain
\begin{align}
\nonumber\frac{|\alpha|^2}{c_r^3}e^{\frac{t_r}2(\alpha^2e^{i\theta}+\alpha^{*2}e^{-i\theta})-|\alpha|^2}&=\frac{|\gamma|^2}{c_r^3}e^{-|\gamma|^2-\frac{t_r}2(\gamma^2+\gamma^{*2})}\\
\nonumber&=\frac{x^2+y^2}{c_r^3}e^{-(1+t_r)x^2}e^{-(1-t_r)y^2}\\
\nonumber&=(1-t_r^2)^{3/2}(x^2+y^2)e^{-(1+t_r)x^2}e^{-(1-t_r)y^2}\\
&=(1-t_r^2)^{3/2}(x^2+y^2)e^{-(1-t_r)(x^2+y^2)}e^{-2t_rx^2}\\
\nonumber&\le(1-t_r^2)^{3/2}(x^2+y^2)e^{-(1-t_r)(x^2+y^2)}\\
\nonumber&\le\frac{(1-t_r^2)^{3/2}}{e(1-t_r)}\\
\nonumber&=\frac1e\sqrt{(1-t_r)(1+t_r)^3},
\end{align}
and this upperbound is attained for $x=0$ and $y=\frac1{\sqrt{1-t_r}}$.
Finally, we have $\max_{u\in[0,1]}{(1-u)(1+u)^3}=\frac{27}{16}$, attained for $u=\frac12$, so we obtain the stellar robustness of a single photon Fock state:
\be
R^\star(1)=\sqrt{1-\frac{3\sqrt3}{4e}}.
\ee

\end{mdframed}
\end{proof}

\noindent Since the stellar robustness inherits the property of invariance under Gaussian unitary operations of the stellar rank, Corollary~\ref{coro:robust} implies the same robustness value for states obtained from a single photon Fock state by unitary Gaussian operations, such as photon-added or photon-subtracted squeezed states, by Eq.~(\ref{Gconvexample}). 

\begin{figure}[h!]
	\begin{center}
		\includegraphics[width=0.51\columnwidth]{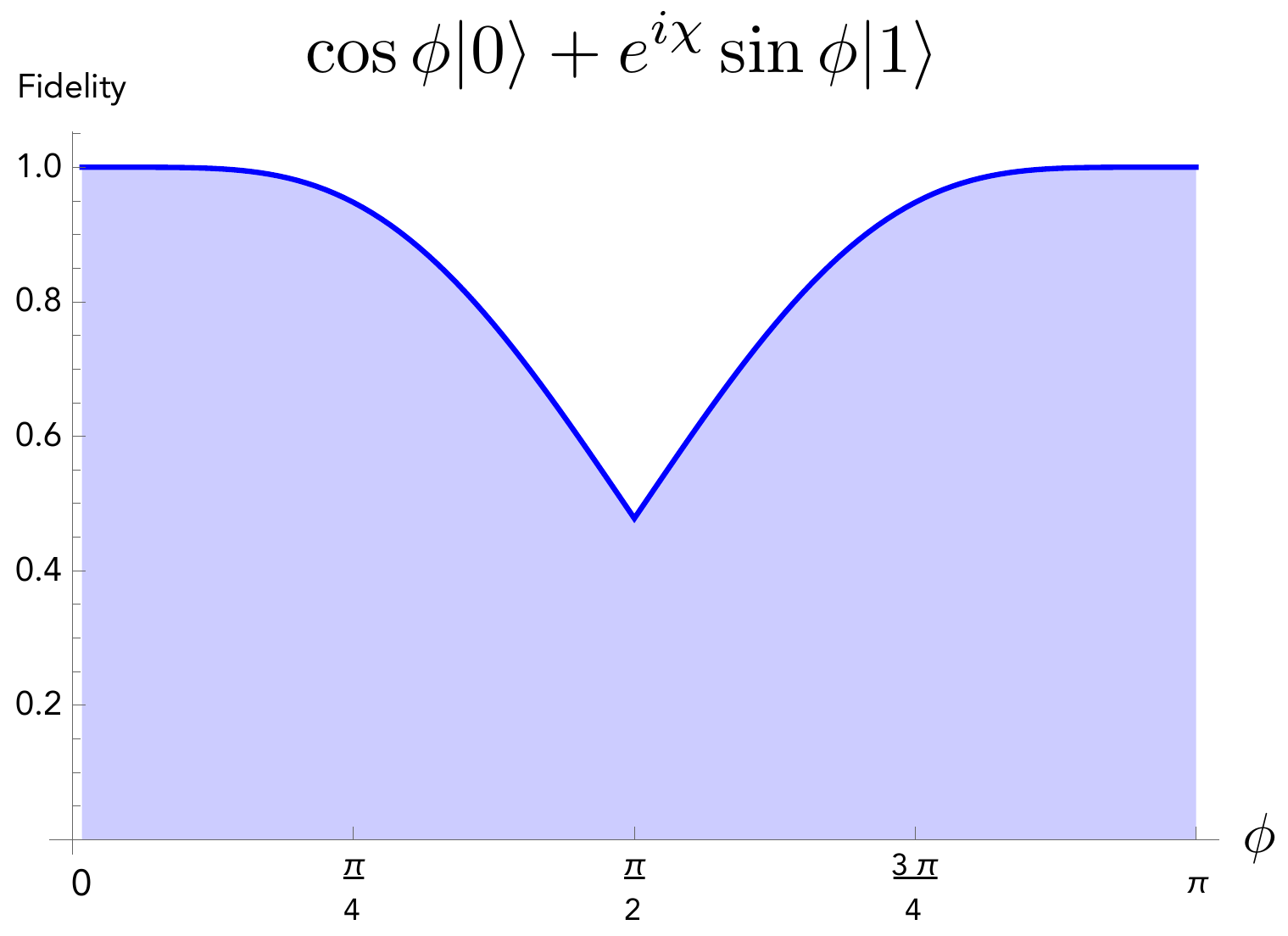}
		\caption{Achievable fidelities with Gaussian states for target core states $\cos\phi\ket0+e^{i\chi}\sin\phi\ket1$, for all $\phi,\chi\in[0,2\pi]$, as a function of $\phi$. The maximum fidelities are independent of $\chi$, and yield the stellar robustness through Lemma~\ref{lem:Rfide}.}
		\label{fig:R01}
	\end{center}
\end{figure}

We have computed numerically the stellar robustness for the states $\cos\phi\ket0+e^{i\chi}\sin\phi\ket1$, for all $\phi,\chi\in[0,2\pi]$, which is independent of $\chi$ (Fig.~\ref{fig:R01}). Setting $\phi=\frac\pi2$ yields the single photon Fock state, which is thus the most robust state of stellar rank $1$, up to Gaussian unitary operations.\\

\begin{figure}[h!]
	\begin{center}
		\includegraphics[width=0.85\columnwidth]{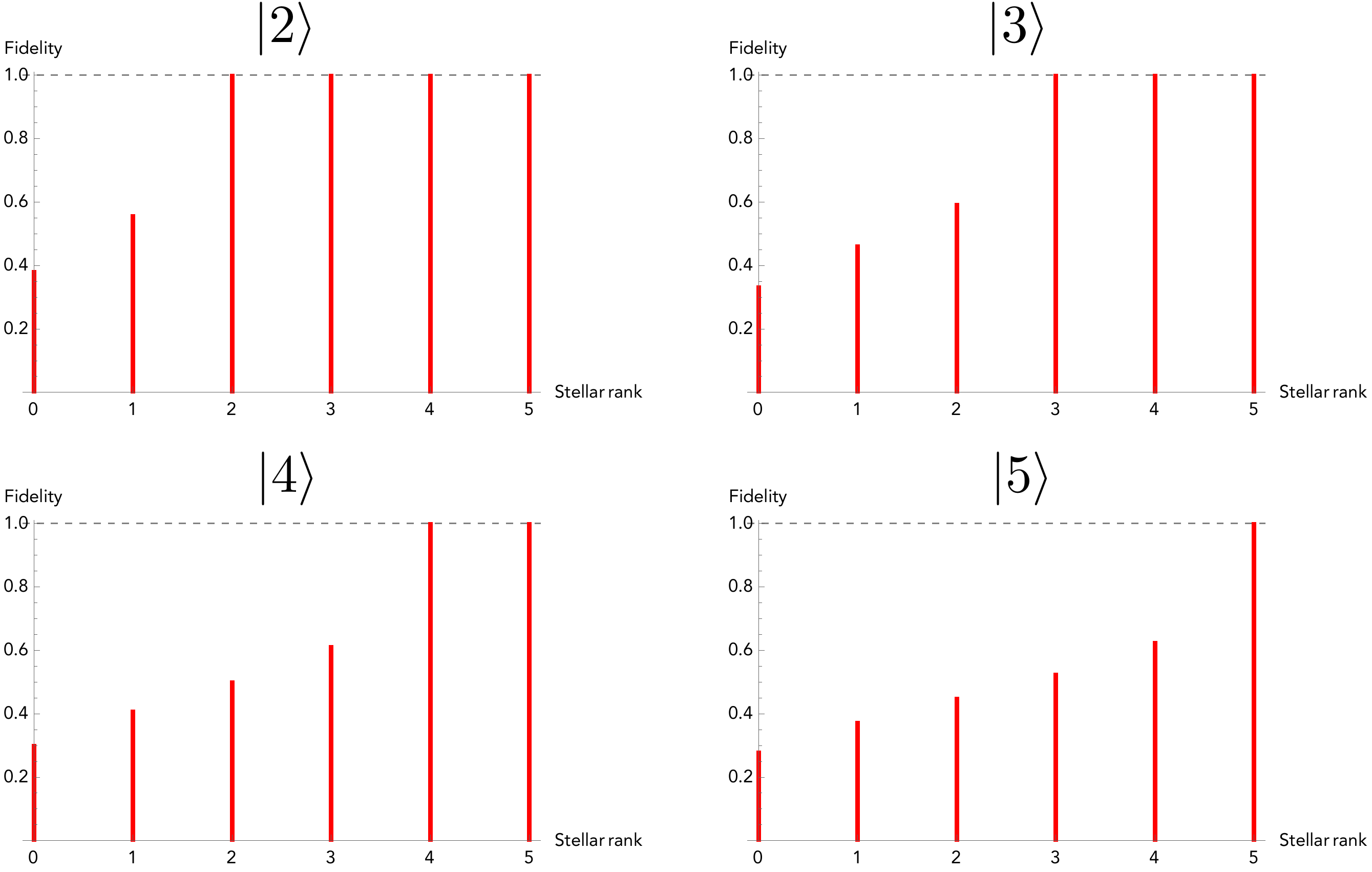}
		\caption{Achievable fidelities for target Fock states $\ket2$, $\ket3$, $\ket4$ and $\ket5$. For each rank $k\in\mathbb N^*$, the vertical line depicts the achievable fidelities between the target state and states of rank $k$. The sequence of maximum fidelities for each rank yields the robustness profile through Lemma~\ref{lem:Rfide}.}
		\label{fig:Rprofiles}
	\end{center}
\end{figure}

\noindent We have also obtained numerically the achievable fidelities with target states $\ket2$, $\ket3$, $\ket4$ and $\ket5$, depicted in Fig.~\ref{fig:Rprofiles}.

\medskip

\noindent As previously mentioned, with Lemma~\ref{lem:Rfide}, preparing a target pure state of finite stellar rank $\ket\psi$ with fidelity better than $1-[R_k^\star(\psi)]^2$, which may be computed using Corollary~\ref{coro:robustcomp}, ensures that the obtained state has a stellar rank equal to or greater than $k$. 

One may obtain the $\epsilon$-smoothed non-Gaussianity of formation of a given target pure state---i.e., the minimal stellar rank of $\epsilon$-close states---from its profile of achievable fidelities as follows: for a given $\epsilon>0$, it is the $x$-coordinate of the leftmost intersection point of the horizontal line of height $1-\epsilon^2$ with the vertical lines of the profile. For example, let $\epsilon=0.7$, so that $1-\epsilon^2=0.51$. With Fig.~\ref{fig:Rprofiles}, the $0.7$-smoothed non-Gaussianity of formation of the Fock state $\ket3$ is equal to $2$. Hence, an experimental (mixed) quantum state which has fidelity greater than $0.51$ (corresponding to $\epsilon=0.7$) with the Fock state $\ket3$ has a stellar rank greater or equal to $2$. 

The stellar hierarchy may thus be certified experimentally by direct fidelity estimation with a target pure state. In chapter~\ref{chap:certif}, we make use of this result and discuss in particular the certification of the stellar rank using Gaussian measurements and heterodyne detection.

\section{Discussion and open problems}

Based on the stellar representation of single-mode continuous variable quantum states, we have defined the stellar rank as the number of zeros of the stellar function, or equivalently of the Husimi $Q$ function. Using the analytic properties of the stellar function, we have shown that this rank is invariant under Gaussian operations and induces a hierarchy over the space of single-mode normalised states. We have characterized the states of finite stellar rank as the states obtained by successive single-photon additions to a Gaussian state, or equivalently as finite superpositions of (equally) displaced and squeezed number states. Additionally, we have given the stellar rank an operational meaning, as the minimal non-Gaussian cost for engineering a state, in terms of single-photon additions and subtractions. 
We have derived the equivalence classes for Gaussian convertibility using the notion of core states, and we have studied in detail the robustness of the ranks of the stellar hierarchy, showing that finite stellar rank states are robust, while infinite stellar rank states are not. In particular, we have shown how to compute the robustness.

\medskip

\noindent While the stellar representation unveils the structure of single-mode non-Gaussian states, various open questions remain:

How to extend the stellar formalism to the case of multimode states? The stellar function for multimode states is a multivariate analytic function, which prevents the use of the factorisation theorem, crucial in the derivation of the results. However, one can consider in the multimode case both the set of states that can be obtained from the vacuum by multimode Gaussian operations and a finite number of photon additions and the set of states that are obtained by multimode Gaussian operations acting on an input with a multivariate polynomial stellar function. We consider both sets of states in the next chapter and investigates their computational power.

Can we interpret geometrically the non-Gaussian properties of quantum states? Can we view the stellar representation as a limit of the Majorana representation? This representation provides a beautiful interpretation of non-entangling operations for symmetric states as a class of transformations of the sphere~\cite{ribeiro2011entanglement,aulbach2011symmetric}. Can we derive an analogous statement to characterize how Gaussian operations affect the roots of the stellar function? The displacement operator simply displaces the sphere on the complex plane, but the action of the squeezing seems nontrivial.

The stellar rank has an interpretation as a non-Gaussianity of formation, i.e., as a cost for quantum state engineering in terms of elementary non-Gaussian operations. Can we identify a computational task for which this rank quantifies how resourceful a state is?

Even though they are not robust, infinite stellar rank states are interesting from a conceptual point of vue, given their use for for error-correction~\cite{cochrane1999macroscopically,Gottesman2001}. Can we classify these states, for example based on the density of the zeros of their stellar function? To that end, what is the precise location of the zeros for GKP states?

\clearemptydoublepage
%
%
\let\textcircled=\pgftextcircled
\chapter{Beyond-classical quantum continuous variable models}
\label{chap:CVS}

\initial{D}ifferent approaches are possible to probe the quantum computational advantage regime and to study the boundary between the quantum systems which are efficiently simulable classically and those universal for quantum computing. 
On the one hand, the regime of classical simulability can be explored~\cite{feynman1982simulating}: being able to efficiently simulate a quantum computational model with a classical computer up to a certain regime may suggest a quantum advantage beyond this regime~\cite{valiant2002quantum,terhal2002classical,Bartlett2002}. 
On the other hand, subuniversal models of quantum computing can be defined: these models lie somewhere in-between classical and universal quantum computing, in the sense that, although not possessing the full computational power of a universal quantum computer, they may outperform classical computational capabilities with respect to specific problems~\cite{Bremmer2010,Aaronson2013,Morimae2014, Bremner2015,farhi2016quantum,Douce2017,boixo2018characterizing}. 
In both cases, one can aim for minimal extensions beyond the classically simulable models which are more likely to be implementable in the near term than universal quantum computers. For concrete applications, however, it is not the quantum computational model but rather the task at hand whose classical simulability matters. Studying application-specific classical simulation regimes is therefore also of great importance. 

We investigate these approaches for different continuous variable quantum computational models. After introducing classical simulation notions, we consider linear optics with input single photons and adaptive photon-number measurement, and study the classial simulation regime for probability estimation and overlap estimation, two computational tasks that are central to machine learning applications~\cite{bravyi2019classical}. Next, we turn to Gaussian quantum circuits with non-Gaussian input states and derive sufficient conditions for an efficient classical strong simulation. Finally, we focus on a specific subclass of Gaussian quantum circuits with non-Gaussian input states, the CVS circuits, which relates to Boson Sampling with continuous variable measurements. We identify the regime for which an efficient classical weak simulation of circuits in this subclass would imply a collapse of the polynomial hierarchy of complexity classes.

This chapter is based on \cite{inprepaMLALO,inprepaSimuNG,chabaud2017continuous}.


\section{Classical simulation of quantum computations}

Depending on the approach used for simulating classically the functioning of quantum devices, several notions of simulability are commonly used. In what follows, we review the ones we will be considering in this chapter.

\subsection{Strong simulation}

To each quantum computation is associated a probability distribution from which classical outcomes are sampled. In the case of continuous variable quantum computations with continuous variable outcomes, the output probability distribution is replaced by an output probability density. This motivates the following (informal) definition~\cite{terhal2002classical,pashayan2020estimation}.

\begin{defi}[Strong simulation]\label{Strongs}
A quantum computation is \textit{strongly simulable} if there exists a classical algorithm which evaluates its output probability distribution (density) or any of its marginals for any outcome in time polynomial in the size of the quantum computation.
\end{defi}

\noindent Various relaxations of this definition are possible, allowing the classical evaluation to be approximate rather than exact, or to abort with a small probability. Hereafter we only consider the definition above. When there exists no efficient classical algorithm for strong simulation, we say that \textit{strong simulation is hard}.

This notion of simulability is referred to as strong because it asks more from the classical simulation algorithm than from the quantum computation. Indeed, the quantum computation is merely sampling from a probability distribution (density), while the classical algorithm has to compute efficiently probabilities.

\subsection{Weak simulation}

A sampling counterpart to the notion of strong simulation is to ask the classical simulation algorithm to mimic the output of the quantum computation~\cite{terhal2002classical,pashayan2020estimation}. Informally:

\begin{defi}[Weak simulation]\label{Weaks}
A quantum computation is \textit{weakly simulable} if there exists a classical algorithm which outputs samples from its output probability distribution (density) in time polynomial in the size of the quantum computation.
\end{defi}

\noindent Akin to strong simulation, various relaxations of this definition are possible, allowing the classical sampling to be approximate rather than exact, or to abort with a small probability. Hereafter we only consider the definition above. When there exists no efficient classical algorithm for weak simulation, we say that \textit{weak simulation is hard}. 

In the case of continuous variable quantum computations with continuous variable outcomes, a weaker requirement is to ask the classical simulation not to sample from the output probability density, but rather from a discretised probability distribution obtained from the probability density by performing an efficient binning of the sample space. Indeed, samples from the output probability density yield samples of such a discretised probability distribution with efficient classical post-processing.

\medskip

\noindent Consider a quantum computation of size $m$ yielding discrete classical outcomes from a probability distribution $P(X_1,\dots,X_m)$, where $X_i$ may take at most $M=\poly m$ values for all $i\in\{1,\dots,m\}$ (the sample space has size $M^m$). Then, weak simulation is weaker than strong simulation, with the following result~\cite{terhal2002classical,pashayan2020estimation}:

\begin{lem}\label{lem:SWsimulation}
An efficient classical algorithm for strong simulation provides an efficient classical algorithm for weak simulation (assuming one can efficiently sample from efficiently computable univariate probability distributions over a polynomial number of samples).
\end{lem}

\noindent We reproduce the proof below for completeness.

\begin{proof}
\begin{mdframed}[linewidth=1.5,topline=false,rightline=false,bottomline=false]

Assuming the existence of an efficient classical algorithm for strong simulation of a quantum computation of size $m$ yielding classical outcomes from a discrete probability distribution $P(X_1,\dots,X_m)$, where $X_i$ may take at most $M=\poly m$ values for all $i\in\{1,\dots,m\}$, one first computes the marginal probabilities $P(X_1)$ for all $M$ possible values of $X_1$. Then, one samples the value $x_1$ from $P(X_1)$ (which is an efficiently computable univariate probability distribution over a polynomial number of samples). With that sample $x_1$, one computes the conditional probability distribution
\be
P(X_2|x_1)=\frac{P(x_1,X_2)}{P(x_1)},
\ee
for all $M$ possible values of $X_2$. Then, one samples the value $x_2$ from $P(X_2|x_1)$ (which is also an efficiently computable univariate probability distribution over a polynomial number of samples). Repeating the same procedure up to
\be
P(X_m|x_1,\dots,x_{m-1})=\frac{P(x_1,\dots,x_{m-1},X_m)}{P(x_1,\dots,x_{m-1})},
\ee
one obtains a sample $(x_1,\dots,x_m)$ from $P(X_1,\dots,X_m)$ efficiently.

\end{mdframed}
\end{proof}

\noindent For quantum computations yielding continuous variable classical outcomes, the result still holds with the same proof for binned discretised probability distributions rather than the corresponding probability density, as long as the discretised probabilities can be computed efficiently from the probability density and have support on a polynomial number of bins for each mode.

\subsection{Probability and overlap estimation}

While the previous two notions of simulation of quantum computations are the most commonly used, other type of simulation may be useful: if the output samples of a quantum computation are used to compute a quantity which may be computed efficiently classically by other means, it is no longer necessary to simulate the whole quantum device. We consider two concrete examples which are prominent for variational quantum algorithms in quantum machine learning: probability estimation and overlap estimation~\cite{havlivcek2019supervised,schuld2019quantum}.

\begin{defi}[Probability estimation]
Let $P$ be a probability distribution over $m$ outcomes. Given any outcome $\bm x$ in the sample space of $P$, \textit{probability estimation} refers to the computational task of outputting an estimate $\tilde P[\bm x]$ such that
\be
P[\bm x]-\frac1{\poly m}\le\tilde P[\bm x]\le P[\bm x]+\frac1{\poly m},
\ee
with probability greater than $1-\frac1{\exp m}$.
\end{defi}

\noindent Probability estimation amounts to outputting a polynomially precise additive estimate of the probability with exponentially small probability of failure. One may use the samples from a quantum computation in order to perform probability estimation for any given outcome: given a quantum device of size $m$ which outputs samples from some probability distribution and a fixed outcome $\bm x$ in the sample space, one may run the device $M=\poly m$ times, recording the value $1$ whenever the outcome $\bm x$ is obtained and the value $0$ otherwise. Then, summing and dividing by $M$, one obtains the frequency of the outcome $\bm x$ over the $M$ uses of the quantum device, which is a polynomially precise additive estimate of the probability of the outcome $\bm x$ with exponentially small probability of failure, by virtue of Hoeffding inequality \cite{hoeffding1963probability}.

Weak simulation is at least as hard as probability estimation, since by the previous reasoning one may obtain polynomially precise additive estimates of probabilities from samples of the probability distribution. Moreover, they are some quantum computations for which weak simulation is hard (assuming widely believed conjectures from complexity theory), but probability estimation can be done efficiently classically. This is the case for IQP circuits~\cite{Bremmer2010,havlivcek2019supervised}, Boson Sampling~\cite{Aaronson2013} and even the period-finding subroutine of Shor's algorithm~\cite{shor1994algorithms}. Let us detail the latter case: if $N$ is an $n$ bits integer to factor, the period-finding subroutine measures the output state
\be
\frac1N\sum_x\sum_ye^{\frac{2i\pi xy}N}\ket y\ket{f(x)}
\ee
in the computational basis, where $f$ is a periodic function over $\{0,\dots,N-1\}$ which can be evaluated efficiently. The probability of obtaining an outcome $y_0,f(x_0)$ is given by
\be
\Pr\,[y_0,f(x_0)]=\left|\frac1N\sum_{f(x)=f(x_0)}e^{\frac{2i\pi xy_0}N}\right|^2.
\ee
Now let
\be
g_{x_0,y_0}:x\mapsto\begin{cases}e^{\frac{2i\pi xy_0}N}\text{ if } f(x)=f(x_0),\\0\text{ otherwise.}\end{cases}
\ee
The function $g_{x_0,y_0}$ can be evaluated efficiently and we have
\be
\Pr\,[y_0,f(x_0)]=\left|\underset{x\leftarrow N}{\mathbb E}[g_{x_0,y_0}(x)]\right|^2,
\ee
where $\underset{x\leftarrow N}{\mathbb E}$ denotes the expected value for $x$ drawn uniformly randomly from $\{0,\dots,N-1\}$. By virtue of Hoeffding inequality, this quantity may be estimated efficiently (in $n$ the number of bits of $N$) classically by sampling uniformly a polynomial number of values in $\{0,\dots,N-1\}$ and computing the modulus squared of the mean of $g_{x_0,y_0}$ for these values.

However, note that probability estimation of quantum circuits is a \textsf{BQP}-complete computational task almost by definition, since given a polynomially precise estimate of the probability of acceptance of an input $x$ to a quantum circuit, one may determine whether it is accepted or rejected by the circuit. In particular, unless factoring is in \textsf{P}, probability estimation for the quantum circuit corresponding to Shor's algorithm \textit{as a whole} is hard and weak simulation of the period-finding subroutine is also hard, since in Shor's algorithm the output samples from the period-finding subroutine are used for a different classical computation than probability estimation (essentially obtaining promising candidates for the period).

\medskip

\noindent A more general computational task than probability estimation in the context of quantum computing is the following:

\begin{defi}[Overlap estimation]
Let $\ket\phi$ and $\ket\psi$ be quantum output states of two quantum computations of size $m$. \textit{Overlap estimation} refers to the computational task of outputting an estimate $\tilde O$ such that
\be
|\braket{\phi|\psi}|^2-\frac1{\poly m}\le\tilde O\le|\braket{\phi|\psi}|^2+\frac1{\poly m},
\ee
with probability greater than $1-1/\text{exp}\,(m)$.
\end{defi}

\noindent The overlap between two quantum states is a measure of their distinguishability~\cite{dieks1988overlap} and overlap estimation thus is related to quantum state discrimination. Several techniques exist to perform quantumly the overlap estimation of two states $\ket\phi$ and $\ket\psi$~\cite{fanizza2020beyond}. One of them is to perform the swap test (see chapter~\ref{chap:prog} or~\cite{buhrman2001quantum}) with various copies of both states. 

Overlap estimation can be done efficiently classically for IQP circuits~\cite{havlivcek2019supervised}. We prove in the next section that it is also the case for Boson Sampling and consider the more general setting of passive linear optical quantum computing with input single-photons and adaptive measurements.

\section{Adaptive linear optics}
\label{sec:ALO}

\begin{figure}[h!]
	\begin{center}
		\includegraphics[width=\columnwidth]{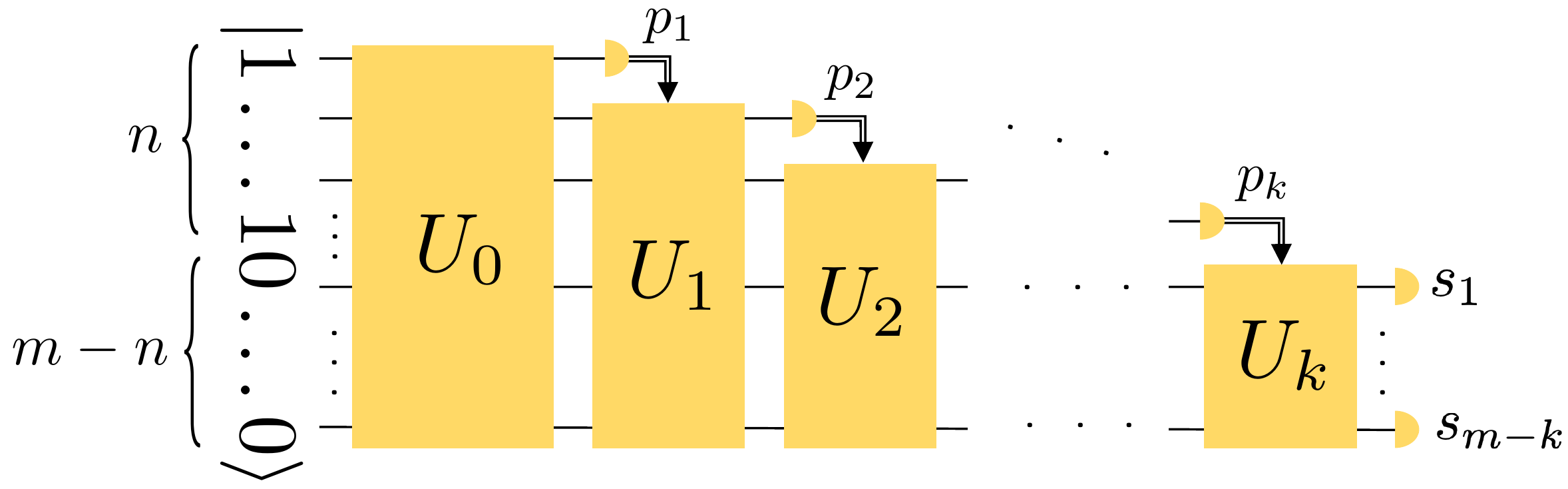}
		\caption{Passive linear optical computing with $k$ adaptive measurements and input state $\ket{1\dots10\dots0}$ with $n$ photons over $m$ modes. The output modes are measured using photon counters. For all $j\in\{1,\dots,k\}$, the unitary interferometer $U_j$, acting on $m-j$ modes, may depend on the measurement outcomes $p_1,\dots,p_j$. The adaptive measurement outcomes $p_1,\dots,p_k$ are used to drive the computation, whose final outcome is $s_1,\dots,s_{m-k}$.}
		\label{fig:adaptive}
	\end{center}
\end{figure}

The complexity of probability estimation and overlap estimation of quantum computations has been well studied in the circuit model~\cite{pashayan2015estimating,bravyi2019classical}. In what follows, we consider the case of passive linear optical quantum computing with adaptive measurements, which we refer to as adaptive linear optics (Fig.~\ref{fig:adaptive}). We use multi-index notations (see section~\ref{sec:multiindex}).

Formally, we consider unitary interferometers of size $m$, described by $m\times m$ unitary matrices (see section~\ref{sec:LO}). We identify the multimode Fock states with $n$ photons over $m$ modes with the elements of $\Phi_{m,n}=\{\bm s\in\mathbb N^m,\text{ }|\bm s|=n\}$, for all $n\in\mathbb N$. We fix the input state $\ket{\bm t}=\ket{\bm1^n\bm0^{m-n}}$, with single photons in the $n$ first modes, where the superscript indicates the size of the string $(0,\dots,0)$ or $(1,\dots,1)$ when there is a possible ambiguity. For $p\in\mathbb N$ and $\bm p\in\Phi_{k,p}$, let us define
\be
U^{\bm p}:=\left[\mathbb 1_k\oplus U_k(p_1,\dots,p_k)\right]\left[\mathbb 1_{k-1}\oplus U_{k-1}(p_1,\dots,p_{k-1})\right]\dots\left[\mathbb 1_1\oplus U_1(p_1)\right]U_0,
\label{Up}
\ee
where $\mathbb 1_j$ is the identity matrix of size $j$. The matrices $U_j$ depend on the measurement outcomes $p_1,\dots,p_j$ for all $j\in\{1,\dots,k\}$. The output state where the adaptive measurement outcome $\bm p$ has been obtained reads
\be
\Tr_k\left[(\ket{\bm p}\!\bra{\bm p}\otimes\mathbb 1_{m-k})U^{\bm p}\ket{\bm t}\!\bra{\bm t}U^{\bm p\dag}\right],
\label{outputALO}
\ee
where the partial trace is over the first $k$ modes and where $\ket{\bm p}$ denotes the $k$-mode Fock state $\ket{p_1\dots p_k}$. The matrix $U^{\bm p}$ describes the interferometer in Fig.~\ref{fig:adaptive}, where the adaptive measurement outcome $\bm p=(p_1,\dots,p_k)$ and the final outcome $\bm s=(s_1,\dots,s_{m-k})$ have been obtained. \\

\noindent Boson Sampling~\cite{Aaronson2013} corresponds to the case $k=0$ and the Knill--Laflamme--Milburn scheme for universal quantum computing~\cite{knill2001scheme} to the case $k=O(m)$. We investigate the transition between these two cases by giving classical algorithms for probability estimation and overlap estimation and identifying various complexity regimes for different numbers of photons $n$ and adaptive measurements $k$.

\subsection{Quantum probability and overlap estimation}

For doing probability estimation with a quantum circuit, one samples the circuit $O(\poly m)$ times, obtaining outcomes, for which the frequency gives a polynomially precise additive estimate of the probability which can be computed efficiently. In the case of a circuit with adaptive measurements, one only looks at the final measurement outcomes and the same holds for adaptive linear optical computations.

\medskip

\noindent For doing overlap estimation with unitary quantum circuits, one may run two circuits $U$ and $V$ in parallel and compare their quantum output states, for example with the swap test. Doing so a polynomial number of times provides a polynomially precise estimate of the overlap. Alternatively, one may build the circuit $UV^\dag$ and project the output quantum state onto the input state. 

In the case of circuits with adaptive measurements, the overlaps are between all possible output states for all possible adaptive measurement results. In particular, if the number of possible adaptive measurement outcomes is exponential, then the probability distribution for these outcomes has to be concentrated on a polynomial number of events for the quantum overlap estimation to be efficient. This is because in order to compute a polynomially precise estimate of the overlap, say, $|\braket{\phi|\psi}|^2$, the states $\ket\phi$ and $\ket\psi$, both corresponding to specific adaptive measurement results, have to be obtained a polynomial number of times.

For adaptive linear optics over $m$ modes with $n$ input photons and $k$ adaptive measurements, the number of possible adaptive measurement outcomes is given by
\be
\ba
\sum_{r=0}^n{|\Phi_{k,r}|}&=\sum_{r=0}^n{\binom{k+r-1}r}\\
&=\binom{k+n}n,
\ea
\ee
where the sum is over the total number of photons detected at the stage of the adaptive measurements. Hence, either the probability distribution for the adaptive measurements outcomes is concentrated on a polynomial number of outcomes, or $\binom{n+k}{n}=O(\poly m)$, which is the case for example when $n=O(1)$ and $k=O(m)$, $n=O(\log m)$ and $k=O(\log m)$, or $n=O(m)$ and $k=O(1)$.
In what follows, we do not assume concentration of the adaptive measurement outcome probability distribution and consider general interferometers with adaptive measurements. The quantum efficient regime for overlap estimation thus corresponds to $\binom{n+k}{n}=O(\poly m)$.

Let $\ket\phi$ and $\ket\psi$ be output states of two adaptive linear interferometers over $m$ modes with $n$ input photons and $k$ adaptive measurements. Let $\bm p$ and $\bm q$ denote the outcomes of the adaptive measurements for $\ket\phi$ and $\ket\psi$, respectively. Let $U^{\bm p}$ in Eq.~(\ref{Up}) be the interferometer for $\ket\phi$, with input Fock state $\ket{\bm t}$. We have
\be
\ba
|\braket{\phi|\psi}|^2&=\Tr\left[\Tr_k[(\ket{\bm p}\!\bra{\bm p}\otimes\mathbb 1_{m-k})U^{\bm p}\ket{\bm t}\!\bra{\bm t}U^{\bm p\dag}]\ket\psi\!\bra\psi\right]\\
&=\Tr\left[(\ket{\bm p}\!\bra{\bm p}\otimes\mathbb 1_{m-k})U^{\bm p}\ket{\bm t}\!\bra{\bm t}U^{\bm p\dag}(\mathbb 1_k\otimes\ket\psi\!\bra\psi)\right]\\
&=\Tr\left[U^{\bm p}\ket{\bm t}\!\bra{\bm t}U^{\bm p\dag}(\ket{\bm p}\!\bra{\bm p}\otimes\ket\psi\!\bra\psi)\right]\\
&=\Tr\left[\ket{\bm t}\!\bra{\bm t}U^{\bm p\dag}\left(\ket{\bm p}\!\bra{\bm p}\otimes\ket\psi\!\bra\psi\right)U^{\bm p}\right],
\ea
\label{overlapALO}
\ee
where we used Eq.~(\ref{outputALO}) in the first line. Because of the conservation of the total number of photons, the overlap between the states $\ket\phi$ and $\ket\psi$ is zero if $|\bm p|\neq|\bm q|$. Otherwise, it can be estimated using a polynomial number of copies of the state $\ket\psi$ as follows: send the input $\ket{\bm p}\otimes\ket\psi$ into the interferometer with unitary matrix $U^{\bm p\dag}$ and mesure the photon number in each output mode. Record the value $1$ if the measurement pattern matches the Fock state $\bm t$ and the value $0$ otherwise. Then, the mean of the obtained values yields a polynomially precise estimate of the overlap $|\braket{\phi|\psi}|^2$ by Eq.~(\ref{overlapALO}) and Hoeffding inequality.
Note that this overlap estimation requires the preparation of the Fock state $\bm p$. By symmetry, one could estimate the overlap alternatively using a polynomial number of copies of the state $\ket\phi$ and preparing the Fock state $\ket{\bm q}$.

\subsection{Classical probability estimation}

In this section, we obtain a classical algorithm for probability estimation of adaptive linear optics over $m$ modes with $n$ input photons and $k$ adaptive measurements.

\medskip

\noindent We first consider the case $k=0$, i.e., Boson Sampling. The probability of the outcome $\bm s\in\Phi_{m,n}$ for the interferometer $U$ given the input $\bm t=(\bm1^n,\bm0^{m-n})\in\Phi_{m,n}$ is given by (see section~\ref{sec:BosonSampling} and~\cite{Aaronson2013})
\be
\text{Pr}_{m,n}[\bm s]=\frac1{\bm s!}|\Per\,(U_{\bm s,\bm t})|^2,
\ee
where $U_{\bm s,\bm t}$ is the $n\times n$ matrix obtained from $U$ by repeating $s_i$ times its $i^{th}$ row for $i\in\{1,\dots,m\}$ and removing its $j^{th}$ column for $j=\{n+1,\dots,m\}$, and where the permanent of an $n\times n$ square matrix $A=(a_{ij}))_{1\le i,j\le n}$ is given by
\begin{equation}
\Per\,(A)=\sum_{\sigma\in S_n}{\prod_{i=1}^n{a_{i\sigma(i)}}},
\end{equation}
where the sum is over the permutations of the set $\{1,\dots,n\}$. When $|\bm s|\neq n$ however, the probability is $0$, since $\bm t$ has $n$ photons and the linear interferometer does not change the total number of photons.
The permanent of a square matrix of size $n$ can be computed exactly in time $O(n2^n)$, thanks to Ryser's formula~\cite{albert1978combinatorial}. However, polynomially precise estimates of the permanent can be obtained in polynomial time~\cite{gurvits2005complexity}, so the probability estimation can be done classically efficiently, which was already noted in~\cite{Aaronson2013}.

\medskip

\noindent We now turn to the case $k>0$, using notations of Eq.~(\ref{Up}) and Fig.~\ref{fig:adaptive}. This case is a direct extension of the case $k=0$. For $p\in\mathbb N$, $\bm p\in\Phi_{k,p}$ and $\bm s\in\Phi_{m-k,n-p}$, the probability of an total outcome $(\bm p,\bm s)\in\Phi_{m,n}$ (adaptive measurement and final outcome) is given by
\be
\text{Pr}^{\text{total}}_{m,n}[\bm p,\bm s]=\frac{1}{\bm p!\bm s!}\left|\Per\left(U^{\bm p}_{(\bm p,\bm s),\bm t}\right)\right|^2.
\ee
Let $p\in\{0,\dots,n\}$ and let $\bm s\in\Phi_{m-k,n-p}$. Then, the probability of obtaining the final outcome $\bm s$ after the adaptive measurements reads
\be
\ba
\text{Pr}_{m,n}^{\text{final}}[\bm s]&=\sum_{\bm p\in\Phi_{k,p}}{\text{Pr}^{\text{total}}_{m,n}[\bm p,\bm s]}\\
&=\frac1{\bm s!}\sum_{\bm p\in\Phi_{k,p}}{\frac{1}{\bm p!}\left|\Per\left(U^{\bm p}_{(\bm p,\bm s),\bm t}\right)\right|^2}.
\ea
\label{probak}
\ee
The sum is taken over the elements of $\Phi_{k,p}$, which has $\binom{k+p-1}{p}\le\binom{k+n-1}{n}$ elements. This last quantity is $O(\poly m)$ when the number of input photons $n$ and the number of adaptive measurements $k$ are small enough compared to $m$. 

\begin{table}[h!]
\centering
\setlength\tabcolsep{20pt}
\bgroup
\def\arraystretch{1.5}
\begin{tabular}{| c | c | c | c |}
\hline
\diagbox[height=8ex,width=9em]{$n$}{$k$} & $O(1)$ & $O(\log m)$ & $O(m)$\\
\hline
$O(1)$ & \cellcolor{blue!15} & \cellcolor{blue!15}  & \cellcolor{blue!15} \\
\hline
$O(\log m)$ & \cellcolor{blue!15} & \cellcolor{blue!15}  & \\
\hline
$O(m)$ & \cellcolor{blue!15}  &  &  \\
\hline
\end{tabular}
\egroup
\caption{Simulability regimes for probability estimation. In blue is the parameter region for which the classical algorithm is efficient.}
\label{tab:probaest}
\end{table}

The simulability regimes are summarised in Table~\ref{tab:probaest}, where the regimes are obtained using Stirling equivalent $n!\sim\sqrt{2\pi n}\left(\frac ne\right)^n$. In particular, as long as both $k$ and $n$ are $O(\log m)$, the output probability can be estimated efficiently (and even computed exactly efficiently).

The universal quantum computing regime corresponds to $n=O(m)$ and $k=O(m)$. The time complexity of the classical algorithm is $O\left(\binom{k+n-1}n\poly m\right)$, so there is a possibility of subuniversal quantum advantage for probability estimation for $n=O(\log m)$ and $k=O(m)$, or $n=O(m)$ and $k=O(\log m)$. However, the runtime of the classical algorithm is subexponential in these cases.

\subsection{Classical overlap estimation}

In this section, we obtain a classical algorithm for overlap estimation of adaptive linear optics over $m$ modes with $n$ input photons and $k$ adaptive measurements.

\medskip

\noindent Once again, we start with $k=0$. The output state of an $m$-mode interferometer $U$ with input state $\bm t\in\Phi_{m,n}$ reads
\be
\ba
\ket\phi&=\sum_{\bm s\in\Phi_{m,n}}{\braket{\bm s|\hat U|\bm t}\ket{\bm s}}\\\
&=\sum_{\bm s\in\Phi_{m,n}}{\frac{\Per\,(U_{\bm s,\bm t})}{\sqrt{\bm s!\bm t!}}\ket{\bm s}},
\ea
\ee
where $U_{\bm s,\bm t}$ is the $n\times n$ matrix obtained from $U$ by repeating $s_i$ times its $i^{th}$ row for $i\in\{1,\dots,m\}$ and repeating $t_j$ times its $j^{th}$ row for $j\in\{1,\dots,m\}$. The composition of two interferometers is another interferometer which unitary representation is the product of the unitary representations of the composed interferometers.
Hence, the inner product of the output states $\ket\phi$ and $\ket\psi$ of two $m$-mode interferometers $U$ and $V$ with the same input state $\bm t\in\Phi_{m,n}$, is equal to the matrix element $\bm t,\bm t$ of $\hat U^\dag\hat V$:
\be
\ba
\braket{\phi|\psi}&=\sum_{\bm u,\bm v\in\Phi_{m,n}}{\braket{\bm t|\hat U^\dag|\bm u}\braket{\bm v|\hat V|\bm t}\braket{\bm u|\bm v}}\\
&=\sum_{\bm s\in\Phi_{m,n}}{\braket{\bm t|\hat U^\dag|\bm s}\braket{\bm s|\hat V|\bm t}}\\
&=\braket{\bm t|\hat U^\dag\hat V|\bm t}\\
&=\frac{\Per\left[(U^\dag V)_{\bm t,\bm t}\right]}{\bm t!},
\ea
\ee
where we used in the third line $\bm t\in\Phi_{m,n}$ and the fact that $\hat U^\dag\hat V$ conserves the space $\Phi_{m,n}$. With the input $\bm t=(\bm1^n,\bm0^{m-n})$ with $n$ photons in $m$ modes, this reduces to
\be
\ba
\braket{\phi|\psi}=\Per\left[(U^\dag V)_n\right],
\ea
\ee
where $(U^\dag V)_n$ is the $n\times n$ top left submatrix of $U^\dag V$. Hence, the inner product and the overlap may be approximated to a polynomial precision efficiently, since this is the case for the permanent~\cite{gurvits2005complexity}. 

\medskip

\noindent We now consider the case $k>0$. Let $p\in\mathbb N$ and let $\bm p\in\Phi_{k,p}$. Writing $\text{Pr}^{\text{adap}}_{m,n}[\bm p]$ the probability of the adaptive measurement outcome $\bm p$, the output state of the interferometer $U^{\bm p}$ with $k$ adaptive measurements with input $\bm t=(\bm1^n,\bm0^{m-n})$ in Fig.~\ref{fig:adaptive}, when the adaptive measurement outcome $\bm p$ is obtained, reads
\be
\frac1{\sqrt{\text{Pr}^{\text{adap}}_{m,n}[\bm p]}}\ket{\psi_{\bm p}},
\ee
where
\be
\ket{\psi_{\bm p}}:=\sum_{\bm s\in\Phi_{m-k,n-p}}{\frac{\Per\left(U^{\bm p}_{(\bm p,\bm s),\bm t}\right)}{\sqrt{\bm p!\bm s!}}\ket{\bm s}}
\ee
and where $\text{Pr}^{\text{adap}}_{m,n}[\bm p]=\braket{\psi_{\bm p}|\psi_{\bm p}}$. More generally, the inner product of two (not normalised) output states $\ket{\psi_{\bm p}}$ and $\ket{\psi_{\bm q}}$ of $m$-mode interferometers $U^{\bm p}$ and $V^{\bm q}$ with $k$ adaptive measurements thus is zero if $|\bm p|\neq|\bm q|$. If $r:=|\bm p|=|\bm q|$, it is given by
\be
\ba
\braket{\psi_{\bm p}|\psi_{\bm q}}&=\frac1{\sqrt{\bm p!\bm q!}}\sum_{\bm s\in\Phi_{m-k,n-r}}{\frac1{\bm s!}\Per\left(U^{\bm p}_{(\bm p,\bm s),\bm t}\right)^*\Per\left(V^{\bm q}_{(\bm q,\bm s),\bm t}\right)}\\
&=\frac1{\sqrt{\bm p!\bm q!}}\sum_{\bm s\in\Phi_{m-k,n-r}}{\frac1{\bm s!}\Per\left(U^{\bm p\dag}_{\bm t,(\bm p,\bm s)}\right)\Per\left(V^{\bm q}_{(\bm q,\bm s),\bm t}\right)}.
\ea
\label{overlapS}
\ee
This expression is a sum of $|\Phi_{m-k,n-r}|$ terms, which is generally exponential in $m$ whenever $n$ is not constant. 
It is reminiscent of the permanent composition formula~\cite{percus2012combinatorial,barvinok2016combinatorics}: for all $m,n,c\in\mathbb N^*$, all $s\in\mathbb N$, all $\bm u\in\Phi_{m,s}$ and all $\bm v\in\Phi_{n,s}$,
\be
\Per\left[(MN)_{\bm u,\bm v}\right]=\sum_{\bm s\in\Phi_{c,s}}{\frac1{\bm s!}\Per\left(M_{\bm u,\bm s}\right)\Per\left(N_{\bm s,\bm v}\right)}
\label{percomp2}
\ee
where $M$ is a $m\times c$ matrix and $N$ is a $n\times c$ matrix.
In what follows, we prove that this expression in Eq.~(\ref{overlapS}) may be rewritten as a sum over fewer terms using the permanent composition formula in Eq.~(\ref{percomp2}). However, this formula is not directly applicable to the expression in Eq.~(\ref{overlapS}). In order to obtain a suitable expression, we first make use of the Laplace formula for the permanent: we expand the permanent of $U^{\bm p\dag}_{\bm t,(\bm p,\bm s)}$ along the columns that are repeated according to $\bm p$ and we expand the permanent of $V^{\bm q}_{(\bm q,\bm s),\bm t}$ along the rows that are repeated according to $\bm q$. 
The general Laplace column expansion formula for the permanent reads: let $n\in\mathbb N^*$, let $W$ be an $n\times n$ matrix, and let $\bm j\in\{0,1\}^n$. Then,
\be
\Per\,(W)=\sum_{\substack{\bm i\in\{0,1\}^n\\|\bm i|=|\bm j|}}{\Per\left(W_{\bm i,\bm j}\right)\Per\left(W_{\bm1^n-\bm i,\bm1^n-\bm j}\right)},
\label{Laplacegen}
\ee
where $W_{\bm i,\bm j}$ is the matrix obtained from $W$ by keeping only the $k^{th}$ rows and $l^{th}$ columns such that $i_k=1$ and $j_l=1$, respectively, and $W_{\bm1^n-\bm i,\bm1^n-\bm j}$ is the matrix obtained from $W$ by keeping only the $k^{th}$ rows and $l^{th}$ columns such that $i_k=0$ and $j_l=0$, respectively. This formula is obtained by applying the Laplace expansion formula for one column various times, for each column with index $l$ such that $j_l=1$, and the same formula holds for rows.

\begin{lem}\label{lem:overlapS}
Let $r\in\mathbb N$. The inner product of two (not normalised) output states $\ket{\psi_{\bm p}}$ and $\ket{\psi_{\bm q}}$ of $m$-mode interferometers $U^{\bm p}$ and $V^{\bm q}$ with adaptive measurements outcome $\bm p,\bm q\in\Phi_{k,r}$ is given by
\be
\braket{\psi_{\bm p}|\psi_{\bm q}}=\frac1{\sqrt{\bm p!\bm q!}}\sum_{\substack{\bm i,\bm j\in\{0,1\}^n\\|\bm i|=|\bm j|=r}}{\Per\left(A^{\bm i}\right)\Per\left(B^{\bm j}\right)\Per\left(C^{\bm i,\bm j}\right)},
\ee
where for all $\bm i,\bm j\in\{0,1\}^n$ such that $|\bm i|=|\bm j|=r$,
\be
A^{\bm i}=U^{\bm p\dag}_{(\bm i,\bm0^{m-n}),(\bm p,\bm0^{m-k})}
\ee
is an $r\times r$ matrix which can be obtained efficiently from $U^{\bm p}$,
\be
B^{\bm j}=V^{\bm q}_{(\bm q,\bm0^{m-k}),(\bm j,\bm0^{m-n})}
\ee
is an $r\times r$ matrix which can be obtained efficiently from $V^{\bm q}$, and
\be
C^{\bm i,\bm j}=U^{\bm p\dag}_{(\bm1^n-\bm i,\bm0^{m-n}),(\bm0^k,\bm1^{m-k})}V^{\bm q}_{(\bm0^k,\bm1^{m-k}),(\bm1^n-\bm j,\bm0^{m-n})}
\ee
is an $(n-r)\times(n-r)$ matrix which can be obtained efficiently from $U^{\bm p}$ and $V^{\bm q}$.
\end{lem}

\begin{proof}
\begin{mdframed}[linewidth=1.5,topline=false,rightline=false,bottomline=false]

We consider the expression for the inner product obtained in Eq.~(\ref{overlapS}):
\be
\braket{\psi_{\bm p}|\psi_{\bm q}}=\frac1{\sqrt{\bm p!\bm q!}}\sum_{\bm s\in\Phi_{m-k,n-r}}{\frac1{\bm s!}\Per\left(U^{\bm p\dag}_{\bm t,(\bm p,\bm s)}\right)\Per\left(V^{\bm q}_{(\bm q,\bm s),\bm t}\right)}.
\label{overlapS1}
\ee

\medskip

\noindent We first apply the general column expansion formula in Eq.~(\ref{Laplacegen}) to the matrix $U^{\bm p\dag}_{\bm t,(\bm p,\bm s)}$ with $\bm j=(\bm1^r,\bm0^{n-r})\in\{0,1\}^n$, obtaining
\be
\Per\left(U^{\bm p\dag}_{\bm t,(\bm p,\bm s)}\right)=\sum_{\substack{\bm i\in\{0,1\}^n\\|\bm i|=r}}{\Per\left[\left(U^{\bm p\dag}_{\bm t,(\bm p,\bm s)}\right)_{\bm i,\bm j}\right]\Per\left[\left(U^{\bm p\dag}_{\bm t,(\bm p,\bm s)}\right)_{\bm1^n-\bm i,\bm1^n-\bm j}\right]}.
\label{expandU}
\ee
Let us consider the matrix $\left(U^{\bm p\dag}_{\bm t,(\bm p,\bm s)}\right)_{\bm i,\bm j}$ appearing in this last expression, for $\bm i\in\{0,1\}^n$. Its rows are obtained by keeping the first $n$ lines of $U^{\bm p\dag}$ since $\bm t=(\bm1^n,\bm0^{m-n})$, then by keeping only the $l^{th}$ rows such that $i_l=1$. Its columns are obtained by repeating $p_l$ times the $l^{th}$ column for $l\in\{1,\dots,k\}$ and $s_l$ times for $l\in\{k+1,\dots,m\}$, then by only keeping the first $r$ columns since $\bm j=(\bm1^r,\bm0^{n-r})$. However, since $|\bm p|=|\bm j|=r$, these are the columes repeated according to $\bm p$. Hence,
\be
\left(U^{\bm p\dag}_{\bm t,(\bm p,\bm s)}\right)_{\bm i,\bm j}=U^{\bm p\dag}_{(\bm i,\bm0^{m-n}),(\bm p,\bm0^{m-k})},
\label{simplerU1}
\ee
where $U^{\bm p\dag}_{(\bm i,\bm0^{m-n}),(\bm p,\bm0^{m-k})}$ is the matrix obtained from $U^{\bm p\dag}$ by keeping only the $l^{th}$ rows such that $i_l=1$ and removing the others, and by repeating $p_l$ times the $l^{th}$ column for $l\in\{1,\dots,k\}$ and removing the others. Similarly, with $|\bm s|=|\bm1^n-\bm j|=n-r$,
\be
\left(U^{\bm p\dag}_{\bm t,(\bm p,\bm s)}\right)_{\bm1^n-\bm i,\bm1^n-\bm j}=U^{\bm p\dag}_{(\bm1^n-\bm i,\bm0^{m-n}),(\bm0^k,\bm s)},
\label{simplerU2}
\ee
where $U^{\bm p\dag}_{(\bm1^n-\bm i,\bm0^{m-n}),(\bm0^k,\bm s)}$  is the matrix obtained from $U^{\bm p\dag}$ by keeping only the $l^{th}$ rows such that $i_l=0$ and removing the others, and by repeating $s_l$ times the $l^{th}$ column for $l\in\{k+1,\dots,m\}$ and removing the others. With Eqs.~(\ref{expandU}), (\ref{simplerU1}) and (\ref{simplerU2}) we obtain
\be
\ba
\Per\left(U^{\bm p\dag}_{\bm t,(\bm p,\bm s)}\right)&=\sum_{\substack{\bm i\in\{0,1\}^n\\|\bm i|=r}}{\Per\left(U^{\bm p\dag}_{(\bm i,\bm0^{m-n}),(\bm p,\bm0^{m-k})}\right)\Per\left(U^{\bm p\dag}_{(\bm1^n-\bm i,\bm0^{m-n}),(\bm0^k,\bm s)}\right)}\\
&=\sum_{\substack{\bm i\in\{0,1\}^n\\|\bm i|=r}}{\Per\left(A^{\bm i}\right)\Per\left(U^{\bm p\dag}_{(\bm1^n-\bm i,\bm0^{m-n}),(\bm0^k,\bm s)}\right)},
\ea
\label{expandU2}
\ee
where we have defined, for all $\bm i\in\{0,1\}^n$ such that $|\bm i|=r$,
\be
A^{\bm i}:=U^{\bm p\dag}_{(\bm i,\bm0^{m-n}),(\bm p,\bm0^{m-k})},
\ee
which is an $r\times r$ matrix independent of $\bm s$ that can be obtained efficiently from $U^{\bm p}$.

\medskip

\noindent The same reasoning with the general row expansion formula for the matrix $V^{\bm q}_{(\bm q,\bm s),\bm t}$ and the rows $\bm i=(\bm1^r,\bm0^{n-r})$ gives
\be
\ba
\Per\left(V^{\bm q}_{(\bm q,\bm s),\bm t}\right)&=\sum_{\substack{\bm j\in\{0,1\}^n\\|\bm j|=r}}{\Per\left[\left(V^{\bm q}_{(\bm q,\bm s),\bm t}\right)_{\bm i,\bm j}\right]\Per\left[\left(V^{\bm q}_{(\bm q,\bm s),\bm t}\right)_{\bm1^n-\bm i,\bm1^n-\bm j}\right]}\\
&=\sum_{\substack{\bm j\in\{0,1\}^n\\|\bm j|=r}}{\Per\left(V^{\bm q}_{(\bm q,\bm0^{m-k}),(\bm j,\bm0^{m-n})}\right)\Per\left(V^{\bm q}_{(\bm0^k,\bm s),(\bm1^n-\bm j,\bm0^{m-n})}\right)},
\ea
\label{expandV}
\ee
where $V^{\bm q}_{(\bm q,\bm0^{m-k}),(\bm j,\bm0^{m-n})}$ is the matrix obtained from $V^{\bm q}$ by repeating $q_l$ times the $l^{th}$ row for $l\in\{1,\dots,k\}$ and removing the others and by keeping only the $l^{th}$ columns such that $j_l=1$, and where $V^{\bm q}_{(\bm0^k,\bm s),(\bm1^n-\bm j,\bm0^{m-n})}$ is the matrix obtained from $V^{\bm q}$ by repeating $s_l$ times the $l^{th}$ row for $l\in\{k+1,\dots,m\}$ and removing the others and by keeping only the $l^{th}$ columns such that $j_l=0$. 
Defining, for all $\bm j\in\{0,1\}^n$ such that $|\bm j|=r$,
\be
B^{\bm j}:=V^{\bm q}_{(\bm q,\bm0^{m-k}),(\bm j,\bm0^{m-n})},
\ee
the expression in Eq.~(\ref{expandV}) rewrites
\be
\Per\left(V^{\bm q}_{(\bm q,\bm s),\bm t}\right)=\sum_{\substack{\bm j\in\{0,1\}^n\\|\bm j|=r}}{\Per\left(B^{\bm j}\right)\Per\left(V^{\bm q}_{(\bm0^k,\bm s),(\bm1^n-\bm j,\bm0^{m-n})}\right)},
\label{expandV2}
\ee
where $B^{\bm j}$ are $r\times r$ matrices independent of $\bm s$ and can be obtained efficiently from $V^{\bm q}$.
 
\medskip

\noindent Plugging Eqs.~(\ref{expandU2}) and (\ref{expandV2}) in Eq.~(\ref{overlapS1}) we obtain
\be
\ba
\braket{\psi_{\bm p}|\psi_{\bm q}}&=\frac1{\sqrt{\bm p!\bm q!}}\sum_{\substack{\bm i,\bm j\in\{0,1\}^n\\|\bm i|=|\bm j|=r}}\Bigg[\Per\left(A^{\bm i}\right)\Per\left(B^{\bm j}\right)\\
&\quad\quad\quad\times\sum_{\bm s\in\Phi_{m-k,n-r}}{\frac1{\bm s!}\Per\left(U^{\bm p\dag}_{(\bm1^n-\bm i,\bm0^{m-n}),(\bm0^k,\bm s)}\right)\Per\left(V^{\bm q}_{(\bm0^k,\bm s),(\bm1^n-\bm j,\bm0^{m-n})}\right)}\Bigg].
\ea
\label{overlapS2}
\ee
The sum appearing in the second line may now be expressed as a single permanent using the permanent composition formula: for all $\bm i,\bm j\in\{0,1\}^n$ such that $|\bm i|=|\bm j|=r$, let us define the $(n-r)\times(m-k)$ matrix
\be
\tilde U^{\bm p,\bm i}:=U^{\bm p\dag}_{(\bm1^n-\bm i,\bm0^{m-n}),(\bm0^k,\bm1^{m-k})},
\ee
and the $(m-k)\times(n-r)$ matrix
\be
\tilde V^{\bm q,\bm j}:=V^{\bm q}_{(\bm0^k,\bm1^{m-k}),(\bm1^n-\bm j,\bm0^{m-n})},
\ee
so that
\be
U^{\bm p\dag}_{(\bm1^n-\bm i,\bm0^{m-n}),(\bm0^k,\bm s)}=\tilde U^{\bm p,\bm i}_{\bm1^{n-r},\bm s}\quad\text{and}\quad V^{\bm q}_{(\bm0^k,\bm s),(\bm1^n-\bm j,\bm0^{m-n})}=\tilde V^{\bm q,\bm j}_{\bm s,\bm1^{n-r}}.
\ee
With the permanent composition formula in Eq.~(\ref{percomp2}) we obtain
\be
\sum_{\bm s\in\Phi_{m-k,n-r}}{\frac1{\bm s!}\Per\left(U^{\bm p\dag}_{(\bm1^n-\bm i,\bm0^{m-n}),(\bm0^k,\bm s)}\right)\Per\left(V^{\bm q}_{(\bm0^k,\bm s),(\bm1^n-\bm j,\bm0^{m-n})}\right)}=\Per\left[\left(\tilde U^{\bm p,\bm i}\tilde V^{\bm q,\bm j}\right)_{\bm1^{n-r},\bm1^{n-r}}\right].
\ee
Since $\tilde U^{\bm p,\bm i}\tilde V^{\bm q,\bm j}$ is an $(n-r)\times(n-r)$ matrix we thus have
\be
\sum_{\bm s\in\Phi_{m-k,n-r}}{\frac1{\bm s!}\Per\left(U^{\bm p\dag}_{(\bm1^n-\bm i,\bm0^{m-n}),(\bm0^k,\bm s)}\right)\Per\left(V^{\bm q}_{(\bm0^k,\bm s),(\bm1^n-\bm j,\bm0^{m-n})}\right)}=\Per\left(\tilde U^{\bm p,\bm i}\tilde V^{\bm q,\bm j}\right).
\ee
Then, Eq.~(\ref{overlapS2}) rewrites
\be
\braket{\psi_{\bm p}|\psi_{\bm q}}=\frac1{\sqrt{\bm p!\bm q!}}\sum_{\substack{\bm i,\bm j\in\{0,1\}^n\\|\bm i|=|\bm j|=r}}{\Per\left(A^{\bm i}\right)\Per\left(B^{\bm j}\right)\Per\left(C^{\bm i,\bm j}\right)},
\label{overlapS3}
\ee
where we have defined
\be
\ba
C^{\bm i,\bm j}&:=\tilde U^{\bm p,\bm i}\tilde V^{\bm q,\bm j}\\
&=U^{\bm p\dag}_{(\bm1^n-\bm i,\bm0^{m-n}),(\bm0^k,\bm1^{m-k})}V^{\bm q}_{(\bm0^k,\bm1^{m-k}),(\bm1^n-\bm j,\bm0^{m-n})},
\ea
\ee
is an $(n-r)\times(n-r)$ matrix which can be obtained efficiently from $U^{\bm p}$ and $V^{\bm q}$.

\end{mdframed}
\end{proof}

\noindent By Lemma~\ref{lem:overlapS}, the overlap is expressed as the modulus squared of a sum over $\binom nr^2$ products of three permanents, of square matrices of sizes $|\bm p|=r$, $|\bm q|=r$ and $(n-r)$, respectively. In the worst case, when $r=n/2$, the sum has at most $O(4^n)$ terms, up to a polynomial factor in $n$. In particular, when $n=O(\log m)$, the overlap reduces to a sum of a polynomial number of terms, which can all be computed in time $O(\poly m)$. Moreover, the cost of computing the overlap is independent of the number $k$ of adaptive measurements, up to the cost of constructing the matrices with repeated lines and columns (which is $O(\poly m)$). The overlap of normalised ouput states is given by
\be
\frac{|\braket{\psi_{\bm p}|\psi_{\bm q}}|^2}{\braket{\psi_{\bm p}|\psi_{\bm p}}\braket{\psi_{\bm q}|\psi_{\bm q}}},
\ee
which may also be computed efficiently when $n=O(\log m)$. The efficiency of the classical algorithm is summarised as a function of $n$ and $k$ in Table~\ref{tab:overlapest} and as a function of $n$ and the number of photons $r$ detected during the adaptive measurements in Table~\ref{tab:overlapest2}, where the regimes are obtained using Stirling equivalent $n!\sim\sqrt{2\pi n}\left(\frac ne\right)^n$.

\begin{table}[h!]
\centering
\setlength\tabcolsep{20pt}
\bgroup
\def\arraystretch{1.5}
\begin{tabular}{| c | c | c | c |}
\hline
\diagbox[height=8ex,width=9em]{$n$}{$k$} & $O(1)$ & $O(\log m)$ & $O(m)$\\
\hline
$O(1)$ & \cellcolor{blue!15} & \cellcolor{blue!15}  & \cellcolor{blue!15} \\
\hline
$O(\log m)$ & \cellcolor{blue!15}  & \cellcolor{blue!15}  &  \cellcolor{blue!15}$\dots$\\
\hline
$O(m)$ &  & $\dots$ & $\dots$ \\
\hline
\end{tabular}
\egroup
\caption{Simulability regimes for overlap estimation as a function of $n$ and $k$. Since the running time is independent of $k$, the columns are the same. In blue is the parameter region for which the classical algorithm is no longer efficient. The symbol $\dots$ indicates regimes where the quantum algorithm is not efficient.}
\label{tab:overlapest}
\end{table}
\begin{table}[h!]
\centering
\setlength\tabcolsep{20pt}
\bgroup
\def\arraystretch{1.5}
\begin{tabular}{| c | c | c | c |}
\hline
\diagbox[height=8ex,width=9em]{$n$}{$r$} & $O(1)$ & $O(\log n)$ & $O(n)$\\
\hline
$O(1)$ &  \cellcolor{blue!15} &  \cellcolor{blue!15} &  \cellcolor{blue!15}\\
\hline
$O(\log m)$ &  \cellcolor{blue!15} &  \cellcolor{blue!15} &  \cellcolor{blue!15}\\
\hline
$O(m)$ &  \cellcolor{blue!15} &  &  \\
\hline
\end{tabular}
\egroup
\caption{Simulability regimes for overlap estimation as a function of $n$ and $r$. In blue is the parameter region for which the classical algorithm is efficient. }
\label{tab:overlapest2}
\end{table}

\noindent Since the quantum efficient regime corresponds to $\binom{k+n}n=O(\poly m)$, there is a possibility of quantum advantage for overlap estimation when $k=O(1)$ and $n=O(m)$. 

\noindent In the case of probability estimation, the possible regimes for quantum advantage do not correspond to near-term implementations: $k$ and $n$ must be both greater than $\log m$. However, for overlap estimation, there is a possiblity of near-term beyond-classical computing with adaptive linear optics using one adaptive measurement, which requires the preparation of photon number states. Note that the interferometer should be concentrating many photons $r$ onto the adaptive measurement in order to obtain possibly hard to estimate overlaps. Using more adaptive measurements does not increase the complexity (apart from polynomial factors in $m$). 

\medskip

\noindent Having characterised these specific simulation regimes, we consider in what follows stronger notions of simulation. In particular, we give classical algorithms for strong simulation of a large class of continuous variable quantum computational models.

\section{The computational power of non-Gaussian states}

Continuous variable systems are being recognized as a promising alternative to the use of qubits, as they allow for the deterministic generation of unprecedented large entangled quantum states, of up to one-million elementary systems~\cite{yokoyama2013optical,Yoshikawa2016} and also offer detection techniques, such as homodyne and heterodyne, with high efficiency and reliability (see section~\ref{sec:heterodynemeasurement}). Any given continuous variable quantum circuit is defined by (i) an input state lying in an infinite-dimensional Hilbert space, (ii) an evolution and (iii) measurements (see section~\ref{sec:CVQI}). An important theorem~\cite{Bartlett2002,mari2012positive} states that if all these elements are described by positive Wigner functions, then there exists a classical algorithm able to efficiently simulate this circuit. Hence, including a negative Wigner function element is mandatory in order to design a continuous variable subuniversal quantum circuit that cannot be efficiently simulated by a classical device. Since Gaussian states and processes have positive Wigner functions, this necessarily corresponds to the use of non-Gaussian resources.

Therefore, if one aims at minimal extensions of Gaussian models, three different families of non trivial quantum circuits can be defined, depending on whether the element yielding the Wigner function negativity is provided by the input state, the unitary evolution, or the measurement. 

In what follows, we analyse the computational power of non-Gaussian states and thus focus on the case where Gaussian circuits and measurements are supplemented with non-Gaussian input states as a computational resource. The results obtained have consequences for all three families of circuits, since non-Gaussian gates and non-Gaussian measurements can be implemented by Gaussian operations together with non-Gaussian ancillary states~\cite{Gottesman2001,ghose2007non,sabapathy2018states}.

\subsection{Gaussian circuits with non-Gaussian inputs}
\label{sec:NGinput}

\noindent We first extend a few definitions from the previous chapter to the multimode case, using multi-index notations (see section~\ref{sec:multiindex}). First, the stellar function, which provides a representation of multimode pure states as multivariate holomorphic functions:

\begin{defi}[Multimode stellar function]
Let $m\in\mathbb N^*$ and let $\ket{\bm\psi}=\sum_{\bm n\ge\bm0}{\psi_{\bm n}\ket{\bm n}}\in\mathcal H^{\otimes m}$ be a normalised pure state over $m$ modes. The \textit{stellar function} of the state $\ket{\bm\psi}$ is defined as
\be
F_{\bm\psi}^\star(\bm z)=e^{\frac12\|\bm z\|^2}\braket{\bm z^*|\bm\psi}=\sum_{\bm n\ge\bm 0}{\frac{\psi_{\bm n}}{\sqrt{\bm n!}}\bm z^{\bm n}},
\label{stellarfuncmulti}
\ee
for all $\bm z\in\mathbb C^m$, where $\ket{\bm z}=e^{-\frac12\|\bm z\|^2}\sum_{\bm n\ge\bm0}{\frac{\bm z^{\bm n}}{\sqrt{\bm n!}}\ket{\bm n}}\in\mathcal H^{\otimes m}$ is the coherent state of amplitude $\bm z$.
\end{defi}

\noindent The following definition also extends naturally from the single-mode case:

\begin{defi}[Multimode core state]
\textit{Multimode core states} are defined as the normalised pure quantum states which have a (multivariate) polynomial stellar function.
\end{defi}

\noindent Like in the single-mode case, these are the states with a finite support over the (multimode) Fock basis. For any $m\in\mathbb N^*$, the set of multimode core states over $m$ modes is dense in the set of normalised states for the trace norm (by considering renormalised cutoff states). We also introduce the following definitions:

\begin{defi}[Degree of a multimode core state]
The \textit{degree} of a multimode core state is defined as the degree-sum of its stellar function.
\end{defi}

\begin{defi}[Support of a multimode core state]
The \textit{support} of a multimode core state is the set of Fock basis elements which have nonzero overlap with the core state.
\end{defi}

\noindent For example, the $3$-mode core state $\frac1{\sqrt2}(\ket{210}+\ket{001})$ is of degree $3$ and has a support of size $2$, and its stellar function  is given by $z_1^2z_2/2+z_3/\sqrt2$, for all $(z_1,z_2,z_3)\in\mathbb C^3$.
 
\medskip

\begin{figure}[h!]
\begin{center}
\includegraphics[width=0.5\columnwidth]{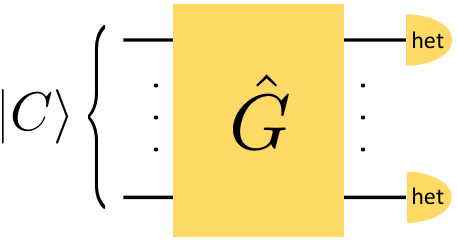}
\caption{Representation of a $G_{\text{core}}$ circuit with multimode core state input $\ket{\bm C}$. The unitary $\hat G$ is Gaussian and the measurement is performed by heterodyne detection.}
\label{fig:GCI}
\end{center}
\end{figure}

\noindent We consider Gaussian circuits with Gaussian measurements, supplemented by non-Gaussian multimode core states in input, which we refer to as $G_{\text{core}}$ circuits. Without loss of generality, a Gaussian measurement may be written as a tensor product of single-mode balanced heterodyne detections preceded by a Gaussian unitary (see section~\ref{sec:heterodynemeasurement}). $G_{\text{core}}$ circuits are thus described by two (multidimensional) parameters: a multimode core state $\ket{\bm C}$ in the input and a Gaussian unitary evolution $\hat G$ (Fig.~\ref{fig:GCI}).\\

\noindent In what follows, we derive a general expression for the output probability density of these circuits. Then, we study the classical simulability of $G_{\text{core}}$ circuits and of various subclasses of circuits.

\medskip

\noindent We first recall a few combinatorial functions related to the permanent, which appear in the expressions of the output probability densities. The hafnian of a square matrix $A=(a_{ij})_{1\le i,j\le2m}$ of size $2m$ is defined as~\cite{caianiello1953quantum}
\begin{equation}
\Haf\,(A):=\sum_{M\in\text{PMP}\,(2m)}{\prod_{\{i,j\}\in M}{a_{ij}}},
\label{Hafnian}
\end{equation}
where the sum is over the perfect matchings of the set $\{1,\dots,2m\}$, i.e., the partitions of $\{1,\dots,2m\}$ in subsets of size $2$. The hafnian of a matrix of odd size is $0$. The hafnian is related to the permanent by
\be
\Haf\begin{pmatrix} \mymathbb0_m & B \\ B^T & \mymathbb0_m \end{pmatrix}=\Per\,(B),
\label{hafper}
\ee
for any $m\times m$ matrix $B$. By convention we set $\Haf\,(\emptyset)=1$, where $\emptyset$ is a square matrix of size $0$.

\medskip

\noindent The loop hafnian of a square matrix $R=(r_{ij})_{1\le i,j\le r}$ of size $r$ is defined as~\cite{bjorklund2019faster}
\be
\lHaf(R):=\sum_{M\in\text{SMP}\,(r)}{\prod_{\{i,j\}\in M}{r_{ij}}},
\label{lHaf}
\ee
where the sum is over the single pair matchings of the set $\{1,\dots,r\}$, defined as the set of perfect matchings of a complete graph with loops with $r$ vertices. This set is isomorphic to the set $\Pi_{1,2}(\{1,\dots,r\})$ of partitions of $\{1,\dots,r\}$ in subsets of size $1$ and $2$ (by mapping a block $\{k\}$ of size $1$ of a partition to the matching $\{k,k\}$ and a block $\{i,j\}$ of size $2$ to the matching $\{i,j\}$). In particular, when $R$ is a matrix whose diagonal entries are all $0$, we have $\lHaf(R)=\Haf\,(R)$.

\medskip

\noindent We obtain a closed expression for the output probability density of Gaussian circuits with multimode core states input in Theorem~\ref{th:Pr}, by adapting proof techniques from~\cite{Hamilton2016,kruse2019detailed,quesada2019franck}. We first state an intermediate technical result.

\begin{lem}\label{lem:efficientC}
Let $m\in\mathbb N^*$, let $V$ be a $2m\times 2m$ symmetric matrix and let $D$ be a column vector of size $2m$. For all $\bm p,\bm q\in\mathbb N^m$, there exists a square matrix $A_{\bm p,\bm q}(V,D)$ of size $|\bm p|+|\bm q|$ such that
\be
\ba
T_{\bm p,\bm q}(V,D)&:=\int_{\bm\beta\in\mathbb C^m}{\exp\left[\frac12\bm{\tilde\beta}^TV\bm{\tilde\beta}+D^T\bm{\tilde\beta}\right]\left(\frac{\partial}{\partial\bm\beta}\right)^{\bm p}\left(\frac{\partial}{\partial\bm\beta^*}\right)^{\bm q}\delta^{2m}(\bm\beta,\bm\beta^*)\,d^m\!\bm\beta\,d^m\!\bm\beta^*}\\
&=(-1)^{|\bm p|+|\bm q|}\lHaf\left[A_{\bm p,\bm q}(V,D)\right],
\ea
\ee
assuming the integral is well defined. The matrix $A_{\bm p,\bm q}(V,D)$ is obtained by repeating the entries of $V$ according to $\bm p$ and $\bm q$ and replacing the diagonal of the matrix obtained by the corresponding elements of $D$ (a detailed example follows the proof).
\end{lem}

\begin{proof}
\begin{mdframed}[linewidth=1.5,topline=false,rightline=false,bottomline=false]

Writing $\bm p=(p_1,\dots,p_m)$ and $\bm q=(q_1,\dots,q_m)$, we first get rid of the integral by successive integration by parts:
\begin{align}
\nonumber T_{\bm p,\bm q}(V,D)&=(-1)^{|\bm p|+|\bm q|}\left(\frac{\partial}{\partial\bm\beta}\right)^{\bm p}\left(\frac{\partial}{\partial\bm\beta^*}\right)^{\bm q}\exp\left[\frac12\bm{\tilde\beta}^TV\bm{\tilde\beta}+D^T\bm{\tilde\beta}\right]\Bigg\rvert_{\bm{\tilde\beta}=\bm0}\\
&=(-1)^{|\bm p|+|\bm q|}\prod_{j=1}^m{\left(\frac{\partial}{\partial\beta_j}\right)^{p_j}\left(\frac{\partial}{\partial\beta_j^*}\right)^{q_j}}\exp\left[\frac12\bm{\tilde\beta}^TV\bm{\tilde\beta}+D^T\bm{\tilde\beta}\right]\Bigg\rvert_{\bm{\tilde\beta}=\bm0}\\
\nonumber&=(-1)^{|\bm p|+|\bm q|}\prod_{j\in\mathcal E_{\bm p,\bm q}}{\left(\frac{\partial}{\partial\tilde\beta_j}\right)}\exp\left[\frac12\bm{\tilde\beta}^TV\bm{\tilde\beta}+D^T\bm{\tilde\beta}\right]\Bigg\rvert_{\bm{\tilde\beta}=\bm0},
\end{align}
where the multiset $\mathcal E_{\bm p,\bm q}$ is defined as the set of size $|\bm p|+|\bm q|$ obtained from $\{1,\dots,2m\}$ by repeating $p_k$ times the index $k$ and $q_k$ times the index $m+k$, for all $k\in\{1,\dots,m\}$.

We make use of Fa\`a di Bruno's formula~\cite{hardy2006combinatorics} in order to expand the product of partial derivatives and we obtain
\be
T_{\bm p,\bm q}(V,D)=(-1)^{|\bm p|+|\bm q|}\sum_{\pi\in\Pi(\mathcal E_{\bm p,\bm q})}\prod_{B\in\pi}{\left(\frac{\partial^{|B|}}{\prod_{j\in B}\partial\tilde\beta_j}\right)}\left[\frac12\bm{\tilde\beta}^TV\bm{\tilde\beta}+D^T\bm{\tilde\beta}\right]\Bigg\rvert_{\bm{\tilde\beta}=\bm0},
\label{Faadi2}
\ee
where $\Pi(\mathcal E_{\bm p,\bm q})$ denotes the set of all partitions of the multiset $\mathcal E_{\bm p,\bm q}$, and where the product runs over the blocks $B$ of the partition $\pi\in\Pi(\mathcal E_{\bm p,\bm q})$, with $|B|$ the size of the block. The function $\bm{\tilde\beta}^\dag V\bm{\tilde\beta}+D^\dag\bm{\tilde\beta}$ is a sum of a quadratic and a linear functions, so all derivatives of order greater than $2$ in the sum vanish. We thus have
\be
\ba
T_{\bm p,\bm q}(V,D)&=(-1)^{|\bm p|+|\bm q|}\sum_{\pi\in\Pi_{1,2}(\mathcal E_{\bm p,\bm q})}\prod_{B\in\pi}{\left(\frac{\partial^{|B|}}{\prod_{j\in B}\partial\tilde\beta_j}\right)}\left[\frac12\bm{\tilde\beta}^TV\bm{\tilde\beta}+D^T\bm{\tilde\beta}\right]\Bigg\rvert_{\bm{\tilde\beta}=\bm0}\\
&=(-1)^{|\bm p|+|\bm q|}\sum_{\pi\in\Pi_{1,2}(\mathcal E_{\bm p,\bm q})}\prod_{\{i,j\}\in\pi}{\left(\frac{\partial^2}{\partial\tilde\beta_i\partial\tilde\beta_j}\right)}\left[\frac12\bm{\tilde\beta}^TV\bm{\tilde\beta}+D^T\bm{\tilde\beta}\right]\Bigg\rvert_{\bm{\tilde\beta}=\bm0}\\
&\quad\quad\quad\quad\quad\quad\quad\quad\quad\quad\quad\quad\times\prod_{\{k\}\in\pi}{\left(\frac{\partial}{\partial\tilde\beta_k}\right)}\left[\frac12\bm{\tilde\beta}^TV\bm{\tilde\beta}+D^T\bm{\tilde\beta}\right]\Bigg\rvert_{\bm{\tilde\beta}=\bm0},
\ea
\ee
where $\Pi_{1,2}(\mathcal E_{\bm p,\bm q})$ denotes the set of all partitions of the multiset $\mathcal E_{\bm p,\bm q}$ in subsets of size $1$ and $2$. All derivatives of order $2$ of the linear term vanish, and all derivatives of order $1$ of the quadratic term vanish when evaluated at $\bm{\tilde\beta}=\bm0$. We thus obtain
\be
T_{\bm p,\bm q}(V,D)=(-1)^{|\bm p|+|\bm q|}\sum_{\pi\in\Pi_{1,2}(\mathcal E_{\bm p,\bm q})}\prod_{\{i,j\}\in\pi}{\left(\frac{\partial^2}{\partial\tilde\beta_i\partial\tilde\beta_j}\right)}\left[\frac12\bm{\tilde\beta}^TV\bm{\tilde\beta}\right]\Bigg\rvert_{\bm{\tilde\beta}=\bm0}\prod_{\{k\}\in\pi}{\left(\frac{\partial}{\partial\tilde\beta_k}\right)}\left[D^T\bm{\tilde\beta}\right]\Bigg\rvert_{\bm{\tilde\beta}=\bm0}.
\ee
Writing $V=(v_{ij})_{1\le i,j\le2m}$, with $V=V^T$, and $D=(d_k)_{1\le k\le2m}$ we obtain
\be
T_{\bm p,\bm q}(V,D)=(-1)^{|\bm p|+|\bm q|}\sum_{\pi\in\Pi_{1,2}(\mathcal E_{\bm p,\bm q})}{\prod_{\{i,j\}\in\pi}{v_{ij}}\prod_{\{k\}\in\pi}{d_k}}.
\label{TmnVDsum}
\ee
We now show that this expression may be rewritten as the loop hafnian of a matrix of size $|\bm p|+|\bm q|$. 
Define $V_{\bm p,\bm q}$ the $(|\bm p|+|\bm q|)\times(|\bm p|+|\bm q|)$ matrix obtained from $V$ by repeating $p_k$ times its $k^{th}$ rows and columns and $q_k$ times its $(m+k)^{th}$ rows and columns, for $k\in\{1,\dots,m\}$. Similarly, define $D_{\bm p,\bm q}$ the column vector of size $|\bm p|+|\bm q|$ obtained from $D$ by repeating $p_k$ times its $k^{th}$ element and $q_k$ times its $(m+k)^{th}$ element, for $k\in\{1,\dots,m\}$. Finally, let $A_{\bm p,\bm q}(V,D)=(a_{ij})_{1\le i,j\le|\bm p|+|\bm q|}$ be the $(|\bm p|+|\bm q|)\times(|\bm p|+|\bm q|)$ matrix obtained from $V_{\bm p,\bm q}$ by replacing its diagonal with the vector $D_{\bm p,\bm q}$. Then, Eq.~(\ref{TmnVDsum}) rewrites
\be
\ba
T_{\bm p,\bm q}(V,D)&=(-1)^{|\bm p|+|\bm q|}\sum_{\pi\in\Pi_{1,2}(\{1,\dots,|\bm p|+|\bm q|\})}{\prod_{\{i,j\}\in\pi}{a_{ij}}\prod_{\{k\}\in\pi}{a_{kk}}}\\
&=(-1)^{|\bm p|+|\bm q|}\sum_{M\in\text{SMP}(|\bm p|+|\bm q|)}{\prod_{\{i,j\}\in M}{a_{ij}}}\\
&=(-1)^{|\bm p|+|\bm q|}\lHaf\left[A_{\bm p,\bm q}(V,D)\right],
\ea
\label{lHafA}
\ee
where the sum in the first line is over the partitions of $\{1,\dots,|\bm p|+|\bm q|\}$ in subsets of size $1$ and $2$, where the sum in the second line is over the single pair matchings of the set $\{1,\dots,|\bm p|+|\bm q|\}$ and where the third line comes from the definition of the loop hafnian in Eq.~(\ref{lHaf}). 

\end{mdframed}
\end{proof}

\noindent Let us illustrate with an example how the matrix $A_{\bm p,\bm q}(V,D)$ appearing in Lemma~\ref{lem:efficientC} is constructed from the matrix $V$ and the vector $D$. Let us set $m=2$, $\bm p=(2,0)$ and $\bm q=(1,0)$. We write
\be
V=\begin{pmatrix}
v_{11}&v_{12}&v_{13}&v_{14}\\
v_{21}&v_{22}&v_{23}&v_{24}\\
v_{31}&v_{32}&v_{33}&v_{34}\\
v_{41}&v_{42}&v_{43}&v_{44}
\end{pmatrix}\quad\text{and}\quad D=\begin{pmatrix}
d_1\\
d_2\\
d_3\\
d_4
\end{pmatrix}.
\ee
We first build the matrix $V_{\bm p,\bm q}$ by repeating $p_k$ times the $k^{th}$ rows and columns of $V$ and $q_k$ times the $(m+k)^{th}$ rows and columns. In that case, $\bm p=(p_1,p_2)=(2,0)$, so we repeat $2$ times the first row and column and discard the second row and column, and $\bm q=(q_1,q_2)=(1,0)$, so we keep the third row and column and discard the fourth row and column, obtaining the $3\times3$ matrix
\be
V_{\bm p,\bm q}=\begin{pmatrix}
v_{11}&v_{11}&v_{13}\\
v_{11}&v_{11}&v_{13}\\
v_{31}&v_{31}&v_{33}
\end{pmatrix}.
\ee
Similarly, we obtain the vector $D_{\bm p,\bm q}$ by repeating $p_k$ times the $k^{th}$ element of $D$ and $q_k$ times the $(m+k)^{th}$ element, as
\be
D_{\bm p,\bm q}=\begin{pmatrix}
d_1\\
d_1\\
d_3
\end{pmatrix}.
\ee
Finally, we replace the diagonal of $V_{\bm p,\bm q}$ by $D_{\bm p,\bm q}$:
\be
A_{\bm p,\bm q}(V,D)=\begin{pmatrix}
d_1&v_{11}&v_{13}\\
v_{11}&d_1&v_{13}\\
v_{31}&v_{31}&d_3
\end{pmatrix}.
\ee
Note that this construction by repeating rows and columns differ from the one encountered in the previous section when dealing with the permanent of matrices, for which the first index denotes which rows are repeated and the second which columns. Here, we are dealing with hafnians of matrices of double size, where the first index denotes which rows and columns are repeated for indices in $\{1,\dots,m\}$, while the second index denotes which rows and columns are repeated for indices in $\{m+1,\dots,2m\}$. However, the two constructions coincide when looking at matrices of the form
\be
\begin{pmatrix} \mymathbb0_m & B \\ B^T & \mymathbb0_m \end{pmatrix},
\ee
through the relation in Eq.~(\ref{hafper}):
\be
\Haf\begin{pmatrix} \mymathbb0_m & B \\ B^T & \mymathbb0_m \end{pmatrix}=\Per\,(B),
\ee
for any $m\times m$ square matrix $B$.

\medskip

\noindent Combining Lemma~\ref{lem:efficientC} with phase space formalism (see section~\ref{sec:phasespace}) and properties of Gaussian states (see section~\ref{sec:Gaussian}), we obtain the following result:

\begin{theo}\label{th:Pr}
Let $m,n\in\mathbb N^*$ and let
\be
\ket{\bm C}=\sum_{\substack{\bm p\in\mathbb N^m\\|\bm p|\le n}}{c_{\bm p}\ket{\bm p}},
\ee
be an $m$-mode core state of degree $n$. Let $\hat G$ be a Gaussian unitary over $m$ modes. For all $\bm\alpha\in\mathbb C^m$, let us write $\bm V$ and $\bm{\tilde d}=(\bm d,\bm d^*)$ the covariance matrix and the displacement vector of the Gaussian state $\hat G^\dag\ket{\bm\alpha}$. Then, the output probability density for the $G_{\text{core}}$ circuit $\hat G$ with input $\ket{\bm C}$ and heterodyne detection, evaluated at $\bm\alpha$, is given by
\be
\text{Pr}_{\text{core}}[\bm\alpha]=\kappa(\bm\alpha,\hat G)\sum_{\substack{\bm p,\bm q\in\mathbb N^m\\|\bm p|\le n,|\bm q|\le n}}{\frac{(-1)^{|\bm p|+|\bm q|}}{\sqrt{\bm p!\bm q!}}c_{\bm p}c_{\bm q}^*\lHaf\left(A_{\bm p,\bm q}\right)},
\label{Pr}
\ee
where $A_{\bm p,\bm q}$ is the square matrix of size $|\bm p|+|\bm q|$ obtained with Lemma~\ref{lem:efficientC} from
\be
V=\begin{pmatrix}\mymathbb0_m & \mathbbm1_m \\ \mathbb1_m & \mymathbb0_m \end{pmatrix}\left[\mathbb1_{2m}-\left(\bm V+\mathbb1_{2m}/2\right)^{-1}\right]\quad\text{and}\quad D=\left[\bm{\tilde d}^\dag\left(\bm V+\mathbb1_{2m}/2\right)^{-1}\right]^T,
\ee
and where
\be
\kappa(\bm\alpha,\hat G)=\frac{\exp\left[-\frac12\bm{\tilde d}^\dag\left(\bm V+\mathbb1_{2m}/2\right)^{-1}\bm{\tilde d}\right]}{\pi^m\sqrt{\Det\,(\bm V+\mathbb1_{2m}/2)}}
\ee
is a Gaussian prefactor.
\end{theo}

\begin{proof}
\begin{mdframed}[linewidth=1.5,topline=false,rightline=false,bottomline=false]

The Gaussian circuit is composed of a Gaussian unitary $\hat G$ and balanced heterodyne detection. The output probability density reads, for all $\bm\alpha=(\alpha_1,\dots,\alpha_m)\in\mathbb C^m$,
\be
\ba
\text{Pr}_{\text{core}}[\bm\alpha]&=\Tr\left[\hat G\ket{\bm C}\!\bra{\bm C}\hat G^\dag\Pi_{\bm\alpha}\right]\\
&=\frac1{\pi^m}\Tr\left[\hat G^\dag\ket{\bm\alpha}\!\bra{\bm\alpha}\hat G\ket{\bm C}\!\bra{\bm C}\right]\\
&=\int_{\bm\beta\in\mathbb C^m}{Q_{\hat G^\dag\ket{\bm\alpha}\!\bra{\bm\alpha}\hat G}(\bm\beta)P_{\ket{\bm C}\!\bra{\bm C}}(\bm\beta)\,d^m\!\bm\beta\,d^m\!\bm\beta^*},
\ea
\label{PrIPAG1}
\ee
where $\Pi_{\bm\alpha}=\frac1{\pi^m}\ket{\bm\alpha}\!\bra{\bm\alpha}$ is the POVM element corresponding to the heterodyne detection of $\bm\alpha=(\alpha_1,\dots,\alpha_m)$. The state $\hat G^\dag\ket{\bm\alpha}$ is a Gaussian state: let $\bm V$ be its covariance matrix and $\bm d$ its displacement vector. For all $\bm\gamma\in\mathbb C^m$, we write $\bm{\tilde\gamma}=(\gamma_1,\dots,\gamma_m,\gamma_1^*,\dots,\gamma_m^*)$. Then, for all $\bm\beta\in\mathbb C^m$,
\be
\ba
Q_{\hat G^\dag\ket{\bm\alpha}\!\bra{\bm\alpha}\hat G}(\bm\beta)&=\frac1{\pi^m\sqrt{\Det\,(\bm V+\mathbb1_{2m}/2)}}\exp\left[-\frac12(\bm{\tilde\beta}-\bm{\tilde d})^\dag\left(\bm V+\mathbb1_{2m}/2\right)^{-1}(\bm{\tilde\beta}-\bm{\tilde d})\right]\\
&=\frac{\exp\left[-\frac12\bm{\tilde d}^\dag\left(\bm V+\mathbb1_{2m}/2\right)^{-1}\bm{\tilde d}\right]}{\pi^m\sqrt{\Det\,(\bm V+\mathbb1_{2m}/2)}}\exp\left[-\frac12\bm{\tilde\beta}^\dag\left(\bm V+\mathbb1_{2m}/2\right)^{-1}\bm{\tilde\beta}+\bm{\tilde d}^\dag\left(\bm V+\mathbb1_{2m}/2\right)^{-1}\bm{\tilde\beta}\right],
\ea
\label{QGaussian}
\ee
i.e., it is a Gaussian function which can be computed efficiently. On the other hand, we have
\be
\ket{\bm C}\!\bra{\bm C}=\sum_{\substack{\bm p,\bm q\in\mathbb N^m\\|\bm p|\le n,|\bm q|\le n}}{c_{\bm p}c_{\bm q}^*\ket{\bm p}\!\bra{\bm q}},
\ee
so that
\be
P_{\ket{\bm C}\!\bra{\bm C}}(\bm\beta)=\sum_{\substack{\bm p,\bm q\in\mathbb N^m\\|\bm p|\le n,|\bm q|\le n}}{c_{\bm p}c_{\bm q}^*P_{\ket{\bm p}\!\bra{\bm q}}(\bm\beta)},
\label{Pcore}
\ee
for all $\bm\beta\in\mathbb C^m$. Moreover we have, for all $\bm p,\bm q\in\mathbb N^m$ and all $\bm\beta\in\mathbb C^m$,
\be
\ba
P_{\ket{\bm p}\!\bra{\bm q}}(\bm\beta)&=\frac{e^{\|\bm\beta\|^2}}{\sqrt{\bm p!\bm q!}}\left(\frac{\partial}{\partial\bm\beta}\right)^{\bm p}\left(\frac{\partial}{\partial\bm\beta^*}\right)^{\bm q}\delta^{2m}(\bm\beta,\bm\beta^*)\\
&=\frac{e^{\frac12\bm{\tilde\beta}^\dag\bm{\tilde\beta}}}{\sqrt{\bm p!\bm q!}}\left(\frac{\partial}{\partial\bm\beta}\right)^{\bm p}\left(\frac{\partial}{\partial\bm\beta^*}\right)^{\bm q}\delta^{2m}(\bm\beta,\bm\beta^*),
\ea
\label{Ppq}
\ee
where $\delta^{2m}(\bm\beta,\bm\beta^*)=\delta(\beta_1)\cdots\delta(\beta_m)\,\delta(\beta_1^*)\cdots\delta(\beta_m^*)$. Combining Eqs.~(\ref{QGaussian}), (\ref{Pcore}) and (\ref{Ppq}) with Eq.~(\ref{PrIPAG1}) we obtain
\be
\ba
\text{Pr}_{\text{core}}[\bm\alpha]&=\kappa(\bm\alpha,\hat G)\sum_{\substack{\bm p,\bm q\in\mathbb N^m\\|\bm p|\le n,|\bm q|\le n}}\frac{c_{\bm p}c_{\bm q}^*}{\sqrt{\bm p!\bm q!}}\int_{\bm\beta\in\mathbb C^m}\Bigg\{\exp\left[-\frac12\bm{\tilde\beta}^\dag\left(\bm V+\mathbb1_{2m}/2\right)^{-1}\bm{\tilde\beta}\right]\\
&\times\exp\left[\bm{\tilde d}^\dag\left(\bm V+\mathbb1_{2m}/2\right)^{-1}\bm{\tilde\beta}\right]e^{\frac12\bm{\tilde\beta}^\dag\bm{\tilde\beta}}\left(\frac{\partial}{\partial\bm\beta}\right)^{\bm p}\left(\frac{\partial}{\partial\bm\beta^*}\right)^{\bm q}\delta^{2m}(\bm\beta,\bm\beta^*)\Bigg\}\,d^m\!\bm\beta\,d^m\!\bm\beta^*,
\ea
\label{PrIPAG2}
\ee
where we have set
\be
\kappa(\bm\alpha,\hat G)=\frac{\exp\left[-\frac12\bm{\tilde d}^\dag\left(\bm V+\mathbb1_{2m}/2\right)^{-1}\bm{\tilde d}\right]}{\pi^m\sqrt{\Det\,(\bm V+\mathbb1_{2m}/2)}}.
\ee
Given that
\be
\bm{\tilde\beta}^\dag=\bm{\tilde\beta}^T\begin{pmatrix} \mymathbb0_m & \mathbbm1_m \\ \mathbb1_m & \mymathbb0_m\end{pmatrix},
\ee
for all $\bm\beta\in\mathbb C^m$, the integral terms in Eq.~(\ref{PrIPAG2}) rewrite as
\be
\int_{\bm\beta\in\mathbb C^m}{\exp\left[\frac12\bm{\tilde\beta}^TV\bm{\tilde\beta}+D^T\bm{\tilde\beta}\right]\left(\frac{\partial}{\partial\bm\beta}\right)^{\bm p}\left(\frac{\partial}{\partial\bm\beta^*}\right)^{\bm q}\delta^{2m}(\bm\beta,\bm\beta^*)\,d^m\!\bm\beta\,d^m\!\bm\beta^*},
\label{PrIPAG3}
\ee
for $|\bm p|\le n$ and $|\bm q|\le n$, where
\be
V=\begin{pmatrix}\mymathbb0_m & \mathbbm1_m\\ \mathbb1_m & \mymathbb0_m \end{pmatrix}\left[\mathbb1_{2m}-\left(\bm V+\mathbb1_{2m}/2\right)^{-1}\right]
\label{V}
\ee
is a $2m\times 2m$ symmetric matrix, due to the initial structure of the covariance matrix, and where
\be
D=\left[\bm{\tilde d}^\dag\left(\bm V+\mathbb1_{2m}/2\right)^{-1}\right]^T
\label{D}
\ee
is a column vector of size $2m$. By Lemma~\ref{lem:efficientC}, the terms in Eq.~(\ref{PrIPAG3}) are equal to
\be
(-1)^{|\bm p|+|\bm q|}\lHaf\left(A_{\bm p,\bm q}\right),
\ee
where the square matrices $A_{\bm p,\bm q}$ of size $|\bm p|+|\bm q|$ are obtained from $V$ by repeating its entries according to $\bm p$ and $\bm q$ and replacing the diagonal by the corresponding elements of $D$ (see the example following Lemma~\ref{lem:efficientC} for a detailed description of the construction). With Eq.~(\ref{PrIPAG2}) we finally obtain
\be
\text{Pr}_{\text{core}}[\bm\alpha]=\kappa(\bm\alpha,\hat G)\sum_{\substack{\bm p,\bm q\in\mathbb N^m\\|\bm p|\le n,|\bm q|\le n}}{\frac{(-1)^{|\bm p|+|\bm q|}}{\sqrt{\bm p!\bm q!}}c_{\bm p}c_{\bm q}^*\lHaf\left(A_{\bm p,\bm q}\right)},
\ee
where
\be
\kappa(\bm\alpha,\hat G)=\frac{\exp\left[-\frac12\bm{\tilde d}^\dag\left(\bm V+\mathbb1_{2m}/2\right)^{-1}\bm{\tilde d}\right]}{\pi^m\sqrt{\Det\,(\bm V+\mathbb1_{2m}/2)}},
\ee
where $\bm V$ and $\bm d$ are the covariance matrix and the diplacement vector of the Gaussian state $\hat G^\dag\ket{\bm\alpha}$, respectively.

\end{mdframed}
\end{proof}

\noindent When the input core state is a multimode Fock state, we refer to the corresponding subclass of $G_{\text{core}}$ circuits as $G_{\text{Fock}}$ circuits. In that case, the sum in Eq.~(\ref{Pr}) reduces to a single term and we obtain the following expression:

\begin{coro}\label{coro:PrFock}
Let $m,n\in\mathbb N^*$ and let $\bm p=(p_1,\dots,p_m)$ with $|\bm p|=n$. Let $\hat G$ be a Gaussian unitary over $m$ modes. For all $\bm\alpha\in\mathbb C^m$, let us write $\bm V$ and $\bm{\tilde d}=(\bm d,\bm d^*)$ the covariance matrix and the displacement vector of the Gaussian state $\hat G^\dag\ket{\bm\alpha}$. Then, the output probability density for the $G_{\text{Fock}}$ circuit $\hat G$ with Fock state input $\ket{\bm p}$ and heterodyne detection, evaluated at $\bm\alpha$, is given by
\be
\text{Pr}_{\text{Fock}}[\bm\alpha]=\frac{\exp\left[-\frac12\bm{\tilde d}^\dag\left(\bm V+\mathbb1_{2m}/2\right)^{-1}\bm{\tilde d}\right]}{\bm p!\pi^m\sqrt{\Det\,(\bm V+\mathbb1_{2m}/2)}}\lHaf(A_{\bm p,\bm p}),
\label{PrFock}
\ee
where $A_{\bm p,\bm p}$ is the square matrix of size $2n$ obtained with Lemma~\ref{lem:efficientC} from
\be
V=\begin{pmatrix}\mymathbb0_m & \mathbbm1_m \\ \mathbb1_m & \mymathbb0_m \end{pmatrix}\left[\mathbb1_{2m}-\left(\bm V+\mathbb1_{2m}/2\right)^{-1}\right]\quad\text{and}\quad D=\left[\bm{\tilde d}^\dag\left(\bm V+\mathbb1_{2m}/2\right)^{-1}\right]^T.
\ee
\end{coro}

\subsection{Strong simulation of weakly non-Gaussian quantum circuits}
\label{sec:StrongsimuNG}

In this section, we use the expression obtained in Theorem~\ref{th:Pr} in order to study strong simulation of Gaussian circuits with few non-Gaussian elements. The first general result deals with general $G_{\text{core}}$ circuits, i.e., Gaussian circuits with multimode core state input.

\begin{theo}\label{th:strong}
Let $m\in\mathbb N^*$ and let $\ket{\bm C}$ be an $m$-mode core state of support size $O(\poly m)$ and degree $n=O(\log m)$. Then, $G_{\text{core}}$ circuits over $m$ modes with input $\ket{\bm C}$ and heterodyne detection can be strongly simulated efficiently classically.
\end{theo}

\begin{proof}
\begin{mdframed}[linewidth=1.5,topline=false,rightline=false,bottomline=false]

By Theorem~\ref{th:Pr}, up to an efficiently computable prefactor, the output probability density is a sum of a polynomial number of loop hafnians, since the support size of the input core state is polynomial. The loop hafnian of a matrix of size $r$ may be computed in time $O(r^32^{r/2})$~\cite{bjorklund2019faster}. For $|\bm p|\le n$ and $\bm q\le n$, the matrices $A_{\bm p,\bm q}$ appearing in Eq.~(\ref{Pr}) are efficiently computable square matrices of size $|\bm p|+|\bm q|\le2n$, so for $n=O(\log m)$, all the loop hafnians may be computed in time $O(\poly m)$. Hence, the output probability density can be evaluated in time $O(\poly m)$.

\medskip

\noindent We now consider the evaluations of the marginal probability densities. Let $k\in\{1,\dots,m-1\}$, for all $\bm\alpha=(\alpha_1,\dots,\alpha_k)\in\mathbb C^k$ we have
\begin{align}
\nonumber \text{Pr}_{\text{core}}[\bm\alpha]&=\Tr\left[\hat G\ket{\bm C}\!\bra{\bm C}\hat G^\dag\left(\Pi_{\bm\alpha}\otimes\mathbb 1_{m-k}\right)\right]\\ 
&=\frac1{\pi^k}\Tr\left[\hat G^\dag\left(\ket{\bm\alpha}\!\bra{\bm\alpha}\otimes\mathbb 1_{m-k}\right)\hat G\ket{\bm C}\!\bra{\bm C}\right]\\
\nonumber &=\pi^{m-k}\int_{\bm\beta\in\mathbb C^m}{Q_{\hat G^\dag\left(\ket{\bm\alpha}\!\bra{\bm\alpha}\otimes\mathbb 1_{m-k}\right)\hat G}(\bm\beta)\,P_{\ket{\bm C}\!\bra{\bm C}}(\bm\beta)\,d^m\!\bm\beta\,d^m\!\bm\beta^*},
\end{align}
where $\Pi_{\bm\alpha}=\frac1{\pi^k}\ket{\alpha_1,\dots,\alpha_k}\!\bra{\alpha_1,\dots,\alpha_k}$ is the POVM element corresponding to the heterodyne detection of $(\alpha_1,\dots,\alpha_k)$ over the first $k$ modes. With Lemma~\ref{lem:efficientC} and the proof of Theorem~\ref{th:Pr}, it is sufficient to show that $Q_{\hat G^\dag\left(\ket{\bm\alpha}\!\bra{\bm\alpha}\otimes\mathbb 1_{m-k}\right)\hat G}$ is an efficiently computable Gaussian function in order to prove that the marginal probability density can be evaluated efficiently. 

For all $(\alpha_1,\dots,\alpha_k)\in\mathbb C^k$ and all $(\gamma_1,\dots,\gamma_{m-k})\in\mathbb C^{m-k}$ we write $\bm\alpha=(\alpha_1,\dots,\alpha_k,0,\dots,0)\in\mathbb C^m$ and $\bm\gamma=(0,\dots,0,\gamma_1,\dots,\gamma_{m-k})\in\mathbb C^m$ so that $\bm\alpha+\bm\gamma=(\alpha_1,\dots,\alpha_k,\gamma_1,\dots,\gamma_{m-k})\in\mathbb C^m$. Using the overcompleteness of coherent states we obtain, for all $(\alpha_1,\dots,\alpha_k)\in\mathbb C^k$ and for all $\bm\beta\in\mathbb C^m$,
\be
\pi^{m-k}Q_{\hat G^\dag\left(\ket{\bm\alpha}\!\bra{\bm\alpha}\otimes\mathbb 1_{m-k}\right)\hat G}(\bm\beta)=\int_{\bm\gamma=(\gamma_1,\dots,\gamma_{m-k})\in\mathbb C^{m-k}}{Q_{\hat G^\dag\ket{\bm\alpha+\bm\gamma}\!\bra{\bm\alpha+\bm\gamma}\hat G}(\bm\beta)\,d^{m-k}\bm\gamma d^{m-k}\bm\gamma^*}.
\label{integralQ}
\ee
Let $S$ and $\bm{\tilde d}=(\bm d,\bm d^*)$ be the symplectic matrix and the displacement vector associated with the Gaussian unitary $\hat G^\dag$. The Gaussian state
\be
\hat G^\dag\ket{\alpha_1,\dots,\alpha_k,\gamma_1,\dots,\gamma_{m-k}}=\hat G^\dag\ket{\bm\alpha+\bm\gamma}
\ee
is described by the covariance matrix $\bm V=\frac12SS^\dag$ and the displacement vector $S(\bm{\tilde\alpha}+\bm{\tilde\gamma})+\bm{\tilde d}$. Its $Q$ function is thus given by
\be
Q_{\hat G^\dag\ket{\bm\alpha+\bm\gamma}\!\bra{\bm\alpha+\bm\gamma}\hat G}(\bm\beta)=\frac{\exp\left[-\frac12(\bm{\tilde\beta}-S(\bm{\tilde\alpha}+\bm{\tilde\gamma})-\bm{\tilde d})^\dag\left(\bm V+\mathbb1_{2m}/2\right)^{-1}(\bm{\tilde\beta}-S(\bm{\tilde\alpha}+\bm{\tilde\gamma})-\bm{\tilde d})\right]}{\pi^m\sqrt{\Det\,(\bm V+\mathbb1_{2m}/2)}},
\ee
for all $(\alpha_1,\dots,\alpha_k)\in\mathbb C^k$, for all $(\gamma_1,\dots,\gamma_{m-k})\in\mathbb C^{m-k}$ and for all $\bm\beta\in\mathbb C^m$. Let us discard the efficiently computable denominator and expand the product in the exponential. Writing $M=\left(\bm V+\mathbb1_{2m}/2\right)^{-1}$, we are left with
\be
\exp\left[-\frac12(\bm{\tilde\beta}-S\bm{\tilde\alpha}-\bm{\tilde d})^\dag M(\bm{\tilde\beta}-S\bm{\tilde\alpha}-\bm{\tilde d})\right]\cdot\exp\left[-\frac12\bm{\tilde\gamma}^\dag S^\dag MS\bm{\tilde\gamma}+(\bm{\tilde\beta}-S\bm{\tilde\alpha}-\bm{\tilde d})^\dag MS\bm{\tilde\gamma}\right],
\ee
The first exponential term is an efficiently computable Gaussian function which factors out of the integral in Eq.~(\ref{integralQ}). Rewriting Eq.~(\ref{integralQ}) up to this efficiently computable Gaussian function we are left with
\be
\ba
\int_{\bm\gamma=(0,\dots,0,\gamma_1,\dots,\gamma_{m-k})\in\mathbb C^m}&{\exp\left[-\frac12\bm{\tilde\gamma}^\dag S^\dag MS\bm{\tilde\gamma}+(\bm{\tilde\beta}-S\bm{\tilde\alpha}-\bm{\tilde d})^\dag MS\bm{\tilde\gamma}\right]d^{m-k}\bm\gamma d^{m-k}\bm\gamma^*}\\
&=\int_{\bm\gamma=(\gamma_1,\dots,\gamma_{m-k})\in\mathbb C^{m-k}}{\exp\left[-\frac12\bm{\tilde\gamma}^TV\bm{\tilde\gamma}+D^T\bm{\tilde\gamma}\right]d^{2(m-k)}\bm{\tilde\gamma}},
\ea
\ee
where $V$ is the $2(m-k)\times2(m-k)$ submatrix of
\be
\begin{pmatrix} \mymathbb0_m & \mathbbm1_m \\ \mathbb1_m & \mymathbb0_m \end{pmatrix}S^\dag MS
\ee
obtained by removing the rows and colums of indices $l$ and $m+l$ for $l\in\{1,\dots,k\}$, and where $D$ is the column vector of size $2(m-k)$ obtained by removing the elements of
\be
\left[(\bm{\tilde\beta}-S\bm{\tilde\alpha}-\bm{\tilde d})^\dag MS\right]^T
\ee
of indices $l$ and $m+l$ for $l\in\{1,\dots,k\}$. The matrix $V$ and the vector $D$ are efficiently computable. Moreover,
\be
\int_{\bm\gamma=(\gamma_1,\dots,\gamma_{m-k})\in\mathbb C^{m-k}}{\exp\left[-\frac12\bm{\tilde\gamma}^TV\bm{\tilde\gamma}+D^T\bm{\tilde\gamma}\right]d^{2(m-k)}\bm{\tilde\gamma}}=\frac{(2\pi)^{m-k}}{\sqrt{\Det\,(V)}}\exp\left[\frac12D^TV^{-1}D\right],
\ee
which is an efficiently computable Gaussian function of $\bm\beta$.

This implies that the value of the marginal probability density $\Pr\,[\alpha_1,\dots,\alpha_k]$ may be computed efficiently. Moreover, it is clear that this does not depent on the choice of $k\in\{1,\dots,m-1\}$ and on the choice of the modes. Hence, all marginal probability densities may be evaluated in time $O(\poly m)$.

\end{mdframed}
\end{proof}

\noindent This result has consequences for the simulability of various continuous variable quantum computing models, in particular those based on Gaussian operations and photon additions or subtractions. We consider three examples in what follows: Interleaved Photon-Added Gaussian circuits (IPAG), Interleaved Photon-Subtracted Gaussian circuits (IPSG) and Gaussian circuits with input Fock states ($G_{\text{Fock}}$).

\medskip

\noindent The stellar hierarchy of single-mode pure quantum states derived in the previous chapter details the engineering of a single-mode quantum state from vacuum using unitary Gaussian operations and single photon addition as a non-Gaussian operation. In particular, the states of finite stellar rank, which corresponds to the states that can be obtained from the vacuum using a finite number of single photon additions or subtractions, are shown to be exactly the states that are obtained by applying a Gaussian unitary operation to a single-mode core state (Theorem~\ref{Gconv}). 

\begin{figure}[h!]
\begin{center}
\includegraphics[width=0.8\columnwidth]{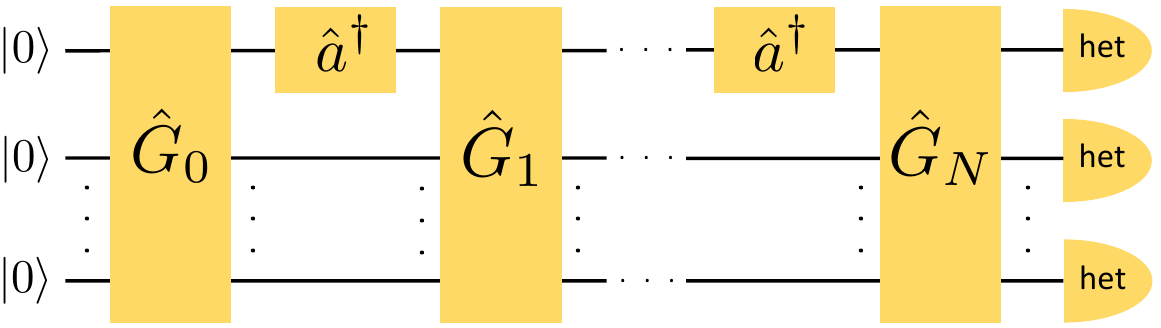}
\caption{Representation of Interleaved Photon-Added Gaussian circuits with $n$ photons additions. The unitaries $\hat G_0,\dots\hat G_n$ are Gaussian and the measurement is performed by balanced heterodyne detection. Note that all photon additions act on the first mode without loss of generality, since swapping two modes is a Gaussian operation.}
\label{fig:GaGa}
\end{center}
\end{figure}

As we will see here, the situation is different in the multimode case: we show that the set of states that can be obtained from a multimode core state with a multimode Gaussian unitary operation is strictly larger than the set of states that can be obtained from the vacuum using a finite number of single photon additions and Gaussian unitary operations (Lemma~\ref{lem:multiGconv}). We also deduce strong simulability results for Gaussian sampling of the latter states. To that end, we consider the family of quantum circuits which sample from states in this set with product unbalanced heterodyne detection, which we refer to as Interleaved Photon-Added Gaussian circuits (IPAG) due to their structure (Fig.~\ref{fig:GaGa}).\\

\noindent Formally, IPAG circuits with $m$ modes and $n$ photon additions are defined as: (i) product vacuum state over $m$ modes in input, (ii) an evolution composed of interleaved multimode Gaussian unitaries $\hat G_0,\dots,\hat G_n$ and $n$ single-mode photon additions, and (iii) product unbalanced heterodyne detection (not necessarily with the same unbalancing for each mode). Without loss of generality, all the photon additions act on the first mode, since swapping two modes is a Gaussian operation. Moreover, up to an added multimode squeezing to the final Gaussian unitary $\hat G_n$, the measurement may be written as a product balanced heterodyne detection.

\medskip

\noindent We first establish a reduction to an equivalent model where the evolution and measurement are Gaussian and only the input state is non-Gaussian. This is done by commuting the photon additions to the input of the circuit. The output state of an IPAG circuit with $m$ modes, $n$ photon additions and Gaussian unitaries $\hat G_0,\dots,\hat G_n$ is given by
\be
\hat G_n\hat a_1^\dag\hat G_{n-1}\hat a_1^\dag\dots\hat G_1\hat a_1^\dag\hat G_0\ket0^{\otimes m},
\label{interleaved1}
\ee
where we have assumed that all the photon additions act on the first mode without loss of generality. Gaussian operations act on annihilation and creation operators through their symplectic representation. They induce affine transformations of the vector of annihilation and creation operators (see section~\ref{sec:Gaussian}). Let us define the column vector of ladder operators
\be
\bm\lambda^\dag=\begin{pmatrix}\hat a_1^\dag\\ \vdots\\ \hat a_m^\dag\\ \hat a_1\\ \vdots\\ \hat a_m\end{pmatrix},
\ee
and let $\hat G$ be an $m$-mode Gaussian operation. Then, there exists a $2m\times2m$ symplectic matrix $S=(s_{ij})_{1\le i,j\le2m}$ and a complex vector $d=(d_1,\dots,d_m)$, such that for all $k\in\{1,\dots,m\}$,
\be
\ba
\hat G\hat a_k^\dag\hat G^\dag&=d_k+(S\bm\lambda^\dag)_k\\
&=d_k+\sum_{l=1}^m{s_{k,l}\hat a_l^\dag+s_{k,m+l}\hat a_l},
\ea
\label{commutG}
\ee
where $(S\bm\lambda^\dag)_k$ indicates the $k^{th}$ element of the column vector $S\bm\lambda^\dag$.
Hence, commuting to the right the creation operators in Eq.~(\ref{interleaved1}), starting by the rightmost one, yields
\be
\ba
\hat G_n\hat a_1^\dag\dots\hat G_1\hat a_1^\dag\hat G_0\ket0^{\otimes m}&=\hat G_n\hat a_1^\dag\hat G_2\dots a_1^\dag\hat G_1\hat G_0\left[d_1^{(0)}+(S^{(0)}\bm\lambda)_1\right]\ket0^{\otimes m}\\
&=\dots\\
&=\hat G_n\dots\hat G_0\left[d_1^{(n-1)}+(S^{(n-1)}\bm\lambda)_1\right]\dots\left[d_1^{(0)}+(S^{(0)}\bm\lambda)_1\right]\ket0^{\otimes m},
\ea
\label{interleaved2}
\ee
where $S^{(k)}$ and $d^{(k)}$ implement the affine transformation corresponding to the action of $(\hat G_k\hat G_{k-1}\dots\hat G_0)^\dag$, for all $k\in\{0,\dots,n-1\}$. Writing $\hat G:=\hat G_n\hat G_{n-1}\dots\hat G_0$, $S^{(k)}=(s^{(k)}_{i,j})_{1\le i,j\le2m}$, and $d^{(k)}=(d_1^{(k)},\dots,d_m^{(k)})$ for $k\in\{0,\dots,n-1\}$, we obtain the output state
\be
\hat G\ket{{\bm C}_{\text{IPAG}}},
\label{IPAGoutput}
\ee
where the state
\be
\ket{{\bm C}_{\text{IPAG}}}:=\left(d_1^{(n-1)}+\sum_{l=1}^m{s_{1,l}^{(n-1)}\hat a_l^\dag+s_{1,m+l}^{(n-1)}\hat a_l}\right)\dots\left(d_1^{(0)}+\sum_{l=1}^m{s_{1,l}^{(0)}\hat a_l^\dag+s_{1,m+l}^{(0)}\hat a_l}\right)\ket0^{\otimes m}
\label{multimodecore}
\ee
is a multimode core state of degree $n$ (and not less, by property of symplectic matrices). Using this characterisation, we obtain the following result:

\begin{lem}\label{lem:multiGconv}
The set of output states of IPAG circuits is strictly included in the set of output states of $G_{\text{core}}$ circuits.
\end{lem}

\begin{proof}
\begin{mdframed}[linewidth=1.5,topline=false,rightline=false,bottomline=false]

The inclusion is immediate with Eq.~(\ref{IPAGoutput}). Up to the Gaussian unitary, it is sufficient to consider core states. To prove the strict inclusion, we show that the $m$-mode core state $(\ket{20}+\ket{01})\otimes\ket0^{\otimes m-2}$ (we omit normalisation), which has degree $2$, is not a core state of the form of Eq.~(\ref{multimodecore}).

\medskip

\noindent By Eq.~(\ref{multimodecore}), all $m$-mode core states of IPAG circuits of degree $2$ have the form
\be
\left(d^{(1)}+\sum_{k=1}^m{s_k^{(1)}\hat a_k^\dag+s_{m+k}^{(1)}\hat a_k}\right)\left(d^{(0)}+\sum_{l=1}^m{s_l^{(0)}\hat a_l^\dag+s_{1,m+l}^{(0)}\hat a_l}\right)\ket0^{\otimes m},
\ee
for some complex numbers $d^{(0)},d^{(1)},s_1^{(0)},\dots,s_{2m}^{(0)},s_1^{(1)},\dots,s_{2m}^{(1)}$. This expression rewrites 
\be
\left(d^{(1)}+\sum_{k=1}^m{s_k^{(1)}\hat a_k^\dag+s_{m+k}^{(1)}\hat a_k}\right)\left(\sum_{l=1}^m{s_l^{(0)}\ket{\bm1_l}}+d^{(0)}\ket{\bm0}\right),
\ee
where for all $l\in\{1,\dots,m\}$, we write $\bm 1_l=(0,\dots,0,1,0\dots,0)$, with a $1$ at the $l^{th}$ position. We finally obtain
\be
\sqrt2\sum_{k=1}^m{s_k^{(0)}s_k^{(1)}\ket{\bm2_k}}+\sum_{\substack{k,l=1\\k\neq l}}^m{s_k^{(0)}s_l^{(1)}\ket{\bm1_k+\bm1_l}}+\sum_{k=1}^m{\left(d^{(1)}s_k^{(0)}+d^{(0)}s_k^{(1)}\right)\ket{\bm1_k}}+\left(d^{(0)}d^{(1)}+\sum_{k=1}^m{s_k^{(0)}s_{m+k}^{(1)}}\right)\ket{\bm0},
\label{IPAGcore}
\ee
where for all $k\in\{1,\dots,m\}$, we write $\bm 2_k=(0,\dots,0,2,0\dots,0)$, with a $2$ at the $k^{th}$ position.
On the other hand we have
\be
(\ket{20}+\ket{01})\otimes\ket0^{\otimes m-2}=\ket{\bm2_1}+\ket{\bm1_2}.
\label{weirdcore}
\ee
In order for this core state to be of the form of Eq.~(\ref{IPAGcore}) we must have
\be
\begin{cases}
s_1^{(0)}s_1^{(1)}\neq0\\
s_k^{(0)}s_l^{(1)}=0,\text{ for }k\neq l,
\end{cases}
\ee
by considering the first and second terms of Eq.~(\ref{IPAGcore}). This implies $s_k^{(0)}=s_k^{(1)}=0$ for all $k\neq1$. Hence, the coefficient of $\ket{\bm1_2}$ in Eq.~(\ref{IPAGcore}) is equal to $0$, while it is nonzero in Eq.~(\ref{weirdcore}). Therefore the core state described by Eq.~(\ref{weirdcore}) cannot be generated by an IPAG circuit.

\end{mdframed}
\end{proof}

\noindent In other words, the set of states that can be obtained from a multimode core state with a multimode Gaussian unitary operation is strictly larger than the set of states that can be obtained from the vacuum using a finite number of single photon additions and Gaussian unitary operations, unlike in the single mode case, where the two sets coincide.

\medskip

\noindent Another consequence of Eqs.~(\ref{IPAGoutput}) and (\ref{multimodecore}) is the following result:

\begin{lem}\label{IPAG}
IPAG circuits over $m$ modes with $n=O(1)$ photon additions can be strongly simulated efficiently classically.
\end{lem}

\begin{proof}
\begin{mdframed}[linewidth=1.5,topline=false,rightline=false,bottomline=false]

When $n=O(1)$, the support size of the core state $\ket{{\bm C}_{\text{IPAG}}}$ in Eq.~(\ref{multimodecore}) is $O(\poly m)$ and its degree is $O(1)$. Then, the result comes from a direct application of Theorem~\ref{th:strong}.

\end{mdframed}
\end{proof}

\noindent When $n=O(\log m)$ however, the support size of the core state is superpolynomial, so the classical algorithm is no longer efficient. 

\medskip

\noindent Similarly, we can define Interleaved Photon-Subtracted Gaussian circuits (IPSG) by replacing photon additions by subtractions in the definition of IPAG circuits. With the same reasoning we obtain the following result:

\begin{coro}\label{coro:IPSG}
IPSG circuits over $m$ modes with $n=O(1)$ photon subtractions can be strongly simulated efficiently classically.
\end{coro}

\begin{proof}
\begin{mdframed}[linewidth=1.5,topline=false,rightline=false,bottomline=false]

We use again the fact that Gaussian operations induce an affine transformation of the vector of annihilation and creation operators. Let $\hat G$ be an $m$-mode Gaussian operation with symplectic matrix $S$ and displacement vector $\bm d$. Writing
\be
\bm\lambda^\dag=\begin{pmatrix}\hat a_1^\dag\\ \vdots\\ \hat a_m^\dag\\ \hat a_1\\ \vdots\\ \hat a_m\end{pmatrix}
\ee
and taking this time the adjoint of Eq.~(\ref{commutG}) we obtain
\be
\ba
\hat G\hat a_k\hat G^\dag&=d_k^*+(S\bm\lambda^\dag)_k^\dag\\
&=d_k^*+\sum_{l=1}^m{s_{k,l}^*\hat a_l+s_{k,m+l}^*\hat a_l^\dag},
\ea
\label{commutG2}
\ee
for all $k\in\{1,\dots,m\}$. The same proof as for IPAG circuits shows that the output state of an IPSG circuit with $n$ photon subtraction and Gaussian evolution $\hat G_0,\dots,\hat G_n$ reads
\be
\hat G\ket{{\bm C}_{\text{IPSG}}},
\ee
where $\hat G=\hat G_n\dots\hat G_0$ and where
\be
\hat G\ket{{\bm C}_{\text{IPSG}}}:=\left(d_1^{*(n-1)}+\sum_{l=1}^m{s_{1,l}^{*(n-1)}\hat a_l+s_{1,m+l}^{*(n-1)}\hat a_l^\dag}\right)\dots\left(d_1^{*(0)}+\sum_{l=1}^m{s_{1,l}^{*(0)}\hat a_l+s_{1,m+l}^{*(0)}\hat a_l^\dag}\right)\ket0^{\otimes m},
\ee
where $S^{(k)}=(s_{i,j}^{(k)})_{1\le i,j\le2m}$ and $\bm d^{(k)}=(d_1^{(k)},\dots,d_m^{(k)})$ are the symplectic matrix and the displacement vector of $(\hat G_k\hat G_{k-1}\dots\hat G_0)^\dag$, for all $k\in\{0,\dots,n-1\}$. When $n=O(1)$, this core state has support size $O(\poly m)$ and degree $O(1)$, and Theorem~\ref{th:strong} concludes the proof.

\end{mdframed}
\end{proof}

\noindent Note that the same reasoning also holds for Gaussian circuits interleaved with both photon additions and subtractions.

\medskip

\noindent A particular subclass of IPAG circuits, where all the photon additions act at the beginning of the circuit, is the class of $G_{\text{Fock}}$ circuits, i.e., Gaussian circuits with Fock state input. In that case, the input is a multimode core state of support size $1$. With Corollary~\ref{coro:PrFock}, we obtain the following result as an immediate consequence of Theorem~\ref{th:strong}:

\begin{lem}\label{th:strongFock}
Let $m\in\mathbb N^*$ and let $\bm p\in\mathbb N^m$, such that $|\bm p|=O(\log m)$. Then, $G_{\text{Fock}}$ circuits over $m$ modes with Fock state input $\ket{\bm p}$ and heterodyne detection can be strongly simulated efficiently classically.
\end{lem}

\noindent In other words, sampling with Gaussian measurements over $m$ modes from $n=O(\log m)$ indistinguishable photons is strongly simulable classically. This contrasts with the case where $m=O(\poly n)$: we show in the next section that strong simulation and even weak simulation of sampling from $n$ photons in $m$ modes with Gaussian measurements is classically hard in that case.

\subsection{Quantum supremacy with non-Gaussian states}
\label{sec:CVS}

In the recent years, there has been an increasing interest in quantum circuits that define subuniversal models of quantum computation~\cite{Bremmer2010,Aaronson2013,Morimae2014, Bremner2015,farhi2016quantum,Douce2017,boixo2018characterizing}. These models may allow for an experimental demonstration of quantum computational supremacy \cite{harrow2017quantum}, i.e., the predicted dramatic speedup of quantum computers over their classical counterparts for some computational tasks~\cite{arute2019quantum}. Subuniversal models for demonstrating quantum supremacy are associated with sampling problems for which the task is to draw random numbers according to a specific probability distribution. Some of these probability distributions are likely to be hard to sample for classical computers, assuming widely accepted conjectures in computer science, such as the fact that the polynomial hierarchy does not collapse, for example with the celebrated Boson Sampling (see section \ref{sec:BosonSampling} and~\cite{Aaronson2013}).

For continuous variable quantum circuits, the classical hardness of circuits with Gaussian input and evolution and non-Gaussian measurement, corresponding to Gaussian Boson Sampling, was proven in~\cite{Lund2014, Hamilton2016}. These circuits are composed of input squeezed states, passive linear optics evolution, and photon counters. In that case, the measurement is a discrete variable measurement.
Subuniversal models with Gaussian input and measurements but non-Gaussian gates are for instance related to the continuous variable implementation of Instantaneous Quantum Computing~\cite{Douce2017,douce2019probabilistic}. Other subuniversal continuous variable circuits that displays non-Gaussian input states together with Gaussian operations and measurements, have been recently considered~\cite{Chakhmakhchyan2017, Lund2017}.

In this section, we define and study a family of continuous variable quantum circuits which we refer to as CVS circuits---for Continuous Variable Sampling---that take non-Gaussian input states and have Gaussian evolution and measurements. The non-Gaussian input states are either single-photons (CVS$_{\text{SP}}$ circuits), single photon-subtracted squeezed vacuum states (CVS$_{\text{PS}}$ circuits), or single photon-added squeezed vacuum states (CVS$_{\text{PA}}$ circuits), and the measurement is unbalanced heterodyne detection (see section~\ref{sec:heterodynemeasurement}), yielding a continuous variable outcome. These models are analog to the Boson Sampling model \cite{Aaronson2013} and the Photon-Added or photon-Subtracted Squeezed Vacuum (PASSV) sampling model \cite{Olson2015}, but with heterodyne detection replacing photon counting. We show in what follows that they allow for the demonstration of quantum computational supremacy with non-Gaussian input states and Gaussian measurements.

The family of CVS circuits is a subclass of IPAG and $G_{\text{core}}$ circuits. Their architecture is inspired by recent experiments performed at Laboratoire Kastler Brossel (LKB), where mode-selective single photon subtraction from a collection of multimode squeezed states has been recently demonstrated~\cite{Ra2017}, and where simultaneous detection of all the optical modes can also be implemented by means of multipixel homodyne detection~\cite{Beck00, ferrini2013compact}.

\medskip

\begin{figure}[h!]
\begin{center}
\includegraphics[width=1\columnwidth]{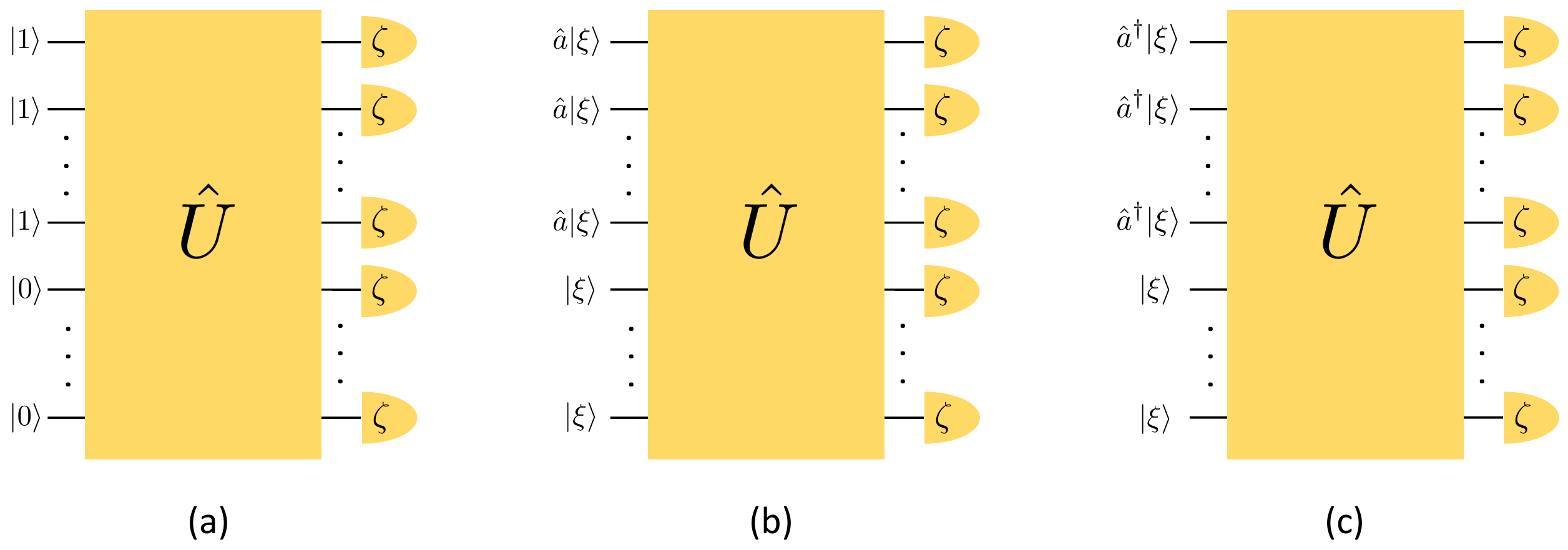}
\caption{Representation of CVS circuits. The passive linear optics evolution is associated with the unitary matrix $U$ defined in Eq.~(\ref{eq:definition-Q}). Measurement is performed by unbalanced heterodyne detection with parameter $\zeta$. (a) CVS$_{\text{SP}}$: in input are vacuum states and single photon states. (b) CVS$_{\text{PS}}$: in input are squeezed vacuum states and single photon-subtracted squeezed vacuum states. (c) CVS$_{\text{PA}}$: in input are squeezed vacuum states and single photon-added squeezed vacuum states.}
\label{fig:CVS}
\end{center}
\end{figure}

\noindent We use for brevity the notations $c_\chi=\cosh\chi$, $s_\chi=\sinh\chi$ and $t_\chi=\tanh\chi$, for all $\chi\in\mathbb R$. CVS$_{\text{PS}}$ circuits are defined formally as follows (see Fig.~\ref{fig:CVS}, CVS$_{\text{PA}}$ and CVS$_{\text{SP}}$ are defined analogously by changing the non-Gaussian input states). Let $m$ be the total number of optical modes. We recall the definition of the squeezing operator with squeezing parameter $\xi\in\mathbb C$: $\hat S(\xi)=e^{\frac12(\xi\hat a^{\dag2}-\xi^*\hat a^2)}$. We restrict to real squeezing parameters in what follows. In that case, $\xi<0$ results in $\hat p$-squeezing while $\xi>0$ in $\hat q$-squeezing.

The first $n$ modes are single photon-subtracted squeezed vacuum states denoted by $\hat a\ket\xi$, where we omit the normalisation factor. The remaining $m-n$ modes are squeezed vacuum states $\ket\xi$. We assume that the real squeezing parameter $\xi$ is uniform over all the modes and does not depend on the number of modes $m$.
We require that $n$ is even and that $m=O(\poly n)\geq2n$.

The input modes undergo a passive linear evolution $\hat U$ that is described by an $m\times m$ unitary matrix $U$ of the form
\begin{equation}
\label{eq:definition-Q}
U=O e^{i\phi\Sigma}
\end{equation}
with $\phi\in\R$, $O\in \mathcal O(m)$ and $\Sigma\in\mathcal O_S(m)$, i.e., $O$ is a real orthogonal matrix, and $\Sigma$ is a real symmetric orthogonal matrix, and hence satisfies $\Sigma^2=1$ (this choice yields a convenient expression for the output probability distribution of CVS circuits).

\noindent Finally, the mode quadratures are measured by unbalanced heterodyne detection with parameter $\zeta\in\mathbb R$, i.e., by projecting the output states onto displaced squeezed vacuum states $\ket{\alpha_j,\zeta} =  \hat D(\alpha_j) \hat S(\zeta) \ket{0}$. The term $\alpha_j=\sqrt{c_\zeta}(e^{-\zeta/2}q_j + i e^{\zeta/2}p_j)$ corresponds to the displacement value of the $j^{th}$ mode, where $q_j$ and $p_j$ are the measured outcomes at the (distinct) output modes of the $j^{th}$-mode heterodyne detector. $\hat D(\alpha)$ is the displacement operator $\hat D(\alpha)= e^{\alpha \hat a^{\dagger}- \alpha^* \hat a}$ (see also section~\ref{sec:heterodynemeasurement}). The squeezing parameter of the detection $\zeta$ is uniform over all the modes and satisfies
\be
|\zeta|=\Omega\left(2^{-\poly m}\right).
\label{zeta}
\ee

\medskip

\begin{figure}[h!]
\begin{center}
\includegraphics[width=0.5\columnwidth]{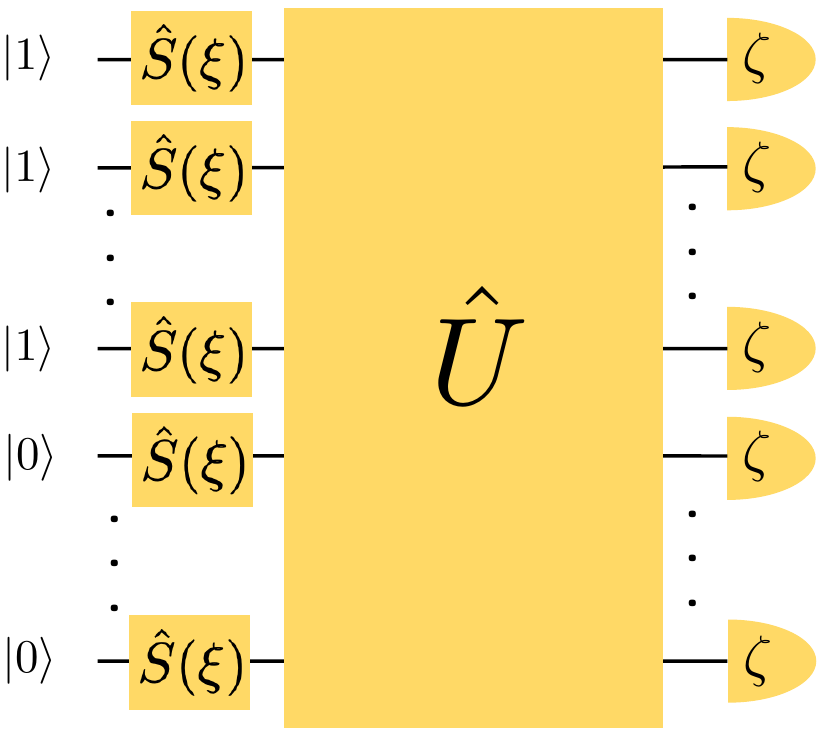}
\caption{An alternative representation of CVS circuits. The input has been rewritten using the mapping of Eq.~(\ref{eq:first-mapping}). $\hat S(\xi)$ is the unitary associated to a squeezing with parameter $\xi\in\mathbb R$, while $\hat U$ is a passive linear optics transformation described by a unitary matrix $U$ defined in Eq.~(\ref{eq:definition-Q}). The output is measured using unbalanced heterodyne detection with parameter $\zeta\in\mathbb R$.}
\label{fig:first-mapping}
\end{center}
\end{figure}
\noindent From the Gaussian convertibility example in Eq.~(\ref{Gconvexample}) of the previous chapter, we know that a photon-subtracted squeezed vacuum state, a photon-added squeezed vacuum state and a squeezed single-photon Fock state, all with the same real squeezing parameter $\xi\in\mathbb R$, are equal:
\begin{equation}
\hat{S}(\xi)\ket1=-\frac1{s_\xi}\hat{a}\ket{\xi} =\frac1{c_\xi}\hat{a}^\dag\ket\xi.
\label{eq:first-mapping}
\end{equation}
By virtue of these identities, the architectures CVS$_{\text{PS}}$ and CVS$_{\text{PA}}$ are in fact identical. Moreover, the architecture CVS$_{\text{SP}}$ is obtained from the first two by letting the squeezing parameter $\xi$ go to $0$. Hereafter, we therefore refer to all three configurations as CVS circuits over $m$ modes, with $n$ non-Gaussian input states, input squeezing $\xi\in\mathbb R$, evolution $U=Oe^{i\phi\Sigma}$ and unbalanced heterodyne detection $\zeta\in\mathbb R$ (Fig.~\ref{fig:first-mapping}): all CVS circuits are therefore specific $G_{\text{Fock}}$ circuits, being also a subclass of IPAG and $G_{\text{core}}$ circuits.

\medskip

\noindent In order to obtain an output probability distribution, we introduce a finite binning of size $\eta>0$ for the output probability density of CVS circuits.
This allows for the definition of a set of indices $\mathbf b=(b^{(q)}_1,\ldots,b^{(q)}_m,b^{(p)}_1,\ldots,b^{(p)}_m)\in\mathbb Z^{2m}$ that corresponds to bins for the $\hat{q}$ and $\hat{p}$ quadratures. We denote $\Pr_{\text{CVS}}^\eta[\mathbf b]$ the discrete probability that, for all $j\in\{1,\ldots,m\}$, the $j^{th}$-mode measurement outcome $(q_j,p_j)$ falls into the boxes $B_j^{(q)}=\left[b^{(q)}_j\eta,(b^{(q)}_j+1)\,\eta\right]$, $B_j^{(p)}=\left[b^{(p)}_j\eta,(b^{(p)}_j+1)\,\eta\right]$. This probability distribution is related to the real-valued probability density associated with CVS circuits, $\text{Pr}_{\text{CVS}}[q_1,p_1,\dots,q_m,p_m]$, by
\begin{equation}
\label{distribution}
\text{Pr}^\eta_{\text{CVS}}[\mathbf b]=\prod _{j=1}^m{\int _{ B^{(q)}_j  }\int _{B^{(p)}_j  }{\text{Pr}_{\text{CVS}}[q_1,p_1,\dots,q_m,p_m]\,dq_j\,dp_j}},
\end{equation}
where $q_1,p_1,\dots,q_m,p_m$ are the continuously distributed measurement outcomes of the product unbalanced heterodyne detection over $m$ modes.
This model of detection is equivalent to perfect heterodyne detection, followed by a binning of the outcome results performed at the stage of post-processing. We assume a resolution scaling with the number of modes as $\eta\sim2^{-\poly m}$. 

\medskip

\noindent We prove that the probability distribution $\Pr_{\text{CVS}}^\eta[\mathbf b]$ is hard to sample for a classical computer, both in the worst case scenario---i.e., weak simulation of all CVS circuits is hard---and in the average case scenario---i.e., weak simulation of a randomly chosen CVS circuit is hard---under the assumption that the polynomial hierarchy does not collapse (see section~\ref{sec:BosonSampling} for a brief review of the complexity classes appearing in this section). The argument adapts proof techniques from~\cite{Aaronson2013,Hamilton2016,Lund2017,Chakhmakhchyan2017} and follows these lines:
\begin{itemize}
\item
We compute the expression $\Pr_{\text{CVS}}[\bm0]$ of the (continuous) probability density evaluated at $\bm0=(0,\dots,0)$ for a given CVS circuit.
\item
We show that for any real matrix $X$, one can find a CVS circuit such that the expression $\Pr_{\text{CVS}}[\bm0]$ is related to the square of the permanent of $X$ by a multiplicative factor.
\item
We show that a classical machine sampling efficiently from the (discrete) probability distribution $\Pr_{\text{CVS}}^\eta[\mathbf b]$ associated to this CVS circuit would allow us to approximate multiplicatively the square of the permanent of $X$ in the third level of the polynomial hierarchy, yielding a contradiction with the widely believed conjecture that the polynomial hierarchy does not collapse.
\end{itemize}

\begin{lem}\label{lem:PrCVS0} We consider a CVS circuit over $m$ modes with $n=2p$ non-Gaussian input states, input squeezing $\xi\in\mathbb R$, evolution $U=Oe^{i\phi\Sigma}$ and unbalanced heterodyne detection $\zeta\in\mathbb R$. Then,
\be
\text{Pr}_{\text{CVS}}[\bm0]=\kappa(\phi,\xi,\zeta)\Haf\,(\Sigma_n)^2,
\ee
where $\Sigma_n$ is the $n\times n$ top left submatrix of $\Sigma$ and where
\be
\kappa(\phi,\xi,\zeta)=\frac{2^{m/2}s_{2\zeta}^n\sin^n(2\phi)}{\pi^m\left[1+c_{2\xi}c_{2\zeta}-s_{2\xi}s_{2\zeta}\cos(2\phi)\right]^{n+m/2}}.
\ee
\end{lem}

\begin{proof}
\begin{mdframed}[linewidth=1.5,topline=false,rightline=false,bottomline=false]

CVS circuits are $G_{\text{Fock}}$ circuits. For a CVS circuit over $m$ modes with $n=2p$ non-Gaussian input states, input squeezing $\xi>0$, evolution $U=Oe^{i\phi\Sigma}$ and unbalanced heterodyne detection $\zeta>0$ (Fig.~\ref{fig:first-mapping}), the multimode Fock state input is $\ket1^{\otimes n}\otimes\ket0^{\otimes m-n}$ and the corresponding Gaussian unitary evolution is given by
\be
\hat G=\hat S^\dag(\zeta)^{\otimes m}\hat U\hat S(\xi)^{\otimes m}.
\ee
Let $\bm V$ be the covariance matrix of the Gaussian state $\hat G^\dag\ket{\bm0}$, its displacement vector being $\bm0$. By Corollary~\ref{coro:PrFock}, the output probability density evaluated at $(0,\dots,0)$ is given by
\be
\text{Pr}_{\text{CVS}}[\bm0]=\frac{\lHaf(A_n)}{\pi^m\sqrt{\Det\,(\bm V+\mathbb1_{2m}/2)}},
\label{PrCVS1}
\ee
where $A_n$ is the square matrix of size $2n$ obtained with Lemma~\ref{lem:efficientC} from
\be
V=\begin{pmatrix}\mymathbb0_m & \mathbbm1_m \\ \mathbb1_m & \mymathbb0_m \end{pmatrix}\left[\mathbb1_{2m}-\left(\bm V+\mathbb1_{2m}/2\right)^{-1}\right]\quad\text{and}\quad D=0,
\label{VDCVS}
\ee
by keeping only the $k^{th}$ and $(m+k)^{th}$ rows and columns of $V$ for $k\in\{1,\dots,n\}$ and by replacing its diagonal entries by the corresponding elements of $D$. Now $D=0$, and for a matrix whose diagonal entries are $0$, the loop hafnian is equal to the hafnian. Hence,
\be
\text{Pr}_{\text{CVS}}[\bm0]=\frac{\Haf\,(A_n)}{\pi^m\sqrt{\Det\,(\bm V+\mathbb1_{2m}/2)}}.
\label{PrCVS2}
\ee
We now derive the expression of the matrix $A_n$ in terms of the CVS circuit parameters:

\begin{lem} \label{lem:Amatrix}
Define
\be
B:=\frac{-s_{2\xi}c_{2\zeta}+c_{2\xi}s_{2\zeta}\cos(2\phi)}{1+c_{2\xi}c_{2\zeta}-s_{2\xi}s_{2\zeta}\cos(2\phi)}\mathbb1_m+i\frac{s_{2\zeta}\sin(2\phi)}{1+c_{2\xi}c_{2\zeta}-s_{2\xi}s_{2\zeta}\cos(2\phi)}\Sigma.
\label{Bmatrix}
\ee
Then,
\be
A_n=\begin{pmatrix} B_n^* & \mymathbb0_n \\ \mymathbb0_n & B_n \end{pmatrix},
\ee
where $B_n$ is the $n\times n$ top left submatrix of $B$.
\end{lem}

\begin{proof}

We show that $V=\begin{pmatrix} B^* & \mymathbb0_m \\ \mymathbb0_m & B \end{pmatrix}$. From Eq.~(\ref{VDCVS}) we have
\be
V=\begin{pmatrix} \mymathbb0_m & \mathbbm1_m \\ \mathbb1_m & \mymathbb0_m \end{pmatrix}\left[\mathbb1_{2m}-\left(\bm V+\mathbb1_{2m}/2\right)^{-1}\right],
\label{VCVS}
\ee
where $\bm V$ is the covariance matrix of the Gaussian state
\be
\hat S^\dag(\xi)^{\otimes m}\hat U^\dag\hat S(\zeta)^{\otimes m}\ket0\bra0^{\otimes m}\hat S^\dag(\zeta)^{\otimes m}\hat U\hat S(\xi)^{\otimes m}.
\ee
This covariance matrix is given by (see section~\ref{sec:symplectic})
\be
\bm V=S_{-\xi}S_{U^\dag}S_\zeta\bm V_{\text{vac}}S_\zeta^\dag S_{U^\dag}^\dag S_{-\xi}^\dag,
\label{sigma1}
\ee
where $\bm V_{\text{vac}}=\mathbb1_{2m}/2$ is the covariance matrix of the vacuum state over $m$ modes, and $S_{-\xi}$,  $S_{U^\dag}$ and $S_\zeta$ are the symplectic matrices describing the action on the covariance matrix of the operators $\hat S^\dag(\xi)^{\otimes m}$,  $\hat U^\dag$ and $\hat S(\zeta)^{\otimes m}$, respectively. Using the notation $c_\chi=\cosh\chi$ and $s_\chi=\sinh\chi$ for all $\chi\in\mathbb R$, we have
\be
S_{-\xi}=\begin{pmatrix} c_\xi\mathbb1_m & -s_\xi\mathbb1_m \\ -s_\xi\mathbb1_m & c_\xi\mathbb1_m \end{pmatrix},\quad S_{U^\dag}=\begin{pmatrix} U^T & \mymathbb0_m \\ \mymathbb0_m & U^\dag \end{pmatrix},\quad S_\zeta=\begin{pmatrix} c_\zeta\mathbb1_m & s_\zeta\mathbb1_m \\ s_\zeta\mathbb1_m & c_\zeta\mathbb1_m \end{pmatrix}.
\ee
With Eq.~(\ref{sigma1}) we obtain
\be
\ba
\bm V&=\frac12\begin{pmatrix} c_\xi\mathbb1_m & -s_\xi\mathbb1_m \\ -s_\xi\mathbb1_m & c_\xi\mathbb1_m \end{pmatrix}\begin{pmatrix} U^T & \mymathbb0_m \\ \mymathbb0_m & U^\dag \end{pmatrix}\begin{pmatrix} c_\zeta\mathbb1_m & s_\zeta\mathbb1_m \\ s_\zeta\mathbb1_m & c_\zeta\mathbb1_m \end{pmatrix}\begin{pmatrix} c_\zeta\mathbb1_m & s_\zeta\mathbb1_m \\ s_\zeta\mathbb1_m & c_\zeta\mathbb1_m \end{pmatrix}\begin{pmatrix} U^* & \mymathbb0_m \\ \mymathbb0_m & U \end{pmatrix}\begin{pmatrix} c_\xi\mathbb1_m & -s_\xi\mathbb1_m \\ -s_\xi\mathbb1_m & c_\xi\mathbb1_m \end{pmatrix}\\
&=\frac12\begin{pmatrix} c_\xi c_\zeta U^T-s_\xi s_\zeta U^\dag & c_\xi s_\zeta U^T-s_\xi c_\zeta U^\dag \\ -s_\xi c_\zeta U^T+c_\xi s_\zeta U^\dag & -s_\xi s_\zeta U^T+c_\xi c_\zeta U^\dag \end{pmatrix}\begin{pmatrix} c_\xi c_\zeta U^*-s_\xi s_\zeta U & -s_\xi c_\zeta U^*+c_\xi s_\zeta U \\ c_\xi s_\zeta U^*-s_\xi c_\zeta U & -s_\xi s_\zeta U^*+c_\xi c_\zeta U \end{pmatrix}\\
&=\frac12\begin{pmatrix} [c_{2\xi}c_{2\zeta}-s_{2\xi}s_{2\zeta}\cos(2\phi)]\mathbb1_m &  \!\!\!\!\!\!\!\![-s_{2\xi}c_{2\zeta}+c_{2\xi}s_{2\zeta}\cos(2\phi)]\mathbb1_m+is_{2\zeta}\sin(2\phi)\Sigma \\ [-s_{2\xi}c_{2\zeta}+c_{2\xi}s_{2\zeta}\cos(2\phi)]\mathbb1_m-is_{2\zeta}\sin(2\phi)\Sigma &  \!\!\!\!\!\!\!\![c_{2\xi}c_{2\zeta}-s_{2\xi}s_{2\zeta}\cos(2\phi)]\mathbb1_m \end{pmatrix},
\ea
\ee
where in the third line we used $c_\chi^2+s_\chi^2=c_{2\chi}$, $2c_\chi s_\chi=s_{2\chi}$, and $c_\chi^2-s_\chi^2=1$, as well as $U=Oe^{i\phi\Sigma}$ with $O^TO=\mathbb1_m$ and $\Sigma^2=\mathbb1_m$, so that $U^\dag U=U^TU^*=\mathbb1_m$, $U^TU=\cos(2\phi)\mathbb1_m+i\sin(2\phi)\Sigma$ and $U^\dag U^*=\cos(2\phi)\mathbb1_m-i\sin(2\phi)\Sigma$. 
The matrix $\bm V+\frac12\mathbb1_{2m}$ may thus be expressed as:
\be
\frac12\begin{pmatrix} [1+c_{2\xi}c_{2\zeta}-s_{2\xi}s_{2\zeta}\cos(2\phi)]\mathbb1_m & \!\!\!\!\!\!\!\![-s_{2\xi}c_{2\zeta}+c_{2\xi}s_{2\zeta}\cos(2\phi)]\mathbb1_m+is_{2\zeta}\sin(2\phi)\Sigma \\ [-s_{2\xi}c_{2\zeta}+c_{2\xi}s_{2\zeta}\cos(2\phi)]\mathbb1_m-is_{2\zeta}\sin(2\phi)\Sigma & \!\!\!\!\!\!\!\![1+c_{2\xi}c_{2\zeta}-s_{2\xi}s_{2\zeta}\cos(2\phi)]\mathbb1_m \end{pmatrix}.
\label{sigma3}
\ee
With Eq.~(\ref{VCVS}), we simply need to show that the inverse of the above matrix is
\be
\mathbb1_{2m}-\begin{pmatrix}\mymathbb0_m & \mathbbm1_m \\ \mathbb1_m & \mymathbb0_m \end{pmatrix}\begin{pmatrix} B^* & \mymathbb0_m \\ \mymathbb0_m & B \end{pmatrix}=\begin{pmatrix} \mathbb1_m & -B \\ -B^* & \mathbb1_m \end{pmatrix}.
\ee
A tedious but straightforward matrix multiplication with Eq.~(\ref{sigma3}) concludes the proof, using $c_\chi^2-s_\chi^2=1$ and $\Sigma^2=\mathbb1_m$.

\end{proof}

\noindent With Lemma~\ref{lem:Amatrix} and Eq.~(\ref{PrCVS2}) we have
\be
\text{Pr}_{\text{CVS}}[\bm0]=\frac1{\pi^m\sqrt{\Det\,(\bm V+\mathbb1_{2m}/2)}}\Haf\begin{pmatrix} B_n^* & \mymathbb0_n \\ \mymathbb0_n & B_n \end{pmatrix},
\label{PrCVS3}
\ee
where
\be
B_n=\frac{-s_{2\xi}c_{2\zeta}+c_{2\xi}s_{2\zeta}\cos(2\phi)}{1+c_{2\xi}c_{2\zeta}-s_{2\xi}s_{2\zeta}\cos(2\phi)}\mathbb1_n+i\frac{s_{2\zeta}\sin(2\phi)}{1+c_{2\xi}c_{2\zeta}-s_{2\xi}s_{2\zeta}\cos(2\phi)}\Sigma_n,
\ee
with $\Sigma_n$ the $n\times n$ top left submatrix of $\Sigma$. Since the hafnian of a matrix does not depend on its diagonal entries, Eq.~(\ref{PrCVS3}) can be rewritten as
\be
\ba
\text{Pr}_{\text{CVS}}[\bm0]&=\frac1{\pi^m\sqrt{\Det\,(\bm V+\mathbb1_{2m}/2)}}\Haf\left[\frac{s_{2\zeta}\sin(2\phi)}{1+c_{2\xi}c_{2\zeta}-s_{2\xi}s_{2\zeta}\cos(2\phi)}\begin{pmatrix} -i\Sigma_n & \mymathbb0_n \\ \mymathbb0_n & i\Sigma_n \end{pmatrix}\right]\\
&=\frac1{\pi^m\sqrt{\Det\,(\bm V+\mathbb1_{2m}/2)}}\left[\frac{s_{2\zeta}\sin(2\phi)}{1+c_{2\xi}c_{2\zeta}-s_{2\xi}s_{2\zeta}\cos(2\phi)}\right]^n\Haf\begin{pmatrix} -i\Sigma_n & \mymathbb0_n \\ \mymathbb0_n & i\Sigma_n \end{pmatrix}.
\ea
\label{PrCVSsigma1}
\ee
Now $\Haf\,(M\oplus N)=\Haf\,(M)\Haf\,(N)$, so the previous expression yields
\be
\text{Pr}_{\text{CVS}}[\bm0]=\frac1{\pi^m\sqrt{\Det\,(\bm V+\mathbb1_{2m}/2)}}\left[\frac{s_{2\zeta}\sin(2\phi)}{1+c_{2\xi}c_{2\zeta}-s_{2\xi}s_{2\zeta}\cos(2\phi)}\right]^n\Haf\,(\Sigma_n)^2.
\label{PrCVSsigma2}
\ee
Finally, we compute $\Det\,(\bm V+\mathbb1_{2m}/2)$:

\begin{lem}\label{lem:detsigma}
\be
\Det\,(\bm V+\mathbb1_{2m}/2)=\frac1{2^m}\left[1+c_{2\xi}c_{2\zeta}-s_{2\xi}s_{2\zeta}\cos(2\phi)\right]^m.
\label{detsigma}
\ee
\end{lem}

\begin{proof}

From the proof of Lemma~\ref{lem:Amatrix} we have
\be
(\bm V+\mathbb1_{2m}/2)^{-1}=\begin{pmatrix} \mathbb1_m & -B \\ -B^* & \mathbb1_m \end{pmatrix},
\ee
so that
\be
\ba
\Det\,(\bm V+\mathbb1_{2m}/2)&=\frac1{\Det\begin{pmatrix} \mathbb1_m & -B \\ -B^* & \mathbb1_m \end{pmatrix}}\\
&=\frac1{\Det\,(\mathbb1_m-BB^*)}.
\ea
\label{detsigma1}
\ee
Using the expression of the matrix $B$ in Eq.~(\ref{Bmatrix}) we obtain
\be
\ba
BB^*&=\left(\frac{-s_{2\xi}c_{2\zeta}+c_{2\xi}s_{2\zeta}\cos(2\phi)}{1+c_{2\xi}c_{2\zeta}-s_{2\xi}s_{2\zeta}\cos(2\phi)}\mathbb1_m+i\frac{s_{2\zeta}\sin(2\phi)}{1+c_{2\xi}c_{2\zeta}-s_{2\xi}s_{2\zeta}\cos(2\phi)}\Sigma\right)\\
&\quad\times\left(\frac{-s_{2\xi}c_{2\zeta}+c_{2\xi}s_{2\zeta}\cos(2\phi)}{1+c_{2\xi}c_{2\zeta}-s_{2\xi}s_{2\zeta}\cos(2\phi)}\mathbb1_m-i\frac{s_{2\zeta}\sin(2\phi)}{1+c_{2\xi}c_{2\zeta}-s_{2\xi}s_{2\zeta}\cos(2\phi)}\Sigma\right)\\
&=\frac{\left[-s_{2\xi}c_{2\zeta}+c_{2\xi}s_{2\zeta}\cos(2\phi)\right]^2+s_{2\zeta}^2\sin^2(2\phi)}{\left[1+c_{2\xi}c_{2\zeta}-s_{2\xi}s_{2\zeta}\cos(2\phi)\right]^2}\mathbb1_m,
\ea
\ee
where we used $\Sigma^2=\mathbb1_m$. Hence, with Eq.~(\ref{detsigma1}) we obtain
\begin{align}
\nonumber\Det\,(\bm V+\mathbb1_{2m}/2)&=\frac1{\Det\,(\mathbb1_m-BB^*)}\\
\nonumber&=\frac1{\left[1-\frac{\left[-s_{2\xi}c_{2\zeta}+c_{2\xi}s_{2\zeta}\cos(2\phi)\right]^2+s_{2\zeta}^2\sin^2(2\phi)}{\left[1+c_{2\xi}c_{2\zeta}-s_{2\xi}s_{2\zeta}\cos(2\phi)\right]^2}\right]^{m}}\\
&=\frac{\left[1+c_{2\xi}c_{2\zeta}-s_{2\xi}s_{2\zeta}\cos(2\phi)\right]^{2m}}{\left[\left[1+c_{2\xi}c_{2\zeta}-s_{2\xi}s_{2\zeta}\cos(2\phi)\right]^2-\left[-s_{2\xi}c_{2\zeta}+c_{2\xi}s_{2\zeta}\cos(2\phi)\right]^2-s_{2\zeta}^2\sin^2(2\phi)\right]^m}\\
\nonumber&=\frac1{2^m}\left[1+c_{2\xi}c_{2\zeta}-s_{2\xi}s_{2\zeta}\cos(2\phi)\right]^m.
\label{detsigma2}
\end{align}

\end{proof}

\noindent Combining Eq.~(\ref{PrCVSsigma2}) and Lemma~\ref{lem:detsigma}, we finally obtain
\be
\text{Pr}_{\text{CVS}}[\bm0]=\frac{2^{m/2}s_{2\zeta}^n\sin^n(2\phi)}{\pi^m\left[1+c_{2\xi}c_{2\zeta}-s_{2\xi}s_{2\zeta}\cos(2\phi)\right]^{n+m/2}}\Haf\,(\Sigma_n)^2,
\ee
where $n=2p$ is the number of single photons in the input.

\end{mdframed}
\end{proof}

\noindent Note that the matrix $O$ appearing in the definition of the CVS circuit Eq.~(\ref{eq:definition-Q}) does not contribute to the output probability distribution. It provides additional degrees of freedom that may be useful for experimental considerations. 

Note also that the expression of the prefactor $\kappa(\phi,\xi,\zeta)$ is left invariant when replacing $\xi$ and $\zeta$ by $-\xi$ and $-\zeta$, which corresponds to changing which quadrature is squeezed both in input and output. 

\medskip

\noindent In the case of CVS$_{\text{SP}}$ circuits---with single photons as non-Gaussian inputs---the squeezing parameter $\xi$ is equal to $0$ and we have the following result, using $1+c_{2\zeta}=2c_\zeta^2$ and $s_{2\zeta}=2c_\zeta s_\zeta$:

\begin{coro}
We consider a CVS$_{\text{SP}}$ circuit over $m$ modes with $n=2p$ non-Gaussian input single photon states, evolution $U=Oe^{i\phi\Sigma}$ and unbalanced heterodyne detection $\zeta\in\mathbb R$. Then,
\be
\text{Pr}_{\text{CVS}_{\text{SP}}}[\bm0]=\kappa_{\text{SP}}(\phi,\zeta)\Haf\,(\Sigma_n)^2,
\ee
where $\Sigma_n$ is the $n\times n$ top left submatrix of $\Sigma$ and where
\be
\kappa_{\text{SP}}(\phi,\zeta)=\frac{t_\zeta^n\sin^n(2\phi)}{\pi^mc_\zeta^m},
\ee
with $t_\zeta=\tanh\zeta$ and $c_\zeta=\cosh\zeta$.
\end{coro}

\noindent Next, we relate the output probability density evaluated at $(0,\dots,0)$ of CVS circuits to the permanent of real matrices. Specifically, we provide an explicit construction holding for any real square matrix $X$.

\begin{lem} \label{lem:Cholesky}
Let $n=2p$ and let $X\in\mathbb R^{p\times p}$. For all $m\geq 2n$ and $\nu\leq 1/\vert\vert X\vert\vert$ there exists a matrix $\Sigma^X\in\mathcal{O}_S(M)$ such that its top left $n\times n$ submatrix is 
\begin{equation}\label{eq:Sigmam}
\Sigma_n^X=\nu\begin{pmatrix} 0 & X \\ X^{ T } & 0 \end{pmatrix}.
\end{equation}
\end{lem}

\begin{proof}
\begin{mdframed}[linewidth=1.5,topline=false,rightline=false,bottomline=false]

Define $Y=\nu X$. The matrix $\mathbb1_p-Y^TY$ is symmetric positive semidefinite since $\left\| Y \right\| \le1$. It thus has a Cholesky decomposition $\mathbb1_p-Y^{ T }Y=Z^{ T }Z$ for some square matrix $Z$. The columns of the $n\times p$ matrix
\be
\begin{pmatrix} Y \\ Z \end{pmatrix}
\ee
form an orthonormal family that can be completed into an orthonormal basis of $\mathbb{R}^n$. The matrix obtained with these columns is orthogonal by construction and reads
\begin{equation}
\begin{pmatrix} Y & C \\ B^{ T } & D \end{pmatrix},
\end{equation}
where $B, C,D$ are $p\times p$ matrices. Finally, with the constraint $m\ge2n$, setting
\begin{equation}
\Sigma^X=\begin{pmatrix} 0 & Y & 0 & C & 0 \\ Y^{ T } & 0 & B & 0 & 0 \\ 0 & B^{ T } & 0 & D & 0 \\ C^{ T } & 0 & D^{ T } & 0 & 0 \\ 0 & 0 & 0 & 0 & \mathbb1_{m-2n}\end{pmatrix}
\end{equation}
yields an $m\times m$ symmetric orthogonal matrix---its columns are orthonormal by construction---which top left $n\times n$ submatrix is precisely given by Eq.~(\ref{eq:Sigmam}).

\end{mdframed}
\end{proof}

\noindent Recall that a specific relation holds between the hafnian and the permanent. Namely, for any square matrix $X$, we have
\begin{equation}
\Per\,(X)=\Haf\begin{pmatrix} 0 & X \\ X^{ T } & 0 \end{pmatrix}.
\label{permhaf}
\end{equation}
Using Lemma~\ref{lem:PrCVS0} with the matrix from Lemma~\ref{lem:Cholesky}, we get that for any square matrix $X$ there exists a CVS circuit CVS$_X$ which probability density at the origin reads:
\be
\ba
\text{Pr}_{\text{CVS}_X}[\bm0]&=\kappa(\phi,\xi,\zeta)\Haf\left(\Sigma_n^X\right)^2\\
&=\nu^n\kappa(\phi,\xi,\zeta)\left[\Haf\begin{pmatrix} 0 & X \\ X^{ T } & 0 \end{pmatrix}\right]^2\\
&=\nu^n\kappa(\phi,\xi,\zeta)\Per\,(X)^2,
\ea
\label{PrPerX}
\ee
where $\nu\le\frac1{\|X\|}$, and where
\be
\kappa(\phi,\xi,\zeta)=\frac{2^{m/2}s_{2\zeta}^n\sin^n(2\phi)}{\pi^m\left[1+c_{2\xi}c_{2\zeta}-s_{2\xi}s_{2\zeta}\cos(2\phi)\right]^{n+m/2}}.
\ee

\medskip

\noindent By Theorem 28 of~\cite{Aaronson2013}, multiplicative approximation of $\Per\,(X)^2$ is a \#\textsf{P}-hard problem for real square matrices. Formally, for any $g\in[1,\poly n]$, the following problem is \#\textsf{P}-hard: given a real matrix $X\in\mathbb R^{n\times n}$ such that $1/\|X\|\ge 2^{-\text{poly(n)}}$, output a nonnegative real number $P_X$ such that
\be
\frac{\Per\,(X)^2}g\le P_X\le g\Per\,(X)^2.
\ee
The multiplying factor $\nu^n\kappa(\phi,\xi,\zeta)$ in Eq.~(\ref{PrPerX}) is finite and non-vanishing for some values of $\xi,\zeta$ and $\phi$, so we obtain the following result:

\begin{coro}\label{coro:P0Phard}
For any $g\in[1,\poly n]$, the following problem is \#\textsf{P}-hard: given a real matrix $X\in\mathbb R^{n\times n}$ such that $1/\|X\|\ge 2^{-\text{poly(n)}}$, output a nonnegative real number $\tilde P_X$ such that
\be
\frac{\text{Pr}_{\text{CVS}_X}[\bm0]}g\le\tilde P_X\le g\text{Pr}_{\text{CVS}_X}[\bm0].
\ee
\end{coro}

\noindent This is because by construction $\frac{\tilde P_X}{\nu^n\kappa(\phi,\xi,\zeta)}$ would then provide a multiplicative approximation of $\Per\,(X)^2$. As it turns out, this problem is easier to solve if one can perform weak simulation of CVS circuits classically:

\begin{lem}\label{lem:PH3}
Given access to a classical oracle which samples from the discretised output probability distribution of CVS circuits $\text{Pr}^\eta_{\text{CVS}}$ of resolution $\eta$, for any $g\in[1,\poly n]$, the following problem can be solved in the third level of the polynomial hierarchy \textsf{PH}$_3$: given a real matrix $X\in\mathbb R^{n\times n}$, output a nonnegative real number $\tilde P_X$ such that
\be
\frac{\text{Pr}_{\text{CVS}_X}[\bm0]}g\le\tilde P_X\le g\text{Pr}_{\text{CVS}_X}[\bm0].
\ee
\end{lem}

\noindent By classical oracle, we mean here an oracle that takes a uniformly random input string as its only source of randomness (it has no built-in randomness as a quantum machine would). Note that we consider a classical oracle sampling from the discretised output probability distribution of CVS circuits $\Pr^\eta_{\text{CVS}}$, rather than from the continuous probability density $\Pr_{\text{CVS}}$. This is a strictly weaker oracle since one may obtain samples from $\Pr^\eta_{\text{CVS}}$ using samples from $\Pr_{\text{CVS}}$, with efficient classical post-processing. 

\begin{proof}
\begin{mdframed}[linewidth=1.5,topline=false,rightline=false,bottomline=false]

With Eq.~(\ref{distribution}), the probability distribution for a CVS circuit with a finite resolution of the heterodyne detection $\eta\sim2^{-\poly m}$, evaluated at $\bm0$ (in a slight abuse of notation we denote both the outcome and the corresponding discretised box by $\bm0$), reads:
\begin{equation}
\label{Preta0}
\text{Pr}^\eta_{\text{CVS}}[\bm0]=\prod _{j=1}^m{\int _{q_j=0}^{\eta}\int _{p_j=0}^{\eta}{dq_jdp_j\text{Pr}_{\text{CVS}}[q_1,p_1,\dots,q_m,p_m]}}.
\end{equation}
Performing a Taylor expansion of the multivariate function $\bm x\mapsto \text{Pr}_{\text{CVS}}[\bm x]$ around the value $\bm0=(0,\dots,0)$, we obtain
\be
\text{Pr}_{\text{CVS}}[\bm x]=\sum_{\bm\gamma\in\mathbb N^{2m}}{\frac{\bm x^{\bm\gamma}}{\bm\gamma!}\partial^{\bm\gamma}\text{Pr}_{\text{CVS}}[\bm0]}.
\ee
Plugging this expression in Eq.~(\ref{Preta0}) and integrating we get
\be
\ba
\text{Pr}^\eta_{\text{CVS}}[\bm0]&=\eta^{2m}\sum_{\bm\gamma\in\mathbb N^{2m}}{\frac{\eta^{|\bm\gamma|}}{(\gamma_1+1)!\dots(\gamma_{2m}+1)!}\partial^{\bm\gamma}\text{Pr}_{\text{CVS}}[\bm0]}\\
&=\eta^{2m}\text{Pr}_{\text{CVS}}[\bm0]+\eta^{2m+1}\sum_{\substack{\bm\gamma\in\mathbb N^{2m}\\|\bm\gamma|>0}}{\frac{\eta^{|\bm\gamma|-1}}{(\gamma_1+1)!\dots(\gamma_{2m}+1)!}\partial^{\bm\gamma}\text{Pr}_{\text{CVS}}[\bm0]},
\ea
\label{Taylor1}
\end{equation}
so that
\be
\frac{\text{Pr}^\eta_{\text{CVS}}[\bm0]}{\eta^{2m}}-\text{Pr}_{\text{CVS}}[\bm0]=\eta\sum_{\substack{\bm\gamma\in\mathbb N^{2m}\\|\bm\gamma|>0}}{\frac{\eta^{|\bm\gamma|-1}}{(\gamma_1+1)!\dots(\gamma_{2m}+1)!}\partial^{\bm\gamma}\text{Pr}_{\text{CVS}}[\bm0]}.
\ee
If $\eta$ is small compared to $\text{Pr}_{\text{CVS}}[\bm0]$, a multiplicative approximation of $\text{Pr}^\eta_{\text{CVS}}[\bm0]\,/\eta^{2m}$ thus yields a multiplicative approximation of $\text{Pr}_{\text{CVS}}[\bm0]$. 

We have $m=\poly n$ and $|\zeta|=\Omega(2^{-\poly m})$, by Eq.~(\ref{zeta}). When considering the circuit CVS$_X$ associated to a real matrix $X$ such that $1/\|X\|\ge 2^{-\poly m}$, we have $\text{Pr}_{\text{CVS}_X}[\bm0]=\Omega(2^{-\poly m})$ by Eq.~(\ref{PrPerX}). Hence, with $\eta\sim2^{-\poly m}$, a multiplicative approximation of $\text{Pr}^\eta_{\text{CVS}_X}[\bm0]\,/\eta^{2m}$ is a multiplicative approximation of $\text{Pr}_{\text{CVS}_X}[\bm0]$. 

We use Stockmeyer's approximate counting algorithm~\cite{stockmeyer1985approximation} in order to conclude the proof: it is a classical algorithm which takes as input the classical description of a circuit sampling from a probability distribution and outputs a multiplicative approximation of the probability of a given outcome (see section \ref{sec:BosonSampling}). This algorithm sits in the third level of the polynomial hierarchy \textsf{PH}$_3$ and works as long as the probability to estimate is not superexponentially small, i.e., $o(2^{-\poly m})$~\cite{Lund2017}.

We have $\text{Pr}^\eta_{\text{CVS}}[\bm0]\,/\eta^{2m}=\Omega(2^{-\poly m})$, so with $\eta\sim2^{-\poly m}$ the probability $\text{Pr}^\eta_{\text{CVS}}[\bm0]$ is not superexponentially small.
Having at our disposal a classical oracle which samples from the probability distribution $\text{Pr}^\eta_{\text{CVS}}$ thus allows us to approximate multiplicatively the probability $\text{Pr}^\eta_{\text{CVS}}[\bm0]$ in the third level of the polynomial hierarchy, by making use of Stockmeyer's algorithm. Dividing the estimate obtained by $\eta^{2m}$ finally yields a multiplicative approximation of $\text{Pr}_{\text{CVS}}[\bm0]$ in \textsf{PH}$_3$ (or rather in the class \textsf{FPH}$_3$ of search problems that may be solved by a \textsf{PH}$_3$ machine).

\end{mdframed}
\end{proof}

\noindent This result holds independently of the value of the squeezing parameter $\xi$, and when the detection parameter $\zeta$ satisfies $|\zeta|=\Omega(2^{-\poly m})$, i.e., even when the detection is very close to a balanced heterodyne detection. When $\zeta=0$, however, the algorithm fails and the circuit is actually weakly simulable classically, because the output probability density factorises into products of single mode output probability densites, due to properties of balanced heterodyne detection. The same property will allow us to derive an efficient verification protocol for Boson Sampling and CVS circuits in the next chapter.

\medskip

\noindent Combining Corollary~\ref{coro:P0Phard} and Lemma~\ref{lem:PH3} gives the main result of this section:

\begin{theo}\label{th:worst}
Sampling from the discretised output probability distribution of CVS circuits is classically hard, or the polynomial hierarchy collapses to its third level.
\end{theo}

\begin{proof}
\begin{mdframed}[linewidth=1.5,topline=false,rightline=false,bottomline=false]

Assuming that sampling from the discretised output probability distribution of CVS circuits can be done efficiently classically, Corollary~\ref{coro:P0Phard} and Lemma~\ref{lem:PH3} imply \textsf{P}$^{\#\textsf{P}}\subset\textsf{PH}_3$ (where \textsf{P}$^{\#\textsf{P}}$ is the class of decision problems that can be solved efficiently using an oracle for the class of counting problems \#\textsf{P}). On the other hand, by Toda's theorem~\cite{toda1991pp}, $\textsf{PH}\subset\textsf{P}^{\#\textsf{P}}$, so that $\textsf{PH}\subset\textsf{PH}_3$, i.e., the polynomial hierarchy collapses to its third level.

\end{mdframed}
\end{proof}

\noindent Theorem~\ref{th:worst} implies that using enough non-Gaussian states as computational resources, weak simulation of Gaussian circuits is no longer classically efficient. This contrast with Theorem~\ref{th:strong} from the previous section, i.e., the fact that strong simulation of Gaussian circuits with few non-Gaussian input states is classically efficient.

\medskip

\noindent This statement is a \textit{worst case} statement, i.e., there exists at least one CVS$_X$ circuits which is hard to sample classically. In order to obtain an \textit{average case} statement and identify a fraction of hard to sample CVS circuits, we define the \emph{Real Gaussian Permanent Estimation} problem:

\begin{pb}[Real Gaussian Permanent Estimation]\label{pb}
Given as input a matrix $X\sim\mathcal{N}(0,1)_\mathbb R^{p\times p}$ of i.i.d.\@ Gaussians together with error bounds $\epsilon,\delta>0$, estimate $\Per\,(X)$ to within error $\pm\epsilon\cdot |\Per\,(X)|$, with probability at least $1-\delta$ over $X$, in $\poly(p,1/\epsilon,1/\delta)$ time.
\end{pb}

\noindent We can use the construction of Lemma~\ref{lem:Cholesky} for the particular case of i.i.d.\@ Gaussian matrices: for any $X\sim\mathcal{N}(0,1)_\mathbb R^{p\times p}$ of i.i.d.\@ Gaussians, we obtain a circuit CVS$_X$ such that Eq.~(\eqref{PrPerX}) holds. Hence every instance of the RGPE is associated with a specific CVS circuit.
In relation to the problem above, we introduce the \emph{Permanent of Real Gaussians} Conjecture: 

\begin{conj}[Permanent of Real Gaussians]
\label{conj1}
\text{RGPE} is \#\textsf{P}-hard.
\end{conj}

\noindent We also introduce a second conjecture:

\begin{conj}[Real Permanent Anti-Concentration]
\label{conj2}
There exists a polynomial $P$ such that for all $p$ and $\delta >0$, 
\be
\Pr_{X\sim \mathcal{N}(0,1)_\mathbb{R}^{p\times p}}\left[\left|\Per\,(X)\right|<\frac{\sqrt {p!}}{P(p,1/\delta)}\right]<\delta.
\ee
\end{conj}

\noindent This problem and these conjectures are precisely the real version of the \emph{Gaussian Permanent Estimation} problem and the \emph{Permanent-of-Gaussians} and \emph{Permanent Anti-Concentration} conjectures introduced in~\cite{Aaronson2013}. This leads us to our average case hardness result.

\begin{theo}\label{th:average}
Assuming Conjecture~\ref{conj2} is true, classical circuits sampling from the (discretised) probability distribution of CVS circuits can be used to solve Real Gaussian Permanent Estimation in the third level of the polynomial hierarchy.
Assuming Conjecture~\ref{conj1} is also true, an efficient classical weak simulation of CVS$_X$ circuits, where $X\sim \mathcal{N}(0,1)_\mathbb{R}^{p\times p}$, would imply a collapse of the polynomial hierarchy to its third level.
\end{theo}

\begin{proof}
\begin{mdframed}[linewidth=1.5,topline=false,rightline=false,bottomline=false]

With the same proof as Lemma~\ref{lem:PH3}, with $\eta=O(2^{-\poly m})$, classical circuits sampling from the (discretised) probability distribution of CVS circuits can be used to obtain a multiplicative approximation of $\text{Pr}_{\text{CVS}}[0,\dots,0]$ in the third level of the polynomial hierarchy \textsf{PH}$_3$ by means of Stockmeyer algorithm. In particular, for $X\sim\mathcal{N}(0,1)_\mathbb{R}^{p\times p}$ a square matrix which entries are i.i.d.\@ Gaussians and considering the circuit CVS$_X$, we obtain multiplicative approximation of $\Per\,(X)^2$.

RGPE however refers to estimating $\Per\,(X)$ rather than $\Per\,(X)^2$. It is easy to see that a multiplicative approximation of $\Per\,(X)^2$ can be turned into a multiplicative approximation $|\Per\,(X)|$ by taking the square root of the estimate. Then, in the case of real matrices, only the sign of the permanent remains to be determined. 

A more general version of this question has been addressed in~\cite{Aaronson2013} where they showed that (the complex version of) Conjecture~\ref{conj2} allowed one to estimate the phase of $\Per\,(X)$ from multiplicative approximation of $|\Per\,(X)|^2$, for $X$ i.i.d.\@ \emph{complex} Gaussian matrix. It implies in particular that Conjecture~\ref{conj2} allows one to determine the sign of $\Per\,(X)$ from $\Per\,(X)^2$ if $X$ is i.i.d.\@ \emph{real} Gaussian matrix. Hence, assuming Conjecture~\ref{conj2}, RGPE can be solved in the third level of the polynomial hierarchy using a classical circuit sampling from the output probability distribution of a CVS circuit as an oracle.

\medskip

\noindent Assuming Conjecture~\ref{conj1} is true, RGPE is \#\textsf{P}-hard. With the above, the existence of an efficient classical algorithm which approximates multiplicatively the output distribution of CVS$_X$ circuits implies the existence of a classical algorithm sitting in the third level of the polynomial hierarchy able to solve a \#\textsf{P}-hard problem. This in turn yields a collapse of the polynomial hierarchy to the third level, thanks to Toda's theorem \cite{toda1991pp}.

\end{mdframed}
\end{proof}

\noindent This result is an average case statement, i.e., it implies that a circuit CVS$_X$, where $X\sim \mathcal{N}(0,1)_\mathbb{R}^{p\times p}$, is hard to sample with high probability over $X$, assuming Conjectures~\ref{conj1} and~\ref{conj2} are true. Once again, we assumed the existence of a classical oracle sampling from the discretised output probability distribution of CVS circuits $\Pr^\eta_{\text{CVS}}$, rather than the continuous probability density $\Pr_{\text{CVS}}$. However, one may obtain samples from $\Pr^\eta_{\text{CVS}}$ using samples from $\Pr_{\text{CVS}}$, with efficient classical post-processing.

\section{Discussion and open problems}

We have considered various notions of classical simulation and have studied the transition from classically simulable models to models that are universal for quantum computing for continuous variables.

\medskip

\noindent We have studied the case of adaptive linear optics, an intermediate model between Boson Sampling~\cite{Aaronson2013} and the Knill--Laflamme--Milburn scheme for universal quantum computing~\cite{knill2001scheme}, obtaining classical algorithms for both probability estimation and overlap estimation and analysing their running times. The conclusion to be drawn from our study is that achieving a quantum advantage for either probability estimation or overlap estimation using linear optics, input single photons and adaptive measurements, is challenging. 

A quantum advantage is not ruled out for probability estimation only if the number of adaptive measurements scale at least logarithmically in the size of the interferometer. The challenge posed by the implementation of a quantum algorithm with adaptive linear optics for probability estimation beyond classical capabilities thus comes from the number of adaptive measurements needed. 

For overlap estimation, a quantum advantage is not ruled out for a constant number adaptive measurements, but many overlaps are easy to estimate classically in that case. It is only when a significant fraction of the input photons is detected at the stage of the adaptive measurements that a quantum advantage becomes possible. The challenge posed by the implementation of a quantum algorithm with adaptive linear optics for overlap estimation beyond classical capabilities thus comes from the need of photon number-resolving detection and the preparation of many photon number states.

\medskip

\noindent For strong simulation, we have considered general Gaussian circuits with Gaussian measurements and non-Gaussian inputs and we have given sufficient conditions in terms of non-Gaussian resources for an efficient classical strong simulation. We have defined the $G_{\text{core}}$ circuits, a broad family of Gaussian circuits supplemented with non-Gaussian input states, where the non-Gaussian states are multimode core states. We have identified various subclasses of these circuits:

\begin{itemize}
\item
The Interleaved Photon-Added Gaussian circuits (IPAG), which are circuits that sample with Gaussian measurements from states which can be engineered from the vacuum using multimode Gaussian unitary operations and a finite number of photon additions.
\item
The $G_{\text{Fock}}$ circuits, which are Gaussian circuits supplemented with Fock states in the input.
\item
The CVS$_{PA}$/CVS$_{PS}$/CVS$_{SP}$ circuits, which are specific interferometers with unbalanced heterodyne detection, supplemented with photon-added squeezed states/photon-added squeezed states/single photons in the input.
\end{itemize}

\noindent The relation between these continuous variable quantum computational models is summarised as
\be
\text{CVS}_{SP}\subset\text{CVS}_{PA}=\text{CVS}_{PS}\subset G_{\text{Fock}}\subset\text{IPAG}\subset G_{\text{core}},
\ee
from the smallest class of circuits to the largest. The tools developped in this chapter also allows us to consider Gaussian circuits supplemented with non-Gaussian states and photon counters, by writing the photon counting POVM element as $\ket n\!\bra n=\frac1{n!}(\hat a^\dag)^n\ket0\!\bra0\hat a^n$, for $n\in\mathbb N$ and commuting the creation operators to the input through the Gaussian computation. Classical algorithms simulating this type of computational model have been derived recently~\cite{quesada2020exact}.

\medskip

\noindent For weak simulation, we have proven the computational hardness of a sampling problem that stems from the family of CVS circuits, relating their discretised output probability density to the permanent of real matrices. Introducing equivalent conjectures to those of~\cite{Aaronson2013} for real matrices, we have extended the hardness result to an average case hardness.

\medskip

\noindent With this collection of results comes various related open problems:

One of the main outstanding problems is to prove the hardness of approximately sampling from CVS circuits. Following~\cite{Aaronson2013}, this may involve making conjectures about anticoncentration and average case hardness of the loop hafnian rather than the permanent, as well as collecting evidence and ultimately proving these conjectures. These conjectures have already been extended from the permanent to the hafnian for the Gaussian Boson Sampling proposal~\cite{Hamilton2016,kruse2019detailed}. 

A related problem is to prove the hardess of sampling from CVS circuits with a binning resolution which either scales as $\frac1{\poly m}$ or is constant with respect to the number of modes, since an exponentially small resolution is not experimentally realistic.

Comparing more precisely IPAG and IPSG circuit families would give insight on the differences between photon addition and photon subtraction in the multimode case. 

Whether the set of output states of IPAG circuits is dense in the set of all multimode states (the multimode equivalent of Lemma~\ref{lem:dense} from the previous chapter) is also an interesting question. In other words, is it possible to approximate with arbitrary precision (in trace distance) any multimode quantum state using only single photon additions and Gaussian unitary operations?

Another main open problem, which we solve in the next chapter, is the verification of the output of CVS circuits and Boson Sampling, necessary to a proper demonstration of quantum supremacy with these computational models.

\clearemptydoublepage
%
%
\let\textcircled=\pgftextcircled
\chapter{Certification of continuous variable quantum states}
\label{chap:certif}

\initial{O}ut of the many properties featured by quantum physics,
the impossibility to perfectly determine an unknown state~\cite{d1996impossibility}
is specially interesting. 
This property is at the heart of quantum cryptography protocols such as quantum key distribution~\cite{BB84}.
On the other hand, it makes certification of the correct functioning of 
quantum devices a challenge, 
since the output of such devices can only be determined approximately, 
through repeated measurements over numerous copies of the output states. The involved configurations spaces have enormous dimensions, a serious burden for any characterization. What is more, certification comes along with an ironic twist: it is highly non-trivial in light of the fact that certain quantum computations are expected to exponentially outperform any attempt at classically solving the same problem. Determining an unknown state is difficult especially for continuous variable quantum states, which are described by possibly infinitely many complex parameters.

In this chapter, after introducing known methods for the characterisation of continuous variable quantum states, we develop new methods using heterodyne measurement in both the trusted and untrusted settings.

Firstly, based on quantum state tomography with heterodyne detection, we introduce a reliable method for continuous variable quantum state certification, which directly yields the elements of the density matrix of the state considered with analytical confidence intervals. This method requires neither mathematical reconstruction of the data nor discrete binning of the sample space, and uses a single Gaussian measurement setting, namely heterodyne detection. 

Secondly, beyond quantum state tomography and without its identical copies assumption, we promote our reliable tomography method to an efficient protocol for verifying single-mode continuous variable pure quantum states with Gaussian measurements against fully malicious adversaries, i.e., making no assumptions whatsoever on the state generated by the adversary.

Thirdly, we generalise the previous protocols to the multimode case and obtain efficient protocols for verifying a large class of multimode continuous variable quantum states, with and without the identical copies assumption. In particular, we show how to efficiently verify the output state of a Boson Sampling experiment with a single-mode Gaussian measurement, thus enabling a proper demonstration of quantum supremacy with Boson Sampling.

This chapter is based on \cite{eisert2020quantum,chabaud2019building,inprepaLKB,chabaud2020efficient}.


\section{Building trust for a continuous variable quantum state}
\label{sec:trust}

With rapidly developing quantum technologies for communication, simulation, computation and sensing, 
the ability to assess the correct functioning of quantum devices is of major importance, for near-term systems, the so-called noisy intermediate-scale quantum devices~\cite{preskill2018quantum}, and for the more sophisticated devices. 
Depending on the desired level of trust and in particular the assumptions one is ready to make, several methods are available for certifying the output of quantum devices~\cite{eisert2020quantum}. A common assumption is that the outcomes of the tested quantum device are \textit{independent and identically distributed} (i.i.d.) over various uses of the device. This implies in particular that the conclusions drawn from test runs are also valid for future computational runs with the same device.

In the following, the task of checking the output state of a quantum device 
is denoted \textit{tomography} for state independent methods, when i.i.d.\@ behaviour is assumed, \textit{certification} for a given a target state, when i.i.d.\@ behaviour is assumed,
and \textit{verification} for a given target state, with no assumption whatsoever, and in particular without the i.i.d.\@ assumption.

\subsection{Tomography, certification and verification}

Quantum state tomography~\cite{d2003quantum} is an important technique which aims at reconstructing a good approximation of the output state of a quantum device by performing multiple rounds of measurements on several copies of said output states. 
Given an ensemble of identically prepared systems, with measurement outcomes from the same
observable, one can build up a histogram, from which a probability density can be estimated.
According to Born's rule, this probability density is the square modulus of the state
coefficients, taken in the basis corresponding to the measurement. 
However, a single measurement setting cannot yield the full state information 
since the phase of its coefficients are then lost. 
Many sets of measurements on many subensembles must be performed and combined to reconstruct the density matrix of the state. 
The data do not yield the state directly, but rather indirectly through data analysis.
Quantum state tomography commonly assumes an i.i.d.\@ behaviour for the device, i.e., that the density matrix of the output state considered is the same at each round of measurement. This assumption may be relaxed with a tradeoff in the efficiency of the protocol~\cite{christandl2012reliable}. 

A certification task corresponds to a setting where one wants to benchmark an industrial quantum device, or check the output of a physical experiment. On the other hand, a verification task corresponds to a cryptographic scenario, where the device to be tested is untrusted, or the quantum data is given by a potentially malicious party, for example in the context of delegated quantum computing.
In the latter case, the task of quantum verification is to ensure that either the device behaved properly, or the computation aborts with high probability.
While delegated computing is a natural platform for the emerging quantum devices, one can provide a physical interpretation to this adversarial setting by emphasising that we aim for deriving verification schemes that make no assumptions whatsoever about the noise model of the underlying systems.
Various methods for verification of quantum devices have been investigated, in particular for discrete variable quantum information~\cite{gheorghiu2017verification}, and they provide different efficiencies and security parameters depending on the computational power of the verifier. The common feature for all these approaches is to utilise some basic obfuscation scheme that allows one to reduce the problem of dealing with a fully general noise model, or a fully general adversarial deviation of the device, to a simple error detection scheme~\cite{vidick2018verification}.
 
For continuous variable quantum devices, checking that the output state is close to a target state may be done with linear optics using optical homodyne tomography~\cite{lvovsky2009continuous}. This method allows one to reconstruct the Wigner function of a generic state using only Gaussian measurements, namely homodyne detection. 
Because of the continuous character of its outcomes,  
one must proceed to a discrete binning of the sample space, in order to build probability histograms. 
Then, the state representation in phase space is determined by a mathematical reconstruction. 

For cases where we have a specific target state, more efficient options are possible. 
For multimode Gaussian states, more efficient certification methods have been derived with Gaussian measurements~\cite{aolita2015reliable}. 
These methods involve the computation of a fidelity witness, i.e., a lower bound on the fidelity, from the measured samples. 
The cubic phase state certification protocol of~\cite{liu2018client} also introduces a fidelity witness, and is an example of certification of a specific non-Gaussian state with Gaussian measurements, which assumes an i.i.d.\@ state preparation. 
The verification protocol for Gaussian continuous variable weighted hypergraph states of~\cite{takeuchi2018resource} removes this assumption, again for this specific family of states.

\subsection{General single-mode protocol}

We address two main issues in what follows. First, existing continuous variable state tomography methods are not reliable in the sense of~\cite{christandl2012reliable}, because errors coming from the reconstruction procedure are indistinguishable from errors coming from the data. Second, there is no Gaussian verification protocol for non-Gaussian states without i.i.d.\@ assumption. \\

\begin{figure}[h!]
	\begin{center}
		\includegraphics[width=3.1in]{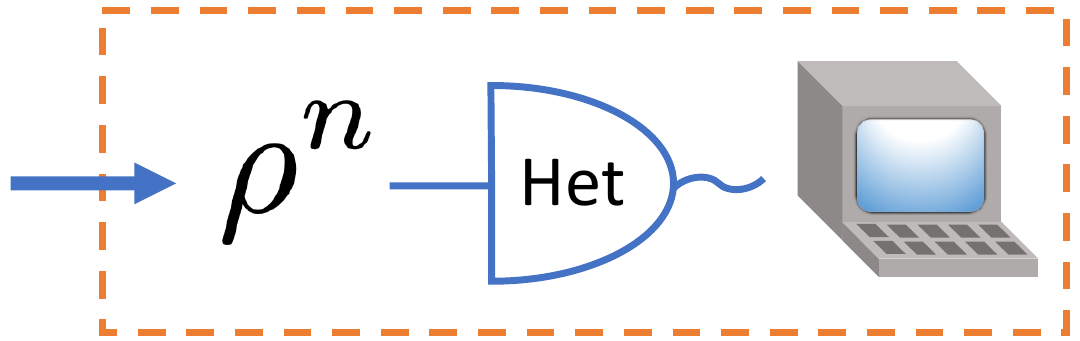}
		\caption{A schematic representation of the protocol. The tester (within the dashed rectangle) receives a continuous variable quantum state $\rho^n$ over $n$ subsystems. This state could be for example the outcome of $n$ successive runs of a physical experiment, the output of a commercial quantum device, or directly sent by some untrusted quantum server. The tester measures with heterodyne detection some of the subsystems of $\rho^n$ and uses the samples obtained and efficient classical post-processing to deduce information about the remaining subsystems.}
		\label{fig:protocol}
	\end{center}
\end{figure}

\noindent We thus introduce a general \textit{receive-and-measure} protocol for building trust for single-mode continuous variable quantum states, using solely Gaussian measurements, namely heterodyne detection (see section~\ref{sec:heterodynemeasurement} and~\cite{ferraro2005gaussian,teo2017heterodyning}). This protocol allows us to perform reliable continuous variable quantum state tomography based on heterodyne detection, which we refer to as \textit{heterodyne tomography} in what follows. This tomography technique only requires a single fixed measurement setting, compared to homodyne tomography. This protocol also provides a means for certifying single-mode continuous variable quantum states, under the i.i.d.\@ assumption. Finally, the same protocol also allows us to verify single-mode continuous variable quantum states, without the i.i.d.\@ assumption. For these three applications, the measurements performed are the same. It is only the selection of subsystems to be measured and the classical post-processing performed that differ from one application to another.

The structure of the protocol is depicted in Fig.~\ref{fig:protocol}: given a quantum state $\rho^n$ over $n$ subsystems, measure some of the subsystems with balanced heterodyne detection. Then, post-process the samples obtained to retrieve information about the remaining subsystems.
We show in the following sections how this protocol may be used to perform reliable tomography, certification and verification of single-mode continuous variable quantum states, and we detail the corresponding choices of subsystems and the classical post-processing for each task.

\section{Heterodyne estimator}
\label{sec:certif1}

In this section, we introduce
a generalisation of the optical equivalence theorem for antinormal ordering~\cite{cahill1969density}, which provides an estimator for the expected value of an operator acting on a state with 
bounded support over the Fock basis, from samples of heterodyne detection of the state. From this result, we derive various protocols in the following sections, ranging from state tomography to state verification.

\medskip

\noindent We denote by $\underset{\alpha\leftarrow D}{\mathbb E}[f(\alpha)]$ the expected value of a function $f$ for samples drawn from a distribution $D$.
Let us introduce for $k,l\ge0$ the polynomials
\be
\mathcal{L}_{k,l}(z)=e^{zz^*}\frac{(-1)^{k+l}}{\sqrt{k!}\sqrt{l!}}\frac{\partial^{k+l}}{\partial z^k\partial z^{*l}}e^{-zz^*},
\label{2DL}
\ee
for $z\in\mathbb C$, which are, up to a normalisation, the Laguerre $2$D polynomials, appearing in particular in the expressions of Wigner function of Fock states~\cite{wunsche1998laguerre}. 
For any operator $A=\sum_{k,l=0}^{+\infty}{A_{kl}\ket k\!\bra l}$ and all $E\in\mathbb N$, we define with these polynomials the function
\be
f_A(z,\eta)=\frac1\eta e^{\left(1-\frac{1}{\eta}\right)zz^*}\sum_{k,l=0}^{E}{\frac{A_{kl}}{\sqrt{\eta^{k+l}}}\mathcal{L}_{k,l}\left(\frac z{\sqrt{\eta}}\right)},
\label{f}
\ee
for all $z\in\mathbb C$, and all $0<\eta<1$. We omit the dependency in $E$ for brevity. The function $z\mapsto f_A(z,\eta)$,
being a polynomial multiplied by a converging Gaussian function, 
is bounded over $\mathbb C$.
With the same notations, we also define the following constant:
\be
K_A=\sum_{k,l=0}^E{|A_{kl}|\sqrt{(k+1)(l+1)}}.
\label{K}
\ee

\begin{theo} \label{thmain}
Let $E\in\mathbb N$ and let \mbox{$0<\eta<\frac2{E}$}. Let also $A=\sum_{k,l=0}^{+\infty}{A_{kl}\ket k\!\bra l}$ be an operator and let \mbox{$\rho=\sum_{k,l=0}^{E}{\rho_{kl}\ket k\!\bra l}$} be a density operator with bounded support.
Then,
\be
\left|\Tr\left(A\rho\right)-\underset{\alpha\leftarrow Q_{\mathrlap\rho}}{\mathbb E}[f_A(\alpha,\eta)]\right|\le\eta K_A,
\label{th1}
\ee
where the function $f$ and the constant $K$ are defined in Eqs.~(\ref{f}) and (\ref{K}).
\end{theo}

\noindent The function $f_A$ defined in Eq.~(\ref{f}) is, up to a numerical factor of $\pi$, a bounded approximation of the Glauber--Sudarshan function $P_A$ of the operator $A$. This approximation is parametrised by a precision $\eta$, and a cutoff value $E$. The optical equivalence theorem for antinormal ordering reads (see section~\ref{sec:phasespace} and~\cite{cahill1969density})
\be
\Tr\,(A\rho)=\pi\int_{\alpha\in\mathbb C}{Q_\rho(\alpha)P_A(\alpha)\,d^2\alpha}.
\ee
Given that
\be
\underset{\alpha\leftarrow Q_{\mathrlap\rho}}{\mathbb E}[f_A(\alpha,\eta)]=\int_{\alpha\in\mathbb C}{Q_\rho(\alpha)f_A(\alpha,\eta)\,d^2\alpha},
\ee
we would expect that $\underset{\alpha\leftarrow Q_{\mathrlap{\rho}}}{\mathbb E}[f_A(\alpha,\eta)]$ is an approximation of $\Tr\,(A\rho)$ parametrised by $\eta$ and $E$. Theorem~\ref{thmain} makes this statement more precise. We prove this theorem in what follows.

\begin{proof}
\begin{mdframed}[linewidth=1.5,topline=false,rightline=false,bottomline=false]

\noindent With Eq.~(\ref{f}) we obtain
\begin{align}
\nonumber\left|\Tr\left(A\rho\right)-\underset{\alpha\leftarrow Q_{\mathrlap{\rho}}}{\mathbb E}[f_A(\alpha,\eta)]\right|&=\left|\sum_{k,l=\mathrlap 0}^{+\infty}{A_{lk}\Tr\left(\ket l\!\bra k\rho\right)}-\smashoperator{\sum_{k,l=0}^E}{A_{lk}\underset{\alpha\leftarrow Q_{\mathrlap{\rho}}}{\mathbb E}[f_{\ket l\!\bra k}(\alpha,\eta)]}\right|\displaybreak\\
&=\left|\sum_{k,l=\mathrlap 0}^{E}{A_{lk}\left(\Tr\left(\ket l\!\bra k\rho\right)-\underset{\alpha\leftarrow Q_{\mathrlap{\rho}}}{\mathbb E}[f_{\ket l\!\bra k}(\alpha,\eta)]\right)}\right|\label{app:triang}\\ 
\nonumber&\le\smashoperator{\sum_{k,l=0}^E}{\left|A_{lk}\right|\left|\Tr\left(\ket l\!\bra k\rho\right)-\underset{\alpha\leftarrow Q_{\mathrlap{\rho}}}{\mathbb E}[f_{\ket l\!\bra k}(\alpha,\eta)]\right|},
\end{align}
where we used in the second line the fact that $\rho$ has a bounded support over the Fock basis. This shows that it is sufficient to prove the Theorem for $A=\ket l\!\bra k$, for all $k,l$ from $0$ to $E$. We first introduce the following result:

\begin{lem} For all $0\le k,l\le E$,
\be
\underset{\alpha\leftarrow Q_{\mathrlap\rho}}{\mathbb E}[f_{\ket l\!\bra k}(\alpha,\eta)]=\rho_{kl}+\smashoperator{\sum_{\substack{m>k,n>l\\m-n=k-l}}^E}{\rho_{mn}\eta^{\frac{m+n-k-l}2}\sqrt{\binom mk\binom nl}}.
\ee
\label{lem:Eflk}
\end{lem}

\begin{proof}

Let us fix $k,l$ in $0,\dots,E$. By Eqs.~(\ref{2DL}) and (\ref{f}) we have, for all $z\in\mathbb C$,
\be
\ba
f_{\ket l\!\bra k}(z)&=\left(\frac{1}{\eta}\right)^{1+\frac{k+l}2}e^{\left(1-\frac{1}{\eta}\right)zz^*}
  \mathop{\mathcal{L}_{l,k}}\left(\frac{z}{\sqrt{\eta}}\right)\\
&=\left(\frac{1}{\eta}\right)^{1+\frac{k+l}2}e^{zz^*}\frac{(-1)^{\mathrlap{k+l}}}{\sqrt{k!}\sqrt{l!}}\left.\frac{\partial^{\mathrlap{k+l}}}{\partial u^{*k}\partial u^l}e^{-uu^*}\right\vert_{u=\frac{z}{\sqrt{\eta}}}\\
&=\frac{1}{\eta}e^{\left(1-\frac{1}{\eta}\right)zz^*}
   \sum_{p=0}^{\min{(k,l)}}{\frac{(-1)^p\sqrt{k!}\sqrt{l!}}{p!(k-p)!(l-p)!}
       \left(\frac1\eta\right)^{k+l-\mathrlap{p}}z^{k-p}z^{*l-p}}.
\ea
\ee
Moreover, for all $\alpha\in\mathbb C$,
\be
\ba
Q_\rho(\alpha)&=\frac1\pi\braket{\alpha|\rho|\alpha}\\
&=\frac1\pi\sum_{m,n=0}^E{\rho_{mn}\braket{\alpha|m}\braket{n|\alpha}}\\
&=\frac1\pi\sum_{m,n=0}^E{\rho_{mn}\frac{\alpha^{*m}\alpha^n}{\sqrt{m!n!}}}e^{-|\alpha|^2}.
\ea
\ee
Combining these expressions we obtain
\begin{align}
\nonumber\underset{\alpha\leftarrow Q_{\mathrlap\rho}}{\mathbb E}[f_{\ket l\!\bra k}(\alpha,\eta)]
  &=\int_{\alpha\in\mathbb C}{Q_\rho(\alpha)f_{\ket l\!\bra k}(\alpha,\eta)\,d^2\alpha}\\
  &=\frac1\pi\sum_{m,n=0}^E{\frac{\rho_{mn}}{\sqrt{m!n!}}}\int_{\alpha\in\mathbb C}{\alpha^{*m}\alpha^ne^{-|\alpha|^2}f_{\ket l\!\bra k}(\alpha,\eta)\,d^2\alpha}\\
 \nonumber=\frac{1}{\pi\eta}\sum_{m,n\mathrlap{=0}}^E\rho_{mn}\frac{\sqrt{k!}\sqrt{l!}}{\sqrt{m!}\sqrt{n!}}
    &\sum_{p=0}^{\min{(\mathrlap{k,l)}}}\frac{(-1)^p}{p!(k-p)!(l-p)!}
    \left(\frac1\eta\right)^{k+l-p}\int_{\alpha\in\mathbb C}{\alpha^{k+n-p}\alpha^{*(l+m-p)}e^{-\frac1\eta|\alpha|^2}d^2\alpha}.
\end{align}
Setting $\alpha=re^{i\theta}$, we have $d^2\alpha=rdrd\theta$ and the integral on the last line may be computed as
\be
\ba
\int_{\alpha\in\mathbb C}{\alpha^{k+n-p}\alpha^{*(l+m-p)}e^{-\frac1\eta|\alpha|^2}d^2\alpha}
  &=\smashoperator{\int_0^{+\infty}}{r^{k+l+m+n-2p+1}e^{-\frac{r^2}\eta}dr}
    \smashoperator{\int_0^{2\pi}}{e^{i(k+n-l-m)\theta}d\theta}\\
&=\begin{cases} \pi\left(\frac{k+l+m+n}{2}-p\right)!\eta^{\frac{k+l+m+n}{2}-p+1}&\text{ for }k-l=m-n,\\&\\0&\text{ for }k-l\neq m-n,\end{cases}
\ea
\ee
where we used $\int_0^{+\infty}{r^{2t+1}e^{-\frac{r^2}\eta}}=\frac12t!\eta^{t+1}$ for $t=\frac{k+l+m+n}{2}-p$, which is obtained directly by induction and integration by parts (note that for $k-l=m-n$, and $p\le\min(k,l)$, we have indeed $t\in\mathbb N$). Hence,
\be
\ba
\underset{\alpha\leftarrow Q_{\mathrlap\rho}}{\mathbb E}[f_{\ket l\!\bra k}(\alpha,\eta)]
  &=\smashoperator{\sum_{\substack{m,n=0\\m-n=k-l}}^E}
      {\rho_{mn}\frac{\sqrt{k!}\sqrt{l!}}{\sqrt{m!}\sqrt{n!}}
      \sum_{p=0}^{\min{(k,l)}}{\frac{(-1)^p\left(\frac{k+l+m+n}{2}-p\right)!}{p!(k-p)!(l-p)!}\eta^{\frac{m+n-k-l}2}}}\\
  &=\smashoperator{\sum_{\substack{m,n=0\\m-n=k-l}}^E}{\rho_{mn}\eta^{\frac{m+n-k-l}2}\frac{\left(\frac{k+l+m+n}{2}\right)!}{\sqrt{m!}\sqrt{n!}\sqrt{k!}\sqrt{l!}}
  \sum_{p=0}^{\min(\mathrlap{k,l)}}{(-1)^p\frac{\binom kp\binom lp}{\binom{\frac{k+l+m+n}{2}}p}}}.
\ea
\label{interE}
\ee
Now for $k\le l$ we have, for all $q\in\mathbb N$ (see, e.g., result 7.1 of~\cite{gould1972combinatorial}),
\be
\sum_{p=0}^k{(-1)^p\frac{\binom kp\binom lp}{\binom qp}}=\begin{cases}\frac{\binom{q-l}{k}}{\binom{q}{k}}&\text{ for }q\ge k+l,\\&\\0&\text{ for }q<k+l.\end{cases}
\ee
When $k\le l$, Eq~(\ref{interE}) thus yields
\begin{align}
\nonumber\underset{\alpha\leftarrow Q_{\mathrlap\rho}}{\mathbb E}[f_{\ket l\!\bra k}(\alpha,\eta)]
 &=\smashoperator{\sum_{\substack{m,n=0\\m-n=k-l\\m+n\ge k+l}}^E}
   {\rho_{mn}\eta^{\frac{m+n-k-l}2}\frac{\left(\frac{k+l+m+n}{2}\right)!}{\sqrt{m!}\sqrt{n!}\sqrt{k!}\sqrt{l!}}\frac{\binom{\frac{k+l+m+n}{2}-l}{k}}{\binom{\frac{k+l+m+n}{2}}{k}}}\\
\nonumber  &=\smashoperator{\sum_{\substack{m\ge k,n\ge l\\m-n=k-l}}^E}
   {\rho_{mn}\eta^{\frac{m+n-k-l}2}\frac1{\sqrt{m!}\sqrt{n!}\sqrt{k!}\sqrt{l!}}\frac{\left(\frac{k-l+m+n}{2}\right)!\left(\frac{-k+l+m+n}{2}\right)!}{\left(\frac{-k-l+m+n}{2}\right)!}}\\
  &=\smashoperator{\sum_{\substack{m\ge k,n\ge l\\m-n=k-l}}^E}
    {\rho_{mn}\eta^{\frac{m+n-k-l}2}\frac{\sqrt{m!}\sqrt{n!}}{\sqrt{k!}\sqrt{l!}\sqrt{(m-k)!}\sqrt{(n-l)!}}}\\
\nonumber  &=\smashoperator{\sum_{\substack{m\ge k,n\ge l\\m-n=k-l}}^E}
    {\rho_{mn}\eta^{\frac{m+n-k-l}2}\sqrt{\binom mk\binom nl}},
\end{align}
where we used that within the summation $m-n=k-l$. This formula is also valid for $l\le k$, with the same reasoning. We finally obtain, for any $k,l$ in $0,\dots,E$
\be
\ba
\underset{\alpha\leftarrow Q_{\mathrlap\rho}}{\mathbb E}[f_{\ket l\!\bra k}(\alpha,\eta)]
  &=\smashoperator{\sum_{\substack{m\ge k,n\ge l\\m-n=k-l}}^E}{\rho_{mn}\eta^{\frac{m+n-k-l}2}\sqrt{\binom mk\binom nl}}\\
&=\rho_{kl}
  +\smashoperator{\sum_{\substack{m>k,n>l\\m-n=k-l}}^E}{\rho_{mn}\eta^{\frac{m+n-k-l}2}\sqrt{\binom mk\binom nl}}.
\ea
\ee

\end{proof}

\noindent Using Lemma~\ref{lem:Eflk}, we obtain
\be
\ba
\left|\Tr\,(\ket l\!\bra k\rho)-\underset{\alpha\leftarrow Q_{\mathrlap\rho}}{\mathbb E}
  [f_{\ket l\!\bra k}(\alpha,\eta)]\right|
&=\left|\rho_{kl}-\underset{\alpha\leftarrow Q_{\mathrlap\rho}}{\mathbb E}[f_{\ket l\!\bra k}(\alpha,\eta)]\right|\\
&=\left|\phantom{{}_{m.}}\smashoperator{\sum_{\substack{m>k,n>l\\m-n=k-l}}^E}
  {\rho_{mn}\eta^{\frac{m+n-k-l}2}\sqrt{\binom mk\binom nl}}\right|\\
&\le\smashoperator{\sum_{\substack{m>k,n>l\\m-n=k-l}}^E}{|\rho_{mn}|\eta^{\frac{m+n-k-l}2}\sqrt{\binom mk\binom nl}}\\
&= \sum_{s=1}^{\mathllap E-\mathrlap{\max{(k,l)}}}
  {|\rho_{s+k,s+l}|\eta^s\sqrt{\binom{s+k}k\binom{s+l}l}}\\
&\le \sum_{s=1}^{\mathllap E-\max\mathrlap{(k,l)}}
  {\eta^s\sqrt{\binom{s+k}k\binom{s+l}l}\sqrt{\rho_{s+k,s+k}}\sqrt{\rho_{s+l,s+l}}},
\ea
\label{interE2}
\ee
where we set $s=m-k=n-l=\frac{m+n-k-l}2$ in the third line, and where we used $|\rho_{s+k,s+l}|\le\sqrt{\rho_{s+k,s+k}}\sqrt{\rho_{s+l,s+l}}$ in the last line, since $\rho$ is a positive semidefinite matrix.
In order to obtain an upper bound independent of $\rho$, we now show for all $s$ that $\eta^s\sqrt{\binom{s+k}k\binom{s+l}l}\le\eta\sqrt{(k+1)(l+1)}$ for $\eta\le\frac2{E}$.
For all $k,l$ in $0,\dots,E$ and for all $s$ in $2,\dots,E-\max({k,l})$, we have
\be
\frac{\sqrt{s+k}\sqrt{s+l}}{s}\le\frac E2.
\ee
This in turn implies that for all $s$ in $2,\dots,E-\max({k,l})$
\begin{align}
\nonumber\eta^s\sqrt{\binom{s+k}k\binom{s+l}l}&=\eta\frac{\sqrt{(s+k)(s+l)}}s\eta^{s-1}\sqrt{\binom{s-1+k}k\binom{s-1+l}l}\\ \displaybreak
&\le\frac{\eta E}2\eta^{s-1}\sqrt{\binom{s-1+k}k\binom{s-1+l}l}\\
\nonumber&\le\eta^{s-1}\sqrt{\binom{s-1+k}k\binom{s-1+l}l},
\end{align}
since we assumed $\eta\le\frac2{E}$. Hence by induction, for all $s$ in $2,\dots,E-\max({k,l})$,
\be
\eta^s\sqrt{\binom{s+k}k\binom{s+l}l}\le\eta^1\sqrt{\binom{1+k}k\binom{1+l}l}=\eta\sqrt{(k+1)(l+1)}.
\ee
Combining this with Eq.~(\ref{interE2}) yields
\be
\ba
\left|\Tr\,(\ket l\!\bra k\rho)-\underset{\alpha\leftarrow Q_{\mathrlap\rho}}{\mathbb E}[f_{\ket l\!\bra k}(\alpha,\eta)]\right|
&\le\eta\sqrt{(k+1)(l+1)}
 \sum_{s=1}^{\mathllap E-\mathrlap{\max(k,l)}}
 {\sqrt{\rho_{s+k,s+k}}\sqrt{\rho_{s+l,s+l}}}\\
&\le\eta\sqrt{(k+1)(l+1)}\sqrt{\sum_{s=1}^{\mathllap E-\mathrlap{\max(k,l)}}
 {\rho_{s+k,s+k}}\sum_{s=1}^{\mathllap E-\mathrlap{\max(k,l)}}{\rho_{s+l,s+l}}}\\ 
&\le\eta\sqrt{(k+1)(l+1)},
\ea
\ee
for all $k,l$ in $0,\dots,E$, where we used Cauchy-Schwarz inequality and the fact that $\Tr\,(\rho)=1$. Note that the above bound still holds when $E\to+\infty$. Together with Eq.~(\ref{app:triang}) we obtain
\be
\ba
\left|\Tr\left(A\rho\right)-\underset{\alpha\leftarrow Q_{\mathrlap\rho}}{\mathbb E}
 [f_A(\alpha,\eta)]\right|
&\le\eta\,\smashoperator{\sum_{k,l=0}^E}{\left|A_{kl}\right|\sqrt{(k+1)(l+1)}}\\
&=\eta K_A,
\ea
\ee
by Eq.~(\ref{K}).

\end{mdframed}
\end{proof}

\noindent This result provides an estimator for the expected value of any operator $A$ acting on a
continuous variable state $\rho$ with bounded support over the Fock basis. 
This estimator is the expected value of a bounded function $f_A$ over samples drawn 
from the Husimi $Q$ function of $\rho$. This probability density corresponds to a Gaussian measurement of $\rho$, 
namely heterodyne detection (see section~\ref{sec:heterodynemeasurement}).
The right hand side of Eq.~(\ref{th1}) is an energy bound, which depends on the operator $A$ and the value $E$.

When the operator $A$ is the density matrix of a continuous variable pure state $\ket\psi$, the previous estimator approximates the fidelity $F(\psi,\rho)=\braket{\psi|\rho|\psi}$ between $\ket\psi\!\bra\psi$ and $\rho$. With the same notations:

\begin{coro} 
Let $E\in\mathbb N$ and let \mbox{$0<\eta<\frac2{E}$}. Let also $\ket\psi\!\bra\psi=\sum_{k,l=0}^{+\infty}{\psi_k\psi_l^*\ket k\!\bra l}$ be a normalised pure state and let \mbox{$\rho=\sum_{k,l=0}^{E}{\rho_{kl}\ket k\!\bra l}$} be a density operator with bounded support.
Then,
\be
\ba
\left|F\left(\psi,\rho\right)-\underset{\alpha\leftarrow Q_{\mathrlap\rho}}{\mathbb E}[f_\psi(\alpha,\eta)]\right|&\le\eta K_\psi\le\frac\eta2(E+1)(E+2),
\ea
\label{co1}
\ee
where the function $f_A$ and the constant $K_A$ are defined in Eqs.~(\ref{f}) and (\ref{K}), for $A=\ket\psi\!\bra\psi$.
\label{corofidelity}
\end{coro}

\begin{proof}
\begin{mdframed}[linewidth=1.5,topline=false,rightline=false,bottomline=false]

We apply Theorem~\ref{thmain} for $A=\ket\psi\!\bra\psi$ a pure state. We obtain
\begin{align}
\nonumber\left|\braket{\psi|\rho|\psi}-\underset{\alpha\leftarrow Q_{\mathrlap\rho}}{\mathbb E}[f_{\psi}(\alpha,\eta)]\right|&\le\eta K_\psi\\ \displaybreak
\nonumber&=\eta\smashoperator{\sum_{k,l=0}^E}{|\psi_k\psi_l|\sqrt{(k+1)(l+1)}}\\
&=\eta\left(\smashoperator{\sum_{n=0}^E}{|\psi_n|\sqrt{n+1}}\right)^2\\
\nonumber&\le\eta\sum_{n=0}^{E}{|\psi_n|^2}\sum_{n=0}^E{(n+1)}\\ 
\nonumber&\le\frac\eta2(E+1)(E+2),
\end{align}
where we used Cauchy-Schwarz inequality, and $\sum_{n=0}^{E}{|\psi_n|^2}\le\Tr\,(\ket\psi\!\bra\psi)=1$. 
Since $\ket\psi$ is a pure state, we have $F(\psi,\rho)=\braket{\psi|\rho|\psi}$, which concludes the proof.

\end{mdframed}
\end{proof}

\noindent This result provides an estimator for the fidelity between any target pure state $\ket\psi$ and any continuous variable (mixed) state $\rho$ with bounded support over the Fock basis. This estimator is the expected value of a bounded function $f_\psi$ over samples 
drawn from the probability density corresponding to heterodyne detection of $\rho$. 
The right hand side of Eq.~(\ref{co1}) is an energy bound, which may be refined depending on the expression of $\ket\psi$. In particular, the second bound is independent of the target state $\ket\psi$. The assumption of bounded support makes sense for tomography, where the energy range of the measured state is known, but not necessarily in a more adversarial setting. 

Given these results, one may choose a target pure state $\ket\psi$ and measure with heterodyne detection various copies of the output (mixed) state $\rho$ of a quantum device with bounded support over the Fock basis. Then, using the samples obtained, one may estimate the expected value of $f_\psi$, thus obtaining an estimate of the fidelity between the states $\ket\psi\!\bra\psi$ and $\rho$. Using this result, we introduce a reliable method for performing continuous variable quantum state tomography using heterodyne detection.

\section{Reliable heterodyne tomography}
\label{sec:certif2}

\noindent Continuous variable quantum state tomography methods usually make two assumptions: firstly that the measured states are independent identical copies (i.i.d.\@ assumption, for \textit{independently and identically distributed}), and secondly that the measured states have a bounded support over the Fock basis~\cite{lvovsky2009continuous}.
With the same assumptions, we present a reliable method for state tomography with heterodyne detection which has the advantage of providing analytical confidence intervals.
Our method directly provides estimates of the elements of the state density matrix, 
phase included. As such, neither mathematical reconstruction of the phase, nor binning of the sample space is needed, since the samples are used only to compute expected values of bounded functions. 
Moreover, only a single fixed Gaussian measurement setting is needed, namely heterodyne detection (Fig.~\ref{fig:2homodyne}).

\medskip

\noindent The law of large numbers ensures that the sample average from independently and identically distributed (i.i.d.\@) random variables converges to the expected value of these random variables, when the number of samples goes to infinity. The following key lemma refines this statement and quantifies the speed of convergence:

\begin{lem}\textbf{(Hoeffding inequality)} Let $\lambda>0$, let $n\ge1$, let $z_1,\dots,z_n$ be i.i.d.\@ complex random variables from a probability density $D$ over $\mathbb R$, and let $f:\mathbb C\mapsto\mathbb R$ such that $|f(z)|\le M$, for $M>0$ and all $z\in\mathbb C$. Then
\be
\Pr\left[\left|\frac{1}{n}\sum_{i=1}^n{f(z_i)}- \underset{z\leftarrow D}{\mathbb E}[f(z)]\right|\ge\lambda\right] \le 2\exp\left[{-\frac{n\lambda^2}{2M^2}}\right].
\ee
\label{lem:HoeffdingR}
\end{lem}

\noindent This comes directly from Hoeffding inequality~\cite{hoeffding1963probability} applied to the real bounded i.i.d.\@ random variables $f(z_1),\dots,f(z_N)$.
When dealing with complex random variables, we use the following result instead:

\begin{lem}\textbf{(Hoeffding inequality for complex random variables)} Let $\lambda>0$, let $n\ge1$, let $z_1,\dots,z_n$ be i.i.d.\@ complex random variables from a probability density $D$ over $\mathbb C$, and let $f:\mathbb C\mapsto\mathbb C$ such that $|f(z)|\le M$, for $M>0$ and all $z\in\mathbb C$. Then
\be
\Pr\left[\left|\frac{1}{n}\sum_{i=1}^n{f(z_i)}- \underset{z\leftarrow D}{\mathbb E}[f(z)]\right|\ge\lambda\right] \le 4\exp\left[{-\frac{n\lambda^2}{4M^2}}\right].
\ee
\label{lem:HoeffdingC}
\end{lem}

\begin{proof} 
\begin{mdframed}[linewidth=1.5,topline=false,rightline=false,bottomline=false]

For all $a>0$ and all $z\in\mathbb C$, $|z|=\sqrt{\Re{(z)}^2+\Im{(z)}^2}\ge a$ implies $|\Re{(z)}|\ge a/\sqrt2$ or $|\Im{(z)}|\ge a/\sqrt2$. Hence,
\be
\Pr\left[|z|\ge a\right]\le\Pr\left[|\Re{(z)}|\ge \frac{a}{\sqrt2}\right]+\Pr\left[|\Im{(z)}|\ge \frac{a}{\sqrt2}\right],
\ee
so applying twice Lemma~\ref{lem:HoeffdingR} for the real random variables $\Re{(f(z))}$ and $\Im{(f(z))}$, respectively, yields Lemma~\ref{lem:HoeffdingC}.
\end{mdframed}
\end{proof}

\noindent For tomographic application, all copies of the state are measured. For $n\ge1$, let $\alpha_1,\dots,\alpha_n\in\mathbb C$ be samples from heterodyne detection
of $n$ copies of a quantum state $\rho$. For $\epsilon>0$ and $k,l\in\mathbb N$, we define
\be
\rho^\epsilon_{kl}=\frac{1}{n}\sum_{i=1}^n{f_{\ket l\!\bra k}\left(\alpha_i,\frac\epsilon{K_{\ket l\!\bra k}}\right)},
\label{Fkleps}
\ee
where the function $f_A$ and the constant $K_A$ are defined in Eqs.~(\ref{f}) and (\ref{K}), for $A=\ket l\!\bra k$, and where $\epsilon>0$ is a free parameter. The quantity $\rho^\epsilon_{kl}$ is the average of the function $f_{\ket l\!\bra k}$ over the samples $\alpha_1,\dots,\alpha_n$. 
The next result shows that this estimator approximates the matrix element $k,l$ of this state with high probability.
We use the notations of Theorem~\ref{thmain}.

\begin{theo}[Reliable heterodyne tomography]
Let $\epsilon,\epsilon'>0$, let $n\ge1$, and let $\alpha_1,\dots,\alpha_n$ be samples obtained by measuring with heterodyne detection $n$ copies of a state $\rho=\sum_{k,l=0}^E{\rho_{kl}\ket k\!\bra l}$ with bounded support, for $E\in\mathbb N$. Then
\be
\left|\rho_{kl}-\rho^\epsilon_{kl}\right|\le\epsilon+\epsilon',
\label{QST}
\ee
for all $0\le k, l\le E$, with probability greater than
\be
1-4\smashoperator{\sum_{0\le k \le l \le E}}\exp\left[{-\frac{n\epsilon^{2+k+l}\epsilon'^2}{4C_{kl}}}\right],
\ee
where the estimate $\rho^\epsilon_{kl}$ is defined in Eq.~(\ref{Fkleps}), and where
\be
C_{kl}:=\left[(k+1)(l+1)\right]^{1+\frac{k+l}2}2^{|l-k|}\binom{\max{(k,l)}}{\min{(k,l)}}
\label{Ckl}
\ee
is a constant independent of $\rho$.
\label{thQST}
\end{theo}

\begin{proof}
\begin{mdframed}[linewidth=1.5,topline=false,rightline=false,bottomline=false]

In order to prove Theorem~\ref{thQST}, we apply Lemma~\ref{lem:HoeffdingC} to the functions $z\mapsto f_{\ket l\!\bra k}(z,\eta)$ defined in Eq.~(\ref{f}). We first bound these functions:

\begin{lem} For all $k,l\ge0$, define
\be
M_{kl}:=\sqrt{2^{|l-k|}\binom{\max{(k,l)}}{\min{(k,l)}}}.
\label{Mkl}
\ee
Then for all $k,l$ and all $z\in\mathbb C$,
\be
\left|f_{\ket k\!\bra l}(z,\eta)\right|\le\frac{M_{kl}}{\eta^{1+\frac{k+l}2}}.
\label{maxfkl}
\ee
\label{lem:boundfkl}
\end{lem}

\begin{proof} For $k$ or $l>E$ the inequality is trivial. For all $k,l\le E$ and all $z\in\mathbb C$,
\be
\ba
\left|f_{\ket k\!\bra l}(z,\eta)\right|&=\left(\frac{1}{\eta}\right)^{1+\frac{k+l}2}e^{\left(1-\frac{1}{\eta}\right)|z|^2}\left|\mathcal{L}_{k,l}\left(\frac{z}{\sqrt\eta}\right)\right|\\
&=\frac{1}{\eta}e^{\left(1-\frac{1}{\eta}\right)|z|^2}\frac{1}{\sqrt{k!}\sqrt{l!}}\left|\sum_{p=0}^{\min\mathrlap{(k,l)}}{\frac{(-1)^pk!l!}{p!(k-p)!(l-p)!}\frac1{\eta^{k+l-p}}z^{l-p}z^{*(k-p)}}\right|,
\ea
\label{maxfkl1}
\ee
where we used Eq.~(\ref{2DL}). Now for all $z\in\mathbb C^*$ and all $a>0$ we have~\cite{wunsche1998laguerre}
\be
\ba
\left|\sum_{p=0}^{\min\mathrlap{(k,l)}}{\frac{(-1)^pk!l!}{p!(k-p)!(l-p)!}a^{k+l-p}z^{l-p}z^{*(k-p)}}\right|&=a^kl!|z|^{k-l}\left|L_l^{(k-l)}\left(a|z|^2\right)\right|\\
&=a^lk!|z|^{l-k}\left|L_k^{(l-k)}\left(a|z|^2\right)\right|,
\ea
\ee
where
\be
L_n^{(\alpha)}(x)=\sum_{q=0}^n{\frac{(-1)^q}{q!}\binom{n+\alpha}{n-q}x^q}
\ee
are the generalised Laguerre polynomials~\cite{abramowitz1965handbook}, defined for $\alpha\in\mathbb R$ and $n\in\mathbb N$. Plugging this relation into Eq.~(\ref{maxfkl1}) we obtain
\be
\ba
\left|f_{\ket k\!\bra l}(z,\eta)\right|&=e^{\left(1-\frac{1}{\eta}\right)|z|^2}\frac{|z|^{l-k}}{\eta^{1+l}}\frac{\sqrt{k!}}{\sqrt{l!}}\left|L_k^{(l-k)}\left(\frac{|z|^2}{\eta}\right)\right|\\
&=e^{\left(1-\frac{1}{\eta}\right)|z|^2}\frac{|z|^{k-l}}{\eta^{1+k}}\frac{\sqrt{l!}}{\sqrt{k!}}\left|L_l^{(k-l)}\left(\frac{|z|^2}{\eta}\right)\right|,
\ea
\label{maxfkl2}
\ee
for all $z\in\mathbb C$. The generalised Laguerre polynomials are bounded as~\cite{rooney1985further}
\be
\left|L_n^{(\alpha)}(x)\right|\le\frac{\Gamma(n+\alpha+1)}{n!\Gamma(\alpha+1)}e^{\frac{x}{2}},
\label{boundLag1}
\ee
for all $x\ge0$, all $\alpha\ge0$ and all $n\in\mathbb N$, and as
\be
\left|L_n^{(\alpha)}(x)\right|\le2^{-\alpha}e^{\frac{x}{2}},
\label{boundLag2}
\ee
for all $x\ge0$, all $\alpha\le-\frac12$ and all $n\in\mathbb N$. 

\medskip

\noindent Let $a>0$. Assuming $k<l$, we have $|z|^{l-k}\le a^{l-k}$ for $|z|\le a$, and $|z|^{k-l}\le a^{k-l}$ for $|z|\ge a$. Thus, the first line of Eq.~(\ref{maxfkl2}), together with Eq.~(\ref{boundLag1}), give
\be
\ba
\left|f_{\ket k\!\bra l}(z,\eta)\right|&\le e^{\left(1-\frac{1}{\eta}\right)|z|^2}\frac{a^{l-k}}{\eta^{1+l}}\frac{\sqrt{k!}}{\sqrt{l!}}\frac{l!}{k!(l-k)!}e^{\frac{|z|^2}{2\eta}}\\
&\le\frac{a^{l-k}}{\eta^{1+l}}\frac{\sqrt{l!}}{(l-k)!\sqrt{k!}},
\ea
\label{maxfkl3}
\ee
for $|z|\le a$ and $k<l$. Similarly, the second line of Eq.~(\ref{maxfkl2}), together with Eq.~(\ref{boundLag2}), give
\be
\ba
\left|f_{\ket k\!\bra l}(z,\eta)\right|&\le e^{\left(1-\frac{1}{\eta}\right)|z|^2}\frac{a^{k-l}}{\eta^{1+k}}\frac{\sqrt{l!}}{\sqrt{k!}}2^{l-k}e^{\frac{|z|^2}{2\eta}}\\
&\le\frac{a^{k-l}}{\eta^{1+k}}\frac{\sqrt{l!}}{\sqrt{k!}}2^{l-k},
\ea
\label{maxfkl4}
\ee
for $|z|\ge a$ and $k<l$. 

\noindent These two last bounds in Eqs.~(\ref{maxfkl3}) and (\ref{maxfkl4}) are equal for $a^{l-k}=(2\eta)^{\frac{l-k}{2}}\sqrt{(l-k)!}$, yielding the bound
\be
\left|f_{\ket k\!\bra l}(z,\eta)\right|\le\sqrt{\frac{2^{l-k}}{\eta^{2+k+l}}\binom{l}{k}},
\label{maxfklle}
\ee
for all $z\in\mathbb C$ and $k<l$. For $l<k$ the same reasoning gives
\be
\left|f_{\ket k\!\bra l}(z,\eta)\right|\le\sqrt{\frac{2^{k-l}}{\eta^{2+k+l}}\binom{k}{l}}.
\label{maxfklge}
\ee
Finally, for $k=l$ the previous bounds also hold, by combining Eqs.~(\ref{maxfkl2}) and (\ref{boundLag1}), and this proves the lemma.

\end{proof}

Let $k,l\ge0$, $n\in\mathbb N$ and $\epsilon'>0$. Applying Lemma~\ref{lem:HoeffdingC} to the function $f_{\ket l\!\bra k}$, with the bound from Lemma~\ref{lem:boundfkl} yields
\be
\Pr\left[\left|\frac{1}{n}\sum_{i=1}^n{f_{\ket l\!\bra k}(\alpha_i,\eta)}- \underset{\alpha\leftarrow Q_{\mathrlap\rho}}{\mathbb E}[f_{\ket l\!\bra k}(\alpha,\eta)]\right|\ge\epsilon'\right] \le 4\exp\left[{-\frac{n\eta^{2+k+l}\epsilon'^2}{4M_{kl}^2}}\right].
\label{Hoeffapp1}
\ee
Applying Theorem~\ref{thmain} for $A=\ket l\!\bra k$ we also obtain
\be
\left|\rho_{kl}-\underset{\alpha\leftarrow Q_{\mathrlap\rho}}{\mathbb E}\left[f_{\ket l\!\bra k}(\alpha,\eta)\right]\right|\le\eta\sqrt{k+1}\sqrt{l+1}.
\label{th1app}
\ee
Let $\alpha_1,\dots,\alpha_n$ be samples from the $Q$ function of $\rho$. Combining Eqs.~(\ref{Hoeffapp1}) and (\ref{th1app}), we obtain with the triangular inequality
\be
\left|\rho_{kl}-\frac{1}{n}\sum_{i=1}^n{f_{\ket l\!\bra k}(\alpha_i,\eta)}\right|\le\eta\sqrt{k+1}\sqrt{l+1}+\epsilon',
\ee
with probability greater than
\be
1-4\exp\left[{-\frac{n\eta^{2+k+l}\epsilon'^2}{4M_{kl}^2}}\right].
\ee
We have $K_{\ket l\!\bra k}=\sqrt{(k+1)(l+1)}$ by Eq.~(\ref{K}). Taking $\eta=\frac\epsilon{K_{\ket l\!\bra k}}$ yields
\be
\left|\rho_{kl}-\frac{1}{n}\sum_{i=1}^n{f_{\ket l\!\bra k}\left(\alpha_i,\frac{\epsilon}{K_{\ket l\!\bra k}}\right)}\right|\le\epsilon+\epsilon',
\ee
with probability greater than
\be
1-4\exp\left[{-\frac{n\epsilon^{2+k+l}\epsilon'^2}{4C_{kl}}}\right],
\label{probaQSTkl}
\ee
where we defined
\be
\ba
C_{kl}&:=\left[(k+1)(l+1)\right]^{1+\frac{k+l}2}M_{kl}^2\\
&=\left[(k+1)(l+1)\right]^{1+\frac{k+l}2}2^{|l-k|}\binom{\max{(k,l)}}{\min{(k,l)}}.
\ea
\label{Cklproof}
\ee
Now this holds for $0\le k,l\le E$. Together with the union bound, this proves the theorem.

\end{mdframed}
\end{proof}

\noindent In light of this result, the principle for performing reliable heterodyne tomography is straightforward and as follows: $n$ identical copies $\rho^{\otimes n}$ of the output quantum state of a physical experiment or quantum device are measured with heterodyne detection, yielding the values $\alpha_1,\dots,\alpha_n$. These values are used to compute the estimates $\rho^\epsilon_{kl}$, defined in Eq.~(\ref{Fkleps}), for all $k,l$ in the range of energy of the experiment. Then, Theorem~\ref{thQST} directly provides confidence intervals for all these estimates of $\rho_{kl}$, the matrix elements of the density operator $\rho$, without the need for a binning of the sample space or any additional data reconstruction, using a single measurement setting. For a desired precision $\epsilon$ and a failure probability $\delta$, the number of samples needed scales as $n=\poly(1/\epsilon,\log(1/\delta))$.

Both homodyne and heterodyne quantum state tomography assume a bounded support over the Fock basis for the output state considered, i.e., that all matrix elements are equal to zero beyond a certain value, and that the output quantum states are i.i.d.\@, i.e., that all measured output states are independent and identical. While these assumptions are natural when looking at the output of a physical experiment, corresponding to a noisy partially trusted quantum device with bounded energy, they may be questionable in the context of untrusted devices. We remove these assumptions in what follows: we first drop the bounded support assumption, deriving a certification protocol for continuous variable quantum states of an i.i.d.\@ device with heterodyne detection ; then, we drop both assumptions, deriving a general verification protocol for continuous variable quantum states against an adversary who can potentially be fully malicious.

\section{Continuous variable quantum state certification protocol}
\label{app:Etesti.i.d.}

\noindent Given an untrusted source of quantum states, the purpose of state certification and state verification protocols is to check whether if its output state is close to a given target state, or far from it. To achieve this, a verifier tests the output state of the source. Ideally, one would like to obtain an upper bound on the probability that the state is not close from the target state, given that it passed a test. However, this is known to be impossible without prior knowledge of the tested state distribution~\cite{gheorghiu2017verification}. Indeed, writing this conditional probability
\be
\Pr\left[\text{incorrect}|\text{accept}\right]=\frac{\Pr\left[\text{incorrect}\cap\text{accept}\right]}{\Pr\left[\text{accept}\right]},
\ee
in a situation where the device always produces a bad output state, it is rejected by the verifier's test most of the time, so the acceptance probability is very small while the conditional probability is equal to $1$. Therefore, the quantity that will always be bounded in certification and verification protocols in which one does not have prior knowledge of the device is the joint probability that the tested state is not close to the target state \textit{and} that it passes the test. Equivalently, we obtain lower bounds on the probability that the tested state is close to the target state or that it fails the test.

\medskip

\noindent We first consider the certification of the output of an i.i.d.\@ quantum device, i.e., which output state is the same at each round. However, we do not assume that the output states of the device have bounded support over the Fock basis anymore. This is instead ensured probabilistically using the samples from heterodyne detection.

Let us define the following operators for $E\ge0$:
\be
U=\smashoperator{\sum_{n=E+1}^{+\infty}}{\ket n\!\bra n}=1-\Pi_E,
\ee
where $\Pi_E=\sum_{n=0}^E{\ket n\!\bra n}$ is the projector onto the Hilbert space $\bar{\mathcal H}$ of states with less than $E$ photons, and
\be
T=\frac{1}{\pi}\smashoperator{\int_{\quad|\alpha|^2\ge E}}{\ket\alpha\!\bra\alpha d^2\alpha},
\ee
where $\ket\alpha$ is a coherent state. We have the following result, proven in~\cite{leverrier2013security} by expanding $T$ in the Fock basis:
\be
U\le2T.
\label{U2T}
\ee
In particular, 
\be
\Tr\,(U\rho)\le2\Tr\,(T\rho).
\label{ineqtrUT}
\ee
The probability $P_r$ that exactly $r$ among $n$ values of $|\alpha_i|^2$ are bigger than $E$ and $n-r$ values are lower, and that the projection $\Pi_E$ of the state $\rho$ onto the Hilbert space $\bar{\mathcal H}$ of states with less than $E$ photons fails is bounded as
\begin{align}
\nonumber P_r&=\binom nr\Tr\left[\left(1-\Pi_E\right)T^r(1-T)^{n-r}\rho^{\otimes n+1}\right]\\
\nonumber&=\binom{n}{r}\Tr\left[UT^r(1-T)^{n-r}\rho^{\otimes n+1}\right]\\
\nonumber&\le2\binom{n}{r}\Tr\left(T\rho\right)^{r+1}\Tr\left[(1-T)\rho\right]^{n-r}\\ 
\nonumber&\le2\binom{n}{r}\max_p{\left|p^{r+1}(1-p)^{n-r}\right|}\\ \displaybreak
&=2\binom{n}{r}\left(\frac{r+1}{n+1}\right)^{r+1}\left(1-\frac{r+1}{n+1}\right)^{n-r}\\ 
\nonumber&\le\frac{2n^r}{r!}\left(\frac{r+1}{n+1}\right)^{r+1}\left(1-\frac{r+1}{n+1}\right)^{n-r}\\
\nonumber&\le\frac{2n^r}{r!}\frac{(r+1)^{r+1}}{n^{r+1}}\exp\left[{-\frac{(n-r)(r+1)}{n+1}}\right]\\
\nonumber&\le\frac{2}{n}\frac{r+1}{\sqrt{2\pi(r+1)}}\exp\left[{\frac{(r+1)^2}{n+1}}\right]\\
\nonumber&\le\frac{\sqrt{r+1}}{n}\exp\left[{\frac{(r+1)^2}{n+1}}\right],
\end{align}
where we used Eq.~(\ref{ineqtrUT}), $1-x\le e^{-x}$ and $(r+1)!\ge\sqrt{2\pi(r+1)}(r+1)^{r+1}e^{-(r+1)}$. For $s\in\mathbb N$, and for all $r\le s$,
\be
\frac{\sqrt{r+1}}{n}\exp\left[{\frac{(r+1)^2}{n+1}}\right]\le\frac{\sqrt{s+1}}{n}\exp\left[\frac{(s+1)^2}{n+1}\right],
\ee
hence the probability that at most $s$ among $n$ values of $|\alpha_i|^2$ are bigger than $E$, and that the projection $\Pi_E$ of the state $\rho$ onto the Hilbert space $\bar{\mathcal H}$ of states with less than $E$ photons fails is bounded by
\be
P^{iid}_{\text{support}}:=\frac{(s+1)^{3/2}}{n}\exp\left[{\frac{(s+1)^2}{n+1}}\right].
\label{Etesti.i.d.}
\ee
For $1\ll s\ll n$, this implies that either $\rho$ is contained in a lower dimensional subspace, or the score at the support estimation step is higher than $s$, with high probability.

\medskip

\noindent Our continuous variable quantum state certification protocol is then as follows: let $\ket\psi$ be a target pure state, of which one wants to certify $m$ copies. The values $s$ and $E$ are free parameters of the protocol. One instructs the i.i.d.\@ device to prepare $n+m$ copies of $\ket\psi$, and the device outputs an i.i.d.\@ (mixed) state $\rho^{\otimes n+m}$. One keeps $m$ copies $\rho^{\otimes m}$, and measures the $n$ others with heterodyne detection, obtaining the samples $\alpha_1,\dots,\alpha_n$. One records the number $r$ of samples such that $|\alpha_i|^2>E$. We refer to this step as \textit{support estimation}. For a given $\epsilon>0$, one also computes with the same samples the estimate
\be
F_\psi(\rho)=\left[\frac{1}{n}\sum_{i=1}^n{f_\psi\left(\alpha_i,\frac\epsilon{mK_\psi}\right)}\right]^m,
\label{tildeF}
\ee
where the function $f_A$ and the constant $K_A$ are defined in Eqs.~(\ref{f}) and (\ref{K}), for $A=\ket\psi\!\bra\psi$, and where $\epsilon>0$ is a free parameter. Note that the support estimation step is no longer necessary if the target state has a bounded support over the Fock basis.

The next result quantifies how close this estimate is from the fidelity between the remaining $m$ copies of the output state $\rho^{\otimes m}$ of the tested device and $m$ copies of the target state $\ket\psi\!\bra\psi^{\otimes m}$.

\begin{theo}[Gaussian certification of continuous variable quantum states]
Let $\epsilon,\epsilon'>0$, let $s\le n$, and let $\alpha_1,\dots,\alpha_n$ be samples obtained by measuring with heterodyne detection $n$ copies of a state $\rho$. Let $E$ in $\mathbb N$, and let $r$ be the number of samples such that $|\alpha_i|^2>E$. Let also $\ket\psi$ be a pure state. Then for all $m\in\mathbb N^*$,
\be
\left|F(\psi^{\otimes m},\rho^{\otimes m})-F_\psi(\rho)\right|\le\epsilon+\epsilon',
\ee
or $r>s$, with probability greater than
\be
1-\left(P_{\text{Support}}^{iid}+P_{\text{Hoeffding}}^{iid}\right),
\ee
\vspace{-5pt}
where
\vspace{-5pt}
\be
P_{\text{Support}}^{iid}=\frac{(s+1)^{3/2}}n\exp\left[{\frac{(s+1)^2}{n+1}}\right],
\ee
\vspace{-10pt}
\be
P_{\text{Hoeffding}}^{iid}=2\exp\left[{-\frac{n\epsilon^{2+2E}\epsilon'^2}{2m^{4+2E}C^2_\psi}}\right],
\ee
where the estimate $F_\psi(\rho)$ is defined in Eq.~(\ref{tildeF}), and where
\be
C_\psi=\sum_{k,l=0}^E{|\psi_k\psi_l|\left(\frac\epsilon m\right)^{E-\frac{k+l}2}K_\psi^{1+\frac{k+l}2}\sqrt{2^{|l-k|}\binom{\max{(k,l)}}{\min{(k,l)}}}}
\label{Cpsi}
\ee
is a constant independent of $\rho$, with the constant $K$ defined in Eq.~(\ref{K}).
\label{thi.i.d.}
\end{theo}

\noindent In order to prove this theorem we make use of the following simple result:

\begin{lem} Let $\eta>0$ and $a,b\in[0,1]$ such that $|a-b|\le\eta$. Then for all $m\ge1$,
\be
\left|a^m-b^m\right|\le m|a-b|\le m\eta.
\ee
\label{lem:simplem}
\end{lem}

\begin{proof}
\begin{mdframed}[linewidth=1.5,topline=false,rightline=false,bottomline=false]
With the notations of the lemma,
\be
\ba
\left|a^m-b^m\right|&=|a-b|\left|\sum_{j=0}^{m-1}{a^jb^{m-j-1}}\right|\\
&\le m|a-b|\\
&\le m\eta.
\ea
\ee
\end{mdframed}
\end{proof}

\noindent We first consider the case of $m=1$ from which we deduce the general case with the lemma.

\begin{proof}
\begin{mdframed}[linewidth=1.5,topline=false,rightline=false,bottomline=false]

Let us write $\ket\psi=\sum_{n\ge0}\psi_n\ket n$. For $\eta>0$, the function $z\mapsto f_\psi(z,\eta)$ is real-valued, since $\ket\psi\!\bra\psi$ is hermitian. It is bounded as
\begin{align}
\nonumber\left|f_{\psi}(\alpha,\eta)\right|&=\left|\sum_{k,l=0}^E{\psi_k\psi_l^*f_{\ket k\!\bra l}(\alpha,\eta)}\right|\\ \displaybreak
\nonumber&\le\sum_{k,l=0}^E{\left|\psi_k\psi_l^*f_{\ket k\!\bra l}(\alpha,\eta)\right|}\\
&\le\sum_{k,l=0}^E{\left|\psi_k\psi_l\right|\frac{M_{kl}}{\eta^{1+\frac{k+l}2}}}\\
\nonumber&=\frac{1}{\eta^{1+E}}\sum_{k,l=0}^E{\left|\psi_k\psi_l\right|\eta^{E-(k+l)/2}M_{kl}}\\
\nonumber&=\frac{M_\psi(\eta)}{\eta^{1+E}},
\end{align}
where we used Lemma~\ref{lem:boundfkl}, and where we defined
\be
M_\psi(\eta):=\sum_{k,l=0}^E{\left|\psi_k\psi_l\right|\eta^{E-(k+l)/2}M_{kl}}.
\label{boundftau}
\ee
Applying Lemma~\ref{lem:HoeffdingR} to the real-valued function $z\mapsto f_\psi(z,\eta)$ thus yields
\be
\Pr\left[\left|\frac{1}{n}\sum_{i=1}^n{f_{\psi}(\alpha_i,\eta)}-\underset{\alpha\leftarrow Q_{\mathrlap\rho}}{\mathbb E}[f_{\psi}(\alpha,\eta)]\right|\ge\epsilon'\right] \le 2\exp\left[{-\frac{n\eta^{2+2E}\epsilon'^2}{2M_\psi^2(\eta)}}\right],
\label{apphoeffding}
\ee
for $\epsilon',\eta>0$, where the probability is over i.i.d.\@ samples from heterodyne detection of $\rho$.

\medskip

\noindent In what follows, we first assume that $\rho\in\bar{\mathcal H}$. By Corollary~\ref{corofidelity} we have
\be
\left|F\left(\psi,\rho\right)-\underset{\alpha\leftarrow Q_{\mathrlap\rho}}{\mathbb E}[f_{\psi}(\alpha,\eta)]\right|\le\eta K_\psi.
\label{appth4}
\ee
Combining Eqs.~(\ref{apphoeffding}) and (\ref{appth4}) yields
\be
\left|F\left(\psi,\rho\right)-\frac{1}{n}\sum_{i=1}^n{f_{\psi}(\alpha_i,\eta)}\right|\le\eta K_\psi+\epsilon',
\ee
with probability greater than $1-2\exp\left[{-\frac{n\eta^{2+2E}\epsilon'^2}{2M_\psi^2(\eta)}}\right]$.
Setting $\eta=\frac\epsilon{K_\psi}$ yields
\be
\left|F\left(\psi,\rho\right)-\frac{1}{n}\sum_{i=1}^n{f_{\psi}\left(\alpha_i,\frac\epsilon{K_\psi}\right)}\right|\le\epsilon+\epsilon',
\label{m1}
\ee
with probability greater than $1-2\exp\left[{-\frac{n\epsilon^{2+2E}\epsilon'^2}{2C_{\psi,1}^2(\epsilon)}}\right]$,
where we defined
\be
\ba
C_{\psi,1}(\epsilon)&:=K_\psi^{1+E}M_\psi\left(\frac\epsilon{K_\psi}\right)\\
&=\sum_{k,l=0}^E{|\psi_k\psi_l|\epsilon^{E-\frac{k+l}2}K_\psi^{1+\frac{k+l}2}\sqrt{2^{|l-k|}\binom{\max{(k,l)}}{\min{(k,l)}}}}.
\ea
\label{CtauE}
\ee
Combining Lemma~\ref{lem:simplem} and Eq.~(\ref{m1}) we obtain
\be
\left|F\left(\psi,\rho\right)^m-\left[\frac{1}{n}\sum_{i=1}^n{f_{\psi}\left(\alpha_i,\frac\epsilon{K_\psi}\right)}\right]^m\right|\le m(\epsilon+\epsilon'),
\ee
with probability greater than $1-2\exp\left[{-\frac{n\epsilon^{2+2E}\epsilon'^2}{2C^2_{\psi,1}(\epsilon)}}\right]$.
Note that we excluded the pathological case $\frac{1}{n}\sum_{i=1}^n{f_{\psi}(\alpha_i,\epsilon/ K_\psi)}>1$:
when that is the case we instead set \mbox{$\frac{1}{n}\sum_{i=1}^n{f_{\psi}(\alpha_i,\epsilon/ K_\psi)}=1$}. 
The target state $\psi$ is pure so $F(\psi^{\otimes m},\rho^{\otimes m})=F\left(\psi,\rho\right)^m$. Hence, replacing $\epsilon$ and $\epsilon'$ by $\epsilon\,/m$ and $\epsilon'/m$, respectively, gives
\be
\left|F\left(\psi,\rho\right)^m-F_\psi(\rho)\right|\le\epsilon+\epsilon',
\ee
with probability greater than
\be
P_{\text{Hoeffding}}^{iid}:=1-2\exp\left[{-\frac{n\epsilon^{2+2E}\epsilon'^2}{2m^{4+2E}C^2_\psi}}\right],
\ee
where
\be
F_\psi(\rho)=\left[\frac{1}{n}\sum_{i=1}^n{f_\psi\left(\alpha_i,\frac\epsilon{mK_\psi}\right)}\right]^m,
\ee
and where
\be
\ba
C_\psi&:=C_{\psi,1}(\epsilon/m)\\
&=\sum_{k,l=0}^E{|\psi_k\psi_l|\left(\frac\epsilon m\right)^{E-\frac{k+l}2}K_\psi^{1+\frac{k+l}2}\sqrt{2^{|l-k|}\binom{\max{(k,l)}}{\min{(k,l)}}}}.
\ea
\label{Cpsiproof}
\ee
Until now we have assumed $\rho\in\bar{\mathcal H}$. By Eq.~(\ref{Etesti.i.d.}), the probability that at most $s$ among $n$ values of $|\alpha_i|^2$ are bigger than $E$, and that the projection $\Pi_E$ of the state $\rho$ onto the Hilbert space $\bar{\mathcal H}$ of states with less than $E$ photons fails is bounded by
\be
P^{iid}_{\text{support}}=\frac{(s+1)^{3/2}}{n}\exp\left[{\frac{(s+1)^2}{n+1}}\right].
\ee
With the union bound we thus obtain
\be
\left|F\left(\psi,\rho\right)^m-F_\psi(\rho)\right|\le\epsilon+\epsilon',
\ee
or $r>s$, with probability greater than $1-\left(P^{iid}_{\text{support}}+P_{\text{Hoeffding}}^{iid}\right)$.

\end{mdframed}
\end{proof}

\noindent This result implies that the quantity $F_\psi(\rho)$ is a good estimate of the fidelity $F(\psi^{\otimes m},\rho^{\otimes m})$, or the score at the support estimation step is higher than $s$, with high probability.
The values of the energy parameters $E$ and $s$ should be chosen to guarantee completeness, i.e., that if the correct state $\ket\psi$ is sent, then $r\le s$ with high probability. 
This theorem is valid for all continuous variable target pure states $\ket\psi$, and the failure probability may be greatly reduced depending on the expression of $\ket\psi$.
The number of samples needed for certifying a given number of copies $m$ with a precision $\epsilon$ and a failure probability $\delta$ scales as \mbox{$n=\poly(m,1/\epsilon,1/\delta)$}. 

This certification protocol is promoted to a verification protocol for single-mode states in the following section, by removing the i.i.d.\@ assumption.

\section{Continuous variable quantum state verification protocol}
\label{sec:certif3}

\noindent We now consider an adversarial setting, where a verifier delegates the preparation of a continuous variable quantum state to a potentially malicious party, called the \textit{prover}. One could see the verifier as the experimentalist in the laboratory and the prover as the noisy device, where we aim not to make any assumptions about its correct functionality or noise model. Given the absence of any direct error correction mechanism that permits a fault tolerant run of the device, the aim of verification is to ensure that a wrong outcome is not being accepted. In the context of state verification, this amounts to making sure that the output state of the tested device is close to an ideal target state.

The prover is not supposed to have i.i.d.\@ behaviour. In particular, when asked for various
copies of the same state, the prover may actually send a large state entangled over all subsystems, 
possibly also entangled with a quantum system on his side.
In that case, the certification protocol derived in the previous section is not reliable. With usual tomography measurements, the number of samples needed for a given precision of the fidelity estimate scales exponentially in the number of copies to verify. This is an essential limitation of quantum tomography techniques, because they check all possible correlations between the different subsystems. 

However we prove that, because of the symmetry of the protocol, the verifier can assume that the prover is sending permutation-invariant states, i.e., states that are invariant under any permutation of their subsystems. After a specific support estimation step, reduced states of permutation-invariant states are close to mixture almost-i.i.d.\@ states, i.e., states that are i.i.d.\@ on almost all subsystems. At the heart of this reduction is a de Finetti theorem for infinite-dimensional systems~\cite{renner2009finetti}, which allows us to restrict to an almost-i.i.d.\@ prover.

\subsection{Description of the protocol}

The verification protocol is as follows: the verifier wants to verify $m$ copies of a target pure state $\ket\psi$. The numbers $n$, $k$, $q$, $s$ and $E$ are free parameters of the protocol.

\begin{itemize}
\item
The prover is instructed to prepare $n+k$ copies of $\ket\psi$ and send them to the verifier. We denote by $\rho^{n+k}$ the state received by the verifier.
\item
The verifier picks $k$ subsystems of the state $\rho^{n+k}$ at random and measures them with heterodyne detection, obtaining the remaining state $\rho^n$ and the samples $\beta_1,\dots,\beta_k$. The verifier records the number $r$ of values $|\beta_i|^2>E$ (support estimation step).
\item
The verifier discards $4q$ subsystems at random, obtaining the remaining state state $\rho^{n-4q}$, and measures all the others subsystems but $m$ chosen at random with heterodyne detection, obtaining the remaining state $\rho^m$ and the samples $\alpha_1,\dots,\alpha_{n-4q-m}$. 
\item
The verifier computes with these samples the estimate
\be
F_\psi(\rho)=\left[\frac{1}{n-4q-m}\smashoperator{\sum_{i=1}^{n-4q-m}}{f_\psi\left(\alpha_i,\frac\epsilon{mK_\psi}\right)}\right]^m,
\label{tildeF2}
\ee
where the function $f_A$ and the constant $K_A$ are defined in Eqs.~(\ref{f}) and (\ref{K}), for $A=\ket\psi\!\bra\psi$ and where $\epsilon>0$ is a free parameter.
\end{itemize}
 
\noindent Note that this estimate is identical to the one defined in Eq.~(\ref{tildeF}) for the certification protocol, replacing $n$ by $n-4q-m$. In order to show that this is a good estimate of the fidelity between the remaining state $\rho^m$ and $m$ copies of the target state $\ket\psi$, we generalise results from~\cite{renner2008security,renner2008finetti,renner2009finetti}. More precisely, we obtain the following results:

\begin{itemize}
\item 
\textit{Support estimation for permutation-invariant states}: with high probability, most of the subsystems of the permutation-invariant state $\rho^{n-4q}$ lie in a lower dimensional subspace, or the score of the state $\rho^{n+k}$ at the support estimation step is high (section~\ref{app:Etestsym}).
\item 
\textit{De Finetti reduction}: any permutation-invariant state with most of its subsystems in a lower dimensional subspace has a purification in the symmetric subspace that still has most of its subsystems in a lower dimensional subspace. This purification is well approximated by a mixture of almost-i.i.d.\@ states (section~\ref{app:deFinetti}).
\item 
\textit{Hoeffding inequality for almost-i.i.d.\@ states}: mixtures of almost-i.i.d.\@ states can be certified in a similar fashion as i.i.d.\@ states (section~\ref{app:almost-i.i.d.}).
\end{itemize} 

\noindent Using these intermediate results, we obtain the following theorem:

\begin{theo}[Gaussian verification of continuous variable quantum states] 
Let $n\ge1$, let $s\le k$, and let $\rho^{n+k}$ be a state over $n+k$ subsystems. Let $\beta_1,\dots,\beta_k$ be samples obtained by measuring $k$ subsystems at random with heterodyne detection and let $\rho^n$ be the remaining state after the measurement. Let $E$ in $\mathbb N$, and let $r$ be the number of samples such that $|\beta_i|^2>E$. Let also $q\ge m$, and let $\rho^m$ be the state remaining after discarding $4q$ subsystems of $\rho^n$ at random, and measuring $n-4q-m$ other subsystems at random with heterodyne detection, yielding the samples $\alpha_1,\dots,\alpha_{n-4q-m}$. Let $\epsilon,\epsilon'>0$ and let $\ket\psi$ be a target pure state. Then,
\be
\left|F\left(\psi^{\otimes m},\rho^m\right)-F_\psi(\rho)\right|\le\epsilon+\epsilon'+P_{deFinetti},
\ee
or $r>s$, with probability greater than
\be
1-\left(P_{\text{support}}+P_{\text{deFinetti}}+P_{\text{choice}}+P_{\text{Hoeffding}}\right),
\ee
where
\be
P_{\text{support}}=8k^{3/2}\exp\left[{-\frac{k}{9}\left(\frac{q}{n}-\frac{2s}{k}\right)^2}\right],
\ee
\be
P_{\text{deFinetti}}=q^{(E+1)^2/2}\exp\left[{-\frac{2q(q+1)}{n}}\right],
\ee
\be
P_{\text{choice}}=\frac{m(4q+m-1)}{n-4q},
\ee
\be
P_{\text{Hoeffding}}=2\binom{n-4q}{4q}\exp\left[{-\frac{n-8q}{2m^{4+2E}}\left(\frac{\epsilon^{1+E}\epsilon'}{C_\psi}-\frac{8qm^{2+E}}{n-4q-m}\right)^2}\right],
\ee
where the estimate $F_\psi(\rho)$ is defined in Eq.~(\ref{tildeF2}), and where
\be
C_\psi=\sum_{k,l=0}^E{|\psi_k\psi_l|\left(\frac\epsilon m\right)^{E-\frac{k+l}2}K_\psi^{1+\frac{k+l}2}\sqrt{2^{|l-k|}\binom{\max{(k,l)}}{\min{(k,l)}}}}
\label{Cpsi}
\ee
is a constant independent of $\rho$, with the constant $K$ defined in Eq.~(\ref{K}).
\label{thVUCVQC}
\end{theo}

\noindent We defer the proof of this result to section~\ref{app:finalproof}.
It implies that either the quantity $F_\psi(\rho)$ is a good estimate of the fidelity $F(\psi^{\otimes m},\rho^m)$, or the score at the support estimation step is higher than $s$, with high probability.
Like for the certification protocol, the values of the energy parameters $E$ and $s$ should be chosen by the verifier to guarantee completeness, i.e., that if the prover sends the correct state $\ket\psi$, then $r\le s$ with high probability. 

For specific choices of the free parameters of the protocol, detailed in the proof of the theorem, either the estimate $F_\psi(\rho)$ is polynomially precise in $m$, or $r>s$, with polynomial probability in $m$, with $n,k,q=\poly m$.
In particular, the efficiency of the protocol may be greatly refined by taking into account the expression of $\ket\psi$ in the Fock basis, and optimizing over the free parameters. 

This verification protocol let the verifier gain confidence about the precision of the estimate of the fidelity in Eq.~(\ref{tildeF2}). If the value of the estimate is close enough to $1$, the verifier may decide to use the state to run a computation. Indeed, statements on the fidelity of a state allow one to infer the correctness of any trusted computation done afterwards using this state.
Let $\beta>0$, and let $\mathcal O$ be the observable corresponding to the result
of the trusted computation performed on $\rho^m$, the reduced state over $m$ subsystems
 instead of $\ket\psi^{\otimes m}$, $m$ copies of the target state $\ket\psi$. In other words, $\mathcal O$ encodes the resources 
which the verifier can perform perfectly (ancillary states, evolution and measurements),
the imperfections being encoded in $\rho$.
Then, $F\left(\psi^{\otimes m},\rho^m\right)\ge1-\beta$ implies the following bound
on the total variation distance between the probability densities of the computation
output of the actual and the target computations:
\be
  \|P_{\psi^{\otimes m}}^{\mathcal O}-P_{\rho^m}^{\mathcal O}\|_{tvd}
    \le D(\psi^{\otimes m},\rho^m)
    \le\sqrt\beta,
    \label{tvdbound}
\ee 
by standard properties of the trace distance $D$ (see section~\ref{sec:DVQI} and~\cite{fuchs1999cryptographic}). What this means is that the distribution of outcomes for the state $\rho^m$ sent by the prover 
is almost indistinguishable from the distribution of outcomes for $m$ copies of the ideal
target state $\ket\psi$, when the fidelity is close enough to one.

\medskip

\noindent In what follows, we detail the intermediate steps described above and prove Theorem~\ref{thVUCVQC}.

\subsection{Support estimation for permutation-invariant states}
\label{app:Etestsym}

We first derive a support estimation step for permutation-invariant states.
We will use in this section the following operators, already introduced in section~\ref{app:Etesti.i.d.}: for $E\ge0$:
\be
U=\smashoperator{\sum_{n=E+1}^{+\infty}}{\ket n\!\bra n}=1-\Pi_E,
\ee
where $\Pi_E=\smashoperator{\sum_{n=0}^E}{\ket n\!\bra n}$ is the projector onto 
the Hilbert space $\bar{\mathcal H}$ of states with at most $E$ photons, and
\be
T=\frac{1}{\pi}\smashoperator{\int_{\quad|\alpha|^2\ge E}}{\ket\alpha\!\bra\alpha d^2\alpha},
\ee
where $\ket\alpha$ is a coherent state. We also recall the following result, from Eq.~(\ref{U2T}), proven in~\cite{leverrier2013security}:
\be
U\le2T.
\label{UleT}
\ee
We recall a few notations and results from~\cite{renner2008finetti}: let $\mathcal A=\{A_0,A_1\},\mathcal B=\{B_0,B_1\}$ be two binary POVMs over $\mathcal H$. Define for $\delta>0$,
\be
\gamma_{A\rightarrow B}(\delta)=\underset{\psi}{\sup}\left\{\Tr\,(B\psi),\text{s.t.}\Tr\,(A\psi)\le\delta\right\}.
\ee
In particular,
\be
\gamma_{T\rightarrow U}(\delta)\le2\delta,
\label{gammaTU}
\ee
by Eq.~(\ref{UleT}). We recall the following result (Lemma~III.1. of~\cite{renner2008finetti}):

\begin{lem} 
Let $n\ge2k$, let $\delta>0$, let $\mathcal A=\{A_0,A_1\}$ and $\mathcal B=\{B_0,B_1\}$ be two binary POVMs over $\mathcal H$, and let $x_1,\dots,x_{n+k}$ the $(n+k)$-partite classical outcome of the measurement $\mathcal{A}^{\otimes n}\otimes \mathcal{B}^{\otimes k}$ applied to any permutation-invariant state $\rho^{n+k}$. Then
\be
\Pr\left[\frac{x_{1}+\dots+x_{n}}{n}>\gamma_{B_1\rightarrow A_1}\left(\frac{x_{n+1}+\dots+x_{n+k}}{k}+\delta\right)+\delta\right]\le8k^{3/2}e^{-k\delta^2}.
\ee
\label{lem:Serf2}
\end{lem}

\noindent This result is a refined version of Serfling's bound~\cite{serfling1974probability}. It relates the outcomes of a measurement on some subsystems of a symmetric state with the outcomes of a related measurement on the rest of the subsystems. With this technical Lemma, we derive in what follows a support estimation step for permutation-invariant states using samples from heterodyne detection.

\medskip

\noindent Let $\rho^{n+k}$ be a state over $n+k$ subsystems. Applying a random permutation to this state and measuring its last $k$ subsystems with heterodyne detection is equivalent to measuring $k$ subsystems at random. We thus assume in the following that the state $\rho^{n+k}$ is a permutation-invariant state, without loss of generality, and that the verifier measures its last $k$ subsystems with heterodyne detection.

Let $\mathcal T=\{1-T,T\}$ and $\mathcal U=\{1-U,U\}$. Let $x_1,\dots,x_{n+k}$ the $(n+k)$-partite classical outcome of the measurement $\mathcal{U}^{\otimes n}\otimes \mathcal{T}^{\otimes k}$ applied to the permutation-invariant state $\rho^{n+k}$ sent by the prover. A value $x_i=1$ for $i\in1,\dots,n$ means that the projection of the $i^{th}$ subsystem onto $\bar{\mathcal H}$ failed, while a value $x_j=1$ for $j\in n+1,\dots,n+k$ means that the value $|\beta|^2$ obtained when measuring the $j^{th}$ subsystem with heterodyne detection was bigger than $E$. In particular, the number of values $\beta_i$ satisfying $|\beta_i|^2>E$, is expressed as $x_{n+1}+\dots+x_{n+k}$.
Let $\mathcal T_{\le s}^k$ be the event that at most $s$ of the $k$ values $\beta_i$ satisfy $|\beta_i|^2>E$, and let $\mathcal F_q^n$ be the event that the projection onto $\bar{\mathcal H}$ fails for more than $q$ subsystems of the remaining state $\rho^n$. Then:

\begin{lem}[Support estimation for permutation-invariant states]
\be
\Pr\left[\mathcal F_q^n\cap\mathcal T_{\le s}^k\right]\le P_{\text{support}}.
\ee
where $P_{\text{support}}=8k^{3/2}\exp\left[{-\frac{k}9\left(\frac qn-\frac{2s}k\right)^2}\right]$.
\label{lem:supportpi}
\end{lem}

\begin{proof} 
\begin{mdframed}[linewidth=1.5,topline=false,rightline=false,bottomline=false]

With Eq.~(\ref{gammaTU}), we have for all $\delta>0$
\be
\gamma_{T\rightarrow U}\left(\frac{x_{n+1}+\dots+x_{n+k}}{k}+\delta\right)+\delta\le2\frac{x_{n+1}+\dots+x_{n+k}}{k}+3\delta.
\ee
Taking $\delta_0=\frac{1}{3}\left(\frac qn-\frac{2s}k\right)$ we obtain
\be
\gamma_{T\rightarrow U}\left(\frac{x_{n+1}+\dots+x_{n+k}}{k}+\delta_0\right)+\delta_0\le\frac qn+2\left(\frac{x_{n+1}+\dots+x_{n+k}}{k}-\frac sk\right),
\ee
so if $x_{1}+\dots+x_{n}>q$ and $x_{n+1}+\dots+x_{n+k}\le s$, then
\be
\gamma_{T\rightarrow U}\left(\frac{x_{n+1}+\dots+x_{n+k}}{k}+\delta_0\right)+\delta_0<\frac{x_{1}+\dots+x_{n}}{n}.
\ee
Hence,
\be
\ba
\Pr\left[\mathcal F_q^n\cap\mathcal T_{\le s}^k\right]&=\Pr\left[\left(x_{1}+\dots+x_{n}>q\right)\cap\left(x_{n+1}+\dots+x_{n+k}\le s\right)\right]\\
&\le\Pr\left[\left(\frac{x_{1}+\dots+x_{n}}{n}>\gamma_{T\rightarrow U}\left(\frac{x_{n+1}+\dots+x_{n+k}}{k}+\delta_0\right)+\delta_0\right)\right]\\
&\le8k^{3/2}e^{-k\delta_0^2}\\
&=8k^{3/2}\exp\left[{-\frac{k}9\left(\frac qn-\frac{2s}k\right)^2}\right],
\ea
\ee
where we used Lemma~\ref{lem:Serf2} for $\mathcal A=\mathcal U$ and $\mathcal B=\mathcal T$.

\end{mdframed}
\end{proof}

\noindent Recall that $\bar{\mathcal H}$ is the Hilbert space of states with at most $E$ photons, of dimension $E+1$.
For $q\le n$, let us define the set of permutation-invariant states over $n$ subsystems, with at most $q$ subsystems out of this lower dimensional subspace (introduced in~\cite{renner2009finetti}):
\be
\mathcal S^n_{\bar{\mathcal H}^{\otimes n-q}}:=\text{span }\underset\pi\bigcup\text{ }\pi\left(\bar{\mathcal H}^{\otimes n-q}\otimes\mathcal H^{\otimes q}\right)\pi^{-1},
\ee
where the union is taken over all permutations. Lemma~\ref{lem:supportpi} then gives
\be
\Pr\left[\mathcal F_q^n\cap\mathcal T_{\le s}^k\right]\le P_{\text{support}},
\ee
where $\mathcal F_q^n$ is the event that the projection of $\rho^n$ (the remaining state after the support estimation step) onto $\mathcal S^n_{\bar{\mathcal H}^{\otimes n-q}}$ fails, and where $P_{\text{support}}=8k^{3/2}\exp\left[{-\frac{k}9\left(\frac qn-\frac{2s}k\right)^2}\right]$. For $1\ll q\ll n$ and $q/n\ll s/k$, this implies that either $\rho^n$ has most of its subsystems in a lower dimensional subspace, or the score at the support estimation step is higher than $s$, with high probability.

\subsection{De Finetti reduction}
\label{app:deFinetti}

We recall in this section two results from~\cite{renner2009finetti}.

\begin{itemize}

\item The first result says that any permutation-invariant state with most of its subsystems in a lower dimensional subspace has a purification in the symmetric subspace that still has most of its subsystems in a lower dimensional subspace. Formally, for $n\in\mathbb N$, and given a Hilbert space $\mathcal K$, let us write Sym$^n(\mathcal K)=\{\phi\in\mathcal K^{\otimes n},\text{ }\pi\phi=\phi\text{ }(\forall \pi)\}$ the symmetric subspace of a Hilbert space $\mathcal K^{\otimes n}$, then (Lemma 3 of~\cite{renner2009finetti}):

\begin{lem} For all $q\le n$, any permutation-invariant state $\rho^n\in\mathcal S^n_{\bar{\mathcal H}^{\otimes n-q}}$ has a purification $\tilde\rho^n$ in \mbox{$\Sym^n(\mathcal H\otimes\mathcal H)\bigcap\mathcal S^n_{(\bar{\mathcal H}\otimes\bar{\mathcal H})^{\otimes n-2q}}$}.
\label{lem:purif}
\end{lem}

\end{itemize}

\noindent The states of the form $\ket v^{\otimes n}$ are the so-called \textit{i.i.d.\@ states}. For all $n,r\ge0$ and all $\ket v\in\bar{\mathcal H}\otimes\bar{\mathcal H}$, the set of \textit{almost-i.i.d.\@ states along $\ket v$}, $\mathcal S^n_{v^{\otimes n-r}}$, is defined as the span of all vectors that are, up to reorderings, of the form $\ket v^{\otimes n-r}\otimes\ket\phi$, for an arbitrary $\phi\in\left(\mathcal H\otimes\mathcal H\right)^{\otimes r}$. In the following, we simply refer to these states as \textit{almost-i.i.d.\@ states} (which becomes relevant when $r\ll n$).

\begin{itemize}

\item The second result is a de Finetti theorem for states in $\Sym^n(\mathcal H\otimes\mathcal H)\bigcap\mathcal S^n_{(\bar{\mathcal H}\otimes\bar{\mathcal H})^{\otimes n-2q}}$, which says that reduced states from them are well approximated by mixtures of almost-i.i.d.\@ states. Formally (Theorem 4 of~\cite{renner2009finetti}, applied to $\mathcal K=\mathcal H\otimes\mathcal H$ and $\bar{\mathcal K}=\bar{\mathcal H}\otimes\bar{\mathcal H}$, with dim$(\bar{\mathcal K})=(E+1)^2$):

\begin{theo}
Let $\tilde\rho^n\in\Sym^n(\mathcal H\otimes\mathcal H)\bigcap\mathcal S^n_{(\bar{\mathcal H}\otimes\bar{\mathcal H})^{\otimes n-2q}}$ and let $\tilde\rho^{n-4q}=\Tr_{4q}(\tilde\rho^n)$. Then, there exist a finite set $\mathcal V$ of unit vectors $\ket v\in\bar{\mathcal H}\otimes\bar{\mathcal H}$, a probability distribution $\{p_v\}_{v\in\mathcal V}$ over $\mathcal V$, and almost-i.i.d.\@ states $\tilde\rho_v^{n-4q}\in\mathcal S^{n-4q}_{v^{\otimes n-8q}}$ such that
\be
F\left(\tilde\rho^{n-4q},\sum_{v\in\mathcal V}{p_v\tilde\rho_v^{n-4q}}\right)>1-q^{(E+1)^2}\exp\left[{-\frac{4q(q+1)}{n}}\right].
\ee
\label{thDeFinetti}
\end{theo}
\end{itemize}

\noindent Given a state $\rho^n\in\mathcal S^n_{\bar{\mathcal H}^{\otimes n-q}}$, applying Theorem~\ref{thDeFinetti} to the purification $\tilde\rho^n$ given by Lemma~\ref{lem:purif} shows that the reduced state $\tilde\rho^{n-4q}$ is close in fidelity to a mixture of states that are i.i.d.\@ on $n-8q$ subsystems.

\subsection{Hoeffding inequality for almost-i.i.d.\@ states}
\label{app:almost-i.i.d.}

We recall here Lemma~\ref{lem:HoeffdingR}, in the context of a product measurement applied to an i.i.d.\@ state $\ket v\!\bra v^{\otimes n}$:

\begin{lem}\textbf{(Hoeffding inequality for i.i.d.\@ states)} 
Let $M>0\in\mathbb R$ and let $f:\mathbb C\mapsto\mathbb R$ be a function bounded as $|f(\alpha)|<M$ for all $\alpha\in\mathbb C$. Let $\lambda>0$, let $p\in\mathbb N^*$, and let $\ket v\in \mathcal H$. Let $\mathcal M=\{\mathcal M_\alpha\}_{\alpha\in\mathbb C}$ be a POVM on $\mathcal H$ and let $D_{\ket v}$ be the probability density function of the outcomes of the measurement $\mathcal M$ applied to $\ket v\!\bra v$. Then
\begin{equation}
\underset{\bm{\alpha}}{\Pr}\left[\left|\frac{1}{p}\sum_{i=1}^p{f(\alpha_i)}- \underset{\beta\leftarrow D_{\mathrlap{\ket v}}}{\mathbb E}[f(\beta)]\right|\ge\lambda\right] 
\le 2\exp\left[{-\frac{p\lambda^2}{2M^2}}\right],
\end{equation}
where the probability is taken over the outcomes $\bm{\alpha}=(\alpha_1,\dots,\alpha_p)$ of the product measurement $\mathcal M^{\otimes p}$ applied to $\ket v\!\bra v^{\otimes p}$.
\label{lem:Hoeffding2}
\end{lem}

\noindent The next result gives an equivalent statement for almost-i.i.d.\@ states along a state $\ket v$, measured with a product measurement. It generalises Theorem 4.5.2 of~\cite{renner2008security}, where the probability distributions over finite sets, corresponding to product measurements with finite number of outcomes, are replaced by continuous variable probability densities, corresponding to product measurements with continuous variable outcomes. Frequencies estimators are also replaced with estimators of expected values of bounded functions. We will use this result for the POVM corresponding to a product heterodyne detection.

\begin{lem}[Hoeffding inequality for almost-i.i.d.\@ states]
Let $M>0\in\mathbb R$ and let $f:\mathbb C\mapsto\mathbb R$ be a function bounded as $|f(\alpha)|\le M$ for all $\alpha\in\mathbb C$. Let $\mu>0$ and $1\le m\le r<t$ such that
\be
(t-m)\mu>2Mr.
\ee
Let also $\ket v\in\bar{\mathcal H}$ and $\ket\Phi\in\mathcal S^t_{v^{\otimes t-r}}$. Let $\mathcal M=\{\mathcal M_\alpha\}_{\alpha\in\mathbb C}$ be a POVM on $\mathcal H$ and let $D_{\ket v}$ be the probability density function of the outcomes of the measurement $\mathcal M$ applied to $\ket v\!\bra v$. Then
\begin{equation}
\underset{\bm{\alpha}}{\Pr}\left[\left|\frac{1}{t-m}\sum_{i=1}^{t-m}{f(\alpha_i)}- \underset{\beta\leftarrow D_{\ket v}}{\mathbb E}[f(\beta)]\right|\ge\mu\right] \le 2\binom{t}{r}\exp\left[{-\frac{t-r}2\left(\frac{\mu}{M}-\frac{2r}{t-m}\right)^2}\right],
\end{equation}
where the probability is taken over the outcomes $\bm{\alpha}=(\alpha_1,\dots,\alpha_{t-m})$ of the product measurement $\mathcal M^{\otimes t-m}$ applied to $\ket\Phi\!\bra\Phi$.
\label{lem:almosti.i.d.}
\end{lem}

\noindent In essence, this lemma says that a product measurement on all but $m$ subsystems of an almost-i.i.d.\@ state along a state $\ket v$ will yield statistics that are similar to the ones that would be obtained by measuring the i.i.d.\@ state $\ket v^{\otimes t-m}$.

\begin{proof} 
\begin{mdframed}[linewidth=1.5,topline=false,rightline=false,bottomline=false]

$\ket\Phi\in\mathcal S^t_{v^{\otimes t-r}}$, so by Lemma 4.1.6 of~\cite{renner2008security}, there exist a finite set $\mathcal S$ of size at most $\binom{t}{r}$, a family of states $\ket{\tilde\Phi^s}\in\mathcal H^{\otimes r}$ for $s\in\mathcal S$, complex amplitudes $\{\gamma_s\}_{s\in\mathcal S}$ and permutations $\{\pi_s\}_{s\in\mathcal S}$ over $[1,\dots,t]$ such that
\be
\ba
\ket\Phi&:=\sum_{s\in\mathcal S}{\gamma_s\ket{\Phi^s}}\\
&=\sum_{s\in\mathcal S}{\gamma_s\pi_s\left(\ket v^{\otimes t-r}\otimes\ket{\tilde\Phi^s}\right)}.
\ea
\label{Psidec}
\ee
With the notations of the Lemma, let us define for $\mu>0$:
\be
\Omega_\mu=\left\{\bm{\alpha}\in\mathbb C^{t-m},\left|\frac{1}{t-m}\sum_{i=1}^{t-m}{f(\alpha_i)}- \underset{\beta\leftarrow D_{\ket v}}{\mathbb E}[f(\beta)]\right|>\mu\right\}.
\ee
We recall here Lemma of 4.5.1 of~\cite{renner2008security}:

\begin{lem}
Let $|\mathcal X|$ be a finite set and $\ket\psi=\sum_{x\in\mathcal X}{\ket{\psi^x}}$, and let $A$ be a non-negative operator. Then
\be
\braket{\psi|A|\psi}\le|\mathcal X|\sum_{x\in\mathcal X}{\braket{\psi^x|A|\psi^x}}.
\ee
\end{lem}

\noindent In particular, using Eq.~(\ref{Psidec}) and this lemma when $A$ is a POVM element of the product measurement \mbox{$\mathcal M_{\bm\alpha}\equiv\mathcal M_{\alpha_1}\otimes\dots\otimes\mathcal M_{\alpha_{t-m}}$}, we obtain:
\be
\ba
\underset{\bm{\alpha}\leftarrow\ket\Phi}{\Pr}[\bm{\alpha}\in\Omega_\mu]&=\int_{\mathrlap{\Omega_\mu}}{\braket{\Phi|\mathcal M_{\bm{\alpha}}|\Phi}d^{2(t-m)}\bm{\alpha}}\\
&\le\int_{\mathrlap{\Omega_\mu}}{|\mathcal S|\sum_{s\in\mathcal S}{|\gamma_s|^2\braket{\Phi^s|\mathcal M_{\bm{\alpha}}|\Phi^s}}d^{2(t-m)}\bm{\alpha}}\\
&\le|\mathcal S|\sum_{s\in\mathcal S}{|\gamma_s|^2\int_{\mathrlap{\Omega_\mu}}{\braket{\Phi^s|\mathcal M_{\bm{\alpha}}|\Phi^s}}d^{2(t-m)}\bm{\alpha}}\\
&=|\mathcal S|\sum_{s\in\mathcal S}{|\gamma_s|^2\underset{\bm{\alpha}\leftarrow\ket{\Phi^s}}{\Pr}[\bm{\alpha}\in\Omega_\mu]},
\ea
\label{PrOmega}
\ee
where we write $\bm{\alpha}\leftarrow\ket\chi$ to indicate that $\bm{\alpha}=(\alpha_1,\dots,\alpha_{t-m})$ is distributed according to the outcomes of the product measurement $\mathcal M^{\otimes t-m}$ applied to $\ket\chi$.

Let $\bm{\alpha}\leftarrow\ket{\Phi^s}$. We have $\ket{\Phi^s}=\pi_s(\ket v^{\otimes t-r}\otimes\ket{\tilde\Phi^s})$, and in particular $(\alpha_{\pi_s(1)},\dots,\alpha_{\pi_s(t-r)})$ is distributed according to the outcomes of the product measurement $\mathcal M^{\otimes t-r}$ applied to $\ket{v}^{\otimes t-r}$. 
We also have, for $|f|\le M$,
\begin{align}
\nonumber\left|\frac{1}{t-r}\sum_{i=1}^{t-r}{f(\alpha_{\pi_s(i)})}-\frac{1}{t-m}\sum_{i=1}^{t-m}{f(\alpha_i)}\right|
&=\left|\frac{1}{t-r}\sum_{i=1}^{t-r}{f(\alpha_{\pi_s(i)})}
  -\frac{1}{t-m}\left(\sum_{i=1}^{t}{f(\alpha_i)}
                     -\smashoperator{\sum_{i=t-m+1}^{t}}{f(\alpha_i)}\right)\right|\\
\nonumber&=\left|\frac{1}{t-r}\sum_{i=1}^{t-r}{f(\alpha_{\pi_s(i)})}
  -\frac{1}{t-m}\left(\sum_{i=1}^{t}{f(\alpha_{\pi_s(i)})}
                     -\smashoperator{\sum_{i=t-m+1}^{t}}{f(\alpha_i)}\right)\right|\\ \displaybreak
\nonumber&=\left|\left(\frac{1}{t-r}-\frac{1}{t-m}\right)\sum_{i=1}^{t-r}{f(\alpha_{\pi_s(i)})}
  +\frac{1}{t-m}\left(\sum_{i=t\mathrlap{-m+1}}^{t}{f(\alpha_i)}
                     -\sum_{i=t\mathrlap{-r+1}}^{t}{f(\alpha_{\pi_s(i)})}\right)\right|\\
\nonumber&\le\left|\frac{1}{t-r}-\frac{1}{t-m}\right|\sum_{i=1}^{t-r}{|f(\alpha_{\pi_s(i)})|}
  +\frac{1}{t-m}\left(\sum_{i=t\mathrlap{-m+1}}^{t}{|f(\alpha_i)|}
                     +\sum_{i=t\mathrlap{-r+1}}^{t}{|f(\alpha_{\pi_s(i)})|}\right)\\
\nonumber&\le\frac{|r-m|}{t-m}M+\frac{(m+r)}{t-m}M\\
&=\frac{2rM}{t-m},
\label{boundfreq}
\end{align}
where we used $r\ge m$.
Now for all $s\in\mathcal S$,  
\be
\ba
\underset{\bm{\alpha}\leftarrow\ket{\Phi^s}}{\Pr}[\bm{\alpha}\in\Omega_\mu]
 &=\underset{\bm{\alpha}\leftarrow\ket{\Phi^s}}{\Pr}\left[
    \left|\frac{1}{t-m}\sum_{i=1}^{t-m}{f(\alpha_i)}
   -\underset{\beta\leftarrow D_{\mathrlap{\ket v}}}{\mathbb E}[f(\beta)]\right|>\mu\right]\\
&\le\underset{\bm{\alpha}\leftarrow\ket{\Phi^s}}{\Pr}\left[
    \left|\frac{1}{t-r}\sum_{i=1}^{t-r}{f(\alpha_{\pi_s(i)})}
    -\underset{\beta\leftarrow D_{\mathrlap{\ket v}}}{\mathbb E}[f(\beta)]\right|
   +\left|\frac{1}{t-m}\smashoperator{\sum_{i=1}^{t-m}}{f(\alpha_i)}
    -\frac{1}{t-r}\sum_{i=1}^{t-r}{f(\alpha_{\pi_s(i)})}\right|>\mu\right]\\
&\le\underset{\bm{\alpha}\leftarrow\ket{\Phi^s}}{\Pr}\left[
    \left|\frac{1}{t-r}\sum_{i=1}^{t-r}{f(\alpha_{\pi_s(i)})}
     -\underset{\beta\leftarrow D_{\mathrlap{\ket v}}}{\mathbb E}[f(\beta)]\right|
    >\mu-\frac{2rM}{t-m}\right]\\
&\le2\exp\left[{-\frac{t-r}2\left(\frac{\mu}{M}-\frac{2r}{t-m}\right)^2}\right],
\ea
\ee
where we used triangular inequality in the second line, Eq.~(\ref{boundfreq}) in the third line and Lemma~\ref{lem:Hoeffding2} in the fourth line with $p=t-r$ and $\lambda=\mu-\frac{2rM}{t-m}>0$. Combining this last equation with Eq.~(\ref{PrOmega}), and using $|\mathcal S|\le\binom{t}{r}$ we finally obtain,
\be
\underset{\bm{\alpha}\leftarrow\ket\psi}{\Pr}\left[\left|\frac{1}{t-m}\sum_{i=1}^{t-m}{f(\alpha_i)}- \underset{\beta\leftarrow D_{\ket v}}{\mathbb E}[f(\beta)]\right|\ge\mu\right] \le 2\binom{t}{r}\exp\left[{-\frac{t-r}2\left(\frac{\mu}{M}-\frac{2r}{t-m}\right)^2}\right].
\ee
\end{mdframed}
\end{proof}

\noindent We recall the bound on $z\mapsto f_\psi(z,\eta)$ obtained in Eq.~(\ref{boundftau}): for all $\alpha\in\mathbb C$,
\be
|f_\psi(\alpha,\eta)|\le\frac{M_\psi(\eta)}{\eta^{1+E}},
\label{boundfM}
\ee
where
\be
M_\psi(\eta)=\sum_{k,l=0}^E{\left|\psi_k\psi_l\right|\eta^{E-(k+l)/2}\sqrt{2^{|l-k|}\binom{\max{(k,l)}}{\min{(k,l)}}}}.
\ee
Let $\mu,\eta>0$, $E\in\mathbb N$, let $\ket v\in\bar{\mathcal H}\otimes\bar{\mathcal H}$, and let $\ket{\Phi_v}^{n-4q}\in\mathcal S^{n-4q}_{v^{\otimes n-8q}}$. Applying Lemma~\ref{lem:almosti.i.d.} for the real-valued function $f_\psi$, for $t=n-4q$, for $r=4q$, for $D_{\ket v}=Q_{\ket v\!\bra v}$, and with the bound from Eq.~(\ref{boundfM}), we obtain 
\be
\ba
\underset{\bm{\alpha}}{\Pr}&\left[
  \left|\frac{1}{n-4q-m}\sum_{i=1}^{n-4q-m}{f_\psi(\alpha_i,\eta)}
  \mathrlap{-}\underset{\quad\beta\leftarrow Q_{\ket v\!\mathrlap{\bra v}}}{\mathbb E}[f_\psi(\beta,\eta)]
  \right|\ge\mu\right] \\
 &\quad\quad\quad\quad\quad\quad\quad\quad\le2\binom{n-4q}{4q}\exp\left[{-\frac{n-8q}2\left(\frac{\eta^{1+E}\mu}{M_\psi(\eta)}-\frac{8q}{n-4q-m}\right)^2}\right],
\ea
\label{Hoeffdingpsiv}
\ee
where the probability is over the outcomes $\bm\alpha$ of a product heterodyne measurement of the first $n-4q-m$ subsystems of $\ket{\Phi_v}^{n-4q}\in\mathcal S^{n-4q}_{v^{\otimes n-8q}}$.

\subsection{Proof of Theorem~\ref{thVUCVQC}}
\label{app:finalproof}

\noindent We introduce the following simple result:

\begin{lem} Let $0<\beta<1$. Let $\rho_1,\rho_2$ be two states such that $F(\rho_1,\rho_2)>1-\beta$. Let $\ket\Phi$ be a pure state, then
\be
\left|F(\Phi,\rho_1)-F(\Phi,\rho_2)\right|\le D(\rho_1,\rho_2)\le\sqrt\beta.
\ee
\label{lem:fidelitytriangular}
\end{lem}

\begin{proof} 
\begin{mdframed}[linewidth=1.5,topline=false,rightline=false,bottomline=false]

Let us write $P^\Phi_{\rho_1}$ and $P^\Phi_{\rho_2}$ the probability distributions associated to the binary measurement $\{\ket\Phi\!\bra\Phi,I-\ket\Phi\!\bra\Phi\}$ of the states $\rho_1$ and $\rho_2$, respectively. Then, $P^\Phi_{\rho_1}(0)+P^\Phi_{\rho_1}(1)=P^\Phi_{\rho_2}(0)+P^\Phi_{\rho_2}(1)=1$, and
\be
\ba
\|P^\Phi_{\rho_1}-P^\Phi_{\rho_2}\|_{tvd}&=\frac12\left(|P^\Phi_{\rho_1}(0)-P^\Phi_{\rho_2}(0)|+|P^\Phi_{\rho_1}(1)-P^\Phi_{\rho_2}(1)|\right)\\
&=|P^\Phi_{\rho_1}(0)-P^\Phi_{\rho_2}(0)|.
\ea
\ee
Hence,
\be
\ba
\left|F(\Phi,\rho_1)-F(\Phi,\rho_2)\right|&=\left|\braket{\Phi|\rho_1|\Phi}-\braket{\Phi|\rho_2|\Phi}\right|\\
&=|P^\Phi_{\rho_1}(0)-P^\Phi_{\rho_2}(0)|\\
&=\|P^\Phi_{\rho_1}-P^\Phi_{\rho_2}\|_{tvd}\\
&\le D(\rho_1,\rho_2)\\
&\le\sqrt{1-F(\rho_1,\rho_2)}\\
&\le\sqrt\beta,
\ea
\ee
where we used Eqs.~(\ref{td1}, \ref{td2}).

\end{mdframed}
\end{proof}

\noindent With these intermediate results, we are now in position to prove Theorem~\ref{thVUCVQC}.

\begin{proof}
\begin{mdframed}[linewidth=1.5,topline=false,rightline=false,bottomline=false]

Let $\ket\psi\!\bra\psi$ be the target pure state, and let $\rho^{n+k}$ be a state sent over $n+k$ subsystems. Let $\beta_1,\dots,\beta_k$ be samples obtained by measuring $k$ subsystems at random of $\rho^{n+k}$ with heterodyne detection. Let $\rho^n$ be the remaining state after the support estimation step. 
In what follows, we first assume that $\rho^n\in\mathcal S^n_{\bar{\mathcal H}^{\otimes n-q}}$.

Let $\rho^{n-4q}$ be the state obtained from $\rho^n$ by tracing over the first $4q$ subsystems. In that case, by section~\ref{app:deFinetti}, there exist a finite set $\mathcal V$ of unit vectors $\ket v\in\bar{\mathcal H}\otimes\bar{\mathcal H}$, a probability distribution $\{p_v\}_{v\in\mathcal V}$ over $\mathcal V$, and almost-i.i.d.\@ states $\tilde\rho_v^{n-4q}\in\mathcal S^{n-4q}_{v^{\otimes n-8q}}$ such that
\be
F\left(\rho^{n-4q},\sum_{v\in\mathcal V}{p_v\rho_v^{n-4q}}\right)>1-q^{(E+1)^2}\exp\left[{-\frac{4q(q+1)}{n}}\right],
\label{DeFinettifide}
\ee
where $\rho_v^{n-4q}$ is the remaining state after tracing over the purifying subsystems, since the fidelity is non-decreasing under quantum operations~\cite{barnum1996noncommuting}. We also obtain 
\be
F\left(\rho^m,\sum_{v\in\mathcal V}{p_v\rho_v^m}\right)>1-q^{(E+1)^2}\exp\left[{-\frac{4q(q+1)}{n}}\right],
\label{DeFinettifidereduced}
\ee
where $\rho^m$ (resp.\@ $\rho_v^m$) is the remaining state after measuring the first $n-4q-m$ subsystems of $\rho^{n-4q}$ (resp.\@ $\rho_v^{n-4q}$) with heterodyne detection.

\medskip

Let $\alpha_1,\dots,\alpha_{n-4q-m}$ be the samples obtained by measuring the first $n-4q-m$ subsystems of $\rho^{n-4q}$ with heterodyne detection. 
The verifier computes the estimate~(\ref{tildeF2})
\be 
F_\psi(\rho)=\left[\frac{1}{n-4q-m}\sum_{i=1}^{n-4q-m}{f_\psi\left(\alpha_i,\frac\epsilon{mK_\psi}\right)}\right]^m,
\label{tildeF3}
\ee
and whenever $F_\psi\ge1$ we instead set $F_\psi=1$. Let us define the completely positive map $\mathcal E$ on $\mathcal H^{n-4q}$ associated to the classical post-processing of the protocol as:
\be
\sigma\mapsto\mathcal E(\sigma)=\sum_e{\Pr\left[F_{\psi}(\sigma)=e\right]\ket e\!\bra e}.
\label{postprocessingmap}
\ee
The sum ranges over the values that the estimate may take. 
With Eq.~(\ref{DeFinettifide}) and Lemma~\ref{lem:fidelitytriangular} we obtain
\be
D\left(\rho^{n-4q},\sum_{v\in\mathcal V}{p_v\rho_v^{n-4q}}\right)\le q^{\frac{(E+1)^2}2}\exp\left[{-\frac{2q(q+1)}{n}}\right],
\label{Drhomrhov}
\ee
The trace distance is non-increasing under quantum operations, so Eq.~(\ref{Drhomrhov}) implies
\be
D\left(\mathcal E\left(\rho^{n-4q}\right),\mathcal E\left(\sum_{v\in\mathcal V}{p_v\rho_v^{n-4q}}\right)\right)\le q^{\frac{(E+1)^2}2}\exp\left[{-\frac{2q(q+1)}{n}}\right].
\ee
Using the definition of the map $\mathcal E$, we obtain a bound in total variation distance:
\be
\left\|P\left[F_\psi(\rho)\right]-P\left[F_\psi\left(\sum_{v\in\mathcal V}{p_v\rho_v^{n-4q}}\right)\right]\right\|_{tvd}\le q^{\frac{(E+1)^2}2}\exp\left[{-\frac{2q(q+1)}{n}}\right],
\ee
where $P$ denotes the probability distributions for the values of the estimates $F_\psi(\rho)$ and $F_\psi\left(\sum_{v\in\mathcal V}{p_v\rho_v^{n-4q}}\right)$. 

\medskip

In particular, this bound implies that for all $\lambda>0$,
\be
\ba
\quad&\left|\Pr\left[\left|F(\psi^{\otimes m},\rho^m)-F_\psi(\rho)\right|>\lambda\right]-\Pr\left[\left|F(\psi^{\otimes m},\rho^m)-F_\psi\left(\sum_{v\in\mathcal V}{p_v\rho_v^{n-4q}}\right)\right|>\lambda\right]\right|\\
&\quad\quad\quad\quad\quad\quad\quad\quad\quad\quad\quad\quad\quad\quad\quad\quad\le q^{\frac{(E+1)^2}2}\exp\left[{-\frac{2q(q+1)}{n}}\right],
\ea
\ee
and thus
\be
\ba
\Pr\left[\left|F(\psi^{\otimes m},\rho^m)-F_\psi(\rho)\right|>\lambda\right]&\le q^{\frac{(E+1)^2}2}\exp\left[{-\frac{2q(q+1)}{n}}\right]\\
&\quad+\Pr\left[\left|F(\psi^{\otimes m},\rho^m)-F_\psi\left(\sum_{v\in\mathcal V}{p_v\rho_v^{n-4q}}\right)\right|>\lambda\right].
\ea
\label{boundprobaeps}
\ee
With Eq.~(\ref{DeFinettifidereduced}) and Lemma~\ref{lem:fidelitytriangular} we obtain
\be
\left|F\left(\psi^{\otimes m},\rho^m\right)-F\left(\psi^{\otimes m},\sum_{v\in\mathcal V}{p_v\rho_v^m}\right)\right|\le q^{\frac{(E+1)^2}2}\exp\left[{-\frac{2q(q+1)}{n}}\right],
\label{boundlemma91}
\ee
where $\psi^{\otimes m}$ is $m$ copies of the target pure state $\ket\psi$. 
With the triangular inequality,
\be
\ba
\,&\left|F(\psi^{\otimes m},\rho^m)-F_\psi\left(\sum_{v\in\mathcal V}{p_v\rho_v^{n-4q}}\right)\right|\\
&\le\left|F(\psi^{\otimes m},\rho^m)-F\left(\psi^{\otimes m},\sum_{v\in\mathcal V}{p_v\rho_v^m}\right)\right|+\left|F\left(\psi^{\otimes m},\sum_{v\in\mathcal V}{p_v\rho_v^m}\right)-F_\psi\left(\sum_{v\in\mathcal V}{p_v\rho_v^{n-4q}}\right)\right|\\
&\le q^{\frac{(E+1)^2}2}\exp\left[{-\frac{2q(q+1)}{n}}\right]+\left|F\left(\psi^{\otimes m},\sum_{v\in\mathcal V}{p_v\rho_v^m}\right)-F_\psi\left(\sum_{v\in\mathcal V}{p_v\rho_v^{n-4q}}\right)\right|,
\ea
\ee
where we used Eq.~(\ref{boundlemma91}) in the last line. With Eq.~(\ref{boundprobaeps}) we obtain, for all $\lambda>0$
\be
\ba
\,&\Pr\left[\left|F(\psi^{\otimes m},\rho^m)-F_\psi(\rho)\right|>\lambda\right]\le q^{\frac{(E+1)^2}2}\exp\left[{-\frac{2q(q+1)}{n}}\right]\\
&\quad\quad\quad\quad\quad\quad+\Pr\left[\left|F\left(\psi^{\otimes m},\sum_{v\in\mathcal V}{p_v\rho_v^m}\right)-F_\psi\left(\sum_{v\in\mathcal V}{p_v\rho_v^{n-4q}}\right)\right|>\lambda-q^{\frac{(E+1)^2}2}e^{-\frac{2q(q+1)}{n}}\right].
\ea
\label{almost}
\ee
By linearity of the probabilities, it suffices to bound $\Pr\left[\left|F(\psi^{\otimes m},\Phi^m)-F_\psi(\Phi)\right|>\mu\right]$, for $\mu=\lambda-q^{\frac{(E+1)^2}2}\exp\left[{-\frac{2q(q+1)}{n}}\right]$, where $\ket\Phi\in\mathcal S^{n-4q}_{v^{\otimes n-8q}}$, for $\ket v\in\bar{\mathcal H}\otimes\bar{\mathcal H}$, and where $\Phi^m$ is the state obtained from $\ket\Phi\!\bra\Phi$ by measuring the first $n-4q-m$ subsystems with heterodyne detection and tracing over the purifying subsystems.\\

\begin{lem}
Let $\ket\Phi\in\mathcal S^{n-4q}_{v^{\otimes n-8q}}$. For all $\epsilon'>0$,
\be
\ba
\Pr\left[\left|F(\psi^{\otimes m},\Phi^m)-F_\psi(\Phi)\right|>\epsilon+\epsilon'\right]&\le2\binom{n-4q}{4q}\exp\left[{-\frac{n-8q}{2m^{4+2E}}\left(\frac{\epsilon^{1+E}\epsilon'}{C_\psi}-\frac{8qm^{2+E}}{n-4q-m}\right)^2}\right]\\
&\quad\quad\quad+\frac{m(4q+m-1)}{n-4q},
\ea
\ee
where
\be
C_\psi=\sum_{k,l=0}^E{|\psi_k\psi_l|\left(\frac\epsilon m\right)^{E-\frac{k+l}2}K_\psi^{1+\frac{k+l}2}\sqrt{2^{|l-k|}\binom{\max{(k,l)}}{\min{(k,l)}}}}\underset{\epsilon\rightarrow0}{\longrightarrow}|\psi_E|^2K_\psi^{1+E}.
\ee
\label{interbound}
\end{lem}

\begin{proof} 
Let $\alpha_1,\dots,\alpha_{n-4q-m}$ be samples obtained by measuring the first $n-4q-m$ subsystems of $\ket\Phi\!\bra\Phi$ with heterodyne detection. We have~(\ref{tildeF2})
\be
F_\psi(\Phi)=\left[\frac{1}{n-4q-m}\smashoperator{\sum_{i=1}^{n-4q-m}}{f_\psi\left(\alpha_i,\frac\epsilon{mK_\psi}\right)}\right]^m,
\ee
and
\be
\ba
\left|F(\psi^{\otimes m},\Phi^m)-F_\psi(\Phi)\right| &\le\left|F(\psi^{\otimes m},{\Phi}^m)-F(\psi^{\otimes m},\ket{v}\!\bra{v}^{\otimes m})\right|\\
  &\quad+\left|F(\psi^{\otimes m},\ket{v}\!\bra{v}^{\otimes m})
   -\left(\underset{\beta\leftarrow Q_{\ket v\!\bra v}}{\mathbb E}%
                                        \left[f_\psi\left(\beta,\frac\epsilon{mK_\psi}\right)\right]\right)^m\right|\\
  &\quad+\left|\left(\underset{\beta\leftarrow Q_{\ket v\!\bra v}}{\mathbb E}%
                                        \left[f_\psi\left(\beta,\frac\epsilon{mK_\psi}\right)\right]\right)^m
        -F_\psi(\Phi)\right|\\
 &=\left|F(\psi^{\otimes m},{\Phi}^m)-F(\psi^{\otimes m},\ket{v}\!\bra{v}^{\otimes m})\right|\\
  &\quad+\left|F(\psi,\ket{v}\!\bra{v})^m
   -\left(\underset{\beta\leftarrow Q_{\ket v\!\bra v}}{\mathbb E}%
                                        \left[f_\psi\left(\beta,\frac\epsilon{mK_\psi}\right)\right]\right)^m\right|\\
  &\quad+\left|\left(\underset{\beta\leftarrow Q_{\ket v\!\bra v}}{\mathbb E}%
                                        \left[f_\psi\left(\beta,\frac\epsilon{mK_\psi}\right)\right]\right)^m
        -\left(\frac{1}{n-4q-m}\sum_{i=1}^{n-4q-m}{f_\psi\left(\alpha_i,\frac\epsilon{mK_\psi}\right)}\right)^m\right|\\
 &\le\left|F(\psi^{\otimes m},{\Phi}^m)-F(\psi^{\otimes m},\ket{v}\!\bra{v}^{\otimes m})\right|\\
  &\quad+m\left|F(\psi,\ket v\!\bra v)
    -\underset{{\beta}\leftarrow Q_{\ket v\!\bra v}}{\mathbb E}\left[f_\psi\left(\beta,\frac\epsilon{mK_\psi}\right)\right]\right|\\
&\quad+m\left|\underset{{\beta}\leftarrow Q_{\ket v\!\bra v}}{\mathbb E}\left[f_\psi\left(\beta,\frac\epsilon{mK_\psi}\right)\right]-\frac{1}{n-4q-m}\sum_{i=1}^{n-4q-m}{f_\psi\left(\alpha_i,\frac\epsilon{mK_\psi}\right)}\right|,
\ea
\label{threebounds}
\ee
where we used Lemma~\ref{lem:simplem}. We bound these three terms in the following.

\medskip

\noindent When selecting at random $m$ subsystems from an almost-i.i.d.\@ state over $n-4q$ subsystems which is i.i.d.\@ on $n-8q$ subsystems, the probability that all of the selected states are from the $n-8q$ i.i.d.\@ subsystems is
\be
\frac{\binom{n-8q}{m}}{\binom{n-4q}{m}}=\frac{(n-8q)(n-8q-1)\dots(n-8q-m+1)}{(n-4q)(n-4q-1)\dots(n-4q-m+1)},
\ee
and we have
\be
\ba
1-\frac{(n-8q)(n-8q-1)\dots(n-8q-m+1)}{(n-4q)(n-4q-1)\dots(n-4q-m+1)}&\le1-\frac{(n-8q-m+1)^m}{(n-4q)^m}\\
&=1-\left(1-\frac{4q+m-1}{n-4q}\right)^m\\
&\le\min{\left(1,\frac{m(4q+m-1)}{n-4q}\right)}\\
&\le\frac{m(4q+m-1)}{n-4q},
\ea
\label{boundmNQ}
\ee
where we used $1-(1-x)^a\le ax$ for all $a\ge1$ and $x\in[0,1]$. In particular, for $\ket\Phi\in\mathcal S^{n-4q}_{v^{\otimes n-8q}}$, and $\Phi^m$ its reduced state over $m$ modes chosen at random, we have
\be
\Phi^m=\ket v\bra v^{\otimes m},
\ee
with probability greater than $1-\frac{m(4q+m-1)}{n-4q}$, where we used the definition of $\mathcal S^{n-4q}_{v^{\otimes n-8q}}$, and Eq.~(\ref{boundmNQ}). Using Lemma~\ref{lem:fidelitytriangular}, the \textit{first term} in Eq.~(\ref{threebounds}) vanishes with probability greater than:
\be
1-\frac{m(4q+m-1)}{n-4q}.
\label{finalbound1}
\ee
The bound for the \textit{second term} is given by Corollary~\ref{corofidelity} applied to the state $\ket v$, for $\eta=\frac\epsilon{mK_\psi}$:
\be
m\left|F(\psi,\ket v\!\bra v)
    -\underset{{\beta}\leftarrow Q_{\ket v\!\bra v}}{\mathbb E}\left[f_\psi\left(\beta,\frac\epsilon{mK_\psi}\right)\right]\right|\le\epsilon.
\label{finalbound2}
\ee

The bound for the \textit{third term} is probabilistic, given by Eq.~(\ref{Hoeffdingpsiv}), for $\eta=\frac\epsilon{mK_\psi}$ and $\mu=\frac{\epsilon'}m$. For all $\epsilon'>0$,
\be
\ba
\quad&\underset{\bm{\alpha}}{\Pr}\left[
  \left|\frac{1}{n-4q-m}\sum_{i=1}^{n-4q-m}{f_\psi\left(\alpha_i,\frac\epsilon{mK_\psi}\right)}
  -\underset{\beta\leftarrow Q_{\ket v\!\bra v}}{\mathbb E}\left[f_\psi\left(\beta,\frac\epsilon{mK_\psi}\right)\right]
  \right|
 \ge\frac{\epsilon'}m\right] \\
&\quad\quad\quad\le2\binom{n-4q}{4q}\exp\left[{-\frac{n-8q}2\left(\frac{\epsilon^{1+E}\epsilon'}{m^{2+E}K_\psi^{1+E}M_\psi(\frac\epsilon{mK_\psi})}-\frac{8q}{n-4q-m}\right)^2}\right].
\ea
\label{finalbound3}
\ee

We now bring together the previous bounds in order to prove Lemma~\ref{interbound}. Combining Eqs.~(\ref{threebounds}), (\ref{finalbound1}), (\ref{finalbound2}) and (\ref{finalbound3}) yields
\begin{align}
\nonumber\quad&\Pr\left[\left|F(\psi^{\otimes m},\Phi^m)-F_\psi(\Phi)\right|>\epsilon+\epsilon'\right]\\
\nonumber&\quad\quad\quad\le\underset{\bm{\alpha}}{\Pr}\left[
  \left|\frac{1}{n-4q-m}\sum_{i=1}^{n-4q-m}{f_\psi\left(\alpha_i,\frac\epsilon{mK_\psi}\right)}
  \mathrlap{-}\underset{\quad\beta\leftarrow Q_{\ket v\!\bra v}}{\mathbb E}\left[f_\psi\left(\beta,\frac\epsilon{mK_\psi}\right)\right]
  \right|
 \ge\frac{\epsilon'}m\right]\displaybreak \\
&\quad\quad\quad\le2\binom{n-4q}{4q}\exp\left[{-\frac{n-8q}2\left(\frac{\epsilon^{1+E}\epsilon'}{m^{2+E}K_\psi^{1+E}M_\psi(\frac\epsilon{mK_\psi})}-\frac{8q}{n-4q-m}\right)^2}\right]\\
\nonumber&\quad\quad\quad\quad\quad\quad+\frac{m(4q+m-1)}{n-4q}\\
\nonumber&\quad\quad\quad=2\binom{n-4q}{4q}\exp\left[{-\frac{n-8q}{2m^{4+2E}}\left(\frac{\epsilon^{1+E}\epsilon'}{C_\psi}-\frac{8qm^{2+E}}{n-4q-m}\right)^2}\right]+\frac{m(4q+m-1)}{n-4q},
\end{align}
where
\be
\ba
C_\psi&=K_\psi^{1+E}M_\psi\left(\frac\epsilon{mK_\psi}\right)\\
&=\sum_{k,l=0}^E{|\psi_k\psi_l|\left(\frac\epsilon m\right)^{E-\frac{k+l}2}K_\psi^{1+\frac{k+l}2}\sqrt{2^{|l-k|}\binom{\max{(k,l)}}{\min{(k,l)}}}}\underset{\epsilon\rightarrow0}{\longrightarrow}|\psi_E|^2K_\psi^{1+E}.
\ea
\ee
\end{proof}

\noindent Combining Eq.~(\ref{almost}) and Lemma~\ref{interbound}, we finally obtain
\be
\ba
\Pr &\left[\left|F(\psi^{\otimes m},\rho^m)-F_\psi(\rho)\right|>\epsilon+\epsilon'+q^{\frac{(E+1)^2}2}e^{-\frac{2q(q+1)}{n}}\right]\\
&\le q^{\frac{(E+1)^2}2}\exp\left[{-\frac{2q(q+1)}{n}}\right]+2\binom{n-4q}{4q}\exp\left[{-\frac{n-8q}{2m^{4+2E}}\left(\frac{\epsilon^{1+E}\epsilon'}{C_\psi}-\frac{8qm^{2+E}}{n-4q-m}\right)^2}\right]\\
&\quad\quad\quad+\frac{m(4q+m-1)}{n-4q}.
\ea
\label{appprobafinale}
\ee
Setting 
\be
P_{\text{Hoeffding}}=2\binom{n-4q}{4q}\exp\left[{-\frac{n-8q}{2m^{4+2E}}\left(\frac{\epsilon^{1+E}\epsilon'}{C_\psi}-\frac{8qm^{2+E}}{n-4q-m}\right)^2}\right],
\ee
\be
P_{\text{choice}}=\frac{m(4q+m-1)}{n-4q},
\ee
and
\be
P_{\text{deFinetti}}=q^{\frac{(E+1)^2}2}\exp\left[{-\frac{2q(q+1)}{n}}\right],
\ee
we obtain
\be
\Pr\left[\left|F(\psi^{\otimes m},\rho^m)-F_\psi(\rho)\right|>\epsilon+\epsilon'+P_{\text{deFinetti}}\right]\le P_{\text{deFinetti}}+P_{\text{choice}}+P_{\text{Hoeffding}}.
\ee
Until now we have assumed $\rho^n\in\mathcal S^n_{\bar{\mathcal H}^{\otimes n-q}}$. By section~\ref{app:Etestsym},
\be
\Pr\left[\mathcal F_q^n\cap\mathcal T_{\le s}^k\right]\le P_{\text{support}}.
\ee
where $\mathcal F_q^n$ is the event that the projection of $\rho^n$ (the remaining state after the support estimation step) onto $\mathcal S^n_{\bar{\mathcal H}^{\otimes n-q}}$ fails, where $\mathcal T_{\le s}^k$ is the event that at most $s$ of the $k$ values $\beta_i$ from the support estimation step satisfy $|\beta_i|^2>E$, and where $P_{\text{support}}=8k^{3/2}\exp\left[{-\frac{k}9\left(\frac qn-\frac{2s}k\right)^2}\right]$. With the union bound we thus obtain
\be
\Pr\left[\left(\left|F(\psi^{\otimes m},\rho^m)-F_\psi(\rho)\right|>\epsilon+\epsilon'+P_{\text{deFinetti}}\right)\cap\mathcal T_{\le s}^k\right]\le P_{\text{support}}+P_{\text{deFinetti}}+P_{\text{choice}}+P_{\text{Hoeffding}},
\ee
where
\begin{align}
\nonumber P_{\text{support}}&=8k^{3/2}\exp\left[{-\frac{k}9\left(\frac qn-\frac{2s}k\right)^2}\right],\\ 
P_{\text{deFinetti}}&=q^{\frac{(E+1)^2}2}\exp\left[{-\frac{2q(q+1)}{n}}\right],\\
P_{\text{choice}}&=\frac{m(4q+m-1)}{n-4q},\\
\nonumber P_{\text{Hoeffding}}&=2\binom{n-4q}{4q}\exp\left[{-\frac{n-8q}{2m^{4+2E}}\left(\frac{\epsilon^{1+E}\epsilon'}{C_\psi}-\frac{8qm^{2+E}}{n-4q-m}\right)^2}\right].
\end{align}
The variables $\epsilon,\epsilon',n,m,q,k,s,E$ are free parameters of the protocol. Let us fix, e.g., $E=O(1)$, $s=O(1)$, \mbox{$n=O\left(m^{19+8E}\right)$}, $k=O(m^{19+8E}), q=O\left(m^{10+4E}\right)$, and $\epsilon=\epsilon'=O(\frac1m)$.
Then, either the estimate $F_\psi(\rho)$ of the fidelity $F(\psi^{\otimes m},\rho^m)$ is polynomially precise (in $m$), or the score at the support estimation step is higher than $s$, with polynomial probability (in $m$), by plugging the different scalings in the above expressions.

\end{mdframed}
\end{proof}

\noindent These general single-mode state certification and verification protocols may be used for various usecases. We present selected applications in the following section, relating to the certification of non-Gaussian properties of quantum states.

\section{Certification of non-Gaussian properties}
\label{sec:applicationscertif}

\subsection{Certifying the stellar rank}
\label{sec:applicationscertif1}

The stellar hierarchy can be certified with the previous protocol using the estimate of the fidelity obtained as a witness for the stellar rank.
We recall a few definitions and results from chapter~\ref{chap:stellar}. The stellar rank of a single-mode normalised pure quantum state corresponds to the minimal number of photon additions necessary to engineer the state from the vacuum, together with Gaussian unitary operations. Moreover, a mixed state which has a stellar rank equal to $n$ cannot be expressed as a mixture of pure states of ranks strictly lower than $n$. Given $k\in\mathbb N^*$ and a target pure state $\ket\psi$, if a mixed state $\rho$ satisfies
\be
F(\psi,\rho)>1-[R^\star_k(\psi)]^2,
\ee
where $R^\star_k(\psi)$ is the $k$-robustness of the state $\ket\psi$, then it has a stellar rank greater or equal to $k$. This in turn can be checked by computing the robustness profile of the state $\ket\psi$.

With Theorem~\ref{thi.i.d.} for $m=1$, we obtain the following protocol for certifying the stellar rank under the i.i.d.\@ assumption, where $E$, $s$, $\epsilon$ and $\epsilon'$ are free parameters:

Let $\ket\psi$ be a target pure state. First, measure with heterodyne detection $n$ copies of the (mixed) state $\rho$, obtaining the samples $\alpha_1,\dots,\alpha_n$. Then, record the number $r$ of samples such that $|\alpha_i|^2>E$. Compute with the same samples the estimate
\be
F_\psi(\rho)=\frac{1}{n}\sum_{i=1}^n{f_\psi\left(\alpha_i,\frac\epsilon{K_\psi}\right)},
\label{tildeF1}
\ee
where the function $f_A$ and the constant $K_A$ are defined in Eqs.~(\ref{f}) and (\ref{K}), for $A=\ket\psi\!\bra\psi$.
Then,
\be
\left|F(\psi,\rho)-F_\psi(\rho)\right|\le\epsilon+\epsilon',
\ee
or $r>s$, with probability greater than
\be
1-\left(\frac{(s+1)^{3/2}}n\exp\left[{\frac{(s+1)^2}{n+1}}\right]+2\exp\left[{-\frac{n\epsilon^{2+2E}\epsilon'^2}{2C^2_\psi}}\right]\right),
\ee
where
\be
C_\psi=\sum_{k,l=0}^E{|\psi_k\psi_l|\epsilon^{E-\frac{k+l}2}K_\psi^{1+\frac{k+l}2}\sqrt{2^{|l-k|}\binom{\max{(k,l)}}{\min{(k,l)}}}}
\label{Cpsi2}
\ee
is a constant independent of $\rho$, with the constant $K$ defined in Eq.~(\ref{K}).
In particular, if the estimate obtained satisfies
\be
F_\psi(\rho)>1-[R^\star_k(\psi)]^2+\epsilon+\epsilon',
\ee
which can be readily checked from the robustness profile of the target state $\ket\psi$, then either the score at the support estimation step is high or the state $\rho$ has stellar rank greater or equal to $k$, with high probability for a large number of samples. An analogous statement holds for the case of verification, without the i.i.d.\@ assumption, with Theorem~\ref{thVUCVQC}.

\subsection{Certifying Wigner negativity}
\label{sec:applicationscertif2}

\noindent In the previous section, we detail how to certify a nonzero stellar rank of any experimental (mixed) state, which implies that this state is non-Gaussian. However, such a mixed state may still have positive Wigner function. Since processes with positive Wigner functions are classically simulable~\cite{mari2012positive}, negativity of the Wigner function is also a crucial property to look for. In this section, we show how our certification protocol with heterodyne detection allows for the certification of Wigner negativity without the need for a full tomography. 

The Wigner function of a state $\rho$ evaluated at $\alpha\in\mathbb C$ is related to the expected value of the parity operator displaced by $\alpha$~\cite{royer1977wigner}:
\be
W_\rho(\alpha)=\frac2\pi\Tr\left[\hat D(\alpha)\hat\Pi\hat D^\dag(\alpha)\rho\right],
\ee
where
\be
\hat\Pi=\sum_{n\ge0}{(-1)^n\ket n\!\bra n}
\ee
is the parity operator. Hence, we can use the certification protocol to obtain mean value estimations of the operator $\frac2\pi\hat D(\alpha)\hat\Pi\hat D^\dag(\alpha)$ and retrieve the value of the Wigner function at $\alpha$. Moreover, since the displacement can be reverted in post-processing by translating the samples by $\alpha$, we can alternatively obtain mean value estimations of the operator $\frac2\pi\hat\Pi$ (which has a simpler expression in Fock basis) using translated samples.

Using either the certification protocol from Theorem~\ref{thi.i.d.} or the verification protocol from Theorem~\ref{thVUCVQC} allows us to witness Wigner negativity under or wihtout the i.i.d.\@ assumption, respectively.

\section{Certifying multimode continuous variable quantum states}
\label{sec:applicationscertif3}

The certification and verification protocols described in the previous sections allow us to obtain efficiently estimates of fidelities of any single-mode continuous variable quantum state with any target single-mode pure state, with analytical confidence intervals, either with i.i.d.\@ assumption or with no assumption whatsoever. These protocols also allow us to estimate efficiently fidelities with multimode i.i.d.\@ pure states. However, translating them directly to efficient protocols for general multimode states seems hopeless, since verifying a multimode state implies accounting for all possible correlations between its subsystems, of which there is an exponential number in the size of the state. 

On the other hand, we show in what follows that being able to estimate single-mode fidelities with heterodyne detection is enough to provide fidelity witnesses for a large class of multimode states.
This result combines the following two observations:

\begin{itemize}
\item
If all the single-mode subsystems $\rho_i$ of a multimode quantum state $\bm\rho$ are close enough to single-mode pure states $\ket{\psi_i}\!\bra{\psi_i}$, then $\bm\rho$ is close to the tensor product of these pure states (Lemma~\ref{lem:product}). In particular, being able to estimate single-mode fidelities is enough to provide fidelity witnesses for product of pure states.
\item
Passive linear transformations followed by single-mode Gaussian unitary operations and product of single-mode balanced heterodyne detections can be simulated by performing unbalanced heterodyne detections first, then post-processing efficiently the samples (Lemma~\ref{lem:hetmagic}). In particular, for such an operation $\hat V$, if the multimode state $\bm\rho$ can be efficiently certified using heterodyne detection, then it is also the case for the state $\hat V\bm\rho\hat V^\dag$.
\end{itemize}

\noindent This allows us to verify efficiently a large class of multimode continuous variable quantum states, with and without the i.i.d.\@ assumption, including the $m$-mode states of the form
\be
\left(\bigotimes_{i=1}^m\hat G_i\right)\hat U\left(\bigotimes_{i=1}^m\ket{\psi_i}\right),
\label{class}
\ee
where $\hat U$ is a passive linear transformation (a unitary transformation of the creation and annihilation operators of the modes) and where, for all $i\in\{1,\dots,m\}$, the state $\ket{\psi_i}$ is a single-mode pure state with constant energy (which does not scale with the number of modes $m$) and the operation $\hat G_i$ is a single-mode Gaussian unitary which may be written as a combination of a single-mode displacement and a single-mode squeezing (see section~\ref{sec:Gaussian}). In particular, these states includes multimode Gaussian states and the output states of Boson Sampling interferometers and of CVS circuits (see sections~\ref{sec:BosonSampling} and~\ref{sec:CVS}).

The fidelity witnesses presented here extend the work of~\cite{aolita2015reliable} in various respects. Their work provides fidelity witnesses for multimode photonic state preparations with Gaussian measurements, under the i.i.d.\@ assumption. However, the witnesses are for a more restricted class of target states and are efficient for Gaussian pure states only. In particular, the number of copies needed to certify with constant precision the output of a Boson Sampling interferometer with $n$ input photons over $m$ modes with their protocol scales as $\Omega(m^{n+4})$, which is worse than exponential in the antibunching regime $n=O(\sqrt m)$, while we show that our protocol provides tight fidelity witnesses with constant precision with $O(m^4\log m)$ copies. Moreover, we are able to remove the i.i.d.\@ state preparation assumption, at the cost of an increased---though still polynomial---number of measurements needed for the same estimate precision and confidence interval.

\medskip

\noindent In the following sections, we present the general protocol and detail its application in the case of Boson Sampling.

\subsection{General multimode protocol}

We present the two versions of the multimode verification protocol, with or without i.i.d.\@ assumption. Under the i.i.d.\@ assumption:

\begin{enumerate}
\item
The verifier chooses an $m$-mode target pure state $\ket{\bm\tau_{U,\bm\xi,\bm\beta}}:=\hat S(\bm\xi)\hat D(\bm\beta)\,\hat U(\bigotimes_{i=1}^m\ket{\psi_i})$, as in Eq.~(\ref{class}), where for all $i\in\{1,\dots,m\}$ the state $\ket{\psi_i}$ has constant energy, where $\hat U$ is an $m$-mode passive linear transformation with $m\times m$ unitary matrix $U$ and where $\bm\xi,\bm\beta\in\mathbb C^m$. The verifier also chooses a precision parameter $0<\eta<1$ and energy cutoff values $E_1,\dots,E_m$.
\item
The verifier asks the prover for $N=O(\poly m)$ copies of the target state $\ket{\bm\tau_{U,\bm\xi,\bm\beta}}$. Let $\bm\rho^{\otimes N}$ be the $(N\times m)$-mode (mixed) state sent by the prover, where $\bm\rho$ is an $m$-mode (mixed) state.
\item
The verifier measures with unbalanced heterodyne detection with unbalancing parameters $\bm\xi$ all the $m$ subsystems of all the $N$ copies of $\bm\rho$, obtaining the $N$ vectors of samples $\bm\gamma^{(1)},\dots,\bm\gamma^{(N)}\in\mathbb C^m$. 
\item
For all $k\in\{1,\dots,N\}$, the verifier computes $\bm\alpha^{(k)}=U^\dag(\bm\gamma^{(k)}-\bm\beta)$. We write $\bm\alpha^{(k)}=(\alpha_1^{(k)},\dots,\alpha_m^{(k)})$.
\item
For all $i\in\{1,\dots,m\}$, the verifier records the number $r_i$ of values among $\alpha_i^{(1)},\dots,\alpha_i^{(N)}$ such that $|\alpha_i^{(k)}|^2>E_i$ (\textit{support estimation}).
\item
For all $i\in\{1,\dots,m\}$, the verifier computes the mean $\tilde F_i$ of the function $z\mapsto f_{\psi_i}(z,\epsilon,E_i)$ over the same values $\alpha_i^{(1)},\dots,\alpha_i^{(N)}$, where the function $f$ is defined in Eq.(~\ref{f}).
\item
The verifier computes $\tilde W=1-\sum_{i=1}^m(1-\tilde F_i)$.
\end{enumerate}

\noindent The cutoff values $E_1,\dots,E_m$ should be chosen by the verifier to guarantee completeness for the estimation of the single-mode fidelities, i.e., that if the prover is sending a near-ideal state it is accepted with high probability. For a sufficiently large number of copies $N=O(\poly m)$, we show in what follows that $\tilde W$ is a tight lower bound on the fidelity with inverse polynomial precision, or one of the scores $r_1,\dots,r_m$ is high, with high probability. 

Without i.i.d.\@ assumption, an equivalent protocol is obtained by using the version of the protocol which does not assume i.i.d.\@ state preparation for estimating the single-mode fidelities in Theorem~\ref{thVUCVQC}. In that case, the final protocol is nearly identical, up to slight differences for the classical post-processing: a small fraction of the measured subsystems have to be discarded at random and the samples used for the support estimation step must be randomly chosen and cannot be used to compute the fidelity estimates. This comes at the cost of an increased number of measurements necessary for the same witness precision and confidence interval, which corresponds however to a polynomial overhead in $m$.

For both protocols, note that the efficiency may be greatly refined by taking into account the expression of the single-mode target pure states $\ket{\psi_i}$ in Fock basis. We give an example of such optimisation in the next section, in the case of Boson Sampling output states, when the single-mode target pure states are either single-photon Fock states or vacuum states. In particular, for the protocol under i.i.d.\@ assumption, if the single-mode target states have a finite support over the Fock basis then the support estimation step is no longer needed.

\medskip

\noindent We now show that the estimate $\tilde W$ is a tight fidelity witness for a number of samples $O(\poly m)$. We first prove the following result:

\begin{lem}\label{lem:product}
Let $\bm\rho$ be an $m$-mode state. For all $i\in\{1,\dots,m\}$, we denote by $\rho_i$ the single-mode reduced state of $\bm\rho$ over the $i^{th}$ mode. Let $\ket{\psi_1},\dots,\ket{\psi_m}$ be single-mode pure states. For all $i\in\{1,\dots,m\}$, we write $F(\rho_i,\psi_i)=1-\epsilon_i$, where $F$ is the fidelity. Then,
\be
1-\sum_{i=1}^m{\epsilon_i}\le F(\bm\rho,\psi_1\otimes\dots\otimes\psi_m)\le\prod_{i=1}^m{(1-\epsilon_i)}.
\label{fideproduct}
\ee
In particular, when $\epsilon_1=\dots=\epsilon_m=\epsilon$, 
\be
1-m\epsilon\le F(\rho,\psi_1\otimes\dots\otimes\psi_m)\le(1-\epsilon)^m.
\label{fideproductequal}
\ee
\end{lem}

\begin{proof}
\begin{mdframed}[linewidth=1.5,topline=false,rightline=false,bottomline=false]

Since $\ket{\psi_1},\dots,\ket{\psi_m}$ are pure states,
\be
F(\bm\rho,\psi_1\otimes\dots\otimes\psi_m)=\Tr\,[\bm\rho\ket{\psi_1}\!\bra{\psi_1}\otimes\dots\otimes\ket{\psi_m}\!\bra{\psi_m}],
\label{fidepure1}
\ee
and
\be
\ba
F(\bm\rho_i,\psi_i)&=\Tr[\bm\rho_i\ket{\psi_i}\!\bra{\psi_i}]\\
&=\Tr[\bm\rho\,\mathbb1_{i-1}\otimes\ket{\psi_i}\!\bra{\psi_i}\otimes\mathbb1_{m-i}]
\ea
\label{fidepure2}
\ee
for all $i\in\{1,\dots,m\}$.
The left hand side of Eq.~(\ref{fideproduct}) is obtained by writing $F(\bm\rho,\psi_1\otimes\dots\otimes\psi_m)$ as a telescopic sum:
\be
\ba
\Tr\,[\bm\rho\ket{\psi_1}\!\bra{\psi_1}\otimes\dots\otimes\ket{\psi_m}\!\bra{\psi_m}]&=\Tr\,[\bm\rho\,\mathbb1_m]\\
&\quad-\Tr\,[\bm\rho(\mathbb1-\ket{\psi_1}\!\bra{\psi_1})\otimes\mathbb1_{m-1}]\\
&\quad-\Tr\,[\bm\rho\ket{\psi_1}\!\bra{\psi_1}\otimes(\mathbb1-\ket{\psi_2}\!\bra{\psi_2})\otimes\mathbb1_{m-2}]\\
&\quad-\Tr\,[\bm\rho\ket{\psi_1}\!\bra{\psi_1}\otimes\ket{\psi_2}\!\bra{\psi_2}\otimes(\mathbb1-\ket{\psi_3}\!\bra{\psi_3})\otimes\mathbb1_{m-3}]\\
&\quad-\dots\\
&\quad-\Tr\,[\bm\rho\ket{\psi_1}\!\bra{\psi_1}\otimes\ket{\psi_2}\!\bra{\psi_2}\otimes\dots\otimes(\mathbb1-\ket{\psi_m}\!\bra{\psi_m})]\\
&\ge1-\sum_{i=1}^m{\left(1-\Tr[\bm\rho\,\mathbb1_{i-1}\otimes\ket{\psi_i}\!\bra{\psi_i}\otimes\mathbb1_{m-i}]\right)},
\ea
\ee
by linearity of the trace, where we used $\Tr\,(\bm\rho)=1$. This gives
\be
F(\bm\rho,\psi_1\otimes\dots\otimes\psi_m)\ge1-\sum_{i=1}^m{\left(1-F(\rho_i,\psi_i)\right)},
\label{lhsprod}
\ee
with Eqs.~(\ref{fidepure1}) and (\ref{fidepure2}).

\medskip

\noindent The right hand side of Eq.~(\ref{fideproduct}) is obtained by Cauchy-Schwarz inequality and a simple induction:
\be
\ba
\Tr\,[\bm\rho\ket{\psi_1}\!\bra{\psi_1}\otimes\dots\otimes\ket{\psi_m}\!\bra{\psi_m}]&=\Tr\left[\left(\sqrt{\bm\rho}\ket{\psi_1}\!\bra{\psi_1}\otimes\mathbb1_{m-1}\right)\left(\mathbb1\otimes\ket{\psi_2}\!\bra{\psi_2}\otimes\dots\otimes\ket{\psi_m}\!\bra{\psi_m}\sqrt{\bm\rho}\right)\right]\\
&\le\Tr\left[\bm\rho\ket{\psi_1}\!\bra{\psi_1}\otimes\mathbb1_{m-1}\right]\Tr\left[\bm\rho\,\mathbb1\otimes\ket{\psi_2}\!\bra{\psi_2}\otimes\dots\otimes\ket{\psi_m}\!\bra{\psi_m}\right]\\
&\le\dots\\
&\le\prod_{i=1}^m{\Tr\,[\bm\rho\,\mathbb1_{i-1}\otimes\ket{\psi_i}\!\bra{\psi_i}\otimes\mathbb1_{m-i}]},
\ea
\ee
where we used the cyclicity of the trace and the fact that $\ket{\psi_1},\dots,\ket{\psi_m}$ are pure states. This gives
\be
F(\bm\rho,\psi_1\otimes\dots\otimes\psi_m)\le\prod_{i=1}^m{F(\rho_i,\psi_i)},
\label{rhsprod}
\ee
with Eqs.~(\ref{fidepure1}) and (\ref{fidepure2}).

\medskip

\noindent Writing $F(\rho_i,\psi_i)=1-\epsilon_i$, we obtain, with Eqs.~(\ref{lhsprod}) and~(\ref{rhsprod}),
\be
1-\sum_{i=1}^m{\epsilon_i}\le F(\bm\rho,\psi_1\otimes\dots\otimes\psi_m)\le\prod_{i=1}^m{(1-\epsilon_i)},
\ee
which concludes the proof.
Additionnally, by the inequality of arithmetic and geometric means,
\begin{align}
\prod_{i=1}^m{(1-\epsilon_i)}&\le\left(1-\frac1m\sum_{i=1}^m{\epsilon_i}\right)^m\displaybreak\\
\nonumber&\le\exp\left(-\sum_{i=1}^m{\epsilon_i}\right),
\end{align}
which gives a looser bound in terms of the total single-mode deviation $\sum_{i=1}^m{\epsilon_i}$:
\be
1-\sum_{i=1}^m{\epsilon_i}\le F(\bm\rho,\psi_1\otimes\dots\otimes\psi_m)\le\exp\left(-\sum_{i=1}^m{\epsilon_i}\right).
\ee
\end{mdframed}
\end{proof}

\noindent Note that Eq.~(\ref{fideproductequal}) is tight for small $\epsilon$, since its right hand side is then equivalent to $1-m\epsilon$. Lemma~\ref{lem:product} implies that if the fidelities of single-mode subsystems of an $m$-mode quantum state with some target pure states are higher than $1-\frac\lambda m$, for some $\lambda>0$, then the $m$-mode state has fidelity at least $1-\lambda$ with the target $m$-mode product state.

Together with the union bound and the single-mode certification and verification protocols from Theorems~\ref{thi.i.d.} and~\ref{thVUCVQC}, this provides a means for obtaining efficiently tight fidelity witnesses with any target tensor product of single-mode pure states with analytical confidence intervals, with and without i.i.d.\@ assumption.

\medskip

\noindent At this point, we can obtain fidelity witnesses only for pure product states, with no entanglement, using a fidelity estimation protocol for each of the single-mode subsystems in parallel. 
We make use of the properties of heterodyne detection in order to extend the class of target states for which fidelity witnesses can be efficiently obtained, from pure product states to the multimode states that are obtained from a pure product state with a passive linear transformation followed by single-mode Gaussian unitary operations, as in Eq.~(\ref{class}).

The POVM elements of product single-mode unbalanced heterodyne detection over $m$ modes with unbalancing parameters $\bm\xi\in\mathbb C^m$ are given by (see section~\ref{sec:heterodynemeasurement})
\be
\Pi_{\bm\alpha}^{\bm\xi}=\frac1{\pi^m}\ket{\bm\alpha,\bm\xi}\!\bra{\bm\alpha,\bm\xi},
\ee
for all $\bm\alpha=(\alpha_1,\dots,\alpha_m)\in\mathbb C^m$, where $\ket{\bm\alpha,\bm\xi}=\bigotimes_{i=1}^m{\ket{\alpha_i,\xi_i}}$ is a product of squeezed coherent states $\hat S(\xi_i)\hat D(\alpha_i)\ket0$.

The POVM elements of product single-mode balanced heterodyne detection are given by $\Pi_{\bm\alpha}^{\bm0}$, for all $\bm\alpha\in\mathbb C^m$, and we have $\Pi_{\bm\alpha}^{\bm\xi}=\hat S(\bm\xi)\Pi_{\bm\alpha}^{\bm0}\hat S^\dag(\bm\xi)$. In particular, a single-mode squeezing followed by a single-mode balanced heterodyne detection can be simulated by performing directly an unbalanced heterodyne detection according to the squeezing parameter. One retrieves balanced heterodyne detection by setting the unbalancing parameter to $0$ and homodyne detection by letting the modulus of the unbalancing parameter go to infinity.

Passive linear transformations correspond to unitary transformations of the creation and annihilation operators of the modes. These transformations, which may be implemented by unitary optical interferometers, map coherent states to coherent states: if $\hat U$ is a passive linear transformation and $U$ is the unitary matrix describing its action on the creation and annihilation operators of the modes, an input coherent state $\ket{\bm\alpha}$ is mapped to an output coherent state $\hat U\ket{\bm\alpha}=\ket{U\bm\alpha}$, where $U\bm\alpha$ is obtained by multiplying the vector $\bm\alpha$ by the unitary matrix $U$. Hence, the POVM elements corresponding to a passive linear transformation $\hat U$ followed by a product of single-mode balanced heterodyne detection are given by $\hat U\Pi^{\bm0}_{\bm\alpha}\hat U^\dag=\Pi^{\bm0}_{U\bm\alpha}$, for all $\bm\alpha\in\mathbb C^m$. This implies that the passive linear transformation $\hat U^\dag$ followed by a product of single-mode heterodyne detections can be simulated by performing the heterodyne detections first, then multiplying the vector of samples obtained by $U$.

A similar property holds with single-mode displacements: since displacements map coherent states to coherent states, up to a global phase, by displacing their amplitude, a single-mode displacement followed by a single-mode heterodyne detection can be simulated by performing the heterodyne detection first, then translating the sample obtained according to the displacement amplitude. In particular we have $\hat D(\bm\beta)\Pi_{\bm\alpha}^{\bm0}\hat D^\dag(\bm\beta)=\Pi_{\bm\alpha+\bm\beta}^{\bm0}$ for all $\bm\alpha,\bm\beta\in\mathbb C^m$, where $\bm\alpha+\bm\beta=(\alpha_1+\beta_1,\dots,\alpha_m+\beta_m)$.

Combining the properties of heterodyne detection we obtain the following result:

\begin{lem}\label{lem:hetmagic} Let $\bm\beta,\bm\xi\in\mathbb C^m$ and let $\hat V=\hat S(\bm\xi)\hat D(\bm\beta)\,\hat U$, where $\hat U$ is an $m$-mode passive linear transformation with $m\times m$ unitary matrix $U$. For all $\bm\alpha\in\mathbb C^m$, let $\bm\gamma=U\bm\alpha+\bm\beta$. Then,
\be
\Pi^{\bm\xi}_{\bm\gamma}=\hat V\Pi^{\bm0}_{\bm\alpha}\hat V^\dag.
\ee
\end{lem}

\begin{proof}
\begin{mdframed}[linewidth=1.5,topline=false,rightline=false,bottomline=false]

\noindent We have $\Pi_{\bm\alpha}^{\bm\xi}=\frac1{\pi^m}\ket{\bm\alpha,\bm\xi}\!\bra{\bm\alpha,\bm\xi}$, for all $\bm\alpha,\bm\xi\in\mathbb C^m$, where $\ket{\bm\alpha,\bm\xi}=\hat S(\bm\xi)\hat D(\bm\alpha)\ket{\bm0}$ is a tensor product of squeezed coherent states. We also have
\be
\begin{cases}
\hat U\Pi^{\bm0}_{\bm\alpha}\hat U^\dag=\Pi^{\bm0}_{U\bm\alpha},\\
\hat D(\bm\beta)\Pi_{\bm\alpha}^{\bm0}\hat D^\dag(\bm\beta)=\Pi_{\bm\alpha+\bm\beta}^{\bm0},\\
\hat S(\bm\xi)\Pi_{\bm\alpha}^{\bm0}\hat S^\dag(\bm\xi)=\Pi_{\bm\alpha}^{\bm\xi},
\end{cases}
\ee
for all $\bm\alpha,\bm\beta,\bm\xi\in\mathbb C^m$ and all $m$-mode passive linear transformations $\hat U$ with $m\times m$ unitary matrix $U$. Writing $\hat V=\hat S(\bm\xi)\hat D(\bm\beta)\hat U$, we obtain
\be
\ba
\hat V\Pi_{\bm\alpha}^{\bm0}\hat V^\dag&=\hat S(\bm\xi)\hat D(\bm\beta)\hat U\Pi_{\bm\alpha}^{\bm0}\hat U^\dag\hat D^\dag(\bm\beta)\hat S^\dag(\bm\xi)\\
&=\hat S(\bm\xi)\hat D(\bm\beta)\Pi_{U\bm\alpha}^{\bm0}\hat D^\dag(\bm\beta)\hat S^\dag(\bm\xi)\\
&=\hat S(\bm\xi)\Pi_{U\bm\alpha+\bm\beta}^{\bm0}\hat S^\dag(\bm\xi)\\
&=\Pi_{U\bm\alpha+\bm\beta}^{\bm\xi}.
\ea
\ee

\end{mdframed}
\end{proof}

\noindent Lemma~\ref{lem:hetmagic} implies that the POVM $\{\hat V\Pi^{\bm0}_{\bm\alpha}\hat V^\dag\}_{\bm\alpha\in\mathbb C^m}$ can be simulated with the POVM $\{\Pi^{\bm\xi}_{\bm\gamma}\}_{\bm\gamma\in\mathbb C^m}$ by computing $\bm\alpha=U^\dag(\bm\gamma-\bm\beta)$, i.e., translating the vector of samples $\bm\gamma$ by the vector of complex amplitudes $-\bm\beta$ and multiplying the vector obtained by the $m\times m$ unitary matrix $U^\dag$.
This means that a passive linear transformation followed by single-mode Gaussian unitary operations before balanced heterodyne detection can be simulated by performing unbalanced heterodyne detection directly, then post-processing efficiently the classical outcomes. In particular, for such a transformation $\hat V$, if a multimode pure product state $\bigotimes_{i=1}^m{\ket{\psi_i}}$ can be efficiently verified using balanced heterodyne detection, then the state $\hat V(\bigotimes_{i=1}^m{\ket{\psi_i}})$ can be efficiently verified using unbalanced heterodyne detection.

Formally, let $\rho$ be an $m$-mode (mixed) state. Let $\ket{\psi_1},\dots,\ket{\psi_m}$ be single-mode pure states and let $\hat V=\hat S(\bm\xi)\hat D(\bm\beta)\hat U$, with $\bm\beta,\bm\xi\in\mathbb C^m$, where $\hat U$ is a passive linear transformation over $m$ modes with an associated $m\times m$ unitary matrix $U$. Then,
\be
\ba
F(\bm\rho,\hat V\ket{\psi_1}\!\bra{\psi_1}\otimes\dots\otimes\ket{\psi_m}\!\bra{\psi_m}\hat V^\dag)&=F(\hat V^\dag\bm\rho\hat V,\ket{\psi_1}\!\bra{\psi_1}\otimes\dots\otimes\ket{\psi_m}\!\bra{\psi_m})\\
&\ge1-\sum_{i=1}^m{\left(1-F(\ket{\psi_i}\!\bra{\psi_i},(\hat V^\dag\bm\rho\hat V)_i)\right)},
\ea
\ee
where we have used Lemma~\ref{lem:product} and where $(\hat V^\dag\bm\rho\hat V)_i$ is the $i^{th}$ single-mode reduced density matrix of the state $\hat V^\dag\bm\rho\hat V$. 

The single-mode fidelities $F(\ket{\psi_i}\!\bra{\psi_i},(\hat V^\dag\bm\rho\hat V)_i)$ can be estimated with analytical confidence intervals by measuring multiple copies of the $m$-mode state $\hat V^\dag\bm\rho\hat V$ with product balanced heterodyne detection and post-processing the samples for individual subsystems according to the protocols from Theorems~\ref{thi.i.d.} and~\ref{thVUCVQC}. By Lemma~\ref{lem:hetmagic}, this is equivalent to measuring the state $\bm\rho$ directly with a product of single-mode unbalanced heterodyne detections with unbalancing parameters $\bm\xi$, translating the vector of samples $\bm\gamma$ obtained by the vector of complex amplitudes $-\bm\beta$ and multiplying the vector obtained by the unitary matrix $U^\dag$. Then, the obtained samples may be post-processed according to the heterodyne certification or verification protocols.

If all the single-mode fidelity estimates obtained are precise to $\frac1{\poly m}$ and greater than $1-\frac1{\poly m}$ with high probability, which can be checked in time $O(\poly m)$, then with the union bound for the failure probabilities, the fidelity between the $m$-mode state $\bm\rho$ and the target state $\hat V\ket{\psi_1}\!\bra{\psi_1}\otimes\dots\otimes\ket{\psi_m}\!\bra{\psi_m}\hat V^\dag$ is greater than $1-\frac1{\poly m}$, with high probability. Hence the single-mode fidelity estimation protocols give a verification protocol for obtaining tight multimode fidelity witnesses, under or without the i.i.d.\@ assumption. 

The single-mode protocols from Theorems~\ref{thi.i.d.} and~\ref{thVUCVQC} are efficient as long as the energy of the single-mode target pure state is constant, i.e., it does not scale with the number of modes. Note that additional displacements may be introduced to reduce the energy of the single-mode target pure states, since by modifying their amplitudes these displacements can be braided through the transformation $\hat V$ and accounted for by translating the heterodyne detection samples. The efficiently verifiable states thus are the pure states of the form $\hat S(\bm\xi)\hat D(\bm\beta)\,\hat U(\bigotimes_{i=1}^m\ket{\psi_i})$, such that for all $i\in\{1,\dots,m\}$, the state $\ket{\psi_i}$ can be displaced onto a state of constant energy. 

In particular, multimode Gaussian pure states with constant squeezing parameter can be efficiently verified, since these can be written as a product of pure single-mode squeezed coherent states followed by a passive linear transformation (see section~\ref{sec:Gaussian}). Note however that under the i.i.d.\@ assumption the witnesses from~\cite{aolita2015reliable} may provide a more efficient certification method for Gaussian states.

Remarkably, the class of efficiently verifiable states also includes the output states of CVS circuits and Boson Sampling interferometers. Our verification protocol may thus be used to verify quantum supremacy, as we detail in the following section.

\subsection{Quantum supremacy with Boson Sampling: from validation to verification}

The experimental demonstration of quantum computational supremacy is regarded as an important milestone in the field of quantum information. It involves a quantum device solving efficiently a computational task which is provably hard for classical computers, together with a verification of its correct functionning~\cite{harrow2017quantum}. While the former has been recently accomplished with superconducting circuits~\cite{arute2019quantum}, the latter is still partial or relying on various computational assumptions. 

Demonstrating quantum supremacy is inherently difficult because the computational task at hand is a sampling task from an anti-concentrating probability distribution over an exponential sample space. For that reason, direct non-interactive verification of quantum computational supremacy with a verifier restricted to classical computations is impossible~\cite{hangleiter2019sample}. 
Possible verification with a classical verifier includes interactive protocols with additional computational assumptions~\cite{regev2009lattices,aaronson2016complexity,mahadev2018classical}, or partial verification, which ultimately relies on making assumptions about the inner functionning of the quantum device.

If one is reluctant to rely on additional assumptions, another way for performing verification is to allow the verifier to have quantum capabilities. However, the computational power of the verifier needs to be as small as possible, as it would not make sense if the verifier had enough computational power to perform the sampling task directly. 
In the context of discrete variable quantum computing, a minimal quantum capability would correspond to being able to prepare only single-qubit states or to perform only simple local measurements. For example, protocols for verification of IQP circuits~\cite{Shepherd2009} with these minimal requirements have been derived under the i.i.d.\@ assumption with single-qubit states~\cite{mills2017information} or with local measurements~\cite{Hangleiter2016} and more recently without the i.i.d.\@ assumption with single-qubit states~\cite{kapourniotis2019nonadaptive} or with local measurements~\cite{takeuchi2018verification}.

In the context of continuous variable quantum computing, this minimal quantum capability would corresponds to being able to prepare only single-mode Gaussian states, or to perform only single-mode Gaussian measurements. An efficient certification protocol exists for verifying multimode Gaussian states~\cite{aolita2015reliable} and thus instances of Gaussian Boson Sampling~\cite{Hamilton2016} with single-mode Gaussian measurements under the i.i.d.\@ assumption~\cite{abrahao2018continuous}.
However, there is no efficient certification nor verification protocol using single-mode Gaussian measurements for Boson Sampling with input single photons: current methods used for validation of Boson sampling are either not scalable, e.g., computing the total variation distance with the ideal probability distribution, or else provide incomplete certificates, e.g., telling apart the tested distribution from classical mock-up distributions such as the uniform distribution~\cite{brod2019photonic,wang2019boson}.

When introducing the Boson Sampling model, Aaronson and Arkhipov importantly showed that even an approximate version of Boson Sampling is hard to sample for classical computers, provided two conjectures on the permanent of random Gaussian matrices hold true (see section~\ref{sec:BosonSampling} and~\cite{Aaronson2013}). More precisely, they showed under these conjectures that sampling from a probability distribution that has small constant total variation distance with an ideal Boson Sampling distribution is classically hard in the so-called antibunching regime $n=O(\sqrt m)$. In particular, verifying a Boson Sampling quantum supremacy experiment amounts to verifying that the experimental quantum device samples from an ideal probability distribution, up to a constant error in total variation distance.

Our verification protocol derived in the previous section can be applied to check efficiently the fidelity of the output state of an experimental Boson Sampling interferometer with the ideal output state, using only balanced heterodyne detection. The fidelity witness gives in turn a certificate of the total variation distance with the ideal probability distribution for any observable by Eq.~(\ref{tvdbound}), therefore allowing for an experimental demonstration of quantum supremacy with Boson Sampling, with a verifier having minimal continuous variable quantum computational power, namely the ability to perform single-mode Gaussian measurements.

Performing verified Boson Sampling with our protocol, even under i.i.d.\@ assumption, would already provide a convincing evidence of quantum supremacy with photonic quantum computing, as the verification without i.i.d.\@ assumption only comes at the cost of an increased number of measurements, still polynomial in the number of modes $m$.

To that end, we optimise the bounds for the multimode certification protocol under i.i.d.\@ assumption. In particular, the support estimation step in the protocol is no longer necessary, because the single-mode target pure states are either single-photon Fock states or vacuum states and thus have finite support over the Fock basis. We show that the number of copies needed for a constant additive precision in the antibunching regime $n=O(\sqrt m)$---which is required for a demonstration of quantum supremacy---scales as $O(m^4\log m)$, making reliable verification of Boson Sampling using single-mode Gaussian measurements within the reach of current experiments. Remarkably, this is only a logarithmic factor harder than verifying multimode Gaussian states~\cite{aolita2015reliable}.

\medskip

\noindent Let $0<\eta<2/3$ and define, for all $z\in\mathbb C$,
\be
f_0(z,\eta)=\frac1\eta\exp\left[\left(1-\frac1\eta\right)|z|^2\right],
\ee
and
\be
f_1(z,\eta)=\frac1{\eta^2}(\frac{|z|^2}\eta-1)\exp\left[\left(1-\frac1\eta\right)|z|^2\right].
\ee
The verification protocol for Boson Sampling with $n$ photons fed into a unitary interferometer $U$ of size $m$ under i.i.d.\@ assumption reads:

\begin{enumerate}
\item
The verifier chooses two precision parameters $0<\eta_0,\eta_1<2/3$.
\item
The verifier asks the prover for $N=O(m^4\log m)$ copies of the target state $U\ket{1\dots1\,0\dots0}$. Let $\bm\rho^{\otimes N}$ be the $(N\times m)$-mode (mixed) state sent by the prover, where $\bm\rho$ is an $m$-mode (mixed) state.
\item
The verifier measures with balanced heterodyne detection all the $m$ subsystems of all the $N$ copies of $\bm\rho$, obtaining the $N$ vectors of samples $\bm\gamma^{(1)},\dots,\bm\gamma^{(N)}\in\mathbb C^m$. 
\item
For all $k\in\{1,\dots,N\}$, the verifier computes $\bm\alpha^{(k)}=U^\dag\bm\gamma^{(k)}$. We write $\bm\alpha^{(k)}=(\alpha_1^{(k)},\dots,\alpha_m^{(k)})$.
\item
For all $i\in\{n+1,\dots,m\}$ the verifier computes the mean $\tilde F_i$ of the function $z\mapsto f_0(z,\eta_0)$ over the values $\alpha_i^{(1)},\dots,\alpha_i^{(N)}$ and for all $j\in\{1,\dots,n\}$ the mean $\tilde F_j$ of the function $z\mapsto f_1(z,\eta_1)$. 
\item
The verifier computes $\tilde W=1-\sum_{i=1}^m(1-\tilde F_i)$.
\end{enumerate}

\begin{theo}[Certification of Boson Sampling using Gaussian measurements]\label{th:verifBS}
$\tilde W$ is an estimate with constant precision of a tight lower bound on the fidelity with the ideal Boson Sampling output state, with probability exponentially close to $1$.
\end{theo}

\noindent The estimate $\tilde W$ thus provides an efficient and reliable certificate of the total variation distance with the ideal probability distribution for any observable by Eq.~(\ref{tvd}).
 
\begin{proof}
\begin{mdframed}[linewidth=1.5,topline=false,rightline=false,bottomline=false]

\noindent Let $0<\eta<2/3$. By Lemma~\ref{lem:Eflk} we have, for any single-mode mixed state $\rho=\sum_{k,l\ge0}{\rho_{kl}\ket k\!\bra l}$,
\be
\underset{\alpha\leftarrow Q_\rho(\alpha)}{\mathbb E}[f_0(\alpha,\eta)]=\Tr\,(\rho\ket0\!\bra0)+\eta\sum_{n=0}^{+\infty}{\eta^n\rho_{n+1,n+1}},
\label{boundEf0}
\ee
and
\be
\underset{\alpha\leftarrow Q_\rho}{\mathbb E}[f_1(\alpha,\eta)]=\Tr\,(\rho\ket1\!\bra1)+\eta\sum_{n=0}^{+\infty}{\eta^n(n+2)\rho_{n+2,n+2}},
\label{boundEf1}
\ee
where $\underset{\alpha\leftarrow Q_\rho}{\mathbb E}[f]$ denotes the expected value of the function $f$ for samples from single-mode balanced heterodyne detection of $\rho$. 

Since $\eta\le2/3$ we have $\eta^{n+1}<\eta^n$ and $\eta^{n+1}(n+3)<\eta^n(n+2)$ for all $n\in\mathbb N$, so by a simple induction $\eta^n\le1$ and $\eta^n(n+2)\le2$, for all $n\in\mathbb N$. In particular, $\sum_{n=0}^{+\infty}{\eta^n\rho_{n+1,n+1}}\le1$ and $\sum_{n=0}^{+\infty}{\eta^n(n+2)\rho_{n+2,n+2}}\le2$, since $\Tr\,(\rho)=1$. With Eqs.~(\ref{boundEf0}) and~(\ref{boundEf1}) we obtain
\be
\left|\Tr\,(\rho\ket0\!\bra0)-\underset{\alpha\leftarrow Q_\rho}{\mathbb E}[f_0(\alpha,\eta)]\right|\le\eta,
\label{approx0}
\ee
and
\be
\left|\Tr\,(\rho\ket1\!\bra1)-\underset{\alpha\leftarrow Q_\rho}{\mathbb E}[f_1(\alpha,\eta)]\right|\le2\eta.
\label{approx1}
\ee

We also have, for all $z\in\mathbb C$, 
\be
0<f_0(z,\eta)\le\frac1\eta,
\label{rangef0}
\ee
and
\be
-\frac1{\eta^2}\le f_1(z,\eta)\le\frac{e^{\eta-2}}{\eta^2(1-\eta)}.
\label{rangef1}
\ee
For $\eta<2/3$, we have $\frac{e^{\eta-2}}{(1-\eta)}<1$. In particular, the range of the function $f_1$ is less than $\frac2{\eta^2}$.

Let $N\in\mathbb N^*$, let $\alpha_1,\dots,\alpha_N$ be i.i.d.\@ samples from single-mode balanced heterodyne detection of a single mode state $\rho_0$ and let $\beta_1,\dots,\beta_N$ be i.i.d.\@ samples from single-mode balanced heterodyne detection of a single mode state $\rho_1$. Let $\epsilon_0,\eta_0,\epsilon_1,\eta_1>0$, by Hoeffding inequality,
\be
\Pr\left[\left|\frac1N\sum_{p=1}^N{f_0(\alpha_p,\eta_0)}-\underset{\alpha\leftarrow Q_{\rho_0}}{\mathbb E}[f_0(\alpha,\eta_0)]\right|\ge\epsilon_0\right]\le2e^{-2N\epsilon_0^2\eta_0^2},
\label{Hoef0}
\ee
and
\be
\Pr\left[\left|\frac1N\sum_{p=1}^N{f_1(\beta_p,\eta_1)}-\underset{\beta\leftarrow Q_{\rho_1}}{\mathbb E}[f_1(\beta,\eta_1)]\right|\ge\epsilon_1\right]\le2e^{-\frac{N\epsilon_1^2\eta_1^4}2}.
\label{Hoef1}
\ee
Let $\bm\sigma$ be an $m$-mode state and let $\sigma_i$ denote its $k^{th}$ single-mode subsystem for all $k\in\{1,\dots,m\}$. Let $\bm\alpha^{(1)},\dots,\bm\alpha^{(N)}\in\mathbb C^m$ be samples from product balanced heterodyne detection of $N$ identical copies of the state $\bm\sigma$. For all $p\in\{1,\dots,N\}$, we write $\bm\alpha_p=(\alpha_1^{(p)},\dots,\alpha_m^{(p)})$. 

Combining Eqs.~(\ref{approx0}) and~(\ref{Hoef0}) we obtain, for any $i\in\{n+1,\dots,m\}$,
\be
\left|\Tr\,(\sigma_i\ket0\!\bra0)-\frac1N\sum_{p=1}^N{f_0\left(\alpha_i^{(p)},\eta_0\right)}\right|\le\epsilon_0+\eta_0,
\label{estimate0}
\ee
with probability greater than $1-2\exp\left[-N\epsilon_0^2\eta_0^2\right]$. Similarly, combining Eqs.~(\ref{approx1}) and~(\ref{Hoef1}) we obtain, for any $j\in\{1,\dots,n\}$,
\be
\left|\Tr\,(\sigma_j\ket1\!\bra1)-\frac1N\sum_{p=1}^N{f_1\left(\alpha_j^{(p)},\eta_1\right)}\right|\le\epsilon_1+2\eta_1,
\label{estimate1}
\ee
with probability greater than $1-2\exp\left[-N\epsilon_1^2\eta_1^4/2\right]$. 

We now choose $\epsilon_0,\eta_0,\epsilon_1,\eta_1$ in order to minimize the error probabilities for a given precision.
Let $\lambda_0>0$. Setting $\epsilon_0+\eta_0=\lambda_0$, the optimal choice, which maximises $\epsilon_0^2\eta_0^2$, is $\epsilon_0=\eta_0=\frac{\lambda_0}2$ and with Eq.~(\ref{estimate0}) we obtain, for any $i\in\{n+1,\dots,m\}$,
\be
\left|\Tr\,(\sigma_i\ket0\!\bra0)-\frac1N\sum_{p=1}^N{f_0\left(\alpha_i^{(p)},\frac{\lambda_0}2\right)}\right|\le\lambda_0,
\label{estimateopti0}
\ee
with probability greater than $1-2\exp\left[-\frac{N\lambda_0^4}8\right]$. 

Let $\lambda_1>0$. Setting $\epsilon_1+2\eta_1=\lambda_1$, the optimal choice, which maximises $\epsilon_1^2\eta_1^4$, is $\epsilon_1=\eta_1=\frac{\lambda_1}3$ and with Eq.~(\ref{estimate1}) we obtain, for any $j\in\{1,\dots,n\}$,
\be
\left|\Tr\,(\sigma_j\ket1\!\bra1)-\frac1N\sum_{p=1}^N{f_1\left(\alpha_j^{(p)},\frac{\lambda_1}{3m}\right)}\right|\le\lambda_1,
\label{estimateopti1}
\ee
with probability greater than $1-2\exp\left[-\frac{N\lambda_1^6}{1458}\right]$.

\medskip

Let us define the fidelity witness
\be
W:=1-\left(\sum_{k=n+1}^m{1-F(\sigma_k,\ket0\!\bra0)}+\sum_{l=1}^n{1-F(\sigma_k,\ket1\!\bra1)}\right),
\ee
as in Lemma~\ref{lem:product}, and the witness estimate
\be
\tilde W(\lambda_0,\lambda_1):=1-\left(\sum_{i=n+1}^m{\left[1-\frac1N\sum_{p=1}^N{f_0\left(\alpha_i^{(p)},\frac{\lambda_0}2\right)}\right]}+\sum_{j=1}^n{\left[1-\frac1N\sum_{p=1}^N{f_1\left(\alpha_j^{(p)},\frac{\lambda_1}{3m}\right)}\right]}\right).
\ee
Taking the union bound of the failure probabilities for $i\in\{n+1,\dots,m\}$ and $j\in\{1,\dots,n\}$, we obtain with Eqs.~(\ref{estimateopti0}) and (\ref{estimateopti1}),
\be
\left|W-\tilde W(\lambda_0,\lambda_1)\right|\le(m-n)\lambda_0+n\lambda_1,
\label{witnessopti}
\ee
with probability greater than
\be
1-2\left((m-n)\exp\left[-\frac{N\lambda_0^4}8\right]+n\exp\left[-\frac{N\lambda_1^6}{1458}\right]\right).
\ee
By Lemma~\ref{lem:product} we have
\be
F(\bm\sigma,\ket{1\dots1\,0\dots0}\!\bra{1\dots1\,0\dots0})\ge W,
\ee
hence with Eq.~(\ref{witnessopti}) we obtain
\be
F(\bm\sigma,\ket{1\dots1\,0\dots0}\!\bra{1\dots1\,0\dots0})\ge\tilde W(\lambda_0,\lambda_1)-(m-n)\lambda_0-n\lambda_1,
\label{Fsigma}
\ee
with probability greater than
\be
1-2\left((m-n)\exp\left[-\frac{N\lambda_0^4}8\right]+n\exp\left[-\frac{N\lambda_1^6}{1458}\right]\right).
\label{Psigma}
\ee
Let $\bm\rho=\hat U^\dag\bm\sigma\hat U$, where $\hat U$ is an $m$-mode passive linear transformation with $m\times m$ unitary matrix $U$. By Lemma~\ref{lem:hetmagic} from the main text, the estimate $\tilde W$ can be computed using samples of product balanced heterodyne detection of $\bm\rho$ multiplied by the unitary matrix $U^\dag$, rather than samples from the balanced heterodyne detection of $\bm\sigma$. 

\noindent With Eqs.~(\ref{Fsigma}) and~(\ref{Psigma}) we obtain
\be
F(\bm\rho,\hat U\ket{1\dots1\,0\dots0}\!\bra{1\dots1\,0\dots0}\hat U^\dag)\ge\tilde W(\lambda_0,\lambda_1)-(m-n)\lambda_0-n\lambda_1,
\ee
with probability greater than
\be
1-2\left((m-n)\exp\left[-\frac{N\lambda_0^4}8\right]+n\exp\left[-\frac{N\lambda_1^6}{1458}\right]\right),
\ee
where $\tilde W(\lambda_0,\lambda_1)$ is computed using samples of product balanced heterodyne detection of $N$ copies of the $m$-mode state $\rho$, each of the $N$ vectors of samples being multiplied by the unitary matrix $U^\dag$.

\medskip 

The values of $\lambda_0$ and $\lambda_1$ must be chosen by the verifier to maximise the above probability for a given precision of the witness. Equivalently, one needs to minimise:
\be
(m-n)\exp\left[-\frac{N\lambda_0^4}8\right]+n\exp\left[-\frac{N\lambda_1^6}{1458}\right],
\label{optithis}
\ee
with the constraint $(m-n)\lambda_0+n\lambda_1=\epsilon$, for $\epsilon>0$. For a given experimental setup, Eq.~(\ref{optithis}) should be minimised depending on the values of $m$ and $n$.

For example, setting $\lambda_0=\frac\epsilon{2(m-n)}$ and $\lambda_1=\frac\epsilon{2n}$ gives
\be
F(\bm\rho,\hat U\ket{1\dots1\,0\dots0}\!\bra{1\dots1\,0\dots0}\hat U^\dag)\ge\tilde W\left(\frac\epsilon{2(m-n)},\frac\epsilon{2n}\right)-\epsilon,
\ee
with probability greater than
\be
1-2\left((m-n)\exp\left[-\frac{2N\epsilon^4}{[4(m-n)]^4}\right]+n\exp\left[-\frac{N\epsilon^6}{2(6n)^6}\right]\right),
\ee
and the estimate $\tilde W$ is $\epsilon$-close to the actual fidelity witness, which by Lemma~\ref{lem:product} is a tight witness of the fidelity. In particular, for a constant precision fidelity witness $W$, the estimate $\tilde W$ yields a constant precision fidelity witness with probability exponentially close (in $m$) to $1$ for $N=O(\max\{(m-n)^4\log(m-n),n^6\log n\})$.
In the antibunching regime $n=O(\sqrt m)$, this means that the estimate has constant precision with exponentially small failure probability already for $N=O(m^4\log m)$.

\end{mdframed}
\end{proof}

\noindent A similar Boson Sampling verification protocol without i.i.d.\@ assumption is obtained by using the version of the protocol which does not assume i.i.d.\@ state preparation for estimating the single-mode fidelities in Theorem~\ref{thVUCVQC}. This comes at the cost of an increased number of samples necessary for the same witness precision and confidence interval, which corresponds to a polynomial overhead in $m$ and a slightly different classical post-processing: a small fraction of the measured subsystems have to be discarded at random and an additional support estimation step is necessary, for which the samples must be randomly chosen and cannot be used to compute the fidelity estimates. More precisely, changing the parameters of the protocol above to those of Theorem~\ref{thVUCVQC}, we get that Theorem~\ref{th:verifBS} holds without the i.i.d.\@ assumption with polynomial confidence.\\

\begin{figure}[h!]
	\begin{center}
		\includegraphics[width=0.4\columnwidth]{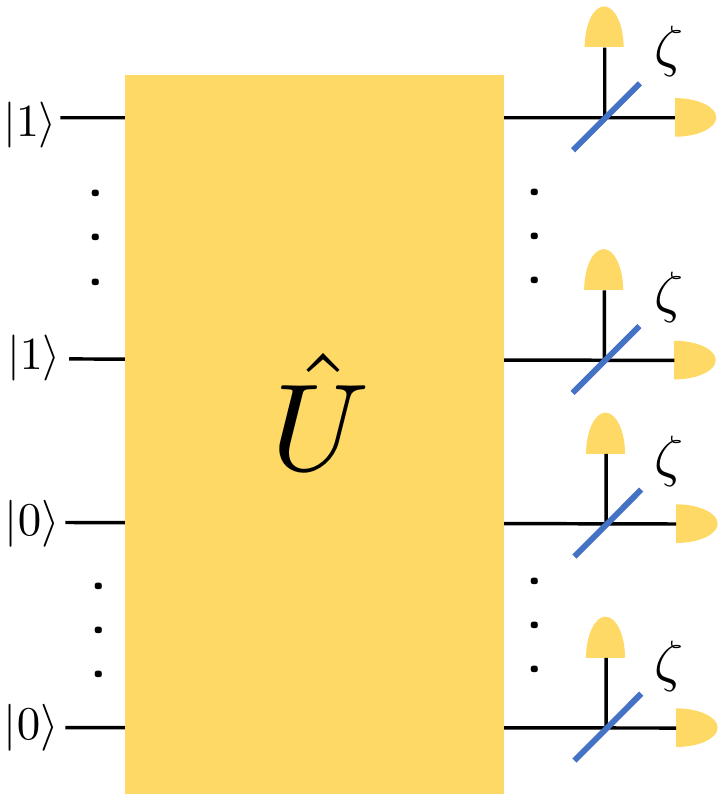}
		\caption{CVS$_{\text{SP}}$ circuit with interferometer $U$ and heterodyne detection with reconfigurable unbalancing parameter $\zeta$. The detectors represented are homodyne detectors. Setting $\zeta=0$ (balanced detection) allows us to certify efficiently the multimode output state with a fidelity witness. Setting $\zeta\neq0$, with $|\zeta|=\Omega(2^{-\poly m})$ allows us to perform efficiently a sampling task which is hard for classical computers, unless the polynomial hierarchy collapses.}
		\label{fig:CVSverif}
	\end{center}
\end{figure}

\noindent An interesting point is that by changing the unbalancing of heterodyne detection of the output modes of a Boson Sampling interferometer, one can switch between verification of Boson Sampling output states and demonstration of quantum sampling supremacy with continuous variable measurements. Indeed, CVS$_{\text{SP}}$ circuits introduced in the previous chapter, which correspond to Boson Sampling with unbalanced heterodyne detection, are hard to sample classically when the unbalancing of the heterodyne detection is not too small (see section~\ref{sec:CVS}), but their output can be efficiently certified simply by switching to balanced heterodyne detection and computing a fidelity witness with the above method. This can be done within the same experimental setup using a reconfigurable beam splitter (Fig.~\ref{fig:CVSverif}) and showing the hardness of approximate CVS circuits sampling is an important step before an experimental demonstration. 

Alternatively, by switching between balanced heterodyne detection and single-photon threshold detection, one can switch between verification of Boson Sampling output states and demonstration of quantum sampling supremacy with discrete variable measurements, for which approximate sampling hardness is demonstrated, assuming two conjectures on the permanent of random Gaussian matrices and the fact that the polynomial hierarchy of complexity classes does not collapse~\cite{Aaronson2013}.

\section{Discussion and open problems}

\noindent Existing methods for building trust for continuous variable quantum states like homodyne quantum state tomography require many different measurement settings, and heavy classical post-processing. For that purpose, we have introduced a reliable method for heterodyne quantum state tomography, which uses heterodyne detection as a single Gaussian measurement setting and allows for the retrieval of the density matrix of an unknown quantum state with analytical confidence intervals, without the need for data reconstruction nor binning of the sample space. 
For data reconstruction methods such as Maximum Likelihood, errors from the reconstruction procedure are usually indistinguishable from errors coming from the tested quantum device. For that reason, such methods do not extend well to the task of verification, unlike our method.

Building on these tomography techniques and with the addition of cryptographic techniques such as the de Finetti theorem, we have derived a protocol for verifying various copies of a continuous variable quantum state, without i.i.d.\@ assumption, with Gaussian measurements. This protocol is robust, as it directly gives a confidence interval on an estimate of the fidelity between the tested state and the target pure state. We emphasize that, while the target state is pure, the tested state is not required to be pure. The general protocol may be tailored to different uses and assumptions, from tomography to verification, simply by changing the classical post-processing.

Our verification protocol is complementary to the approach of~\cite{takeuchi2018resource}, in which a verifier performs continuous variable quantum computing by delegating the preparation of Gaussian cluster states to a prover and has to perform non-Gaussian measurements. In our approach, the measurement-only verifier may perform continuous variable quantum computing by delegating the preparation of non-Gaussian states to the prover and has to perform Gaussian measurement, which are much easier to perform experimentally.

Importantly, we have promoted our single-mode protocols for fidelity estimation to multimode protocols yielding fidelity witnesses, showing in particular how to verify output states of a Boson Sampling interferometer efficiently, either under the i.i.d.\@ assumption or with no assumption whatsoever. These protocols open the way for the most rigourous experimental demonstration of quantum computational supremacy so far, with Boson Sampling.

An exciting open problem is whether the technique employed in this chapter for promoting single-mode fidelity estimation protocols to protocols providing multimode fidelity witnesses can be applied in other contexts, for example discrete variable quantum computing. This technique crucially relies on being able to revert efficiently, at the stage of classical post-processing, specific quantum operations (passive linear transformations in this case) after a specific measurement (heterodyne detection, i.e., sampling from the Husimi $Q$ function in this case).

\clearemptydoublepage
%
%
\let\textcircled=\pgftextcircled
\chapter{Quantum-programmable measurements with linear optics}
\label{chap:prog}

\initial{D}istinguishing two unknown quantum states is central to many quantum applications~\cite{montanaro2013survey}, notably for entanglement testing~\cite{mintert2005concurrence,walborn2006experimental,harrow2013testing}, quantum communication~\cite{buhrman2001quantum,de2004one,kumar2017efficient} and quantum machine learning~\cite{ekert2002direct,lloyd2013quantum}. This task is referred to as \textit{unknown quantum state discrimination}. 

The ability to program a fixed computer to perform a variety of computations is especially important: we do not want to build a new physical device for every different computation. In particular, quantum-programmable devices are quantum machines that take additional quantum states in input as a program, which dictates the rest of the computation.  It is not possible to build a fixed quantum computer which can be programmed to perform any quantum computation~\cite{nielsen1997programmable}, but we can design quantum-programmable devices for a restricted set of computations, such as projective measurements. 

In this chapter, we show a correspondence between unknown quantum state discrimination and quantum-programmable measurements, by generalising the celebrated swap test~\cite{buhrman2001quantum} for quantum state discrimination to an unbalanced setting where multiple copies of only one of the two tested states are available. 

Next, we also generalise a known link between the Hong--Ou--Mandel effect for partially distinguishable photons and the swap test~\cite{garcia2013swap}: we present the Hadamard interferometer and show that it provides a scheme for performing unknown quantum state discrimination and quantum-programmable measurements with linear optics and single photons. 

In order to reduce the experimental requirements for implementation, we consider the case of projective measurements onto coherent states and simplify the previously derived scheme. In this case, we perform a simple analysis of the consequences of experimental imperfections.

This chapter is based on \cite{chabaud2018optimal,inprepa1}.


\section{Testing quantum states}
\label{sec:testing}

In the previous chapter, we discussed the efficient characterization of a continuous variable quantum state, either by full tomographic reconstruction, by fidelity estimation with a target state, or by obtaining a fidelity witness with a target state. In some cases, however, one merely wants to test simple properties of quantum systems. Given two unknown quantum states, one of the simplest questions one may ask is whether these states are equal or not. In this section, we make a connection between unknown quantum state discrimination schemes and quantum-programmable projective measurement devices.

\subsection{Quantum state discrimination: the swap test}

The swap test~\cite{buhrman2001quantum} provides a simple probabilistic tool to compare two unknown quantum states. It takes as input two quantum states $\ket{\phi}$ and $\ket{\psi}$ that are not entangled and outputs $0$ with probability $\frac{1}{2}+\frac{1}{2}|\braket{\phi|\psi}|^2$ and $1$ with probability $\frac{1}{2}-\frac{1}{2}|\braket{\phi|\psi}|^2$, where $\braket{\phi|\psi}$ is the inner product of the states $\ket{\phi}$ and $\ket{\psi}$. When the measurement outcome is $0$ (resp.\@ $1$), we conclude that the states were identical (resp.\@ different), up to a global phase.\\

\begin{figure}[h!]
	\begin{center}
		\includegraphics[width=2.8in]{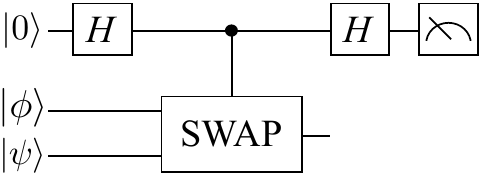}
		\caption{Circuit representation of a swap test. The ancilla qubit is measured in the computational basis.}
		\label{fig:SWAP}
	\end{center}
\end{figure}

\noindent A circuit implementing the swap test for qubits is represented in Fig.~\ref{fig:SWAP}, where an ancilla is first prepared in the $|+\rangle$ state by a Hadamard gate
\be
H=\frac{1}{\sqrt2}\begin{pmatrix} 1 & 1 \\ 1 & -1 \end{pmatrix},
\label{matrixH}
\ee
and controls a swap between the two systems being tested. 

The swap test meets the so-called \textit{one-sided error requirement}~\cite{buhrman1999one}, i.e., if the input states are identical, the test will always declare them as identical. On the other hand, if the input states are different, the test can obtain a wrong conclusion by declaring the states identical. The probability that this happens is strictly less than $1$, hence by repeating the test various times, the probability that the sequence of tests never answers $1$ can be brought down arbitrarily close to zero, exponentially fast. However, the swap test is destructive, in the sense that the output states of a previous test cannot be reused for a new test because they become maximally entangled during the test~\cite{garcia2013swap}. This means that in order to boost the correctness of the test in this manner, multiple copies of both states must be available.

\subsection{Quantum state identity testing: generalised swap test}

Let $m\ge2$. We introduce the following generalisation of the swap test, in the context where one has access to various copies of a reference state $\ket{\psi}$ but to only a single copy of the other tested state $\ket{\phi}$:

\begin{defi}[Swap test of order $m$]
The swap test of order $m$ is a binary test that takes as input a state $\ket{\phi}$ and $m-1$ copies of a state $\ket{\psi}$, and outputs $0$ with probability $\frac1m+\frac{m-1}m|\braket{\phi|\psi}|^2$ and $1$ with probability $(\frac{m-1}m)(1-|\braket{\phi|\psi}|^2)$. If the outcome $0$ (resp.\@ $1$) is obtained, the test concludes that the states $\ket{\phi}$ and $\ket{\psi}$ were identical (resp.\@ different).
\label{defswap}
\end{defi}

\noindent Such a test clearly satisfies the one-sided error requirement.\\
\begin{figure}[h!]
\begin{center}
\includegraphics[width=4in]{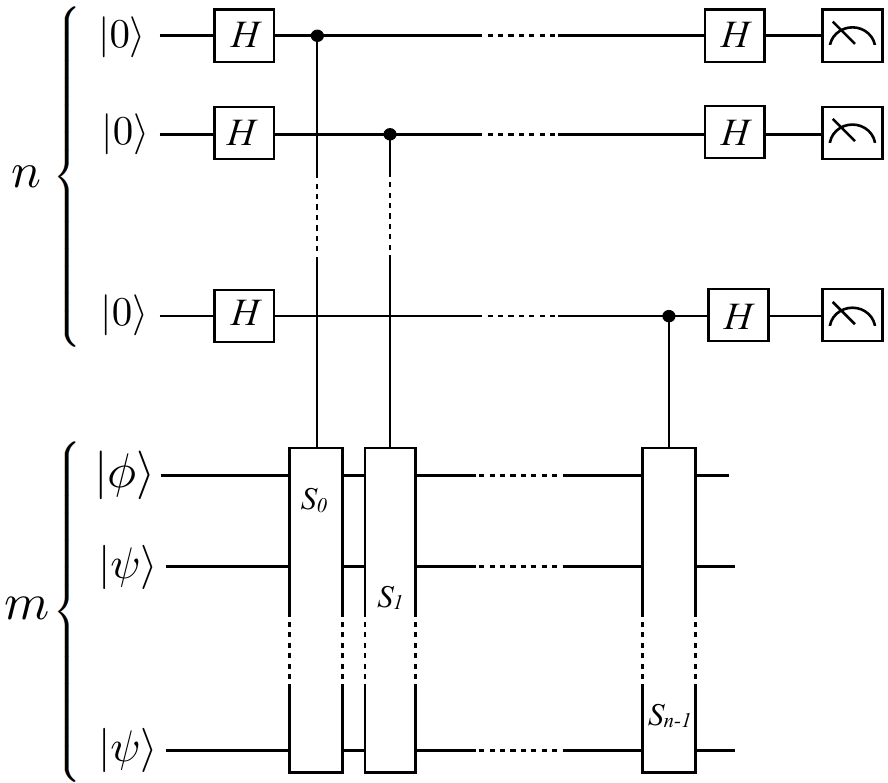}
\caption{Swap circuit of order $m$. The unitaries $S_k$ are tensor products of swap gates described in the main text~(\ref{Sk}). The $n=\log m$ ancilla qubits are measured in the computational basis at the end of the computation. The probability of obtaining $0$ for all measurement outcomes is $\frac1m+\frac{m-1}m|\braket{\phi|\psi}|^2$.}
\label{fig:swap_circuit}
\end{center}
\end{figure}

\noindent In the following, we restrict to the swap test of order $m$ when $m$ is a power of $2$, writing $n=\log m$. We introduce the swap circuit of order $m$ (Fig.~\ref{fig:swap_circuit}), that acts on $m$ input qubits by applying $n$ consecutive layers of products of swap gates controlled by $n$ ancilla qubits. These ancilla qubits are first initialised in the $\ket+$ state using Hadamard gates. Then, they are used as control qubits for the gates $S_0,\dots,S_{n-1}$, which can be applied in any order, where for all \mbox{$k\in\{0,\dots,n-1\}$}
\begin{equation}
S_k=\underset {\substack{ i\in \left[ 0,{ 2 }^{ k }-1 \right] ,\\j\in \left[ 0,{ 2 }^{ n-k-1 }-1 \right] } }{ \bigotimes} \text{SWAP}\left[j { 2 }^{ k+1 }+i , j { 2 }^{ k+1 }+i+{ 2 }^{ k }\right],
\label{Sk}
\end{equation}
with SWAP$[i,j]$ being the unitary operation that swaps the $i^{th}$ and $j^{th}$ qubits for $i,j\in\{0,\dots,m-1\}$. These controlled gates are applied to the input states $\ket{\phi},\ket{\psi},\dots,\ket{\psi}$ (one copy of a state $\ket{\phi}$ and $m-1$ copies of a state $\ket{\psi}$). Finally, a Hadamard gate is applied to each ancilla, which is then measured in the computational basis.
By a simple induction, we obtain that the probability of obtaining the outcome $0$ for all ancilla qubits is the squared norm of the following state:
\begin{equation}
\frac1m(\ket{\phi\psi\dots\psi}+\ket{\psi\phi\dots\psi}+\dots+\ket{\psi\dots\psi\phi}),
\label{perror}
\end{equation}
which only depends on the overlap between the states $\ket{\phi}$ and $\ket{\psi}$. More precisely,
\begin{equation}
\Pr\,[0,\dots,0]=\frac1m+\frac{m-1}m|\braket{\phi|\psi}|^2.
\label{sswap}
\end{equation}
The swap circuit of order $m$ thus implements the swap test of order $m$. Indeed, if the outcome $(0,\dots,0)$ is obtained, the test outputs $0$ and we conclude that the states were identical, while for any other outcome the test outputs $1$ and we conclude that the states were different. Note that in the case where $m=2$, the scheme reduces to the original swap test.

\begin{figure}[h!]
\begin{center}
\includegraphics[width=4.5in]{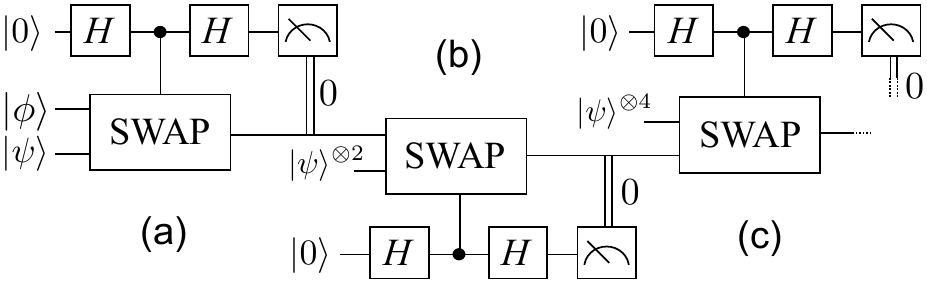}
\caption{The simplified swap circuit of order $m$ consisting in $n=\log m$ consecutive swap tests. (a) The first swap test compares the input states $\ket{\phi}$ and $\ket{\psi}$. (b) If this test is not able to tell apart the input states, i.e., if its outcome is $0$, then the second swap test compares the bipartite output state of the first test with the state $\ket{\psi}^{\otimes2}$. (c) If this test outcome is again $0$, then the third swap test compares the quadripartite output state of the second test with the state $\ket{\psi}^{\otimes4}$, and so on. If the $n$ outcomes are $0$, the test concludes that the states $\ket{\phi}$ and $\ket{\psi}$ were identical.}
\label{fig:swap_circuit_2}
\end{center}
\end{figure}

Because the $m-1$ last input states are identical, swapping them acts as the identity. This can be used to simplify the swap circuit of order $m$ by replacing the $n=\log m$ layers of swap gates in Eq.~(\ref{Sk}) by the following $n$ layers $S_0',\dots,S_{n-1}'$, which have to be applied in this order:
\begin{equation}
S_k'=\overset { { 2 }^{ k }-1 }{ \underset { l=0 }{ \bigotimes  }  } \text{SWAP}\left[ l,l+2^k \right].
\label{Sk2}
\end{equation}
This reduces the total number of swap gates from $\frac{m\log m}{2}$ to $m-1$ without changing the number of ancilla qubits. This circuit has a simple structure of $n=\log m$ consecutive swap tests (Fig.~\ref{fig:swap_circuit_2}). \\

\noindent For $k\in\{0,\dots,n-1\}$, conditioned on all the previous outputs being $0$, the $k^{th}$ swap test compares the output state of the previous test and the state $\ket{\psi}^{\otimes2^{k}}$. Here, the swap test of two multipartite quantum states consists in applying a swap test to each of their corresponding subsystems. However, this multipartite swap test uses only a single ancilla qubit controlling the product of swap gates, as in Eq.~(\ref{Sk2}), instead of an ancilla qubit for each pair of subsystems.

We now prove the optimality of the swap test of order $m$ under the one-sided error requirement, i.e., we show that it achieves the lowest error probability in comparing states $|\phi\rangle$ and $\ket{\psi}$ given $m-1$ copies of $|\phi\rangle$ and one copy of $\ket{\psi}$ such that the one-sided error requirement is satisfied.

For this purpose, we first derive a more general result. In~\cite{kada2008efficiency}, the authors consider the problem of testing if $m$ quantum states are identical or not (the so-called \textit{identity test}), with the promise that all the states are pairwise identical or orthogonal. 
In particular, they show that the optimal value for the error probability of any identity test with these assumptions satisfying the one-sided error requirement is $\frac1m$. We extend this result to the case where the states to be compared are no longer assumed pairwise identical or orthogonal:

\begin{theo} Under the one-sided error requirement, any identity test of $m$ unknown quantum states $\ket{\psi_1},\dots,\ket{\psi_m}$ has an error probability at least
\be
\frac1{m!}\sum_{\sigma\in S_m}{\prod_{k=1}^m{\braket{\psi_k|\psi_{\sigma(k)}}}},
\ee
where $S_m$ is the symmetric group over $\{1,\dots,m\}$.
\label{th:sopt}
\end{theo}

\begin{proof}
\begin{mdframed}[linewidth=1.5,topline=false,rightline=false,bottomline=false]

An identity test satisfying the one-sided error requirement can only be wrong when declaring identical (outputting $0$) states that are not identical. Hence, to prove Theorem~\ref{th:sopt}, it suffices to lower bound the probability of outputting $0$ for any identity test. This is done by showing that the optimal identity test consists in a projection onto the symmetric subspace of the input states Hilbert space. 

An identity test on a Hilbert space $\mathcal{H}$ is a binary test which can be written as a positive-operator valued measure $\{\Pi_0,\Pi_1\}$, with $\Pi_0+\Pi_1=\mathbb 1$. Such a test takes as input a pure tensor product state $\ket{\psi_1\dots \psi_m}\in\mathcal{H}^{\otimes m}$ and outputs $0$ with probability
\be
\Pr[0]=\Tr\,[\Pi_0\ket{\psi_1\dots \psi_m}\bra{\psi_1\dots \psi_m}], 
\ee
and $1$ with probability
\be
\Pr[1]=1-\Pr[0]=\Tr\,[\Pi_1\ket{\psi_1\dots \psi_m}\bra{\psi_1\dots \psi_m}].
\ee
If the output $0$ is obtained we conclude that we had $\ket{\psi_1}=\dots=\ket{\psi_m}$, whereas if the output $1$ is obtained we conclude that the states were not all identical. The one-sided error requirement can thus be written as
\be
\forall\ket{\psi},\Tr\,[\Pi_1\ket{\psi}\bra{\psi}^{\otimes m}]=0.
\label{oser}
\ee
Following~\cite{harrow2013church}, the symmetric subspace of $\mathcal{H}^{\otimes m}$ is characterised as
\be
S=\text{span}\{\ket{\psi}^{\otimes m}:\ket{\psi}\in\mathcal{H}\},
\ee
and the orthogonal projector onto this space can be written as
\be
P_S=\frac{1}{m!}\sum_{\sigma\in S_m}{P_\sigma},
\ee
where for all $\sigma\in S_m$ and all $\ket{\psi_1\dots\psi_m}\in\mathcal{H}^{\otimes m}$ we have $P_\sigma\ket{\psi_1\dots\psi_m}=\ket{\psi_{\sigma(1)}\dots\psi_{\sigma(m)}}$. Given the characterisation of the symmetric subspace, the one-sided error requirement in Eq.~(\ref{oser}) implies that the supports of $P_S$ and $\Pi_1$ are disjoint. The support of $P_S$ is thus included in the support of $\Pi_0$, given that $\Pi_0+\Pi_1=\mathbb 1$ and this implies in turn that $\Pi_0\ge P_S$ by positivity of $\Pi_0$. 

The error probability of the identity test under the one-sided error requirement is given by the probability of outputting the result $0$ while the states were not all identical:
\begin{align}
\nonumber\Pr[0]&=\Tr\,[\Pi_0\ket{\psi_1\dots \psi_m}\bra{\psi_1\dots \psi_m}]\displaybreak\\
\nonumber&\ge\Tr\,[P_S\ket{\psi_1\dots \psi_m}\bra{\psi_1\dots \psi_m}]\\
&\ge\frac1{m!}\sum_{\sigma\in S_m}{\Tr\,[P_\sigma\ket{\psi_1\dots \psi_m}\bra{\psi_1\dots \psi_m}]}\\
\nonumber&\ge\frac1{m!}\sum_{\sigma\in S_m}{\Tr\,[\ket{\psi_{\sigma(1)}\dots\psi_{\sigma(m)}}\bra{\psi_1\dots \psi_m}]}\\
\nonumber&\ge\frac1{m!}\sum_{\sigma\in S_m}{\prod_{k=1}^{m}{\braket{\psi_k|\psi_{\sigma(k)}}}},
\end{align}
where in the third line we used the expression of the orthogonal projector $P_S$ onto the symmetric subspace.

\end{mdframed}
\end{proof}

\noindent Applying Theorem~\ref{th:sopt} with $\ket{\psi_1\dots\psi_{k+l}}=\ket{\phi}^{\otimes k}\otimes\ket{\psi}^{\otimes l}$, we obtain the following lower bound for the error probability of any identity test of $k+l$ states $\ket{\phi}^{\otimes k}\otimes\ket{\psi}^{\otimes l}$:
\be
\frac1{(k+l)!}\sum_{p=0}^{\min{(k,l)}}{\binom{k}{p}\binom{l}{p}k!l!|\braket{\phi|\psi}|^{2p}}=\sum_{p=0}^{\min{(k,l)}}{\frac{\binom{k}{p}\binom{l}{p}}{\binom{k+l}{k}}|\braket{\phi|\psi}|^{2p}},
\ee
where $\binom{k}{p}\binom{l}{p}k!l!$ is the number of partitions of $\{1,\dots,k+l\}$ which map exactly $k-p$ elements of $\{1,\dots,k\}$ to elements of $\{1,\dots,k\}$.
Testing quantum state identity with the input state $\ket{\phi}^{\otimes k}\otimes\ket{\psi}^{\otimes l}$ amounts to comparing the states $\ket\phi$ and $\ket\psi$ using $k$ copies of $\ket{\phi}$ and $l$ copies of $\ket{\psi}$. 

In the case where $k=1$ and $l=m-1$, we have $\ket{\psi_1\dots\psi_m}=\ket{\phi\psi\dots\psi}$ and Theorem~\ref{th:sopt} shows that the value $\frac1m+\frac{m-1}m|\braket{\phi|\psi}|^2$ is a lower bound for the error probability of any identity test of $m$ states $\ket{\phi},\ket{\psi},\dots,\ket{\psi}$, i.e., one copy of a state $\ket{\phi}$ and $m-1$ copies of a state $\ket{\psi}$. With Definition~\ref{defswap} we directly obtain the following result:

\begin{coro} The swap test of order $m$ has optimal error probability $\frac1m+\frac{m-1}m|\braket{\phi|\psi}|^2$ under the one-sided error requirement.
\label{co:0}
\end{coro}

\noindent The swap circuit of order $m$ is thus optimal for quantum state identity testing with an input $\ket{\phi},\ket{\psi},\dots,\ket{\psi}$, under the one-sided error requirement, since it implements the swap test of order $m$.
In the next section, we show that the swap circuit of order $m$ can be used to implement a programmable projective measurement.

\subsection{Universal programmable measurements}

In a typical experiment performing a quantum measurement, the choice of measurement is encoded in macroscopic, classical, information in the experimental setup.
For example it can be encoded into the reflectance of a beam splitter, the phase in the branch of an interferometer or the spacial direction of a Stern Gerlach device. 
Often these choices are made beforehand and fixed. In some cases they can be programmed in a single set up (for example using thermo-optic phase shifters \cite{carolan2015universal}).
In all these cases, however, the choice of measurement basis is effectively programmed classically. 

We consider the case where the choice of measurement is instead controlled by a quantum state.
There are several reasons why one may consider a quantum state to control the choice of measurement. 
This state may be an output of a quantum computer, or a communication protocol, for example, which is not known before hand and only accessible as a quantum state. For example, in the cryptographic setting, non-orthogonal states can be used to remotely program a measurement which allows one to test the behaviour of a remote party. This is the essence behind the delegated blind verified quantum computation in~\cite{fitzsimons2017unconditionally}. 
At a fundamental level quantum programmable measurements separate as much as possible the choice of measurement basis and the bulk of the physical measurement apparatus, which could be interesting in probing
foundational questions, for example in tests of contextuality where information about which measurements are being carried out leads to loopholes \cite{meyer1999finite,clifton2000simulating,winter2014does}.

A related and, in a sense, more general problem is that of a programmable quantum computer, where a quantum program state is used to encode a unitary to be run on a generic quantum computing device (gate array), first proposed by Nielsen and Chuang \cite{nielsen1997programmable}. 
There it was shown that to do so deterministically requires orthogonal program states for every different unitary. To use the continuous parameters available in quantum states to encode more computations, the best one can do is probabilistic.
In principle these techniques can be used to program quantum measurements.
Indeed since the original proposal there have been several alternative schemes, extensions and applications, including programmable quantum state discriminators and measurements \cite{vidal2000storage,duvsek2002quantum,rovsko2003generalized,ziman2005realization,bergou2006programmable}.
These results, however, are either too general to consider the type of efficiency we show here, or specialized to tasks which are different from our setting. \\

\begin{figure}[h!]
	\begin{center}
		\includegraphics[width=4.2in]{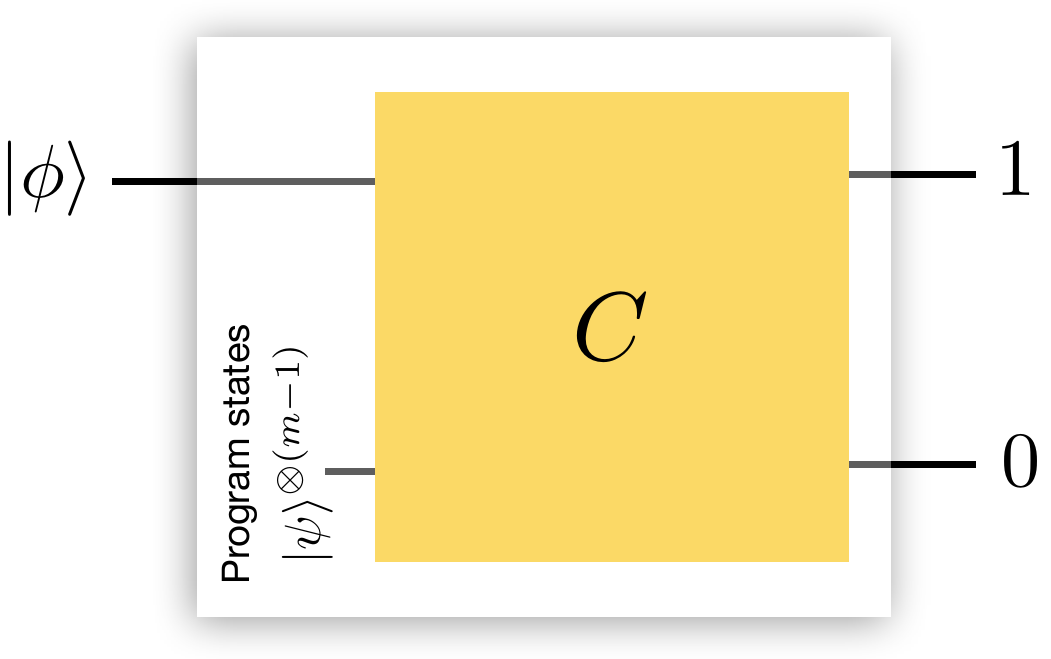}
		\caption{Programmable projective measurement. Given an input $\ket{\phi}$ and $m-1$ program registers $|\psi\rangle^{\otimes m-1}$ and allowing for possible ancillas (not pictured here), we apply some circuit $C$, independent of $|\psi\rangle$, and output a binary result where $0$ is associated to projecting onto $|\psi\rangle$ and $1$ to its complement.}
		\label{fig:programmable projective measurement definition}
	\end{center}
\end{figure}

\noindent We cast our problem as follows, illustrated in Fig.~\ref{fig:programmable projective measurement definition}. 
One has $m-1$ program registers each prepared in the state $|\psi\rangle$ corresponding to the choice of measurement basis, and a single input register prepared in some state $\ket{\phi}$. 
Our aim is to output a classical bit corresponding to a projective measurement, where $0$ represents the outcome $|\psi\rangle$ and $1$ represents its complement.
In an ideal measurement the result $0$ would occur with probability $|\langle  \phi |\psi\rangle|^2$. 
However, this is impossible for finite $m$.
This follows from standard arguments based on the linearity of quantum mechanics, in analogy to necessity of orthogonal program states for computation mentioned above (see for example \cite{nielsen1997programmable} for the case of programmable universal quantum computation, which easily extends to our case).
We can thus only ever approximate perfect measurements. In our case we parametrise this approximation by $\epsilon$, requiring that the result $0$ is returned with probability $\epsilon$-close to $|\langle  \phi |\psi\rangle|^2$.

We present a scheme which achieves this optimally in terms of how $\epsilon$ scales with $m$, under the condition that if the input is $\ket{\psi}$, the measurement always returns $0$. 
This so-called one-sided error requirement~\cite{buhrman1999one} makes sense for various potential applications where it is important not to be wrong for this answer. One such example is the link between our scheme and the swap test~\cite{buhrman2001quantum}. 

In the swap test, two unknown quantum states are compared using a controlled-swap operation. This test is especially relevant for the task of state discrimination. 
The general task of assessing if a set of $m$ arbitrary states are identical has been addressed in~\cite{chefles2004unambiguous,kada2008efficiency}. 
To solve this in generality requires controlled permutations for all possible permutations and therefore scales exponentially in circuit size.
If one restricts oneself to the case where one has $m/2$ copies of one state and $m/2$ copies of the other, one can apply the construction in~\cite{kada2008efficiency} to get an optimal result.
However, this scaling is not much better than simply doing the original swap test $m/2$ times, yet the corresponding test is much more difficult to implement.

From this point of view, the interesting cases of two states comparison is if one has an asymmetric number of one compared state compared to the other. In the most extreme case one would have just one copy of one state and $m-1$ copies of the other, which is exactly the case we consider for our programmable projective measurement, viewing the program state as the one we have many copies of.
In particular, the $m=2$ case reduces to the swap test.

Given that a projective measurement with respect to a state $\ket{\psi}$ is a process that takes as input a state $\ket{\phi}$ and outputs $0$ with probability $|\braket{\phi|\psi}|^2$ and $1$ with probability $1-|\braket{\phi|\psi}|^2$, we introduce the notion of projective measurement with finite error:

\begin{defi}[Approximate projective measurement]
Given a quantum state $\ket{\psi}$ and $\epsilon>0$, a projective measurement with error $\epsilon$ with respect to the reference state $\ket{\psi}$ is a process that takes as input a quantum state $\ket{\phi}$ and outputs $0$ with probability $\Pr[0]$ and $1$ with probability $\Pr[1]$, such that $|\Pr[0]-(|\braket{\phi|\psi}|^2)|\le\epsilon$ and $|\Pr[1]-(1-|\braket{\phi|\psi}|^2)|\le\epsilon$.
\label{def:proj}
\end{defi}

\noindent Note that the two conditions in the previous definition are equivalent, since $\Pr[0]+\Pr[1]=1$. It will thus suffice to consider, e.g., the first condition. In this context, under the one-sided error requirement, a projective measurement with any error $\epsilon$ always outputs $0$ if the input state is equal to the reference state.

\begin{figure}[h!]
\begin{center}
\includegraphics[width=4.4in]{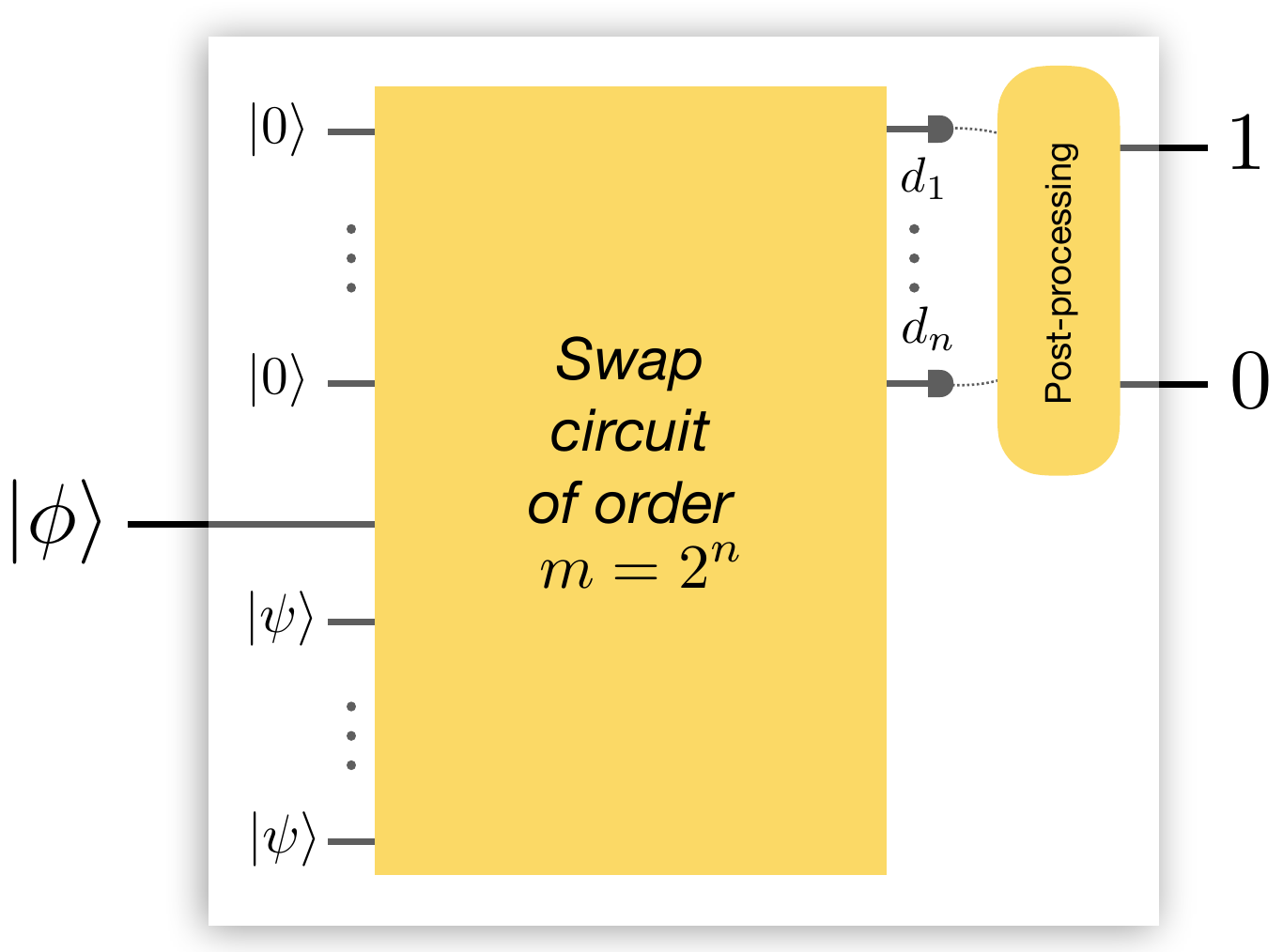}
\caption{The swap circuit of order $m$ used as a programmable projective measurement device. It takes as input a state $\ket{\phi}$ and the internal measurement outcomes are post-processed such that the device outputs $0$ with probability $\frac1m+\frac{m-1}m|\braket{\phi|\psi}|^2$ and $1$ with probability $\frac{m-1}m(1-|\braket{\phi|\psi}|^2)$. The programmable resource is the state $\ket{\psi}$ and the process uses $m-1$ copies of this state as well as $n=\log m$ ancillas.}
\label{fig:setupc}
\end{center}
\end{figure}

\begin{theo} A swap circuit of order $m$ can be used to perform a projective measurement with error $\frac1m$ under the one-sided error requirement. Moreover, it is optimal in the sense that it uses the minimum number of copies of the reference state for achieving such an error.
\label{th:circuit}
\end{theo}

\begin{proof}
\begin{mdframed}[linewidth=1.5,topline=false,rightline=false,bottomline=false]

For the swap circuit of order $m$, we have $\Pr\,[0,\dots,0]=\frac1m+\frac{m-1}m|\braket{\phi|\psi}|^2$ by Eq.~(\ref{sswap}), so we can consider the whole circuit except the state $\ket{\phi}$ as a black box in Fig.~\ref{fig:swap_circuit}, and post-process the measurement outcomes $\bm d$ as follows: if $\bm d=(0,\dots,0)$, output $0$, and output $1$ otherwise. The setup now takes a single state $\ket{\phi}$ in input and outputs $0$ with probability $\Pr[0]=\frac1m+\frac{m-1}m|\braket{\phi|\psi}|^2$, and $1$ with probability $\Pr[1]=1-\Pr[0]$. We have $|\Pr[0]-(|\braket{\phi|\psi}|^2)|\le\frac1m$ and when $\ket{\phi}=\ket{\psi}$, we have $\Pr[0]=1=|\braket{\phi|\psi}|^2$, hence this device performs a projective measurement with error $\frac1m$ and meets the one-sided error requirement. 

We now prove the optimality of this device in terms of resources, i.e., we show that any device implementing a projective measurement with error $\frac1m$ and meeting the one-sided error requirement cannot use less than \mbox{$m-1$} copies of the reference state. 

We consider a device that implements a projective measurement with error $\epsilon$, with respect to a reference state $\ket{\psi}$, using $p$ copies of this reference state. This device takes as input a quantum state $\ket{\phi}$ and outputs $0$ with probability $\Pr_\phi[0]$ and $1$ with probability $\Pr_\phi[1]=1-\Pr_\phi[0]$. By Definition~\ref{def:proj}, the probability of outputting $0$ satisfies $|\Pr_\phi[0]-(|\braket{\phi|\psi}|^2)|\le\epsilon$. When the input state $\ket{\phi}$ is orthogonal to the reference state $\ket{\psi}$, the probability $\Pr_{\phi,\bot}[0]$ of outputting $0$ thus satisfies
\be
\text{Pr}_{\phi,\bot}[0]\le\epsilon. 
\label{Pbot1}
\ee
On the other hand, we can use this device to perform an identity test of $p+1$ states $\ket{\phi},\ket{\psi},\dots,\ket{\psi}$ (one copy of the state $\ket{\phi}$ and $p$ copies of the state $\ket{\psi}$): if the output $0$ (resp.\@ $1$) is obtained we conclude that the states were identical (resp.\@ different). This device meets the one-sided error requirement, so by Theorem~\ref{th:sopt} it has error probability at least $\frac{1}{p+1}+\frac p{p+1}|\braket{\phi|\psi}|^2$. This error probability corresponds to the probability of outputting $0$ when the input states are different. In particular, when the input state $\ket{\phi}$ is orthogonal to the reference state $\ket{\psi}$, the probability $\Pr_{\phi,\bot}[0]$ of outputting $0$ thus satisfies
\be
\text{Pr}_{\phi,\bot}[0]\ge\frac{1}{p+1}. 
\label{Pbot2}
\ee
Combining both inequalities~(\ref{Pbot1}) and (\ref{Pbot2}) we obtain $\frac{1}{p+1}\le\epsilon$ or equivalently $p\ge\frac1\epsilon-1$. For $\epsilon=\frac1m$, this amounts to $p\ge m-1$, which completes the proof.

\end{mdframed}
\end{proof}

\noindent Theorem~\ref{th:circuit} implies that given a large enough swap circuit and the ability to produce many copies of a state $\ket{\psi}$, one can projectively measure any state with respect to the state $\ket{\psi}$ up to arbitrary small error. This error scales as the inverse of the number of copies. The circuit can thus be used as a programmable projective measurement device, where the programmable resource is the reference state $\ket{\psi}$ whose number of copies can be adjusted to control the precision of the measurement (Fig.~\ref{fig:setupc}).

The implementation of the swap circuit of order $m$ is however challenging, due to the presence of many controlled-swap gates. In order to lower the implementation requirements, we study in the next section the linear optical Hadamard interferometer~\cite{crespi2015suppression,crespi2016suppression} and show that its statistics can be efficiently post-processed to reproduce those of a swap circuit of order $m$, without the need for ancillas. This comes at the cost that the device no longer has a quantum output, which however does not matter for most applications. In particular, we show that the Hadamard interferometer provides a simple linear optical platform for implementing the programmable projective measurement that we have described.

\section{Universal programmable projective measurements with linear optics}
\label{sec:univ1}

The swap test has been shown equivalent to the linear optical Hong-Ou-Mandel effect~\cite{garcia2013swap} (see section~\ref{sec:HOM}), in the sense that one can use Hong-Ou-Mandel effect to perform a state discrimination test between two partially distinguishable photons, whose statistics reproduce those of a swap test. Generalising this equivalence, we present a practical solution to our problem with linear optics, using the Hadamard interferometer~\cite{crespi2015suppression,crespi2016suppression}.

\subsection{The Hadamard interferometer}

In what follows, we consider optical unitary interferometers of size $m$ which take as input one single photon in a quantum state $\ket{\phi}$ in the first mode and $m-1$ indistinguishable single photons in a state $\ket{\psi}$, one in each other spatial mode (the spatial modes of the interferometers are indexed from $1$ to $m$). These states should be thought of as encoded in additional degrees of freedom of the photons (e.g., polarisation, time bins). The output modes are measured using photon number-resolving detection. 

There exist complex amplitudes $\alpha$ and $\beta$ and a state $\ket{\psi^\bot}$ with $\braket{\psi|\psi^\bot}=0$ such that
\begin{equation}
\ket\phi=\alpha\ket\psi+\beta\ket{\psi^\bot},
\end{equation}
where $\alpha=\braket{\psi|\phi}$ and $|\alpha|^2+|\beta|^2=1$. We have the following homomorphism property for single photon states:
\begin{equation}
\ket{1_\phi}=\ket{1_{\alpha\psi+\beta\psi^\bot}}=\alpha\ket{1_\psi}+\beta\ket{1_{\psi^\bot}},
\end{equation}
where for any state $\ket{\chi}$, $\ket{1_\chi}$ is the state of a single photon encoding the state $\ket{\chi}$. The single photon encoding maps identity of quantum states to distinguishability of single photons. In order to test the distinguishability of the photons, we look for detection events that do not occur when the photons are indistiguishable. In that case, it suffices to compute the output statistics separately when $\ket{\phi}=\ket{\psi}$ (\textit{indistinguishable case}) and when $\ket{\phi}=\ket{\psi^\bot}$ (\textit{distinguishable case}) to obtain the output statistics in the general case by linearity. The probability of detecting a photon number pattern $\bm d=(d_1,\dots,d_m)$ which does not occur in the indistiguishable case, or equivalently that the $k^{th}$ detector detects $d_k$ photons for all $k\in\{1,\dots,m\}$, is then 
\begin{equation}
\begin{aligned}
\text{Pr}\,(\bm d)&=|\alpha|^2\text{Pr}_i(\bm d)+|\beta|^2\text{Pr}_d(\bm d) \\
&=\left(1-|\braket{\phi|\psi}|^2\right)\text{Pr}_d(\bm d),
\end{aligned}
\label{partialud}
\end{equation}
where $\text{Pr}_i(\bm d)=0$ is the probability in the indistinguishable case and $\text{Pr}_d(\bm d)$ is the probability in the distinguishable case. We thus have
\be
\sum_{\substack{\bm d\\ \text{Pr}_i(\bm d)=0}}{\text{Pr}\,(\bm d)}=\left(1-|\braket{\phi|\psi}|^2\right)\sum_{\substack{\bm d\\ \text{Pr}_i(\bm d)=0}}{\text{Pr}_d(\bm d)}.
\label{partialud2}
\ee
Note that for any measurement outcome $\bm d=(d_1,\dots,d_m)$, we have $d_1+\dots+d_m=m$ since an interferometer is a passive device that does not change the total number of photons. For any interferometer of size $m$, we also obtain the following result:

\begin{lem}\label{lem:PrdPri}
For any detection pattern $\bm d$,
\begin{equation}
\text{Pr}_d(\bm d)\ge\frac{\text{Pr}_i(\bm d)}m,
\label{general}
\end{equation}
\end{lem}

\begin{proof}
\begin{mdframed}[linewidth=1.5,topline=false,rightline=false,bottomline=false]

We consider optical unitary interferometers of size $m$ which take as input one single photon in a quantum state $\ket{\phi}$ and $m-1$ indistinguishable single photons in a state $\ket{\psi}$, one in each spatial mode, indexed from $1$ to $m$. The output modes are measured using photon number detection. A measurement outcome thus has the form $\bm d=(d_1,\dots,d_m)$, with \mbox{$d_1+\dots+d_m=m$}.

Recall that the permanent of an $m\times m$ matrix $A=(a_{ij})_{1\le i,j\le m}$ is defined by
\begin{equation}
\Per\,(A)=\sum_{\sigma\in S_m}{\prod_{k=1}^m{a_{k\sigma(k)}}},
\end{equation}
where $S_m$ is the symmetric group over $\{1,\dots,m\}$. We now compute Pr$_i(\bm d)$ and Pr$_d(\bm d)$ for all detection patterns $\bm d$.

In the indistinguishable case, $m$ indistinguishable photons, one in each mode, are sent through a linear optical network described by an $m\times m$ unitary matrix $U=(u_{ij})_{1\le i,j\le m}$. The probability of a detection event $\bm d$ can be computed as (see, section~\ref{sec:BosonSampling} and~\cite{Aaronson2013})
\begin{equation}
\text{Pr}_i(\bm d)=\frac{|\Per\,(U_{\bm d})|^2}{\bm d!},
\label{pru}
\end{equation}
where $\bm d!=d_1!\dots d_m!$ and where $U_{\bm d}$ is the matrix obtained from $U$ by repeating $d_k$ times the $k^{th}$ column for $k\in\{1,\dots,m\}$. 

In the distinguishable case, $m-1$ indistinguishable photons are sent in modes $2,\dots,m$ through a linear optical network described by an $m\times m$ unitary matrix $U=(u_{ij})_{1\le i,j\le m}$, along with one additional photon in the first mode in an orthogonal state. Since it is fully distinguishable from the others, the additional photon behaves independently, hence the probability of detecting the photon number pattern $\bm d$ for one distinguishable photon and $m-1$ indistinguishable photons in input is
\begin{equation}
\text{Pr}_d(\bm d)=\sum _{ \substack{k=1\\d_k\neq0} }^m{ \text{Pr}_i(\bm d-\bm1_k)\cdot\text{Pr}_i(\bm1_k) },.
\end{equation}
This last expression formalises the fact that the $m-1$ indistinguishable photons give a detection pattern \mbox{$\bm d-\bm1_k$} which, completed by the additional distinguishable photon in the $k^{th}$ output mode, forms the pattern $\bm d$. Developing this expression with Eq.~(\ref{pru}) yields
\begin{equation}
\text{Pr}_d(\bm d)=\frac { 1 }{ \bm d! } \sum _{ \substack{k=1\\d_k\neq0} }^{ m }{ d_{ k }|u_{ 1k }\Per\,(U_{ 1,\bm d-\bm1_{ k } })|^{ 2 } } 
\label{prd}
\end{equation}
where $U_{1,\bm d-\bm1_k}$ is the matrix obtained from $U$ by removing the first row, then by repeating $d_l$ times the $l^{th}$ column for $l\neq k$ and by repeating $d_k-1$ times the $k^{th}$ column.

In order to obtain more readable expressions, we define for all $k\in\{1,\dots,m\}$ and for any detection pattern $\bm d$,
\begin{equation}
p_k(\bm d)=\begin{cases}\frac{u_{1k}\Per\,(U_{1,\bm d-\bm1_k})}{\sqrt{\bm d!}} \text{ if } d_k\neq0, \\ 0\text{ otherwise.} \end{cases}
\end{equation}
Using the Laplace expansion of the permanent, the previous equations~(\ref{pru},~\ref{prd}) rewrite
\begin{equation}
\text{Pr}_i(\bm d)=\left|\sum_{k=1}^m{d_kp_k(\bm d)}\right|^2,
\label{pru2}
\end{equation}
and
\begin{equation}
\text{Pr}_d(\bm d)=\sum_{k=1}^m{d_k|p_k(\bm d)|^2}.
\label{prd2}
\end{equation}
Since $\sum_{k=1}^m{d_k}=m$, we obtain, using Cauchy-Schwarz inequality with the complex vectors $\left\{\sqrt{d_k}\right\}_{1\le k\le m}$ and $\left\{\sqrt{d_k}p_k(\bm d)\right\}_{1\le k\le m}$,
\begin{equation}
\text{Pr}_d(\bm d)\ge\frac{\text{Pr}_i(\bm d)}m,
\label{general}
\end{equation}
for any detection pattern $\bm d$. 

\end{mdframed}
\end{proof}

\noindent For all $\bm d$ we have
\be
\sum_{\substack{\bm d\\ \text{Pr}_i(\bm d)=0}}{\text{Pr}_d(\bm d)}+\sum_{\substack{\bm d\\ \text{Pr}_i(\bm d)\neq0}}{\text{Pr}_d(\bm d)}=1.
\label{completePrd}
\ee
Combining Lemma~\ref{lem:PrdPri} with Eqs.~(\ref{partialud2}) and (\ref{completePrd}) yields
\begin{equation}
\sum_{\substack{\bm d\\ \text{Pr}_i(\bm d)\neq0}}{\text{Pr}\,(\bm d)}\ge\left(\frac{1}m+\frac{m-1}m|\braket{\phi|\psi}|^2\right).
\label{boundpr}
\end{equation}
This last expression is valid for any interferometer and can be used it to retrieve, in the context of linear optics, the error probability bound for state identity testing under the one-sided error requirement obtained in Corollary~\ref{co:0}. Indeed, assume that $\bm d$ is a detection event such that Pr$_i(\bm d)\neq0$, which could be a disjoint union of multiple detection events, used for an identity test: if $\bm d$ is obtained we conclude that the states were identical (or equivalently that the photons were indistinguishable), otherwise we assume that the states were different (or equivalently that the first photon was distinguishable from the others). The one-sided error requirement can thus be written as $\sum_{{\bm d,\text{Pr}_i(\bm d)\neq0}}{\text{Pr}_i(\bm d)}=1$: indistinguishable photons always pass the test. For different input states $\ket{\phi}$ and $\ket{\psi}$, the error probability of the corresponding test is then given by $\sum_{{\bm d,\text{Pr}_i(\bm d)\neq0}}{\text{Pr}\,(\bm d)}$, which by Eq.~(\ref{boundpr}) is lower bounded by $\frac1m+\frac{m-1}m|\braket{\phi|\psi}|^2$.

\begin{figure}[h!]
\begin{center}
\includegraphics[width=2.3in]{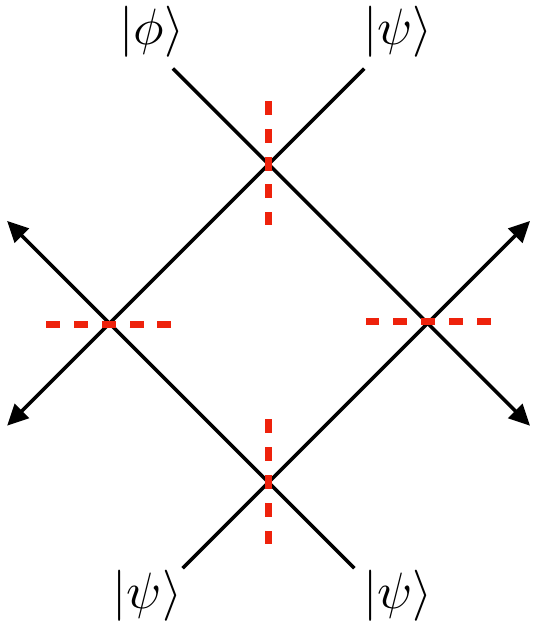}
\caption{Hadamard interferometer with 4 input modes. The dashed red lines represent balanced beam splitters. The input states are one single photon in state $\ket{\phi}$ and three single photons in state $\ket{\psi}$, one in each mode.}
\label{fig:4swap}
\end{center}
\end{figure}

We now study a particular unitary interferometer, when the size $m$ is a power of $2$: the Hadamard interferometer~\cite{crespi2015suppression,crespi2016suppression}. We show that it provides a simple implementation of the swap test of order $m$. For $m=4$ spatial modes (Fig.~\ref{fig:4swap}), this interferometer is described by the Hadamard-Walsh transform of order $2$:
\begin{equation}
\frac{1}{\sqrt{2}}\begin{pmatrix} H & H \\ H & -H \end{pmatrix} 
\end{equation}
where $H$ is a Hadamard matrix, see Eq.~(\ref{matrixH}).\\

\noindent In the general case, the Hadamard interferometer of order $m$ is described by the Hadamard-Walsh transform of order $n=\log m$, which is defined by induction:
\begin{equation}
H_{k+1}=\frac{1}{\sqrt{2}}\begin{pmatrix} H_k & H_k \\ H_k & -H_k \end{pmatrix},
\label{HWtransform}
\end{equation}
with $H_0=1$ and $H_1=H$. We can now state our main result linking the Hadamard interferometer and the swap test of order $m$.

\begin{theo}\label{th:Hn}
The output statistics of the Hadamard interferometer of order $m$ can be classically post-processed in time $O(m\log m)$ to reproduce those of the swap test of order $m$.
\end{theo}

\begin{proof}
\begin{mdframed}[linewidth=1.5,topline=false,rightline=false,bottomline=false]

Let us define
\begin{equation}
S=(s_{ij})_{1\le i,j\le m}=\sqrt{m} H_n,
\end{equation}
thus omitting the normalisation factor. We have
\begin{equation}
S=\underbrace { \sqrt{2}H\otimes \dots \otimes \sqrt{2}H }_{ n\text{ }times },
\end{equation}
where $H$ is a Hadamard matrix. The rows of $\sqrt{2}H$, together with the element-wise multiplication, form a group isomorphic to $\mathbb{Z}/2\mathbb{Z}$, thus the rows of $S$ together with the element-wise multiplication form a group isomorphic to $\left(\mathbb{Z}/2\mathbb{Z}\right)^n$. As a consequence, multiplying element-wise all the rows of $S$ by its $i^{th}$ row for a given $i$ amounts to permuting the rows of $S$. 

Let $\bm d=(d_1,\dots,d_m)$ and $k\in\{1,\dots,m\}$ such that $d_k\neq0$. Let also $S_{\bm d-\bm1_k}$ be the matrix obtained from $S$ by repeating $d_l$ times the $l^{th}$ column for $l\neq k$ and $d_k-1$ the $k^{th}$ column. For all $i\in\{1,\dots,m\}$, one can obtain the matrix $S_{1,\bm d-\bm1_k}$ (with the first row removed) from the matrix $S_{i,\bm d-\bm1_k}$  (with the $i^{th}$ row removed) by multiplying element-wise all rows by the $i^{th}$ row and permuting the rows. Since the permanent is invariant by row permutation we obtain, for all $i\in\{1,\dots,m\}$ and all $k\in\{1,\dots,m1\}$ such that $d_k\neq0$,
\begin{equation}
\Per\,(S_{i,\bm d-\bm1_k})=\epsilon_{ik}(\bm d)\Per\,(S_{1,\bm d-\bm1_k}),
\label{pmper}
\end{equation}
where $\epsilon_{ik}(\bm d)=s_{ik}\prod_{j=1}^m{\left(s_{ij}\right)^{d_j}}$. Let us define for all $\bm d=(d_1,\dots,d_m)$
\be
\pi(\bm d)=\sum_{i=1}^m{\prod_{j=1}^m{\left(s_{ij}\right)^{d_j}}}.
\label{pi}
\ee
We use the Laplace row expansion formula for the permanent of $S_{\bm d}$ to obtain, for all $\bm d=(d_1,\dots,d_m)$ and all $k\in\{1,\dots,m\}$ such that $d_k\neq0$,
\begin{equation}
\begin{aligned}
\Per\,(S_{\bm d})&=\sum_{i=1}^m{s_{ik}\Per\,(S_{i,\bm d-\bm d1_k})}\\
&=\left(\sum_{i=1}^m{s_{ik}\epsilon_{ik}(\bm d)}\right)\Per\,(S_{1,\bm d-\bm1_k})\\
&=\left(\sum_{i=1}^m{\prod_{j=1}^m{\left(s_{ij}\right)^{d_j}}}\right)\Per\,(S_{1,\bm d-\bm1_k})\\
&=\pi(\bm d)\Per\,(S_{1,\bm d-\bm1_k}),
\end{aligned}
\label{perpi}
\end{equation}
where we used Eq.~(\ref{pmper}) in the second line. With the general expressions of Pr$_i(\bm d)$~(\ref{pru}) and Pr$_d(\bm d)$~(\ref{prd}), this equation implies 
\begin{equation}
m\text{Pr}_i(\bm d)=\pi(\bm d)^2\text{Pr}_d(\bm d).
\label{prupiprd}
\end{equation}
With the Laplace column expansion formula for the permanent of $S_{\bm d}$ and the last line of Eq.~(\ref{perpi}), we also obtain
\begin{equation}
m^2\text{Pr}_i(\bm d)=\pi(\bm d)^2\text{Pr}_i(\bm d).
\label{pruprd}
\end{equation}
In particular, combining Eqs.~(\ref{prupiprd}) and (\ref{pruprd}),
\begin{equation}
m^2\pi(\bm d)^2\text{Pr}_d(\bm d)=\pi(\bm d)^4\text{Pr}_d(\bm d).
\end{equation}
Now Pr$_d(\bm d)$ is non-zero for all $\bm d$, since by Eq.~(\ref{prd}) it is a sum of moduli squared of permanents of $(2^n-1)\times(2^n-1)$ matrices, which in turn cannot vanish by a result of~\cite{simion1983+}. Hence the previous equation rewrites
\begin{equation}
m\pi(\bm d)=\pi(\bm d)^2.
\label{piM0}
\end{equation}
As a consequence, $\pi(\bm d)=m$ or $\pi(\bm d)=0$ for all $\bm d$. Combining Eqs.~(\ref{prupiprd}) and (\ref{piM0}) we obtain
\be
\ba
\pi(\bm d)\neq0&\Leftrightarrow\pi(\bm d)=m\\
&\Leftrightarrow \text{Pr}_i(\bm d)\neq0\\
&\Leftrightarrow\text{Pr}_d(\bm d)=\frac{\text{Pr}_i(\bm d)}m,
\ea
\ee
and thus 
\begin{equation}
\begin{aligned}
\text{Pr}_i[\pi(\bm d)=m]&=\sum_{\pi(\bm d)=m}{\text{Pr}_i(\bm d)}\\
&=\sum_{\text{Pr}_i(\bm d)\neq0}{\text{Pr}_i(\bm d)}\\
&=1.
\end{aligned}
\label{pru2}
\end{equation}
We also obtain
\begin{equation}
\begin{aligned}
\text{Pr}_d[\pi(\bm d)=m]&=\sum_{\pi(\bm d)=m}{\text{Pr}_d(\bm d)}\\
&=\frac1m\sum_{\pi(\bm d)=}{\text{Pr}_i(\bm d)}\\
&=\frac1m.
\end{aligned}
\label{prd2}
\end{equation}
We finally conclude by combining Eqs.~(\ref{pru2}), (\ref{prd2}) and (\ref{partialud2}):
\begin{equation}
\begin{aligned}
\Pr\,[\pi(\bm d)=m]&=\sum_{\pi(\bm d)=m}{\text{Pr}\,(\bm d)}\\
&=\frac1m+\frac{m-1}m\left|\braket{\phi|\psi}\right|^2.
\end{aligned}
\end{equation}
The post-processing (i.e., computing $\pi(\bm d)$) can be done efficiently in time $O(m\log m)$ for any detection pattern $\bm d=(d_1,\dots,d_m)$. Indeed, let $S_{\bm d}$ be the $m\times m$ matrix obtained from $S$ by repeating $d_k$ times the $k^{th}$ column for \mbox{$k\in\{1,\dots,m\}$}. The expression $\pi(\bm d)$ in Eq.~(\ref{pi}) is the sum of the product of the elements of each row of $S_{\bm d}$. Since the entries of the matrix $S$ are only $+1$ and $-1$, $\pi(\bm d)=m$ if and only if the number of $-1$ on the rows of $S_{\bm d}$ is even for all rows. The condition $\pi(\bm d)=m$ can thus be written as a system of $m$ linear equations modulo $2$. Since $\left(\mathbb{Z}/2\mathbb{Z}\right)^n$ is finitely generated by $n$ elements, the $m$ rows of $S_{\bm d}$ can be generated with at most $n$ rows using element-wise multiplication, for any measurement outcome $\bm d$. Hence, computing the parity of the number of $-1$ on each row of $S_{\bm d}$, which is equivalent to testing $\pi(\bm d)=m$, can be done by computing at most $n=\log m$ parity equations, with a number of terms in each equation which is at \mbox{most $m$.} 

A simple induction shows that a possible choice for the rows whose parity has to be tested is the rows with index $2^k+1$ for $k\in\{0,\dots,n-1\}$ (the rows of the matrix being indexed from $1$ to $m$).

\end{mdframed}
\end{proof}

\noindent Note that the group structure invoked in the proof is preserved under permutations, so Theorem~\ref{th:Hn} also applies to the unitary interferometers described by permutations of the Hadamard-Walsh transform.

The conclusion to be drawn from Theorem~\ref{th:Hn} is that as long as a state $\ket{\psi}$ can be encoded using single photons, then one can perform a swap test of order $m$ with respect to the state $\ket{\psi}$ using the Hadamard interferometer of order $m$ and an efficient classical post-processing of the measurement outcomes. The post-processing consists in the following parity test: given the measurement outcome $\bm d=(d_1,\dots,d_m)$, where $d_1+\dots+d_m= m $, construct the matrix $S_{\bm d}$ from the matrix $S=\sqrt{m} H_n$ by keeping the $k^{th}$ column only if $d_k$ is odd. If the rows $(2,3,5,\dots,2^{n-1}+1)$ of $S_{\bm d}$ all have an even number of $-1$, output $0$. Output $1$ otherwise. This means that the post-processing only requires the parity of the photon number in each output mode.

In particular, the photon number-resolving detectors can be replaced by detecting the parity of the number of photons in each output mode. Detecting this parity can for example be achieved with microwave technology~\cite{haroche2007measuring,vlastakis2013deterministically,sun2014tracking}. Also only $m-1$ detectors are necessary, since the parity of the number of photon in the remaining mode can be deduced from the parities of the other modes, given that the total number of photons is $m$. If only the parity is measured, the discrimination test is non-destructive and the remaining single-mode state is a mixture of either even or odd photon-number states, depending on the measured parity and the total number of photons.

\begin{figure}[h!]
\begin{center}
\includegraphics[width=5in]{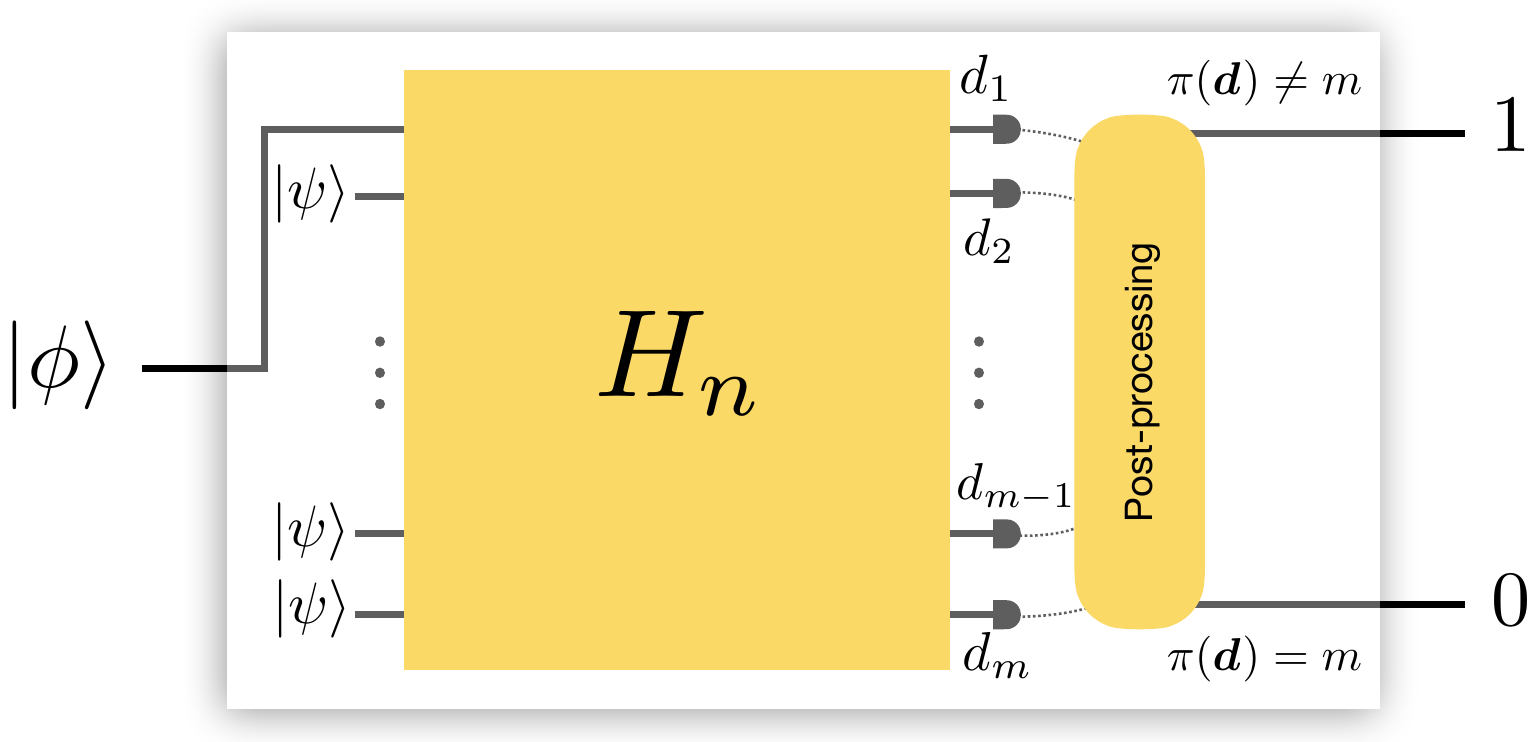}
\caption{The Hadamard interferometer of order $m$ used as a programmable projective measurement device. A single photon in the state $\ket{\phi}$ goes through a linear interferometer along with $m-1$ indistinguishable single photons in the state $\ket{\psi}$. The parity of the number of photons in each output mode is measured and efficiently post-processed, such that the device outputs $0$ with probability $\frac1m+\frac{m-1}m|\braket{\phi|\psi}|^2$ and $1$ with probability $\frac{m-1}m(1-|\braket{\phi|\psi}|^2)$.}
\label{fig:setup}
\end{center}
\end{figure}

Using the argument developed in the proof of Theorem~\ref{th:circuit}, by considering the $m-1$ photons and the interferometer as a black box (Fig.~\ref{fig:setup}) whose outcomes are post-processed as described above, we also deduce the following result from Theorem~\ref{th:Hn}:
\begin{coro} The Hadamard interferometer of order $m$ can be used to perform a projective measurement with error $\frac1m$, using a classical post-processing of its measurement outcomes that takes time $O(m\log m)$.
\label{co:2}
\end{coro}

\noindent Interestingly, the unitary interferometers described by the Hadamard-Walsh transform and its permutations are not the only unitary interferometers which can reproduce the statistics of a swap test with efficient post-processing, and indeed we present a generalisation in the next section. However, it is the simplicity of the Hadamard interferometer in terms of experimental implementation that motivates our interest towards this interferometer. In particular, this interferometer can be simply implemented with a few balanced beam splitters. A result by Reck \textit{et al.}~\cite{reck1994experimental} states that any $m\times m$ unitary interferometer can be implemented using phase shifters and at most $\frac{m(m-1)}{2}$ beam splitters, possibly unbalanced. For the Hadamard interferometer, less beam splitters are needed and no phase shifters:

\begin{lem}\label{lem:fewBS}
The $m\times m$ Hadamard interferometer can be implemented using $\frac{m\log m}2$ balanced beam splitters.
\end{lem}

\begin{proof}
\begin{mdframed}[linewidth=1.5,topline=false,rightline=false,bottomline=false]

The size $m$ is a power of $2$, with $n=\log m$. We prove by induction over $n$ that there exist $P_0(n),\dots,P_{n-1}(n)$ permutation matrices of order $m/2$, such that
\begin{equation}
H_n=\prod _{ k=0 }^{ n-1 }{ { P }_{ k }(n)\left( { \mathbb1 }_{ m/2 }\otimes { H } \right) { { P } }_{ k }(n)^{ T } }.
\label{hyprec}
\end{equation}
Since multiplying matrices is equivalent to setting up experimental devices in sequence, and given that $H$ is the matrix describing a balanced beam splitter, Eq.~(\ref{hyprec}) implies the result we want to prove.

\medskip

\noindent For $n=1$, we have $m=2$ and Eq.~(\ref{hyprec}) is true with $P_0(1)=\mathbb1_1$.  For brevity, we define for all $k$
\begin{equation}
H^{(k)} ={ \mathbb1 }_{ k }{ \otimes} H.
\end{equation}
Assuming that Eq.~(\ref{hyprec}) is true for $n$, we use the recursive definition of the Hadamard-Walsh transform
\begin{equation}
H_{n+1} = H\otimes H_n,
\end{equation}
along with properties of the tensor product of matrices in order to obtain
\begin{align}
\nonumber H_{n+1} &= \left( { H }_{ n }\otimes { \mathbb1}_{ 2 } \right) H^{(m)}\\ 
\nonumber&= Q\left( { \mathbb1 }_{ 2 }\otimes { H }_{ n } \right) { Q }^{ T }H^{(m)}\displaybreak\\ 
&= Q\left[ { \mathbb1 }_{ 2 }\otimes \prod _{ k=0 }^{ n-1 }{ { P }_{ k }(n)H^{(m/2)} { { P } }_{ k }(n )^{ T } }  \right] { Q }^{ T }H^{(m)}\\ 
\nonumber&= Q\left[ \prod _{ k=0 }^{ n-1 }{ \left( { \mathbb1 }_{ 2 }\otimes { P }_{ k }(n) \right) H^{(m)} \left( { \mathbb1 }_{ 2 }\otimes { { P } }_{ k }(n)^{ T } \right)  }  \right] { Q }^{ T }H^{(m)} \\ 
\nonumber&= \prod _{ k=0 }^{ n-1 }{ \left[ Q\left( { \mathbb1 }_{ 2 }\otimes { P }_{ k }(n) \right)  \right] H^{(m)} { \left[ { Q }\left( { \mathbb1 }_{ 2 }\otimes { { P } }_{ k }(n) \right)  \right]  }^{ T } }H^{(m)},
\end{align}
where $Q$ is a permutation matrix of order $m$ and where in the third line we have used Eq.~(\ref{hyprec}). Setting $P_k(n+1)=Q\left( { \mathbb1 }_{ 2 }\otimes { P }_{ k }(n)\right)$ for $k\in\{0,\dots,n-1\}$ and $P_{n}(n+1)=\mathbb1_m$ proves Eq.~(\ref{hyprec}) for $n+1$, since these matrices are permutation matrices of order $m$. This completes the induction and the proof of the result.

\end{mdframed}
\end{proof}

\subsection{Group generalisation}

Using the Hadamard interferometer requires the size parameter $m$ to be a power of $2$. This requirement can be relaxed, possibly raising the experimental requirements at the same time. Indeed, for any value of $m$, one can associate to any abelian group of order $m$ an interferometer of size $m$ which gives the desired statistics. This is the object of the following result that uses the invariant factor decomposition of an abelian group:
\begin{theo} Let $\mathcal G$ be an abelian group of order $m$. Then, there exists $n\in\mathbb{N}^*$ and $a_1,\dots,a_n\in\mathbb{N}^*$, where $a_i|a_{i+1}$ for $i\in\{1,\dots,n-1\}$ and $a_1\cdots a_n=m$, such that the interferometer described by the $m\times m$ unitary matrix
\begin{equation}
U_{\mathcal G}:=\frac{1}{\sqrt{m}}F_{a_1}\otimes\dots\otimes F_{a_n},
\end{equation}
where $F_a=(e^{\frac{2i\pi}{a}(k-1)(l-1)})_{1\le k,l\le a}$ is the quantum Fourier transform of order $a$ for all $a\in\mathbb{N}^*$, can perform a $\frac1m$-approximate projective measurement with a post-processing of its measurement outcomes that takes time at most $mn$. The rows of $F_{\mathcal G}:=\sqrt{m}U_{\mathcal G}$ together with the element-wise multiplication form a group isomorphic to $\mathcal G$.
\label{th:group}
\end{theo}

\begin{proof}
\begin{mdframed}[linewidth=1.5,topline=false,rightline=false,bottomline=false]

We use the notations of the theorem. The invariant factor decomposition of $\mathcal G$ gives
\begin{equation}
\mathcal G\simeq \left(\mathbb{Z}/a_1\mathbb{Z}\right)\otimes\dots\otimes\left(\mathbb{Z}/a_n\mathbb{Z}\right),
\end{equation}
where $n\in\mathbb{N}^*$ and $a_1,\dots,a_n\in\mathbb{N}^*$ are unique, satisfying $a_i|a_{i+1}$ for $i\in\{1,\dots,n-1\}$ and $a_1\cdots a_n=m$. Given that the rows of $F_a$ together with the element-wise multiplication form a group isomorphic to $\left(\mathbb{Z}/a\mathbb{Z}\right)$ for all $a\in\mathbb{N}^*$, the rows of $F_{\mathcal G}=(f_{ij})_{1\le i,j\le m}=\sqrt{m}U_{\mathcal G}$ together with the element-wise multiplication form a group isomorphic to $\mathcal G$.

Since the group structure was the only argument invoked in the proof of Theorem~\ref{th:Hn}, the same conclusion can be drawn here, by following the same argument: for any detection event $\bm d=(d_1,\dots,d_m)$,
\begin{equation}
\Pr\,[\pi(\bm d)=m]=\frac1m+\frac{m-1}m\left|\braket{\phi|\psi}\right|^2,
\end{equation}
where
\begin{equation}
\pi(\bm d)=\sum_{i=1}^m{\prod_{j=1}^m{\left(f_{ij}\right)^{d_j}}}.
\end{equation}
The group $\mathcal G$ is finitely generated by $n$ elements, so $n$ rows of $F_{\mathcal G}$ are sufficient to generate all its rows by element-wise multiplication. The condition $\pi(\bm d)=m$ can thus be checked in time $O(mn)$. 

\end{mdframed}
\end{proof}

\noindent In particular, for $\mathcal G\simeq(\mathbb{Z}/m\mathbb{Z})$, the corresponding interferometer is described by the (normalised) quantum Fourier transform of order $m$, while for $\mathcal G\simeq(\mathbb{Z}/2\mathbb{Z})^n$, we retrieve Theorem~\ref{th:Hn} and the Hadamard interferometer.


\section{Programmable projective measurements with coherent states}
\label{sec:univ2}

The previous scheme for performing programmable measurements with linear optics requires creation and manipulation of high-dimensional superposition states. In order to simplify the experimental requirements, we adapt this scheme to an encoding of quantum states in coherent states of light. Since coherent states are natural realisations of states produced by lasers, they can be efficiently produced and manipulated experimentally. The coherent state scheme takes as input a generic single-mode continuous variable quantum state, the test state, and $m-1$ copies of a coherent state in the program registers, and approximates the projective measurement $\{\ket{\beta}\!\bra{\beta}, \mathbb1-\ket{\beta}\!\bra{\beta}\}$ on the input state in a single run, using only threshold detectors. In particular, we obtain a more faithful projective measurement using coherent states than using a single-photon encoding.

In what follows, we introduce three different schemes for performing state discrimination and programmable projective measurement with coherent states: the \emph{Hadamard scheme}, the \emph{merger scheme}, and the \emph{looped merger scheme}. Further, we give the proof for the optimality of the coherent state projective measurement performed by all three schemes, under the one-sided error requirement.

\subsection{Coherent state discrimination}

\begin{figure}[h!]
\includegraphics[width=3in]{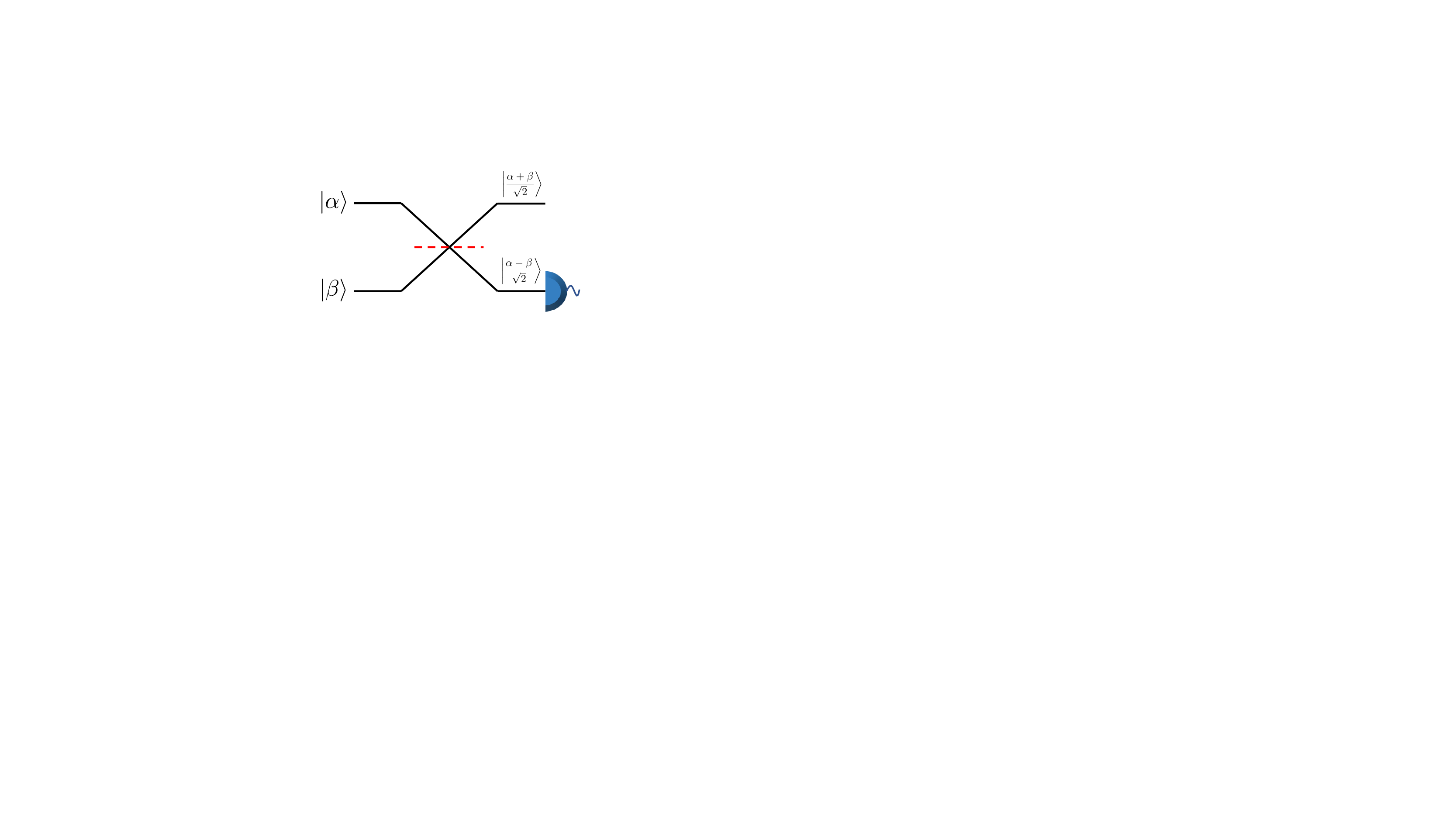}
\centering
\caption{Balanced beam splitter operation on input states $\ket{\alpha}$ and $\ket{\beta}$. The second output mode is measured with a single-photon threshold detector. The probability of obtaining a click relates to the distinguishability of the two unknown coherent states.}
\label{fig:OpticalSwaptest}
\end{figure}

The swap test discriminates between two unknown states. If the unknown states are coherent states instead, then an analogous test can be performed by mixing the states on a balanced beam splitter and measuring the lower output branch with a single-photon threshold detector (Fig.~\ref{fig:OpticalSwaptest}).\\

\noindent A beam splitter maps the input modes creation operators $\{\hat{a}^{\dagger},\hat{b}^{\dagger}\}$ to the output modes creation operators $\{\hat{c}^{\dagger},\hat{d}^{\dagger}\}$ as
\begin{equation}
\begin{split}
\hat{a}^{\dagger} & \rightarrow \frac{1}{\sqrt{2}}(\hat{c}^{\dagger}+\hat{d}^{\dagger}), \\
\hat{b}^{\dagger} & \rightarrow \frac{1}{\sqrt{2}}(\hat{c}^{\dagger}-\hat{d}^{\dagger}).
\end{split}
\end{equation} 
The input state at the beam splitter is
\begin{equation}
\ket{\alpha}_a\otimes \ket{\beta}_b,
\end{equation}
where the subscripts denote the mode in which the coherent states enter the beam splitter.
In the absence of experimental imperfections, this yields the output state
\begin{equation}\label{output}
\left|\frac{\alpha + \beta}{\sqrt{2}}\right\rangle_c\otimes \left|\frac{\alpha - \beta}{\sqrt{2}}\right\rangle_d,
\end{equation}
for a balanced beam splitter. The probability of obtaining a click in the detector on the second mode is given by
\begin{equation}
1-\exp\left(-\frac{|\alpha - \beta|^2}{2}\right)=1-|\langle\alpha|\beta\rangle|.
\label{D1}
\end{equation}
We now consider the scenario when one receives a single copy of an unknown coherent state $\ket{\alpha}$ and the objective is to check wether if the test state is equal to the reference coherent state $\ket{\beta}$. Here one has access to multiple copies of the reference state but is limited to just a single copy of the test state. In the simpler case, the state discrimination can be performed with a single copy of the reference state, like in the previous section. This succeeds with a probability given by Eq.~(\ref{D1}). In this section, we prove that having multiple copies of the reference state $\ket{\beta}$ increases the success probability of discriminating with the test state $\ket{\alpha}$. For this, we first provide a generalised interferometer construction, the \emph{Hadamard scheme}, based on Hadamard transformations, following the previous section. We then show how this interferometer can be simplified, thanks to coherent state encoding and we introduce the \emph{merger scheme} and the \emph{looped merger scheme}.

\subsection{The Hadamard scheme}

We consider the Hadamard interferometer over $m$ modes, where $m$ is a power of $2$. The input is now composed of coherent states: 
\begin{equation}
\ket{\alpha}_1\otimes\ket{\beta}_2\otimes\dots\otimes\ket{\beta}_{m},
\end{equation}
where the subscript denotes the mode in which the coherent state enters the generalised interferometer. For brevity, we address this state as $\ket{\alpha\beta...\beta}$. All the output modes but the first are measured with single-photon threshold detectors.

With Eq.~(\ref{HWtransform}), the Hadamard interferometer of order $m$ is described by the Hadamard-Walsh transform of order $n=\log m$, which is defined by:
\begin{equation}
H_{n}= H^{\otimes n},
\label{interferometern}
\end{equation}
with $H_0=(1)$ and $H_1=H$.
The input coherent states $\ket{\alpha\beta\dots\beta}$ upon interaction with the interferometer of order $m$ transform as,
\be
\ket{\alpha\beta\dots\beta}\mapsto H_n\ket{\alpha\beta\dots\beta}=\ket{\delta_1\delta_2\dots\delta_m},
\ee
where, with a simple induction, we obtain $\delta_1=\frac{\alpha+(m-1)\beta}{\sqrt m}$ and $\delta_k=\frac{\alpha-\beta}{\sqrt m}$ for $k>1$. Thus the last $m-1$ modes have the same probability of a click when measured with single-photon threshold detectors. The probability $\Pr\,[\emptyset]$ that none of the $m-1$ detectors clicks is,
\be
\begin{aligned}
\text{Pr}_{\alpha,\beta,m}[\emptyset]&=\prod_{k=1}^{m-1}{(1-\Pr\,[\text{click in $k^{\text{th}}$ mode}])}\\
&=\prod_{k=1}^{m-1}{[1-(1 - \exp(-|\delta_k|^2))]}\\
&=\exp\left(-\frac{m-1}m|\alpha-\beta|^2\right)\\
&=(|\braket{\alpha|\beta}|^2)^{1-\frac{1}{m}}.
\end{aligned}
\label{Eq:Noclick}
\ee
In particular, for all $\alpha,\beta\in\mathbb{C}$, $\Pr_{\alpha,\beta,+\infty}[\emptyset]=|\braket{\alpha|\beta}|^2$, which corresponds to a perfect projective measurement of the states $\ket{\alpha}$ and $\ket{\beta}$.  Writing $x=|\braket{\alpha|\beta}|^2$ the overlap of the test and reference states, we obtain
\be
\text{Pr}_{x,m}[\emptyset]=x^{1-\frac{1}{m}}.
\ee
Assigning to the event `none of the detectors clicks' the value $0$ and to other detection events (`at least one of the $m-1$ detectors clicks') the value $1$, we obtain a device whose statistics approach those of a projective measurement, with
\be
\text{Pr}_{x,m}[0]=1-\text{Pr}_{x,m}[1]=x^{1-\frac{1}{m}}.
\ee
With Theorem~\ref{th:Hn} and Eq.~(\ref{sswap}), for an $m$-mode input state $\ket{\phi\psi\dots\psi}$ the corresponding statistics with single-photon encoding are
\be
\text{Pr}_{x,m}[0]=1-\text{Pr}_{x,m}[1]=\frac{1}{m}+\left(1-\frac{1}{m}\right)x.
\ee
The single-photon encoding implies having $m-1$ number-resolving or parity detectors. On contrary, the encoding with coherent states requires $m-1$ single-photon threshold detectors. Experimentally, this is relatively easier to implement. The test based on coherent state also satisfies the one-sided error requirement:
if the states are the same, then their trace distance is $0$ and hence the probability of having the detection event $1$ is $0$.
Moreover, coherent state encoding provides a more faithful projective measurement than single-photon encoding. Indeed, the statistics produced by coherent state encoding are closer to the ones of a perfect projective measurement. For any given value of the overlap $x$:
\be
\forall x\in[0,1],\quad x\le x^{1-\frac{1}{m}}\le\frac{1}{m}+\left(1-\frac{1}{m}\right)x.
\ee
In particular, for a given size $m$, the maximal statistical gap with a perfect projective measurement is,
\be
\begin{aligned}
e_{\text{SP}}(m)&=\max_{x\in[0,1]}{\left|\left[\frac{1}{m}+\left(1-\frac{1}{m}\right)x\right]-x\right|}\\
&=\frac{1}{m},
\end{aligned}
\ee
for the single-photon encoding, and
\be
\begin{aligned}
e_{\text{CS}}(m)&=\max_{x\in[0,1]}{\left|\left( x^{1-\frac{1}{m}}\right)-x\right|}\\
&=\frac{(m-1)^{m-1}}{m^m}\\
&\sim\frac{1}{e}\cdot\frac{1}{m},
\end{aligned}
\ee
for the coherent state encoding, which is lower than the single-photon encoding gap.
This happens because for the single-photon encoding no assumption is made about the states $\ket\phi$ and $\ket\psi$, while the states $\ket\alpha$ and $\ket\beta$ are assumed to be coherent states. This additional information about the states allows us to better approximate a perfect projective measurement with the same number of input states. We show in the next section that there exists a simpler measurement setting than the Hadamard interferometer, achieving the same performance in the test, due to coherent state encoding.

\subsection{The merger scheme} 
\label{ampscheme}

\begin{figure}[h!]
\begin{center}
\includegraphics[width=5in]{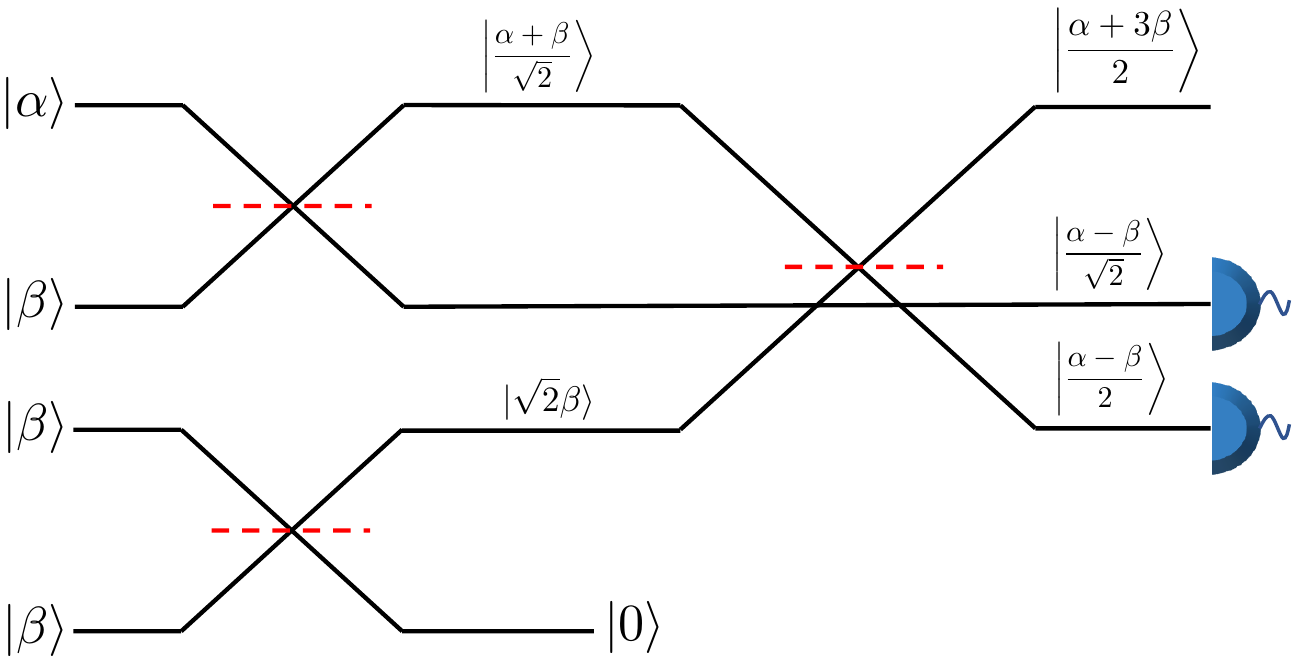}
\caption{Merger scheme with 4 input modes. The input states are one tested state $\ket{\alpha}$ and three local states $\ket{\beta}$, one in each mode. The detectors are single-photon threshold detectors.}
\label{fig:simpl}
\end{center}
\end{figure}

\noindent By Lemma~\ref{lem:fewBS}, the Hadamard scheme of size $m$ uses $\frac{m\log m}2$ balanced beam splitters and $m-1$ single-photon threshold detectors. We introduce a simplified scheme over the same number of modes $m$, which only uses $m-1$ balanced beam splitters and $\log m$ detectors, and show that it achieves the same performance than the Hadamard scheme. We refer to this scheme as the \textit{merger} scheme, since it merges identical input coherent states into an amplified coherent state in the first output mode and the vacuum in all other modes.

For $m=4$ spatial modes, this interferometer acting on modes $\{1,2,3,4\}$ is described by the following unitary matrix:
\be
\ba
U_4&=\begin{pmatrix}\frac{1}{2}&\frac{1}{2}&\frac{1}{2}&\frac{1}{2}\\ \frac{1}{\sqrt2}&-\frac{1}{\sqrt2}&0&0\\ \frac{1}{2}&\frac{1}{2}&-\frac{1}{2}&-\frac{1}{2}\\0&0&\frac{1}{\sqrt2}&-\frac{1}{\sqrt2}\end{pmatrix}\\
&=\begin{pmatrix}\frac{1}{\sqrt2}&0&\frac{1}{\sqrt2}&0\\0&1&0&0\\ \frac{1}{\sqrt2}&0&-\frac{1}{\sqrt2}&0\\0&0&0&1\end{pmatrix}\times\begin{pmatrix}\frac{1}{\sqrt2}&\frac{1}{\sqrt2}&0&0\\ \frac{1}{\sqrt2}&-\frac{1}{\sqrt2}&0&0\\0&0&\frac{1}{\sqrt2}&\frac{1}{\sqrt2}\\0&0&\frac{1}{\sqrt2}&-\frac{1}{\sqrt2}\end{pmatrix}\\
&=H_{1,3}\times(H_{1,2}\oplus H_{3,4}),
\ea
\label{U2}
\ee
where $H_{i,j}$ corresponds to the balanced beam splitter operation acting on modes $i$ and $j$ (where the modes are indexed from $1$ to $m$) and identity on the other modes (Fig.~\ref{fig:simpl}). 

\begin{figure}[h!]
\begin{center}
\includegraphics[width=5.3in]{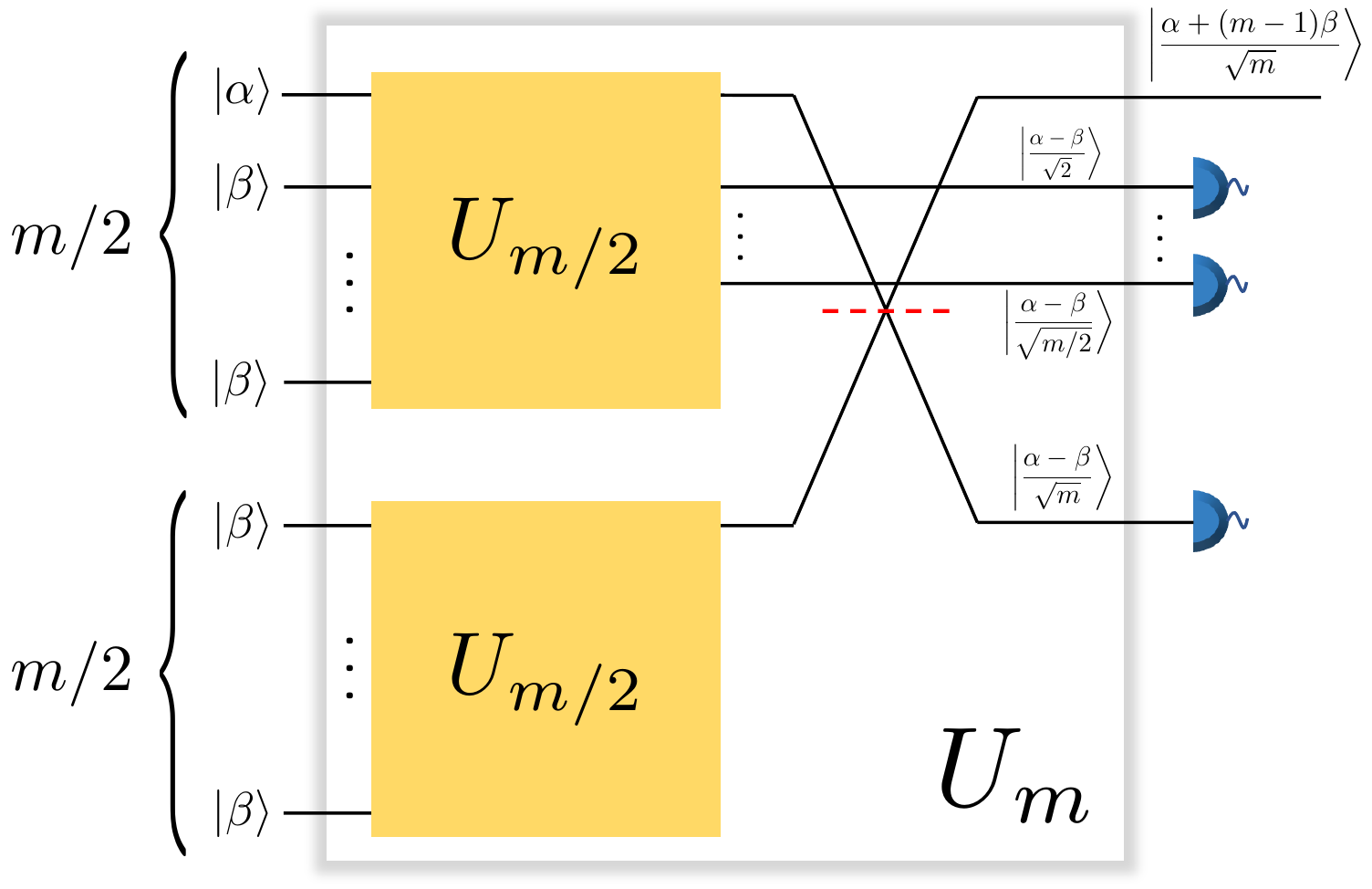}
\caption{General merger scheme of size $m$, with one copy of $\ket{\alpha}$ and $m - 1$ copies of $\ket{\beta}$: the first output modes of two interferometers described by $U_{m/2}$ are mixed on a balanced beam splitter. Indexing the spatial modes from $1$ to $m$, the $2^k$+1 output modes are measured with single-photon threshold detectors, for $k = 0\dots n-1$. }
\label{fig:induction}
\end{center}
\end{figure}

The generalised merger interferometer is defined by induction:
\be
U_m=H_{1,m/2+1}\times(U_{m/2}\oplus U_{m/2}),
\label{U}
\ee
where $U_1=H_{0,1}=H$ is a Hadamard matrix. This induction relation is illustrated in Fig.~\ref{fig:induction}.

Indexing the spatial modes from $1$ to $m$, the $2^k+1$ output modes are measured with single-photon threshold detectors, for $k=0\dots n-1$. A simple induction shows that the output state in the $2^k+1$ output mode is $\left|\frac{\alpha-\beta}{\sqrt{2^{k+1}}}\right\rangle$.
Hence, the probability that none of the $n=\log m$ detectors clicks is given by
\be
\begin{aligned}
\text{Pr}_{\alpha,\beta,m}[\emptyset]&=\prod_{k=0}^{n-1}{(1-\Pr\,[\text{click in the }2^kth\text{ mode}])}\\
&=\prod_{k=0}^{n-1}{\left[1-\left(1 - \exp\left(-\left|\frac{\alpha-\beta}{2^{\frac{k+1}{2}}}\right|^2\right)\right)\right]}\\
&=\exp\left(-\sum_{k=0}^{n-1}{\left(\frac{1}{2}\right)^{k+1}}|\alpha-\beta|^2\right)\\
&=\exp\left(-\frac{m-1}{m}|\alpha-\beta|^2\right)\\
&=(|\braket{\alpha|\beta}|^2)^{1-\frac{1}{m}},
\end{aligned}
\label{Eq:Noclick2}
\ee
thus retrieving the statistics obtained with the Hadamard scheme, using only $n=\log m$ detectors. Moreover, a simple induction shows that the merger interferometer can be implemented with only $m-1$ balanced beam splitters.

\medskip

\begin{figure}[h!]
\begin{center}
\includegraphics[width=3.3in]{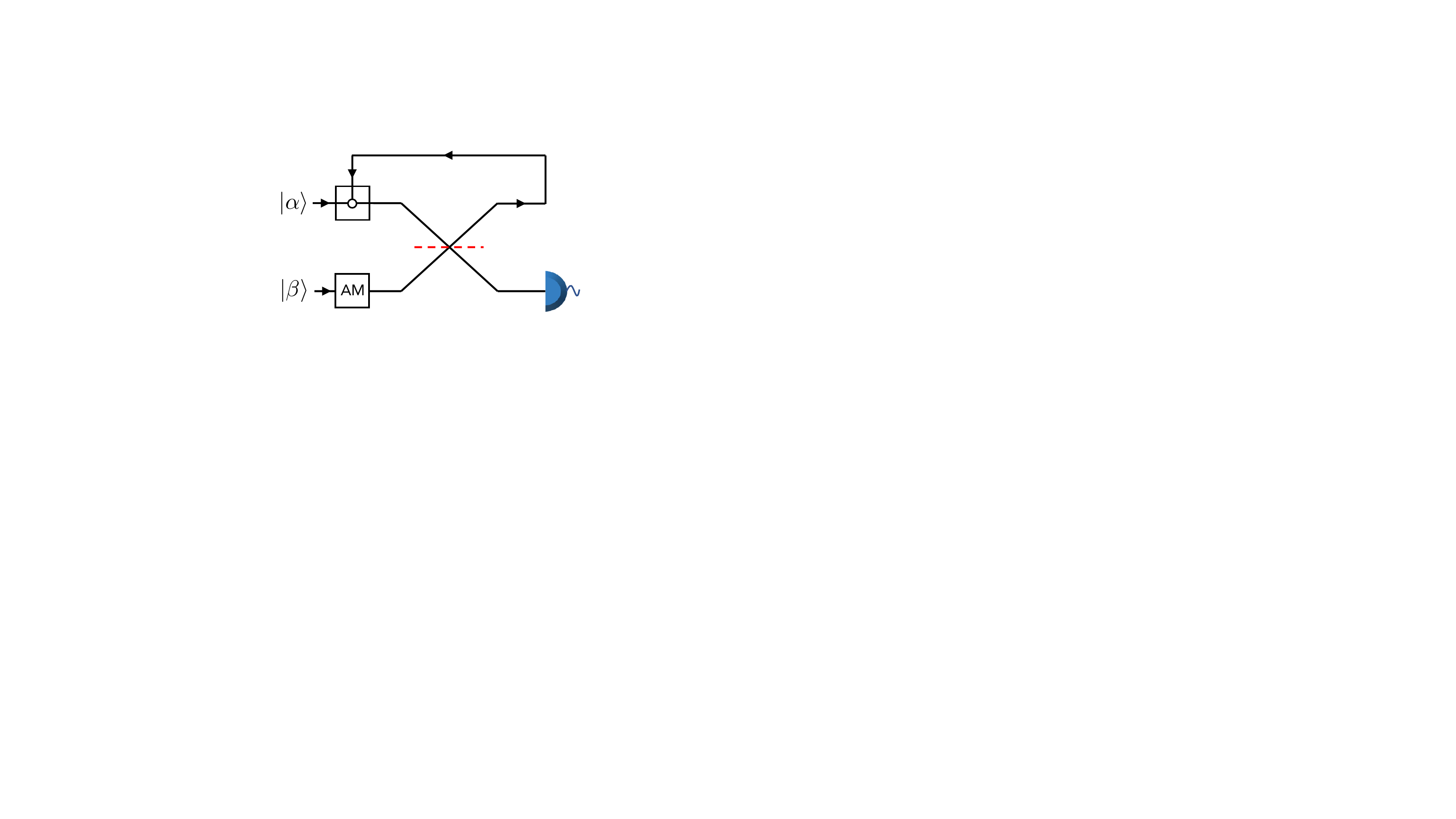}
\caption{The looped merger scheme. The coherent pulse $\ket\alpha$ is sent once, while at each loop a new coherent pulse $\ket\beta$ is sent. An optical switch ensures a closed loop after the first pulse $\ket\alpha$ passes through. An amplitude modulator (AM) transforms the $k$-th pulse $\ket\beta$ to $\ket{\sqrt{2^k}\beta}$, $k$ starting at $0$.}
\label{fig:loop}
\end{center}
\end{figure}

\noindent Noting the recursive character of the merger scheme, we present another possible implementation of the merger scheme using a looped beam splitter interaction, one single-photon threshold detector and an active optical element, namely an active amplitude modulator (Fig.~\ref{fig:loop}).\\

\noindent This setup now uses an active optical element and a constant number of passive linear optical elements and approximates a perfect projective measurement up to arbitrary precision. By construction, the statistics of the setup after $m-1$ pulses $\ket\beta$ sent reproduce those of the merger scheme of size $m$.

\medskip

\noindent The three schemes discussed provide experimentally-friendly devices to perform a variety of quantum information processing tasks using coherent states, ranging from state discrimination to programmable projective measurements, in a non-destructive manner. These schemes are also optimal for coherent state discrimination:

\begin{theo}
The Hadamard interferometer and the merger interferometer are optimal for coherent states discrimination, under the one-sided error requirement.
\end{theo}

\begin{proof}
\begin{mdframed}[linewidth=1.5,topline=false,rightline=false,bottomline=false]

The proof extends results from~\cite{sedlak2007unambiguous}. We first start by deriving the optimal POVM for discriminating coherent states under the one-sided error requirement.

\medskip

\noindent Let $\{\Pi_0,\Pi_1\}$ be a POVM for discriminating coherent states $\ket\alpha$ and $\ket\beta$ under the one-sided error requirement, when provided a single copy of $\ket\alpha$ and $m-1$ copies of $\ket\beta$ (the proof of~\cite{sedlak2007unambiguous} assumes $m=2$). The operator $\Pi_0$ corresponds to saying that the states $\ket\alpha$ and $\ket\beta$ are the same, while the operator $\Pi_1$ corresponds to saying that they are different. These operators thus verify the following conditions:
\begin{equation}
\Pi_0,\Pi_1\ge\mymathbb0,\text{ }\Pi_0+\Pi_1=\mathbb 1,
\label{cond0}
\end{equation}
and
\be
\forall\alpha\in\mathbb{C},\Tr\,\left[\Pi_1\ket\alpha\bra\alpha^{\otimes m}\right]=0,
\label{cond1}
\ee
where the last condition is the one-sided error requirement. Integrating this condition over $\mathbb{C}$ yields
\begin{equation}
0=\int{d^2\alpha\Tr\left[\Pi_1\ket\alpha\bra\alpha^{\otimes m}\right]}=\Tr\left[\Pi_1\Delta_m\right],
\label{cond2}
\end{equation}
where we have defined
\begin{equation}
\Delta_m=\int{d^2\alpha\ket\alpha\bra\alpha^{\otimes m}}\ge0.
\label{Delta1}
\end{equation}
Note that the condition in~(\ref{cond2}) is equivalent to the one-sided requirement in~(\ref{cond1}) because the operators $\Pi_1$ and $\ket\alpha\bra\alpha^{\otimes m}$ are positive. 

The operator $\frac{m}{\pi}\Delta_m$ is actually a projector. This result can be obtained by writing the state $\ket\alpha$ in the Fock basis and an integration in polar coordinates, where $\alpha = re^{i\theta}$, as follows:
writing
\be
\ket\alpha=e^{-\frac{|\alpha|^2}{2}}\sum_{k=0}^{+\infty}{\frac{\alpha^k}{\sqrt{k!}}\ket k},
\ee
we obtain
\begin{equation}
\begin{aligned}
\Delta_m&=\int{d^2\alpha\text{ } \exp[-m|\alpha|^2]\sum_{\substack{k_j,l_j=0\\\forall j\in [m]}}^{\infty}{\frac{\alpha^{\sum_j k_j}(\alpha^*)^{\sum_j l_j}}{\sqrt{k_1!\dots k_m!l_1!\dots l_m!}}\ket{k_1\dots k_m}\bra{l_1\dots l_m}}}\\
&=\sum_{\substack{k_j,l_j=0}}^{\infty}\frac{\ket{k_1\dots k_m}\bra{l_1\dots l_m}}
{\sqrt{k_1!\dots k_m!l_1!\dots l_m!}}\int_{r=0}^{\infty}{dr\text{ }\exp[{-mr^2}]r^{1+\sum_j k_j+l_j}}\int_{\theta=0}^{2\pi}{d\theta\text{ }\exp[{i\theta\sum_j (k_j-l_j)}]}\\
&=\frac{\pi}{m}\sum_{\substack{k_j,l_j=0}}^{\infty}\frac{\delta_{\sum_j k_j,\sum_j l_j}}{m^{\frac{\sum_j k_j}{2}}m^{\frac{\sum_j l_j}{2}}}\sqrt{\frac{(\sum_j k_j)!(\sum_j l_j)!}{k_1!\dots k_m!l_1!\dots l_m!}}\ket{k_1\dots k_m}\bra{l_1\dots l_m}\\
&=\frac{\pi}{m}\sum_{p=0}^{\infty}{\sum_{\substack{\sum_j k_j=p\\\sum_j l_j=p}}{m^{-p}\sqrt{\frac{p!}{k_1!\dots k_m!}}\sqrt{\frac{p!}{l_1!\dots l_m!}}\ket{k_1\dots k_m}\bra{l_1\dots l_m}}}\\
&=\frac{\pi}{m}\sum_{p=0}^{\infty}{\ket{\chi_p^m}\bra{\chi_p^m}},
\end{aligned}
\label{Delta2}
\end{equation}
where we have defined for all $p\ge0$,
\begin{equation}
\ket{\chi_p^m}=m^{-p/2}\sum_{\sum_j k_j=p}{\sqrt{\frac{p!}{k_1!\dots k_m!}}\ket{k_1\dots k_m}}.
\label{chi}
\end{equation}
With the multinomial formula, we obtain $\braket{\chi_p^m|\chi_p^m}=1$ for all $p\ge0$, and since the states $\ket{\chi_p^m}$ have exactly $p$ photons, we have $\braket{\chi_p^m|\chi_{q}^m}=\delta_{pq}$ for all $p,q\ge0$. The states $\ket{\chi_p^m}$ thus are orthonormal and with Eq.~(\ref{Delta2}), the operator $\frac{m}{\pi}\Delta_m$ is a projector.

By Eq.~(\ref{cond2}), the supports of $\Pi_1$ and $\frac{m}{\pi}\Delta_m$ are disjoint, and by Eq.~(\ref{cond1}) we have $\Pi_0+\Pi_1=\mathbb 1$, so the support of $\frac{m}{\pi}\Delta_m$ is included in the support of $\Pi_0$. The optimal POVM $\{\Pi_0^{opt},\Pi_1^{opt}\}$ for state discrimination minimises the error probability, hence with the one-sided error requirement $\Pi_0^{opt}$ must have minimal support, meaning that
\begin{equation}
\Pi_0^{opt}=\frac{m}{\pi}\Delta_m=\sum_{p=0}^{+\infty}{\ket{\chi_p^m}\bra{\chi_p^m}}\quad\text{and}\quad\Pi_1^{opt}=\mathbb 1-\Pi_0^{opt}.
\label{POVMopt}
\end{equation}
Note that, with the same proof, this choice of POVM is also optimal in the generalised setting where one is given one unknown generic state and $m-1$ unknown coherent states, and is asked to test if all the states are identical or not.

\medskip

\noindent We now show that the POVM $\{\Pi_0^h,\Pi_1^h\}$ corresponding to the Hadamard interferometer with a threshold detection of the last $m-1$ modes is optimal for coherent state discrimination under the one-sided error requirement, i.e., that
\begin{equation}
\Pi_0^h=\Pi_0^{opt},
\end{equation}
where $\Pi_0^{opt}$ is defined in Eq.~(\ref{POVMopt}). We have
\begin{equation}
\Pi_0^h=\hat H_n^\dag\Pi_0^d\hat H_n,
\end{equation}
where $\hat H_n$ is the unitary evolution corresponding to the action of the interferometer of order $m$ defined in Eq.~(\ref{interferometern}), with $n=\log m$, and $\Pi_0^d=\mathbb1\otimes\ket0\bra0^{\otimes m-1}$ is the POVM operator corresponding to the event where none of the $m-1$ threshold detectors clicks. We obtain
\begin{equation}
\begin{aligned}
\Pi_0^h&=\hat H_n^\dag\left(\mathbb1\otimes\ket0\bra0^{\otimes m-1}\right)\hat H_n\\
&=\sum_{p=0}^{+\infty}{\tilde H_n^\dag\left(\ket{p}\bra{p}\otimes\ket0\bra0^{\otimes m-1}\right)\hat H_n}.
\end{aligned}
\label{Pi0h}
\end{equation}
For $k=1\dots m$, we write $a_k^\dag$ the creation operator for the $k^{th}$ mode. For all $p\ge0$ we have
\begin{align}
\nonumber\hat H_n^\dag\left(\ket{p}\otimes\ket0^{\otimes m-1}\right)&=\frac{1}{\sqrt{p!}}\hat H_n^\dag (\hat a_1^\dag)^p\ket0^{\otimes m}\\
\nonumber&=\frac{1}{\sqrt{p!}}(\hat H_n^\dag\hat a_1^\dag \hat H_n)^p\ket0^{\otimes m}\\
\nonumber&=\frac{m^{-p/2}}{\sqrt{p!}}(\hat a_1^\dag+\dots+\hat a_m^\dag)^p\ket0^{\otimes m}\displaybreak\\
&=\frac{m^{-p/2}}{\sqrt{p!}}\sum_{k_1+\dots+k_m=p}{\frac{p!}{k_1!\dots k_m!}(\hat a_1^\dag)^{k_1}\dots(\hat a_m^\dag)^{k_m}}\ket0^{\otimes m}\\
\nonumber&=m^{-p/2}\sum_{k_1+\dots+k_m=p}{\sqrt{\frac{p!}{k_1!\dots k_m!}}\ket{k_1\dots k_m}}\\
\nonumber&=\ket{\chi_p^m},
\end{align}
where we have used $\hat H_n\ket0^{\otimes m}=\ket0^{\otimes m}$, $\hat H_n^\dag \hat H_n=\mathbb 1$, $\hat H_n^\dag\hat a_1^\dag \hat H_n=\frac{\hat a_1^\dag+\dots+\hat a_m^\dag}{\sqrt m}$, the multinomial formula, and Eq.~(\ref{chi}). With Eqs.~(\ref{POVMopt}) and (\ref{Pi0h}), this concludes the proof.

Given that the statistics obtained with the merger scheme and the looped merger scheme mimic those of the Hadamard scheme, these schemes are also optimal for the same discrimination task.

\end{mdframed}
\end{proof}

\noindent While these devices are relatively easy to implement, any implementation will suffer from experimental imperfections. In the next section, we investigate how such imperfections affect the performance of the merger scheme, for $m=4$ modes.

\subsection{Experimental imperfections}

In this section, we analyse the performance of the merger scheme in presence of experimental imperfections. Our error model is the following, with three major sources of error: $(i)$ the limited detector efficiency and channel transmission loss, characterized by a parameter $0\leq\eta\leq1$, which changes the coherent state $\ket\alpha$ to $\ket{\sqrt{\eta}\alpha}$ thus reducing the probability of obtaining a click using a single-photon threshold detector by a factor $\eta$; $(ii)$ the limited beam-splitter visibility $0\leq\nu\leq1$, which may lead to a click in the wrong detector, and $(iii)$ the dark count in the detectors characterized by a probability $p_{dark}$. For our analysis, the click probability due to the coherent states is of $O(1)$ and thus significantly larger than the dark count probability $p_{dark}\sim 10^{-8}$. The dark counts can thus be safely ignored.\\

\begin{table}[h!]
\centering
\setlength{\tabcolsep}{8pt}
\begin{tabular}{||c | c c c ||}
\hline
 & $\eta$ & $\nu$ & $p_{dark}$ \\
\hline
Exp. & $0.9$ & $(98.8 \pm 0.3)\%$ & $(1 \pm 0.1)*10^{-8}$\\
\hline
\end{tabular}
\caption{Table illustrating the experimental parameters used in simulation of our results. The dark-count rate, $p_{dark}$, achievable with super-conducting detectors~\cite{schweickert2018demand}. The standard values are set-up efficiency, $\eta$, and beam splitter visibility $\nu$ are from \cite{kumar2019experimental}.}
\label{TableExpSWAP}
\end{table}

\noindent For $m=2$, when the input $\ket{\alpha}$, $\ket{\beta}$ is fed in an imperfect beam splitter, the transformation from input modes $\{\hat{a}^{\dagger},\hat{b}^{\dagger}\}$ into the output modes $\{\hat{c}^{\dagger},\hat{d}^{\dagger}\}$, is the following:
\begin{equation}
\ket{\alpha}_a\otimes\ket{\beta}_b\mapsto\Ket{\sqrt\nu\frac{\alpha+\beta}{\sqrt2}+\sqrt{1-\nu}\frac{\alpha-\beta}{\sqrt2}}_c\otimes\Ket{\sqrt\nu\frac{\alpha-\beta}{\sqrt2}+\sqrt{1-\nu}\frac{\alpha+\beta}{\sqrt2}}_d.
\end{equation}
The corresponding unitary transformation is
\begin{equation}
H'=\frac{1}{\sqrt2}\begin{pmatrix} A&B\\A&-B \end{pmatrix},
\label{Eq:hadamard}
\end{equation}
where $A = \sqrt{\nu} + \sqrt{1 - \nu}$, and $B = \sqrt{\nu} - \sqrt{1 - \nu}$.

We consider the case of $m = 4$ spatial modes (Fig~\ref{fig:simpl}), indexed from $1$ to $4$. We apply the imperfect transformation on the input $\ket{\alpha\beta\beta\beta}$. This results in
\be
\begin{aligned}
\ket{\alpha\beta\beta\beta}\mapsto U'_4\ket{\alpha\beta\beta\beta}&=\ket{\delta_1\delta_2\delta_3\delta_4},
\end{aligned}
\ee
where from Eq.~(\ref{U2}) we derive
\be
\ba
U'_2&=H'_{1,3}\times(H'_{1,2}\oplus H'_{3,4})\\
&=\begin{pmatrix}\frac{1}{2}A^2&\frac{1}{2}AB&\frac{1}{2}AB&\frac{1}{2}B^2\\ \frac{1}{\sqrt2}A&-\frac{1}{\sqrt2}B&0&0\\ \frac{1}{2}A^2&\frac{1}{2}AB&-\frac{1}{2}AB&-\frac{1}{2}B^2\\0&0&\frac{1}{\sqrt2}A&-\frac{1}{\sqrt2}B\end{pmatrix},
\ea
\ee
with $A=\sqrt\nu+\sqrt{1-\nu}$ and $B=\sqrt\nu-\sqrt{1-\nu}$.
We obtain $\delta_2=\frac{A\alpha-B\beta}{\sqrt2}$, and $\delta_3 = \frac{A^2\alpha-B^2\beta}{2}$. 
Adding the channel and detector losses $\eta$, the output is mapped as $\delta_k \mapsto\sqrt{\eta}\delta_k$, for all $k$.

Similar to the analysis without experimental imperfection, we detect the output modes $1$ and $2$ of the imperfect merger interferometer, with the coherent state input being $\ket{\alpha\beta\beta\beta}$. The probability that none of the two detectors clicks is given by
\be
\exp\big(-\eta(|\delta_2|^2 + |\delta_3|^2)\big).
\ee
Assigning to the detection event \textit{no detector clicks} the value $0$, and to other detection events, i.e., \textit{at least one of the detectors clicks}, the value $1$, we obtain a device whose statistics approximate those of a projective measurement.

When the states are the same, the completeness, which is the probability of not obtaining the detection event $1$ is
\begin{equation}
c_4^{exp} =  \exp(-2\eta(1-\nu)(1+2\nu)|\alpha|^2).
\label{4completeexp}
\end{equation}
We observe that if $\nu = 1$ (no imperfections), then $c_4^{exp} = 1$, thus we obtain perfect completeness. For the imperfection values of Table~\ref{TableExpSWAP}, the value of $c_4^{exp}$ is close to $1$ for small $|\alpha|^2$ values. 

The analogous completeness for $m = 2$ is
\begin{equation}
c_2^{exp} =  \exp(-2\eta(1 - \nu)|\alpha|^2).
\label{2completeexp}
\end{equation}
From Eq.~(\ref{2completeexp}) and Eq.~(\ref{4completeexp}), we observe that $c_2^{exp} \le c_4^{exp}$, which implies that the completeness for the $m = 4$ scheme is less than the completeness for the $m = 2$ scheme. The reduction in completeness probability for the $m=4$ scheme is precisely what accounts for a lower failure probability when the local and reference states are different, which we detail in the next paragraph.

\medskip

\noindent If the states are different, the probability of obtaining the detection event $1$ (soundness) is given by:

\begin{lem}
\be
\ba
s_4^{exp}=1-\exp\Big[(4\nu^2-1)|\alpha-\beta|^2&+4\left[(1+2\nu)(1-\nu)+2\sqrt{\nu(1-\nu)}\right]|\alpha|^2\\
&+4\left[(1+2\nu)(1-\nu)-2\sqrt{\nu(1-\nu)}\right]|\beta|^2\bigg)\Big].
\ea
\label{Eq:Soundness}
\ee
\end{lem}

\begin{proof}
\begin{mdframed}[linewidth=1.5,topline=false,rightline=false,bottomline=false]

When the states are different, the probability of obtaining the detection event $0$ (failure probability) is
\be
1-s_4=\exp{\left(-\frac{\eta}{4}A\right)},
\ee
where
\be
\ba
A&=2\left|\sqrt\nu(\alpha-\beta)+\sqrt{1-\nu}(\alpha+\beta)\right|^2+\left|\alpha-\beta+2\sqrt{\nu(1-\nu)}(\alpha+\beta)\right|^2\\
&=(1+2\nu)|\alpha-\beta|^2+2(1+2\nu)(1-\nu)|\alpha+\beta|^2+8\sqrt{\nu(1-\nu)}(|\alpha|^2-|\beta|^2),
\ea
\ee
where we used $(\alpha-\beta)(\alpha+\beta)^*+(\alpha-\beta)^*(\alpha+\beta)=2|\alpha|^2-2|\beta|^2$. Using $|\alpha+\beta|^2=2|\alpha|^2+2|\beta|^2-|\alpha-\beta|^2$ we obtain
\be
A=(4\nu^2-1)|\alpha-\beta|^2+4\left[(1+2\nu)(1-\nu)+2\sqrt{\nu(1-\nu)}\right]|\alpha|^2+4\left[(1+2\nu)(1-\nu)-2\sqrt{\nu(1-\nu)}\right]|\beta|^2.
\ee

\end{mdframed}
\end{proof}

\noindent The analogous soundness in $m=2$ experimental imperfection scheme is
\begin{equation}
s_2^{exp} =1 - \exp\bigg[-\eta\bigg(\nu - \frac{1}{2}\bigg)|\alpha-\beta|^2 -\eta\bigg(1 - \nu + \sqrt{\nu(1 - \nu)}\bigg)|\alpha|^2-\eta\bigg(1 - \nu - \sqrt{\nu(1 - \nu)}\bigg)|\beta|^2\bigg].
\end{equation}
We then obtain:

\begin{lem}\label{lem:s2s4} For all experimental parameters,
\be
s_2^{exp}\le s_4^{exp}.
\ee
\end{lem}

\begin{proof}
\begin{mdframed}[linewidth=1.5,topline=false,rightline=false,bottomline=false]

We have
\begin{equation}
\begin{aligned}
s_2^{exp} ={} & 1 - \exp\bigg[-\eta\bigg(\nu - \frac{1}{2}\bigg)|\alpha-\beta|^2 -\eta\bigg(1 - \nu + \sqrt{\nu(1 - \nu)}\bigg)|\alpha|^2 \\
&\hspace{17mm} -\eta\bigg(1 - \nu - \sqrt{\nu(1 - \nu)}\bigg)|\beta|^2\bigg]\\
&:=1-\exp[-\eta A_2],
\end{aligned}
\end{equation}
and
\begin{equation}
\begin{aligned}
s_4^{exp} ={} & 1 - \exp\bigg[-\eta\bigg(\nu^2-\frac{1}{4}\bigg)|\alpha-\beta|^2 \\
&\hspace{15mm} -\eta\bigg((1+2\nu)(1-\nu)+2\sqrt{\nu(1-\nu)}\bigg)|\alpha|^2 \\
&\hspace{15mm} -\eta\bigg((1+2\nu)(1-\nu)-2\sqrt{\nu(1-\nu)}\bigg)|\beta|^2\bigg]\\
&:=1-\exp[-\eta A_4].
\end{aligned}
\end{equation}
Since the function $x\mapsto 1-e^{-x}$ is increasing, it is sufficient to show that $A_2\le A_4$ for all $\alpha,\beta$. Writing $\alpha=re^{i\phi}$ and $\beta=te^{i\psi}$, where $r,t\ge0$ and $\phi,\psi\in[0,2\pi]$, we obtain
\begin{equation}
\begin{aligned}
A_4-A_2&=\bigg(\frac{1}{4}+\nu(1-\nu)+\sqrt{\nu(1-\nu)}\bigg)r^2 \\
&\hspace{5mm} + \bigg(\frac{1}{4}+\nu(1-\nu)-\sqrt{\nu(1-\nu)}\bigg)t^2\\
&\hspace{5mm} -2rt\bigg(\frac{1}{4}-\nu(1-\nu)\bigg)\cos(\phi-\psi).
\end{aligned}
\end{equation}
This last expression is a polynomial of degree $2$ in $r$, with a positive leading coefficient. Thus if its discriminant is negative, then the expression is always positive. The discriminant is
\be
\ba
\Delta&=4t^2\bigg[\bigg(\frac{1}{4}-\nu(1-\nu)\bigg)^2\cos(\phi-\psi)^2-\bigg(\frac{1}{4}+\nu(1-\nu)+\sqrt{\nu(1-\nu)}\bigg)\bigg(\frac{1}{4}+\nu(1-\nu)-\sqrt{\nu(1-\nu)}\bigg)\bigg]\\
&\le-6t^2\nu(1-\nu)\\
&\le0,
\ea
\ee
where the second line is obtained by using $\cos(\phi-\psi)\le1$.
Hence for all experimental parameters within the error model we consider, we have $s_2^{exp}\le s_4^{exp}$.

\end{mdframed}
\end{proof}

\noindent Hence, the experimental $m=4$ scheme outperforms the $m=2$ scheme in soundness for all values of the noise parameters. On the other hand, the completeness of the scheme suffers from experimental imperfections.


\section{Discussion and open problems}

We have identified a connection between unknown quantum state discrimination and quantum-programmable measurements. We have presented an optimal scheme for a programmable projective measurement device, and a linear optical implementation, with the Hadamard interferometer and single-photon encoding, which is straightforward and efficient. 
This could for example be used to design a photonic circuit which would act as a universal projective measurement device for a broad range of potential applications from quantum information and cryptography to tests of contextuality.

Our scheme can also be interpreted as an optimal swap test when one has a single copy of one state, and $m-1$ of the other. 
We have chosen to phrase the problem in terms of $m-1$ copies of the state $|\psi\rangle$. In principle we could have chosen any other encoding of the quantum input into $m-1$ registers. 
The reason for this choice is twofold. Firstly it is part of the envisaged problem setting---we imagine a device producing states encoding our measurement, for example these could be the output of a computation. Secondly we do so in order to separate as much as possible the resource of $m-1$ program systems and the process of translating them into a measurement. In particular if one had any other encoding, for example into some entangled states, this encoding process could be incorporated into the circuit representing the generic measurement apparatus. In this sense the most quantum information that can be contained about the state $|\psi\rangle$ in $m-1$ systems is $m-1$ copies of the state $|\psi\rangle$---anything more can be done afterwards.
This result also provides a natural interpretation of the notion of projective measurement in quantum mechanics, as a comparison between one state and several copies of another state using an interferometer: in the macroscopic limit, when many copies of a reference eigenstate are available, we retrieve a macroscopic classically programmable quantum measurement set up.

In order to reduce the experimental requirements, we have also presented an optimal programmable measurement scheme that projects the incoming single mode state in the test register into a local coherent state basis of the program registers. Our scheme is implemented using balanced beam splitters and single-photon threshold detectors. Threshold detectors with high efficiency and ultra low dark counts are commercially available \cite{schweickert2018demand}. Additionally, the numbers of detectors needed is logarithmic in the size of the interferometer, which itself is composed only of a linear number of balanced beam splitters.
This implementation using coherent states can act as a backbone in improving the performance of a range of quantum protocols in communication complexity \cite{buhrman2001quantum, de2004one}, cryptography and computational regimes \cite{aaronson2008power, mintert2005concurrence, walborn2006experimental, harrow2013testing, ekert2002direct, lloyd2013quantum}.

For completeness, it would be interesting to characterise the full class of interferometers that are optimal for state identity testing under the one-sided error requirement, as we only gave a broad class of such interferometers using a group construction. It would be also interesting to consider the influence of real experimental conditions in a more general setting.


\clearemptydoublepage
%
%
\let\textcircled=\pgftextcircled
\chapter{Quantum weak coin flipping with linear optics}
\label{chap:WCF}

\initial{W}eak coin flipping is among the fundamental cryptographic primitives which ensure the security of modern communication networks. It allows two mistrustful parties to remotely agree on a random bit when they favor opposite outcomes. Unlike other two-party computations, one can achieve information-theoretic security using quantum mechanics only: both parties are prevented from biasing the flip with probability higher than $1/2+\epsilon$, where $\epsilon$ is arbitrarily low. Classically, the dishonest party can always cheat with probability $1$ unless computational assumptions are used. Despite its importance, no physical implementation has been proposed so far for quantum weak coin flipping. 

In this chapter, we present a practical protocol for quantum weak coin flipping that requires a single photon and linear optics only. We show that it is secure even when threshold single-photon detectors are used, and reaches a bias as low as $\epsilon=1/\sqrt{2}-1/2\approx 0.207$. We further show that the protocol may display quantum advantage over a few hundred meters with state-of-the-art technology.

This chapter is based on \cite{bozzio2020quantum}.


\section{Weak coin flipping protocol with linear optics}

Compared to weak coin flipping, where two mistrustful parties wish to remotely agree on the outcome of a coin flip when they favor different outcomes, the cryptographic task of strong coin flipping corresponds to the case where they want to agree on an unbiaised random bit when they do not necessarily favor a particular outcome. Despite its name, strong coin flipping is less general than weak coin flipping in the sense that optimal strong coin flipping protocols may be designed which use weak coin flipping protocols as a subroutine~\cite{IEEE:CK09}.

While quantum strong coin flipping protocols have been experimentally demonstrated \cite{MTV:PRL05,BBB:NC11,PJL:NC14}, no implementation has been proposed for quantum weak coin flipping. This may be explained by two reasons. First, it is difficult to find an encoding and implementation which is robust to losses: a dishonest party may always declare an abort when they are not satisfied with the flip's outcome. Second, none of the proposed quantum weak coin flipping protocols~\cite{SR:PRL02,KN:IPL04,M:FOCS04,M:PRA05,M:arx07,ACG:SIAM16,ARW:arx18,ARV:arx19} translate trivially into a simple experiment: they all involve performing single-shot generalized measurements or generating beyond-qubit states.

We introduce a family of quantum weak coin flipping protocols, inspired by \cite{SR:PRL02}, which achieve biases as low as $\epsilon = 1/\sqrt{2}-1/2\approx 0.207$. Our protocols involve simple projective measurements instead of generalized ones, require a single photon and linear optics only, and need at most three rounds of communication between the parties. The information is encoded by mixing a single photon with vacuum on an unbalanced beam splitter, which generates entanglement \cite{MBH:PRL13}: both parties may then agree on a random bit, while the entanglement is simultaneously verified. This encoding is very robust to noise, as the single photon need not be pure or indistinguishable from other photons in any degree of freedom, save photon number.
We also use a version of our schemes to construct a quantum strong coin flipping protocol with bias $\approx 0.31$. We further derive a practical security proof for both number-resolving and threshold single-photon detectors, considering the extension to infinite-dimensional Hilbert spaces. Since the presence of losses may enable classical protocols to reach lower cheating probabilities than quantum protocols, we finally show that our quantum protocol bears no classical equivalent over a few hundred meters of lossy optical fiber and non-unit detection efficiency.

\medskip

\begin{figure}[h!]
\begin{center}
\includegraphics[width=1\columnwidth]{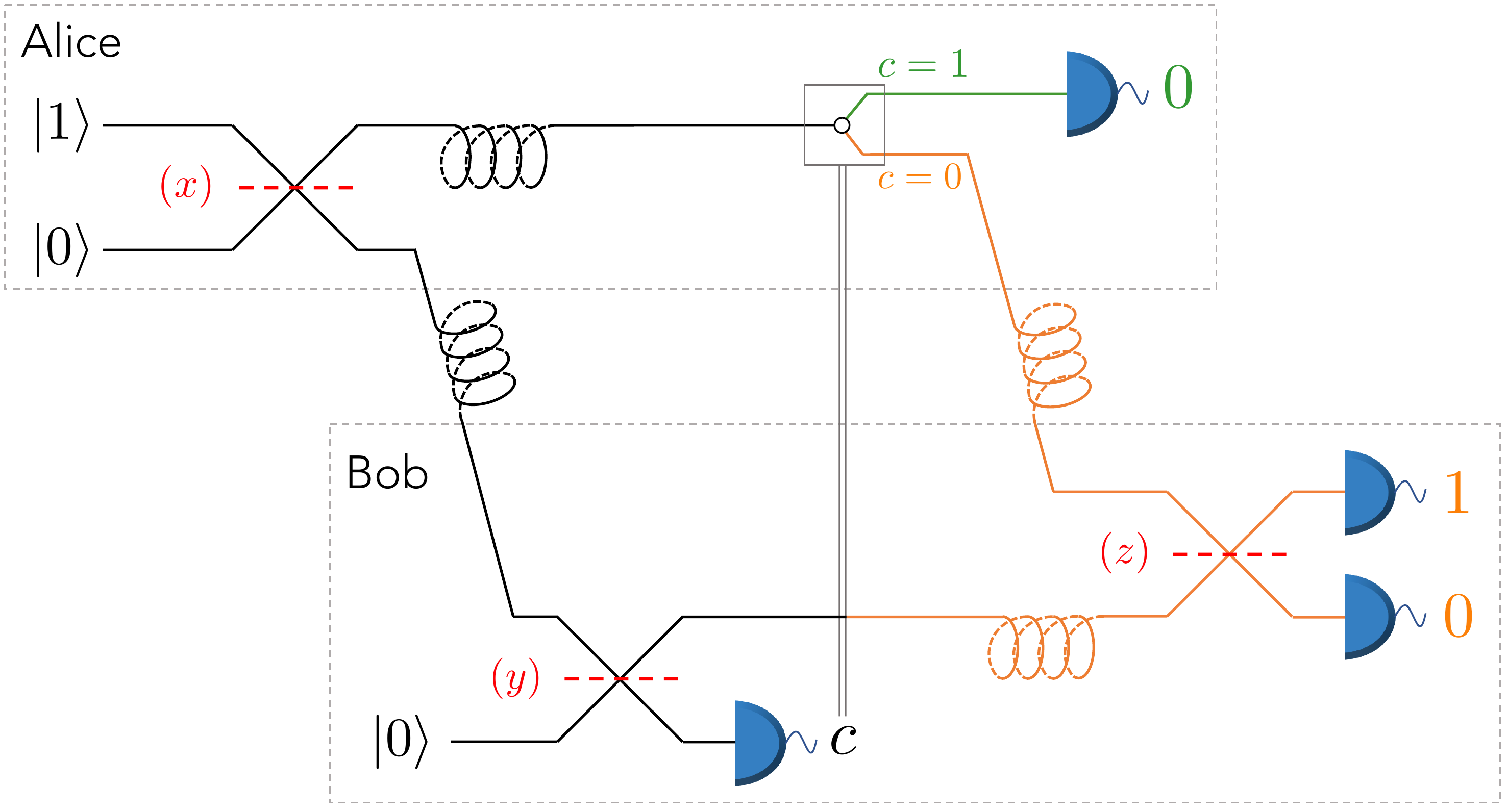}
\caption{\textbf{Representation of the honest protocol.} The dashed boxes indicate Alice and Bob's laboratories, respectively. Dashed red lines represent beam splitters, with the reflectance indicated in red. $\ket{0}$ and $\ket{1}$ are the vacuum and single photon Fock states, respectively. Curly lines represent fiber used for quantum communication from Alice to Bob, or delay lines within Alice's or Bob's laboratory, when waiting for the other party's communication. Bob broadcasts the classical outcome $c$, which controls an optical switch on Alice's side. The protocol when Bob declares $c=0/1$ is represented in orange/green. The final outcomes are the expected outcomes when both parties are honest.}
\label{fig:protocol}
\end{center}
\end{figure}

\noindent In the honest protocol, Alice and Bob wish to toss a fair coin, with a priori knowledge that they each favor opposite outcomes. Fig.~\ref{fig:protocol} represents the implementation of the honest protocol, which follows five distinct steps. Defining $x\in[0,\frac12]$ as a free protocol parameter, these read:
\begin{itemize}
\item 
Alice mixes a single photon with the vacuum on a beam splitter of reflectance $x$.
\item 
Alice keeps the first half of the state obtained, and sends the second half to Bob.
\item 
Bob mixes the half he receives with the vacuum on a beam splitter of reflectance $y=1-\frac1{2(1-x)}$.
\item 
Bob measures the second register of his state with a single-photon detector, and broadcasts the outcome $c\in\{0,1\}$.
\item 
The last step is a verification step, which splits into two cases. If $c=0$, Alice sends her half of the state to Bob, who mixes it with his half on a beam splitter of reflectance $z=2x$. He then measures the two output modes with single-photon detectors. He declares Alice the winner if the outcome $(1,0)$ is obtained. If $c=1$: Bob discards his half, and Alice measures her half with a single-photon detector. If the outcome is $(0)$, Bob is declared winner.
\end{itemize}

\subsection{Completeness}

In what follows, we let the parameters $x,y,z$ vary freely, and derive the relations these parameters need to satisfy to enforce a honest protocol without abort cases. We show that for the specific relations indicated above the protocol is fair, i.e., the probability of winning for each party is $\frac12$ when they are both honest.

Single photons are quantized excitations of the electromagnetic field, which are described by the action of the creation operator onto the vacuum. Beam splitters act linearly on creation operators and leave invariant the vacuum. More precisely, a beam splitter of reflectance $r$ acting on modes $k,l$ maps the creation operators $\hat a_k^\dag$ and $\hat a_l^\dag$ of the input modes onto $\hat b_k^\dag$ and $\hat b_l^\dag$, where
\be
\begin{pmatrix}\hat b_k^\dag\\\hat b_l^\dag\end{pmatrix}=H_{kl}^{(r)}\begin{pmatrix}\hat a_k^\dag\\\hat a_l^\dag\end{pmatrix},
\ee
with
\be
H_{kl}^{(r)}=\begin{pmatrix}\sqrt r&\sqrt{1-r}\\\sqrt{1-r} & -\sqrt r\end{pmatrix}.
\ee
Hence, the evolution of the quantum state over the three modes up to Bob's first measurement reads:
\be
\ba
\ket{100}&\underset{(x),12}\to\sqrt{x}\ket{100}+\sqrt{1-x}\ket{010}\\
&\underset{(y),23}\to\sqrt{x}\ket{100}+\sqrt{(1-x)y}\ket{010}\\
&\quad\quad\quad\quad+\sqrt{(1-x)(1-y)}\ket{001},
\ea\label{eq:evolution2}
\ee
where the notation $(r),kl$ indicates the reflectance of the beam splitter and the corresponding spatial modes.
The probability that Bob obtains outcome $c=1$ when measuring the third register thus is $\Pr\,[1]=(1-x)(1-y)$, while the probability of outcome $c=0$ is $\Pr\,[0]=1-\Pr\,[1]$. Setting $y=1-\frac1{2(1-x)}$ ensures $\Pr\,[0]=\Pr\,[1]=\frac12$. 

When $c=1$, the state on modes $1$ and $2$ is projected onto $\ket{00}$, while $c=0$ projects the state onto $\sqrt{2x}\ket{10}+\sqrt{1-2x}\ket{01}$. In the first case, the measurement performed by Alice outputs $(0)$ with probability $1$. In the second case, the measurement performed by Bob outputs $(1,0)$ with probability $1$ when 
\be
z=\frac{x}{1-(1-x)(1-y)}=2x.
\label{eq:z}
\ee
In that case, the probability that Alice (resp.\@ Bob) wins is directly given by $P_h^{(A)} = \Pr\,[0]$ (resp.\@ $P_h^{(B)} = \Pr\,[1]$). This shows that the protocol is fair, since $\Pr\,[0]=\Pr\,[1]=\frac12$.

\medskip

\noindent In the following, we make use of a simple reduction which allows us to simplify calculations in the proofs:

\begin{lem}\label{lem:reductionUV}
Let $U=(H^{(z)}\otimes \mathbb{1})(\mathbb{1}\otimes H^{(y)})$, with $z>0$. For all density matrices $\tau$,
\begin{equation}
\Tr\,[(\tau\otimes\ket0\bra0)U^\dag(\mathbb{1}\otimes\ket{00}\bra{00})U]=\Tr\,[(\tau\otimes\ket0\bra0)V^\dag(\ket{0}\bra{0}\otimes \mathbb{1}\otimes\ket{0}\bra{0})V],
\end{equation}
where $V=(\mathbb{1}\otimes H^{(b)})(H^{(a)}\otimes \mathbb{1})(\mathbb{1}\otimes R(\pi)\otimes \mathbb{1})$, with $a=\frac{y(1-z)}{1-(1-y)(1-z)}$ and $b=1-(1-y)(1-z)$, and $R(\pi)$ a phase shift of $\pi$ acting on mode $2$.
\end{lem}

\begin{proof}
\begin{mdframed}[linewidth=1.5,topline=false,rightline=false,bottomline=false]

The action of $U$ on the creation operators is given by
\be
\ba
U&=\begin{pmatrix}\sqrt z&\sqrt{1-z}&0\\\sqrt{1-z}&-\sqrt z&0\\0&0&1\end{pmatrix}\begin{pmatrix}1&0&0\\0&\sqrt y&\sqrt{1-y}\\0&\sqrt{1-y}&-\sqrt y\end{pmatrix}\\
&=\begin{pmatrix}\sqrt z&\sqrt{y(1-z)}&\sqrt{(1-y)(1-z)}\\\sqrt{1-z}&-\sqrt{yz}&-\sqrt{(1-y)z}\\0&\sqrt{1-y}&-\sqrt y\end{pmatrix}.
\ea
\ee
Linear interferometers map product coherent states onto product coherent states, and, for all $\alpha\in\mathbb C$, we have that $U^\dag\ket{\alpha00}=\ket{\beta_1\beta_2\beta_3}$, where
\be
\begin{pmatrix}\beta_1\\\beta_2\\\beta_3\end{pmatrix}=\begin{pmatrix}\alpha\sqrt z\\\alpha\sqrt{y(1-z)}\\\alpha\sqrt{(1-y)(1-z)}\end{pmatrix}.
\ee
We have $V=(\mathbb{1}\otimes H^{(b)})(H^{(a)}\otimes \mathbb{1})(\mathbb{1}\otimes R(\pi)\otimes \mathbb{1})$, with $a,b\in[0,1]$, and $R(\pi)$ a phase shift of $\pi$ acting on mode $2$. The action of $V$ on the creation operators is given by
\be
\ba
V&=\begin{pmatrix}1&0&0\\0&\sqrt b&\sqrt{1-b}\\0&\sqrt{1-b}&-\sqrt b\end{pmatrix}\begin{pmatrix}\sqrt a&\sqrt{1-a}&0\\\sqrt{1-a}&-\sqrt a&0\\0&0&1\end{pmatrix}\begin{pmatrix}1&0&0\\0&-1&0\\0&0&1\end{pmatrix}\\
&=\begin{pmatrix}\sqrt a&-\sqrt{1-a}&0\\\sqrt{b(1-a)}&\sqrt{ab}&\sqrt{1-b}\\\sqrt{(1-a)(1-b)}&\sqrt{a(1-b)}&-\sqrt b\end{pmatrix}.
\ea
\ee
For all $\alpha\in\mathbb C$, $V^\dag\ket{0\alpha0}=\ket{\gamma_1\gamma_2\gamma_3}$, where
\be
\begin{pmatrix}\gamma_1\\\gamma_2\\\gamma_3\end{pmatrix}=\begin{pmatrix}\alpha\sqrt{b(1-a)}\\\alpha\sqrt{ab}\\\alpha\sqrt{1-b}\end{pmatrix}.
\ee
Since $a=\frac{y(1-z)}{1-(1-y)(1-z)}$ and $b=1-(1-y)(1-z)$, we have $b(1-a)=z$, $ab=y(1-z)$, and $1-b=(1-y)(1-z)$, so $(\beta_1,\beta_2,\beta_3)=(\gamma_1,\gamma_2,\gamma_3)$.

\noindent Then,
\be
\ba
\Tr\,[(\tau\otimes\ket0\bra0)U^\dag(\mathbb{1}\otimes\ket{00}\bra{00})U]&=\frac1\pi\int_{\mathbb C}{d^2\alpha\Tr\,[(\tau\otimes\ket0\bra0)U^\dag\ket{\alpha00}\bra{\alpha00}U]}\\
&=\frac1\pi\int_{\mathbb C}{d^2\alpha\Tr\,[(\tau\otimes\ket0\bra0)V^\dag\ket{0\alpha0}\bra{0\alpha0}V]}\\
&=\Tr\,[(\tau\otimes\ket0\bra0)V^\dag(\ket{0}\bra{0}\otimes \mathbb{1}\otimes\ket{0}\bra{0})V],
\ea
\ee
where we used the completeness relation of coherent states $\mathbb{1}=\frac1\pi\int_{\mathbb C}{\ket\alpha\bra\alpha d^2\alpha}$.

\end{mdframed}
\end{proof}

\subsection{Soundness}

We now derive the soundness of the protocol. Namely, we obtain the maximal winning probabilities when Bob is dishonest and Alice is honest, and vice versa.

\begin{lem}\label{lem:PdB} 
Bob's optimal cheating probability is given by
\be
P_d^{(B)}=1-x.
\label{pdb1}
\ee
\end{lem}

\begin{proof}
\begin{mdframed}[linewidth=1.5,topline=false,rightline=false,bottomline=false]

Dishonest Bob should always declare the outcome $c=1$ in order to maximize his winning probability. The outcome of the coin flip is then confirmed if Alice obtains the outcome $0$ upon verification. Bob thus needs to maximize the probability of the outcome $0$, applying a general quantum operation to his half of the state. However, the probability that the detector clicks is independent of Bob's action. It is given by $x$, so that Bob's winning probability is upper bounded by $(1-x)$. This upper bound is reached if Bob discards his half of the state and broadcasts $c=1$. Bob's optimal cheating probability thus is $P_d^{(B)}=1-x$.

\end{mdframed}
\end{proof}

\noindent Alice wins when Bob declares $c=0$ and the outcome of his quantum measurement is $(1,0)$. The most general strategy of dishonest Alice is to send a (mixed) state $\sigma$, while Bob performs the rest of the protocol honestly. Assuming honest Bob has number-resolving detectors, we obtain the following result:

\begin{lem}\label{lem:PdA1} 
Alice's optimal cheating probability when Bob has number resolving detectors is given by
\be
P_d^{(A)}=1-(1-y)(1-z).
\ee
\end{lem}

\begin{figure}[h!]
	\begin{center}
		\includegraphics[width=0.6\columnwidth]{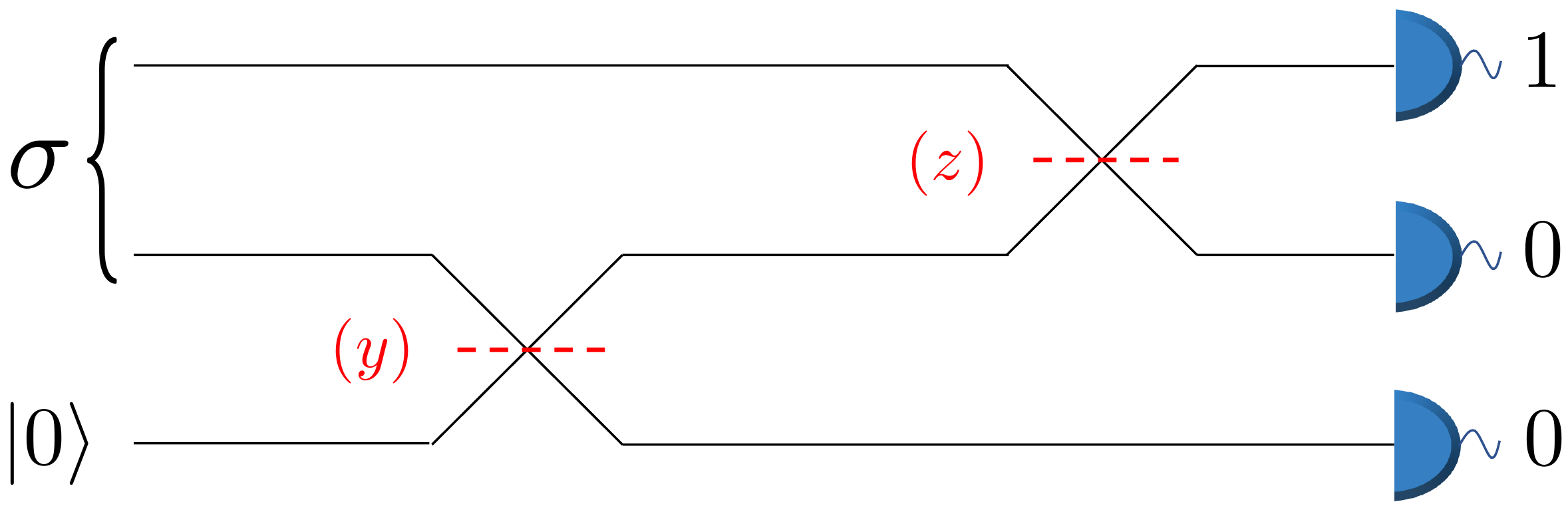}
		\caption{\textbf{Dishonest Alice.} Alice aims to maximize the outcome $(1,0,0)$: an outcome $0$ on the third mode means that Bob declared Alice the winner, while an outcome $(1,0)$ for modes $1$ and $2$ means that Alice passed Bob's verification. The reflectances of the beam splitter are given by $y=1-\frac1{2(1-x)}$ and $z=2x$.}
		\label{fig:Bhonest}
	\end{center}
\end{figure}

\begin{proof}
\begin{mdframed}[linewidth=1.5,topline=false,rightline=false,bottomline=false]

When using number-resolving single-photon detectors, any projection onto the $n>1$ photon subspace leads to Alice getting caught cheating. Alice must therefore maximize the overlap with the projective measurement $\ket{100}\bra{100}$ only (Fig.~\ref{fig:Bhonest}).

Let $\sigma$ be the state sent by Alice. Let $U=(H^{(z)}\otimes \mathbb{1})(\mathbb{1}\otimes H^{(y)})$, with $z=\frac{x}{1-(1-x)(1-y)}$.
Alice needs to maximize the probability of the overall outcome $(1,0,0)$, which is given by
\begin{equation}
    P_d^{(A)}=\Tr\,[U(\sigma\otimes\ket0\bra0)U^\dag\ket{100}\bra{100}],
\end{equation}
since Bob uses number-resolving detectors. By convexity of the probabilities, we may assume without loss of generality that Alice sends a pure state $\sigma=\ket\psi\bra\psi$, which allows us to write:
\be
\ba
P_d^{(A)}&=\Tr\,[U(\ket\psi\bra\psi\otimes\ket0\bra0)U^\dag\ket{100}\bra{100}]\\
&=\Tr\,[(\ket\psi\bra\psi\otimes\ket0\bra0)U^\dag\ket{100}\bra{100}U]\\
&=\Tr\,[\bra{\psi}\otimes\bra{0}U^\dag\ket{100}\bra{100}U\ket{\psi}\otimes\ket{0}].
\ea\label{pwin}
\ee
We have:
\be
\ba
U^\dag\ket{100}&=(\mathbb{1}\otimes H^{(y)})(H^{(z)}\otimes \mathbb{1})\ket{100}\\
&=(\mathbb{1}\otimes H^{(y)})(\sqrt{z}\ket{100}+\sqrt{1-z}\ket{010})\\
&=\sqrt{z}\ket{100}+\sqrt{y(1-z)}\ket{010}+\sqrt{(1-y)(1-z)}\ket{001},
\ea
\ee
and therefore:
\be
\ba
U^\dag\ket{100}\bra{100}U=&z\ket{100}\bra{100}+y(1-z)\ket{010}\bra{010}+(1-y)(1-z)\ket{001}\bra{001}\\
&+\sqrt{yz(1-z)}\left(\ket{100}\bra{010}+\ket{010}\bra{100}\right)\\
&+\sqrt{z(1-y)(1-z)}\left(\ket{100}\bra{001}+\ket{001}\bra{100}\right)\\
&+(1-z)\sqrt{y(1-y)}\left(\ket{010}\bra{001}+\ket{001}\bra{010}\right).
\ea
\ee
Substituting back into Eq. (\ref{pwin}) then reduces to:

\be
\ba
P_d^{(A)}&=\bra{\psi}\left(z\ket{10}\bra{10}+y(1-z)\ket{01}\bra{01}+\sqrt{yz(1-z)}(\ket{10}\bra{01}+\ket{01}\bra{10})\right)\ket{\psi}\\
&=\bra{\psi}\left(\sqrt{z}\ket{10}+\sqrt{y(1-z)}\ket{01}\right)\left(\sqrt{z}\bra{10}+\sqrt{y(1-z)}\bra{01}\right)\ket{\psi}\\
&=\left|\bra{\psi}\left(\sqrt{z}\ket{10}+\sqrt{y(1-z)}\ket{01}\right)\right|^2.
\ea
\ee
Using Cauchy-Schwarz inequality then allows us to upper bound $P_d^{(A)}$ as:
\begin{equation}
\begin{aligned}
    P_d^{(A)} &\leq \|\psi\|^2\left\|\left(\sqrt{z}\ket{10}+\sqrt{y(1-z)}\ket{01}\right)\right\|^2 \leq(1-(1-y)(1-z))\|\psi\|^2,
\end{aligned}
\label{cauchy}
\end{equation}
which is maximized for $\|\psi\|=1$. Hence we finally get:

\begin{equation}
P_d^{(A)} \leq1-(1-y)(1-z).
\end{equation}

In order to find Alice's optimal cheating strategy (i.e., the optimal pure state $\ket{\phi}$ that she must send to achieve this bound), we remark that the unnormalized state $\sqrt{z}\ket{10}+\sqrt{y(1-z)}\ket{01}$ maximizes the expression in Eq. (\ref{cauchy}). Normalizing this state then provides Alice's optimal strategy, which is to prepare the state
\begin{equation}
\ket{\phi}:=\sqrt{\frac z{1-(1-y)(1-z)}}\ket{10}+\sqrt{\frac{y(1-z)}{1-(1-y)(1-z)}}\ket{01}.
\label{eq:phi}
\end{equation}
Hence,
\begin{equation}
P_d^{(A)}=1-(1-y)(1-z).
\end{equation}
In the case of a fair protocol, $y=1-\frac1{2(1-x)}$ and $z=2x$, so
\begin{equation}
P_d^{(A)}=\frac1{2(1-x)},
\end{equation}
and Alice's optimal strategy is to prepare the state
\begin{equation}
\ket{\phi_x}:=2\sqrt{x(1-x)}\ket{10}+(1-2x)\ket{01}.
\end{equation}

\end{mdframed}
\end{proof}

\noindent Remarkably, the protocol is still secure even when Bob only uses single photon threshold detectors, which is essential to the practicality of the protocol. Moreover, Alice's optimal cheating probability remains the same:

\begin{lem}\label{lem:PdA2} 
Alice's optimal cheating probability when Bob has threshold detectors is given by
\be
P_d^{(A)}=1-(1-y)(1-z).
\ee
\end{lem}

\begin{figure}[h!]
	\begin{center}
		\includegraphics[width=0.5\columnwidth]{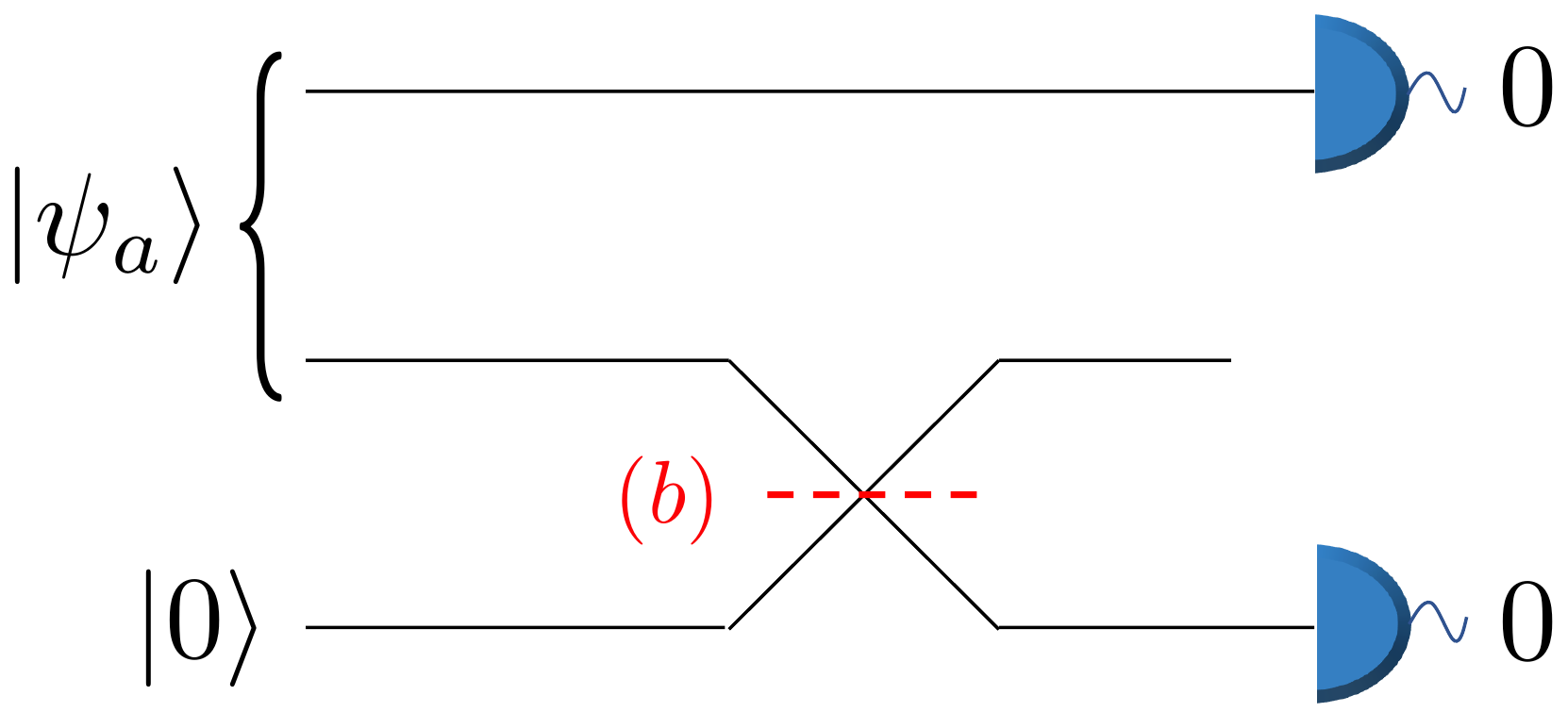}
		\caption{\textbf{Equivalent picture for dishonest Alice.} In the original dishonest setup of Fig.~\ref{fig:Bhonest}, Alice aims to maximize the outcome $(1,0,0)$. This is equivalent to Alice maximizing outcome $0$ on spatial modes $1$ and $3$, independently of what is detected on mode $2$. The outcomes indicated correspond to Alice winning. The reflectance is $b=1-(1-y)(1-z)$.}
		\label{fig:equivproof}
	\end{center}
\end{figure}

\begin{proof}
\begin{mdframed}[linewidth=1.5,topline=false,rightline=false,bottomline=false]

Unlike the previous case, incorrect outcomes with higher photon number could still pass the test: for $n\ge1$, the threshold detectors cannot discriminate between a $\ket{100}$ and $\ket{n00}$ projection. We show in the following that this doesn't help a dishonest Alice, and that the strategy described previously for the case of number resolving detectors is still optimal in the case of threshold detectors.

With the same notations as in the previous proof, Alice needs to maximize the probability of the overall outcome $(1,0,0)$, hence the overlap with the projector $\sum_{n=1}^{\infty}\ket{n00}\bra{n00}=\left(\mathbb{1}-\ket{0}\bra{0}\right)\otimes\ket{00}\bra{00}$. This allows us to write:
\begin{equation}
    P_d^{(A)}=\Tr\,[U(\ket\psi\bra\psi\otimes\ket0\bra0)U^\dag((\mathbb{1}-\ket0\bra0)\otimes\ket{00}\bra{00})],
    \label{eq:PwinAT}
\end{equation}
since Bob uses threshold detectors, where $U=(H^{(z)}\otimes \mathbb{1})(\mathbb{1}\otimes H^{(y)})$, with $z=\frac{x}{1-(1-x)(1-y)}$.

Linear optical evolution conserves photon number. Hence if Alice sends the vacuum state, the detectors will never click. Removing the two-mode vacuum component of the state prepared by Alice and renormalizing therefore always increases her winning probability. Since we are looking for the maximum winning probability, we can assume without loss of generality that $\braket{\psi|00}=0$, i.e.,
\be
\Tr\,[U(\ket\psi\bra\psi\otimes\ket0\bra0)U^\dag\ket{000}\bra{000}]=|\braket{\psi|00}|^2,
\ee
So maximizing the winning probability in Eq.~(\ref{eq:PwinAT}) is equivalent to maximizing
\be
\tilde P_d^{(A)}=\Tr\,[U(\ket\psi\bra\psi\otimes\ket0\bra0)U^\dag(\mathbb{1}\otimes\ket{00}\bra{00})],
\ee
given the constraint $\braket{\psi|00}=0$. We have
\be
\ba
\tilde P_d^{(A)}&=\Tr\,[U(\ket\psi\bra\psi\otimes\ket0\bra0)U^\dag(\mathbb{1}\otimes\ket{00}\bra{00})]\\
&=\Tr\,[(\ket\psi\bra\psi\otimes\ket0\bra0)U^\dag(\mathbb{1}\otimes\ket{00}\bra{00})U].
\label{eq:completeness}
\ea
\ee
With Lemma~\ref{lem:reductionUV} and Eq.~(\ref{eq:completeness}), we may thus write:
\be
\tilde P_d^{(A)}=\Tr\,[(\ket\psi\bra\psi\otimes\ket0\bra0)V^\dag(\ket{0}\bra{0}\otimes \mathbb{1}\otimes\ket{0}\bra{0})V],
\label{eq:PwinAtilde}
\ee
where $V=(\mathbb{1}\otimes H^{(b)})(H^{(a)}\otimes \mathbb{1})(\mathbb{1}\otimes R(\pi)\otimes \mathbb{1})$, with $a=\frac{y(1-z)}{1-(1-y)(1-z)}$ and $b=1-(1-y)(1-z)$. Let us now define:
\be
\ket{\psi_a}:=H^{(a)}(\mathbb{1}\otimes R(\pi))\ket\psi.
\label{eq:psix}
\ee
The constraints $\braket{\psi|00}=0$ and $\braket{\psi_a|00}=0$ are equivalent, because the above transformation leaves the total number of photons invariant. With Eq.~(\ref{eq:PwinAtilde}) we obtain
\be
\tilde P_d^{(A)}=\Tr\,[(\ket{\psi_a}\bra{\psi_a}\otimes\ket0\bra0)(\mathbb{1}\otimes H^{(b)})(\ket{0}\bra{0}\otimes \mathbb{1}\otimes\ket{0}\bra{0})(\mathbb{1}\otimes H^{(b)})],
\label{eq:PwinAtilde2}
\ee
with the constraint $\braket{\psi_a|00}=0$.
Maximizing this expression thus corresponds to maximizing the probability of the outcome $(0,0)$ when measuring modes $1$ and $3$ of the state obtain by mixing the second half of $\ket{\psi_a}$ with the vacuum on a beam splitter of reflectance $b=1-(1-y)(1-z)$ (Fig.~\ref{fig:equivproof}).

We now show that an optimal strategy for Alice is to ensure that $\ket{\psi_a}=\ket{01}$. Let us write
\be
\ket{\psi_a}=\sum_{p+q>0}{\psi_{pq}\ket{pq}},
\ee
where we take into account the constraint $\braket{\psi_x|00}=0$. Then, with Eq.~(\ref{eq:PwinAtilde2}) we obtain
\begin{align}
\nonumber\tilde P_d^{(A)}&=\sum_{p+q>0,p'+q'>0}{\psi_{pq}\psi_{p'q'}^*\Tr\,[\ket{pq0}\bra{p'q'0}(\ket{0}\bra{0}\otimes H^{(b)}(\mathbb{1}\otimes\ket{0}\bra{0})H^{(b)})]}\\ \displaybreak
\nonumber&=\sum_{q>0,q'>0}{\psi_{0q}\psi_{0q'}^*\Tr\,[\ket{q0}\bra{q'0} H^{(b)}(\mathbb{1}\otimes\ket{0}\bra{0})H^{(b)}]}\\
&=\sum_{n\ge0,q>0,q'>0}{\psi_{0q}\psi_{0q'}^*\Tr\,[\ket{q0}\bra{q'0} H^{(b)}\ket{n0}\bra{n0}H^{(b)}]}\\
\nonumber&=\sum_{n>0}{|\psi_{0n}|^2|\braket{n0|H^{(b)}|n0}|^2}\\
\nonumber&=\sum_{n>0}{|\psi_{0n}|^2b^n},
\end{align}
where we used in the fourth line the fact that $H^{(b)}$ doesn't change the number of photons. Since $b\in[0,1]$, this shows that
\be
\ba
\tilde P_d^{(A)}&\leq b\sum_{n>0}{|\psi_{0n}|^2}\\
&=b,
\ea
\ee
since $\ket{\psi_a}$ is normalized, and this bound is reached for $|\psi_{01}|^2=1$, i.e., $\ket{\psi_a}=\ket{01}$. With Eq.~(\ref{eq:psix}), this implies that an optimal strategy for Alice is to prepare the state
\begin{equation}
\begin{aligned}
\ket\psi&=(\mathbb{1}\otimes R(\pi))H^{(a)}\ket{01}\\
&=\sqrt{1-a}\ket{10}+\sqrt{a}\ket{01}\\
&=\sqrt{\frac z{1-(1-y)(1-z)}}\ket{10}+\sqrt{\frac{y(1-z)}{1-(1-y)(1-z)}}\ket{01}\\
&=\ket{\phi},
\end{aligned}
\end{equation}
where $\ket\phi$ is the state that dishonest Alice needs to send to maximize her winning probability when Bob uses number-resolving detectors (Eq. (\ref{eq:phi})). Her winning probability is then
\be
P_d^{(A)}=1-(1-y)(1-z). 
\label{pda1}
\ee
We therefore recover the same result as for number-resolving detectors. Once again, if the protocol is fair then $y=1-\frac1{2(1-x)}$ and $z=2x$, so
\be
P_d^{(A)}=\frac1{2(1-x)},
\ee
and an optimal strategy for Alice is to prepare the state
\be
\ket{\phi_x}:=2\sqrt{x(1-x)}\ket{10}+(1-2x)\ket{01}.
\ee

\end{mdframed}
\end{proof}

\noindent Alice's cheating probability equals $\frac{1}{2(1-x)}$ for $y=1-\frac{1}{2(1-x)}$ and $z=2x$. In particular, for all values of $x$, we retrieve the property shared by the protocols of~\cite{SR:PRL02}: $P_d^{(A)}P_d^{(B)}=\frac12$. 
Setting $x=1-1/\sqrt2$, we obtain a version of the protocol which is balanced, i.e., both players have the same cheating probability $1/\sqrt2$. The protocol bias is then $\epsilon=1/\sqrt2-1/2\approx0.207$.

\subsection{Strong coin flipping protocol}

Following~\cite{IEEE:CK09}, we show that our family of quantum weak coin flipping protocols allows us to construct a quantum strong coin flipping protocol:

\begin{lem}\label{lem:SCF}
There exists a quantum strong coin flipping protocol achieving bias $\epsilon\approx0.31$ which uses an unbalanced linear optical weak coin flipping protocol as a subroutine.
\end{lem}

\begin{proof}
\begin{mdframed}[linewidth=1.5,topline=false,rightline=false,bottomline=false]

An unbalanced quantum weak coin flipping protocol can be turned into a quantum strong coin flipping protocol using an additional classical protocol, as described in~\cite{IEEE:CK09}. In particular, let us consider a weak coin flipping protocol such that:
\be
\ba
P_h^{(A)}&=p\\
P_h^{(B)}&=1-p\\
P_d^{(A)}&=p+\epsilon\\
P_d^{(B)}&=1-p+\epsilon,
\ea
\label{pepsSCF}
\ee
for $p\in[0,1]$ and $\epsilon>0$. Then, the corresponding strong coin flipping protocol has bias~\cite{IEEE:CK09}
\be
\max{\left(\frac12-\frac12(p-\epsilon),\frac1{2-(p+\epsilon)}-\frac12\right)}.
\ee
For our weak coin flipping protocol, we have:
\be
\ba
P_h^{(A)}&=1-(1-x)(1-y)\\
P_h^{(B)}&=(1-x)(1-y)\\
P_d^{(A)}&=1-(1-y)(1-z)\\
P_d^{(B)}&=1-x,
\ea
\ee
with the constraint $z=\frac x{1-(1-x)(1-y)}$ (so that the protocol does not abort in the honest case, Eq.~(\ref{eq:z})).
Enforcing the conditions in Eq.~(\ref{pepsSCF}), and optimizing over the corresponding strong coin flipping bias implies
\be
\ba
x&=\frac{y^2}{(1-y)(1-2y)}\\
z&=\frac y{(1-y)^2}\\
1-\frac x2&=\frac1{2-y-z+yz},
\ea
\ee
which in turn give the values
\be
\ba
x&\approx0.38\\
y&\approx0.31\\
z&\approx0.66,
\ea
\ee
by enforcing $x,y,z\in[0,1]$, and a bias of $\approx0.31$, which is a lower bias than the best implemented strong coin flipping protocol so far \cite{PJL:NC14}.

\end{mdframed}
\end{proof}

\section{Experimental imperfections}

\subsection{Noisy protocol}

We investigate how imperfect state generation, non-ideal beam splitters and single-photon detector dark counts affect the correctness and security of the protocol. While we fixed the parameter values to $y=1-\frac1{2(1-x)}$ and $z=2x$ in the ideal setting, we now allow the three parameters $x$, $y$, $z$ to vary freely.

The vacuum/single-photon encoding is very robust to noise, in comparison to polarization or phase encoding for instance: the only property that must be preserved through propagation is photon number. This implies that photon indistinguishability and purity are not required in any degree of freedom other than photon number. In this case, Alice may simply produce a heralded single photon via spontaneous parametric down-conversion (SPDC) \cite{C:CP18}, which generates a photon pair: one may be used for the flip, while the other may herald the presence of the first one. Given the photon-pair emission probability $p$, accidentally emitting two pairs at the same time using SPDC occurs with probability $p^2$. Since $p$ may be arbitrarily tuned by changing the pump power, $p^2$---and therefore the probability of two photons being accidentally generated by Alice at once---may then be decreased to negligible values.

Note that, in the case where Alice's single photon source is probabilistic but heralded (as in SPDC), she may always inform Bob of a successful state generation prior to his announcement of $c$ without compromising security. In what follows, we may therefore assume that both parties have agreed on the presence of an initial state, and hence know when the protocol occurs.

Noise will therefore stem from the non-ideal reflectances of the beam splitters, and the non-zero detector dark count probability $p_{dc}$. For each party, these may affect the protocol correctness in two ways: an undesired bias of the flip, and an added abort probability during the verification process.

Deviations on the beam splitter reflectances $x$, $y$, and $z$ will first change the honest winning probabilities: these may be re-calculated by replacing the ideal reflectance $r\in\{x,y\}$ with an imperfect $r'$. As regards to honest aborts, a beam splitter with reflectance $z'$ instead of $z$ may be applied on the resulting state when $c=0$. Noisy detectors may cause an unwanted abort corresponding to a click because of dark counts. However, with superconducting nanowire single-photon detectors, this probability is typically very low, of the order of $p_{dc}<10^{-8}$ \cite{H:NP09}. 

We can therefore conclude that any source of noise may be incorporated in the security analysis by simply replacing parameters $x$, $y$, and $z$ with $x'$, $y'$, and $z'$. Furthermore, this source of error will most likely be negligible with current technology. We therefore solely focus on the more consequential effects of losses.

\subsection{Losses: completeness}

\begin{figure}[h]
\begin{center}
\includegraphics[width=1\columnwidth]{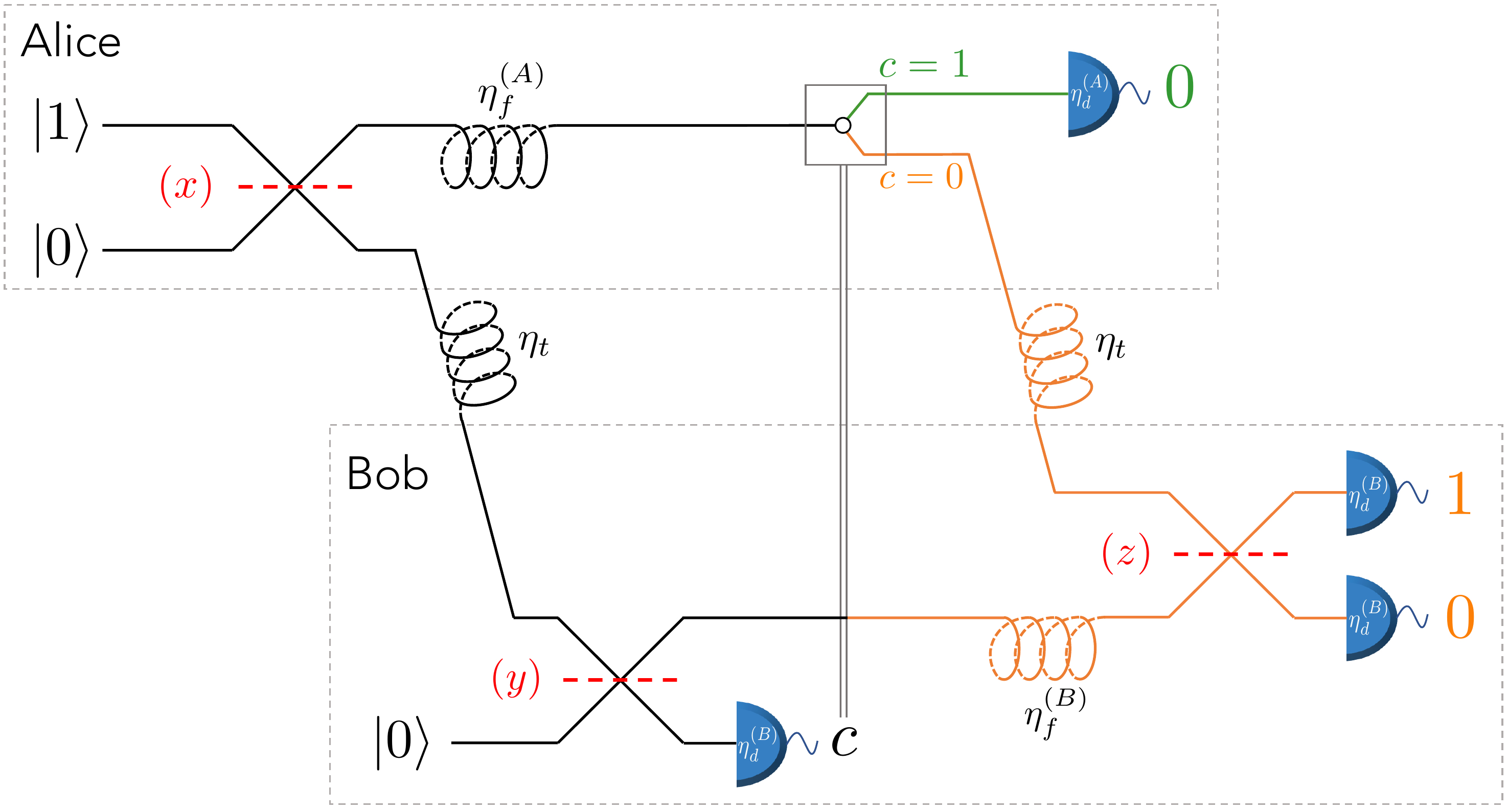}
\caption{\textbf{Representation of the honest protocol with losses.} The dashed boxes indicate Alice and Bob's laboratories, respectively. Dashed red lines represent beam splitters, with the reflectance indicated in red. The efficiencies of the detectors, are indicated in white. Curly lines represent fiber used for quantum communication from Alice to Bob, or delay lines within Alice's or Bob's laboratory. $\ket{0}$ and $\ket{1}$ are the vacuum and single photon Fock states, respectively. Bob broadcasts the classical outcome $c$, which controls an optical switch on Alice's side. The protocol when Bob declares $c=0/1$ is represented in orange/green. The final outcomes are the expected outcomes when both parties are honest.}
\label{fig:protocolapp}
\end{center}
\end{figure}

Losses can be due to the channel transmission and to non-unit delay line transmission and detection efficiencies. We label $\eta_t$ the transmission efficiency of the quantum channel from Alice to Bob. We also define as $\eta_f^{(i)}$ the transmission of party $i$'s fiber delay, while $\eta_d^{(i)}$ denotes the detection efficiency of party $i$'s single-photon detectors. Here, we assume the efficiencies of Bob's detectors to be the same, and that each party introduces a fiber delay whenever they are waiting for the other party's communication. The delay time therefore depends on the distance between the two parties. 
We give a representation of the honest protocol with losses, in Fig.~\ref{fig:protocolapp}.

We recall a useful simple property, which we will use extensively in the following:

\begin{lem} Equal losses on various modes can be commuted through passive linear optical elements acting on these modes.
\label{lem:commut}
\end{lem}

\noindent This result was proven, e.g., in~\cite{berry2010linear}, and we give hereafter a quick proof.

\begin{proof}
\begin{mdframed}[linewidth=1.5,topline=false,rightline=false,bottomline=false]

One way to prove this statement is to use the fact that any interferometer may be decomposed as beam splitters and phase shifters~\cite{reck1994experimental}. Then, losses trivially commute with phase shifters, and are easily shown to commute with beam splitters. Indeed, consider a beam splitter of reflectance $t$ acting on modes $1$ and $2$. Its action on the creation operators of the modes is given by
\be
\hat a_1^\dag,\hat a_2^\dag\rightarrow\sqrt t\hat a_1^\dag+\sqrt{1-t}\hat a_2^\dag,\sqrt{1-t}\hat a_1^\dag-\sqrt t\hat a_2^\dag,
\ee
while equal losses $\eta$ on both modes act as
\be
\hat a_1^\dag,\hat a_2^\dag\rightarrow\sqrt\eta \hat a_1^\dag,\sqrt\eta \hat a_2^\dag.
\ee
Hence, the action of the beam splitter followed by losses is given by
\be
\hat a_1^\dag,\hat a_2^\dag\rightarrow\sqrt\eta(\sqrt t\hat a_1^\dag+\sqrt{1-t}\hat a_2^\dag),\sqrt\eta(\sqrt{1-t}\hat a_1^\dag-\sqrt t\hat a_2^\dag),
\ee
while losses followed by the beam splitter act as
\be
\hat a_1^\dag,\hat a_2^\dag\rightarrow\sqrt t(\sqrt\eta \hat a_1^\dag)+\sqrt{1-t}(\sqrt\eta \hat a_2^\dag),\sqrt{1-t}(\sqrt\eta \hat a_1^\dag)-\sqrt t(\sqrt\eta \hat a_2^\dag),
\ee
which is equal to the previous evolution.

\end{mdframed}
\end{proof}

\noindent In the presence of losses, the protocol may also abort when both parties are honest, when the photon is lost. We obtain the expressions for the honest winning probabilities $P_h^{(A)}$ and $P_h^{(B)}$, and hence the probability $P_{ab}$ of abort, in the presence of losses:

\begin{lem}\label{lem:honestloss}
\begin{equation}
    \begin{aligned}
    P_h^{(A)} &=\eta_t\eta_d^{(B)}\left(\sqrt{xz\eta_f^{(A)}}+\sqrt{(1-x)y(1-z)\eta_f^{(B)}}\right)^2\\
    P_h^{(B)} &=\eta_t\eta_d^{(B)}(1-x)(1-y)\\
    P_{ab}& = 1-P_h^{(A)}-P_h^{(B)}.
    \end{aligned}
\label{eq:loss_correct}
\end{equation}
\end{lem}

\begin{proof}
\begin{mdframed}[linewidth=1.5,topline=false,rightline=false,bottomline=false]

The honest winning probability for Bob is directly given by his chance of detecting the photon (the photon gets to his detector and doesn't get lost):
\be
P_h^{(B)}=\eta_t\eta_d^{(B)}(1-x)(1-y).
\label{eq:phb}
\ee
On the other hand, Alice wins if the photon, starting from her first input mode, is detected by Bob in the last step. 

The evolution of the creation operator of the first mode during the lossy honest protocol is given by:
\begin{align}
\nonumber\hat a_1^\dag&\rightarrow\sqrt x\hat a_1^\dag+\sqrt{1-x}\hat a_2^\dag\\
\nonumber&\rightarrow\sqrt{x\eta_f^{(A)}}\hat a_1^\dag+\sqrt{(1-x)\eta_t}\hat a_2^\dag\\
\nonumber&\rightarrow\sqrt{x\eta_f^{(A)}}\hat a_1^\dag+\sqrt{(1-x)\eta_ty}\hat a_2^\dag+\sqrt{(1-x)(1-y)\eta_t}\hat a_3^\dag\\
\nonumber&\rightarrow\sqrt{x\eta_f^{(A)}}\hat a_1^\dag+\sqrt{(1-x)\eta_ty}\hat a_2^\dag+\sqrt{(1-x)(1-y)\eta_t\eta_d^{(B)}}\hat a_3^\dag\\
&\rightarrow\sqrt{x\eta_f^{(A)}\eta_t}\hat a_1^\dag+\sqrt{(1-x)\eta_ty\eta_f^{(B)}}\hat a_2^\dag+\sqrt{(1-x)(1-y)\eta_t\eta_d^{(B)}}\hat a_3^\dag\\
\nonumber&\rightarrow\left(\sqrt{x\eta_f^{(A)}\eta_tz}+\sqrt{(1-x)\eta_ty\eta_f^{(B)}(1-z)}\right)\hat a_1^\dag+\left(\sqrt{x\eta_f^{(A)}\eta_t(1-z)}-\sqrt{(1-x)\eta_ty\eta_f^{(B)}z}\right)\hat a_2^\dag\\
\nonumber&\qquad+\sqrt{(1-x)(1-y)\eta_t\eta_d^{(B)}}\hat a_3^\dag\\
\nonumber&\rightarrow\left(\sqrt{x\eta_f^{(A)}\eta_tz\eta_d^{(B)}}+\sqrt{(1-x)\eta_ty\eta_f^{(B)}(1-z)\eta_d^{(B)}}\right)\hat a_1^\dag+\left(\sqrt{x\eta_f^{(A)}\eta_t(1-z)\eta_d^{(B)}}-\sqrt{(1-x)\eta_ty\eta_f^{(B)}z\eta_d^{(B)}}\right)\hat a_2^\dag\\
\nonumber&\qquad+\sqrt{(1-x)(1-y)\eta_t\eta_d^{(B)}}\hat a_3^\dag.
\end{align}
In particular, the photon reaches Bob's uppermost detector with probability
\be
\ba
P_h^{(A)}&=\left(\sqrt{x\eta_f^{(A)}\eta_tz\eta_d^{(B)}}+\sqrt{(1-x)\eta_ty\eta_f^{(B)}(1-z)\eta_d^{(B)}}\right)^2\\
&=\eta_t\eta_d^{(B)}\left(\sqrt{xz\eta_f^{(A)}}+\sqrt{(1-x)y(1-z)\eta_f^{(B)}}\right)^2.
\ea
\label{eq:pha}
\ee
Finally, the protocol aborts for all other detection events:
\be
P_{ab}=1-P_h^{(A)}-P_h^{(B)}.
\ee

\end{mdframed}
\end{proof}

\noindent Note that the overall correctness does not depend on Alice's detection efficiency $\eta_d^{(A)}$, since the declaration of outcome $c$ depends solely on Bob's detector and the verification step on Alice's side involves detecting vacuum.

\subsection{Losses: soundness}

The soundness of the protocol is also affected by the presence of losses.

Dishonest Bob's best strategy is to perform the same attack as in the lossless case, because he has no control over Alice's half of the subsystem. His winning probability is then given by the following result:

\begin{lem}
Dishonest Bos's maximum winning probability is given by:
\be
P_d^{(B)}=1-x\eta_f^{(A)}\eta_d^{(A)}.
\label{dishonestBlossy}
\ee
\end{lem}

\noindent In a more general game-theoretic scenario, Bob's best strategy will in fact depend on the rewards and sanctions associated with honest aborts and `getting caught cheating' aborts. In other words, Bob has to minimize his risk-to-reward ratio. Maximizing his winning probability makes him run the risk of getting caught cheating with probability $ x\eta_f^{(A)}\eta_d^{(A)}$.

\medskip

\noindent Dishonest Alice must still generate the state which maximizes the $(1,0,0)$ outcome on Bob's detectors after his honest transformations have been applied. However, the expression for Bob's corresponding projector now changes, as there is a finite probability $(1-\eta_d^{(B)})^n$ that the $n$-photon component is projected onto the vacuum. The $0$ outcome on one spatial mode is therefore triggered by the projection $\Pi_0=\sum_{n=0}^{\infty}(1-\eta_d^{(B)})^n\ket{n}\bra{n}$. The total projector responsible for the $(1,0,0)$ outcome then reads $\Pi_{100}=\left(\mathbb{1}-\Pi_0\right)\otimes\Pi_0\otimes\Pi_0$. 

\begin{lem}
Dishonest Alice's maximum winning probability is given by:
\be
\ba
P_d^{(A)}&=\max_{l>0}\left[\left(1-(1-y\eta_f^{(B)})(1-z)\eta_d^{(B)}\right)^l-\left(1-\eta_d^{(B)}\right)^l\right]\\
&\leq 1-(1-y)(1-z).
\ea
\label{dishonestAlossy}
\ee
\end{lem}

\noindent The value of the upper bound in the second line is Alice's cheating probability in the lossless case. This shows that Alice cannot take advantage of Bob's imperfect detectors or his lossy delay line in order to increase her cheating probability. 

\begin{figure}[h!]
	\begin{center}
		\includegraphics[width=3.5in]{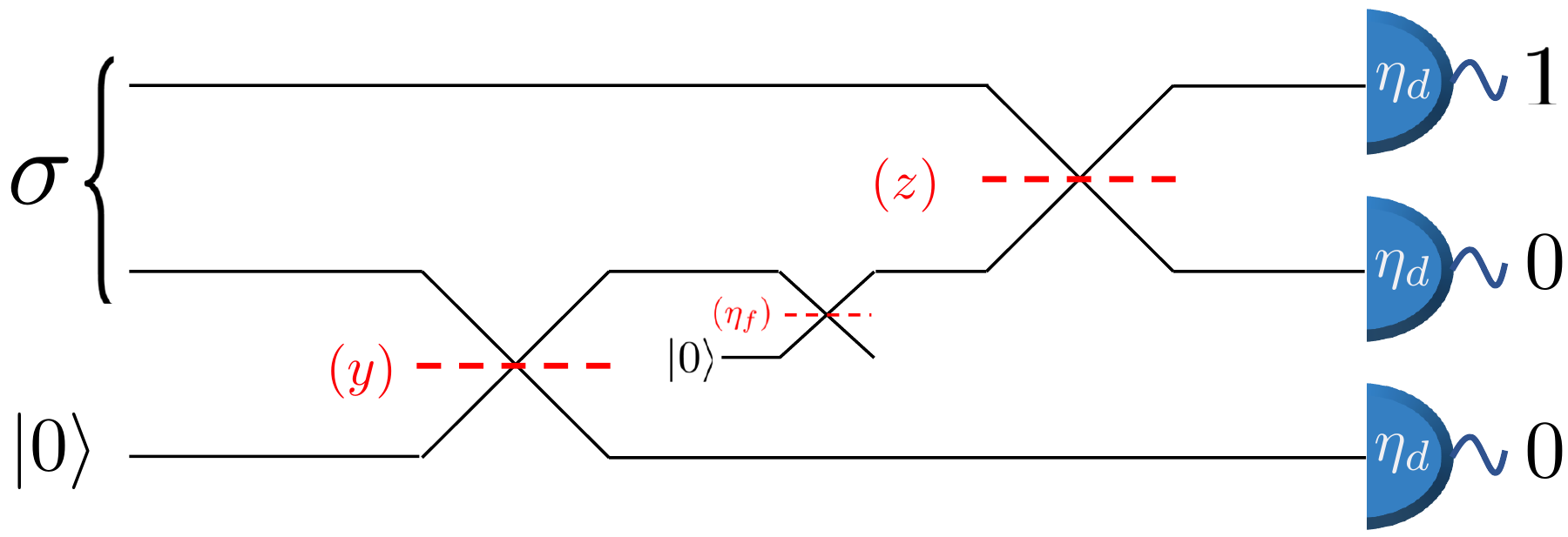}
		\caption{Alice aims to maximize the outcome $(1,0,0)$ by sending the state $\sigma$. The lossy delay line is represented by a mixing with the vacuum on a  beam splitter of transmission amplitude $\eta_f$. The quantum efficiency of the detectors is indicated in white.}
		\label{fig:lossy1}
	\end{center}
\end{figure}
\begin{figure}[h!]
	\begin{center}
		\includegraphics[width=3.6in]{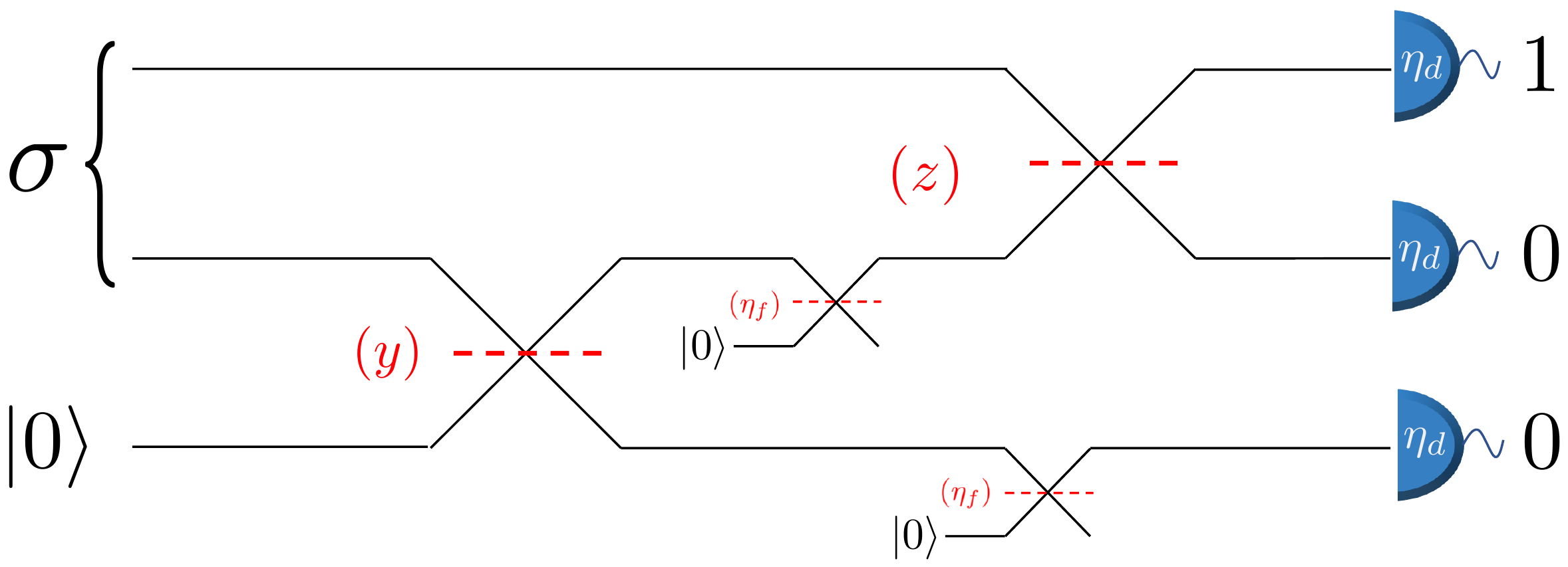}
		\caption{Adding losses on the third mode increases Alice's winning probability.}
		\label{fig:lossy2}
	\end{center}
\end{figure}
\begin{figure}[h!]
	\begin{center}
		\includegraphics[width=3.8in]{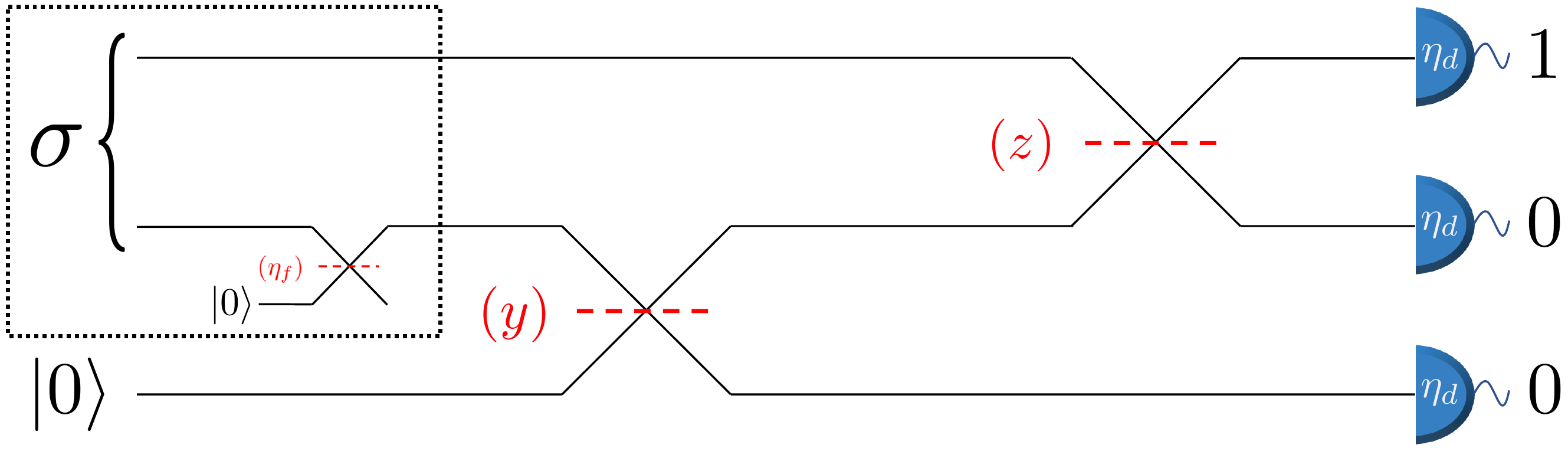}
		\caption{The losses $\eta_f$ are commuted back to Alice's state preparation. The losses on input mode $3$ can be omitted since the input state is the vacuum.}
		\label{fig:lossy3}
	\end{center}
\end{figure}
\begin{figure}[h!]
	\begin{center}
		\includegraphics[width=3.6in]{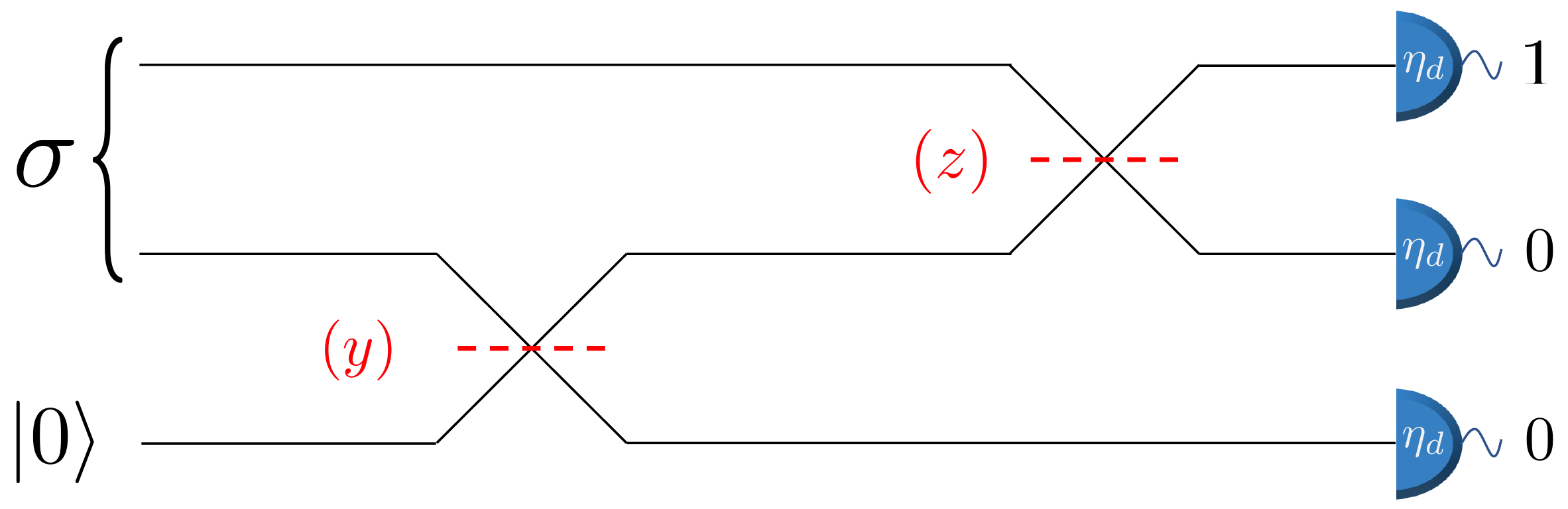}
		\caption{Alice aims to maximize the outcome $(1,0,0)$ by sending the state $\sigma$. The delay line efficiency $\eta_f$ is equal to $1$.}
		\label{fig:lossy4}
	\end{center}
\end{figure}

\begin{proof}
\begin{mdframed}[linewidth=1.5,topline=false,rightline=false,bottomline=false]

The losses $\eta$ correspond to a probability $1-\eta$ of losing a photon. These can be modelled as a mixing with the vacuum on a beam splitter of reflectance $\eta$.
We first show that we can obtain Alice's cheating probability by solving the case with perfect delay line, and replacing the parameter $y$ by $y\eta_f$, independently of the efficiency $\eta_d$ of his detectors. 

\medskip

\noindent The lossy delay line of efficiency $\eta_f$ may be modelled as a mixing with the vacuum on a beam splitter of transmission $\eta_f$.

Alice prepares a state $\sigma$, which goes through the interferometer depicted in Fig.~\ref{fig:lossy1}, and wins if the measurement outcome obtained by Bob is $(1,0,0)$.

In particular, note that the outcome $0$ must be obtained for the third mode. Hence Alice's winning probability is always lower than if the third mode was mixed with the vacuum on a beam splitter of transmission amplitude $\eta_f$ just before the detection (Fig.~\ref{fig:lossy2}), since this increases the probability of the outcome $0$ for this mode.

Let us assume that this is the case. Then, by Lemma~\ref{lem:commut}, the losses $\eta_f$ on output modes $2$ and $3$ may be commuted back through the beam splitter of reflectance $y$, acting on modes $2$ and $3$. 

Since the input state on mode $3$ is the vacuum, the losses on this mode may then be removed (Fig.~\ref{fig:lossy3}).
In that case, the probability of winning is clearly lower than when the delay line is perfect (Fig.~\ref{fig:lossy4}), because Alice is now restricted to lossy state preparation instead of ideal state preparation.

This reduction shows that Alice’s maximum winning probability when Bob is using a lossy delay line is always lower than when Bob’s delay line is perfect, independently of the efficiency $\eta_d$ of his detectors.

\medskip

\noindent Moreover, Alice's maximum cheating probability and optimal cheating strategy may be inferred from the case where Bob has a perfect delay line, as we show in what follows.

By convexity of the probabilities, Alice's best strategy is to send a pure state $\ket\psi=\sum_{k,l\ge0}{\psi_{kl}\ket{kl}}$. Let us denote by $W$ the interferometer depicted in Fig.~\ref{fig:lossy1}, including the detection losses. Let us consider the evolution of Alice's state and the vacuum on the third input mode through the interferometer $W$. The creation operator for the first mode evolves as
\be
\ba
\hat a_1^\dag&\rightarrow\sqrt z\hat a_1^\dag+\sqrt{1-z}\hat a_2^\dag\\
&\rightarrow\sqrt{z\eta_d}\hat a_1^\dag+\sqrt{(1-z)\eta_d}\hat a_2^\dag\\
&=W\hat a_1^\dag W^\dag,
\ea
\ee
while the creation operator for the second mode evolves as
\begin{align}
\nonumber\hat a_2^\dag&\rightarrow\sqrt y\hat a_2^\dag+\sqrt{1-y}\hat a_3^\dag\\
\nonumber&\rightarrow\sqrt{y\eta_f}\hat a_2^\dag+\sqrt{1-y}\hat a_3^\dag\\
&\rightarrow\sqrt{y(1-z)\eta_f}\hat a_1^\dag-\sqrt{yz\eta_f}\hat a_2^\dag+\sqrt{1-y}\hat a_3^\dag\\
\nonumber&\rightarrow\sqrt{y(1-z)\eta_f\eta_d}\hat a_1^\dag-\sqrt{yz\eta_f\eta_d}\hat a_2^\dag+\sqrt{(1-y)\eta_d}\hat a_3^\dag\\
\nonumber&=W\hat a_2^\dag W^\dag.
\end{align}
Hence, the output state (before the ideal threshold detection) is given by
\begin{align}
\nonumber W\ket{\psi0}&=W\sum_{k,l\ge0}{\psi_{kl}\ket{kl0}}\\
\nonumber&=W\left[\sum_{k,l\ge0}{\frac{\psi_{kl}}{\sqrt{k!l!}}(\hat a_1^\dag)^k(\hat a_2^\dag)^l}\right]\ket{000}\displaybreak\\
&=\left[\sum_{k,l\ge0}{\frac{\psi_{kl}}{\sqrt{k!l!}}(W\hat a_1^\dag W^\dag)^k(W\hat a_2^\dag W^\dag)^l}\right]\ket{000}\\
\nonumber&=\Bigg[\sum_{k,l\ge0}\frac{\psi_{kl}}{\sqrt{k!l!}}\left(\sqrt{z\eta_d}\hat a_1^\dag+\sqrt{(1-z)\eta_d}\hat a_2^\dag\right)^k\\
\nonumber&\quad\times\left(\sqrt{y(1-z)\eta_f\eta_d}\hat a_1^\dag-\sqrt{yz\eta_f\eta_d}\hat a_2^\dag+\sqrt{(1-y)\eta_d}\hat a_3^\dag\right)^l\Bigg]\ket{000}.
\end{align}
Now Alice's maximum cheating probability is given by
\be
P_d^{(A)}=\Tr\,[W\ket{\psi0}\bra{\psi0}W^\dag(\mathbb1-\ket0\bra0)\ket{00}\bra{00}].
\ee
Hence, the state after a successful projection $(\mathbb1-\ket0\bra0)\ket{00}\bra{00}$, which has norm $P_d^{(A)}$, reads
\be
\left[\sum_{k+l>0}{\frac{\psi_{kl}}{\sqrt{k!l!}}(z\eta_d)^{k/2}[y(1-z)\eta_f\eta_d]^{l/2}(\hat a_1^\dag)^{k+l}}\right]\ket{000}.
\ee
When Bob has a perfect delay line ($\eta_f=1$) this state reads
\be
\left[\sum_{k+l>0}{\frac{\psi_{kl}}{\sqrt{k!l!}}(z\eta_d)^{k/2}[y(1-z)\eta_d]^{l/2}(\hat a_1^\dag)^{k+l}}\right]\ket{000},
\ee
and its norm is the winning probability of Alice in that case. Hence,
\be
P_d^{(A)}[\eta_f,\eta_d,y,z]=P_d^{(A)}[1,\eta_d,y\eta_f,z],
\label{etamPd}
\ee
i.e., we can obtain Alice's cheating probability by solving the case with perfect delay line, and replacing the parameter $y$ by $y\eta_f$. In the following, we thus derive Alice's optimal strategy in that case.

\end{mdframed}

\medskip

\medskip

\medskip

\begin{figure}[h!]
	\begin{center}
		\includegraphics[width=5in]{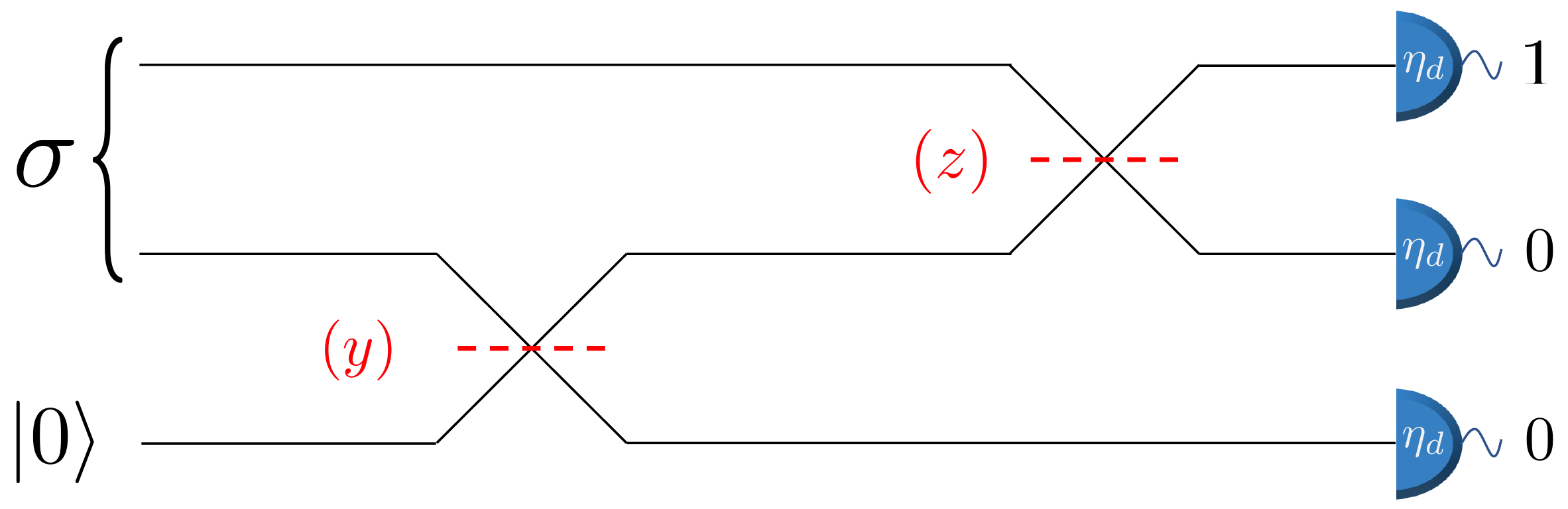}
		\caption{Alice aims to maximize the outcome $(1,0,0)$ by sending the state $\sigma$. The quantum efficiency of the detectors is indicated in white.}
		\label{fig:eta1}
	\end{center}
\end{figure}
\begin{figure}[h!]
	\begin{center}
		\includegraphics[width=5in]{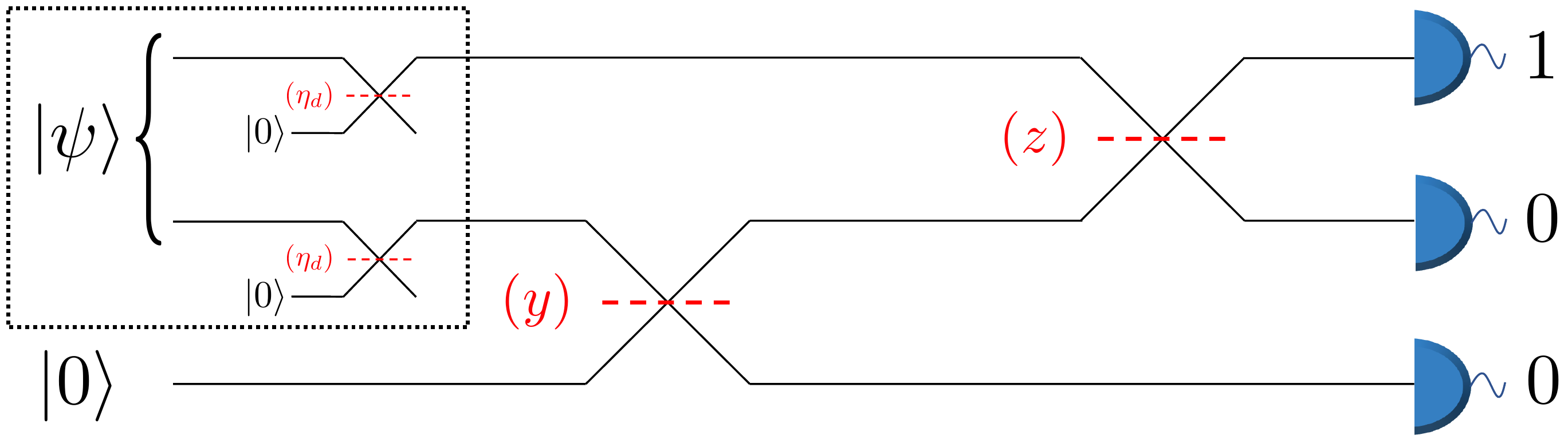}
		\caption{The quantum efficiency are modelled as losses $\eta_d$ on modes $1$, $2$, and $3$, which are then commuted through the interferometer, back to Alice's state preparation. The losses on input mode $3$ can be omitted since the input state is the vacuum.}
		\label{fig:eta2}
	\end{center}
\end{figure}
\begin{figure}[h!]
	\begin{center}
		\includegraphics[width=4.5in]{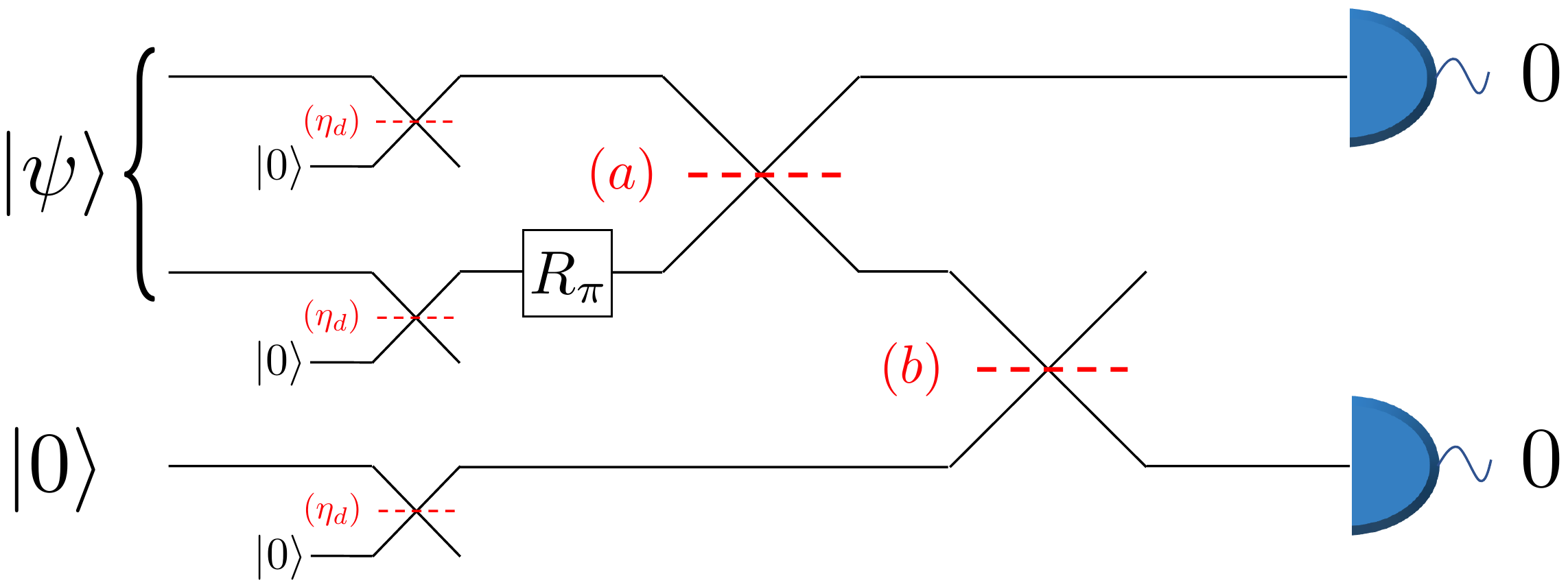}
		\caption{An equivalent picture for the first term $P_1$ of Eq.~(\ref{P1P2}). The term $P_1$ is the probability of the simultaneous outcomes $0$ for modes $1$ and $3$.}
		\label{fig:eta3}
	\end{center}
\end{figure}

\begin{mdframed}[linewidth=1.5,topline=false,rightline=false,bottomline=false]

Let $\sigma$ be the state sent by Alice, and $\eta_d$ the detector efficiency. Alice needs to maximize the probability of the overall outcome $(1,0,0)$ at the output of the interferometer depicted in Fig.~\ref{fig:eta1}, hence the overlap with the projector:

\begin{equation}
\Pi_{(1,0,0)}^{\eta_d}=\left[\mathbb{1}-\sum_m(1-\eta_d)^m\ket{m}\bra{m}\right]\otimes\left[\sum_{n,p}{(1-\eta_d)^{n+p}\ket{n}\bra{n}\otimes\ket{p}\bra{p}}\right].
\label{Pi_100eta}
\end{equation}

By convexity of the probabilities, we may assume without loss of generality that Alice sends a pure state $\sigma=\ket\psi\bra\psi$.
Moreover, the imperfect threshold detectors of quantum efficiency $\eta_d$ can be modelled by mixing the state to be measured with the vacuum on a beam splitter of transmission amplitude $\eta_d$ followed by an ideal threshold detection~\cite{ferraro2005gaussian}. In that case, this corresponds to losses $\eta_d$ on modes $1$, $2$, and $3$, followed by ideal threshold detections. By Lemma~\ref{lem:commut}, commuting the losses back through the interferometer leads to the equivalent picture depicted in Fig.~\ref{fig:eta2}, where the losses on input mode $3$ have been omitted, since the input state is the vacuum.

In that case, Alice's probability of winning is clearly lower than when the threshold detectors are perfect (Fig.~\ref{fig:Bhonest}), because she is restricted to lossy state preparation instead of ideal state preparation. Let $\ket{\tilde\psi}$ be the lossy state obtained by applying losses $\eta_d$ on both modes of Alice's prepared state $\ket\psi$. Alice's winning probability may then be written:
\begin{equation}
\ba
    P_d^{(A)}&=\Tr\,[U(\ket{\tilde\psi}\bra{\tilde\psi}\otimes\ket0\bra0)U^\dag(\mathbb{1}-\ket0\bra0)\otimes\ket{00}\bra{00}]\\
    &=\Tr\,[U(\ket{\tilde\psi}\bra{\tilde\psi}\otimes\ket0\bra0)U^\dag(\mathbb{1}\otimes\ket{00}\bra{00})]-\Tr\,[U(\ket{\tilde\psi}\bra{\tilde\psi}\otimes\ket0\bra0)U^\dag\ket{000}\bra{000}],
\ea
\end{equation}
where $U=(H^{(z)}\otimes \mathbb{1})(\mathbb{1}\otimes H^{(y)})$ is the unitary corresponding to the general interferometer of the lossless protocol. By Lemma~\ref{lem:reductionUV}, we have
\begin{equation}
\Tr\,[(\tau\otimes\ket0\bra0)U^\dag(\mathbb{1}\otimes\ket{00}\bra{00})U]=\Tr\,[(\tau\otimes\ket0\bra0)V^\dag(\ket{0}\bra{0}\otimes \mathbb{1}\otimes\ket{0}\bra{0})V],
\end{equation}
for any density matrix $\tau$, where $V=(\mathbb{1}\otimes H^{(b)})(H^{(a)}\otimes \mathbb{1})(\mathbb{1}\otimes R(\pi)\otimes \mathbb{1})$, with $a=\frac{y(1-z)}{y+z-yz}$ and $b=y+z-yz$, and $R(\pi)$ a phase shift of $\pi$ acting on mode $2$. Hence,
\begin{equation}
    P_d^{(A)}=\Tr\,[V(\ket{\tilde\psi}\bra{\tilde\psi}\otimes\ket0\bra0)V^\dag(\ket{0}\bra{0}\otimes \mathbb{1}\otimes\ket{0}\bra{0})]-\Tr\,[\ket{\tilde\psi}\bra{\tilde\psi}\ket{00}\bra{00}],
\end{equation}
where we used $U^\dag\ket{000}=\ket{000}$ for the second term. Setting $\ket{\tilde\psi_x}=(H^{(a)}\otimes \mathbb{1})(\mathbb{1}\otimes R(\pi))\ket{\tilde\psi}$ yields
\begin{equation}
    P_d^{(A)}=\underbrace{ \Tr\,[(\ket{\tilde\psi_x}\bra{\tilde\psi_x}\otimes\ket0\bra0)(\mathbb{1}\otimes H^{(b)})(\ket{0}\bra{0}\otimes \mathbb{1}\otimes\ket{0}\bra{0})(\mathbb{1}\otimes H^{(b)})]}_{\equiv P_1}-\underbrace{\Tr\,[\ket{\tilde\psi_x}\bra{\tilde\psi_x}\ket{00}\bra{00}]}_{\equiv P_2},
    \label{P1P2}
\end{equation}
where we used $\ket{00}=(\mathbb{1}\otimes R(\pi))H^{(a)}\ket{00}$ for the second term $P_2$.

Let us consider the first term $P_1$. Since $\ket{\tilde\psi}$ is the state obtained by applying losses $\eta_d$ on both modes of the state $\ket\psi$, we obtain the equivalent picture in Fig.~\ref{fig:eta3}, where we have added losses $\eta_d$ also on mode $3$, since the input state is the vacuum.

Let $\ket{\psi_x}=H^{(a)}(\mathbb{1}\otimes R(\pi))\ket\psi$. With Lemma~\ref{lem:commut}, commuting the losses $\eta_d$ to the output of the interferometer in Fig.~\ref{fig:eta3}, and combining the losses on mode $2$ and $3$ yields
\begin{equation}
    P_1=\Tr\,[\ket{\psi_x}\bra{\psi_x}\Pi_{(0)}^{\eta_d}\otimes\Pi_{(0)}^{\eta_d(1-b)}],
\end{equation}
where $\Pi_{(0)}^\eta$ is the POVM element corresponding to no click for a threshold detector of quantum efficiency $\eta$ (recall that this is the same as an ideal detector preceded by a mixing with the vacuum on a beam splitter of transmission amplitude $\eta$). The same reasoning for the second term $P_2$ gives
\begin{equation}
    P_2=\Tr\,[\ket{\psi_x}\bra{\psi_x}\Pi_{(0)}^{\eta_d}\otimes\Pi_{(0)}^{\eta_d}],
\end{equation}
and we finally obtain with Eq.~(\ref{P1P2}),
\begin{equation}
    P_d^{(A)}=\Tr\,[\ket{\psi_x}\bra{\psi_x}\Pi_{(0)}^{\eta_d}\otimes(\Pi_{(0)}^{\eta_d(1-b)}-\Pi_{(0)}^{\eta_d})].
\end{equation}
Let us write $\ket{\psi_x}=\sum_{k,l\ge0}^{+\infty}{\psi_{kl}\ket{kl}}$. With the expression of the POVM in Eq.~(\ref{Pi_100eta}) the last equation reads
\begin{align}
\nonumber P_d^{(A)}&=\sum_{k,l\ge0}{|\psi_{kl}|^2(1-\eta_d)^k[(1-\eta_d(1-b))^l-(1-\eta_d)^l]}\\
\nonumber&\leq\max_{k,l\ge0}{(1-\eta_d)^k[(1-\eta_d(1-b))^l-(1-\eta_d)^l]}\sum_{k,l\ge0}{|\psi_{kl}|^2}\\
&=\max_{k,l\ge0}{(1-\eta_d)^k[(1-\eta_d(1-b))^l-(1-\eta_d)^l]}\\
\nonumber&=\max_{l\ge1}{[(1-\eta_d(1-b))^l-(1-\eta_d)^l]}\\
\nonumber&=\max_{l\ge1}{[(1-\eta_d(1-y)(1-z))^l-(1-\eta_d)^l]},
\end{align}
where we used $b=y+z-yz$. Let $l_0\in\mathbb N^*$ such that $\max_{l\ge1}{[(1-\eta_d(1-b))^l-(1-\eta_d)^l]}=(1-\eta_d(1-b))^{l_0}-(1-\eta_d)^{l_0}$. This last expression is an upperbound for $P_d^{(A)}$, which is attained for $\psi_{kl}=\delta_{k,0}\delta_{l,l_0}$, i.e., $\ket{\psi_x}=\ket{0l_0}$.
Thus, the best strategy for Alice is to send the state
\be
\ba
\ket\psi&=(\mathbb{1}\otimes R(\pi))H^{(a)}\ket{\psi_x}\\
&=(\mathbb{1}\otimes R(\pi))H^{(a)}\ket{0l_0},
\ea
\ee
where $a=\frac{y(1-z)}{y+z-yz}$, and her winning probability is then
\be
P_d^{(A)}=(1-\eta_d(1-y)(1-z))^{l_0}-(1-\eta_d)^{l_0},
\ee
when Bob has a perfect delay line.
Recalling Eq.~(\ref{etamPd}), the best strategy for Alice when Bob has a lossy delay line of efficiency $\eta_f$ is to send the state
\be
\ba
\ket\psi&=(\mathbb{1}\otimes R(\pi))H^{(a)}\ket{\psi_x}\\
&=(\mathbb{1}\otimes R(\pi))H^{(a)}\ket{0l_1},
\ea
\ee
where $a=\frac{y(1-z)\eta_f}{y\eta_f+z-yz\eta_f}$, and $l_1\in\mathbb N^*$ maximizes $(1-\eta_d(1-y\eta_f)(1-z))^l-(1-\eta_d)^l$. Her winning probability is then
\begin{align}
\nonumber P_d^{(A)}&=\max_{l>0}\left[\left(1-(1-y\eta_f)(1-z)\eta_d\right)^l-\left(1-\eta_d\right)^l\right]\\
\nonumber&=(1-\eta_d(1-y\eta_f)(1-z))^{l_1}-(1-\eta_d)^{l_1}\\
\nonumber&=\eta_d[1-(1-y\eta_f)(1-z)]\sum_{j=0}^{l_1-1}{(1-\eta_d)^j(1-\eta_d(1-y\eta_f)(1-z))^{l_1-j-1}}\\
&\leq\eta_d[1-(1-y\eta_f)(1-z)]\sum_{j=0}^{l_1-1}{(1-\eta_d)^j}\\
\nonumber&=\eta_d[1-(1-y\eta_f)(1-z)]\frac{1-(1-\eta_d)^{l_1}}{1-(1-\eta_d)}\displaybreak\\
\nonumber&=[1-(1-y\eta_f)(1-z)][1-(1-\eta_d)^{l_1}]\\
\nonumber&\leq1-(1-y\eta_f)(1-z)\\
\nonumber&\leq 1-(1-y)(1-z),
\end{align}
and this last expression is her winning probability when there are no losses.

\end{mdframed}
\end{proof}

\noindent Let us derive the value of $l_1$ for which the maximum is achieved in Eq.~(\ref{dishonestAlossy}). For this, we define:
\be
\ba
&r = 1-\eta_d(1-y\eta_f)(1-z)\\
&s = 1-\eta_d.\\
\ea
\ee
We then consider a $\lambda_1\in\mathbb{R}^{*+}$ which maximizes $(r^\lambda-s^\lambda)$ for $\lambda\in\mathbb{R}^{*+}$. We have that:
\be
\ba
&\frac{d}{d \lambda_1}(r^{\lambda_1}-s^{\lambda_1})=0 \Leftrightarrow \lambda_1 = \frac{\log{\log{s}}-\log{\log{r}}}{\log{r}-\log{s}},
\ea
\ee
for strictly non-zero $r$ and $s$. This allows us to deduce:
\begin{equation}
   l_1 = \left\{
    \begin{array}{ll}
     &\lfloor\lambda_1\rfloor \:\:\:\:\:\:\:\: \text{if} \:\:\:\:\:\:\:\: r^{\lfloor\lambda_1\rfloor}-s^{\lfloor\lambda_1\rfloor}\ge r^{\lceil\lambda_1\rceil}-s^{\lceil\lambda_1\rceil}\\
     &\lceil\lambda_1\rceil\:\:\:\:\:\:\:\:\:\: \text{if} \:\:\:\:\:\:\:\: r^{\lceil\lambda_1\rceil}-s^{\lceil\lambda_1\rceil} \ge r^{\lfloor\lambda_1\rfloor}-s^{\lfloor\lambda_1\rfloor}.\\
    \end{array}
\right.
\end{equation}

\subsection{Quantum advantage}

We now analyze the performance of our protocol in a practical setting, by enforcing three conditions on the free parameters: the protocol must be fair, balanced, and perform strictly better than any classical protocol. The latter condition is not required in an ideal implementation, since quantum weak coin flipping always provides a security advantage over classical weak coin flipping. Allowing for abort cases, however, may enable some classical protocols to perform better than quantum ones. This is because increasing the abort probability effectively decreases Alice and Bob's cheating probabilities. We say that the protocol allows for quantum advantage when it provides a strictly lower cheating probability than any classical protocol with the same abort probability. This is obtained using the bounds from~\cite{HW:TCC11}, which yield the best classical cheating probability $P_d^{C}=1-\sqrt{P_{ab}}$ for our protocol. 

\medskip

\noindent\textbf{Condition (i):} the first condition enforces a fair protocol, i.e., $P_{h}^{(A)}=P_{h}^{(B)}$. With Eq.~(\ref{eq:loss_correct}), we aim to solve for $y$ as a function of $x$ and $z$:
\begin{equation}
    \begin{aligned}
   (i) \Leftrightarrow &\:\:\eta_t\eta_d^{(B)}\left(\sqrt{xz\eta_f^{(A)}}+\sqrt{(1-x)y(1-z)\eta_f^{(B)}}\right)^2 =  \eta_t\eta_d^{(B)}(1-x)(1-y)\\ 
        \Leftrightarrow &\:\: (1-x)\left[(1-z)\eta_f^{(B)}+1\right]y+2\sqrt{x(1-x)z(1-z)\eta_f^{(A)}\eta_f^{(B)}}\sqrt{y}+xz\eta_f^{(A)}-(1-x)=0.
    \end{aligned}
    \label{eqeq}
\end{equation}
We make the substitution $Y = \sqrt{y}$ in order to transform Eq. (\ref{eqeq}) into a second-order polynomial equation. We then take only the positive solution (since $y$ must be positive) which reads:
\begin{equation}
    Y = \frac{\sqrt{xz(1-z)\eta_f^{(A)}\eta_f^{(B)}-\left[(1-z)\eta_f^{(B)}+1\right]\left[xz\eta_f^{(A)}-(1-x)\right]}-\sqrt{xz(1-z)\eta_f^{(A)}\eta_f^{(B)}}}{\sqrt{1-x}\left[(1-z)\eta_f^{(B)}+1\right]}.
    \label{eq:bigY}
\end{equation}
We may finally write:
\begin{equation}
    (i) \Leftrightarrow \:\: y=f\left(x,z,\eta_f^{(i)},\eta_d,\eta_t\right),
\end{equation}
where
\be
f\left(x,z,\eta_f^{(i)},\eta_d,\eta_t\right)=\frac{\left(\sqrt{(1-x)\left[(1-z)\eta_f^{(B)}+1\right]-xz\eta_f^{(A)}}-\sqrt{xz(1-z)\eta_f^{(A)}\eta_f^{(B)}}\right)^2}{(1-x)\left[(1-z)\eta_f^{(B)}+1\right]^2}.
\ee
Note that $y$ should be a real number, and hence we require that the expression under the first square root of $f\left(x,z,\eta_f^{(i)},\eta_d,\eta_t\right)$ is positive, i.e.,
\begin{equation}
    z\leq \frac{(1-x)(1+\eta_f^{(B)})}{x\eta_f^{(A)}+(1-x)\eta_f^{(B)}}.
\end{equation}
Furthermore, note that, for $\eta_f^{(A)}=\eta_f^{(B)}=\eta_f$, $y$ should be an increasing function of $\eta_f$, and therefore a decreasing function of $d$ when $\eta_f=10^{-\frac{0.2}{10}2d}$. Mathematically speaking, this is to prevent $y'(d)\rightarrow \infty$ and $y(d)>1$. Physically speaking, this condition ensures that, as the probability of transmitting the photon (and of preserving it for verification) gets smaller, Bob should encourage a detection on the third mode, which evens out the honest probabilities of winning.

\medskip

\noindent\textbf{Condition (ii):} the second condition enforces a balanced protocol, i.e., $P_d^{(A)}=P_d^{(B)}$. With Eqs.~(\ref{dishonestBlossy}) and (\ref{dishonestAlossy}), this translates into the following expression for $x$:
\begin{equation}
    (ii) \Leftrightarrow  x=g\left(y,z,\eta_f^{(i)},\eta_d^{(i)}\right),
\end{equation}
where
\begin{equation}
g\left(y,z,\eta_f^{(i)},\eta_d^{(i)}\right)=\frac1{\eta_f^{(A)}\eta_d^{(A)}}\left[1-\max_{l\ge1}{[(1-\eta_d^{(B)}(1-y\eta_f^{(B)})(1-z))^l-(1-\eta_d^{(B)})^l]}\right].
\end{equation}

\medskip

\noindent\textbf{Condition (iii):} we recall the general coin flipping formalism from \cite{HW:TCC11}, in which any classical or quantum coin flipping protocol may be expressed as:

\begin{equation}
    CF\left(p_{00},p_{11},p_{*0},p_{*1},p_{0*},p_{1*}\right),
\end{equation}
where $p_{ii}$ is the probability that two honest players output value $i\in\{0,1\}$, $p_{*i}$ is the probability that Dishonest Alice forces Honest Bob to declare outcome $i$, and $p_{i*}$ is the probability that Dishonest Bob forces Honest Alice to declare outcome $i$. In this formalism, a perfect strong coin flipping protocol can then be expressed as  $CF\left(\frac{1}{2},\frac{1}{2},\frac{1}{2},\frac{1}{2},\frac{1}{2},\frac{1}{2}\right)$, while a perfect weak coin flipping may be expressed as  $CF\left(\frac{1}{2},\frac{1}{2},\frac{1}{2},1,1,\frac{1}{2}\right)$. We may now express our quantum weak coin flipping protocol in the lossless setting as:
\begin{equation}
    CF\left(\frac{1}{2},\frac{1}{2},\left[\frac{1}{2(1-x)}\right],1,1,[1-x]\right).
\end{equation}
In the lossy setting, note that the probabilities that Alice and Bob each choose to lose (i.e., $p_{*1}$ and $p_{0*}$, respectively), both remain $1$. When Dishonest Bob chooses to lose, he may always declare outcome $0$ regardless of what he detects, which yields $p_{0*}=1$. When Dishonest Alice chooses to lose, she may send a state $\ket n$ to Bob, and so:
\be
\ba
p_{*1}&=\Tr\left[H^{(y)}\ket{n0}\bra{n0}H^{(y)}I\otimes(I-\Pi_0)\right]\\
&=1-\Tr\left[H^{(y)}\ket{n0}\bra{n0}H^{(y)}(I\otimes\Pi_0)\right],
\ea
\ee
where $\Pi_0=\sum_{l\ge0}(1-\eta)^l\ket l\bra l$ and $H^{(y)}=\begin{pmatrix}
\sqrt y & \sqrt {1-y}\\\sqrt{1-y} & -\sqrt y
\end{pmatrix}$. 

\medskip

\noindent Now,
\be
\ba
H^{(y)}\ket{n0}&=H^{(y)}\frac{(\hat a_1^\dag)^n}{\sqrt{n!}}\ket{00}\\
&=\frac1{\sqrt{n!}}(\sqrt y\hat a_1^\dag+\sqrt{1-y}\hat a_2^\dag)^n\ket{00}\\
&=\frac1{\sqrt{n!}}\sum_{k=0}^n\binom nky^{\frac k2}(1-y)^{\frac{n-k}2}\hat a_1^{\dag k}\hat a_2^{\dag(n-k)}\ket{00}\\
&=\sum_{k=0}^n\sqrt{\binom nky^k(1-y)^{n-k}}\ket{k\text{ }(n-k)}.
\ea
\ee
We thus obtain, by linearity of the trace:
\be
\ba
p_{*1}&=1-\sum_{l,l'\ge0}(1-\eta)^l\sum_{k,k'=0}^n\sqrt{\binom nky^k(1-y)^{n-k}}\sqrt{\binom n{k'}y^{k'}(1-y)^{n-k'}}\Tr\left[\ket{k\text{ }(n-k)}\bra{k'\text{ }(n-k')}\ket{l'l}\bra{l'l}\right]\\
&=1-\sum_{k=0}^n(1-\eta)^{n-k}\binom nky^k(1-y)^{n-k}\\
&=1-\left[y+(1-\eta)(1-y)\right]^n,
\ea
\ee
which goes to $1$ when $n$ goes to infinity, for $y<1$. Hence, in the lossy setting, the protocol becomes a:
\begin{equation}
    CF\left(P_{h}^{(A)},P_{h}^{(B)},P_d^{(A)},1,1,P_d^{(B)}\right),
\end{equation}
where $P_d^{(A)}=\max_{l>0}\left(1-(1-y\eta_f^{(A)})(1-z)\eta_d^{(B)}\right)^l-\left(1-\eta_d^{(B)}\right)^l$ and $P_d^{(B)} = 1-x\eta_f^{(A)}\eta_d^{(A)}$.

\medskip

\noindent Using Theorem $1$ from \cite{HW:TCC11}, there exists a classical protocol that implements an information-theoretically secure coin flip with our parameters if and only if the following conditions hold:
\be
\begin{cases}
P_h^{(A)}\le P_d^{(A)}\\
P_h^{(B)}\le P_d^{(B)}\\
P_{ab}=1-P_h^{(A)}-P_h^{(B)}\ge(1-P_d^{(A)})(1-P_d^{(B)}).
\label{system_qc}
\end{cases}
\ee
Our quantum protocol therefore presents an advantage over classical protocols if at least one of these conditions \textit{cannot} be satisfied. Since we are interested in fair and balanced protocols, setting
\be
P_{h}=P_{h}^{(A)}=P_{h}^{(B)},\quad\text{and}\quad P_{d}=P_{d}^{(A)}=P_{d}^{(B)}
\ee
allows us to rewrite (\ref{system_qc}) as: 
\be
\begin{cases}
P_h\le P_d\\
P_{ab}=1-2P_h\ge(1-P_d)^2 \Leftrightarrow P_h\le\frac12[1-(1-P_d)^2].
\end{cases}
\ee
Let us finally remark that for all $x$ we have $\frac12[1-(1-x)^2]=x-\frac{x^2}2\le x$, so the first inequality above is implied by the second. The system is thus equivalent to the second inequality:
\be
P_{ab}=1-2P_h\ge(1-P_d)^2,
\ee
provided that $P_h^{(A)}=P_h^{(B)}=P_h$ and $P_d^{(A)}=P_d^{(B)}=P_d$.

\medskip

\noindent In order to get a clearer insight into the meaning of quantum advantage, we express this condition in terms of cheating probability: our protocol displays quantum advantage if and only if the lowest classical cheating probability

\be
P_d^{C}=1-\sqrt{1-2P_h}=1-\sqrt{P_{ab}}
\ee
exceeds our quantum cheating probability $P_d^{Q}$. 

\medskip

\begin{figure}[h!]
	\begin{center}
		\includegraphics[width=0.7\columnwidth]{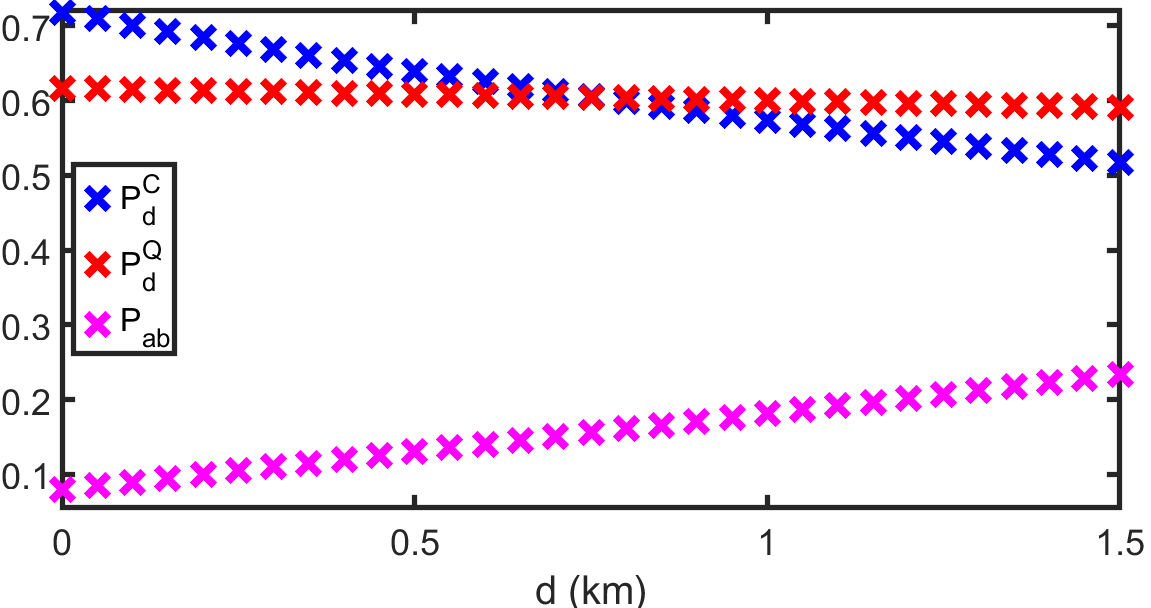}
		\caption{Practical quantum advantage for a fair and balanced protocol: numerical values for the lowest classical and quantum cheating probabilities, $P_d^{C}$ and $P_d^{Q}$, are plotted as a function of distance $d$ in blue and red, respectively. Honest abort probability $P_{ab}$ (responsible for $P_d^{Q}$ being lower than our ideal quantum cheating probability $1/\sqrt{2}$) is plotted in magenta. Our quantum protocol performs strictly better than any classical protocol when $P_d^{Q}<P_d^{C}$. We set $\eta_f=\eta_s\eta_t^2$, where $\eta_s$ is the fiber delay transmission corresponding to $500$ns of optical switching time, and $\eta_t^2=(10^{-\frac{0.2}{10}d})^2$ is the fiber delay transmission associated with travelling distance $d$ twice (once for quantum, once for classical) in single-mode fibers with attenuation $0.2$ dB/km. We have $\eta_d=0.95$ and $z=0.57$.}
		\label{fig:performance}
	\end{center}
\end{figure}

\noindent The three conditions may then be translated into the following system of equations, where we define $P_d^{Q}=P_d^{(A)}=P_d^{(B)}$:
\begin{equation}
   \left\{
    \begin{array}{ll}
     (i) \:\:\:\: &P_h^{(A)}=P_h^{(B)}\:\:\:\:\:\: \text{fairness} \\
     (ii) \:\:\:\: &P_d^{(A)}=P_d^{(B)}\:\:\:\:\: \text{balance} \\
     (iii) &P_d^{Q}< P_d^{C} \:\:\text{quantum advantage} \\
    \end{array}
\right.
\label{eq:systemz}
\end{equation}
Fig.~\ref{fig:performance} shows a choice of parameters obtained numerically for which the system in Eq.~(\ref{eq:systemz}) is satisfied, up to a distance of $d$ km.\\

\section{Discussion and open problems}

By noticing a non-trivial connection between the early protocol from \cite{SR:PRL02} and linear optical transformations, we answer the question of the implementability of quantum weak coin flipping, and show that it is achievable with current technology over a few hundred meters. As the distance increases, the issue of stability of the interferometric setup should also be taken into account. Both parties require a set of beam splitters and single photon threshold detectors. State generation on Alice's side can be performed with any heralded probabilistic single-photon source, for which photon indistinguishability and state purity do not matter. Only Alice requires an optical switch, which is commercially available. Although short-term quantum storage is needed, a spool of optical fiber with twice the length of the quantum channel suffices, and provides the required storage/retrieval efficiency.

On the fundamental level, our results also raise the question of a potentially deeper connection between the large family of protocols from \cite{M:FOCS04,M:PRA05,M:arx07}---which achieves biases as low as $1/6$---and linear optics. Recalling that the protocol from \cite{SR:PRL02}, and hence our protocol, is conjectured optimal for this family, its extension to many rounds should be necessary in order to lower the bias. The optimality of the one-round protocol is crucial, as a recent result shows that the weak coin flipping bias decreases very inefficiently with the number of rounds \cite{M:arx19}.

\clearemptydoublepage
%
%
\let\textcircled=\pgftextcircled
\chapter*{Conclusion and outlook}
\addcontentsline{toc}{chapter}{Conclusion and outlook}  
\chaptermark{Conclusion and outlook} 
\pagestyle{concl}

\initial{G}uided by three general questions about the use of quantum information in existing and upcoming technologies, this thesis has provided some answers in the context of continuous variable quantum information theory and linear quantum optics.\\

\noindent Firstly, \textit{what leads to a quantum advantage?}

\medskip

We have considered the case of non-Gaussian states as a resource for outperforming classical computing capabilities. Introducing the stellar formalism, we have characterised single-mode non-Gaussian states by the number of elementary non-Gaussian operations needed to engineer them \cite{chabaud2020stellar}. Apart from providing insights about the structure of these states, we have seen direct consequences of the use of our formalism for Gaussian convertibility of states, comparing photon addition and photon subtraction, and cat state engineering \cite{inprepaLKB}.

We have studied classical simulation regimes for a variety of continuous variable and optical quantum models \cite{inprepaMLALO,inprepaSimuNG}. Our conclusions provide the minimum requirements necessary for the development of beyond-classical quantum applications with these models. Bridging the gap between classically simulable models and models universal for quantum computing, we have shown that CVS circuits, which form a subuniversal family of optical interferometers relating to Boson Sampling with Gaussian measurements, are hard to simulate classically \cite{chabaud2017continuous}.\\

\noindent Secondly, \textit{how do we check the correct functioning of a quantum device?}

\medskip

We have developped a variety of certification and verification protocols for continuous variable quantum states with single-mode Gaussian measurements. 
As a first step, we showed how to perform efficient reliable tomography and fidelity estimation for any single-mode continuous variable quantum state, under the assumption of identical copies, and with no assumption whatsoever \cite{chabaud2019building}.
Next, we showed how to obtain tight fidelity witnesses for a large class of multimode continuous variable quantum states with analytical confidence intervals \cite{chabaud2020efficient}. 
These fidelity witnesses in turn allowed us to derive a verification protocol for multimode states, including the output states of Boson Sampling experiments, which was missing so far \cite{eisert2020quantum}, thus enabling an experimental demonstration of quantum supremacy with photonic quantum computing.\\

\noindent Thirdly, \textit{what useful advantages can we obtain from the use of quantum information?}

\medskip

We have considered various applications of quantum information and their implementation with linear optics. We have analysed the task of discriminating two unknown quantum states in an unbalanced setting, showing a connection with the concept of universal quantum-programmable measurement. For these two tasks, we have introduced an optimal implementation with linear optics and single-photon encoding \cite{chabaud2018optimal}. To obtain a more practical setup, we have discussed coherent state encoding which simplifies considerably the corresponding scheme \cite{inprepa1}.

Turning to quantum cryptography, we have proposed, using once again linear optics, the first practical implementation of quantum weak coin flipping with information-theoretic security \cite{bozzio2020quantum}, a building block for a variety of cryptographic applications. We have analysed the robustness of our proposal to experimental imperfections and losses and showed that an experimental implementation could display quantum advantage already with current technology.\\

\noindent We have given various open problems at the end of each chapter. Let us finish by considering more general perspectives.

\medskip

We believe that demonstrating verified quantum supremacy with photonic quantum computing using our fidelity witness protocol is a fascinating prospect, either for Boson Sampling or CVS circuits. Given its efficiency, the verification protocol is already within experimental reach. This would represent a milestone in the development of quantum technologies and fundamentally demonstrate the different nature of quantum and classical computations. 

Other immediate perspectives are the ongoing implementations of proof-of-concept experiments for demonstrating the stellar rank with Gaussian measurements, performing quantum-programmable measurements with coherent states and performing quantum weak coin flipping with a single photon.

A natural outlook is to extend the answers to the three general questions above to contexts other than continuous variable quantum information theory and to experimental platforms other than linear quantum optics. The first half of the thesis makes extensive use of phase space formalism. It would be interesting to see if similar results can be obtained for discrete variable quantum information theory by considering analogous discrete phase space methods.

Ultimately, the interaction of various research topics relating to quantum information theory leads to a deeper understanding of quantum advantages and enables the development of exciting quantum technologies.

\clearemptydoublepage
\backmatter
\pagestyle{myvf}
%
\bibliographystyle{alpha}
\refstepcounter{chapter}
\bibliography{thesisbiblio}
\clearemptydoublepage
\clearpage
\end{document}